\documentclass[]{raa}            
\usepackage{graphicx,times}
\usepackage{natbib}

\providecommand{\bm}[1]{\mbox{\boldmath$#1$\unboldmath}}

\usepackage{longtable}
\usepackage{subfigure}
\usepackage[fleqn]{amsmath}
\usepackage{multirow}

\begin{document}

\title{A survey of CO and its isotope lines for possible cloud-cloud collision candidates}

 \volnopage{ {\bf 20xx} Vol.\ {\bf 9} No. {\bf XX}, 000--000}
   \setcounter{page}{1}

   \author{Nan Li
      \inst{1}
   \and Jun-Jie Wang
      \inst{1}
   }

   \institute{National Astronomical Observatories, Chinese Academy of Sciences,
             Beijing 100012, China; {\it linan@bao.ac.cn}\\
\vs \no
   {\small Received [year] [month] [day]; accepted [year] [month] [day] }
}

\abstract {In the ${}^{12}\!$CO ($J$=1-0) survey for the 1331 cold
{\it IRAS\/} sources \citep{b14}, 214 sources show profiles with
multiple-peak profiles and are selected as cloud-cloud collision
candidates. In January 2005, 201 sources are detected with
${}^{12}\!$CO(1-0), ${}^{13}\!$CO(1-0), and C${}^{18}\!$O(1-0)
emission by the 13.7m telescope at Purple Mount Observatory. This is
the first CO and its isotope lines directed toward possible
cloud-cloud collision regions. According to the statistics of the
201 sources in Galactic distribution, the 201 sources show a similar
distribution to the parent sample (1331 cold {\it IRAS\/} sources).
These sources are located over a wide range of the Galactocentric
distances, and are partly associated with the star formation region.
Based on preliminary criteria which describe the spectrum properties
of the possible cloud-cloud collision region \citep{b9,b10}, the 201
sources are classified into four types by the fit of the spectral
profiles between the optically thick and thin lines toward each
source. The survey is focused on the possible cloud-cloud collision
regions, and gives some evidences to help us with selecting the
target region. Then we will carry on the mapping and
multi-wavelength study for the selected region in future.
 }
   \authorrunning{Nan Li \& Jun-Jie Wang }            
   \titlerunning{A survey of CO and its isotope lines for possible cloud-cloud collision candidates}  

   \maketitle


%

\section{Introduction}           
\label{sect:intro}
 In the star forming region, the collapse could
be triggered by self-gravitation or some external interaction
mechanism. Beside the other external shocks (e.g. supernova shocks,
and the stellar winds around a previous generation of OB stars etc.
), the cloud-cloud collision was also proposed as the effective
external influence mechanism in star formation. These processes have
been calculated with numerical simulation \citep{b7,b4,b8,b6}. When
CO and its isotope molecules are used to be a sensitive probe to
detect the cool molecular regions by astronomers, many CO survey
have been made to describe the the molecular cloud distribution
\citep{b5,b3}. For cloud-cloud collision region, CO and its isotope
line are also the good probes to describe the cloud colliding region
\citep{b12,b13}. The {\it IRAS\/} point-source database is an
effective indicator of the star forming regions which are hidden by
dust in visible wavelength. In this paper, we choose the cold {\it
IRAS\/} source according to our criteria to make the CO and its
isotope line survey. This is a relatively complete {\it
IRAS\/}-based CO survey for the the possible cloud-cloud collision
candidates. Although there are some observations were done in the in
earlier times \citep{b23}, it only cover a little part (42 sources)
of our sample in ${}^{12}\!$CO line. There are no such results with
${}^{12}\!$CO(1-0), ${}^{13}\!$CO(1-0), and C${}^{18}\!$O(1-0)
emission for the possible cloud-cloud collision candidates ever
before. And we have selected two possible cloud-cloud collision
candidates ( IRAS 02459+6029 and 22528+5936 ) from our survey to
observe with ${}^{12}\!$CO(2-1), ${}^{13}\!$CO(2-1), and
${}^{12}\!$CO(3-2)$, {}^{13}\!$CO(3-2) by KOSMA in previous work (Li
\& Wang, accepted by RAA, 2012). The CO mapping and multi-wavelength
study show the IRAS 02459+6029 could be a cloud-cloud collision
sample and the IRAS 22528+5936 could be two separate clouds without
colliding.

\section{Sample selection}
We build the sample list on the basis of the ${}^{12}\!$CO(1-0)
survey for the cold {\it IRAS\/} sources \citep{b14}. These 1331
sources are selected from the cold {\it IRAS\/} sources according to
the criteria as follows:

1.(a) $\alpha$(B1950) $\leq 8^h$, $\alpha$(B1950) $\geq 16^h$,
$\delta$(B1950) $\geq -35^\circ$; (b) $|b|$ $\leq 25^\circ$; or (c)
$|b|$ $\geq 1^\circ$ when $l\leq 60^\circ$ or $l\geq 300^\circ$.
Sources coordinates should follow the conditions.(a) and (b) cover
the sources in the northern sky near the galactic plane and avoid
the source confusion in CO from the the Galactic central region.

2. Sources should have been detected at least in three wave bands to
specify their infrared properties.

3.Sources colors over the 12, 25, and 60 $\mu$m bands satisfy with
$\log$(F12/ F25) $\leq -$0.4 and $\log$(F25/ F60) $\leq -$0.4 or
$\log$(F12/ F60) $\leq -$0.4 when the FQUAL25 = 1.These cold {\it
IRAS\/} colors range are derived from statistic data for the
embedded sources associated with recent star formation \citep{b1}.

4. These sources are no association with late-type stars, planetary
nebula, extra galaxies, and other kinds of sources unrelated with
Galactic star formation identified by {\it IRAS\/} PSC.

Then we pick up our sample list based on the criteria for possible
cloud-cloud collision candidate from the 1331 cold {\it IRAS\/}
sources. According to a portion of criteria given by Vallee (1995)
which describe the spectrum properties of the possible cloud-cloud
collision region, we choose the samples which the line profile
features show the double peak or multipeak and the velocity of peaks
is adjacent. It means this kind of sources may have several velocity
components and it might include several colliding cloud clumps. The
Figure 1 shows the CO(1-0) spectrum that we selected from 1331 cold
{\it IRAS\/} sources.

\begin{figure}
\centering
  \vspace{-1pt}
\includegraphics [height=70mm,  angle=-90]{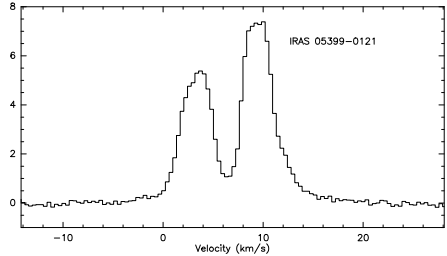}
  \caption{ The ${}^{12}\!$CO(1-0) line profile of initially selected sample for possible cloud-cloud collision from 1331 {\it IRAS\/}
sources (Yang et al. 2002). }
 \label{Fig1}
\end{figure}

After this step, there are 214 sources selected finally. But only
201 sources are detected in the ${}^{12}\!$CO(1-0),
${}^{13}\!$CO(1-0), and C${}^{18}\!$O(1-0) lines. The galactic
distribution of the 201 sources are described in the Figure 2. Most
of the sources are located around the Galactic plane. A little part
of sources are concentrated toward the well-known local star-forming
regions, particularly toward the Perseus, Taurus, and Orion regions
at $l= 160^\circ-200^\circ$ and $b= -10^\circ $ to $-25^\circ$.

\section[]{Observations}

 These sources were already observed in
${}^{12}\!$CO(1-0), ${}^{13}\!$CO(1-0) and  C${}^{18}\!$O(1-0) with
the 13.7m (45 foot) Millimeter Telescope at the Qinghai Station of
Purple Mountain Observatory, CAS (Chinese Academy of Sciences), in
January 2005. The beam size is 50$\pm$7 arcsec in azimuth and
54$\pm$3 arcsec in elevation. The pointing accuracy is around 5
arcsec. We used the cooled mixer SIS receiver which have three
acousto-optical spectrometers (AOSs) working at 110.20, 109.78 and
115.27GHz to get the three CO(1-0) lines simultaneously. The AOS has
1024 channels, and its band width for
${}^{12}\!$CO(1-0),${}^{13}\!$CO(1-0) and C${}^{18}\!$O(1-0) lines
are 145.330, 42.672 and 43.097 MHz. The system temperature is
200-300K during observations. The noise level of antenna temperature
was 0.4K for ${}^{12}\!$CO(1-0), 0.3K for ${}^{13}\!$CO(1-0) and
0.2K for C${}^{18}\!$O(1-0) typically. In this paper, we don't apply
main-beam efficiency corrections to any data, and just show the data
simply in the original antenna temperature scale $T^*_A$.

We observed these sources in position-switching mode in order to
obtain a better baseline. Each source was scanned two times at least
except some sources are measured many times for the bad SNR (signal
to noise ratio) in few cases. The integration time was 2 minutes on
every source. We identified a source to be detected if its peak
antenna temperature is larger than $3\sigma$ rms noise level. We
observed the standard sources, such as W51D, S140, W3(OH), NGC2264,
etc., every 2 hours during observation. These standard sources
served as the secondary temperature standard. We use it to calibrate
target sources within a specific period of time for reducing the
influence of any possible short-period variation of the system.

The data reduction based on the software packages, CLASS (Continuum
and Line Analysis Single-dish Software) and GREG (Grenoble Graphic).

 \begin{figure}[h!!!]
  \centering
 \vspace{-1em}
\includegraphics [height=70mm,  angle=0]{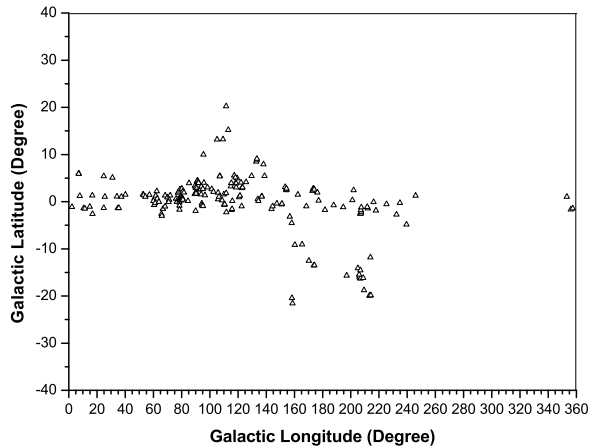}
 \vspace{-1em}
\begin{minipage}[]{150 mm}
  \caption{ Galactic distribution of the 201 sources which are selected as cloud-cloud collision candidates. }
   \end{minipage}
\label{Fig2}
\end{figure}

\section[]{RESULTS}
\subsection{The statistic study of the selected CO sources}
The 201 sources are observed in ${}^{12}\!$CO(1-0),
${}^{13}\!$CO(1-0) and C${}^{18}\!$O(1-0). Table 1 shows the basic
parameters of 201 sources as possible cloud-cloud collision
candidates. Column (1) shows {\it IRAS\/} name of the sources, (2)
and (3) denote equatorial coordinates in the J2000 epoch. Columns
(4) and (5) are their Galactic coordinates, and (6) shows the peak
temperature of the antenna to each sources. Column (7) is the
corresponding rms noise level of the CO spectrum, and (8) contains
the radial velocity of the sources relative to the local standard of
rest. In column (9), the FWHM of each line derived from Gaussian
fitting is presented. Column (10) lists the line span measured at
the 0 K level, column (11) gives the kinematic distance of each
source which derived from other paper \citep{b14}.

The kinematic distances of the CO sources were plotted in Figure 3.
The lack of the inner part of the Galaxy is obvious. Because the
serious contamination in the CO line, we do not observe these
sources in this region ( $|b|$ $< 1^\circ$ when $l\leq 60^\circ$ or
$l\geq 300^\circ$ ). We could investigate the molecular emission
such as high density tracers around these sources in further
observation.

\begin{figure}[h!!!]
 \centering
 \vspace{-1em}
\includegraphics [height=70mm,  angle=0]{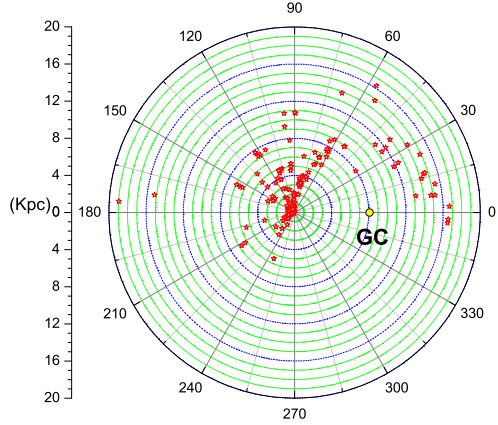}
  \begin{minipage}[]{100 mm}
   \caption{Projected Galactocentric distribution of the 201 candidates. }
   \end{minipage}
  \label{Fig3}
\end{figure}

Figure 4 demonstrates the $L-V_{LSR}$ diagram for the all CO sources
that we selected. The range of the $V_{LSR}$ from 91.73 to -80.09
$kms^{-1}$. The distribution of $V_{LSR}$ conforms to the
distribution of the 1331 sources sample given by Yang \citep{b14}.
It basically follows the Galactic rotation. A significant part of
sources which associated with the star-forming regions distributes
over a large fraction of the Galactic plane. A large portion of the
sources are located in the second Galactic quadrant. Some of them
with the minus sign of the radial velocities and large amplitudes
are located in the Perseus arm possibly.

\begin{figure}[h!!!]
\centering
 \vspace{-1em}
\includegraphics [height=70mm,  angle=0]{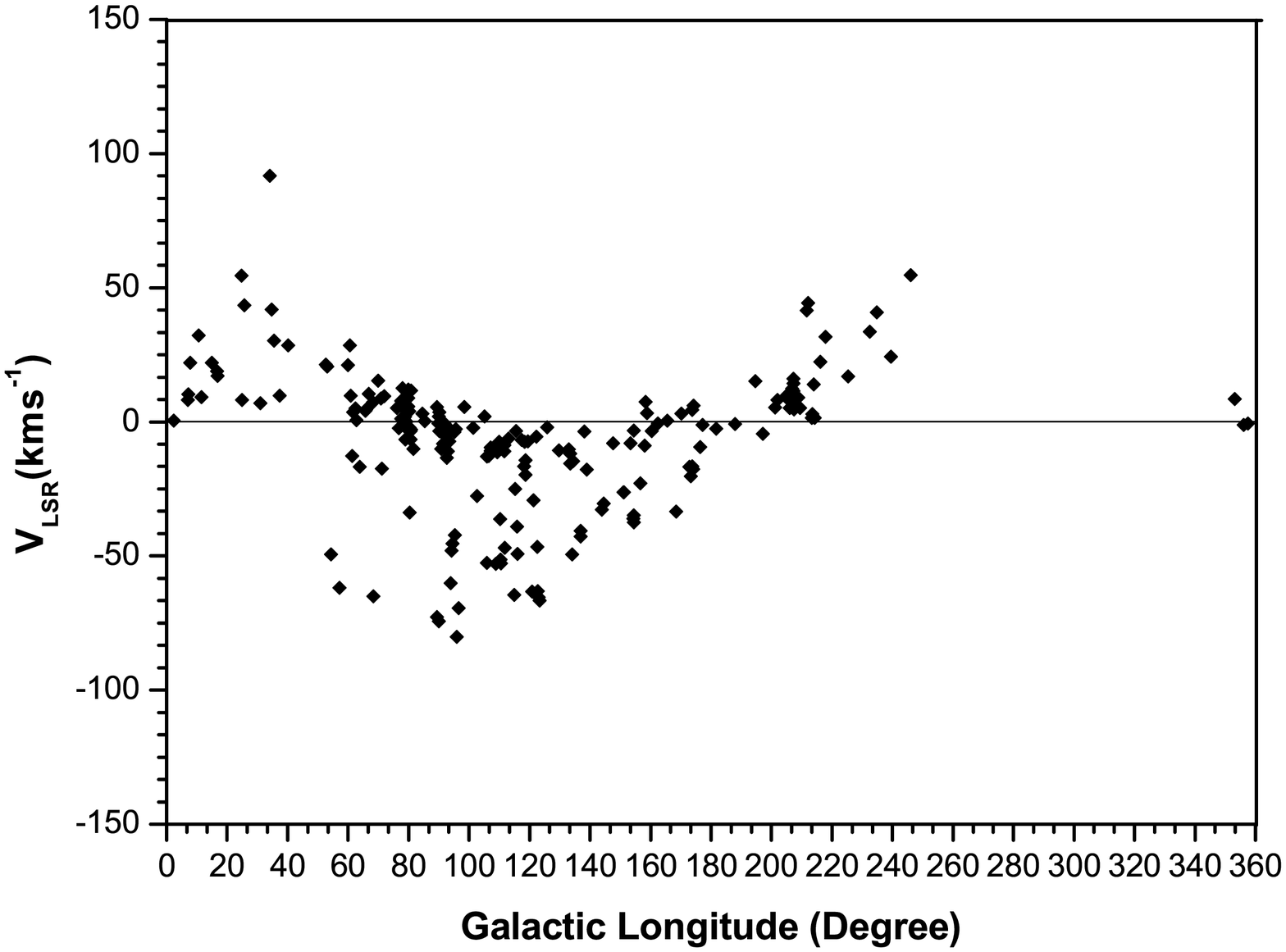}
 \begin{minipage}[]{65 mm}
  \caption{L-$V$ distribution of our survey sample }
   \end{minipage}
     \label{Fig4}
\end{figure}

The statistic distribution of the peak antenna temperature of the
201 sources is illustrated in Figure 5. The range of antenna
temperature covers from 0.73 to 18.4 K and the average temperature
is 3.8 K. The distribution of the line intensity is monotonic
generally from 2 K to 10 K, and the peak lies on the 2-3 K. The
source {\it IRAS\/} 05394-0151 is detected the highest antenna
temperature of 18.4 K.

\begin{figure}[h!!!]
\centering
 \vspace{-1em}
\includegraphics [height=70mm,  angle=0]{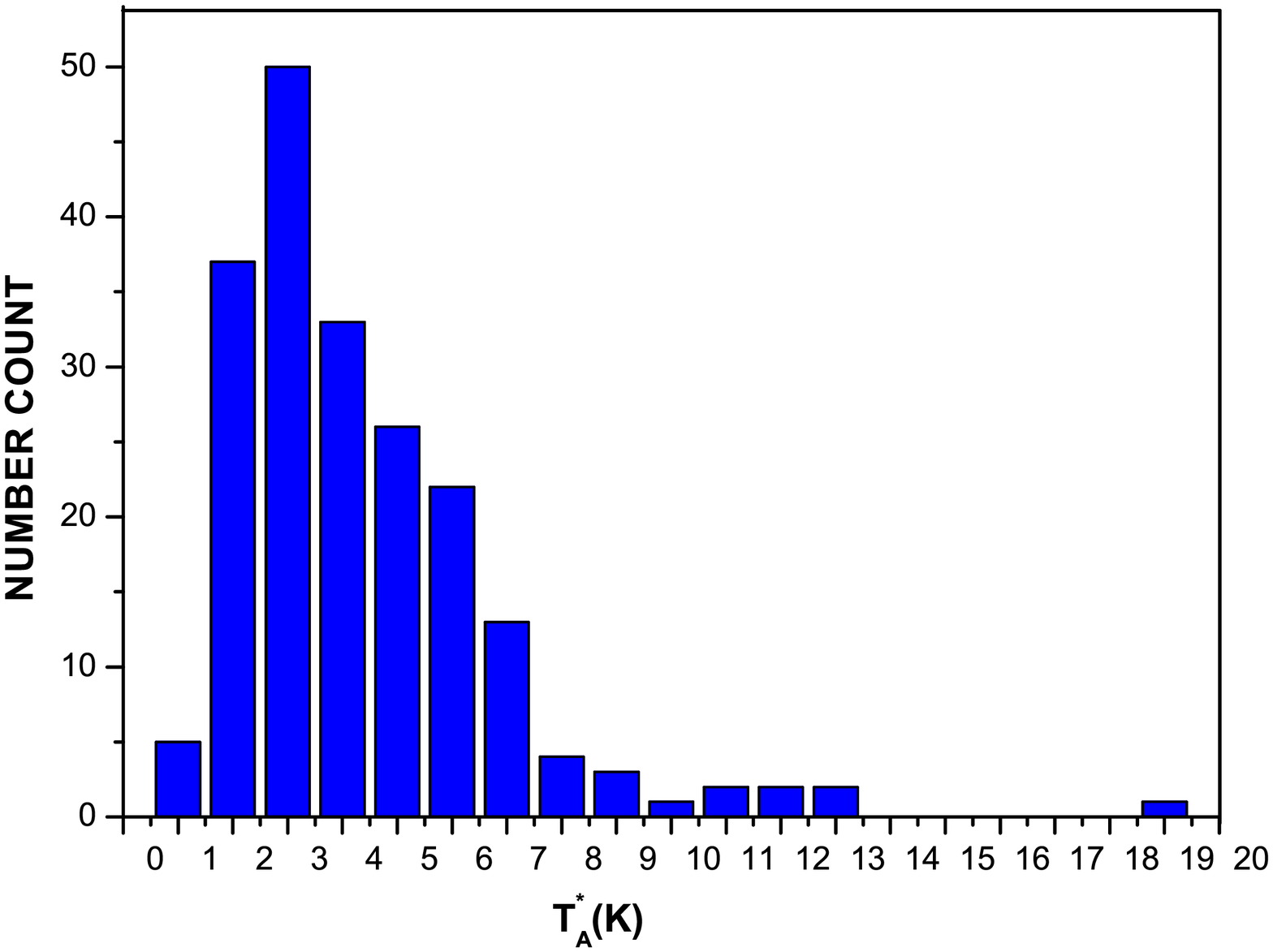}
\begin{minipage}[]{80 mm}
  \caption{Statistical distribution of the $T^*_A$ of the 201 sources }
\end{minipage}
\label{Fig4}
\end{figure}

\small
  \setlength{\tabcolsep}{1mm}
\begin{longtable}{cllllrrrrrr}

 \caption {\label{tab_a1} Basic parameters
of the 201 selected {\it IRAS\/} sources. In this table, 42 sources
with asterisk were also observed by the FCRAO Outer Galaxy Survey
\citep{b23}.
} \\

 \hline \hline
 Name  &   $\alpha (2000)$    &  $\beta(2000)$  &
$l$ & $b$ &
   $T_A^*$  &  $T_\sigma$  & $V_{LSR}$ &  $\delta V$   &  $V_{span}$
    &  $D_k$    \\
    &   &   & (deg) &  (deg) & (K) & (K) & ($kms^{-1}$)&   & ($kms^{-1}$)&  (kpc)  \\
  (1) & (2) & (3)& (4) & (5) & (6) & (7) & (8) & (9) & (10) & (11)
  \\
\hline
\endfirsthead
\caption{$Continued$}\\
\hline\hline
  Name     &   $\alpha (2000)$    &  $\beta(2000)$  &
$l$ & $b$ &
   $T_A^*$  &  $T_\sigma$  & $V_{LSR}$ &  $\delta V$   &  $V_{span}$
    &  $D_k$    \\
        &   &   & (deg) &  (deg) & (K) & (K) & ($kms^{-1}$)&   & ($kms^{-1}$)&  (kpc)  \\
  (1) & (2) & (3)& (4) & (5) & (6) & (7) & (8) & (9) & (10) & (11)
  \\
\hline
\endhead
\hline
\endfoot
$^\ast$00021+6604 & 00 04 43.60 & +66 20 58.0 & 118.26373   &   3.901982    & 2.53  &  0.15  & -7.53   & 2.94   &  5.65   &  0.45            \\
$^\ast$00056+6605 & 00 08 18.80 & +66 21 53.0 & 118.621328  &   3.854562    & 2.89  &  0.12  & -14.34  & 1.47   &  3.47   &  0.89            \\
$^\ast$00070+6516 & 00 09 44.20 & +65 33 21.9 & 118.631671  &   3.033102    & 5.11  &  0.19  & -19.81  & 3.05   &  6.52   &  1.26            \\
$^\ast$00116+6716 & 00 14 20.10 & +67 33 22.0 & 119.390751  &   4.939818    & 1.95  &  0.22  & -7.24   & 1.84   &  8.68   &  0.42            \\
$^\ast$00289+6327 & 00 31 51.50 & +63 43 58.0 & 120.767534  &   0.943341    & 1.14  &  0.12  & -63.37  & 1.78   &  4.8    &  7.19            \\
$^\ast$00342+6347 & 00 37 11.02 & +64 04 01.0 & 121.374013  &   1.238323    & 4.32  &  0.2   & -29.18  & 11.68  &  30.32  &  1.87            \\
$^\ast$00412+6638 & 00 44 15.34 & +66 54 41.1 & 122.226176  &   4.049512    & 2.95  &  0.2   & -5.46   & 2.43   &  7.39   &  0.29            \\
$^\ast$00455+6137 & 00 48 33.27 & +61 53 36.8 & 122.592344  &   -0.97629    & 1.73  &  0.17  & -46.63  & 6.34   &  9.56   &  3.77            \\
\hline\hline
\end{longtable}

\normalsize
 We also use {\it IRAS\/} PSC and {\it MSX \/} PSC to
investigate if there are infrared sources around these CO sources
within 5 arcmin. Table 2 demonstrates the number of {\it IRAS\/}
point source and {\it MSX \/} point source around the 201 sources,
and notes the association of the {\it IRAS\/} point source and {\it
MSX \/} point source by n or a tags. Column (1) is the {\it IRAS\/}
name. In the columns (2) and (3) give the number of the {\it IRAS\/}
point source and {\it MSX \/} point source in the region within 5
arcmin centered with these sources sample. Column (4) show the note
for the association between columns (2) and (3). Label a represents
{\it MSX \/} point source overlap our selected CO source. Label n
means There is no {\it MSX \/} source in the position of central
source. The Columns (4) and (5) denotes the type and identification
of these 201 observed sources.

\small

  \setlength{\tabcolsep}{0.2mm}
\begin{longtable}{cccccc}

\caption{\label{tab_a1}
Detected {\it IRAS\/} and {\it MSX\/} sources  around the selected sources } \\
\hline \hline

  Name     &  {\it IRAS\/} Point Sources & {\it MSX\/} Point Sources  &  Note  &
 Type &  Identification     \\
    & (Sky coverage within 5 arcmin)  &  (Sky coverage within 5 arcmin)  &   &   &   \\
  (1) & (2) & (3)& (4) & (5) & (6)  \\
\hline
\endfirsthead
\caption{$Continued$}\\
\hline\hline
   Name     & {\it IRAS\/} Point Source & {\it MSX\/} Point Source  & Note &
 Type &  Identification     \\
    & (Sky coverage within 5 arcmin)  &  (Sky coverage within 5 arcmin)  &   &   &   \\
  (1) & (2) & (3)& (4) & (5) & (6)  \\
\hline
\endhead
\hline
\endfoot
00021+6604 &    2   &       0   &      &   IR       &                                     \\                                          
00056+6605 &    1   &       1   &  n   &   IR       &                                     \\                    
00070+6516 &    1   &       1   &  a   &   IR       &                                     \\                 
00116+6716 &    1   &       0   &      &   IR       &                                     \\                                          
00289+6327 &    1   &       1   &  a   &   IR       &                                     \\                     
00342+6347 &    2   &       1   &  a   &   IR       &                                     \\                       
00412+6638 &    1   &       2   &  n   &   G        &   2MASX J00441534+6654411           \\                                        
00455+6137 &    2   &       2   &  n   &   IR       &                                     \\
\hline\hline
\end{longtable}

\normalsize
\subsection{The CO and its isotope molecules emission from sources}

A complete set of the ${}^{12}\!$CO(1-0),${}^{13}\!$CO(1-0) and
C${}^{18}\!$O(1-0) line emissions toward 201 sources is published on
line. The ${}^{12}\!$CO(1-0),${}^{13}\!$CO(1-0) and
C${}^{18}\!$O(1-0) spectrum are plotted in blue, green and red
colors respectively. These sources are classified into four types of
the possible cloud-cloud collision region candidates. Type 1
illustrates the lines with different optical thickness of each
source have the similar profile. The ${}^{12}\!$CO, ${}^{13}\!$CO
and C${}^{18}\!$O($J$=1-0) spectra have double-peaked or
multi-peaked main line and the peaks are almost located at the same
velocity. It contains 38 sources. Several reasons could induce the
double-peaked line profile in optically thick ${}^{12}\!$CO lines,
which includes self-absorption, outflow, infall, rotation and two
clouds configuration, etc. People have already compared and
contrasted the spectral profile of ${}^{12}\!$CO with the profile of
${}^{13}\!$CO and C${}^{18}\!$O, and they summarize the feature in
different situation \citep{b12,b2,b11,b13}. Toward this type of
source, the profile of the optically thick ${}^{12}\!$CO lines fits
for the profiles of the optically thin ${}^{13}\!$CO and
C${}^{18}\!$O very well. So the possibility of the self-absorption,
outflow, infall and rotation is remote. It could be the two
separated cloud in double-peaked profile region. We need to
investigate in multi-wavelength toward these possible cloud-cloud
collision regions in future. In type 2, the C${}^{18}\!$O($J$=1-0)
spectra have the bad SNR (signal to noise ratio) toward these 71
sources. But the profile of the optically thick ${}^{12}\!$CO lines
fits for the profiles of the optically thin ${}^{13}\!$CO well, also
show the double-peaked or multi-peaked profile. This type of the
sources are still treated as possible cloud-cloud collision
candidates. Type 3 contains the spectra with ${}^{12}\!$CO
double-peaked or multi-peaked main lines, the profiles of the
${}^{13}\!$CO and C${}^{18}\!$O line fit the profile of the
${}^{12}\!$CO line partly. Considered the uncertain factors, these
sources can't be made a decision as the non cloud-cloud collision
candidates, we need further observation toward these 58 sources to
confirm it. The last 34 sources are plotted in type 4. One part of
these sources all show the ${}^{13}\!$CO, C${}^{18}\!$O and
${}^{12}\!$CO lines with single peak. The other sources present the
${}^{13}\!$CO and C${}^{18}\!$O lines with single peak, and the peak
is located between the double peak of the optically thick
${}^{12}\!$CO lines. It suggests that this type of optical thick
profile might be induced by self-absorption, infall, etc
\citep{b12,b15}. So we treat these sources as non cloud-cloud
collision candidates.

We also check the galactic distribution for the 4 types of sources
in Figure 6. The 4 types of sources are located around the Galactic
plane mostly and the distribution have similarity with the whole
sample. It means the distribution toward the 4 types of the sources
are uniform in our sample. Figure 7 plots the kinematic distances of
the 4 types of sources. The sources of type 1 locate in the first
and second Galactic quadrant. Some sources of type 1 are associated
with the Perseus arm possibly. The sources of type 2 and 3 have the
similar configuration of the distribution. The sources in Type 4
just locate around 5 kpc. Figure 8 demonstrates the $L-V_{LSR}$
diagram for the 4 types. The distributions for 4 types are similar
with the distribution of 201 sources roughly. And the last statistic
data for the peak antenna temperature of the sources in 4 types show
some different distribution in Figure 9. The sources in Type 1 show
a bimodal distribution. From 2 K to 7 K, there are large dispersion.
The $T_A^*$ range in type 2 are concentrated from 2 K to 5 K. The
peak of distribution in Type 3 lies on the 1-2 K. Type 4 shows a
wide dispersion during 0-7 K.

Our next work will carry on mapping observations toward these
possible cloud collision region. We have selected some sources from
our survey list to observe recently. We want to develop new methods
to identify the cloud-cloud collision region and get some new
results of the star formation triggered by cloud-cloud collision.

\begin{figure}[h!!!]
\centering

\includegraphics [height=50mm,  angle=0]{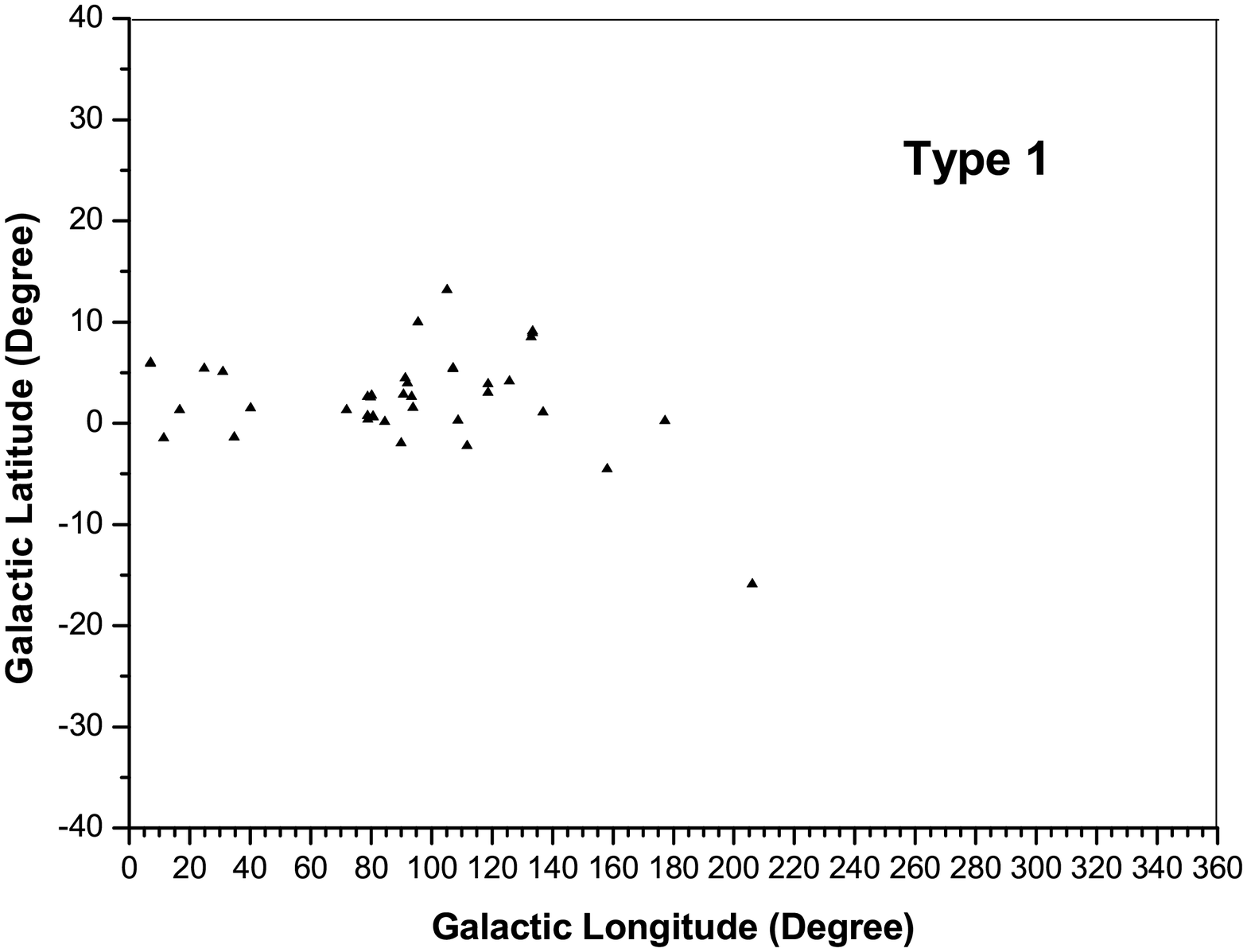}
\includegraphics [height=50mm,  angle=0]{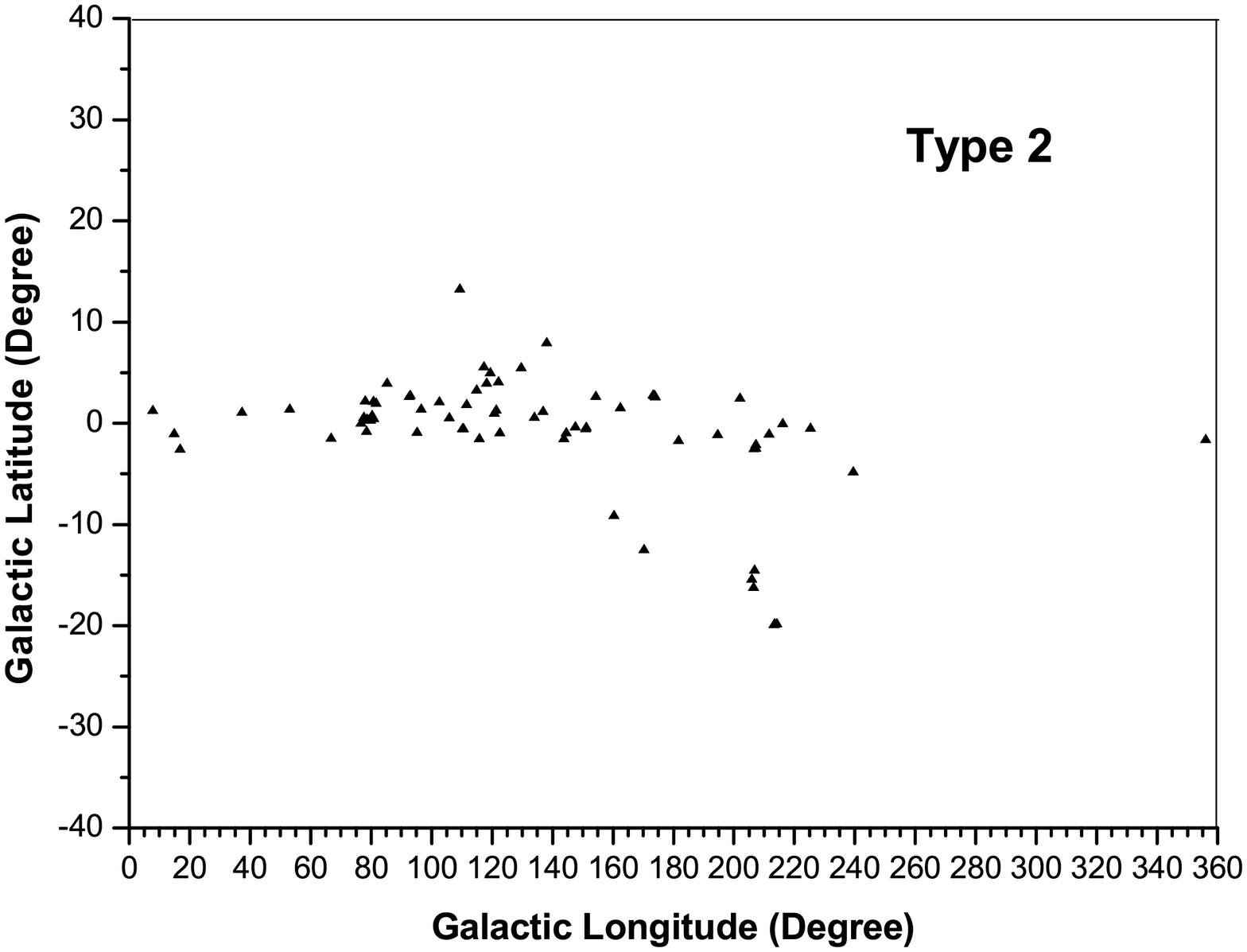}
\includegraphics [height=50mm,  angle=0]{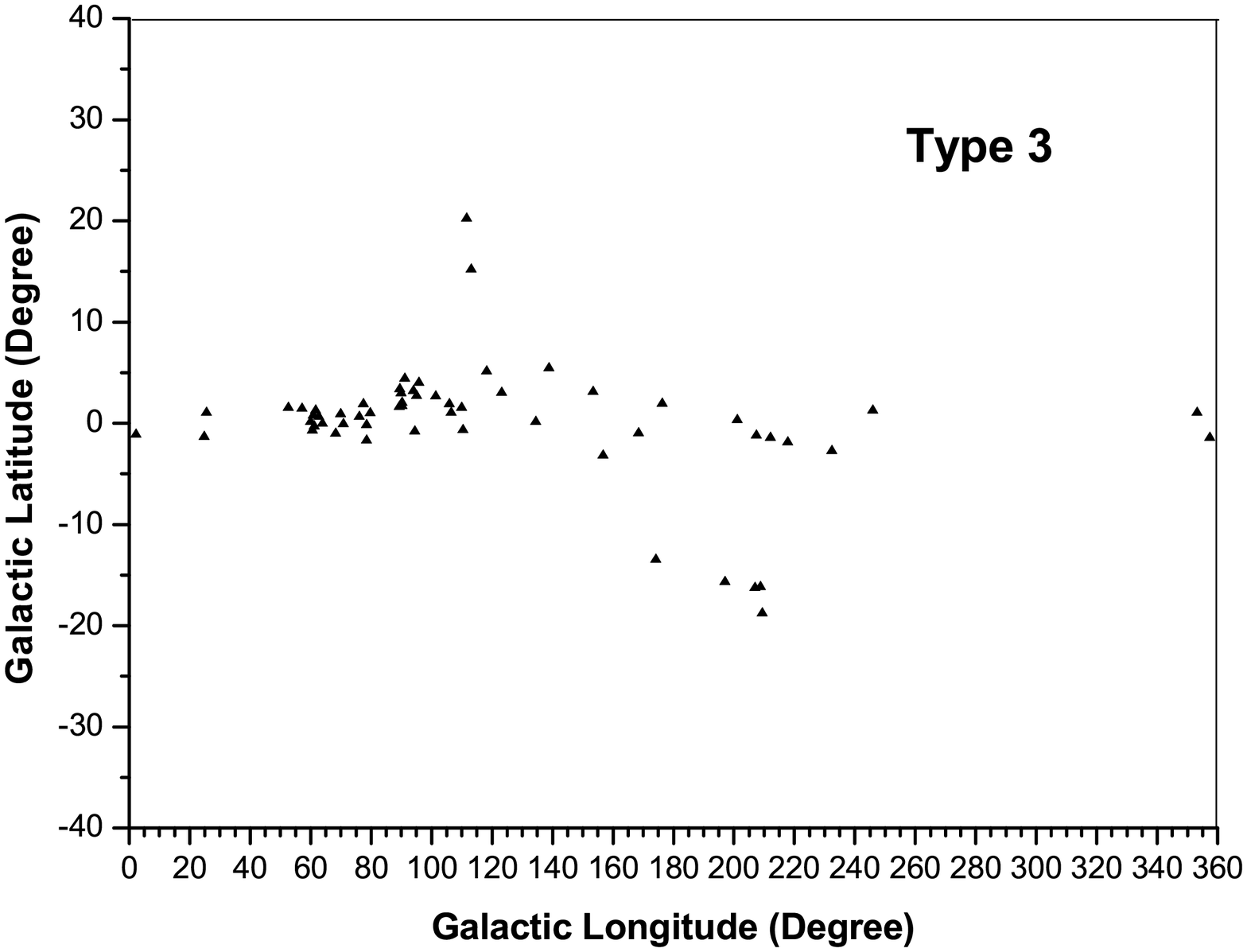}
\includegraphics [height=50mm,  angle=0]{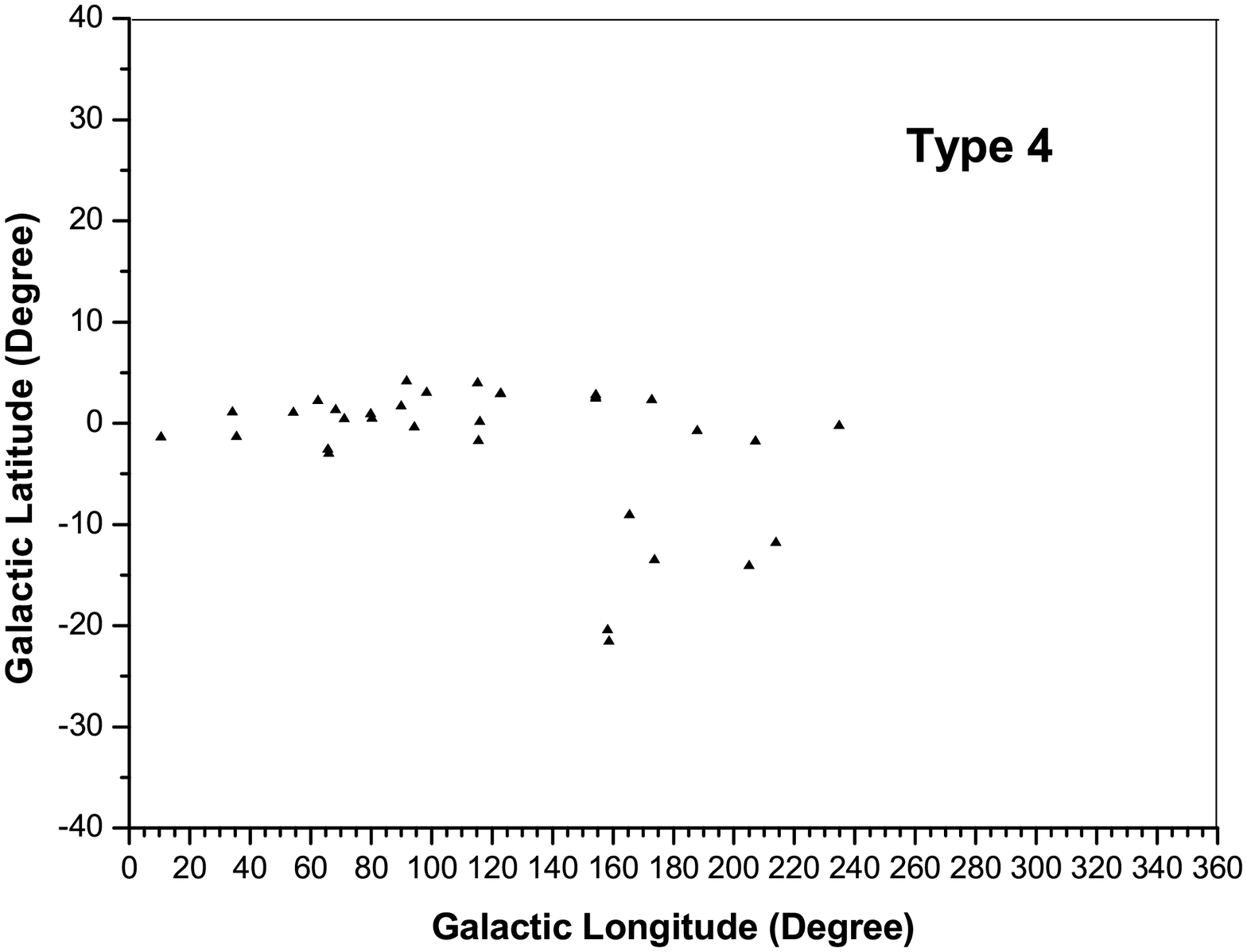}
\begin{minipage}[]{130 mm}
  \caption{Galactic distribution for 4 types of sources which classified  by the fit of the spectrum
profiles between the optically thick and thin lines . }
\end{minipage}
\label{Fig6}
\end{figure}

\begin{figure}[h!!!]
\centering

\includegraphics [height=50mm,  angle=0]{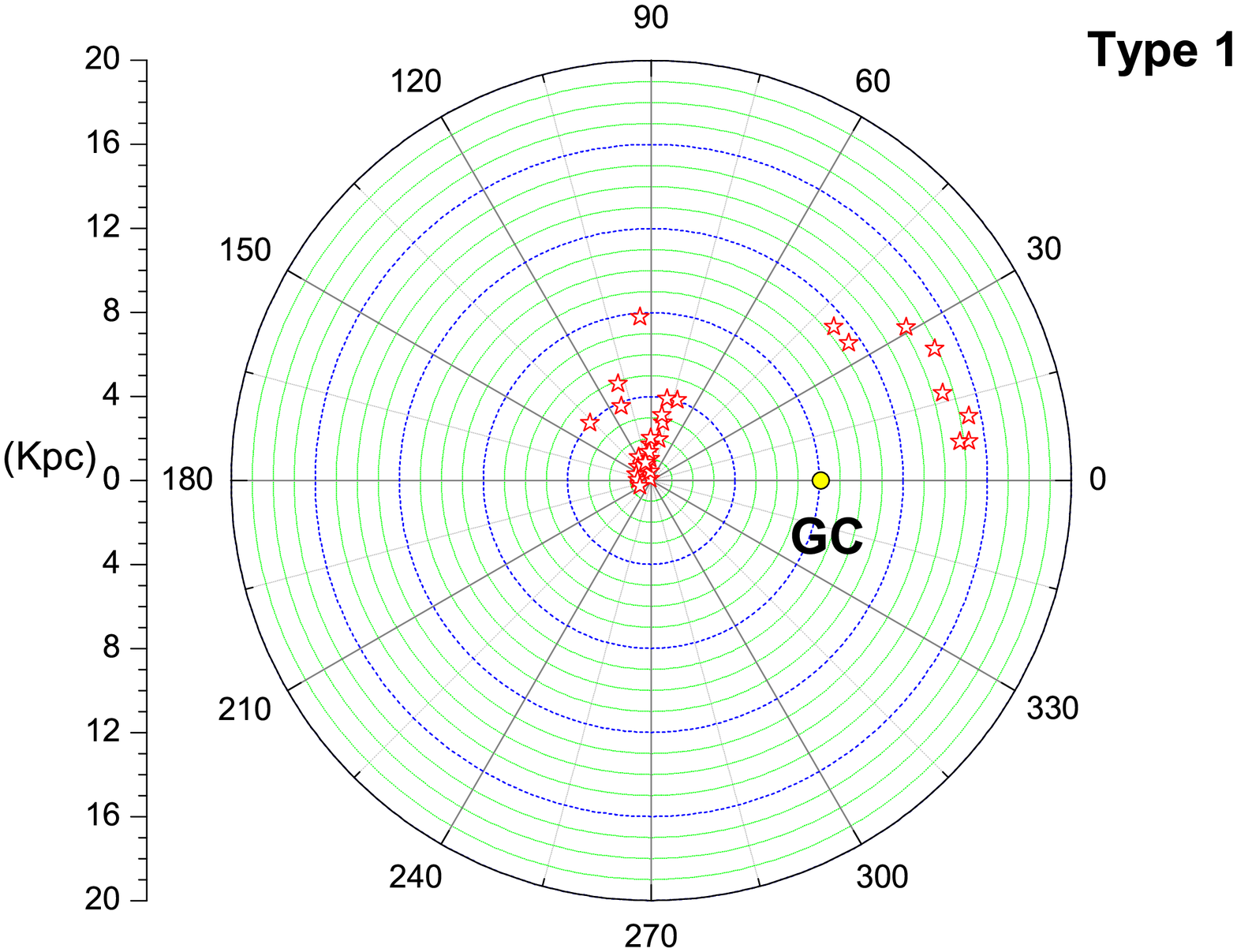}
\includegraphics [height=50mm,  angle=0]{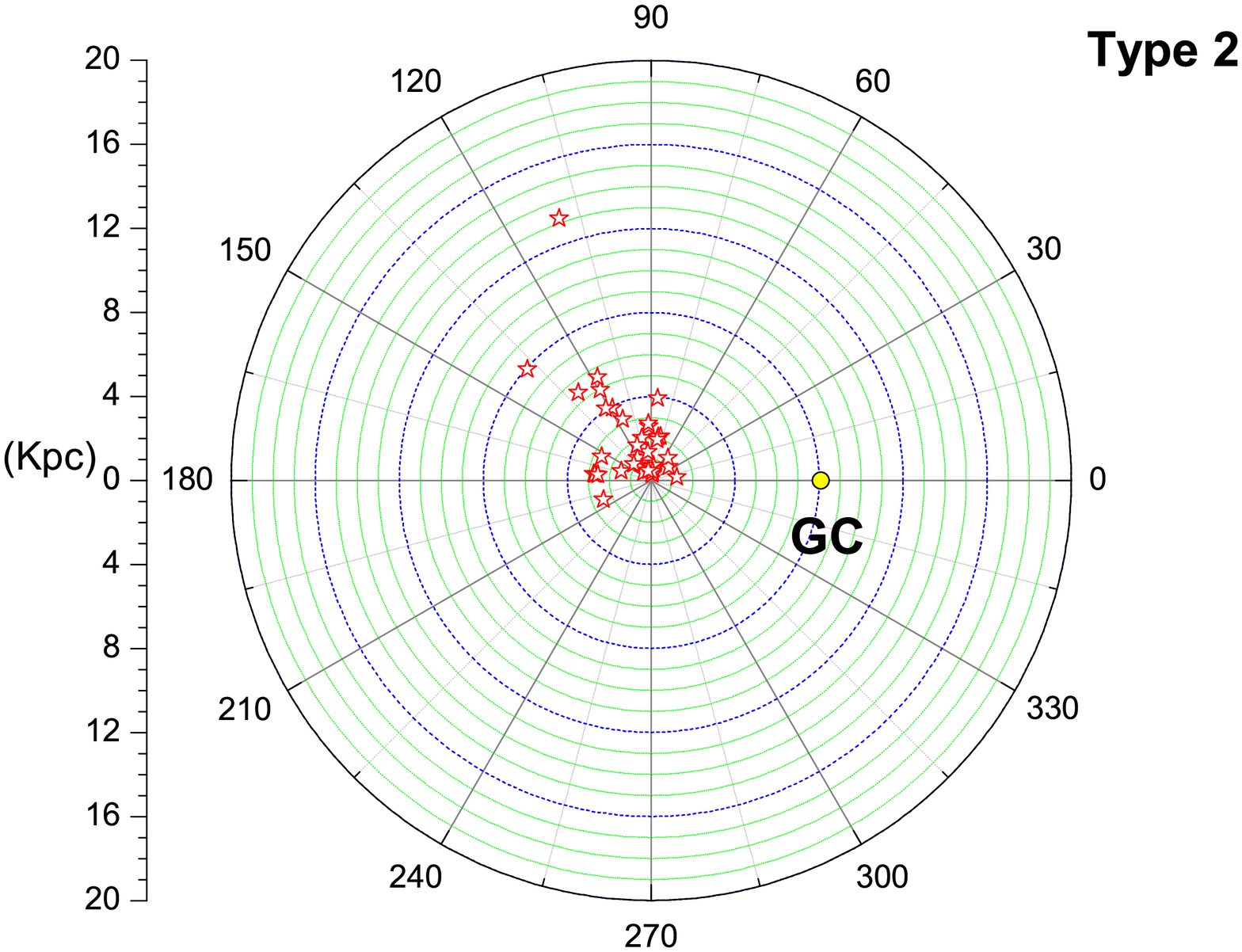}
\includegraphics [height=50mm,  angle=0]{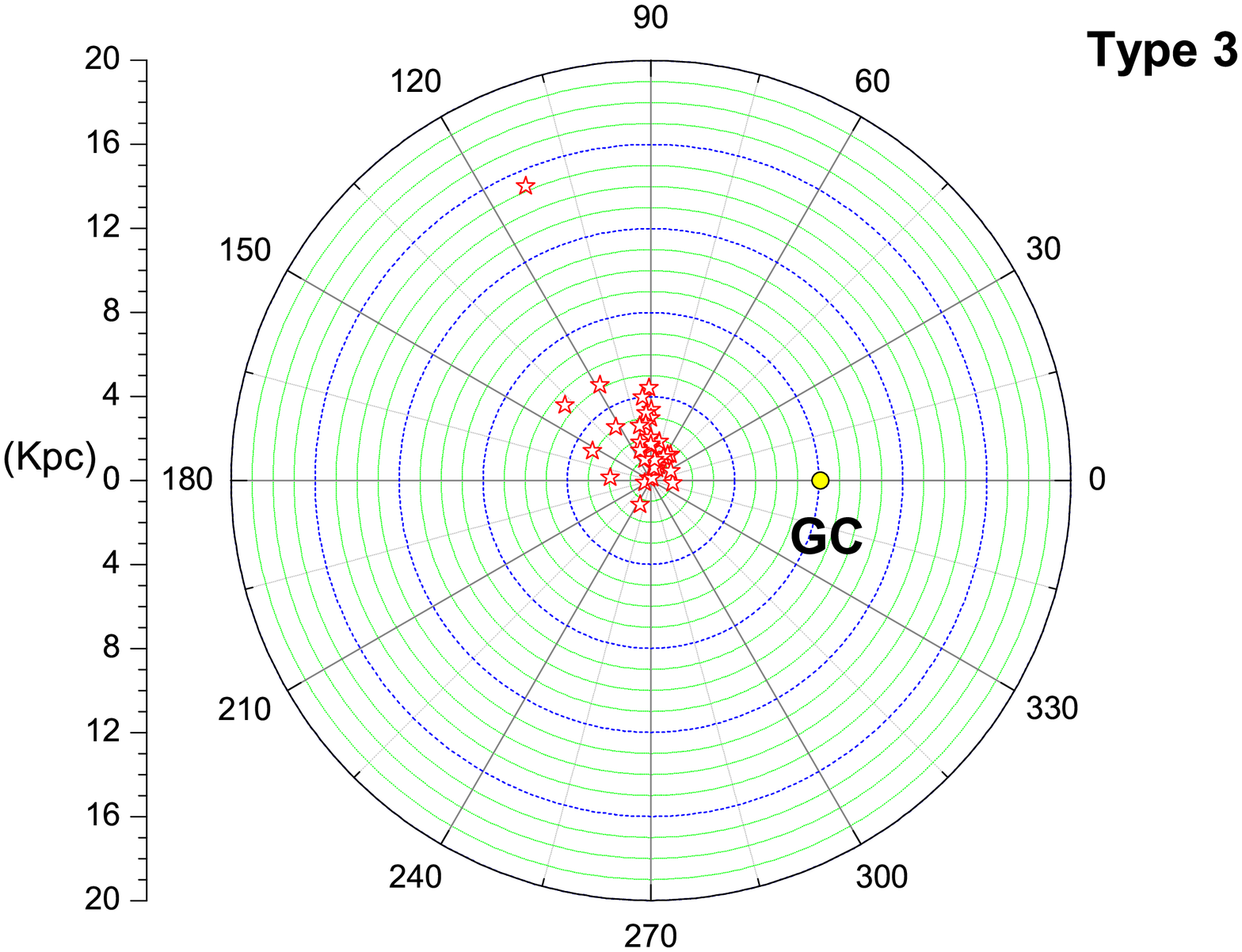}
\includegraphics [height=50mm,  angle=0]{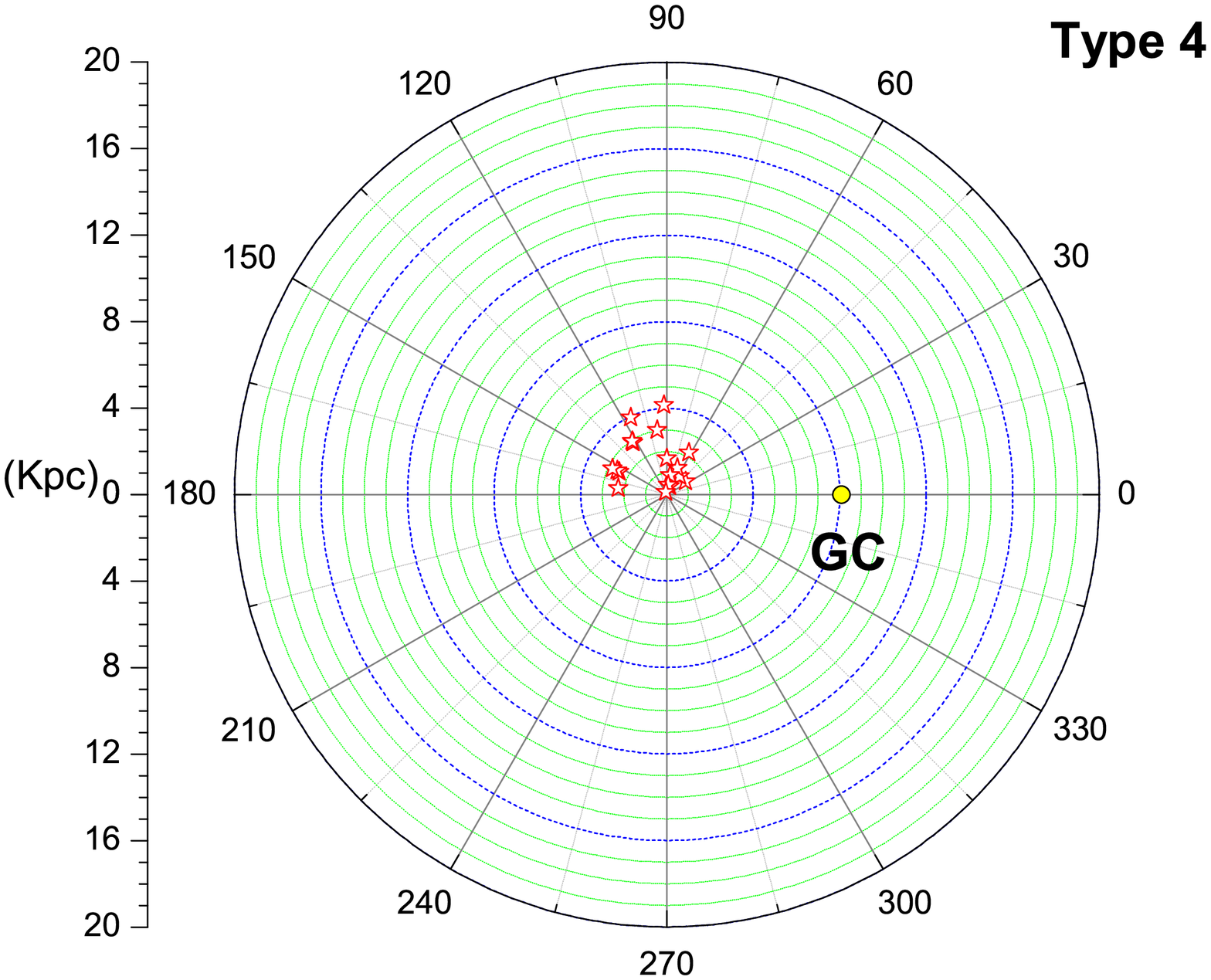}
\begin{minipage}[]{110 mm}
  \caption{Projected Galactocentric distribution toward 4 types of CO sources . }
\end{minipage}
\label{Fig7}
\end{figure}

\begin{figure}[h!!!]
\centering

\includegraphics [height=50mm,  angle=0]{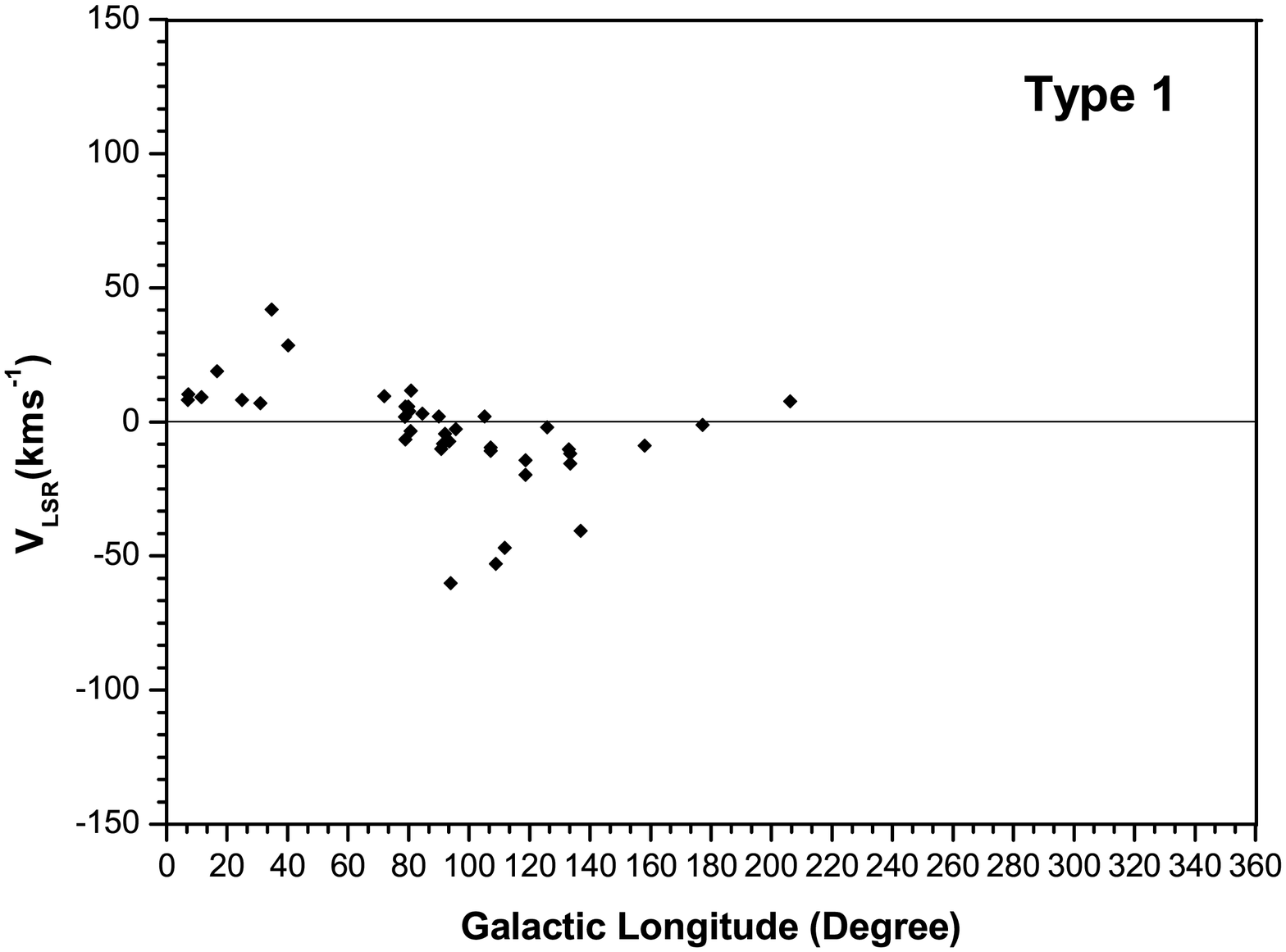}
\includegraphics [height=50mm,  angle=0]{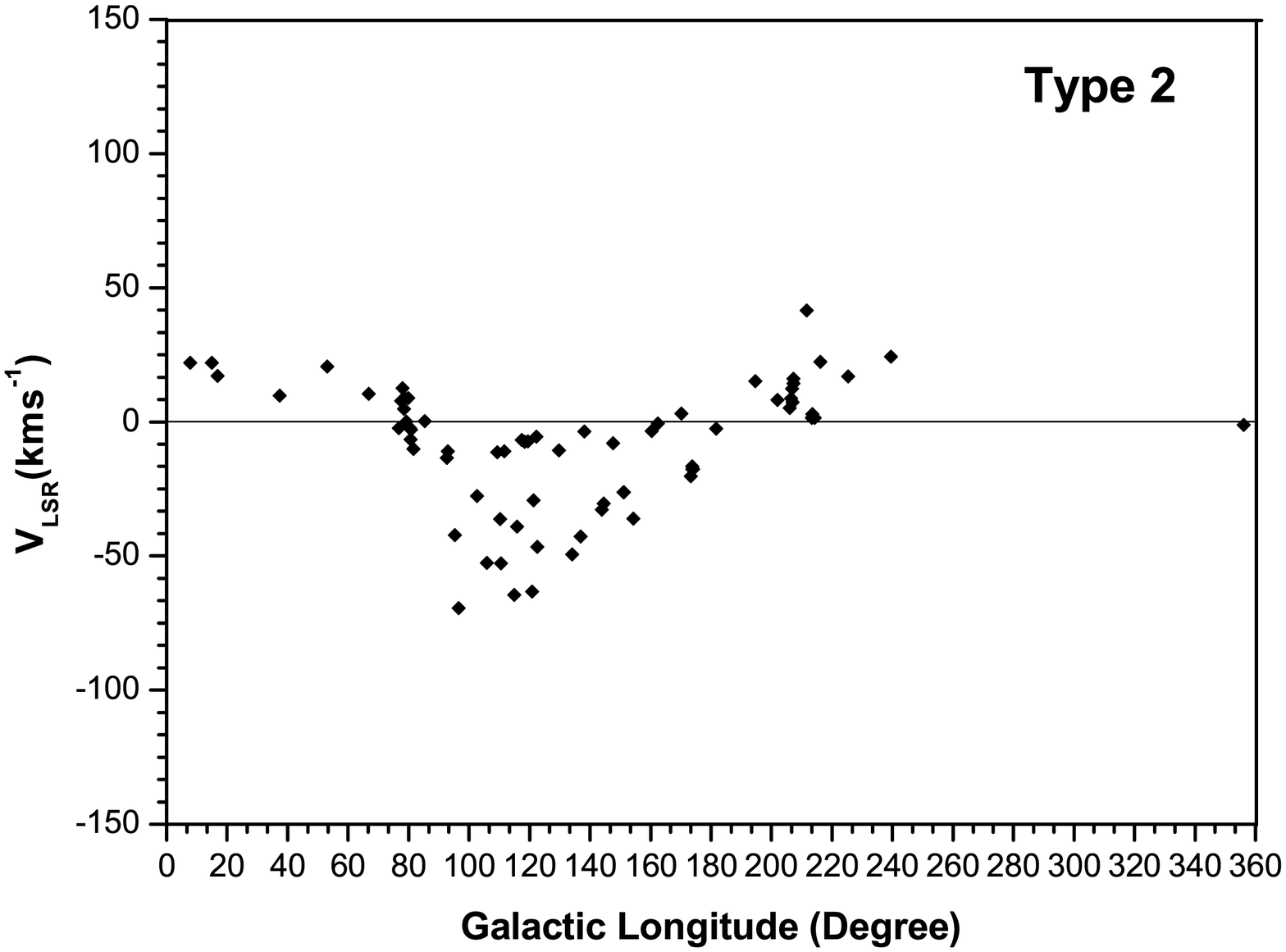}
\includegraphics [height=50mm,  angle=0]{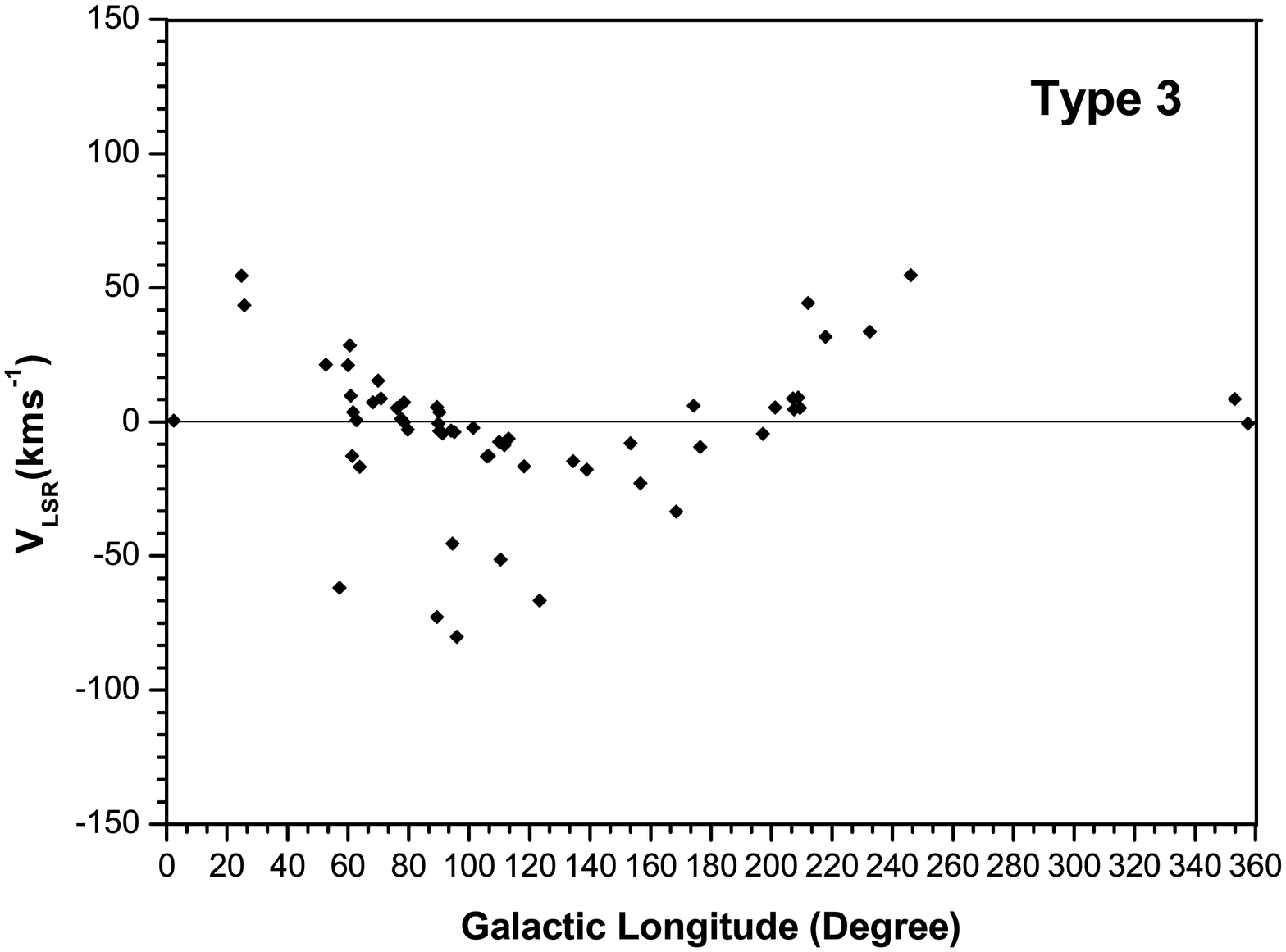}
\includegraphics [height=50mm,  angle=0]{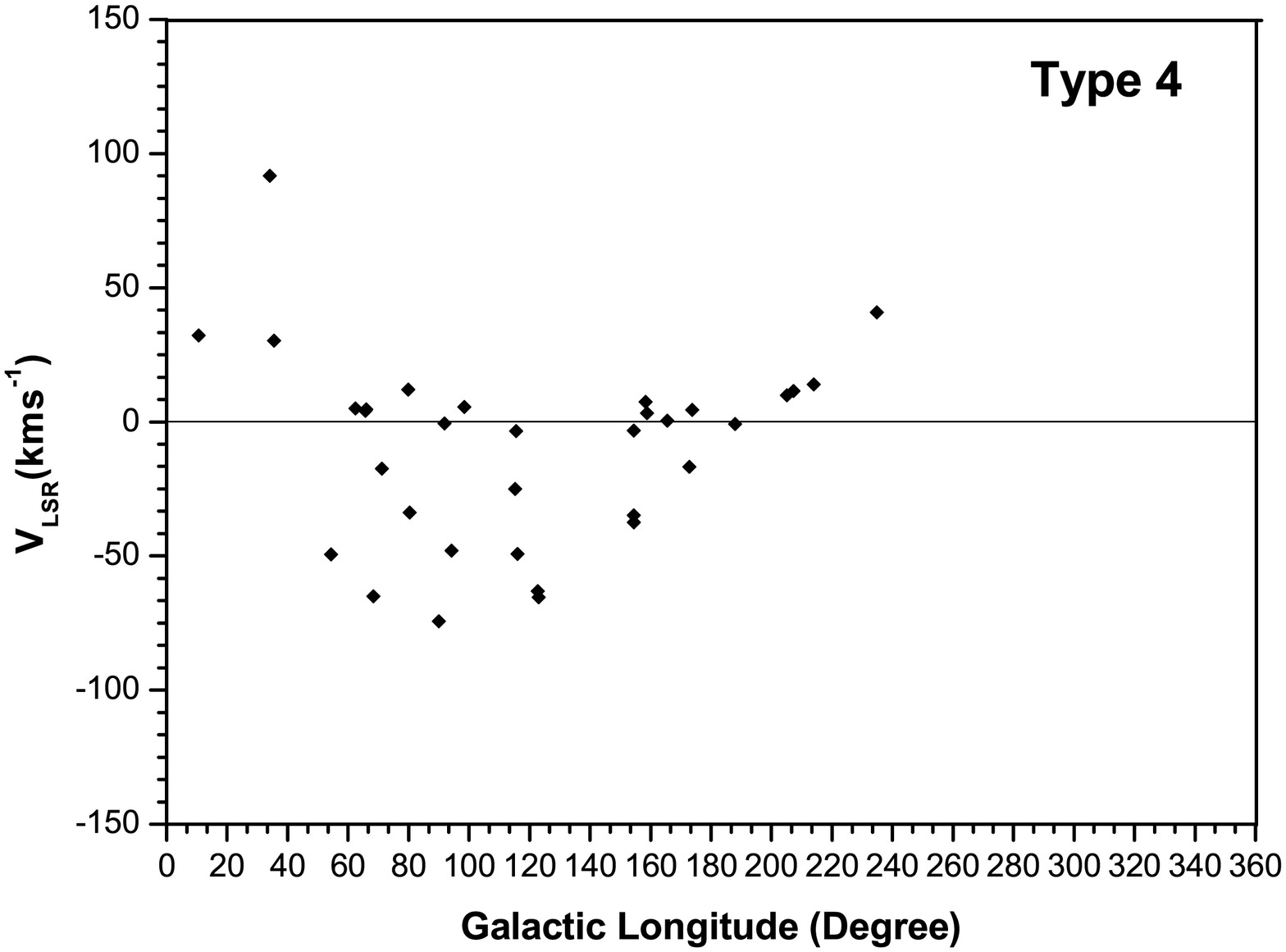}
\begin{minipage}[]{100 mm}
  \caption{L-$V$ distribution about 4 types of sources in our survey sample . }
\end{minipage}
\label{Fig8}
\end{figure}

\begin{figure}[h!!!]
\centering

\includegraphics [height=50mm,  angle=0]{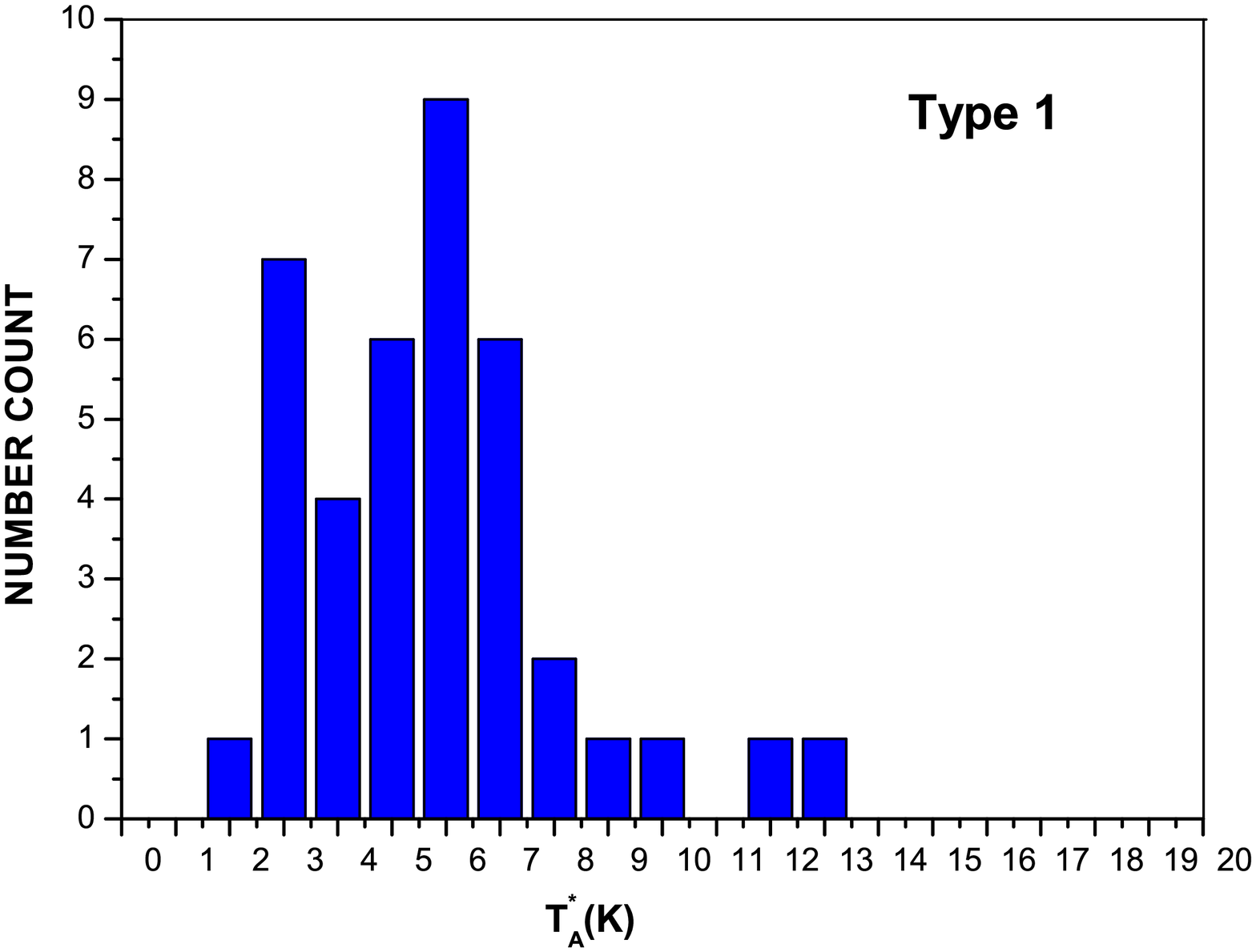}
\includegraphics [height=50mm,  angle=0]{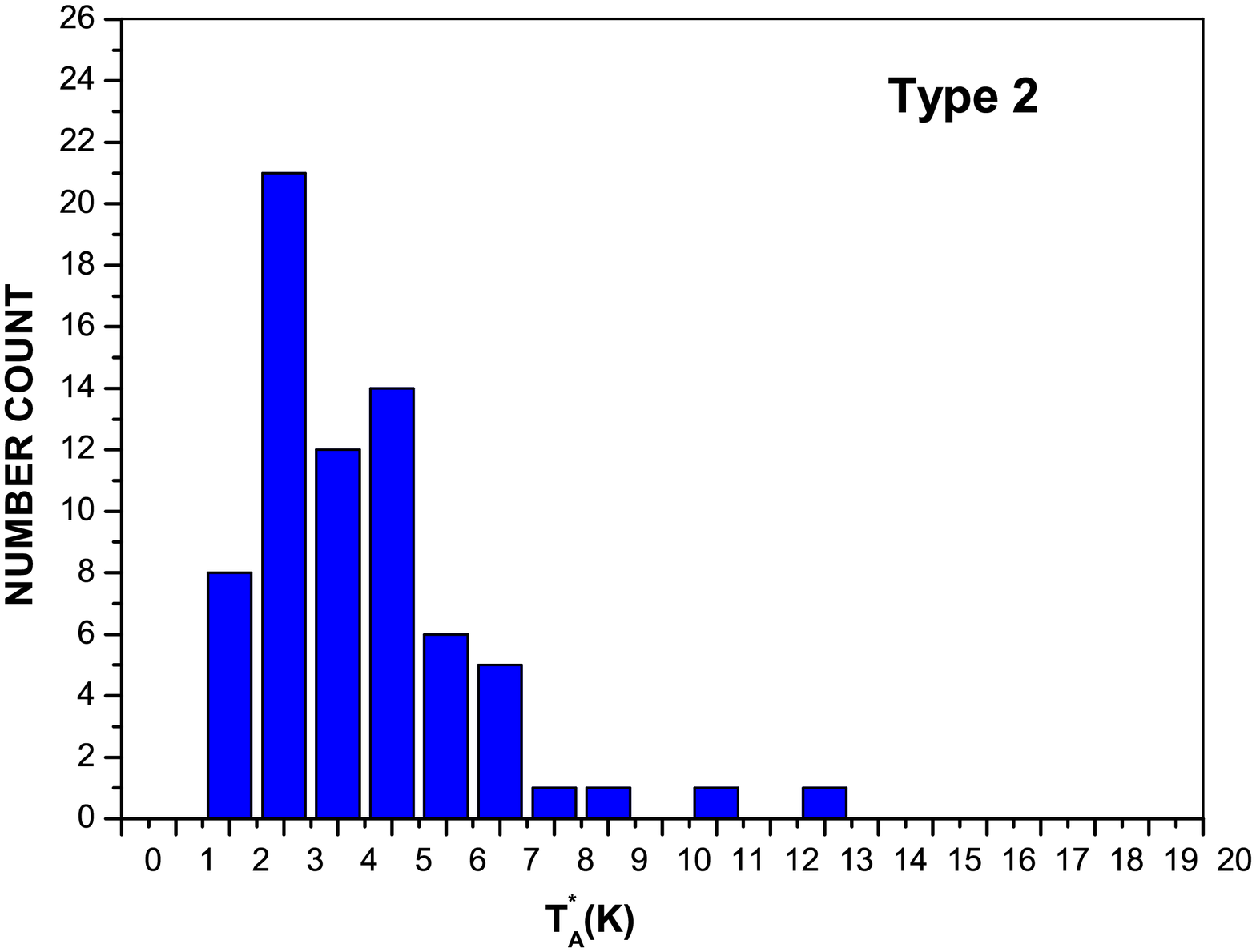}
\includegraphics [height=50mm,  angle=0]{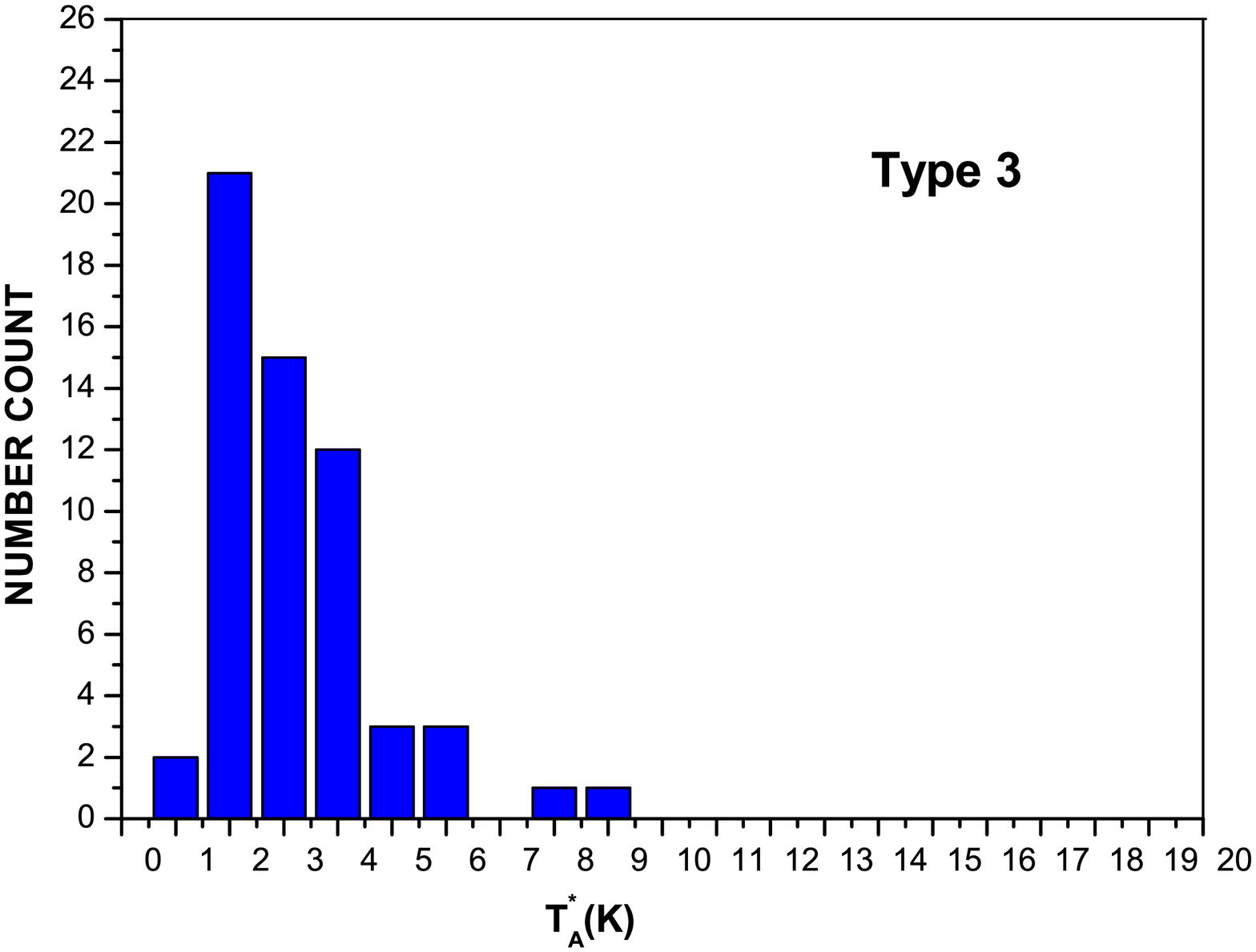}
\includegraphics [height=50mm,  angle=0]{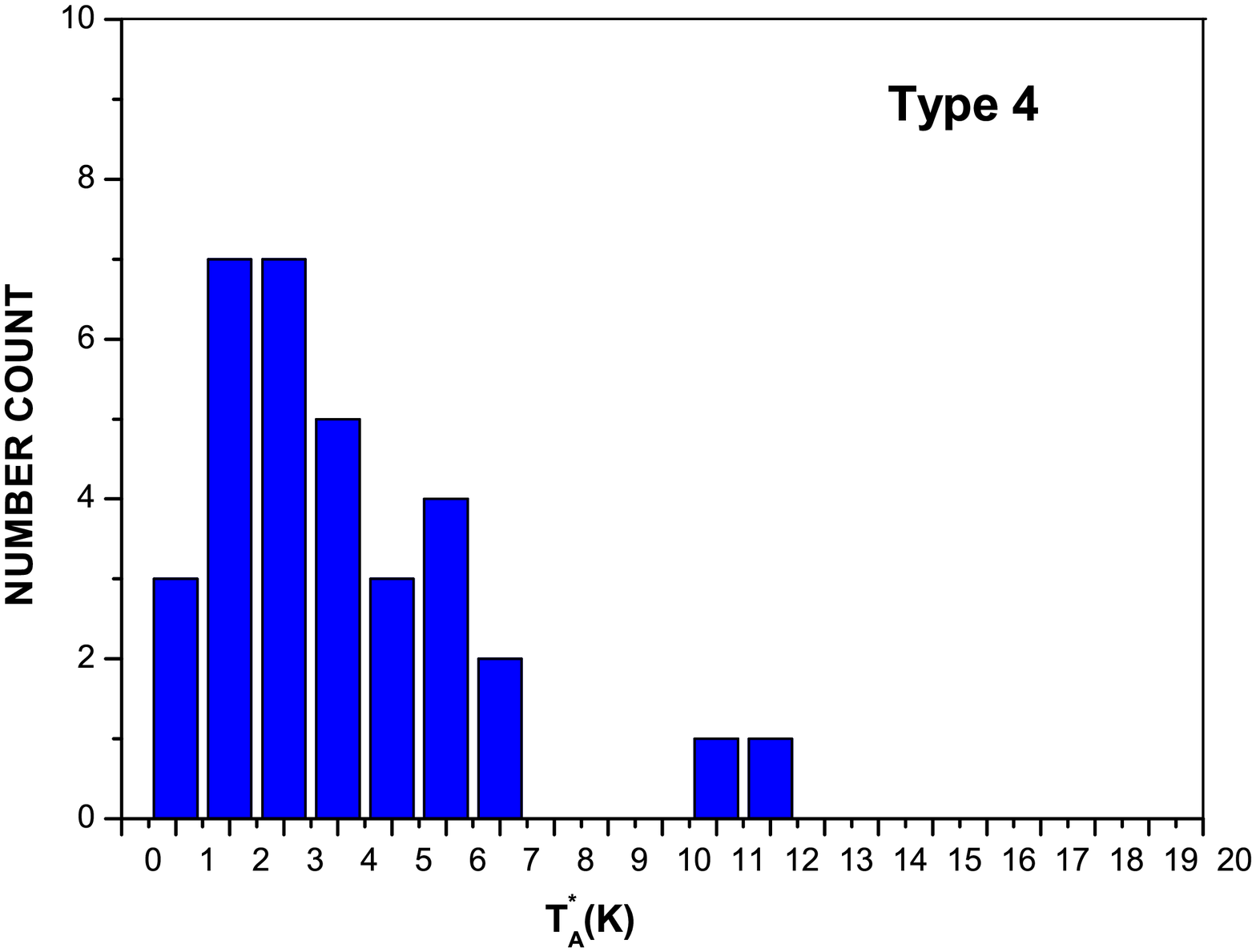}

\begin{minipage}[]{130 mm}
  \caption{Statistical distribution of $T_A^*$ about 4 types of sources in our survey sample . }
\end{minipage}
\label{Fig9}
\end{figure}

\subsection{Derived parameters}

Because of bad SNR (signal to noise ratio) with  the
${}^{13}\!$CO(1-0) line toward some sources in type 2, 3 and 4, we
only calculate the parameters of the 38 sources in type 1. Derived
parameters of the sources in type 1 have been listed in Table 3.
$T_R^* (^{12}CO)$ and $T_R^* (^{13}CO)$ are the ${}^{12}\!$CO and
${}^{13}\!$CO(1-0) radiation temperature: $ T^*_R=T^*_A/ \eta_{mb}$,
where T$_A^*$ is the antenna temperature corrected with atmospheric
attenuation and other losses (these are done by the observatory ),
and $\eta_{mb}$ is the main beam efficiency. $\upsilon_{13} $ is the
Gaussian fit of the $V_{LSR}$(${}^{13}\!$CO) and
$\Delta\upsilon_{13}$ is its full width at half-maximum (FWHM). The
cores' distances from us are obtained from literatures \citep{b17}.
The ${}^{12}\!$CO(1-0) line is usually considered as optically thick
in massive CO cores. Assuming local thermal equilibrium (LTE), the
excitation temperature $T_{ex}$ of the ${}^{13}\!$CO(1-0) transition
is the same as that of the ${}^{12}\!$CO \citep{b16}.

\begin{equation}
 T_{ex}=\frac{h\nu}{k}\{\ln[1+(\frac{kT_R^*(^{12}CO)}{h\nu}+\frac{1}{\exp(
\frac{h\nu}{kT_{bg}}-1)})^{-1} ] \}^{-1} \label{eq:LebsequeI}
\end{equation}

Where the $T_{bg}$=2.7K. The ${}^{13}\!$CO(1-0) transition is
usually optically thin. On the LTE assumption, the opacity of
${}^{13}\!$CO(1-0) is approximately \citep{b22}

\begin{equation}
 \tau(^{13}CO)\approx -\ln[1-\frac{T_R^* (^{13}CO )}{T_R^* (^{12}CO) }]
\label{eq:LebsequeI}
\end{equation}

Sometimes the  ${}^{12}\!$CO spectra are self-reversed by absorption
from the cold part of the cloud core outer. In that case the
${}^{13}\!$CO opacity will be underestimated. The  ${}^{13}\!$CO
column density is given by \citep{b16}

\begin{equation}
N(^{13}CO)=\frac{3k}{8\pi^3 B\mu^2 } \frac{\exp[
\frac{hBJ(J+1)}{kT_{ex}}
]}{J+1}\times\frac{T_{ex}+hB/3k}{[1-exp(\frac{-h\nu}{kT_{ex}}
)]}\int\tau_{\nu} d\nu
 \label{eq:LebsequeI}
\end{equation}

 Where B is the rotational constant of ${}^{13}\!$CO, $\mu$ is
 the permanent dipole moment and $J$ is the rotational quantum number
of the lower state.

 We adopt the ${}^{12}\!$CO and ${}^{13}\!$CO abundances to be

\begin{equation}
\frac{N(H_2)}{N(^{12}CO)}=10000 ,   \frac{N(^{12}CO)}{N(^{13}CO)}=89
\label{eq:LebsequeI}
\end{equation}

\subsection{Infrared results}

 We do the color cuts used for the IRAS sources in our survey for cloud-cloud collision candidates.
 The color cuts used here were derived from values in the literature and were taken from studies of several
types of objects: T Tauri stars, low-mass and intermediate-mass
YSOs, and massive YSOs. Table 4 lists the color cuts we used. We use
the criteria to select the YSOs candidates, but it's not a unique
classification method. The choice of color cuts is merely meant to
be as inclusive as possible. All the results plotted by color-color
diagram are represented in Figure 10. The sources of the 3 types
which are considered as possible cloud-cloud collision candidates
are marked on the map.

\begin{table}

\small \centering

\begin{minipage}[]{70mm}

\caption[]{The parameters of the sources in type 1}\label{Table
3}\end{minipage}
\tabcolsep 3mm
 \begin{tabular}{cccccccc}
  \hline\hline\noalign{\smallskip}
 \it IRAS\ & $T_R^* (^{12}CO)$ &$T_R^* (^{13}CO)$ &$\upsilon_{13}$ &$\Delta\upsilon_{13}$ &$T_{ex}$ &$N_{H_2 }$  &$\tau_{13} $  \\
       & K &K  &$kms^{-1}$ &$kms^{-1}$ & K   &$10^{22}cm^{-2}$  &   \\
  \hline\noalign{\smallskip}

\multirow{2}{*}{00056+6605} &     4.06    &   1.33   &    -6.24     &     0.91    &    	8.47 	   &    0.17    &     0.40         \\
	            &     4.18    &   0.43   &    -2.58     &     1.65    &    	8.59 	   &    0.05    &     0.11         \\
	            \cline{2-8}
\multirow{2}{*}{00070+6516} &     8.54    &   5.06   &    -19.12    &     1.95    &    	13.06    &    0.81    &     0.90         \\
	            &     3.15    &   1.22   &    -5.23     &     0.89    &    	7.52 	   &    0.17    &     0.49         \\
	            \cline{2-8}
\multirow{2}{*}{01166+6635} &     4.33    &   2.10   &    -2.51     &     0.93    &    	8.75 	   &    0.30    &     0.66         \\
	            &     4.64    &   4.57   &    1.78      &     0.78    &    	9.07 	   &    1.96    &     4.12         \\
	            \cline{2-8}
\multirow{2}{*}{02425+6851} &     8.18    &   5.95   &    -10.83    &     1.57    &    	12.69    &    1.12    &     1.30         \\
	            &     5.36    &   1.88   &    -4.08     &     0.77    &    	9.81 	   &    0.23    &     0.43         \\
	            \cline{2-8}
\multirow{2}{*}{02459+6029} &     13.07   &   9.54   &    -40.91    &     1.83    &    	17.64    &    2.06    &     1.31         \\
	            &     18.22   &   5.57   &    -36.61    &     2.00    &    	22.83    &    0.93    &     0.36         \\
	            \cline{2-8}
\multirow{3}{*}{02485+6902} &     3.40    &   1.61   &    -14.96    &     1.75    &    	7.78 	   &    0.23    &     0.64         \\
	            &     8.56    &   4.63   &    -10.57    &     1.94    &    	13.08    &    0.71    &     0.78         \\
	            &     5.36    &   2.22   &    -3.59     &     0.62    &    	9.81 	   &    0.29    &     0.54         \\
	            \cline{2-8}
\multirow{3}{*}{02499+6911} &     9.54    &   6.22   &    -15.57    &     3.05    &    	14.07    &    1.09    &     1.06         \\
	            &     6.35    &   1.61   &    -10.29    &     2.58    &    	10.83    &    0.19    &     0.29         \\
	            &     5.55    &   3.10   &    -3.30     &     0.69    &    	10.01    &    0.46    &     0.82         \\
	            \cline{2-8}
04173+4328    &     4.37    &   3.04   &    -8.61     &     0.85    &    	8.79 	   &    0.53    &     1.19         \\
              \cline{2-8}
\multirow{2}{*}{05363+3127} &     5.19    &   3.14   &    -2.11     &     1.48    &    	9.63 	   &    0.49    &     0.93         \\ 
	            &     9.17    &   5.14   &    1.20      &     1.23    &    	13.69    &    0.81    &     0.82         \\ 
	            \cline{2-8}
\multirow{2}{*}{05399-0121} &     10.66   &   2.65   &    2.52      &     2.33    &    	15.21    &    0.34    &     0.29         \\ 
	            &     14.82   &   12.50  &    8.88      &     1.77    &    	19.41    &    3.48    &     1.85         \\ 
	            \cline{2-8}
\multirow{2}{*}{17364-1946} &     5.81    &   2.08   &    5.17      &     1.01    &    	10.27    &    0.26    &     0.44         \\ 
	            &     4.87    &   6.69   &    10.50     &     1.64    &    	9.31 	   &    $\ldots$    &     $\ldots$     \\ 
	            \cline{2-8}   
\multirow{2}{*}{17369-1945} &     4.52    &   2.81   &    5.27 	    &     0.81    &    	8.95 	   &    0.45    &     0.97         \\ 
	            &     4.76    &   4.55   &    10.43     &     0.88    &    	9.19 	   &    1.51    &     3.11         \\ 
	            \cline{2-8}
\multirow{2}{*}{18134-1942} &     17.96   &   11.40  &    10.67     &     2.70    &    	22.58    &    2.51    &     1.01         \\ 
	            &     7.75    &   5.61   &    21.49     &     1.08    &    	12.25    &    1.04    &     1.29         \\ 
	            \cline{2-8}
\multirow{2}{*}{18136-1347} &     11.50   &   4.55   &    19.36     &     2.08    &    	16.06    &    6.65    &     0.50         \\ 
	            &     14.82   &   9.07   &    20.93     &     1.51    &    	19.41    &    1.78    &     0.95         \\ 
	            \cline{2-8}
\multirow{2}{*}{18273+0034} &     5.52    &   5.50   &    8.08      &     1.80    &    	9.97     &    3.46    &     6.19         \\ 
	            &     4.68    &   5.97   &    9.93      &     1.46    &    	9.11     &    $\ldots$    &     $\ldots$     \\ 
	            \cline{2-8}   
\multirow{2}{*}{18567+0700} &     17.75   &   11.27  &    29.20     &     2.63    &    	22.36    &    2.47    &     1.01         \\ 
	            &     5.96    &   1.39   &    32.97     &     2.64    &    	10.43    &    0.16    &     0.26         \\ 
	            \cline{2-8}
\multirow{2}{*}{18572+0057} &     6.02    &   2.39   &    37.53     &     2.84    &    	10.49    &    0.31    &     0.50         \\ 
	            &     8.66    &   5.18   &    45.68     &     4.51    &    	13.18    &    0.84    &     0.91         \\ 
	            \cline{2-8}
20051+3435    &     9.48    &   5.59   &    11.11     &     3.25    &    	14.01    &    0.92    &     0.89         \\ 
\cline{2-8}
20190+4102    &     16.27   &   8.77   &    2.84      &     1.35    &    	20.88    &    1.67    &     0.77         \\ 
\cline{2-8}
\multirow{2}{*}{20227+4154} &     12.86   &   11.19  &    5.58      &     2.23    &    	17.43    &    3.14    &     2.04         \\ 
	            &     5.65    &   1.24   &    10.48     &     2.42    &    	10.11    &    0.14    &     0.25         \\ 
	            \cline{2-8}
\multirow{2}{*}{20228+4215} &     23.44   &   15.47  &    5.50 	    &     1.68    &    	28.08    &    4.08    &     1.08         \\ 
	            &     5.96    &   1.28   &    8.00 	    &     4.71    &    	10.43    &    0.15    &     0.24         \\ 
	            \cline{2-8}
\multirow{2}{*}{20275+4001} &     14.80   &   9.28   &    -5.81     &     3.41    &    	19.39    &    1.85    &     0.99         \\ 
	            &     5.75    &   2.32   &    0.45 	    &     1.87    &    	10.21    &    0.30    &     0.52         \\ 
	            \cline{2-8}
20293+3952    &     10.91   &   9.54   &    5.78 	    &     3.04    &    	15.47    &    2.56    &     2.07         \\ 
\cline{2-8}
\multirow{2}{*}{20332+4124} &     15.32   &   9.66   &    -2.02     &     3.39    &    	19.92    &    1.96    &     1.00         \\ 
	            &     5.46    &   1.92   &    4.57 	    &     1.92    &    	9.91     &    0.24    &     0.43         \\ 
	            \cline{2-8}
\multirow{2}{*}{20343+4129} &     3.24    &   1.24   &    -2.72     &     1.58    &    	7.62     &    0.17    &     0.48         \\ 
	            &     11.69   &   4.42   &    11.62     &     2.48    &    	16.25    &    0.64    &     0.48         \\ 
	            \cline{2-8}
\multirow{2}{*}{20436+5849} &     5.77    &   3.16   &    -2.29     &     1.39    &    	10.23    &    0.47    &     0.79         \\ 
	            &     4.35    &   0.96   &    3.60 	    &     1.13    &    	8.77 	   &    0.11    &     0.25         \\ 
	            \cline{2-8}
\multirow{2}{*}{20489+4410} &     17.89   &   10.80  &    -2.16     &     2.20    &    	22.50    &    2.30    &     0.93         \\ 
	            &     10.18   &   1.92   &    2.47 	    &     4.77    &    	14.72    &    0.24    &     0.21         \\
	            \cline{2-8} 
\multirow{2}{*}{20553+5208} &     7.09    &   5.48   &    -8.46     &     2.01    &    	11.58    &    1.08    &     1.49         \\ 
	            &     5.34    &   1.81   &    -3.21     &     2.31    &    	9.79 	   &    0.23    &     0.42         \\ 
	            \cline{2-8}
\multirow{2}{*}{21005+5217} &     4.08    &   1.26   &    -10.77    &     1.70    &    	8.49 	   &    0.16    &     0.37         \\ 
	            &     5.22    &   3.53   &    -1.54     &     3.67    &    	9.67 	   &    0.60    &     1.12         \\ 
	            \cline{2-8}
\multirow{2}{*}{21007+5036} &     10.64   &   7.44   &    -10.33    &     1.91    &    	15.19    &    1.43    &     1.20         \\ 
	            &     3.61    &   0.96   &    -1.83     &     4.90    &    	8.00 	   &    0.12    &     0.31         \\ 
	            \cline{2-8}
\multirow{2}{*}{21131+5221} &     4.89    &   3.10   &    -13.49    &     1.57    &    	9.33 	   &    0.50    &     1.00         \\ 
	            &     8.10    &   5.81   &    -5.73     &     3.67    &    	12.61    &    1.07    &     1.26         \\ 
	            \cline{2-8}
21169+6804    &     5.50    &   5.48   &    3.02 	    &     0.97    &    	9.95 	   &    3.40    &     6.10         \\ 
\cline{2-8}
\multirow{2}{*}{21182+4641} &     11.01   &   9.28   &    1.62 	    &     1.25    &    	15.56    &    2.31    &     1.85         \\ 
	            &     4.51    &   0.82   &    5.57 	    &     1.16    &    	8.93 	   &    0.09    &     0.20         \\ 
	            \cline{2-8}
\multirow{2}{*}{21202+5157} &     13.34   &   3.63   &    -59.20    &     2.39    &    	17.92    &    0.51    &     0.32         \\ 
	            &     2.64    &   0.90   &    -54.12    &     1.17    &    	6.99 	   &    0.13    &     0.41         \\ 
	            \cline{2-8}
\multirow{2}{*}{22199+6322} &     16.41   &   9.42   &    -10.64    &     1.70    &    	21.01    &    1.86    &     0.85         \\ 
	            &     8.16    &   2.69   &    -7.05     &     1.09    &    	12.67    &    0.34    &     0.40         \\ 
	            \cline{2-8}
\multirow{2}{*}{22202+6317} &     11.56   &   4.69   &    -10.88    &     1.40    &    	16.11    &    0.69    &     0.52         \\ 
	            &     6.35    &   4.08   &    -7.36     &     1.56    &    	10.83    &    0.67    &     1.03         \\ 
	            \cline{2-8}
22528+5936    &     14.08   &   7.46   &    -53.25    &     2.05    &    	18.66    &    1.32    &     0.75         \\ 
\cline{2-8}
23215+5826    &     9.63    &   3.04   &    -46.59    &     1.44    &    	14.17    &    0.40    &     0.38         \\ 
\cline{2-8}

  \noalign{\smallskip}\hline\hline
\end{tabular}
\end{table}

\begin{table}[h!!!]

\bc

\begin{minipage}[]{80mm}

\caption[]{ {\it IRAS\/} Color Cuts Used For Selection Of YSOs
}\label{Table 4}\end{minipage}
 \small
\tabcolsep 4.5mm
 \begin{tabular}{lcccc}
  \hline\noalign{\smallskip}
 Object Type &[25-12] &[60-25]  &[100-60]   &Reference    \\
  \hline\noalign{\smallskip}
   T Tauri stars &0-0.6 &$<$0.5 &\ldots & 1, 2  \\
Intermediate-mass YSOs &0-1.0 &0.4-1.0 &\ldots &3 \\
High-mass YSOs &0.17-1.5 &0.23-1.13 &0.08-0.57 &4 \\

  \noalign{\smallskip}\hline
\end{tabular}
\ec
\tablecomments{0.86\textwidth}{Where
$[\lambda1-\lambda2]=\log(F1/F2)$. } \tablerefs{0.86\textwidth}{
(1)Harris, Clegg \& Hughes 1988.(2)Prusti,Adorf, \& Meurs, 1992.
(3)Beichman et.al 1986. (4) Chan,Henning, \& Schreyer 1996.
 }
\end{table}

\begin{figure}[h!!!]
\centering

\includegraphics [height=75mm,  angle=0]{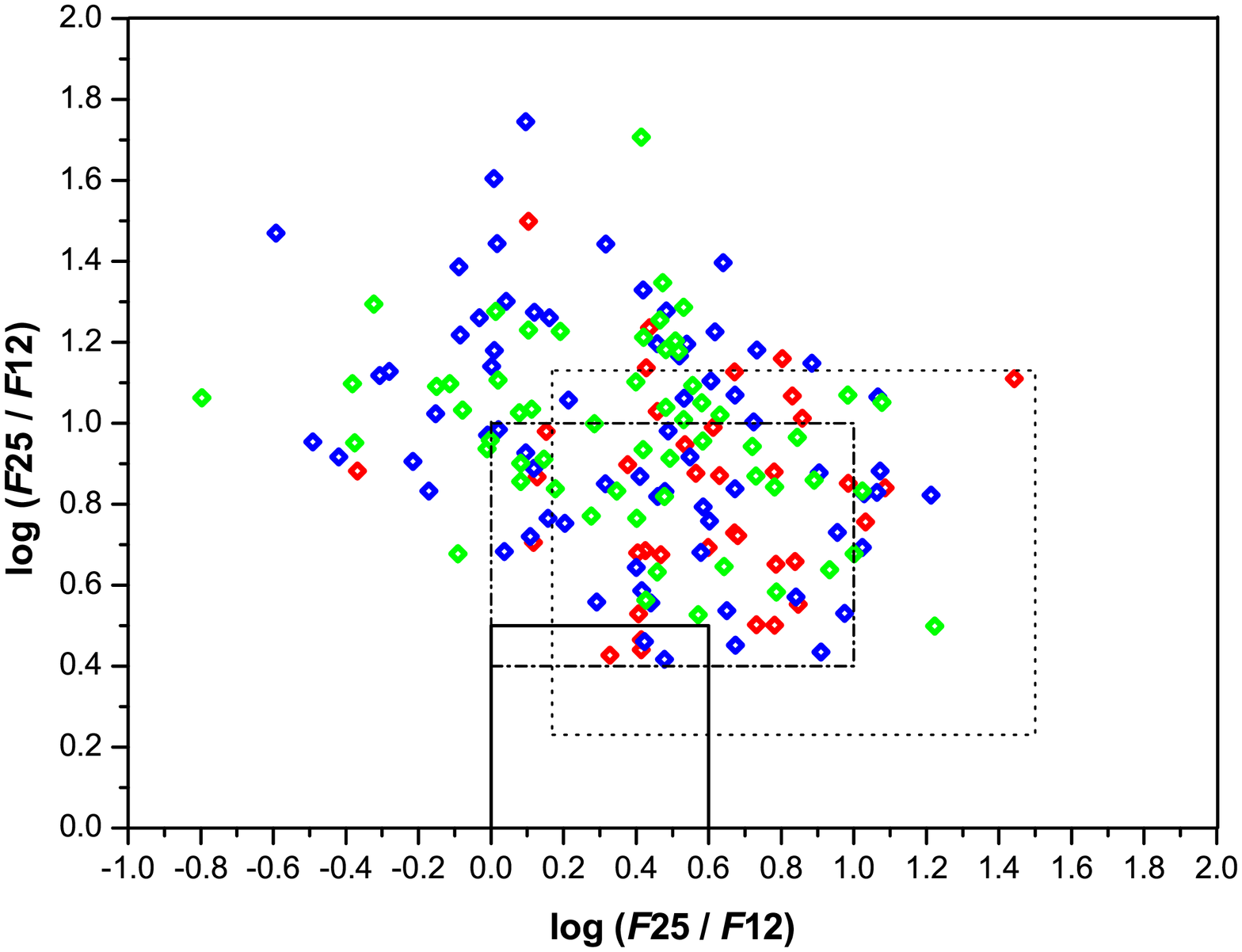}

\begin{minipage}[]{150 mm}
  \caption{The figure describes the color-color distribution of the sources in type 1, 2 and 3.
  The red, blue and green diamonds represent the sources in type 1, 2 and 3, respectively.
  The color box of T Tauri stars is indicated by the solid line. The dashed line box covers the Intermediate-mass YSOs and
   the dotted line box shows the location of the High-mass YSOs.   }
\end{minipage}
\label{Fig10}
\end{figure}

According to the range of the IRAS cuts given by Table 4, we
summarize that 3 T Tauri stars, 25 Intermediate-mass YSOs and 24
High-mass YSOs are identified from the sources of type 1. There are
2 T Tauri stars, 30 Intermediate-mass YSOs and 22 High-mass YSOs
identified in type 2. In type 3, 23 Intermediate-mass YSOs and 27
High-mass YSOs  are identified from the 58 sources and no T Tauri
stars.




\section{Conclusions}

In this paper, according to the criteria for cloud-cloud collision
sample, we observed the ${}^{12}\!$CO, ${}^{13}\!$CO and
C${}^{18}\!$O($J$=1-0) lines in 201 {\it IRAS\/} sources which are
selected as cloud-cloud collision candidates. The criteria are as
follows :

1.A portion of criteria given by Vallee (1995) which describe the
spectrum properties of the possible cloud-cloud collision region.

2.The line profile features show the double peak or multipeak and
the velocity of peaks are adjacent.

These sources are located over a wide range of the Galactocentric
distances, and associated with the star formation region partly.
Then the 201 sources are classified into 4 types by the close fit
between the profile of optically thick ${}^{12}\!$CO lines and the
profiles of the optically thin ${}^{13}\!$CO and C${}^{18}\!$O. The
{\it IRAS\/} flux in four bands have been used to identify sources
with the colors of YSOs. Considering the association of the {\it
IRAS\/} and the {\it MSX\/} PSC, the sources in type 1, 2 and 3
could be selected to do the multi-wavelength mapping. To see if
there is star formation triggered by cloud-cloud collision. Toward
two possible cloud-cloud collision candidates ( {\it IRAS\/}
02459+6029 and 22528+5936 ) selected from our survey, combining the
CO mapping and multi-wavelength study  we prove that the {\it
IRAS\/} 02459+6029 could be a cloud-cloud collision sample and the
{\it IRAS\/} 22528+5936 could be two separate clouds without
colliding (Li \& Wang, accepted by RAA 2012). And we also develop
our new criteria to identify the cloud-cloud collision region in
that paper. This survey is very foundational and efficient for
selected a possible cloud-cloud collision region.


\begin{thebibliography}{99}
\bibitem[\protect\citeauthoryear{Beichman}{1986}]{b1} Beichman C. A., 1986, in Light on Dark Matter, ed. F. P. Israel
(Dordrecht: Reidel), 279

\bibitem[\protect\citeauthoryear{Beichman et~al.}{1986}]{b21} Beichman C. A., Myers P. C. et al., 1986, ApJ, 307, 337

\bibitem[\protect\citeauthoryear{Chan. \& Schreyer}{1996}]{b20} Chan H., Schreyer, 1996, A\&AS, 115, 81

\bibitem[\protect\citeauthoryear{Dobashi}{1993}]{b2} Dobashi K., Onishi T. et al., 1993, AJ, 105, 1487


\bibitem[\protect\citeauthoryear{Frerking \& Langer}{1982}]{b3} Frerking M. A., Langer W. D., 1982, ApJ, 256, 523

\bibitem[\protect\citeauthoryear{Garden et al.}{1991}]{b16} Garden R. P., Hayashi M., 1991, \apj, 374, 540

\bibitem[\protect\citeauthoryear{Gilden}{1984}]{b4} Gilden D. L., 1984, ApJ, 279, 335

\bibitem[\protect\citeauthoryear{Goldreich \& Kwan}{1974}]{b5} Goldreich P., Kwan, J., 1974, ApJ, 189, 441

\bibitem[\protect\citeauthoryear{Habe \& Ohta}{1992}]{b6} Habe A., Ohta K., 1992, PASJ, 44, 203

\bibitem[\protect\citeauthoryear{Harris, Clegg \& Hughes}{1988}]{b18} Harris S., Clegg P., Hughes J., 1988, MNRAS, 235,
441


\bibitem[\protect\citeauthoryear{Hausman}{1981}]{b7} Hausman M. A., 1981, ApJ, 245, 72

\bibitem[\protect\citeauthoryear{Kerton et al.}{2003}]{b23} Kerton C. R., Brunt C. M., 2003, A\&A, 399, 1083

\bibitem[\protect\citeauthoryear{Lattanzio et al.}{1985}]{b8} Lattanzio J. C., Monaghan J. J., Pongracic H., Schwarz M. P., 1985, MNRAS, 215, 125

\bibitem[\protect\citeauthoryear{Prusti et~al.}{1992}]{b19} Prusti T., Adorf H.-M., Meurs E. J. A., 1992, A\&A, 261, 685

\bibitem[\protect\citeauthoryear{Vallee}{1995a}]{b9} Vallee J. P., 1995, AJ, 109, 1724


\bibitem[\protect\citeauthoryear{Vallee}{1995b}]{b10} Vallee J. P., 1995, AJ, 110, 2256


\bibitem[\protect\citeauthoryear{Walsh et al.}{2002}]{b11} Walsh W., Beck R., et al., 2002, A\&A, 388, 7


\bibitem[\protect\citeauthoryear{Wang et al.}{2004}]{b12} Wang J. J., Chen W. P., Miller M., Qin S. L., Wu Y. F., 2004, ApJ, 614L, 105

\bibitem[\protect\citeauthoryear{Wouterloot et al.}{1989}]{b17} Wouterloot J. G. A., Brand J., 1989, A\&AS, 80, 149

\bibitem[\protect\citeauthoryear{Xin \& Wang}{2008}]{b13}  Xin B., Wang J. J., 2008, ChJAA, 8, 433

\bibitem[\protect\citeauthoryear{Guan et al.}{2008}]{b22} Guan X., Wu Y. F., Ju B. G., 2008, MNRAS, 391, 869

\bibitem[\protect\citeauthoryear{Yang et al.}{2002}]{b14} Yang J., Jiang Z. B., Wang M., Ju B. G., Wang H. C., 2002, ApJS, 141, 157


\bibitem[\protect\citeauthoryear{Zhou}{1993}]{b15} Zhou S. D., Evans N. J., Koempe C., Walmsley C. M., 1993, ApJ, 404, 232



\end{thebibliography}
\end{document}


\begin{figure}

\includegraphics[height=70mm,  angle=-90, clip, viewport=150 10 500 750]{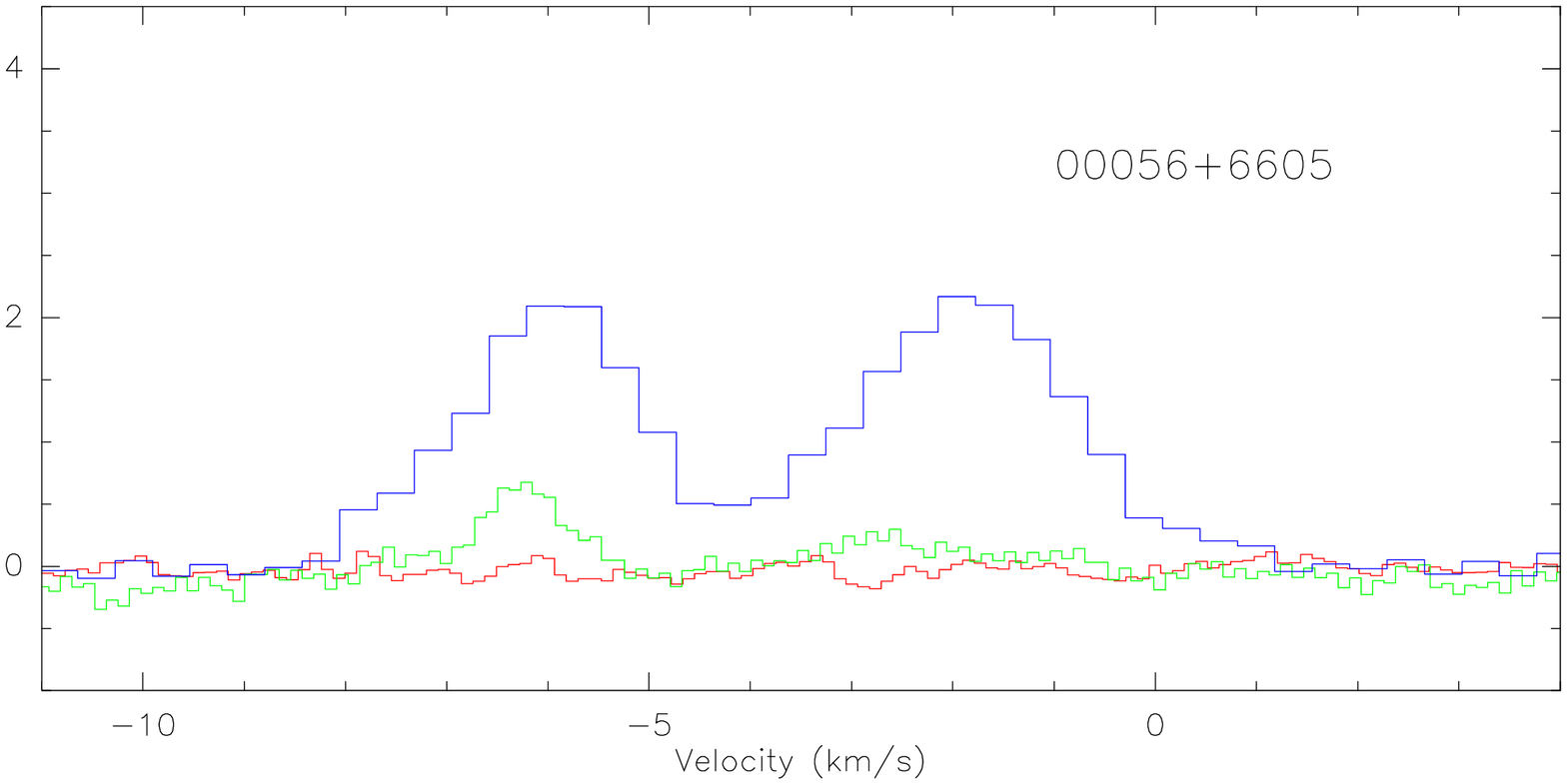}
\includegraphics[height=70mm,  angle=-90, clip, viewport=150 10 500 750]{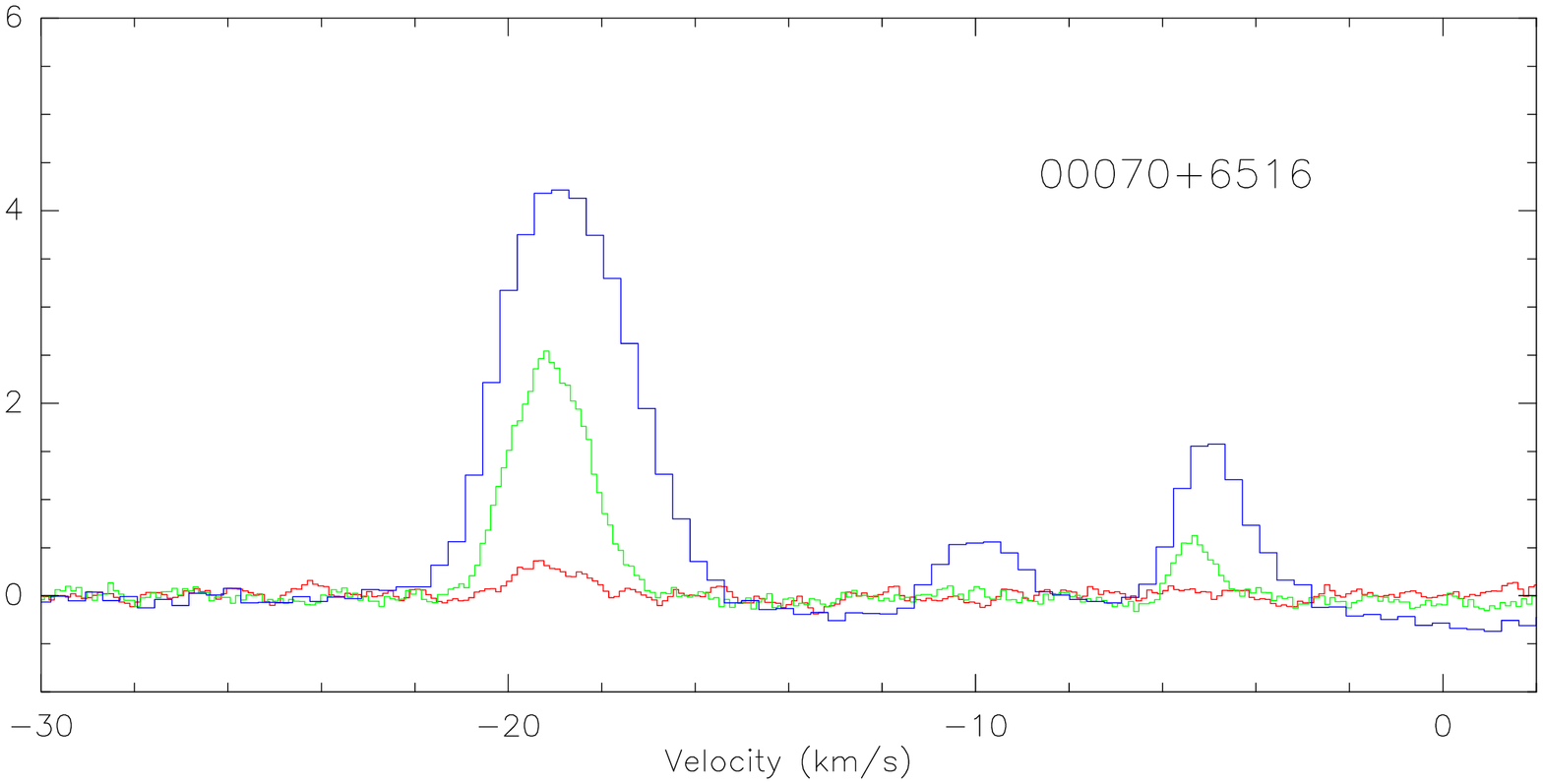}
\includegraphics[height=70mm,  angle=-90, clip, viewport=150 10 500 750]{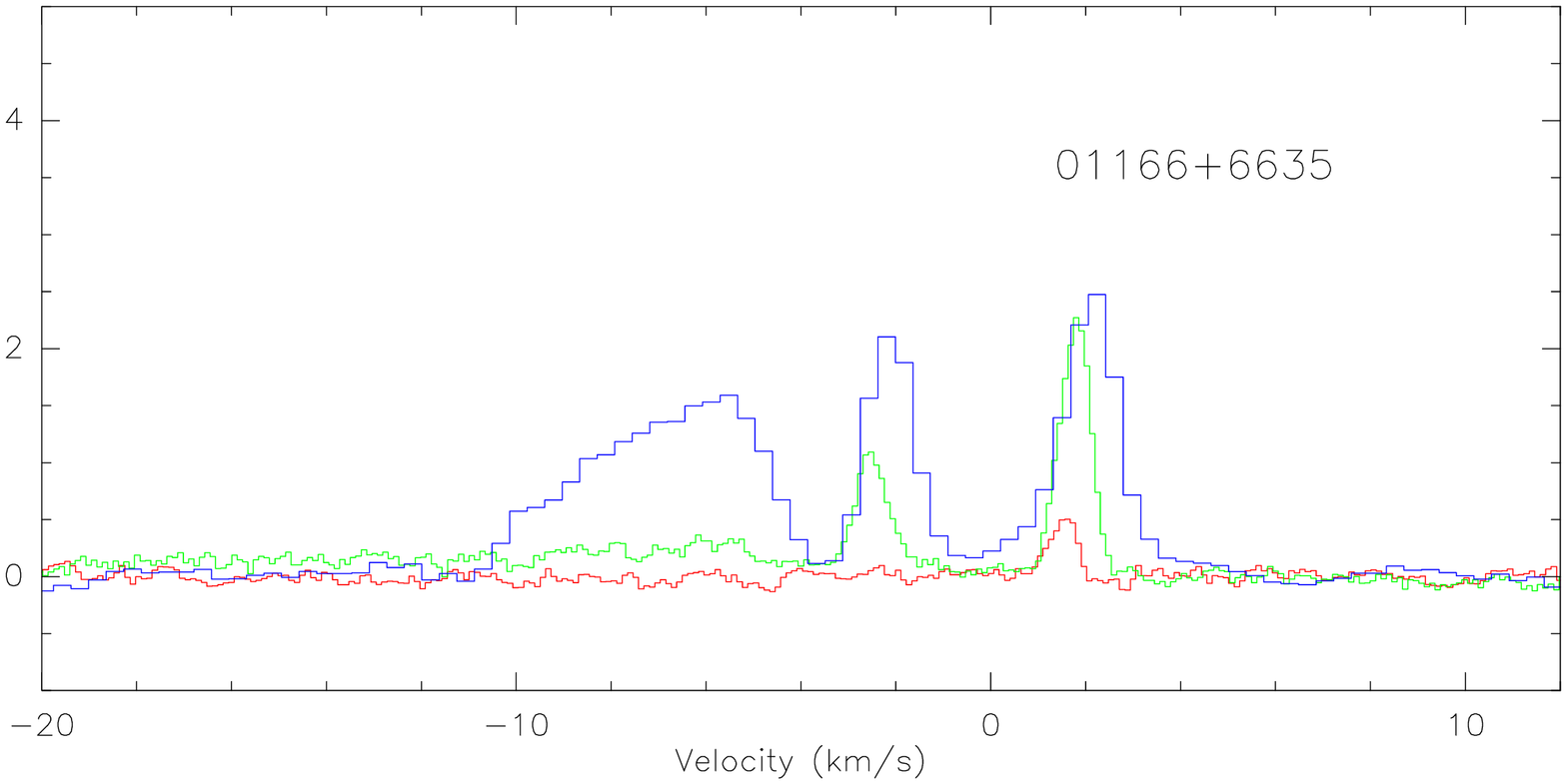}
\includegraphics[height=70mm,  angle=-90, clip, viewport=150 10 500 750]{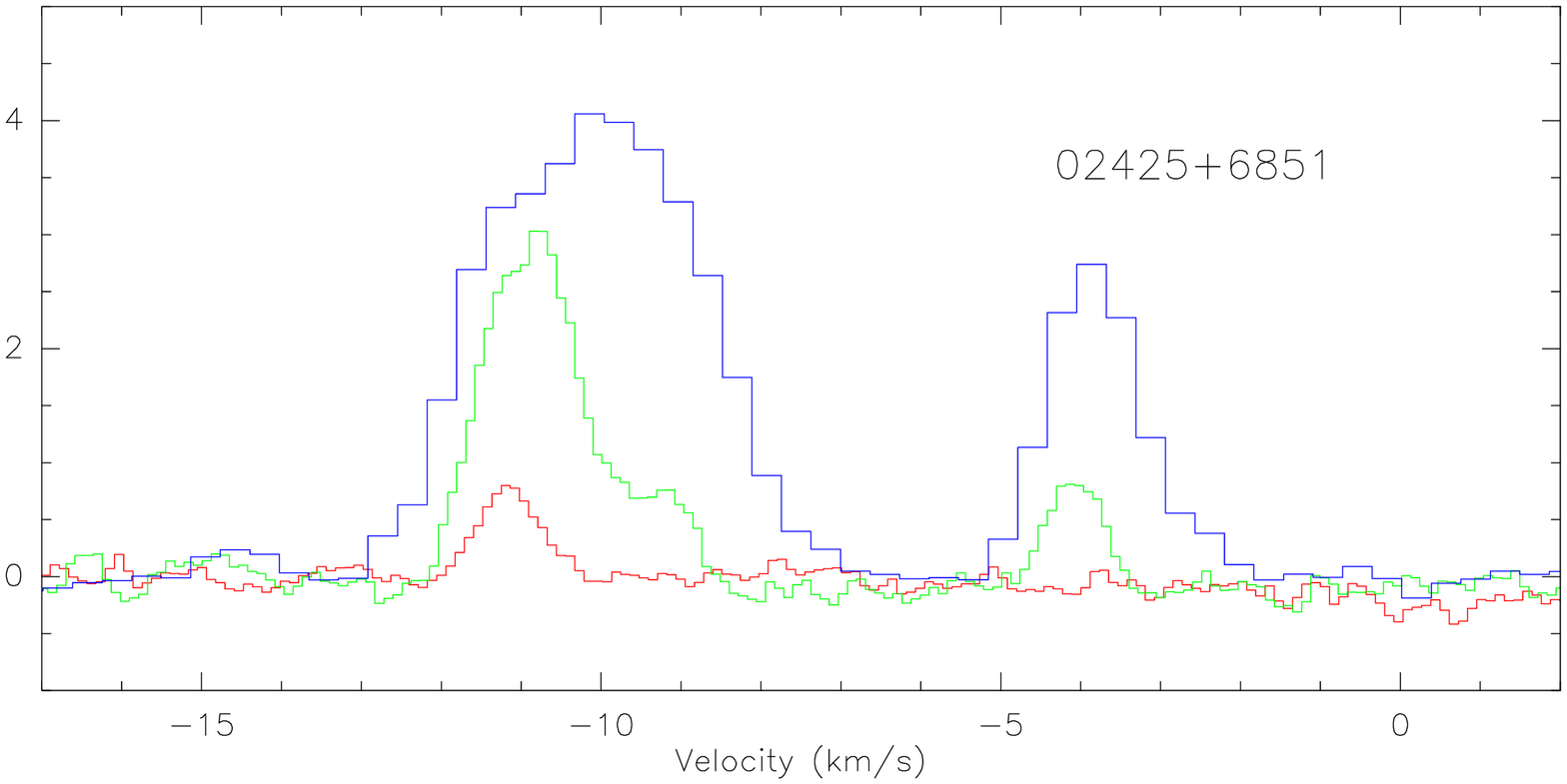}
\includegraphics[height=70mm,  angle=-90, clip, viewport=150 10 500 750]{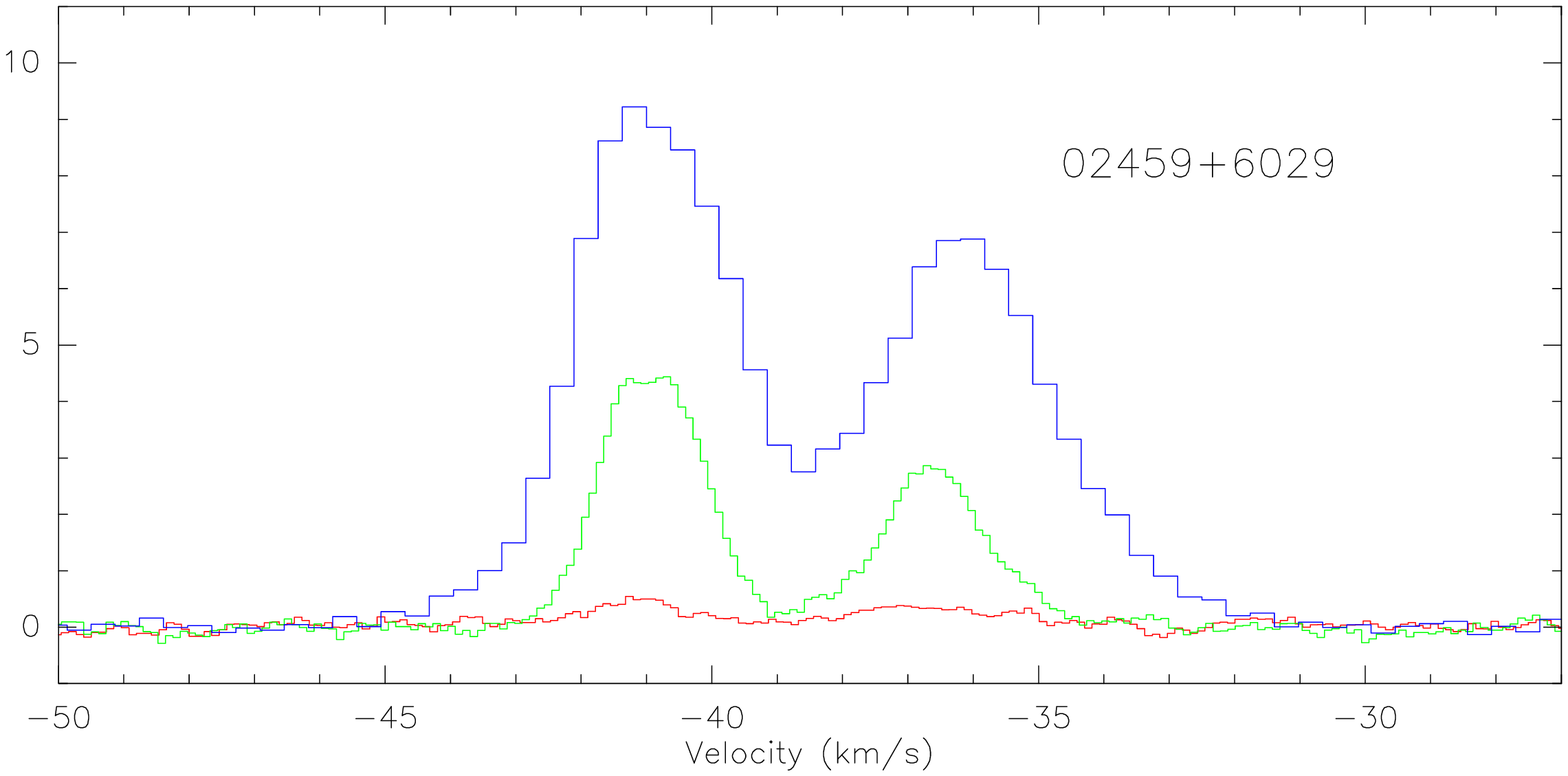}
\includegraphics[height=70mm,  angle=-90, clip, viewport=150 10 500 750]{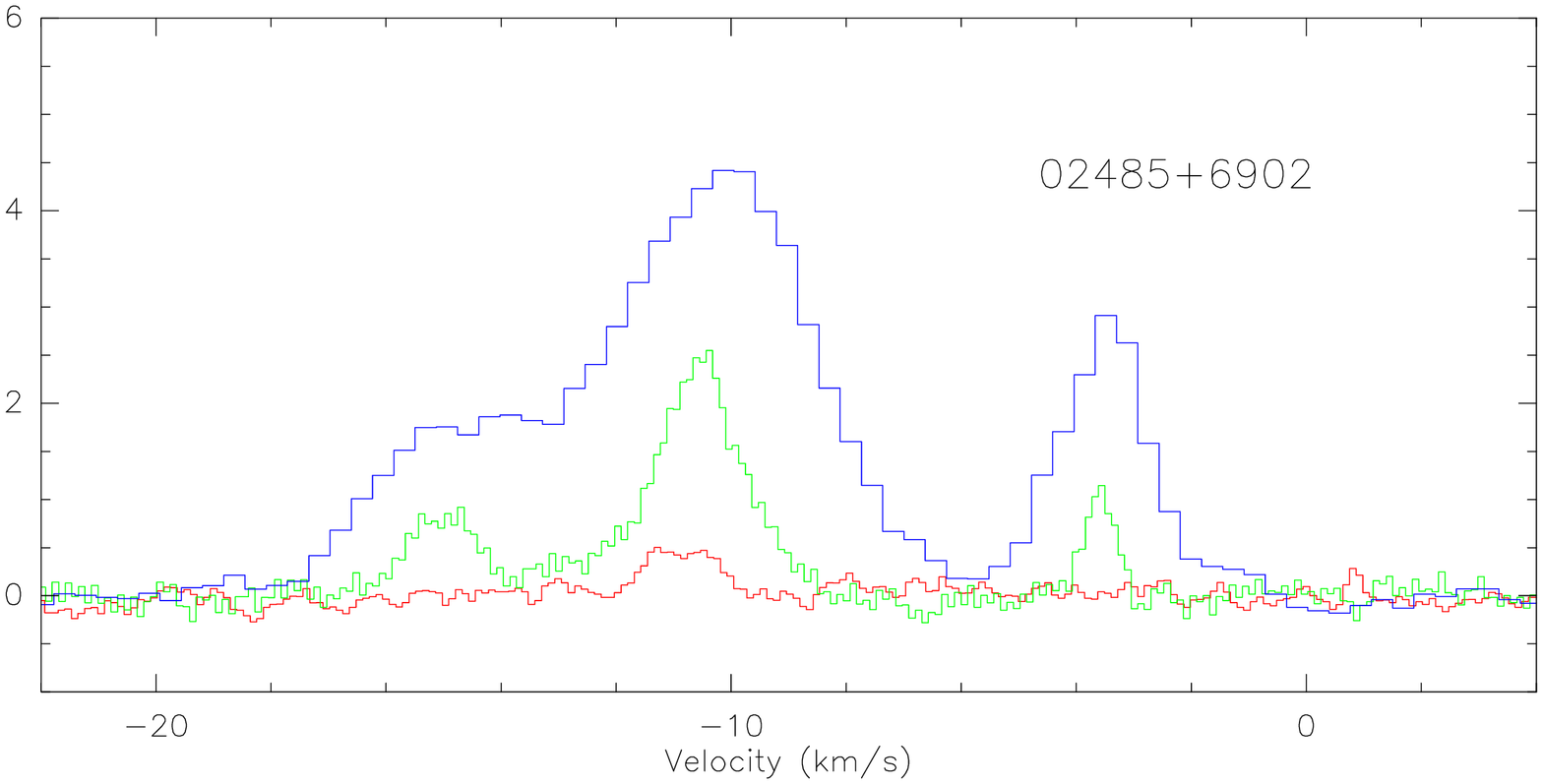}
\includegraphics[height=70mm,  angle=-90, clip, viewport=150 10 500 750]{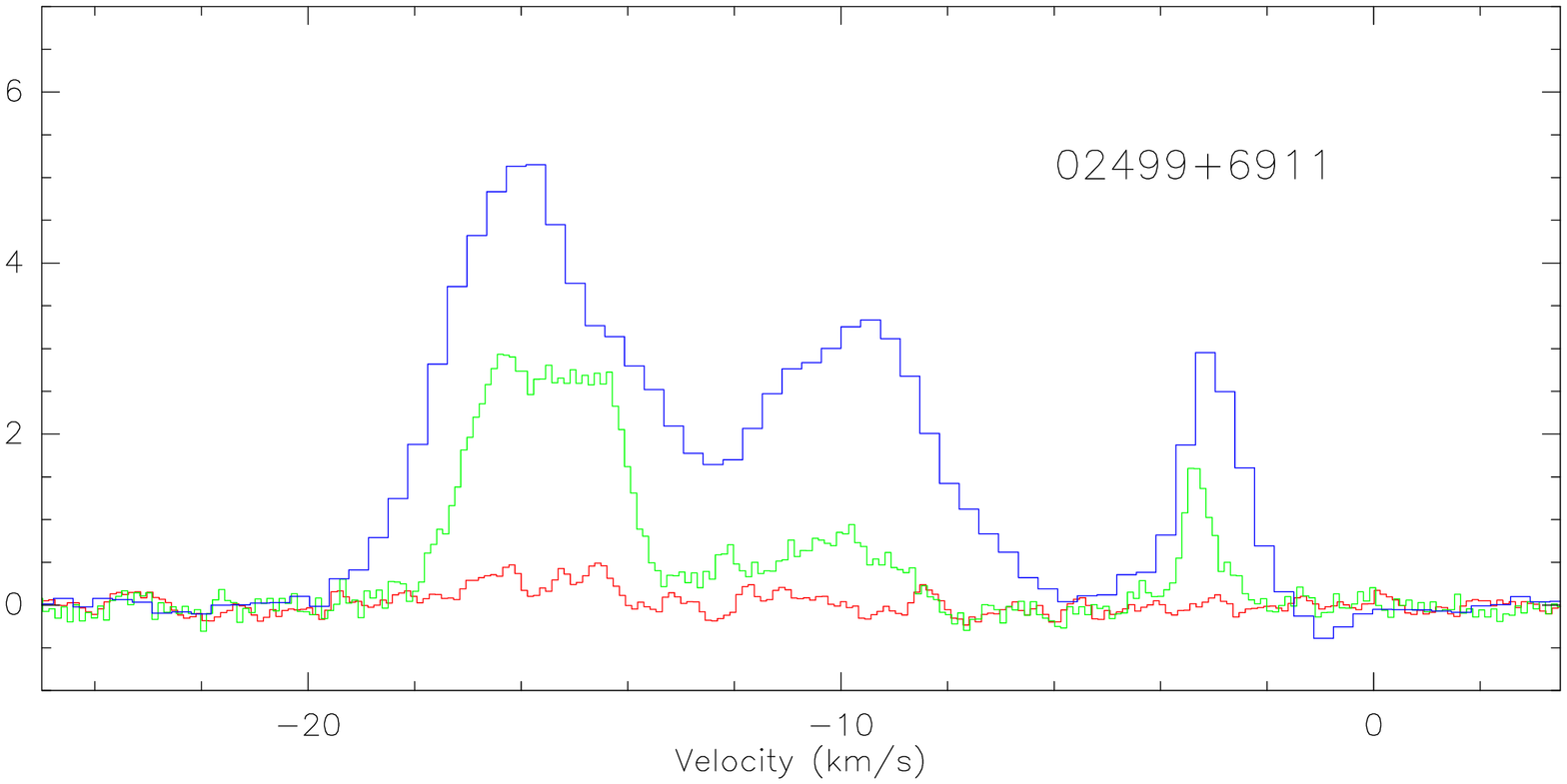}
\includegraphics[height=70mm,  angle=-90, clip, viewport=150 10 500 750]{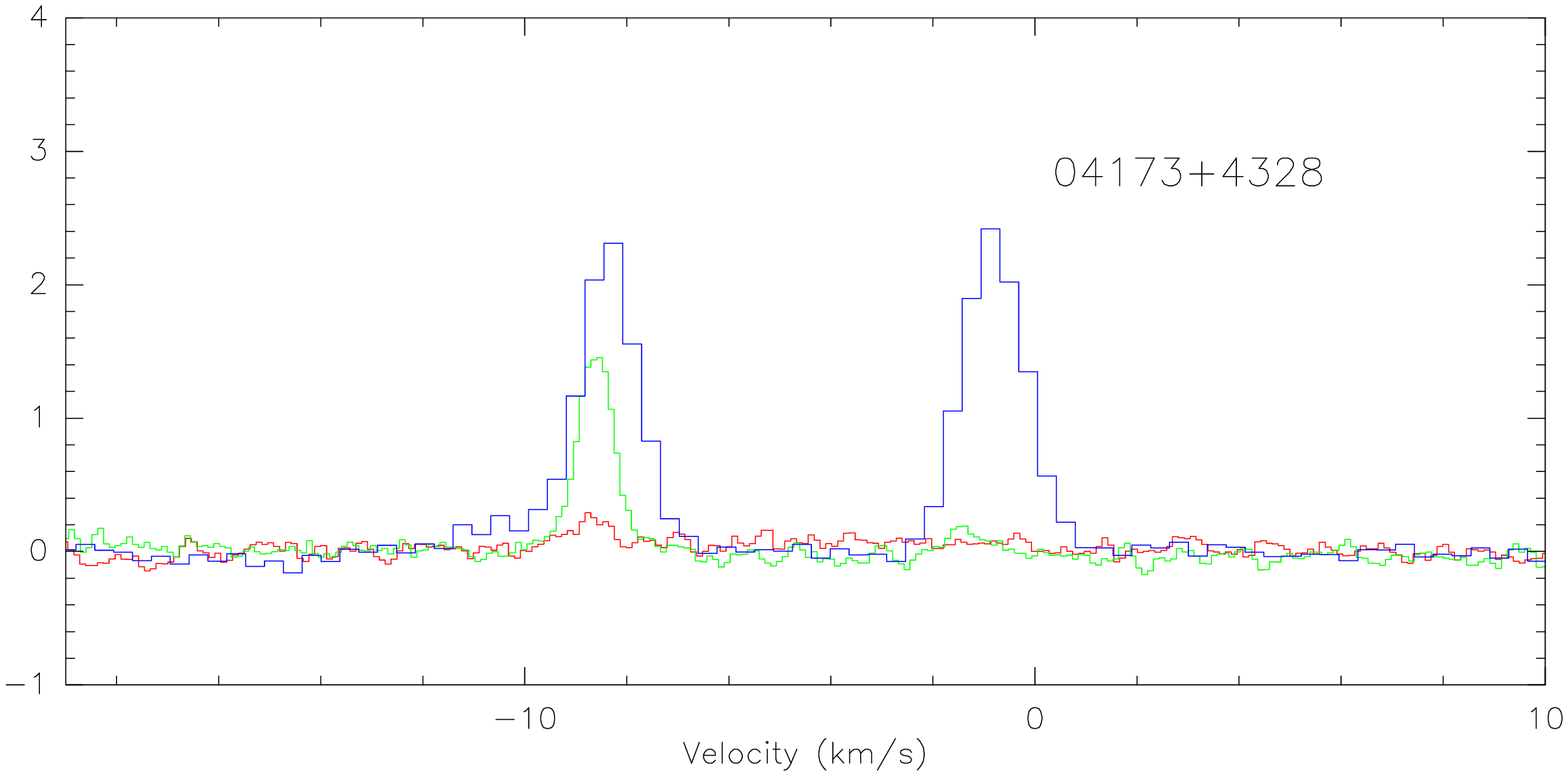}
\includegraphics[height=70mm,  angle=-90, clip, viewport=150 10 500 750]{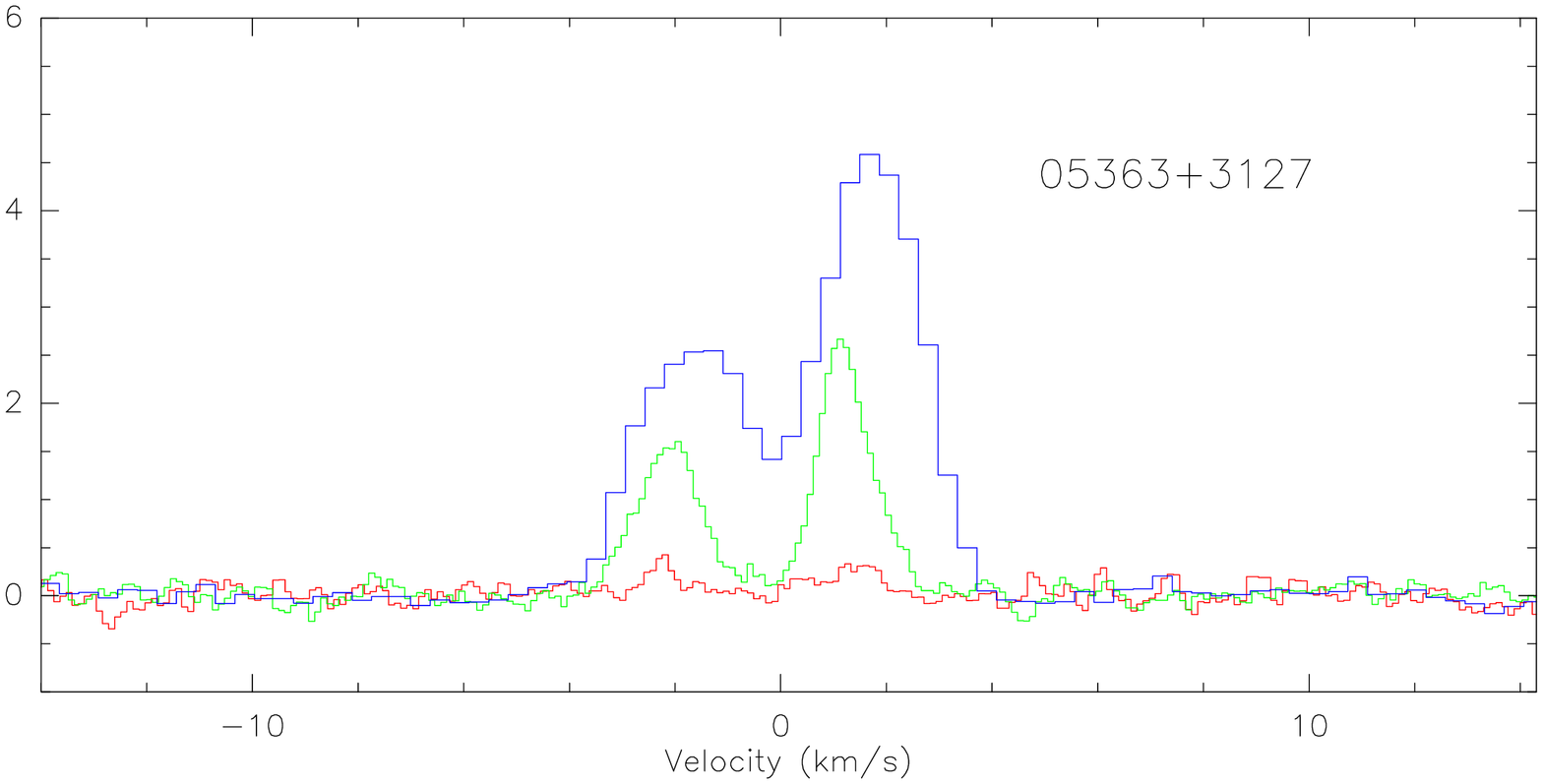}
\includegraphics[height=70mm,  angle=-90, clip, viewport=150 10 500 750]{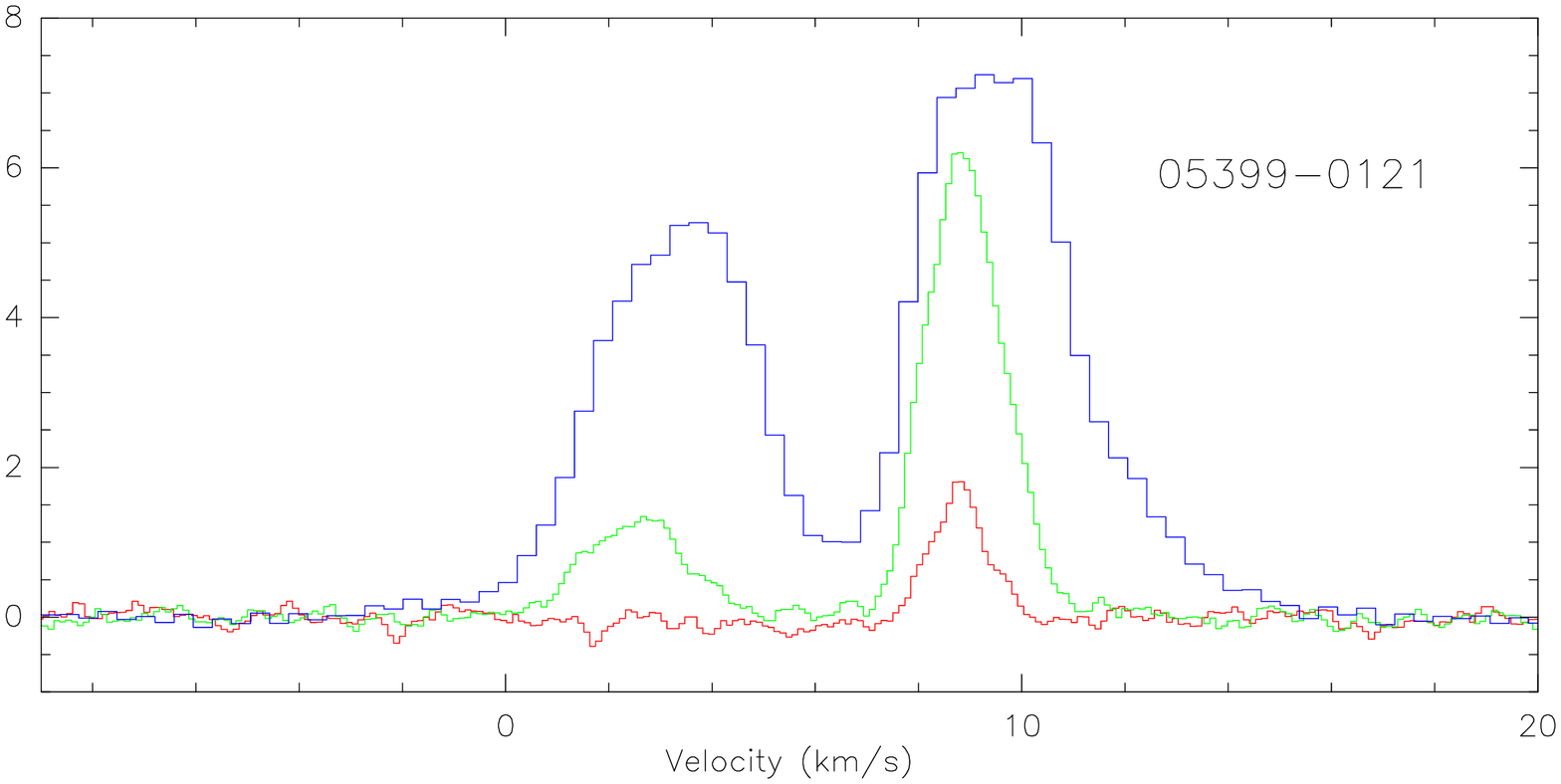}
\includegraphics[height=70mm,  angle=-90, clip, viewport=150 10 500 750]{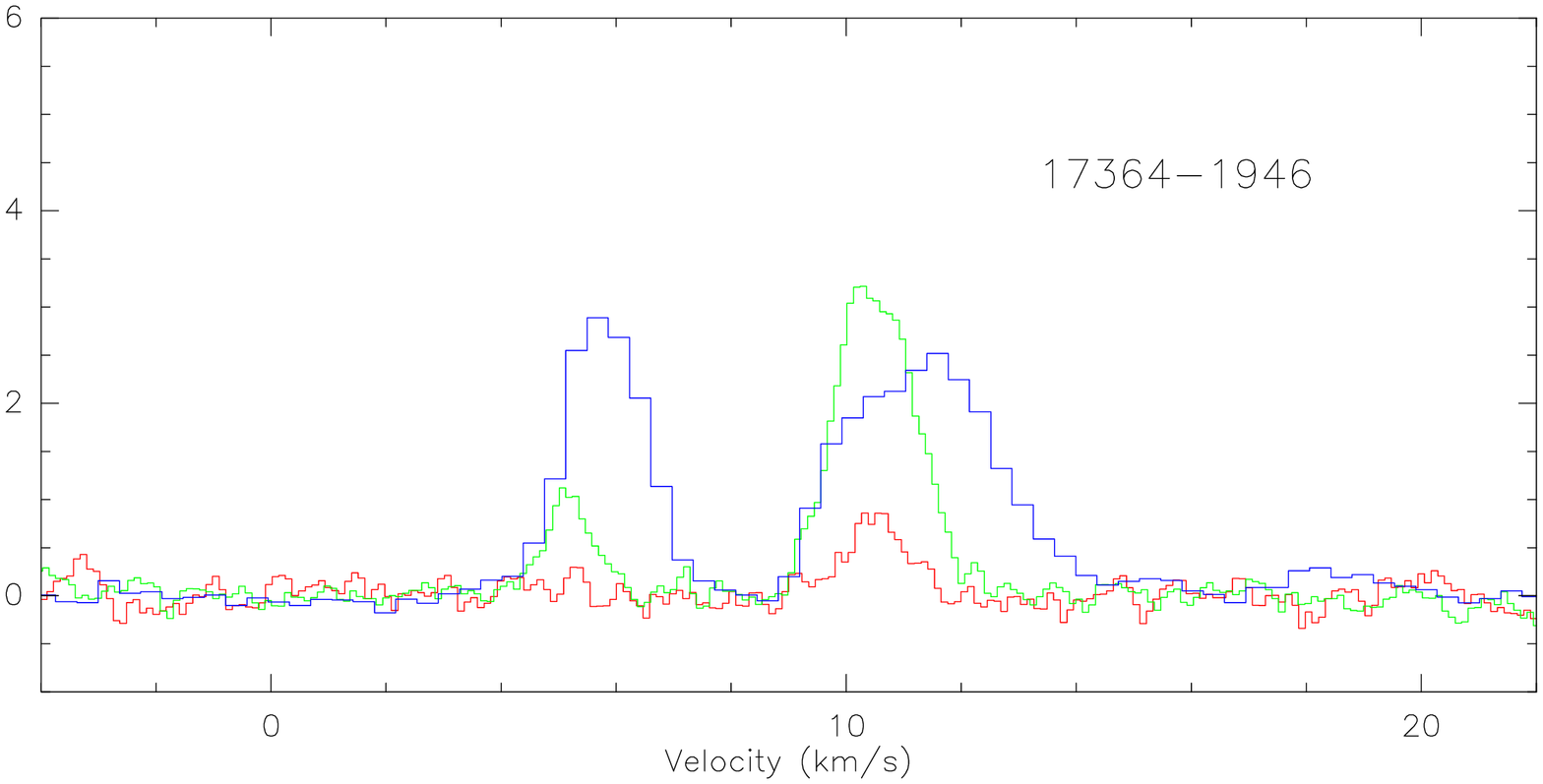}
\includegraphics[height=70mm,  angle=-90, clip, viewport=150 10 500 750]{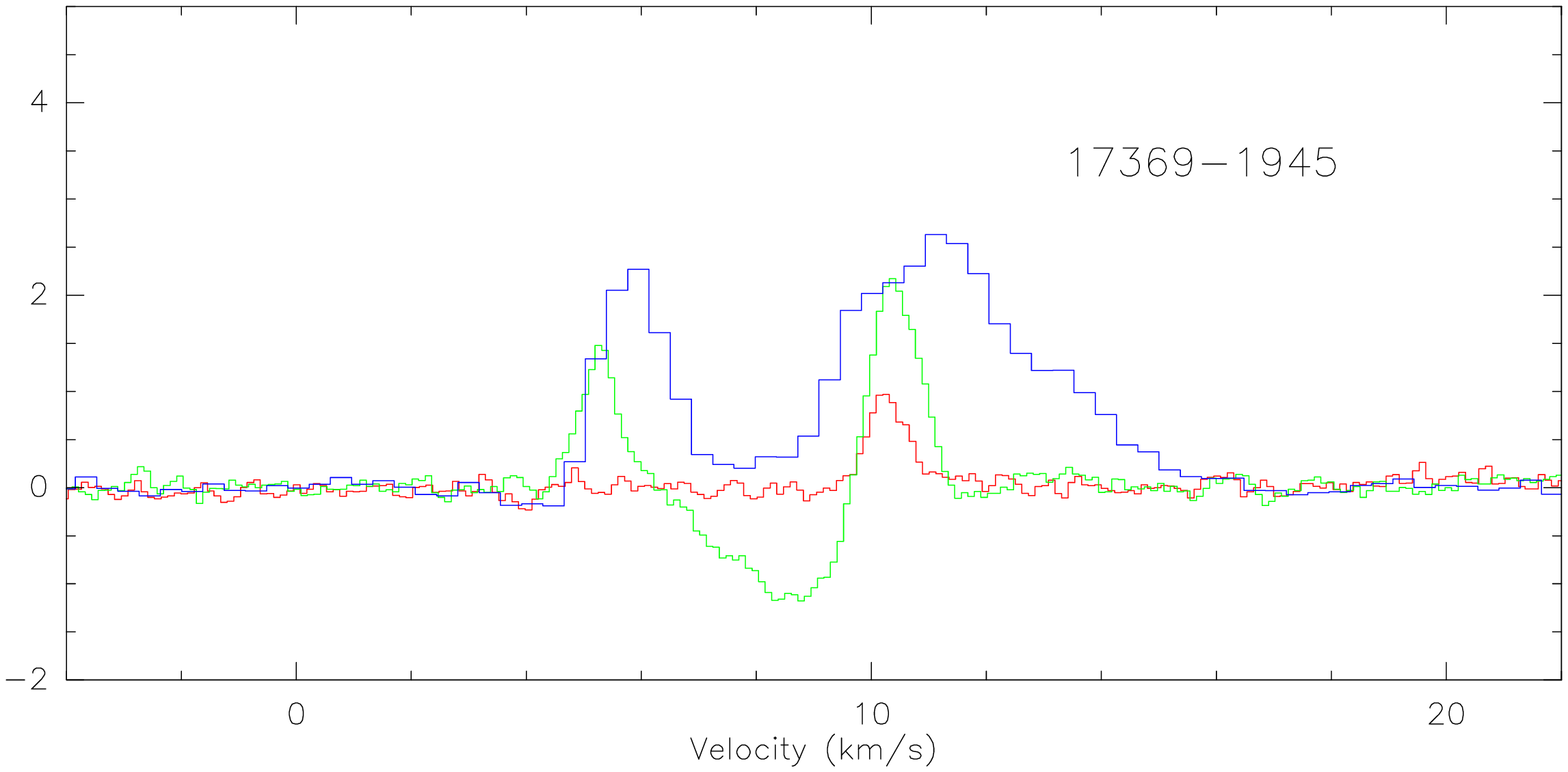}
\includegraphics[height=70mm,  angle=-90, clip, viewport=150 10 500 750]{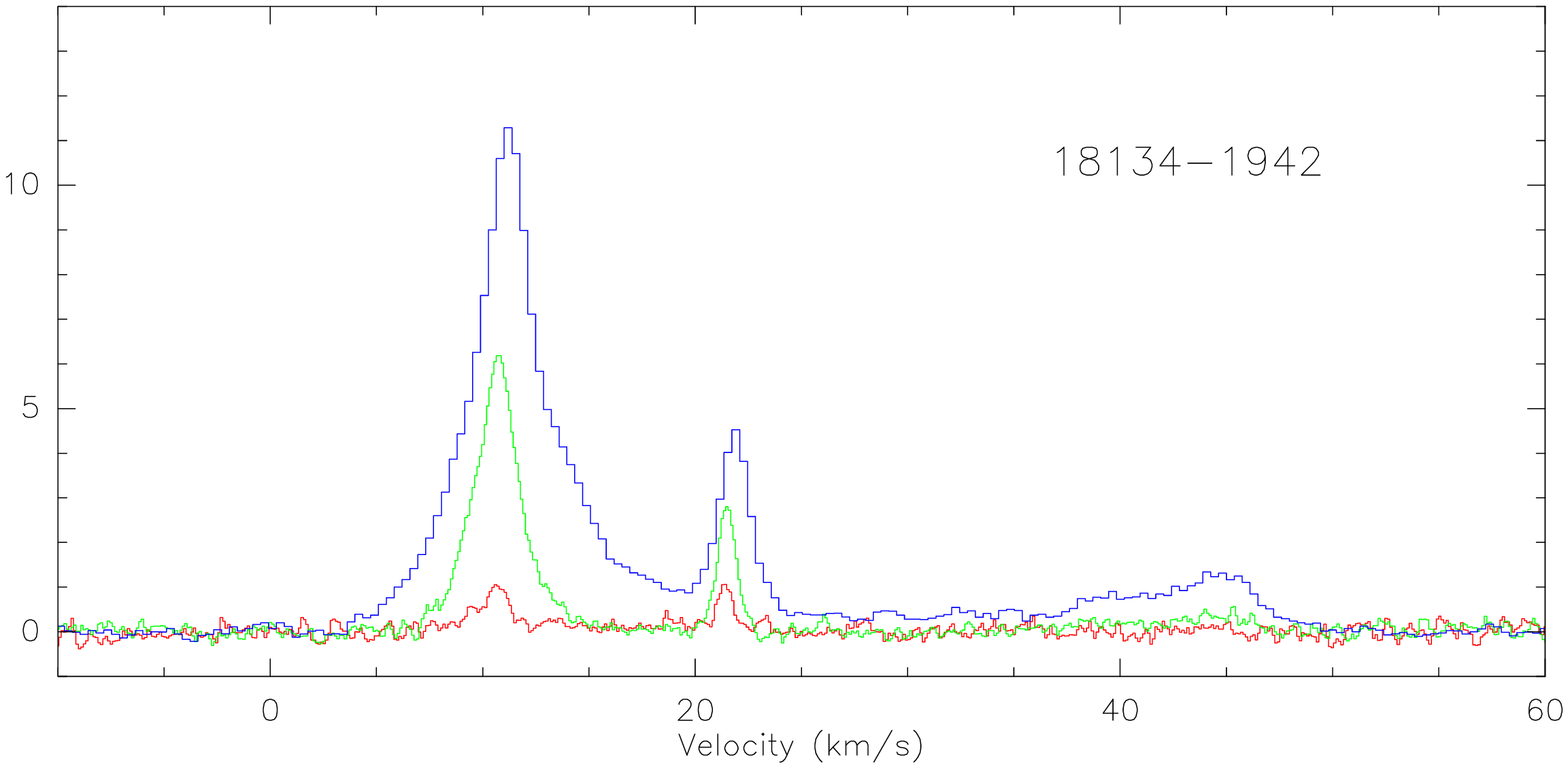}
\includegraphics[height=70mm,  angle=-90, clip, viewport=150 10 500 750]{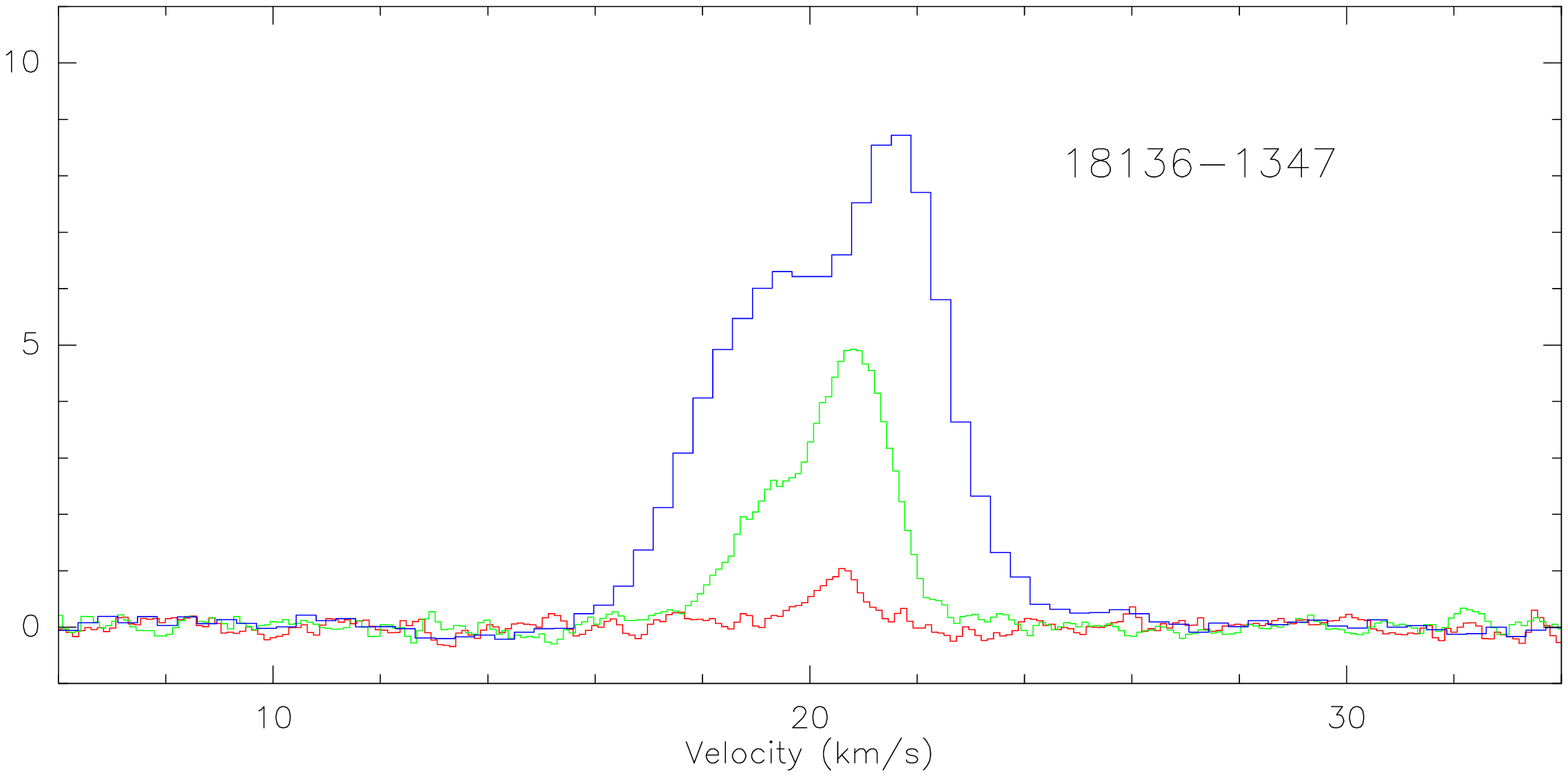}
\centering
\begin{minipage}[]{60mm}
   \caption{The sources of type 1
  }\end{minipage}
   \label{Fig6}
   \end{figure}

\addtocounter{figure}{-1}
\begin{figure}

\includegraphics[height=70mm,  angle=-90, clip, viewport=150 10 500 750]{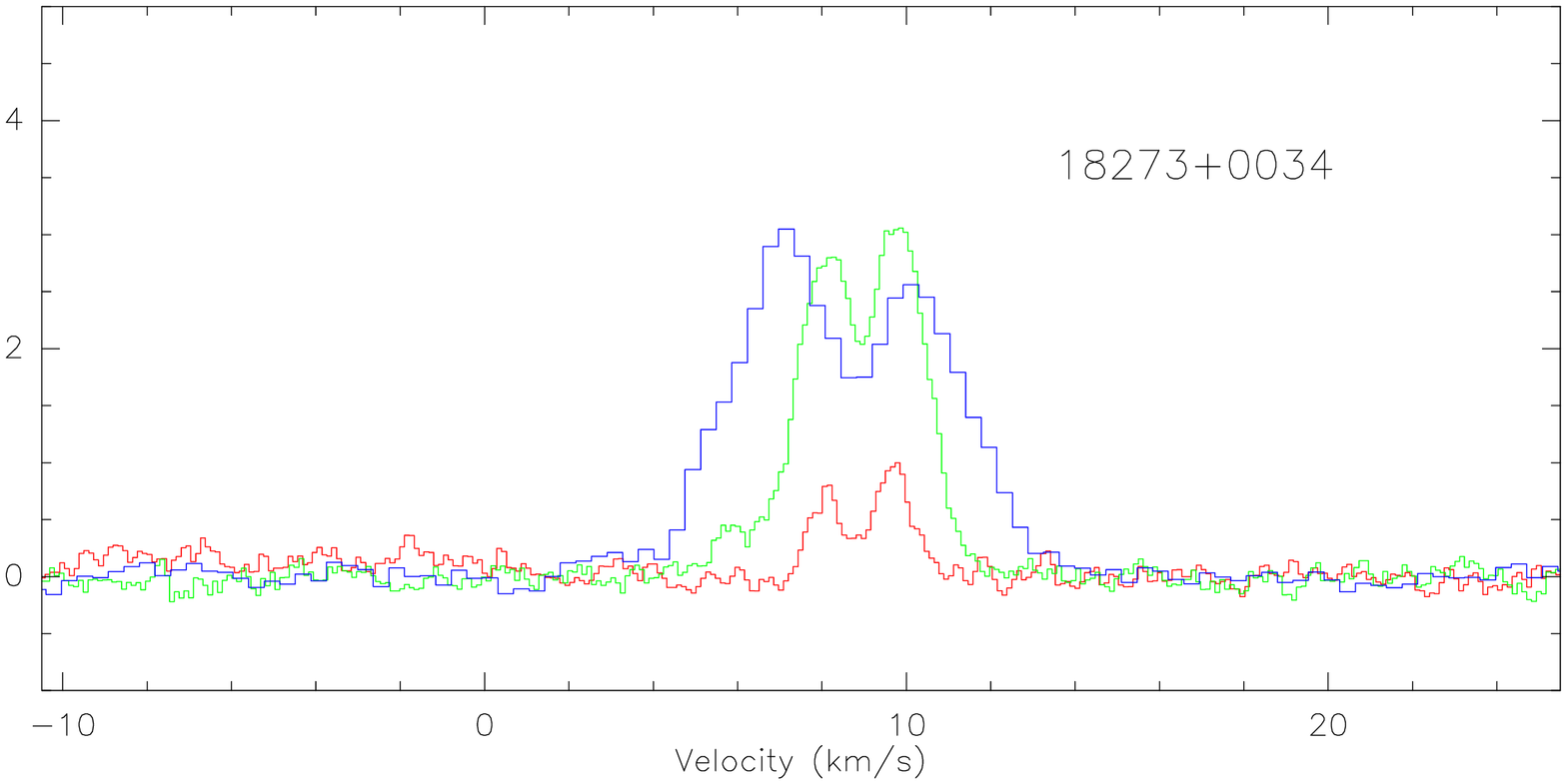}
\includegraphics[height=70mm,  angle=-90, clip, viewport=150 10 500 750]{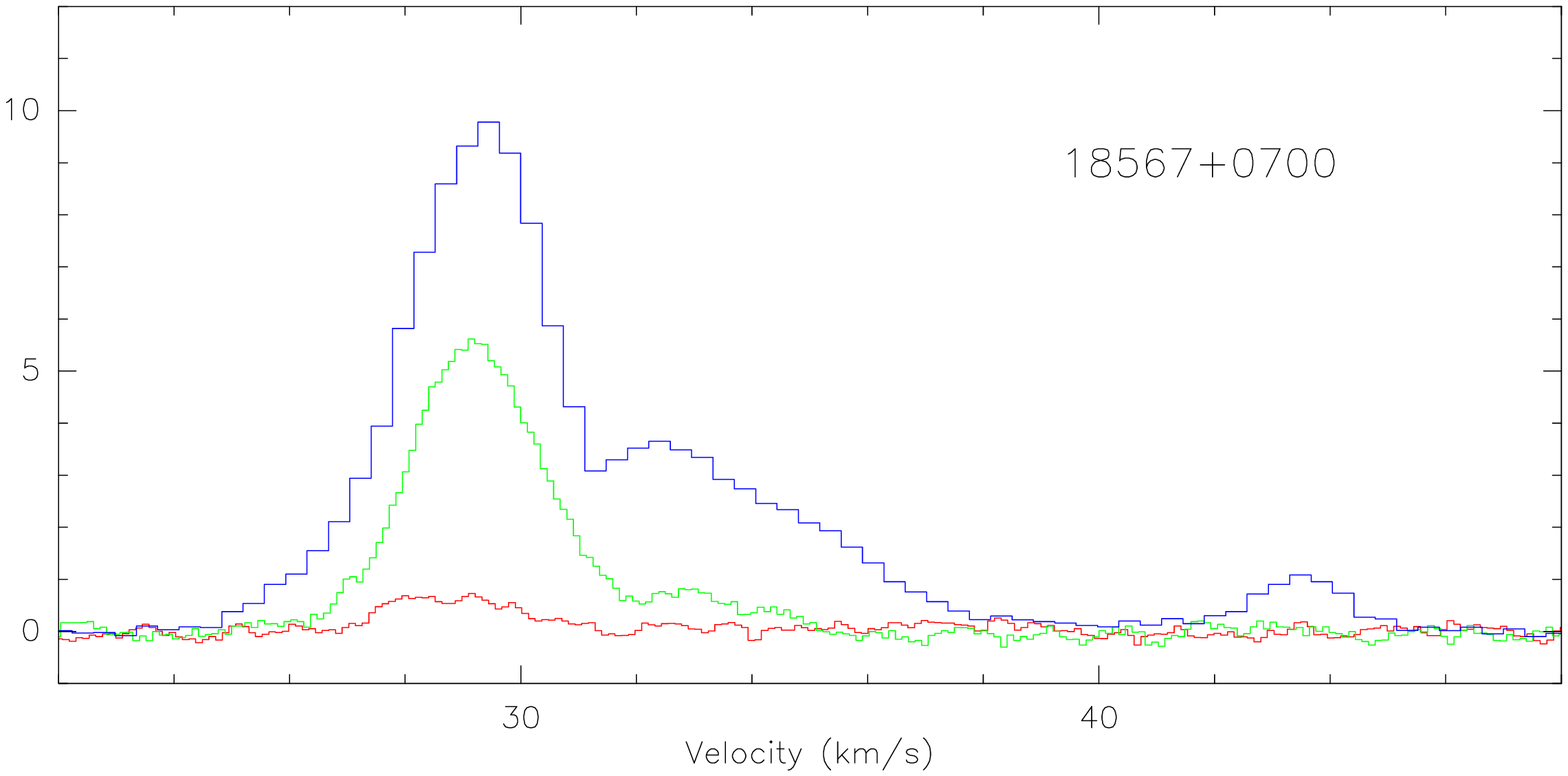}
\includegraphics[height=70mm,  angle=-90, clip, viewport=150 10 500 750]{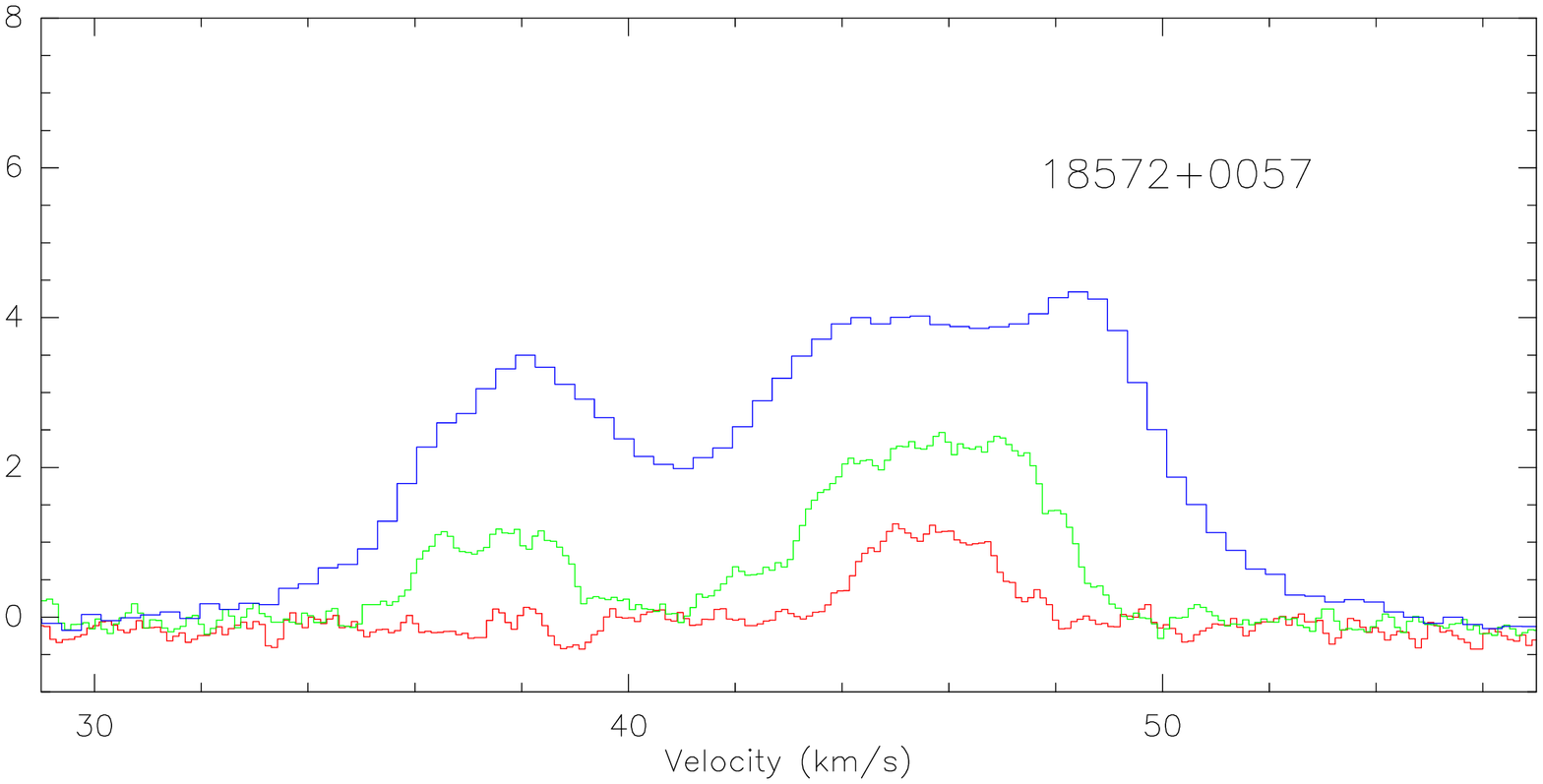}
\includegraphics[height=70mm,  angle=-90, clip, viewport=150 10 500 750]{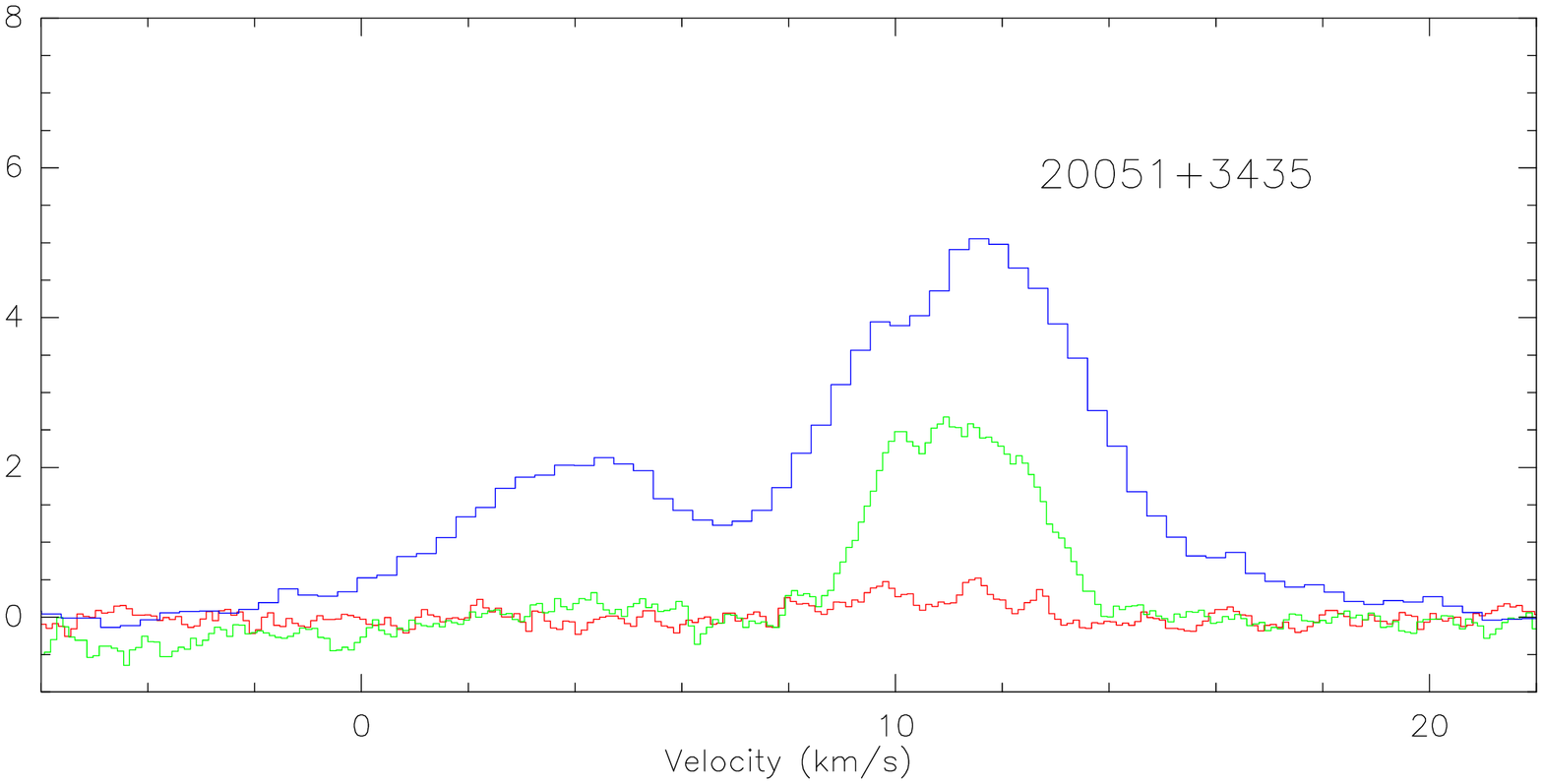}
\includegraphics[height=70mm,  angle=-90, clip, viewport=150 10 500 750]{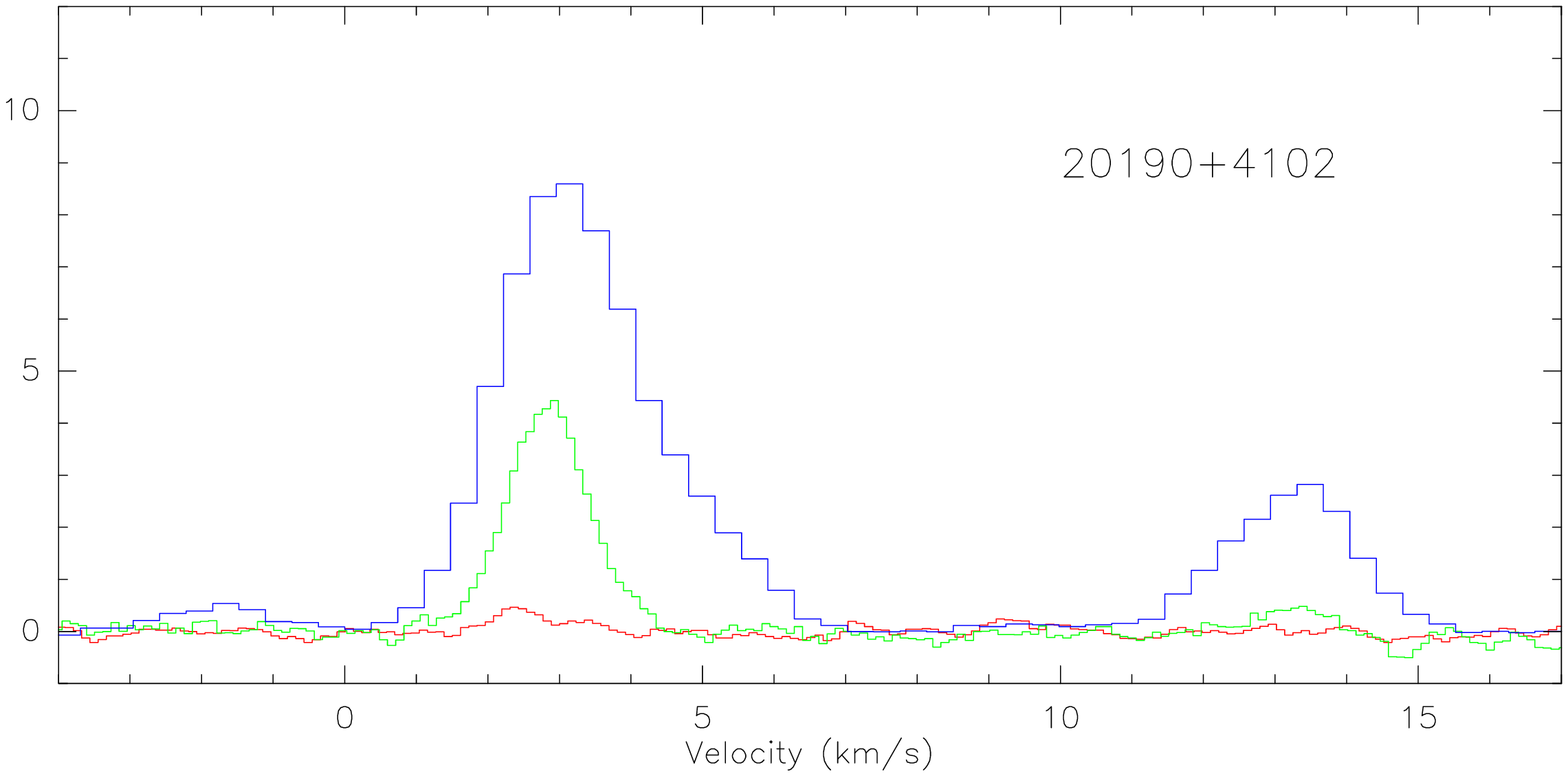}
\includegraphics[height=70mm,  angle=-90, clip, viewport=150 10 500 750]{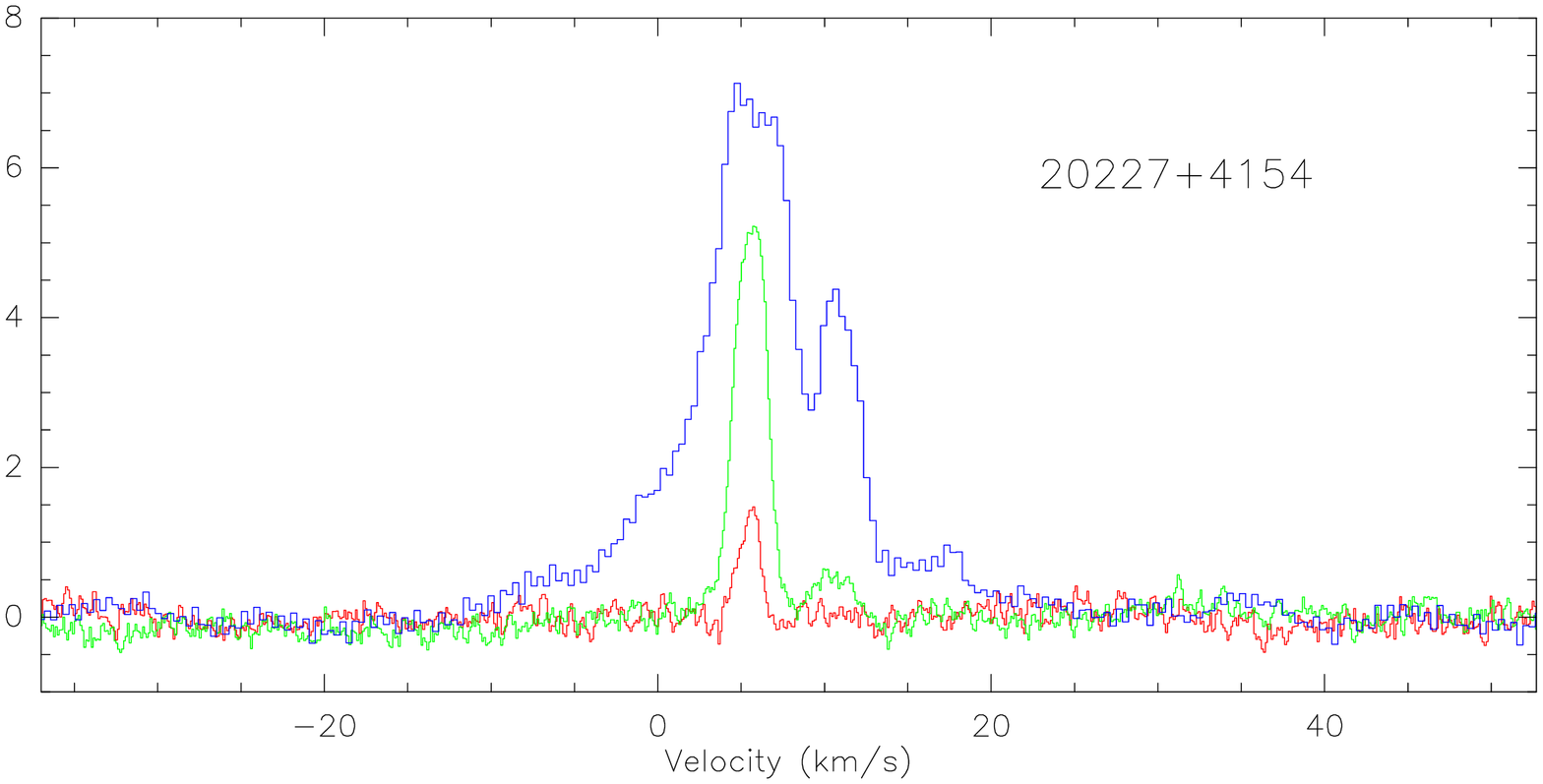}
\includegraphics[height=70mm,  angle=-90, clip, viewport=150 10 500 750]{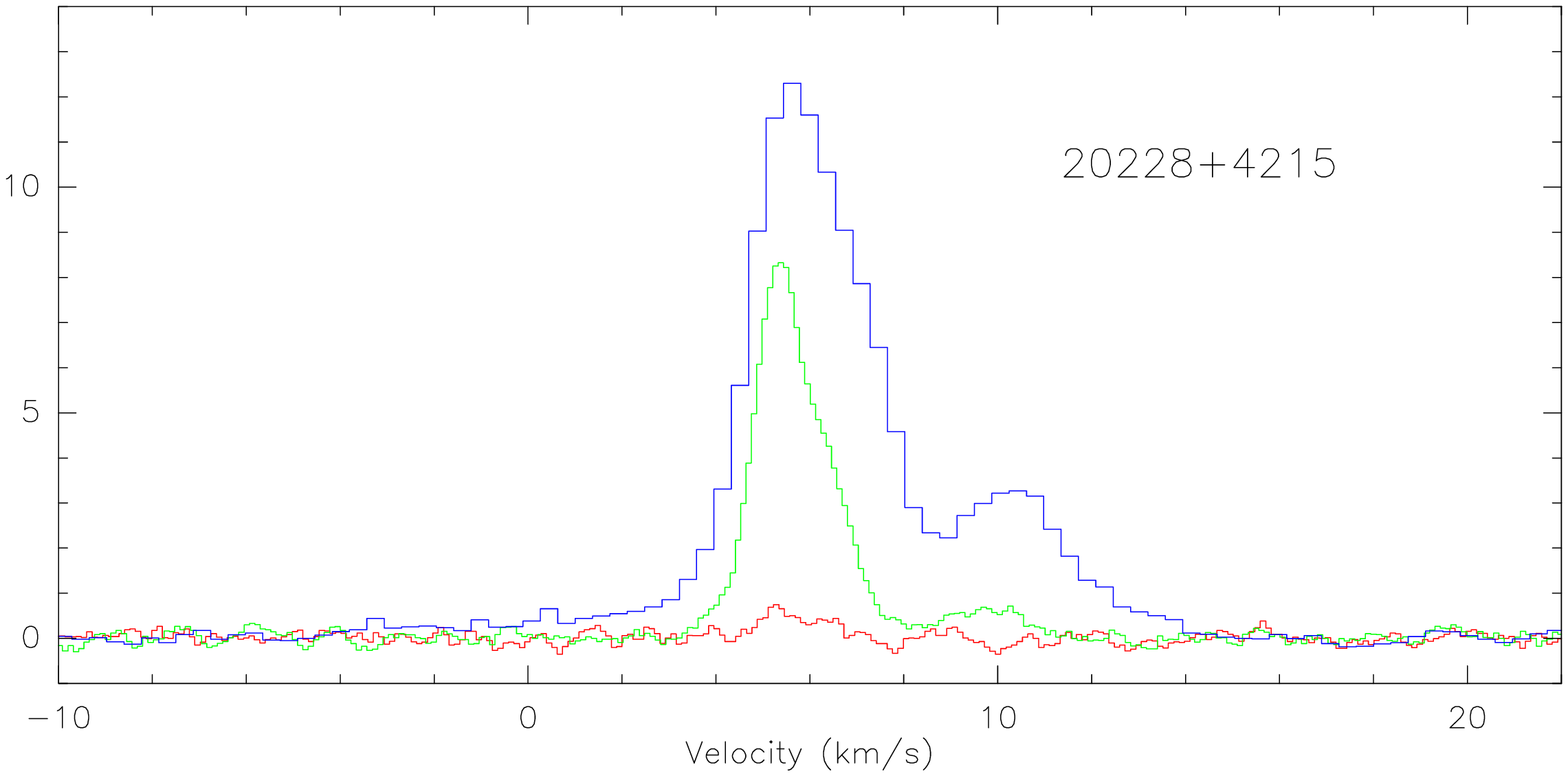}
\includegraphics[height=70mm,  angle=-90, clip, viewport=150 10 500 750]{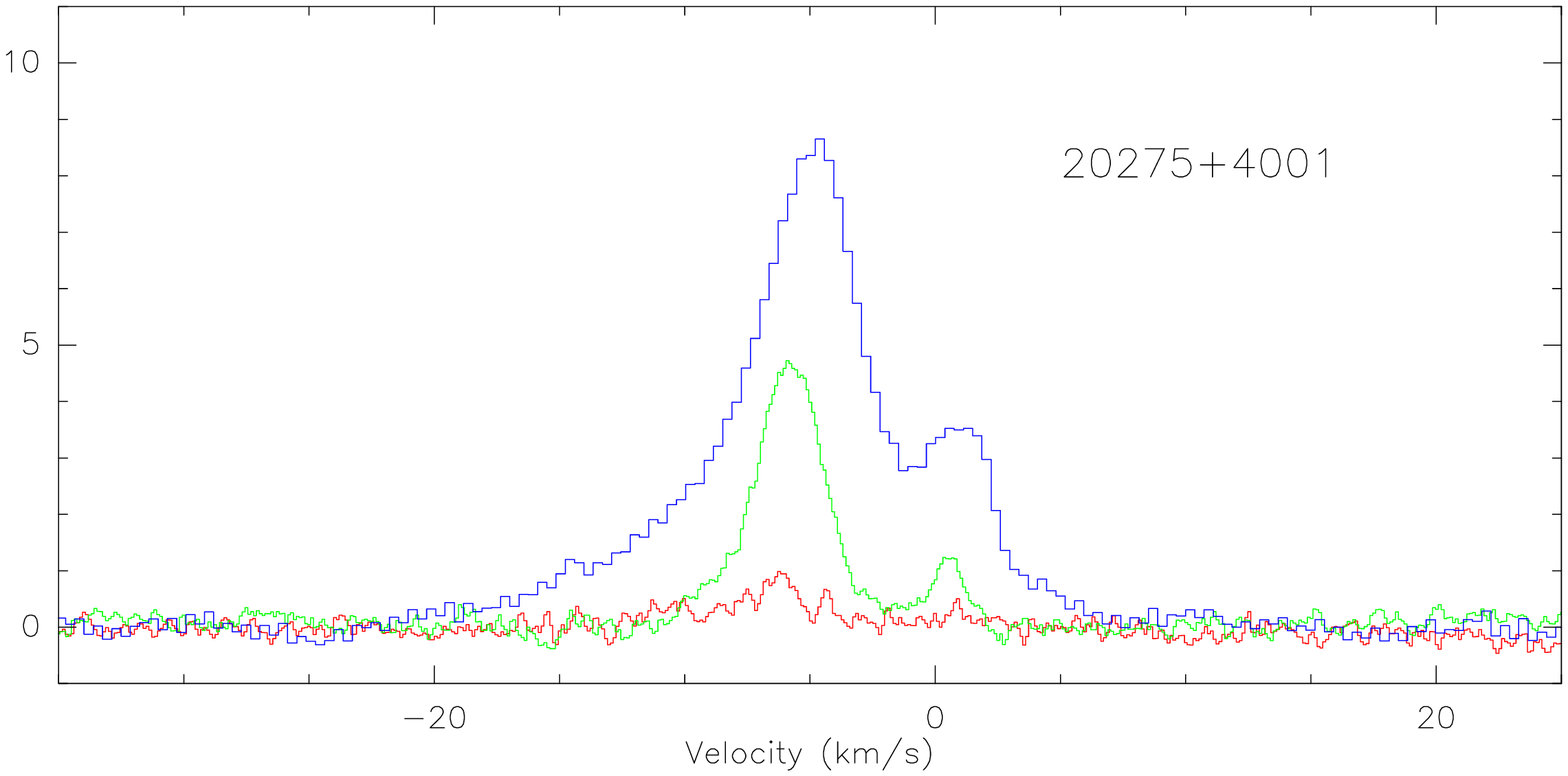}
\includegraphics[height=70mm,  angle=-90, clip, viewport=150 10 500 750]{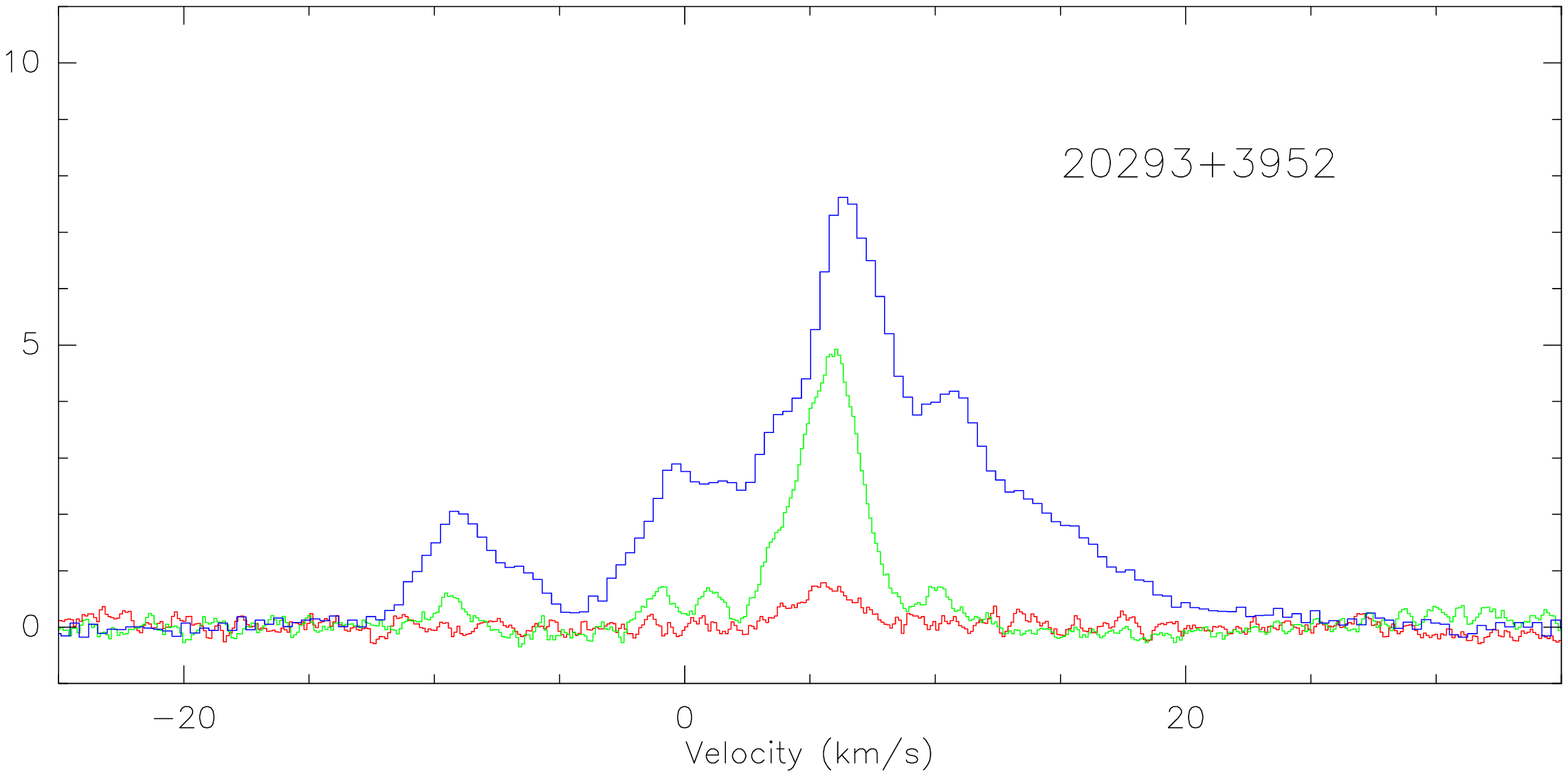}
\includegraphics[height=70mm,  angle=-90, clip, viewport=150 10 500 750]{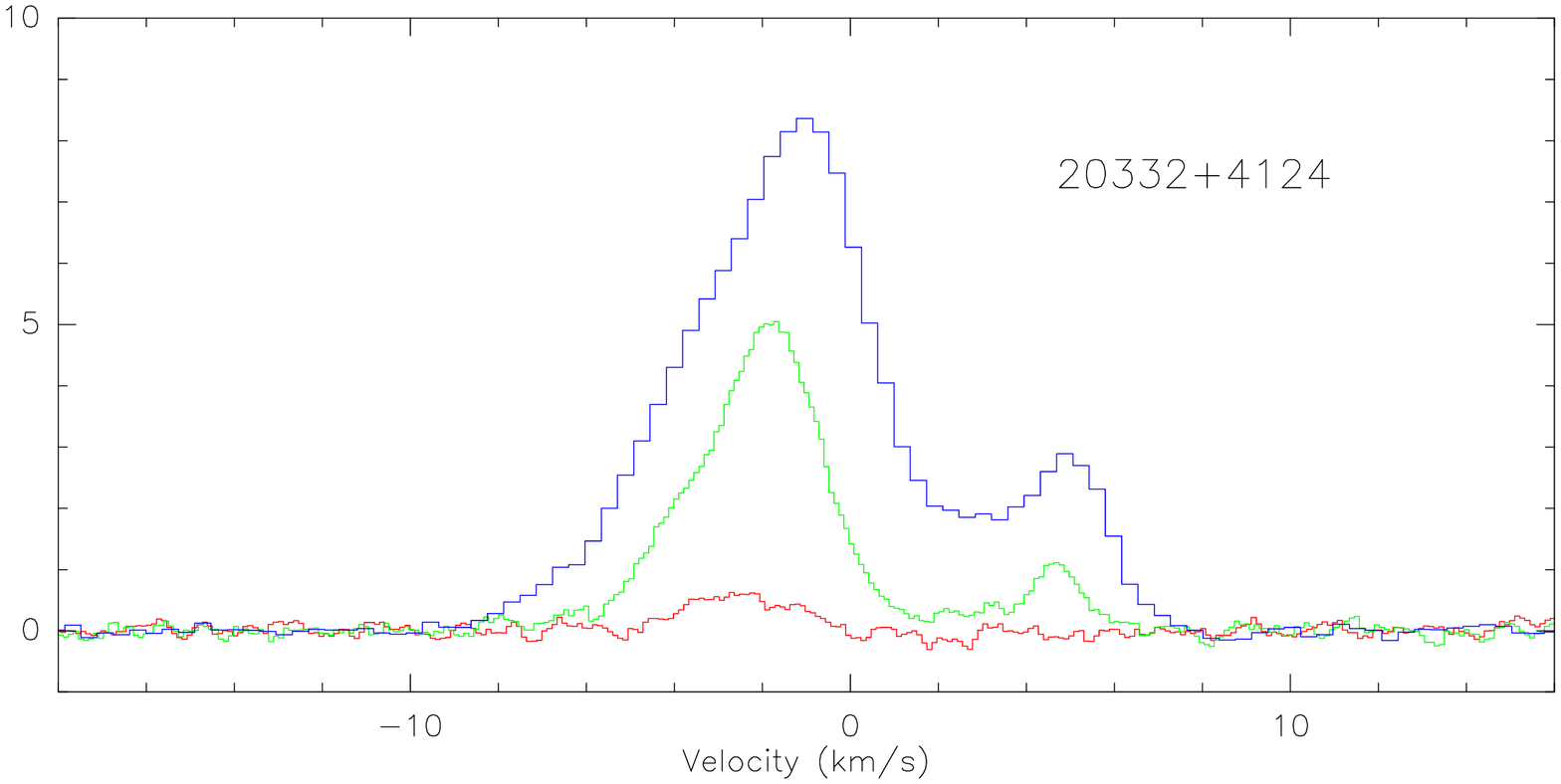}
\includegraphics[height=70mm,  angle=-90, clip, viewport=150 10 500 750]{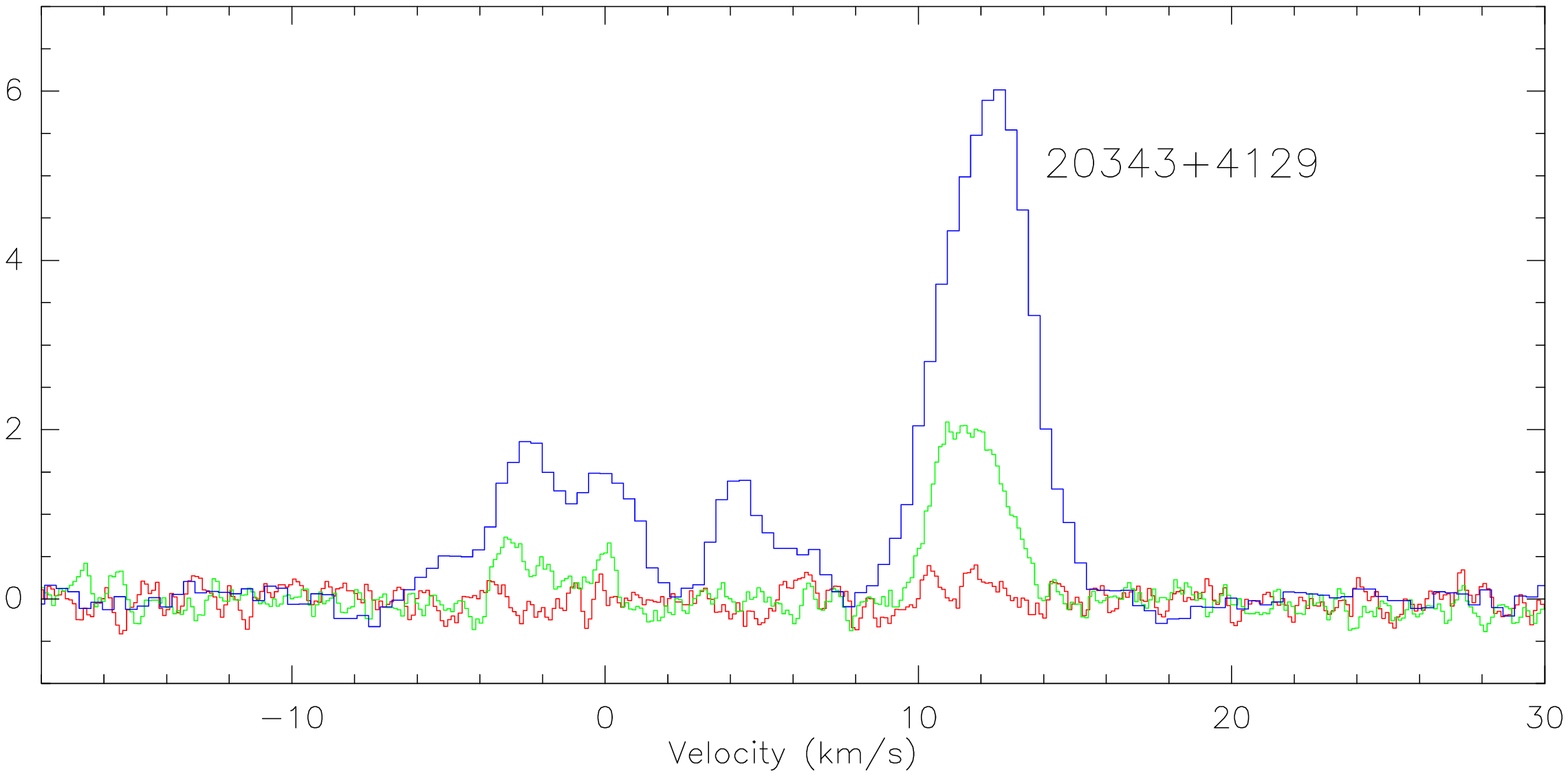}
\includegraphics[height=70mm,  angle=-90, clip, viewport=150 10 500 750]{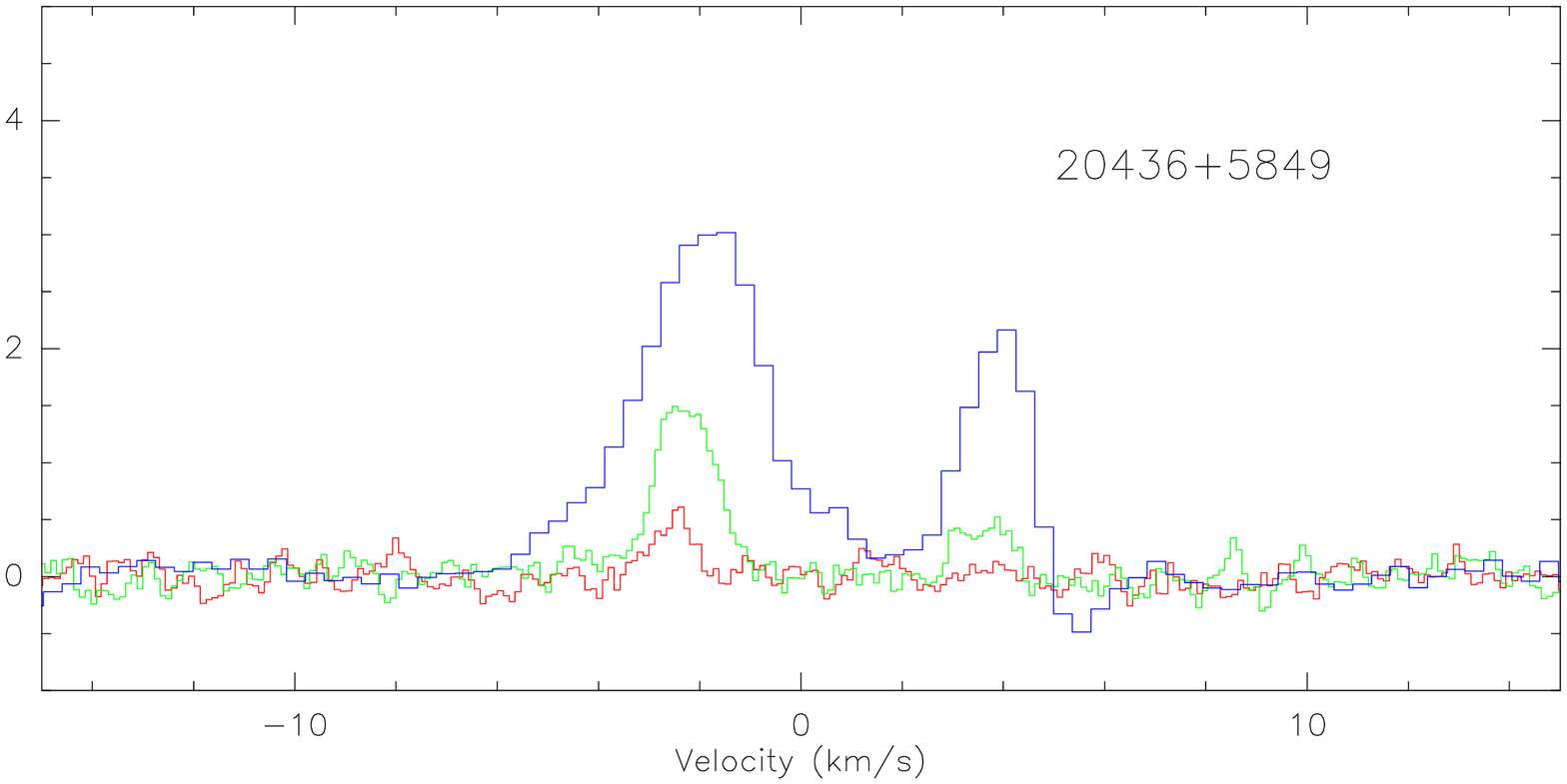}
\includegraphics[height=70mm,  angle=-90, clip, viewport=150 10 500 750]{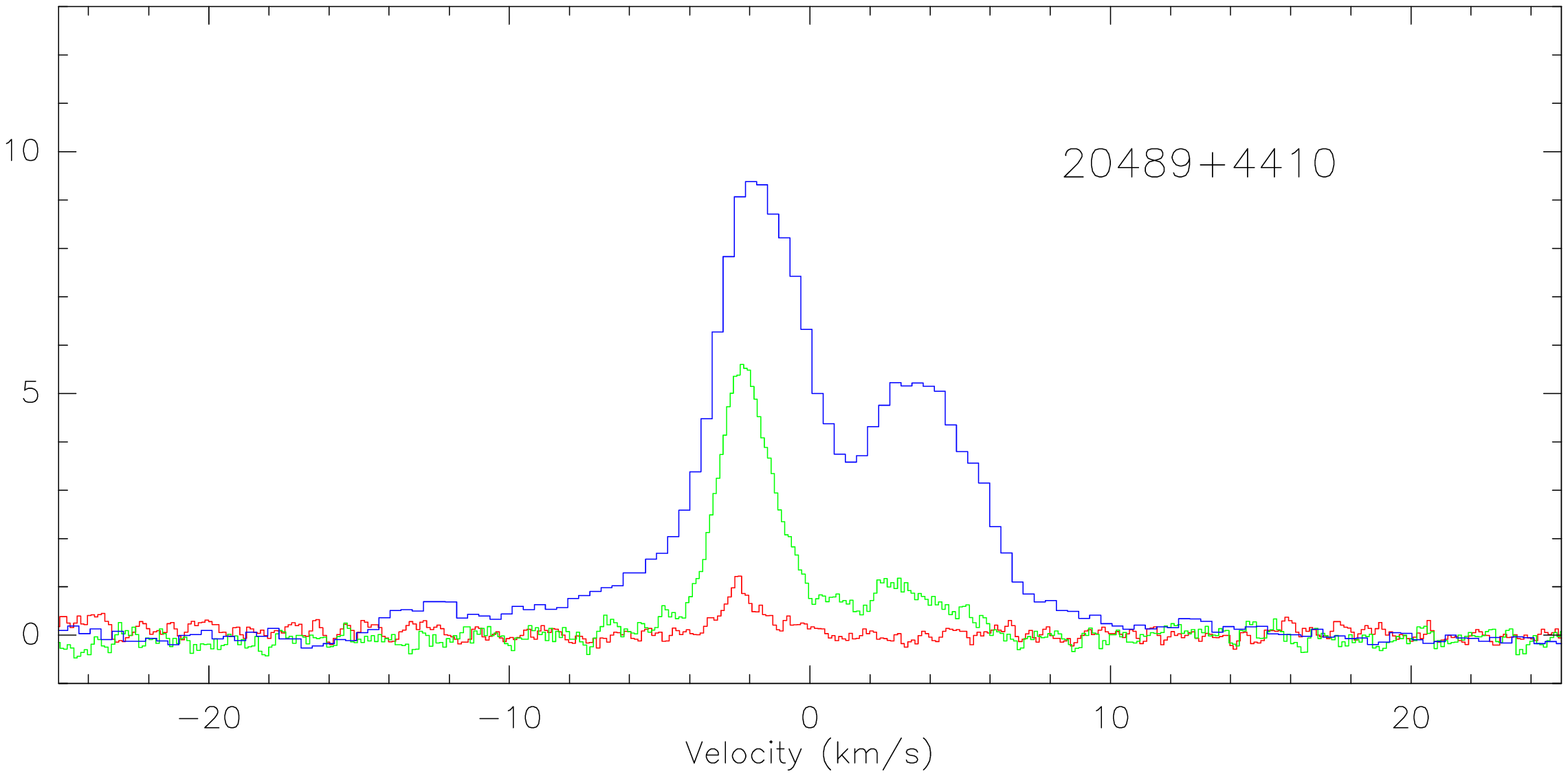}
\includegraphics[height=70mm,  angle=-90, clip, viewport=150 10 500 750]{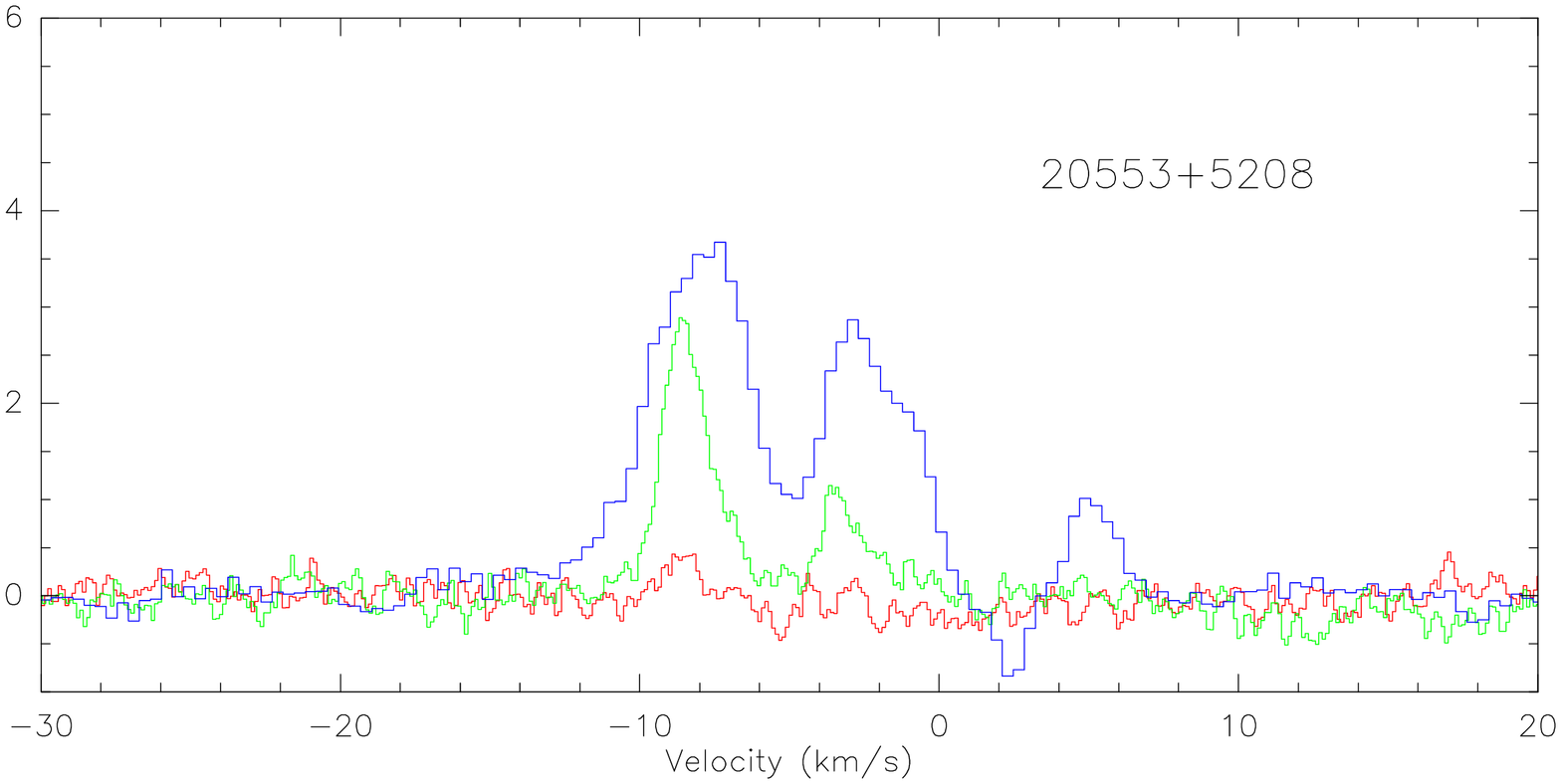}
\centering
\begin{minipage}[]{60mm}
   \caption{The sources of type 1
  }\end{minipage}
   \label{Fig6}
   \end{figure}

\addtocounter{figure}{-1}
\begin{figure}

\includegraphics[height=70mm,  angle=-90, clip, viewport=150 10 500 750]{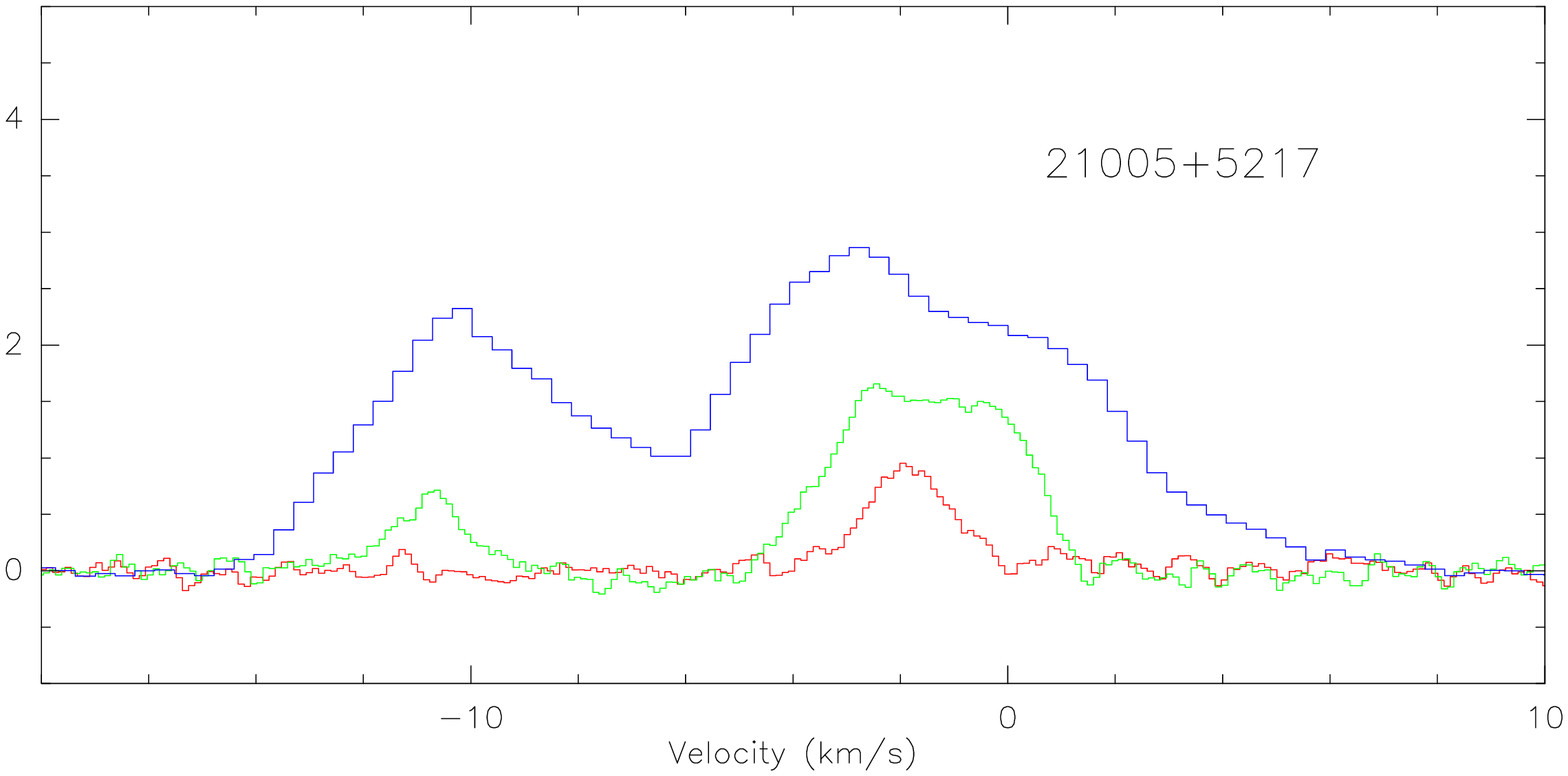}
\includegraphics[height=70mm,  angle=-90, clip, viewport=150 10 500 750]{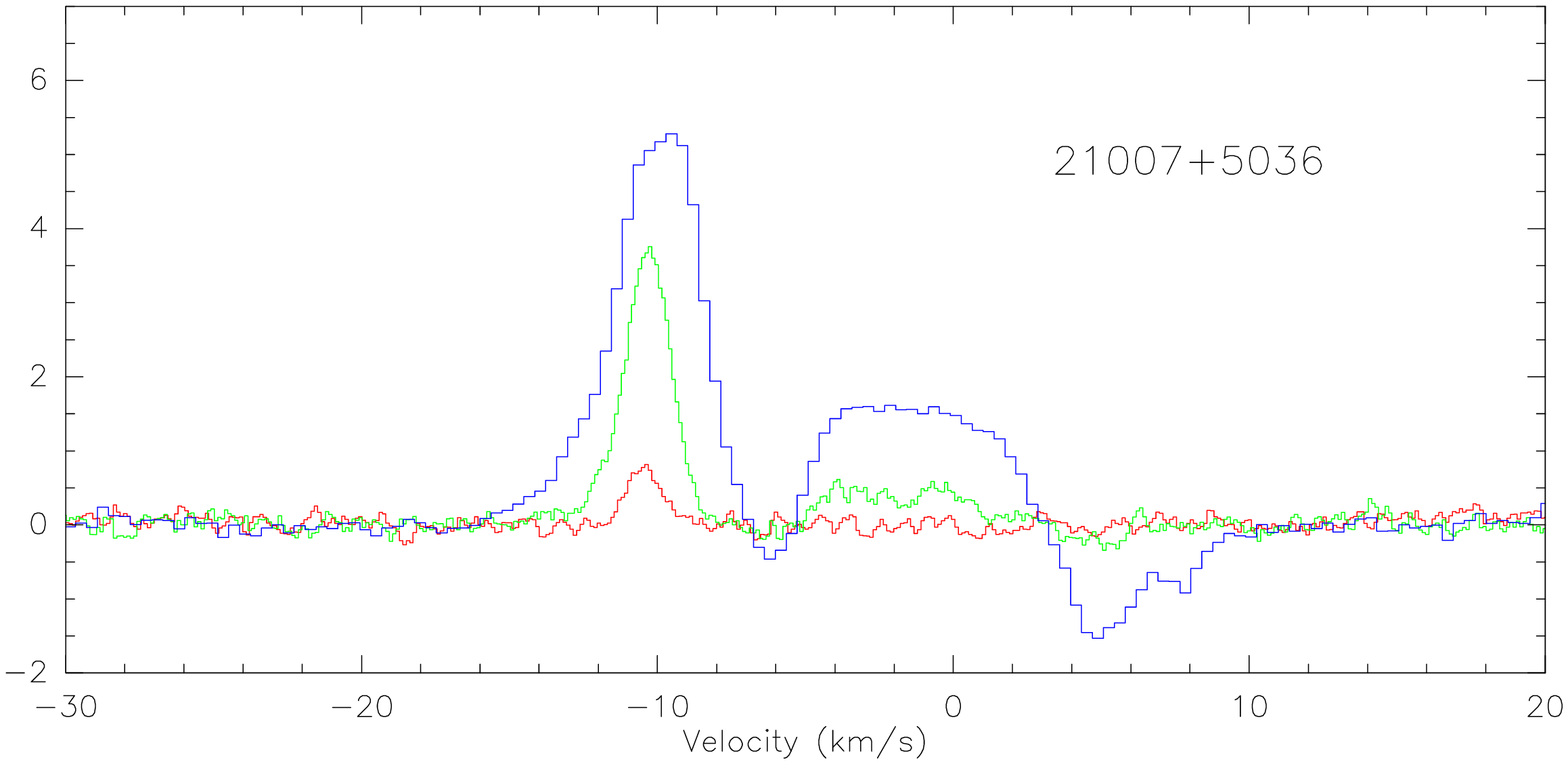}
\includegraphics[height=70mm,  angle=-90, clip, viewport=150 10 500 750]{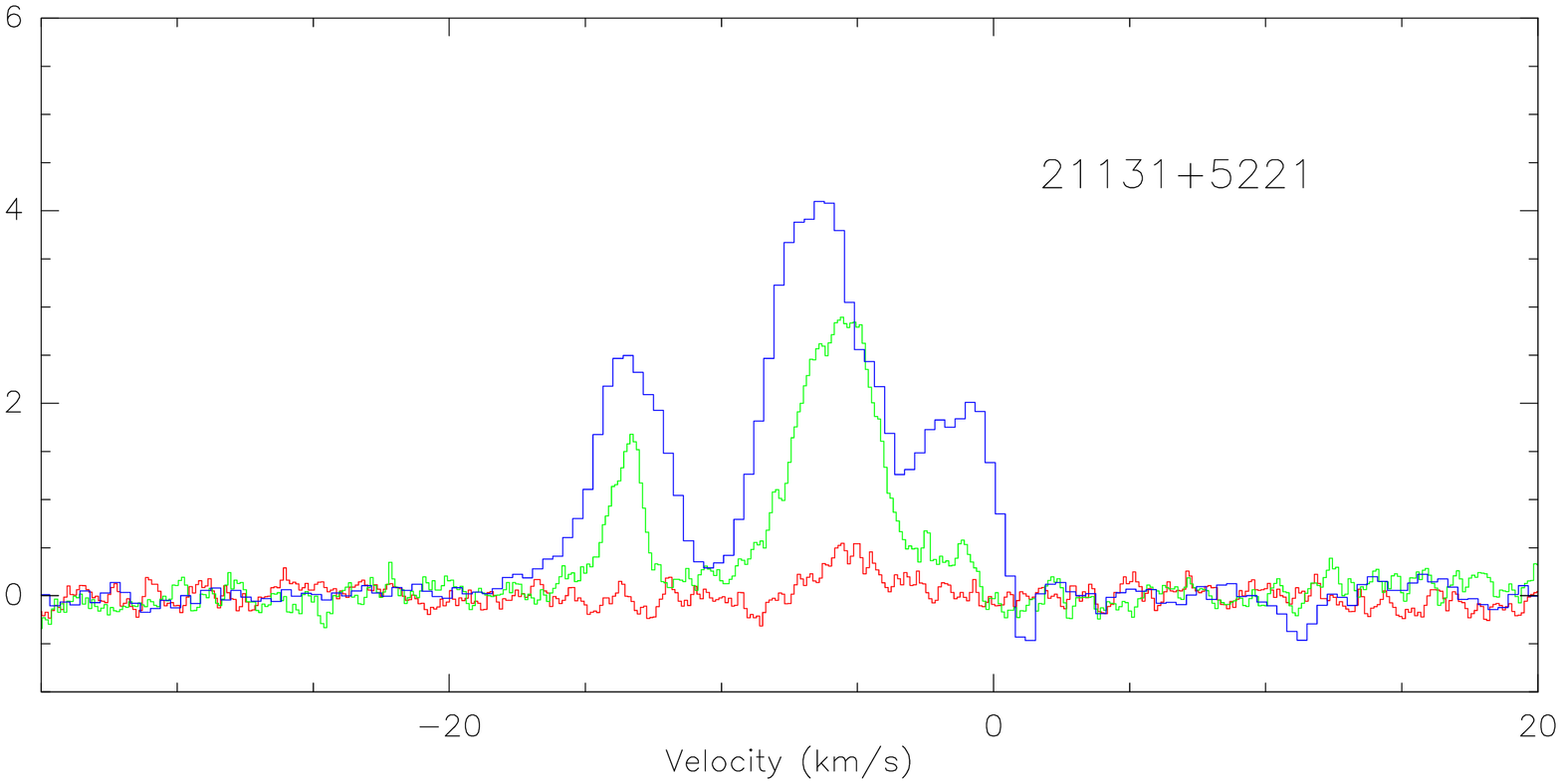}
\includegraphics[height=70mm,  angle=-90, clip, viewport=150 10 500 750]{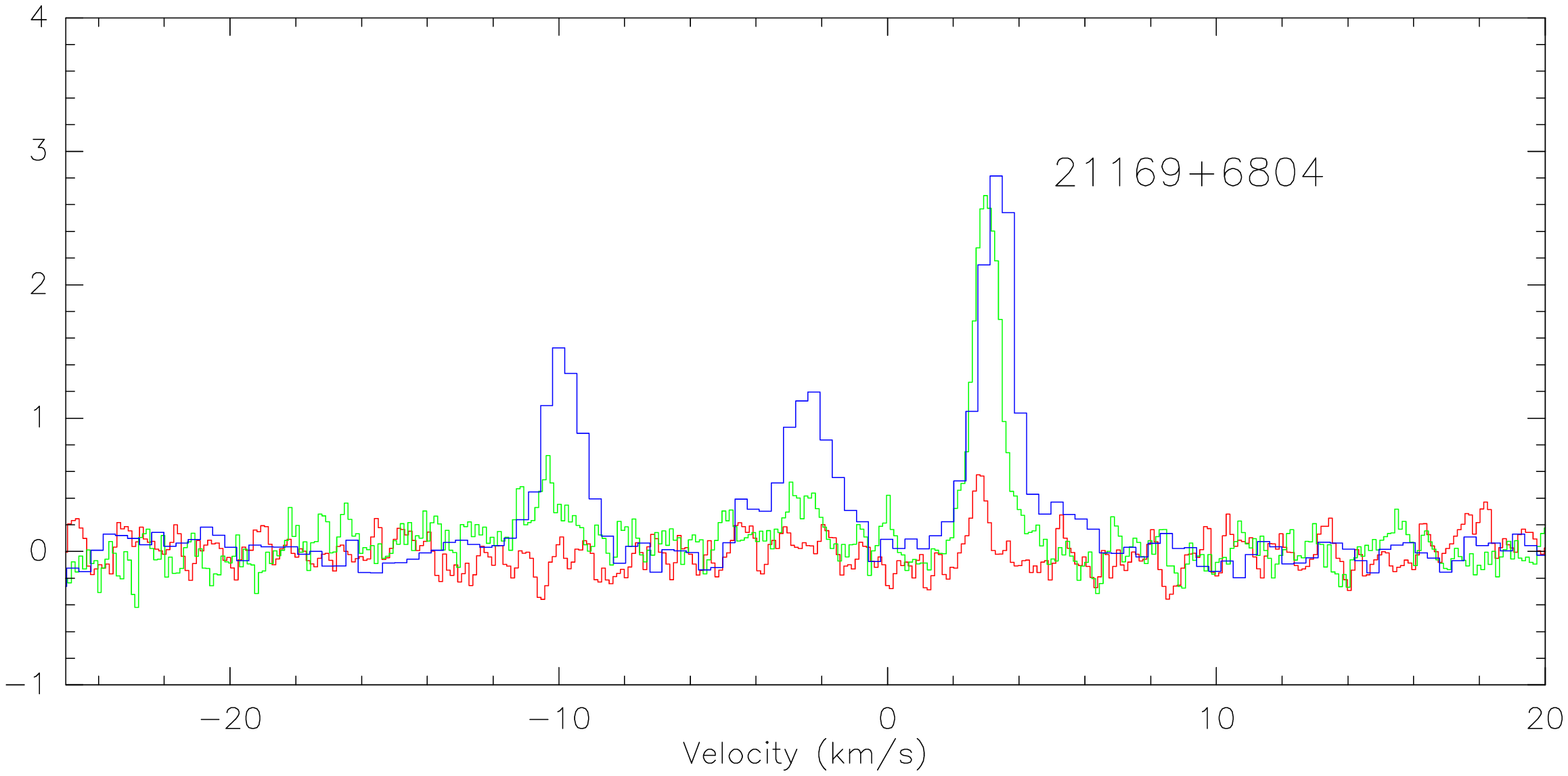}
\includegraphics[height=70mm,  angle=-90, clip, viewport=150 10 500 750]{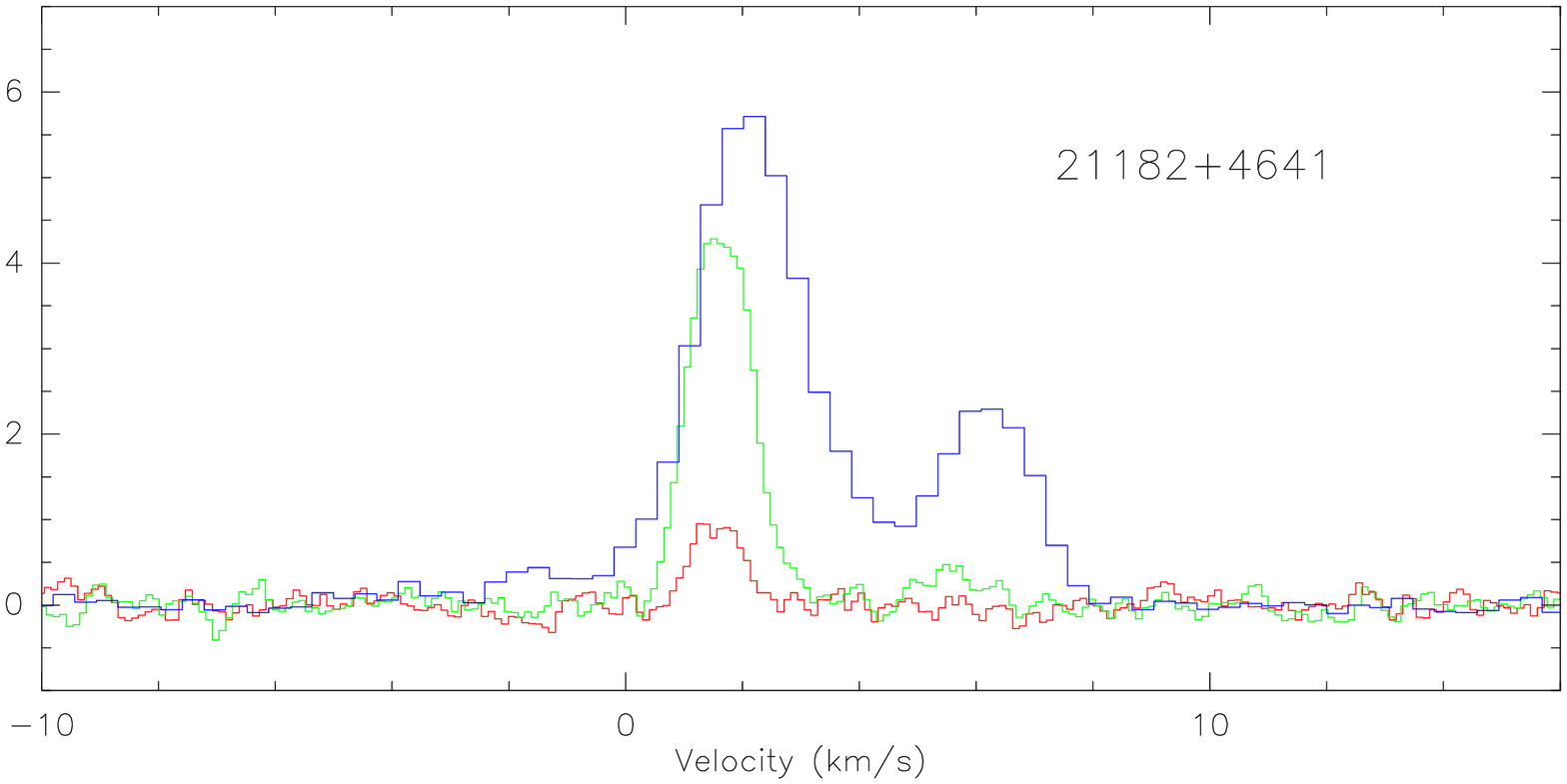}
\includegraphics[height=70mm,  angle=-90, clip, viewport=150 10 500 750]{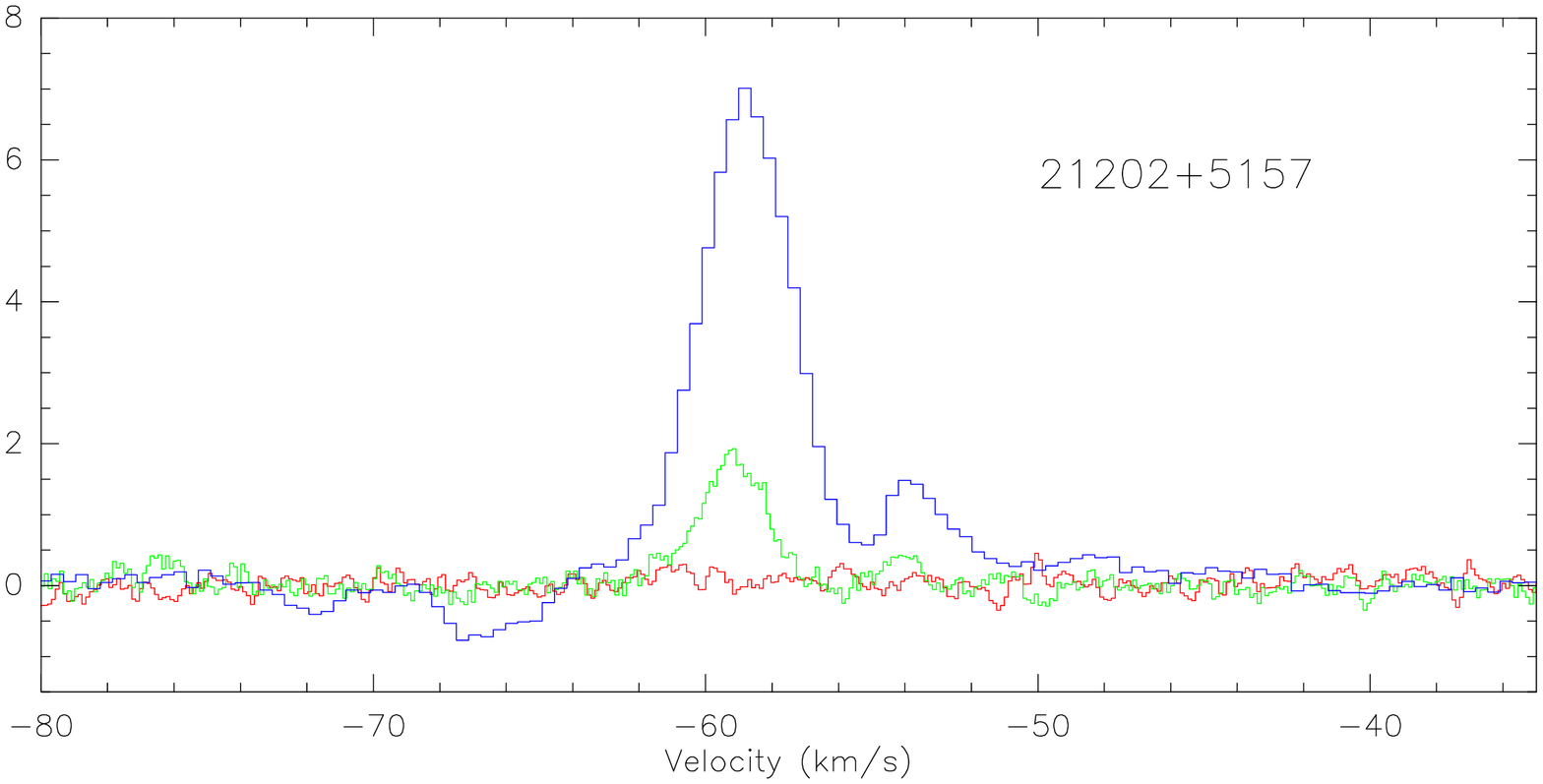}
\includegraphics[height=70mm,  angle=-90, clip, viewport=150 10 500 750]{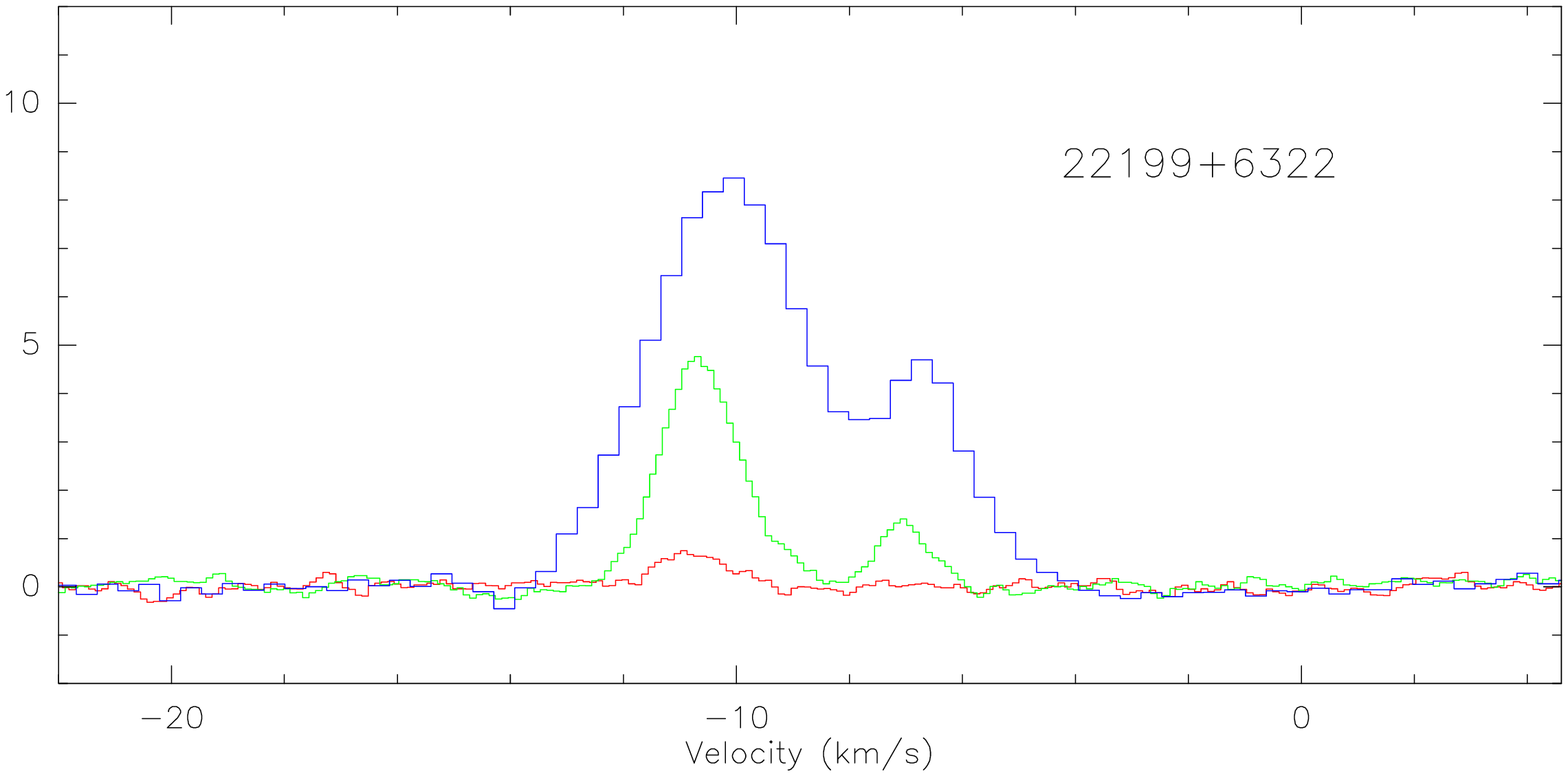}
\includegraphics[height=70mm,  angle=-90, clip, viewport=150 10 500 750]{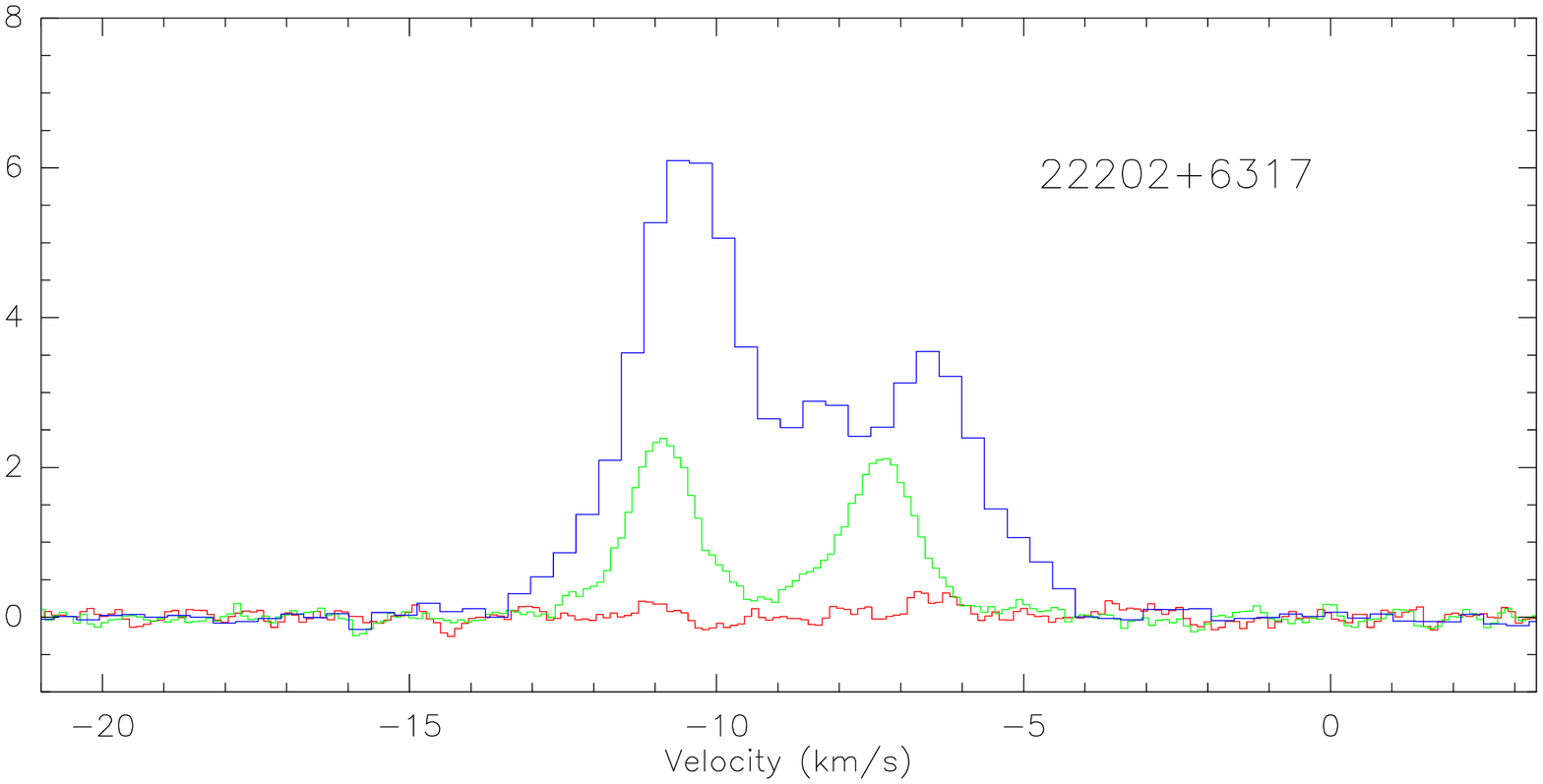}
\includegraphics[height=70mm,  angle=-90, clip, viewport=150 10 500 750]{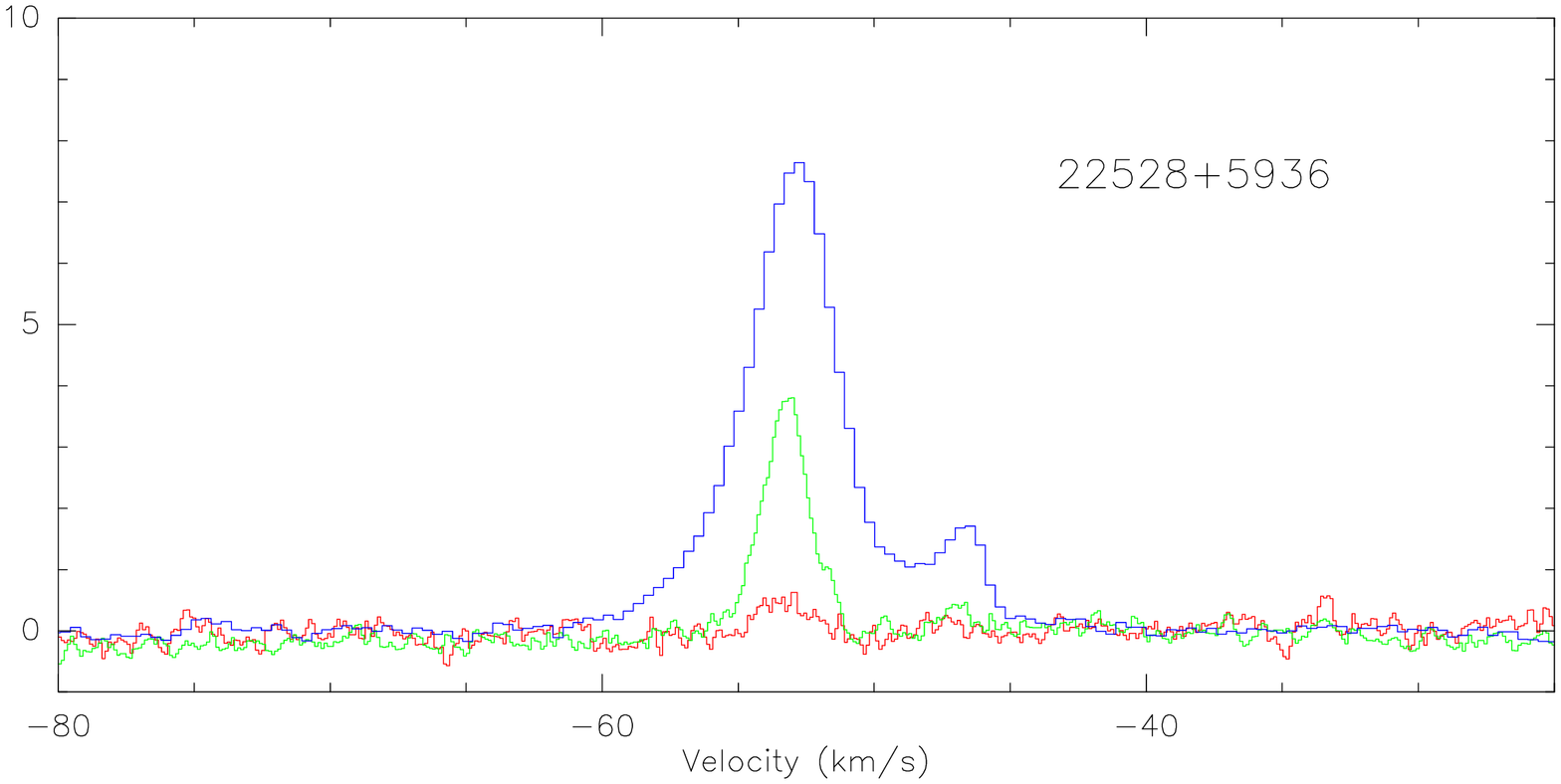}
\includegraphics[height=70mm,  angle=-90, clip, viewport=150 10 500 750]{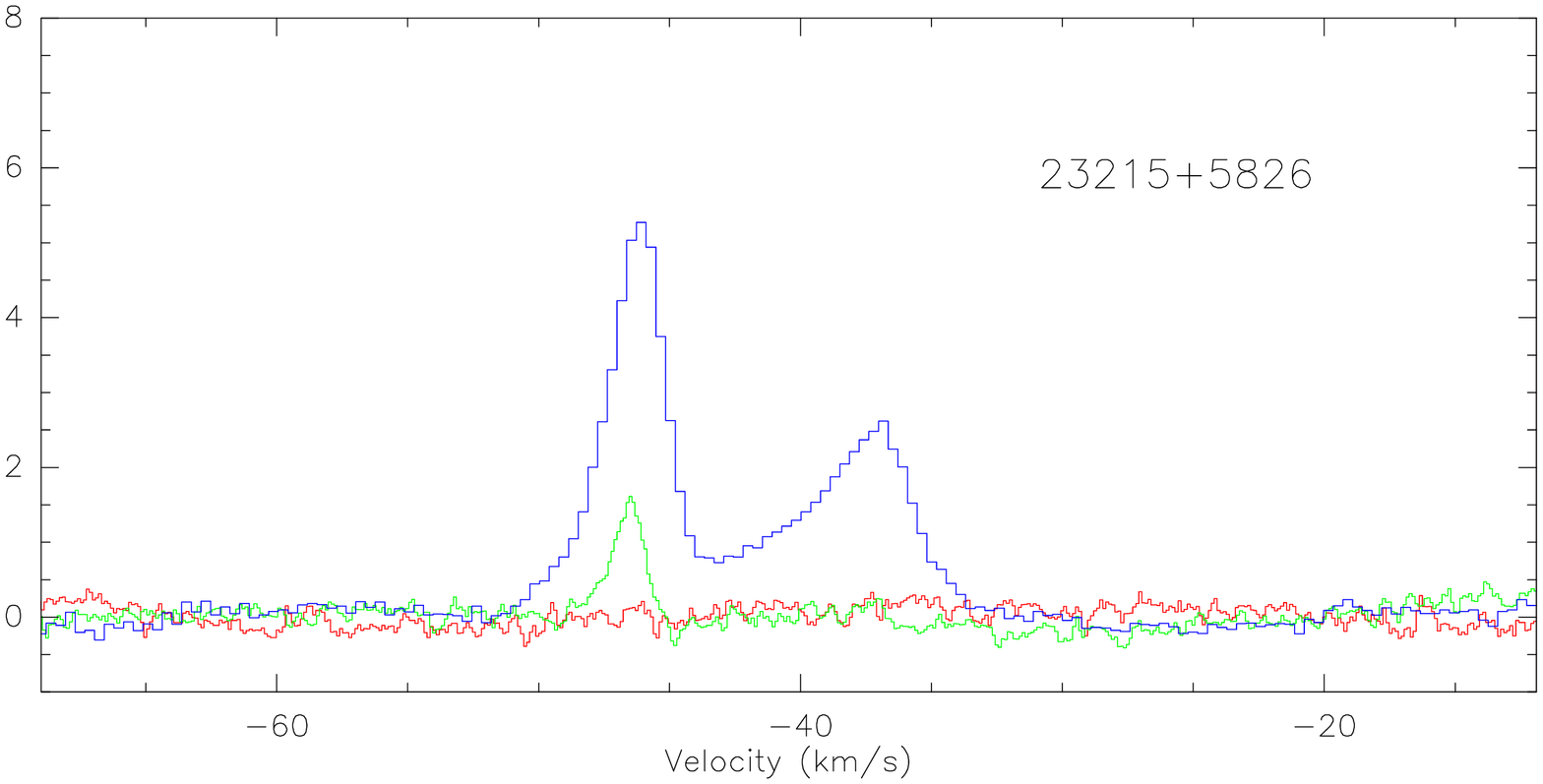}
\centering
\begin{minipage}[]{60mm}
   \caption{The sources of type 1
  }\end{minipage}
   \label{Fig6}
   \end{figure}

\begin{figure}

\includegraphics[height=70mm,  angle=-90, clip, viewport=150 10 500 750]{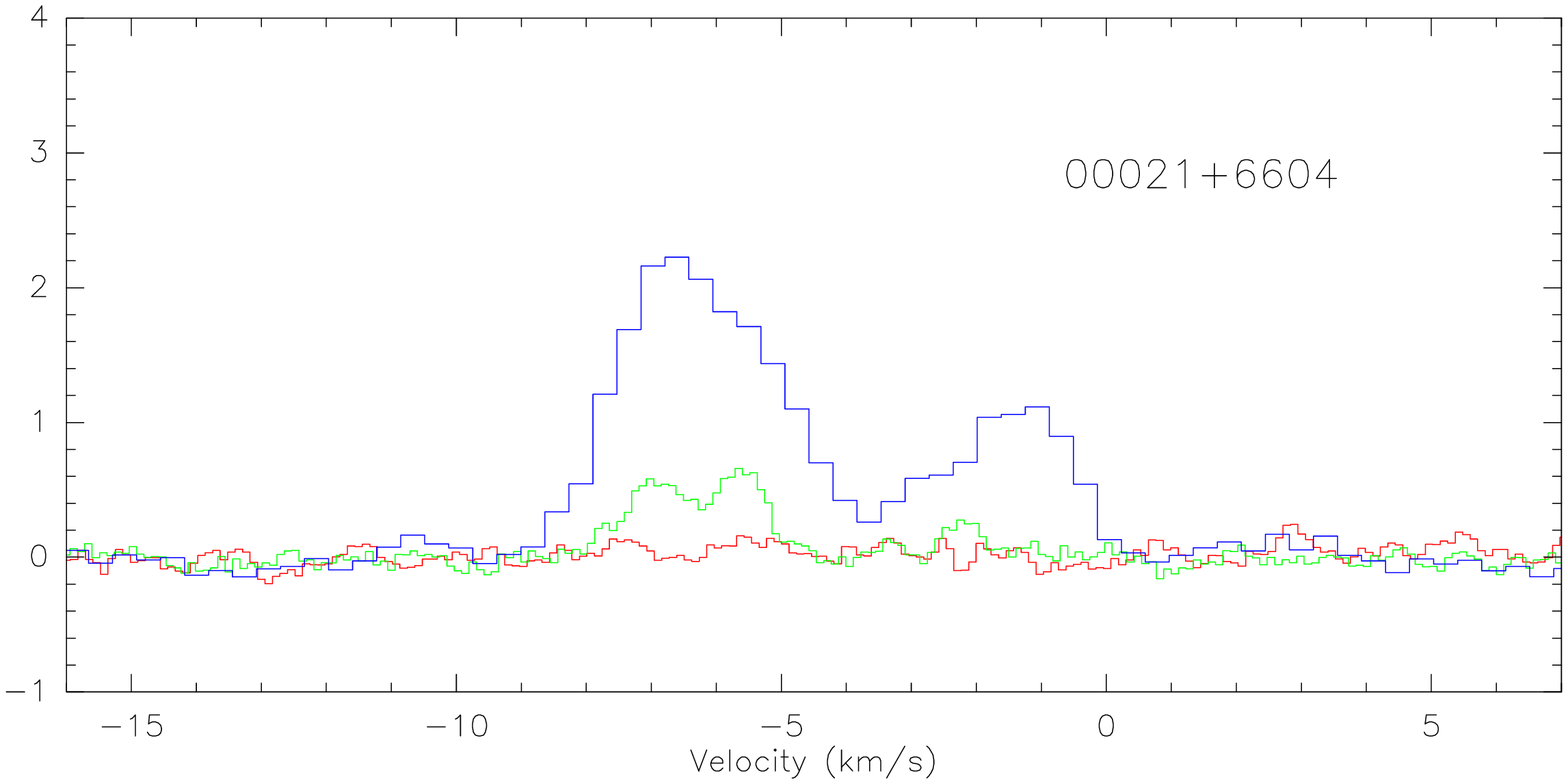}
\includegraphics[height=70mm,  angle=-90, clip, viewport=150 10 500 750]{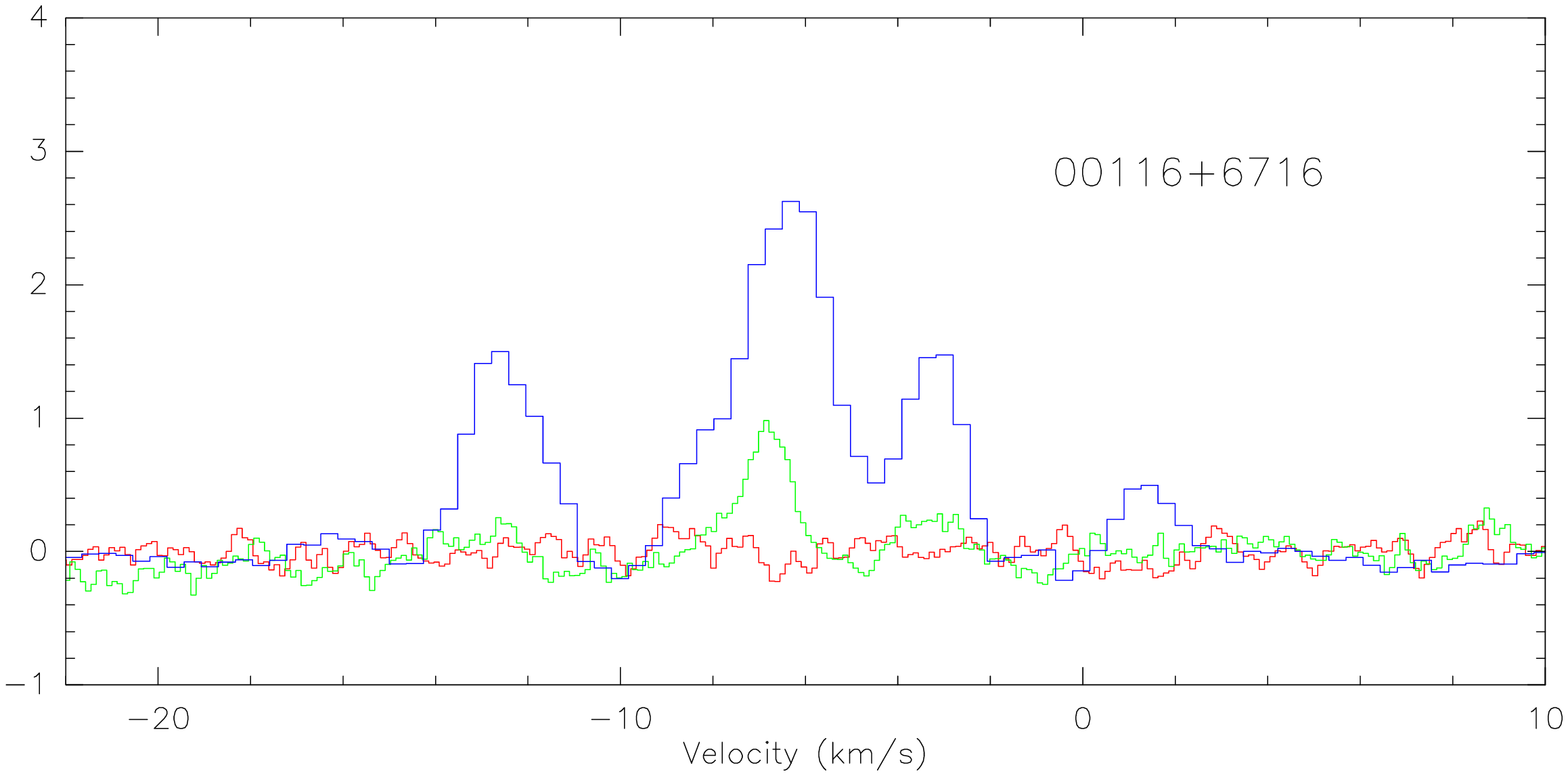}
\includegraphics[height=70mm,  angle=-90, clip, viewport=150 10 500 750]{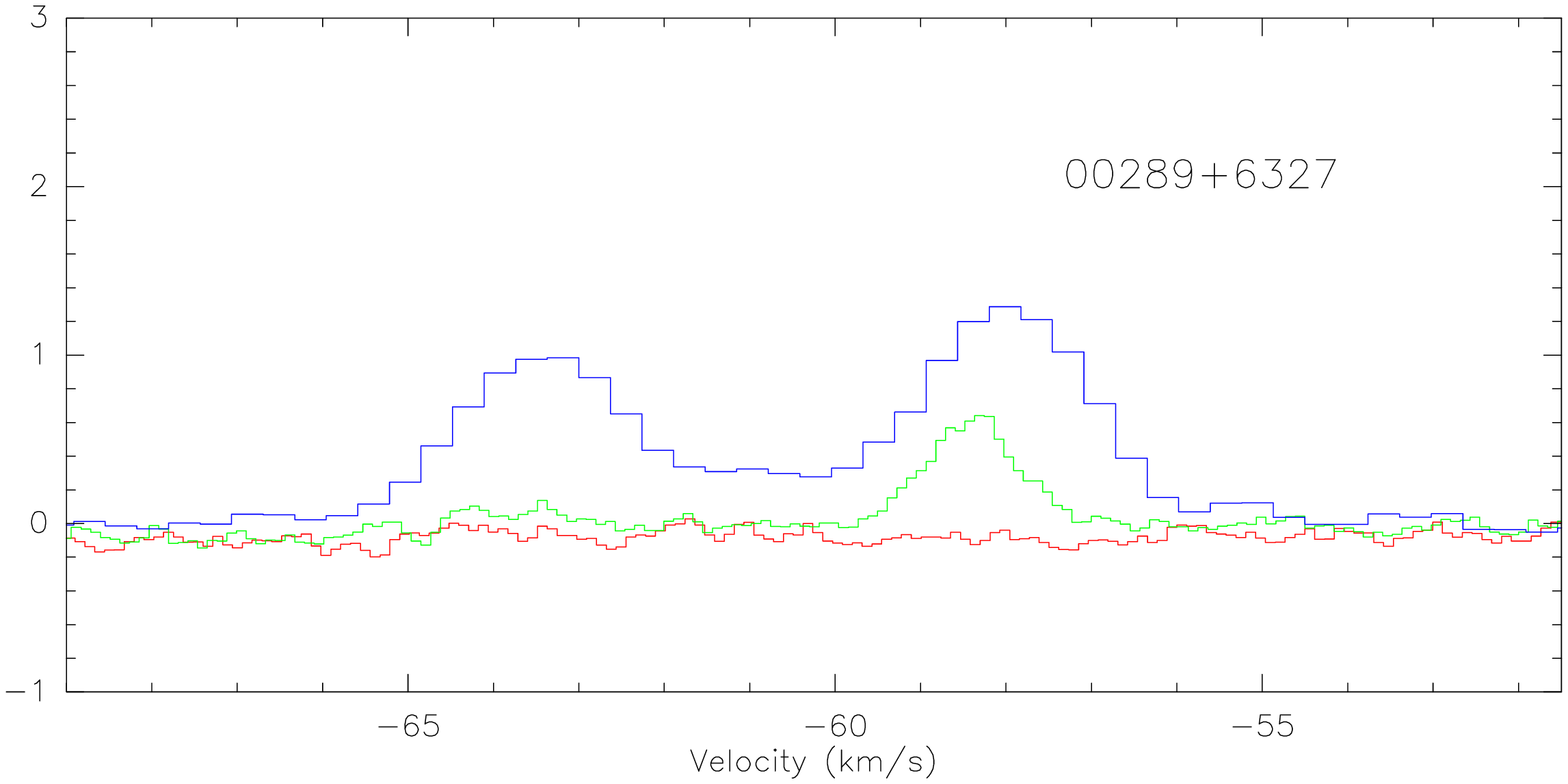}
\includegraphics[height=70mm,  angle=-90, clip, viewport=150 10 500 750]{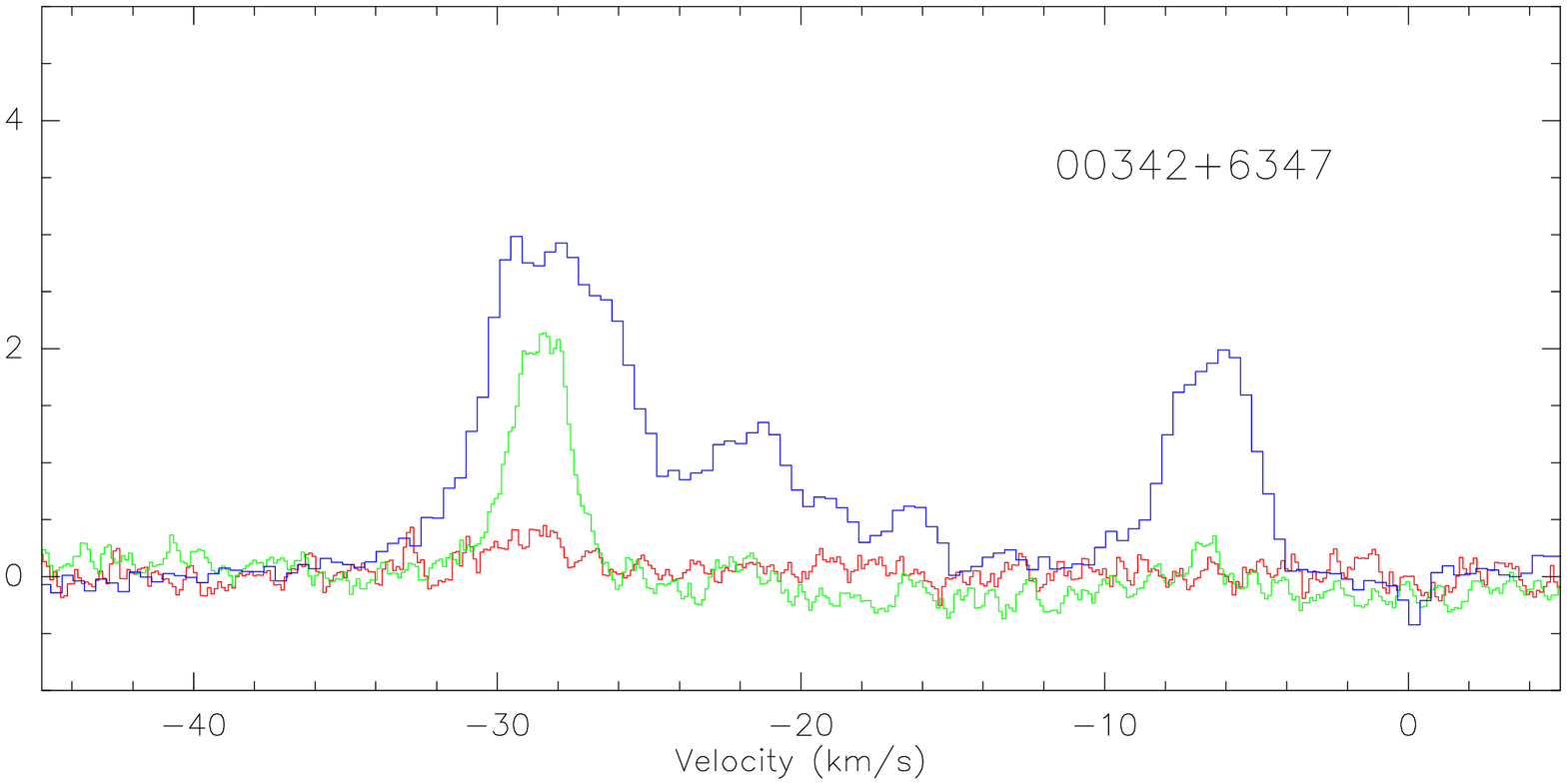}
\includegraphics[height=70mm,  angle=-90, clip, viewport=150 10 500 750]{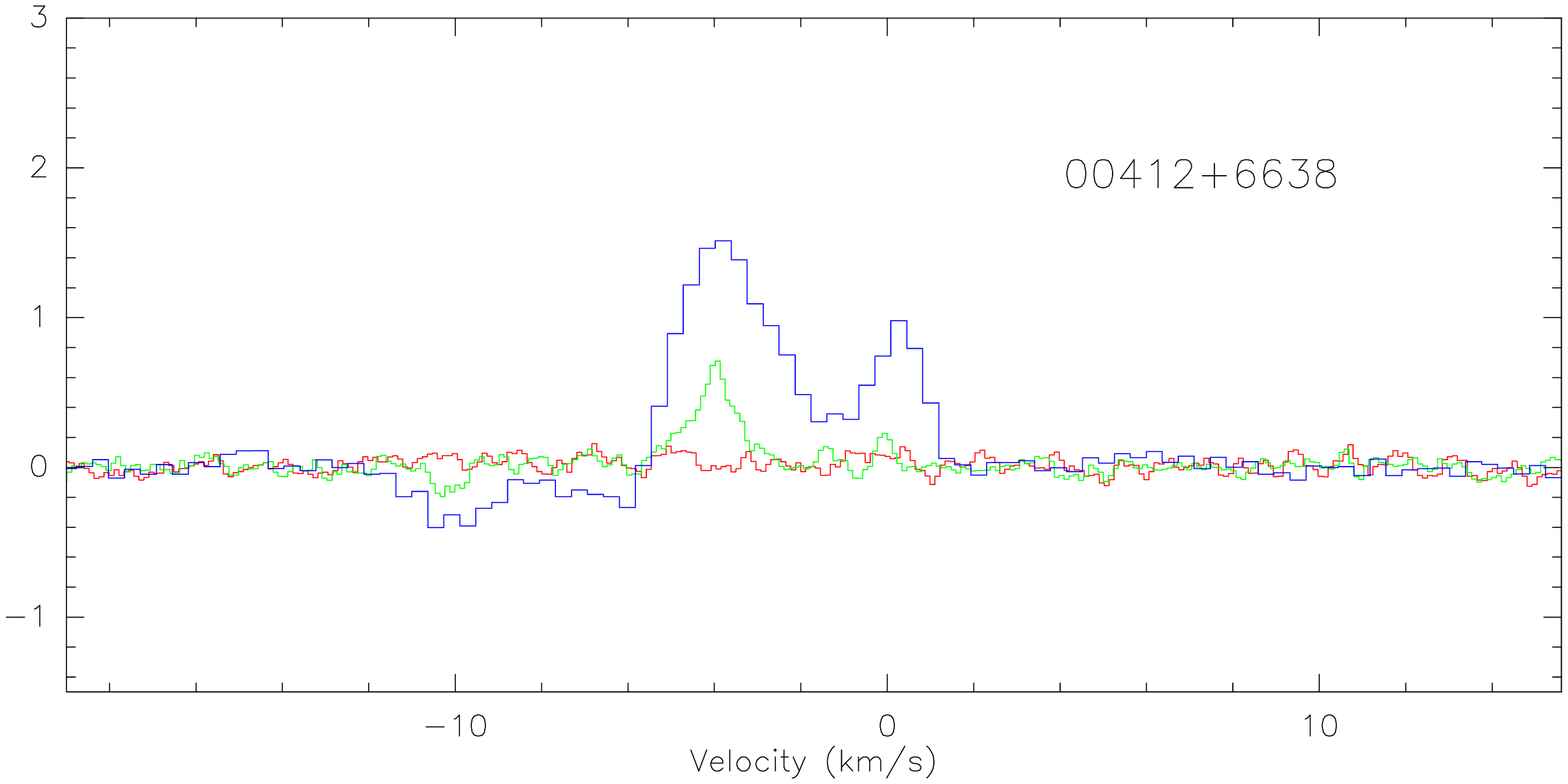}
\includegraphics[height=70mm,  angle=-90, clip, viewport=150 10 500 750]{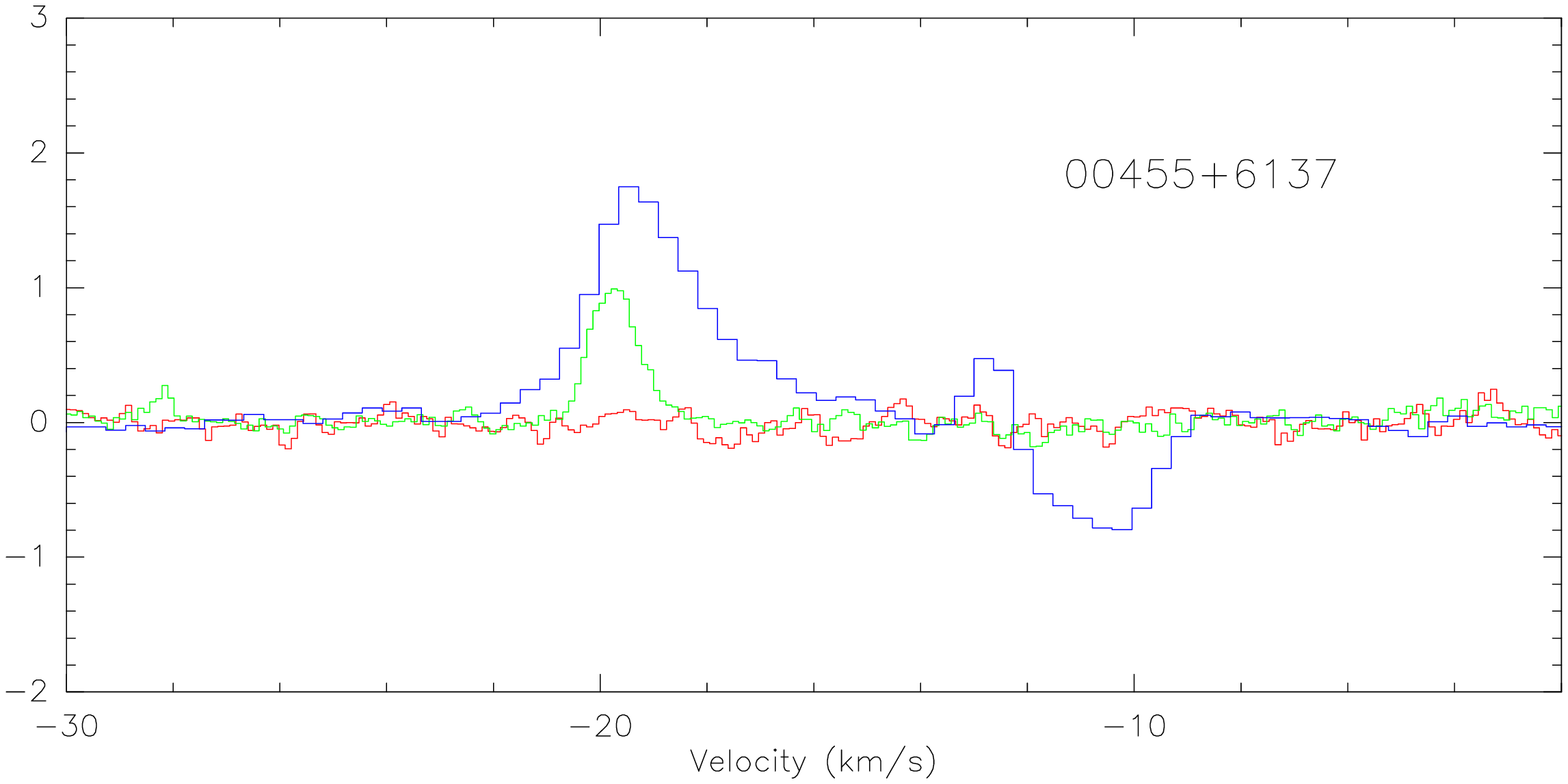}
\includegraphics[height=70mm,  angle=-90, clip, viewport=150 10 500 750]{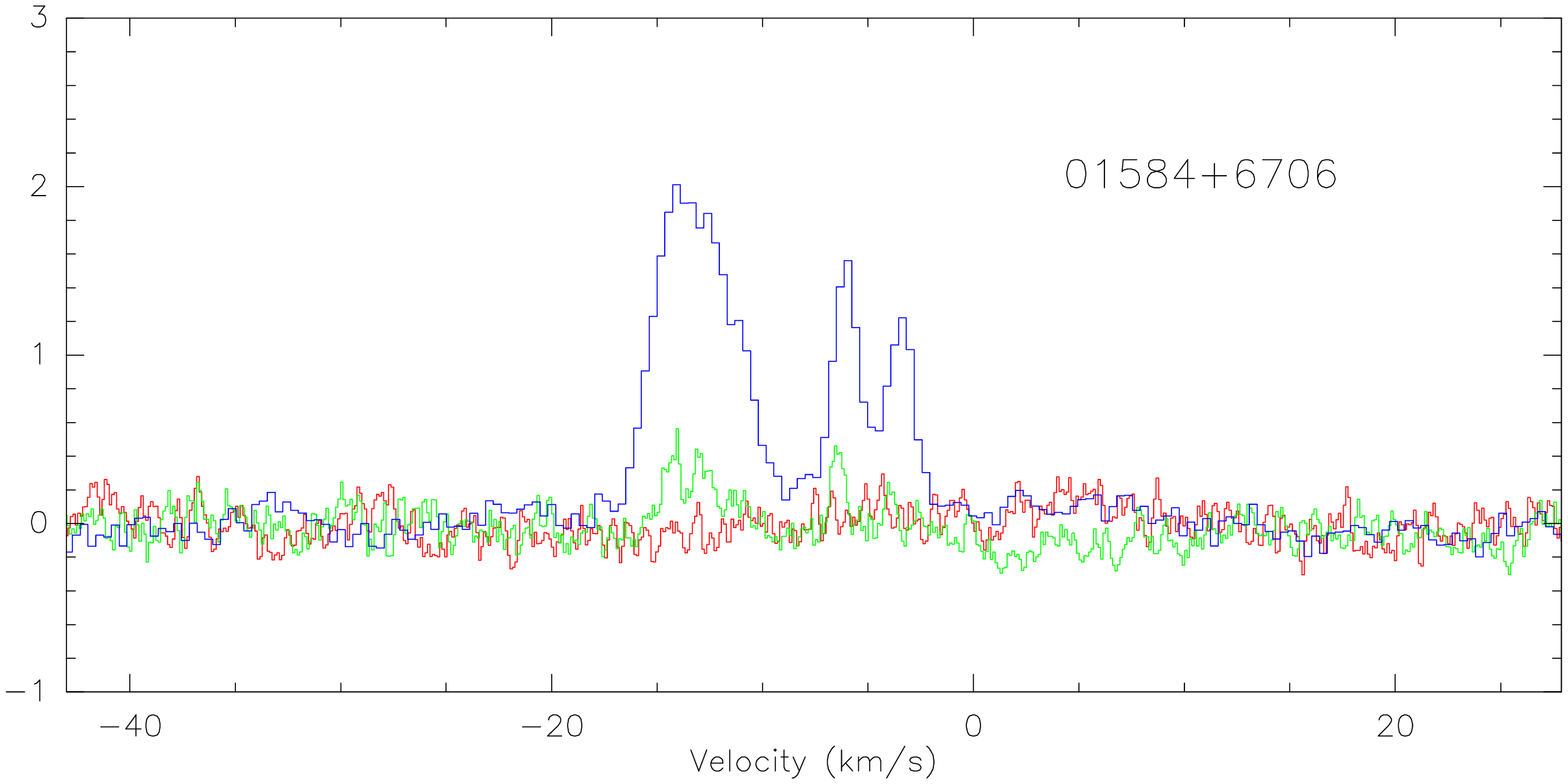}
\includegraphics[height=70mm,  angle=-90, clip, viewport=150 10 500 750]{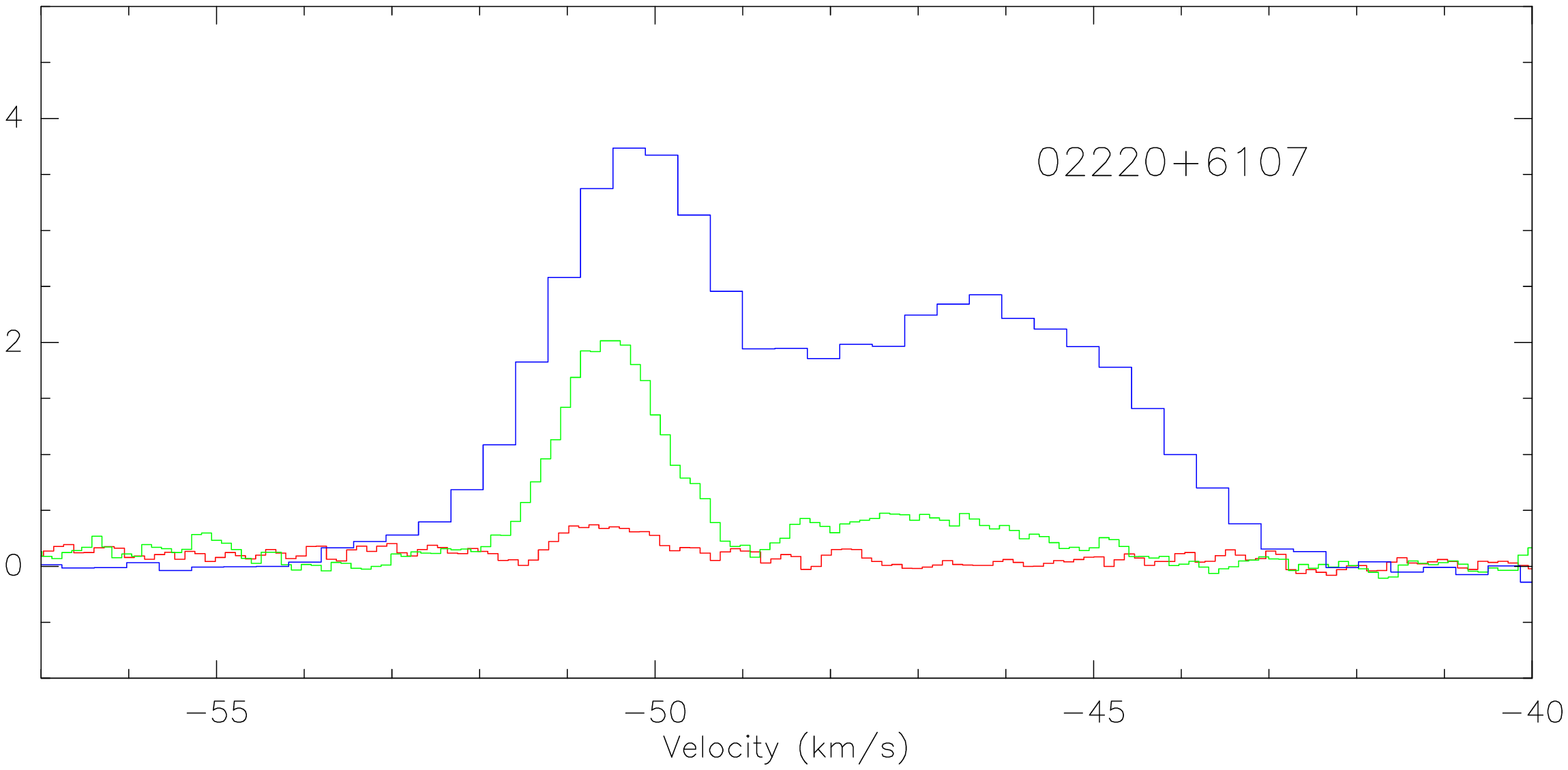}
\includegraphics[height=70mm,  angle=-90, clip, viewport=150 10 500 750]{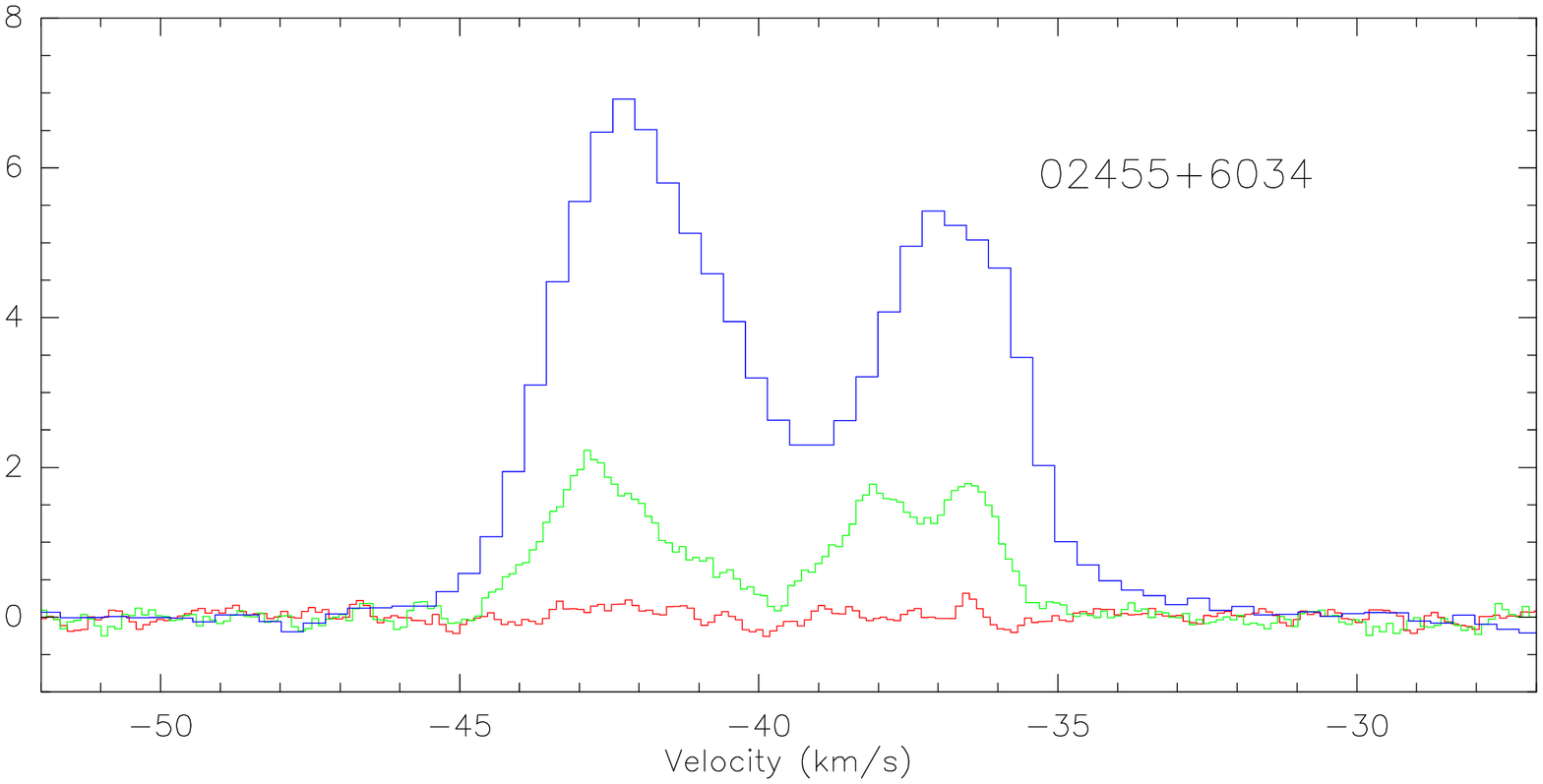}
\includegraphics[height=70mm,  angle=-90, clip, viewport=150 10 500 750]{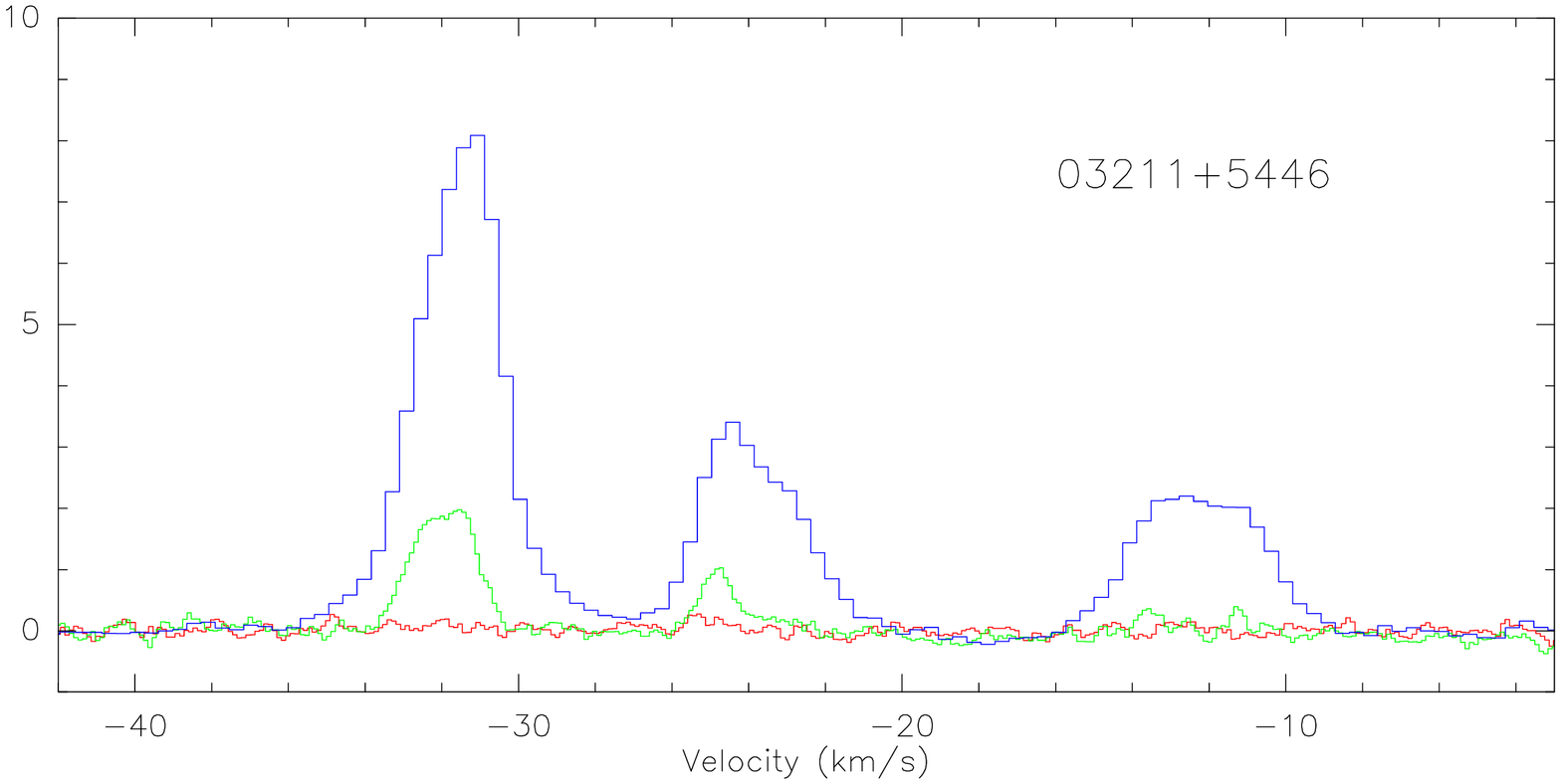}
\includegraphics[height=70mm,  angle=-90, clip, viewport=150 10 500 750]{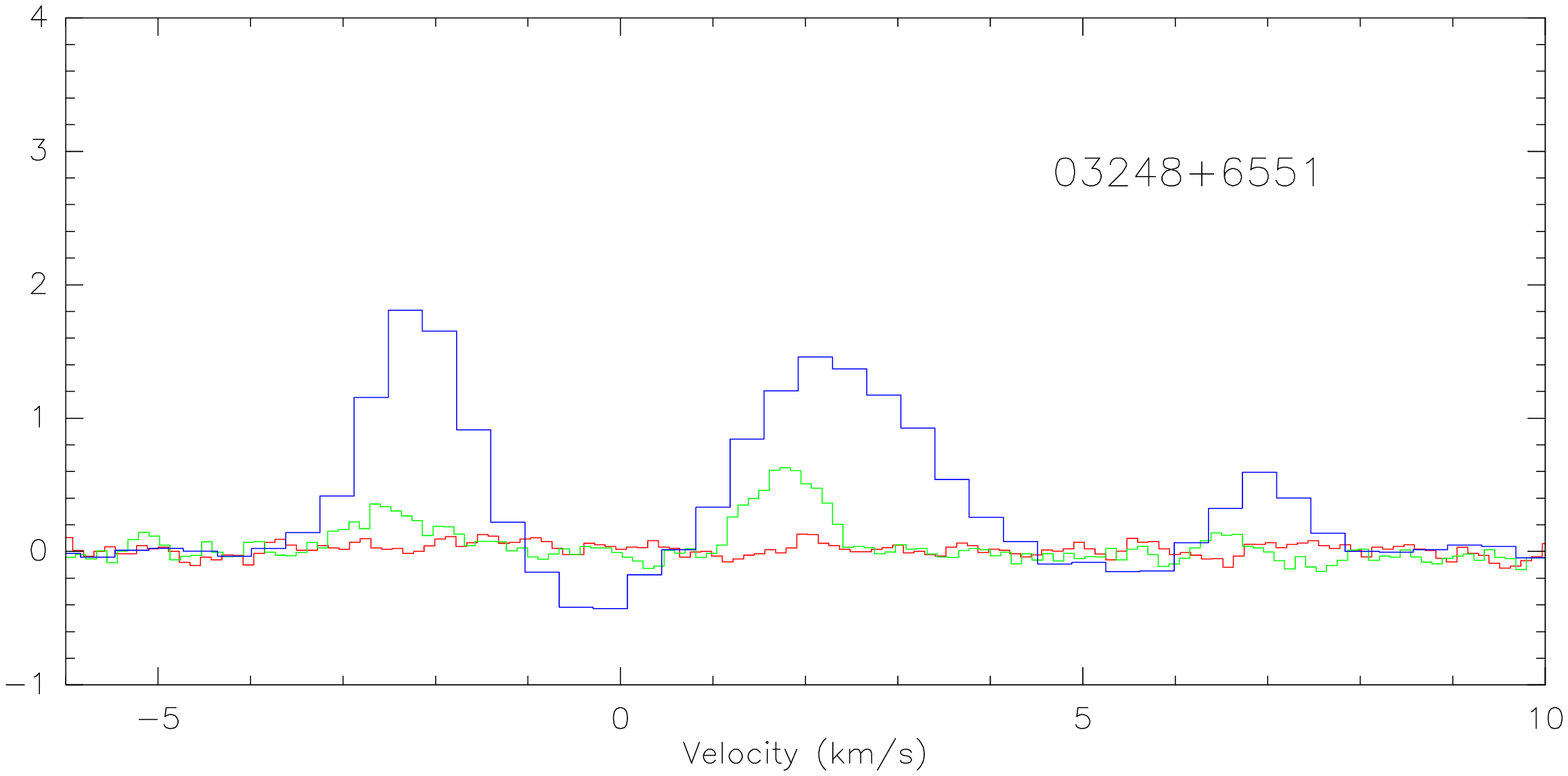}
\includegraphics[height=70mm,  angle=-90, clip, viewport=150 10 500 750]{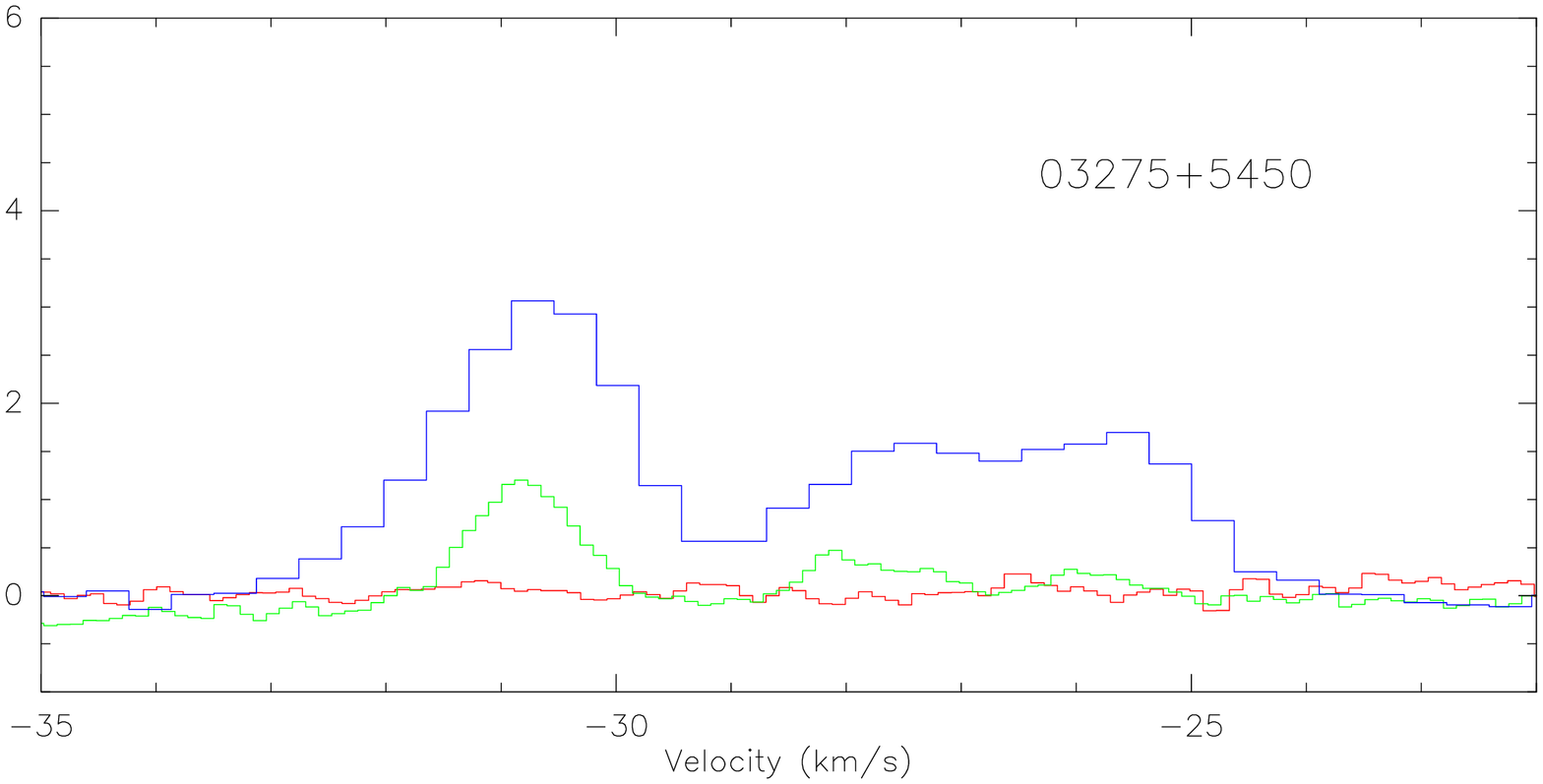}
\includegraphics[height=70mm,  angle=-90, clip, viewport=150 10 500 750]{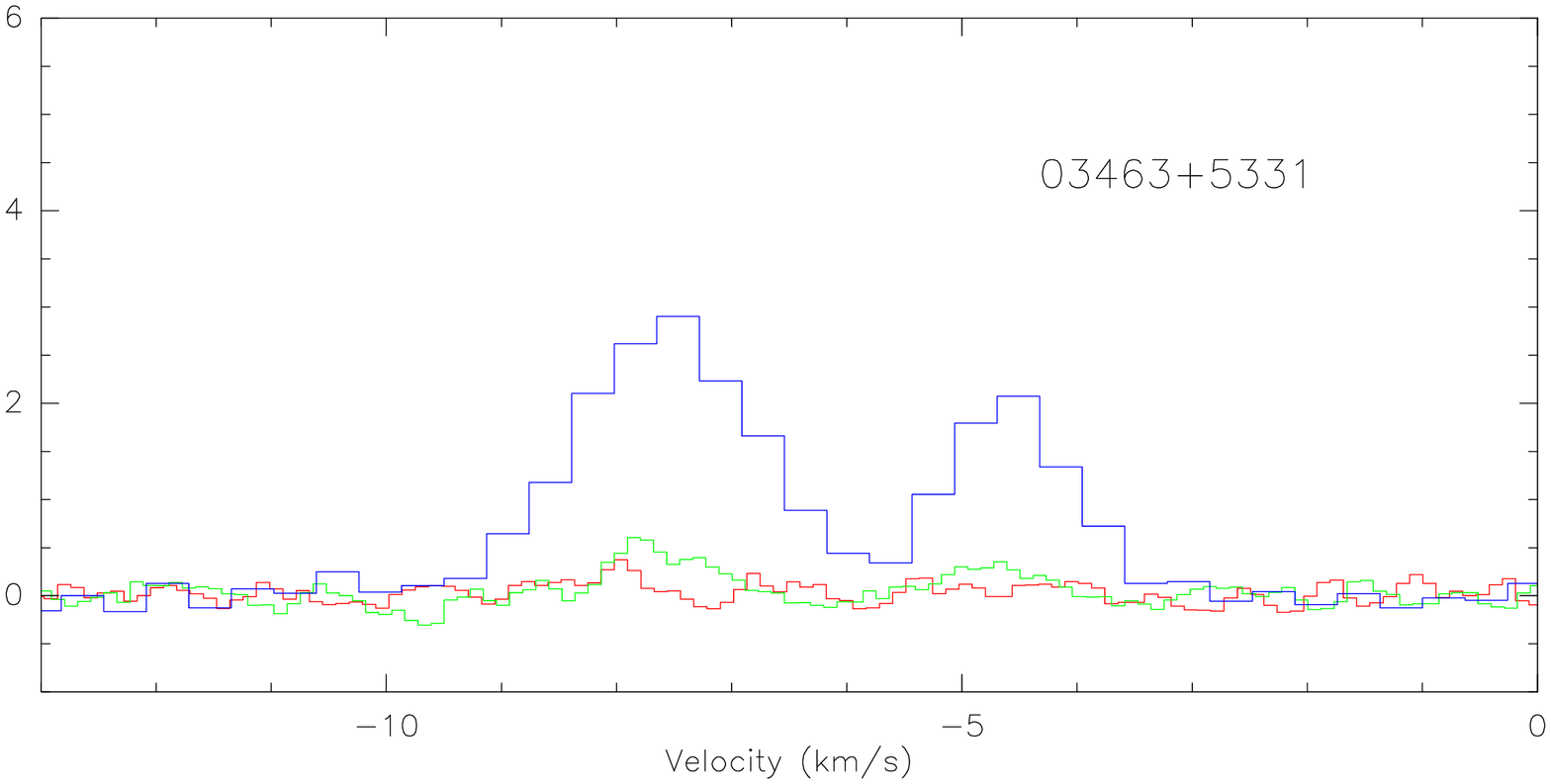}
\includegraphics[height=70mm,  angle=-90, clip, viewport=150 10 500 750]{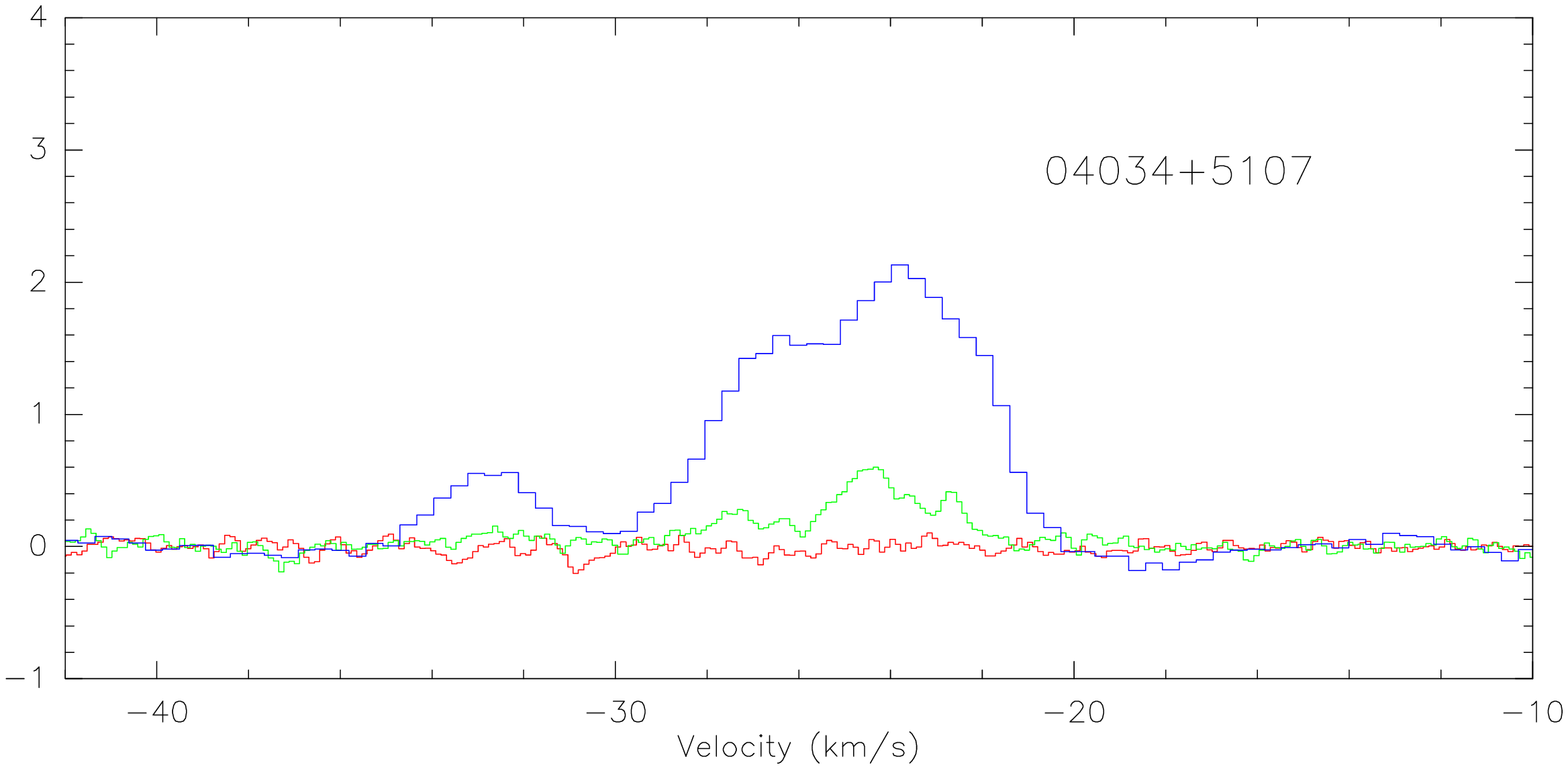}
 \centering
\begin{minipage}[]{60mm}

   \caption{The sources of type 2
  }\end{minipage}
   \label{Fig7}
   \end{figure}

\addtocounter{figure}{-1}
\begin{figure}

 \centering
\includegraphics[height=70mm,  angle=-90, clip, viewport=150 10 500 750]{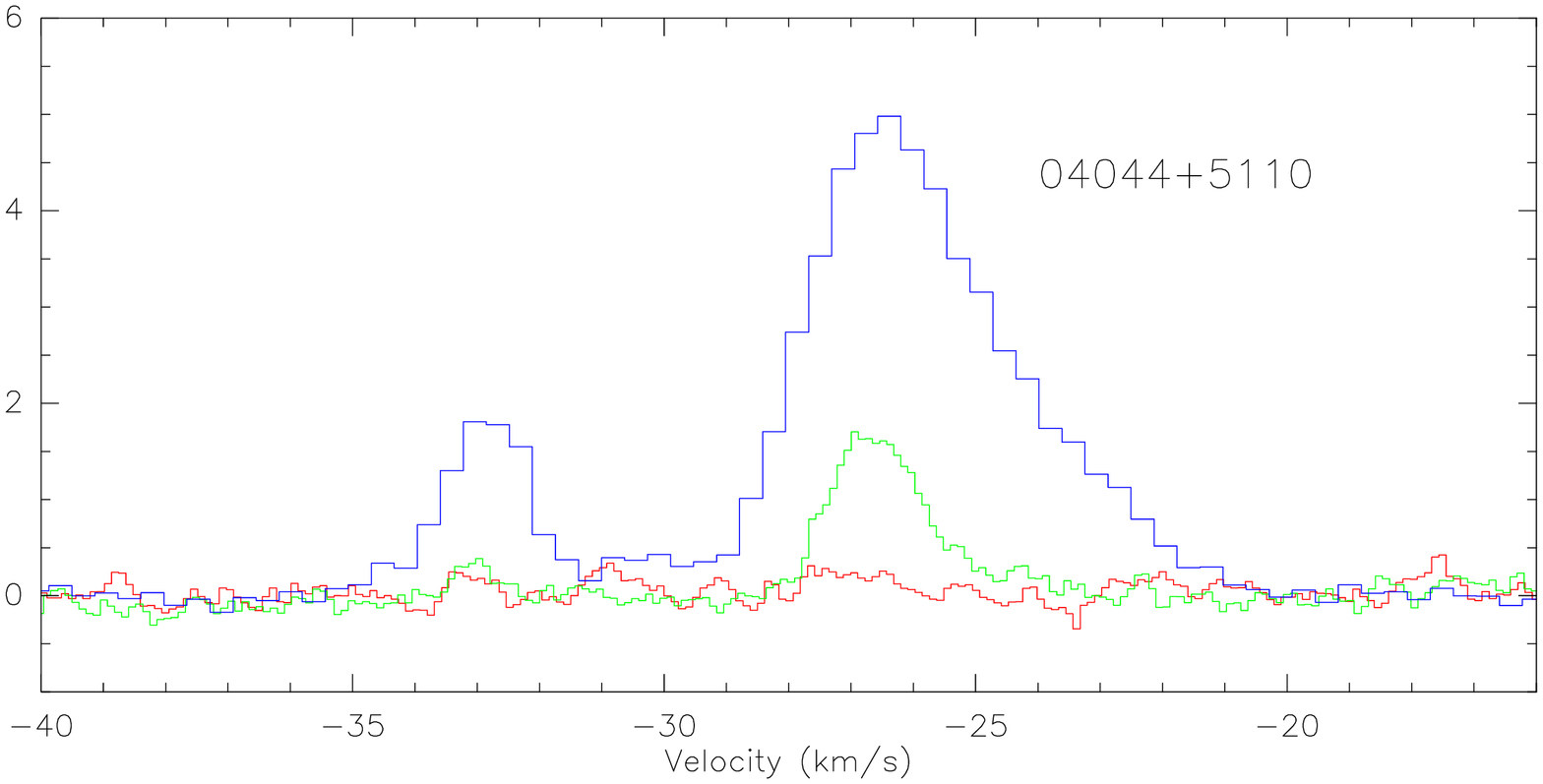}
\includegraphics[height=70mm,  angle=-90, clip, viewport=150 10 500 750]{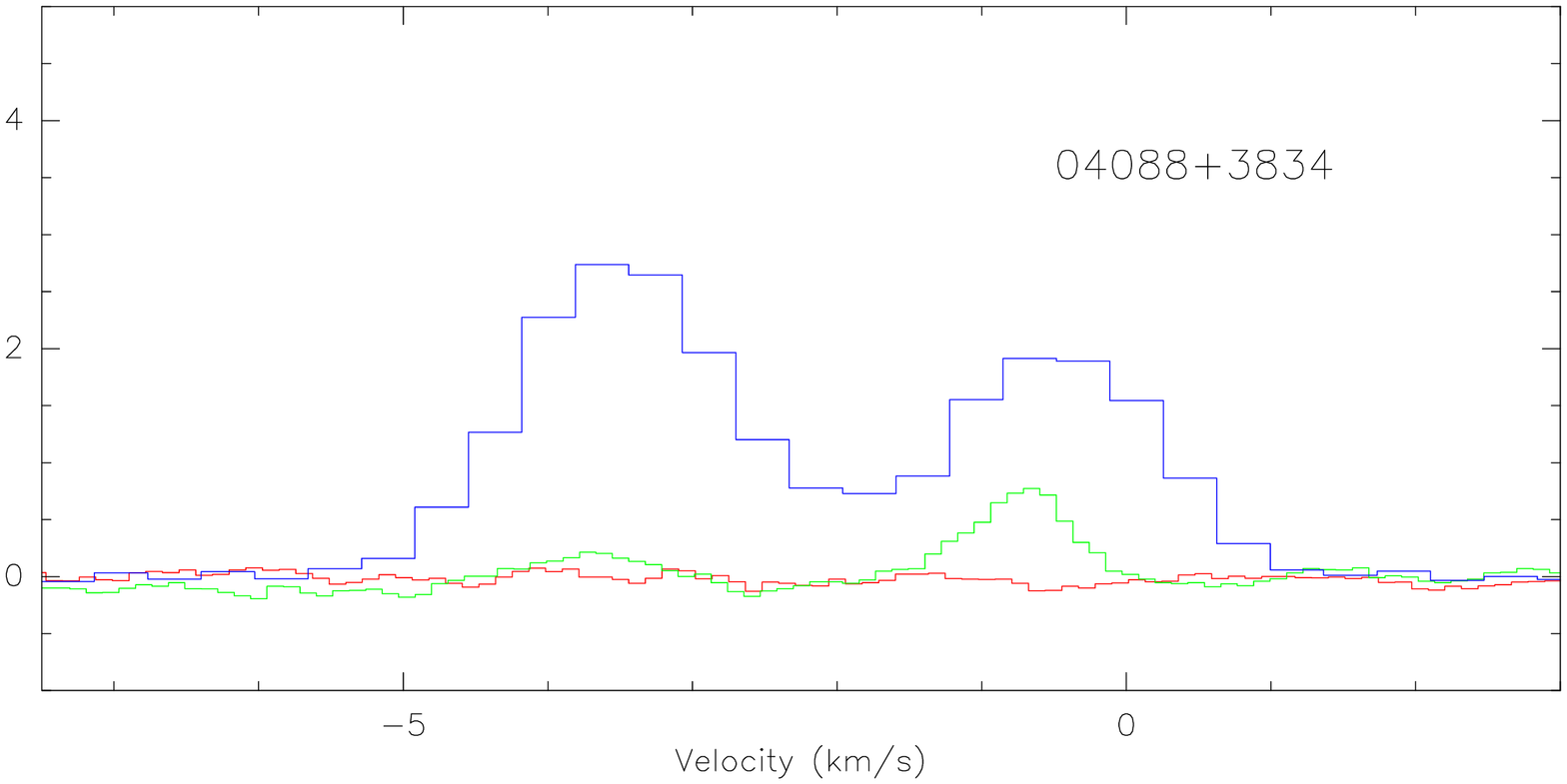}
\includegraphics[height=70mm,  angle=-90, clip, viewport=150 10 500 750]{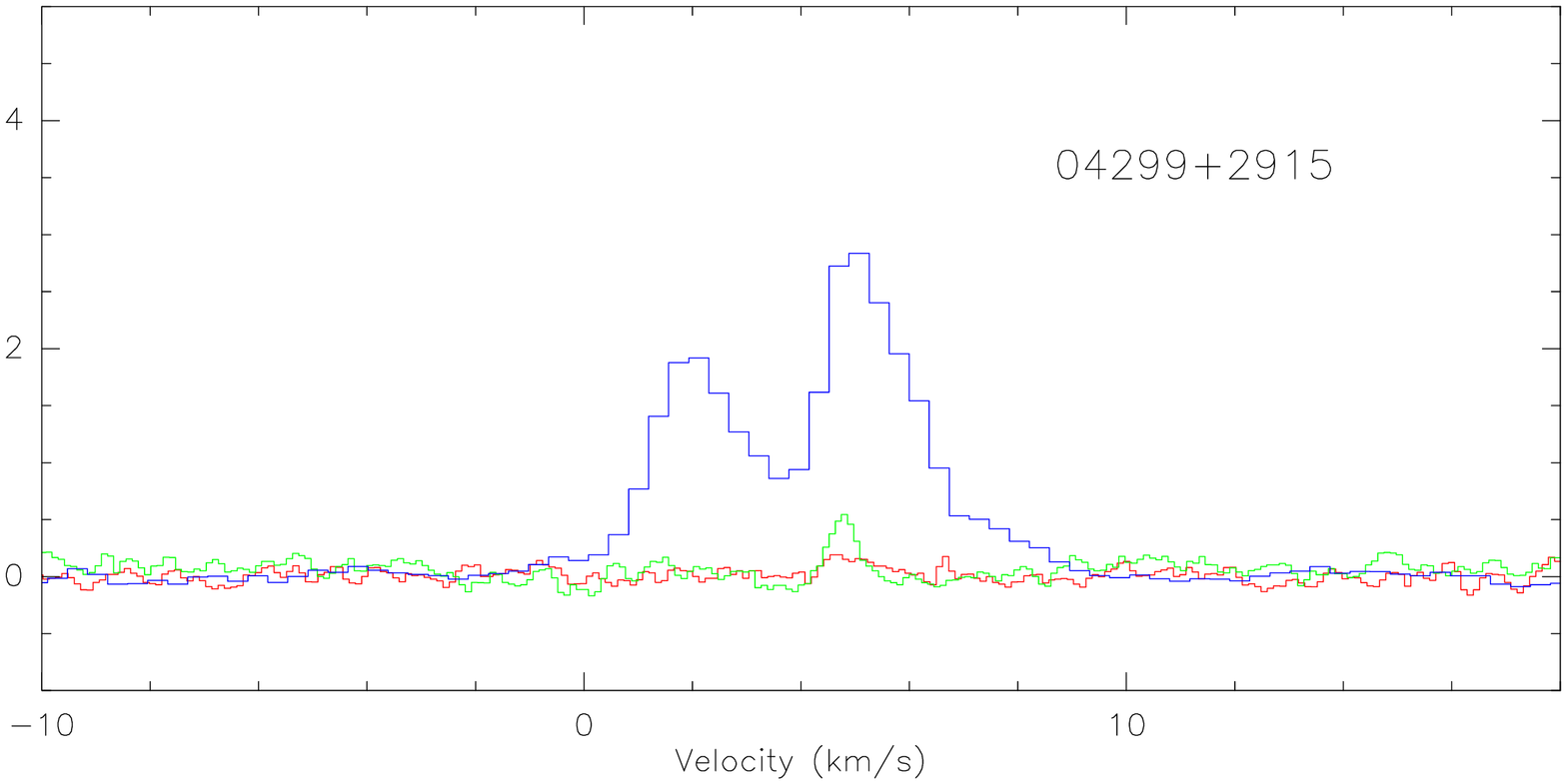}
\includegraphics[height=70mm,  angle=-90, clip, viewport=150 10 500 750]{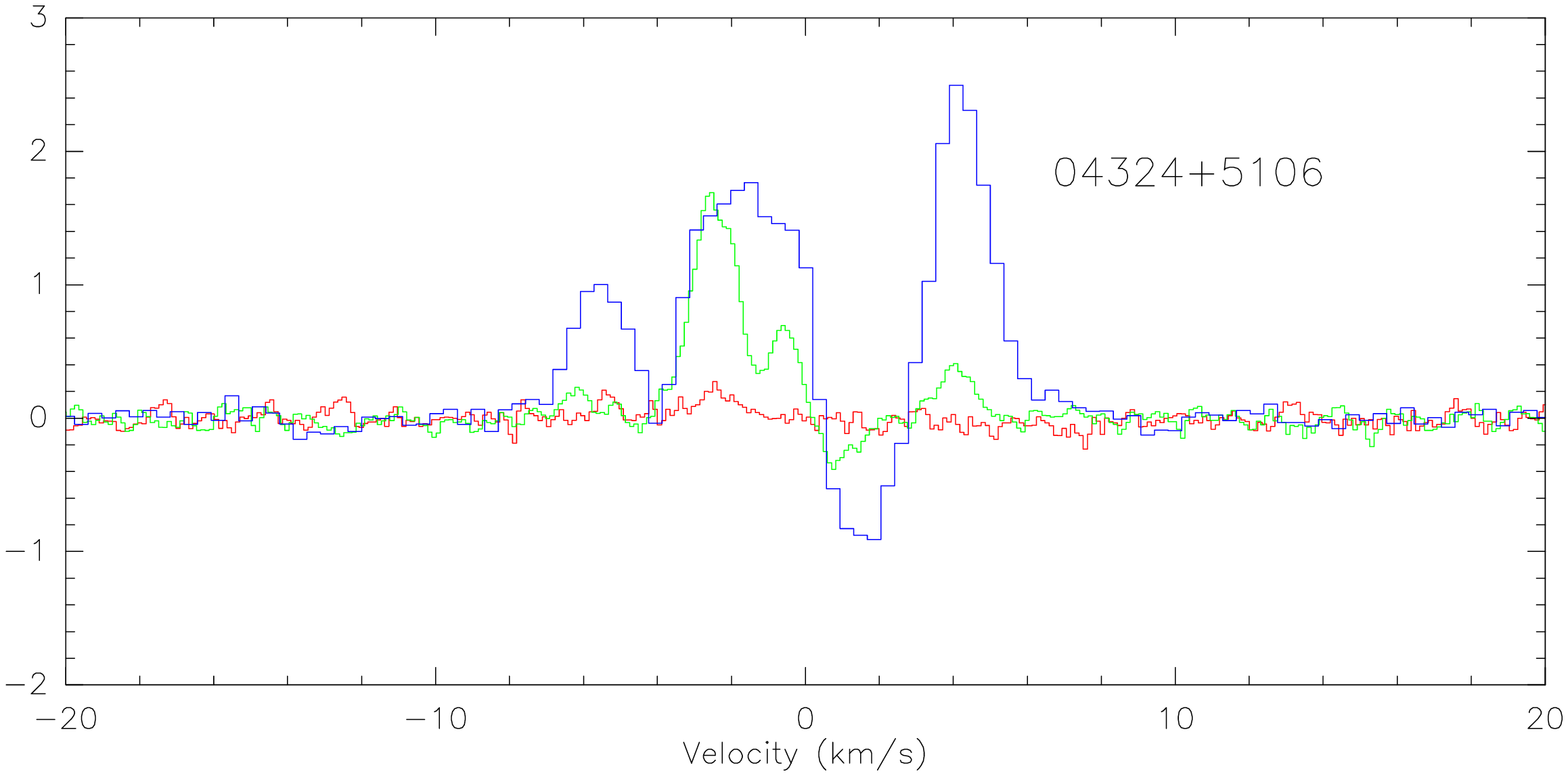}
\includegraphics[height=70mm,  angle=-90, clip, viewport=150 10 500 750]{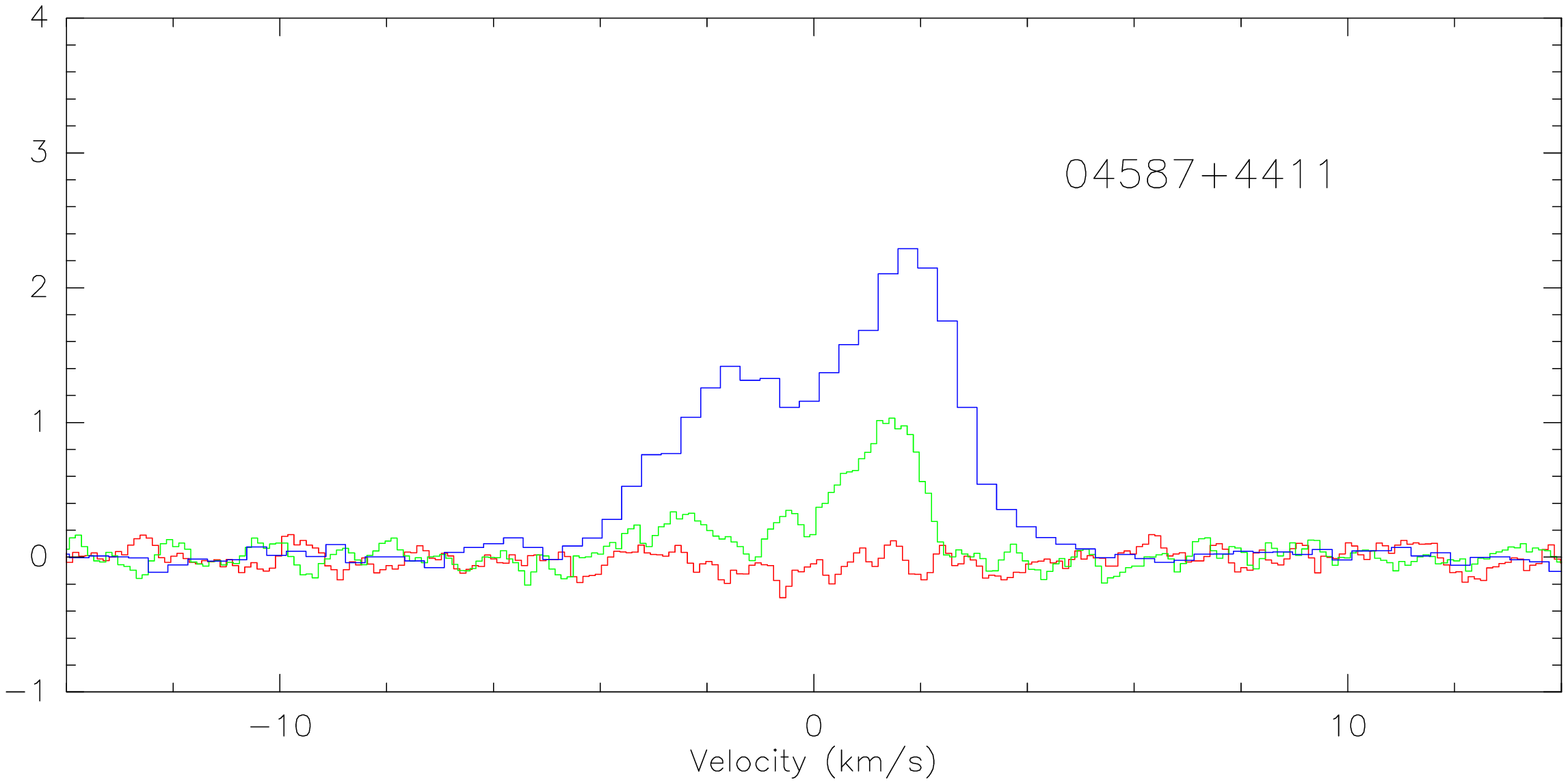}
\includegraphics[height=70mm,  angle=-90, clip, viewport=150 10 500 750]{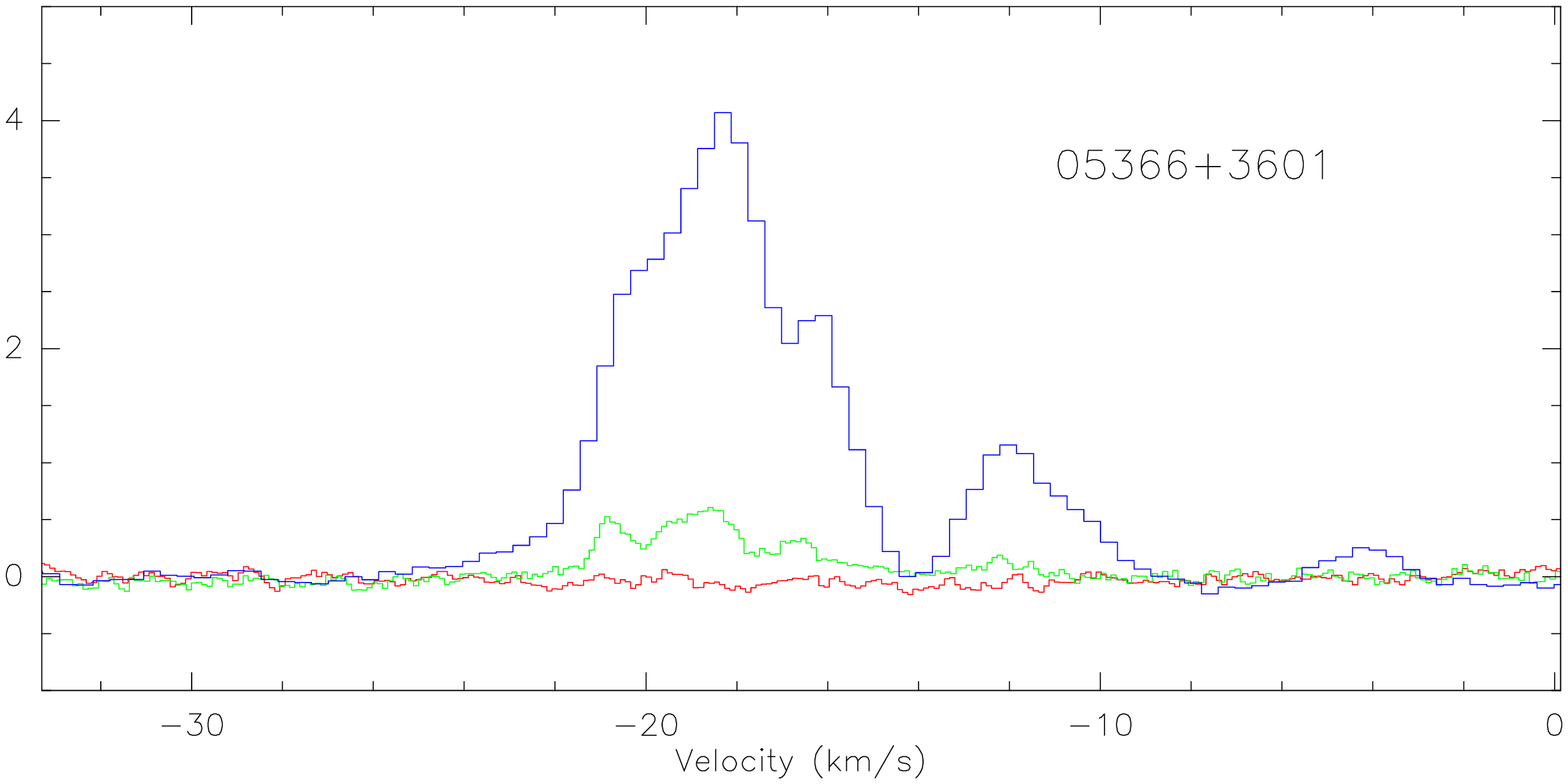}
\includegraphics[height=70mm,  angle=-90, clip, viewport=150 10 500 750]{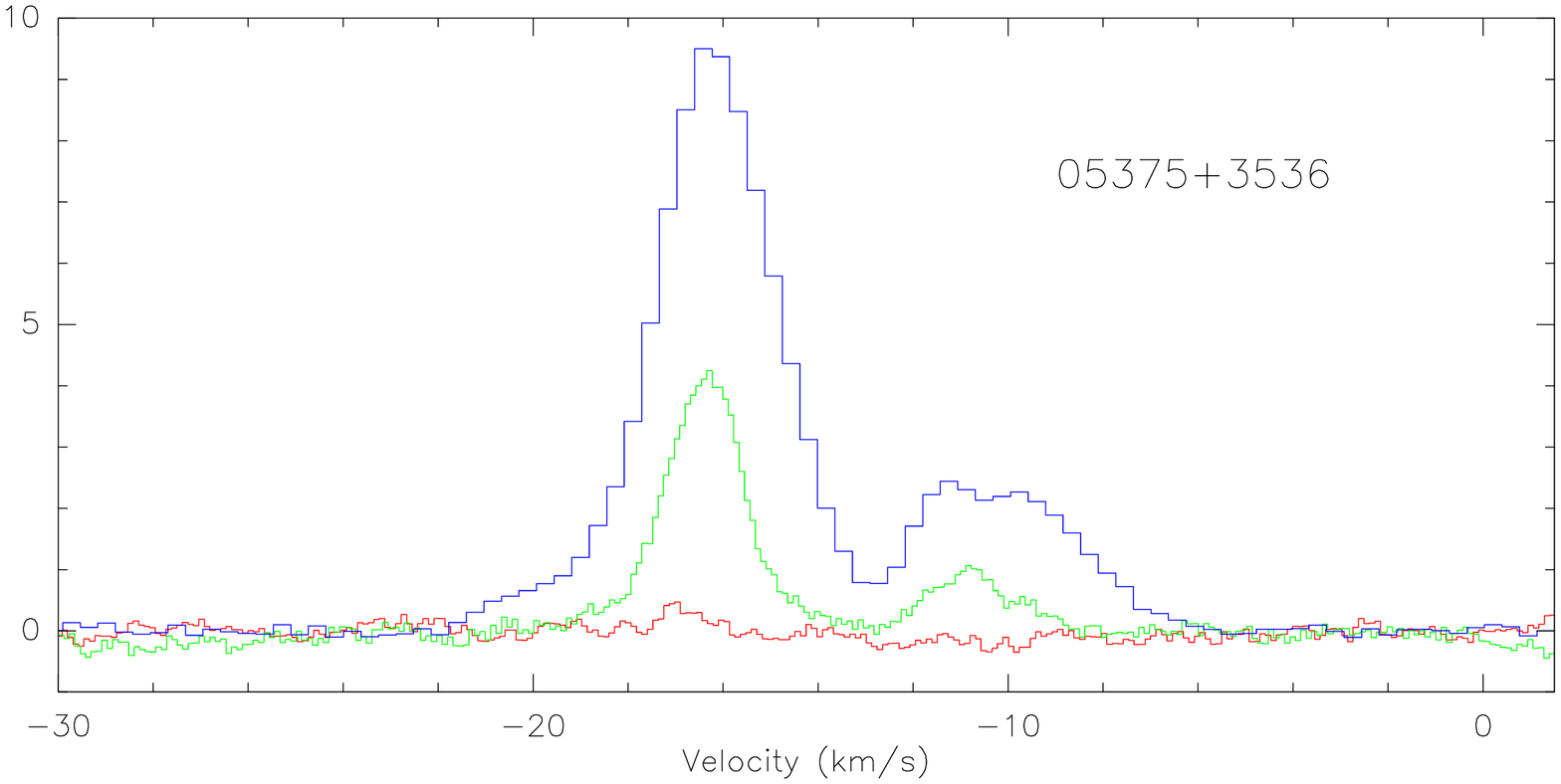}
\includegraphics[height=70mm,  angle=-90, clip, viewport=150 10 500 750]{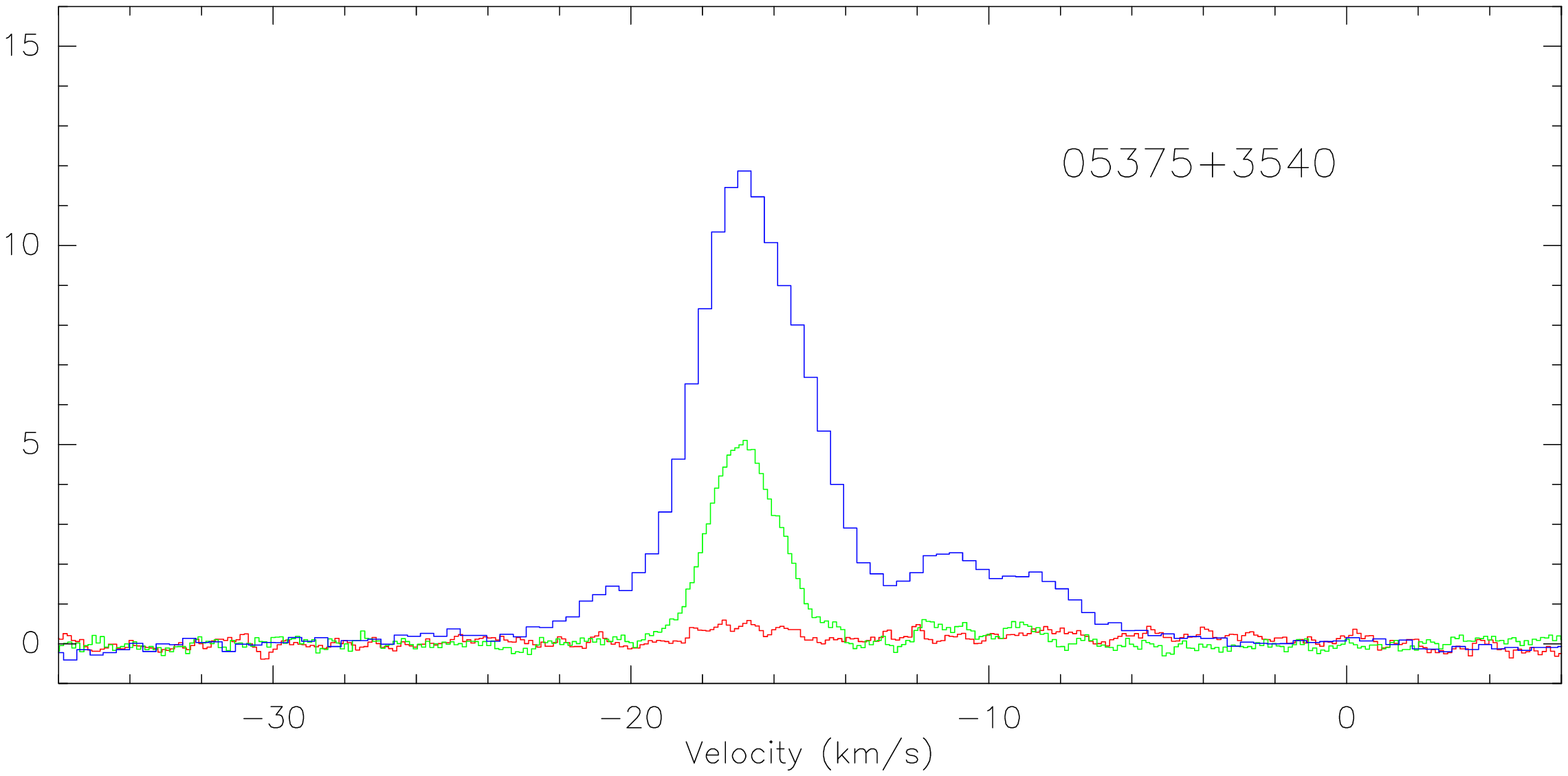}
\includegraphics[height=70mm,  angle=-90, clip, viewport=150 10 500 750]{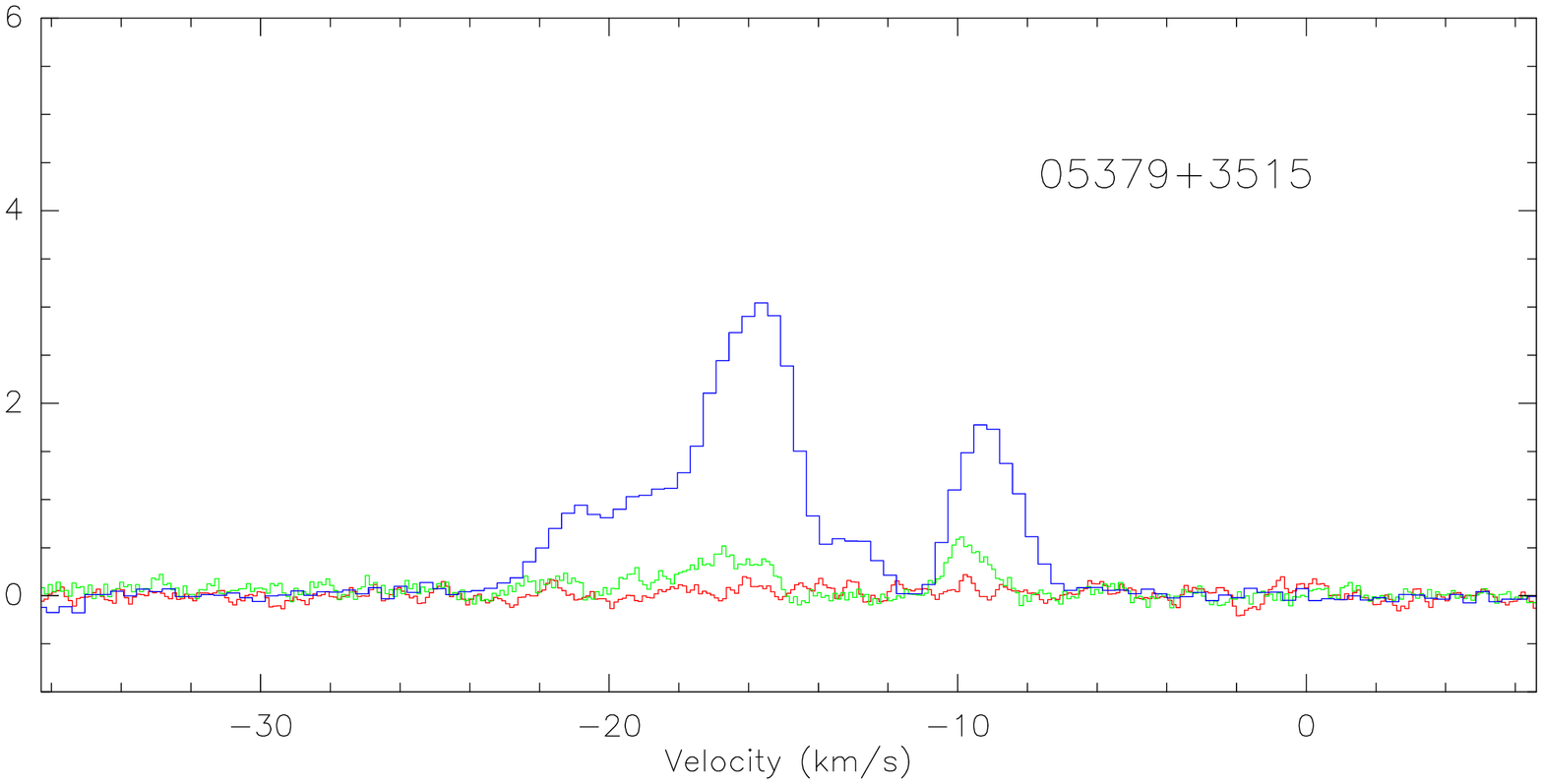}
\includegraphics[height=70mm,  angle=-90, clip, viewport=150 10 500 750]{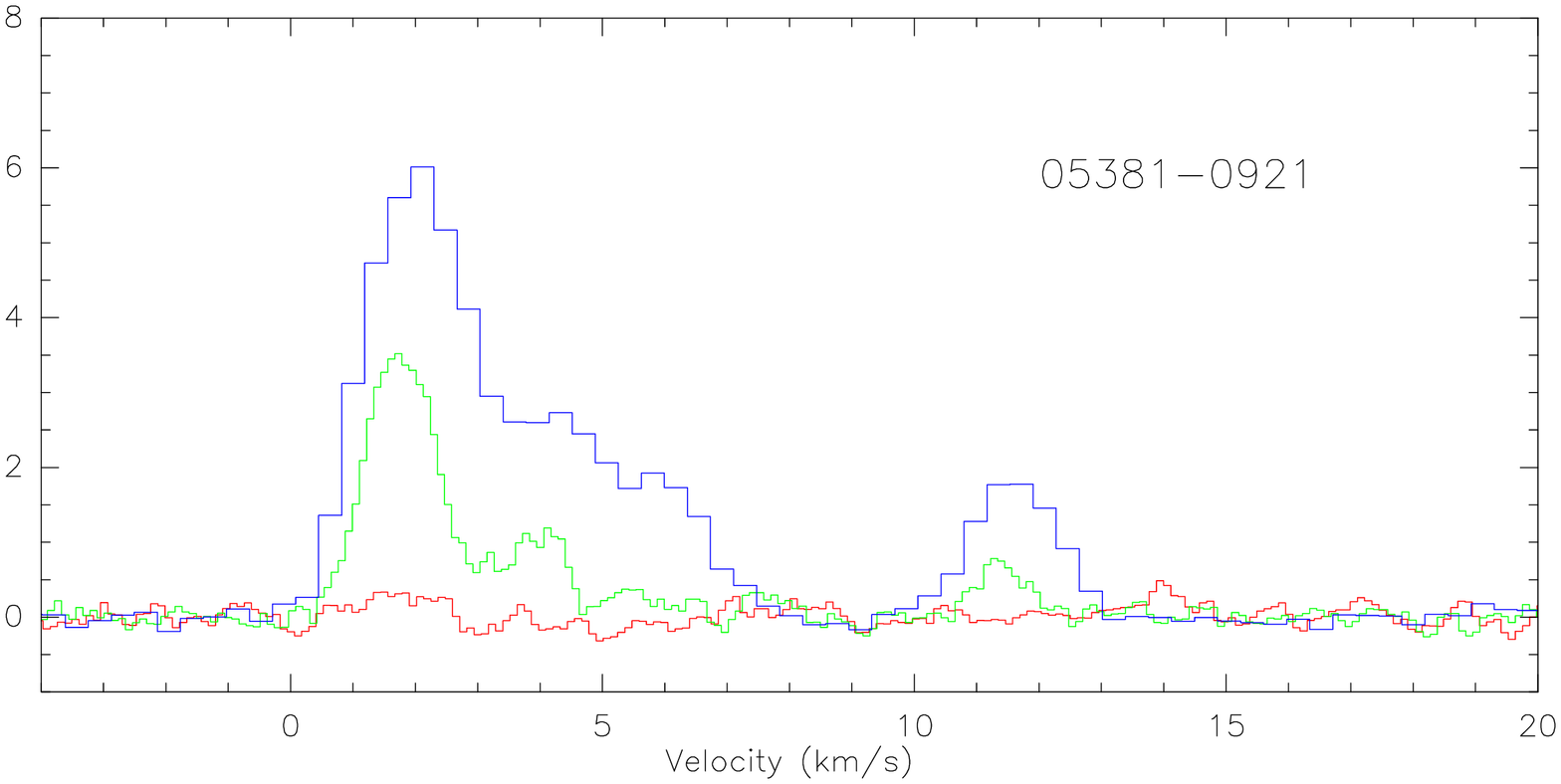}
\includegraphics[height=70mm,  angle=-90, clip, viewport=150 10 500 750]{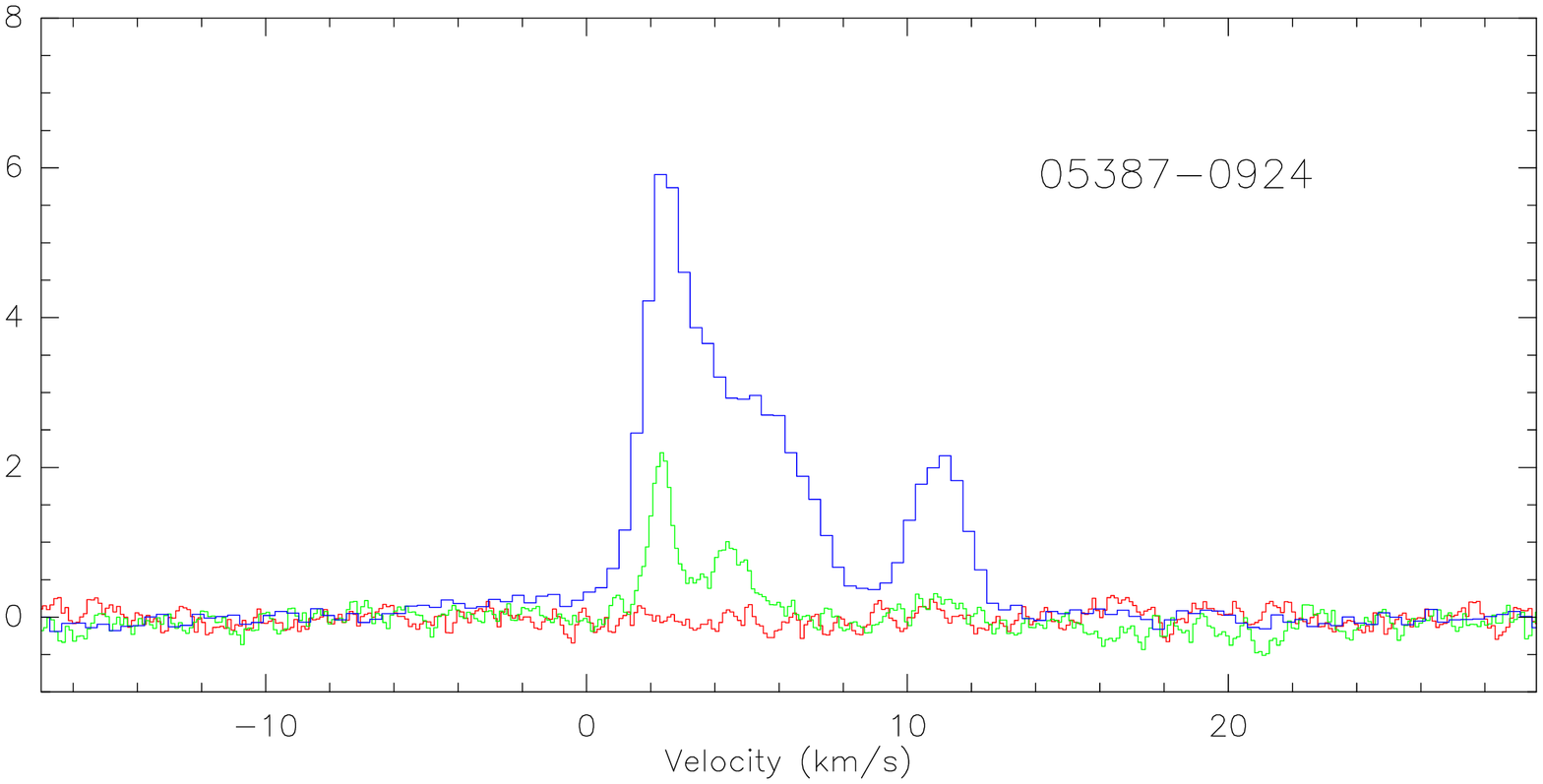}
\includegraphics[height=70mm,  angle=-90, clip, viewport=150 10 500 750]{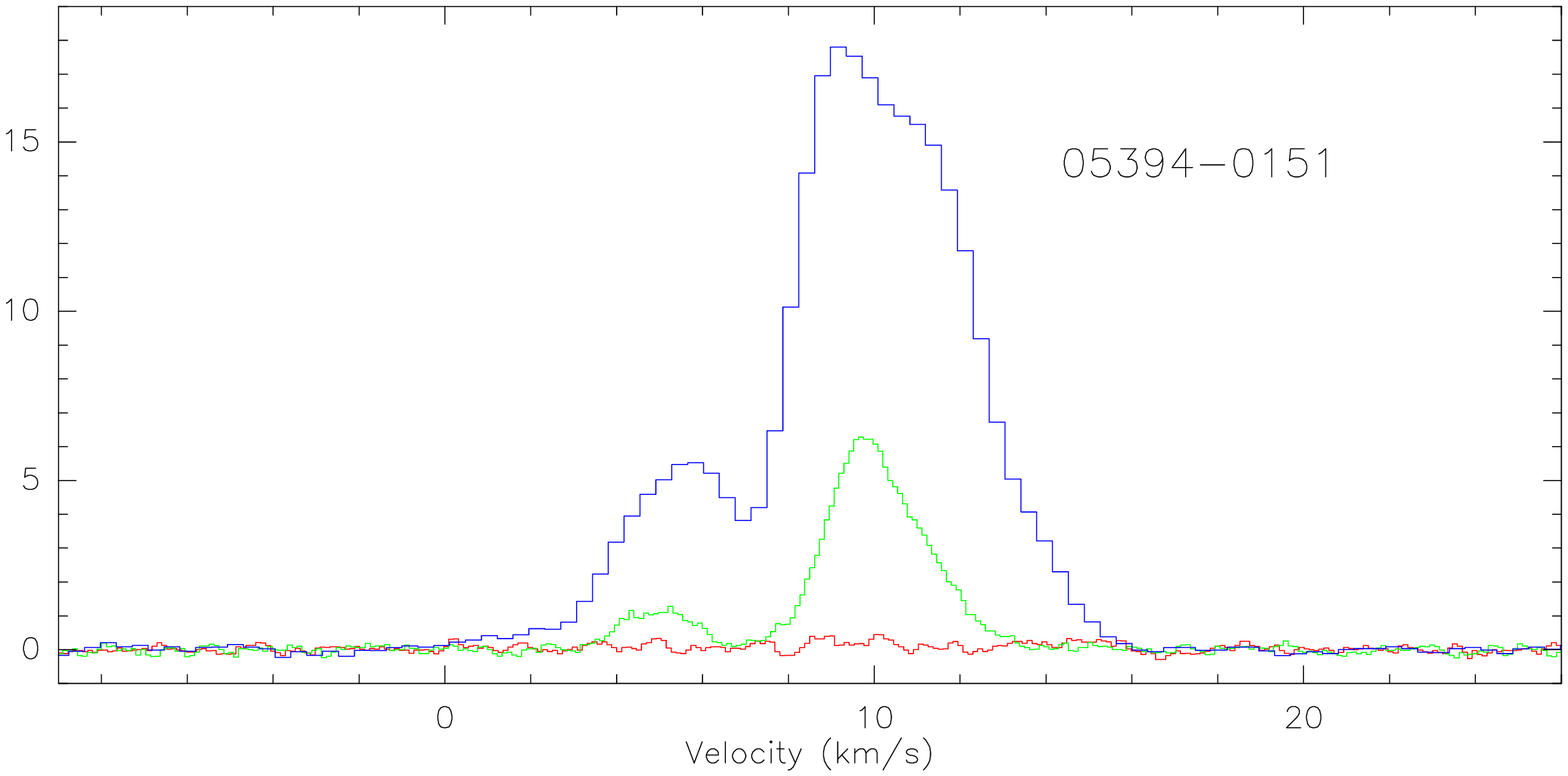}
\includegraphics[height=70mm,  angle=-90, clip, viewport=150 10 500 750]{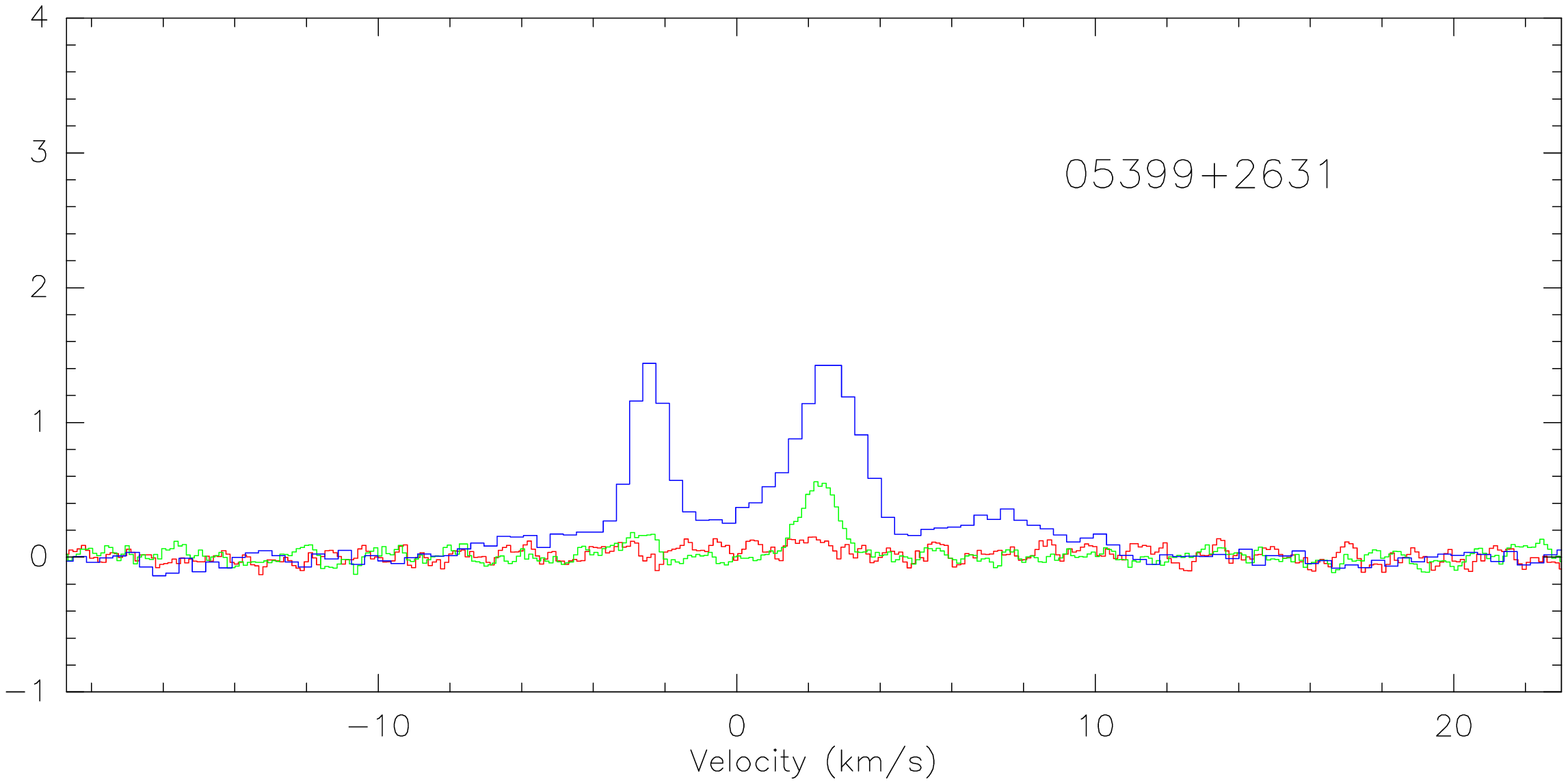}
\includegraphics[height=70mm,  angle=-90, clip, viewport=150 10 500 750]{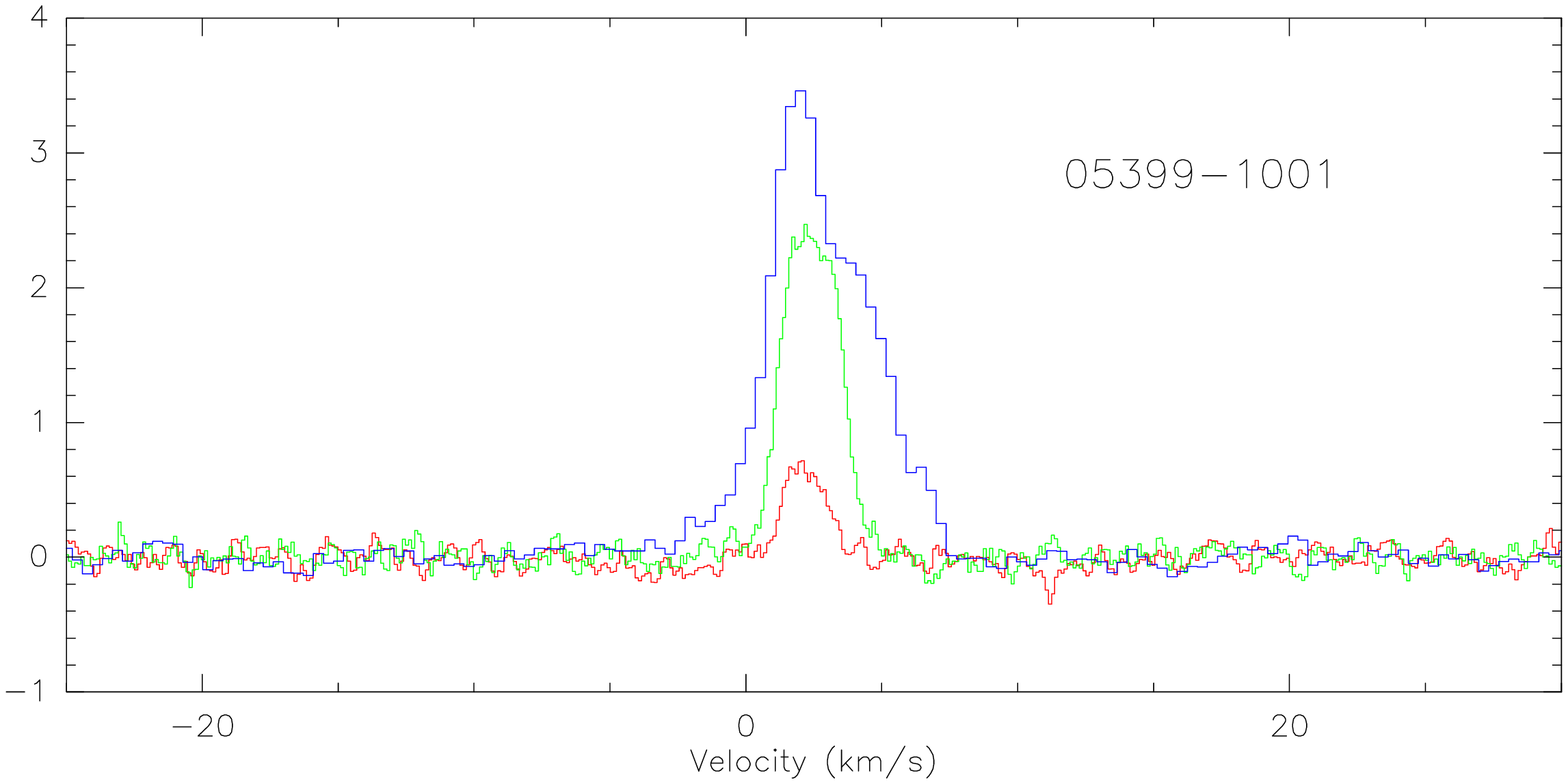}

\centering
\begin{minipage}[]{60mm}

   \caption{The sources of type 2
  }\end{minipage}
   \label{Fig7}
   \end{figure}

\addtocounter{figure}{-1}
\begin{figure}
\centering
\includegraphics[height=70mm,  angle=-90, clip, viewport=150 10 500 750]{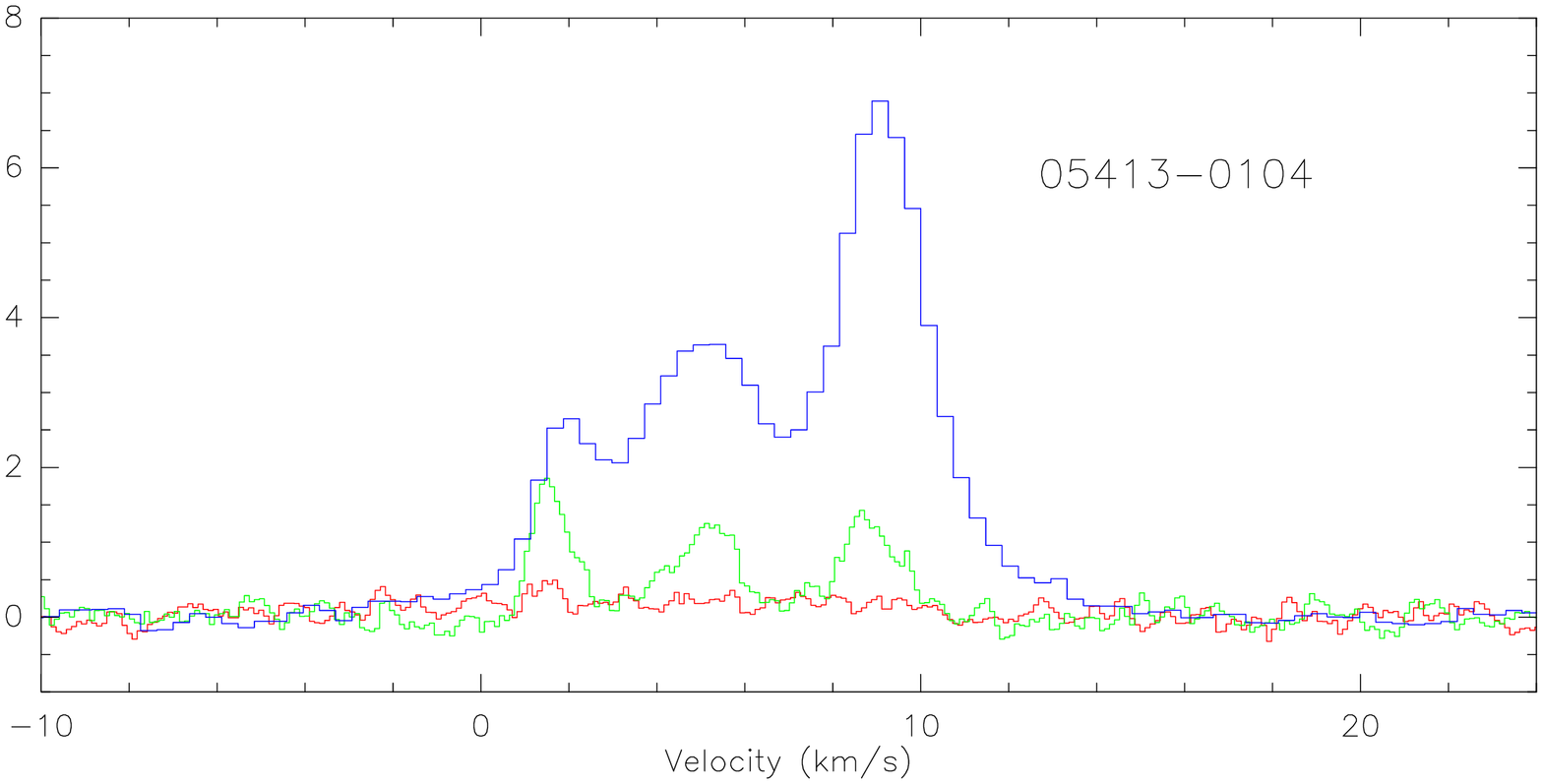}
\includegraphics[height=70mm,  angle=-90, clip, viewport=150 10 500 750]{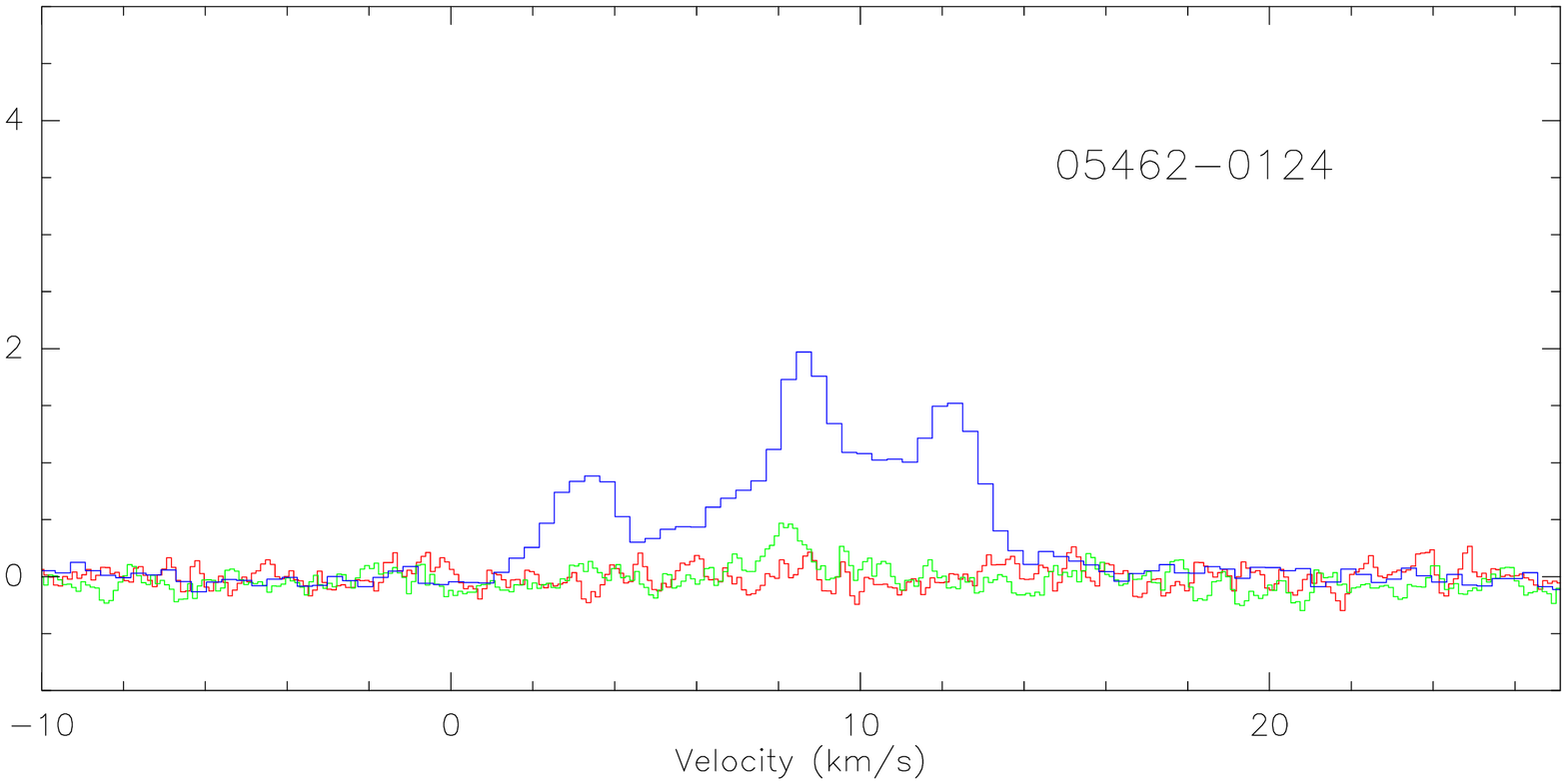}
\includegraphics[height=70mm,  angle=-90, clip, viewport=150 10 500 750]{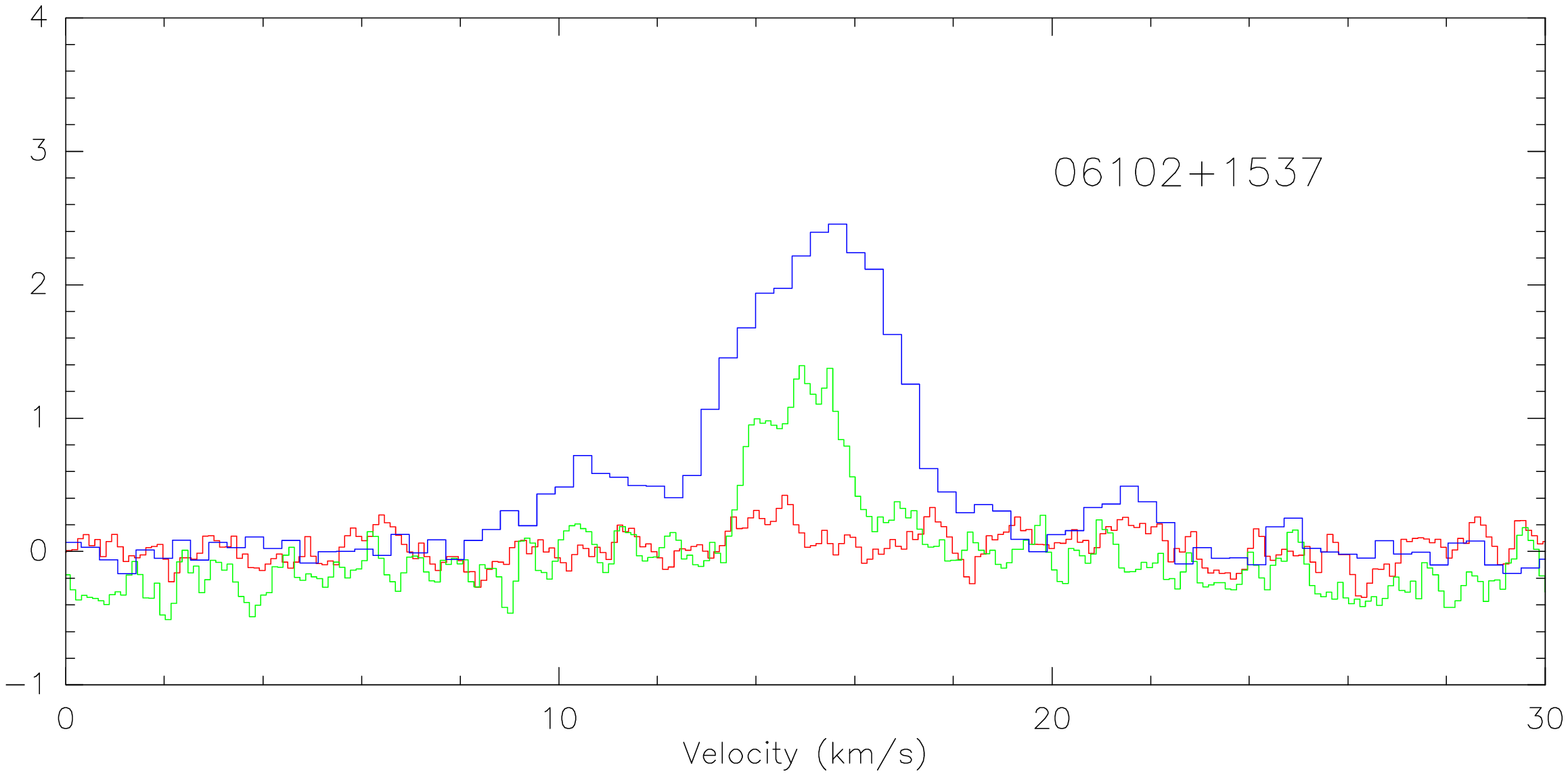}
\includegraphics[height=70mm,  angle=-90, clip, viewport=150 10 500 750]{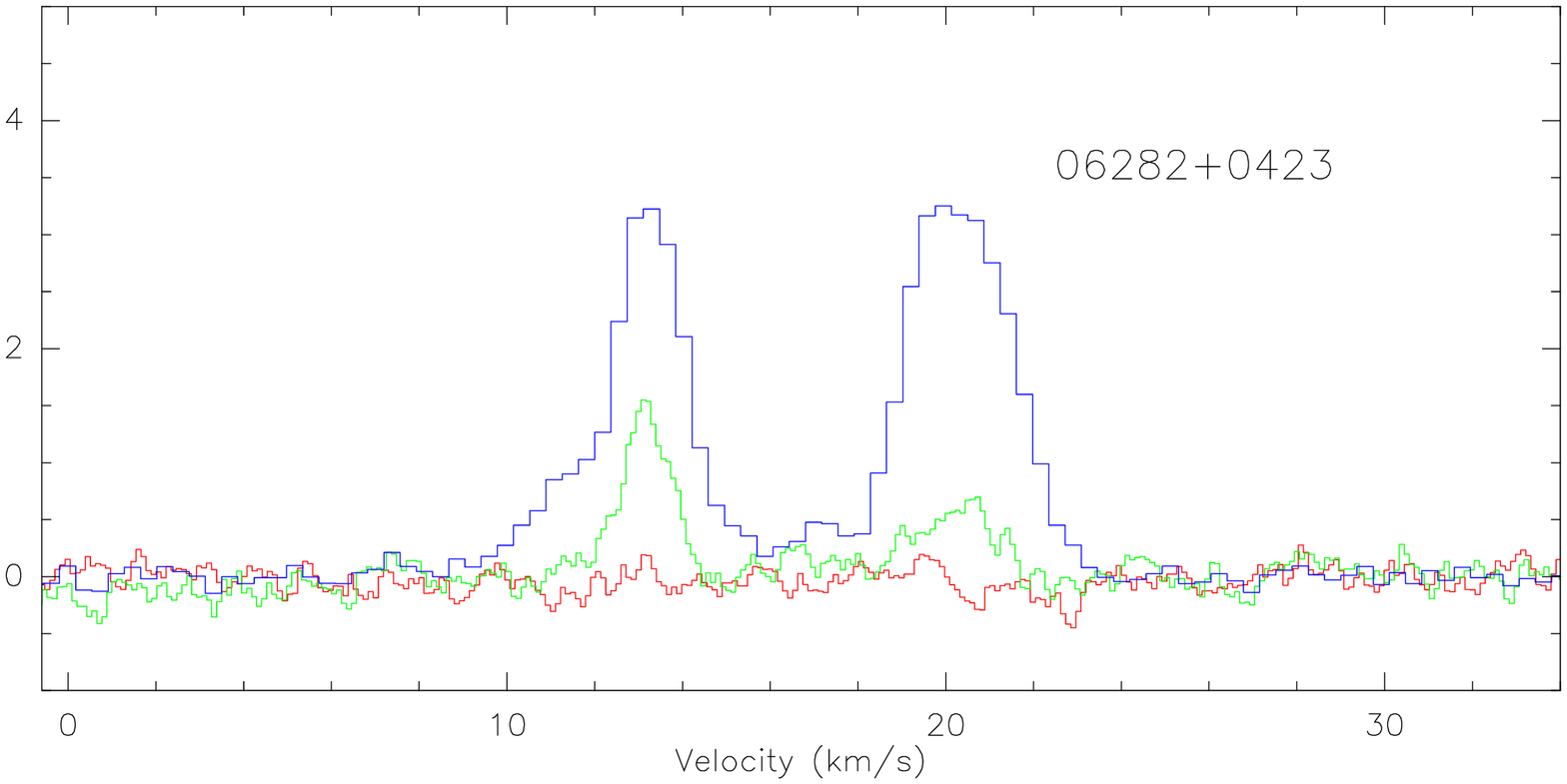}
\includegraphics[height=70mm,  angle=-90, clip, viewport=150 10 500 750]{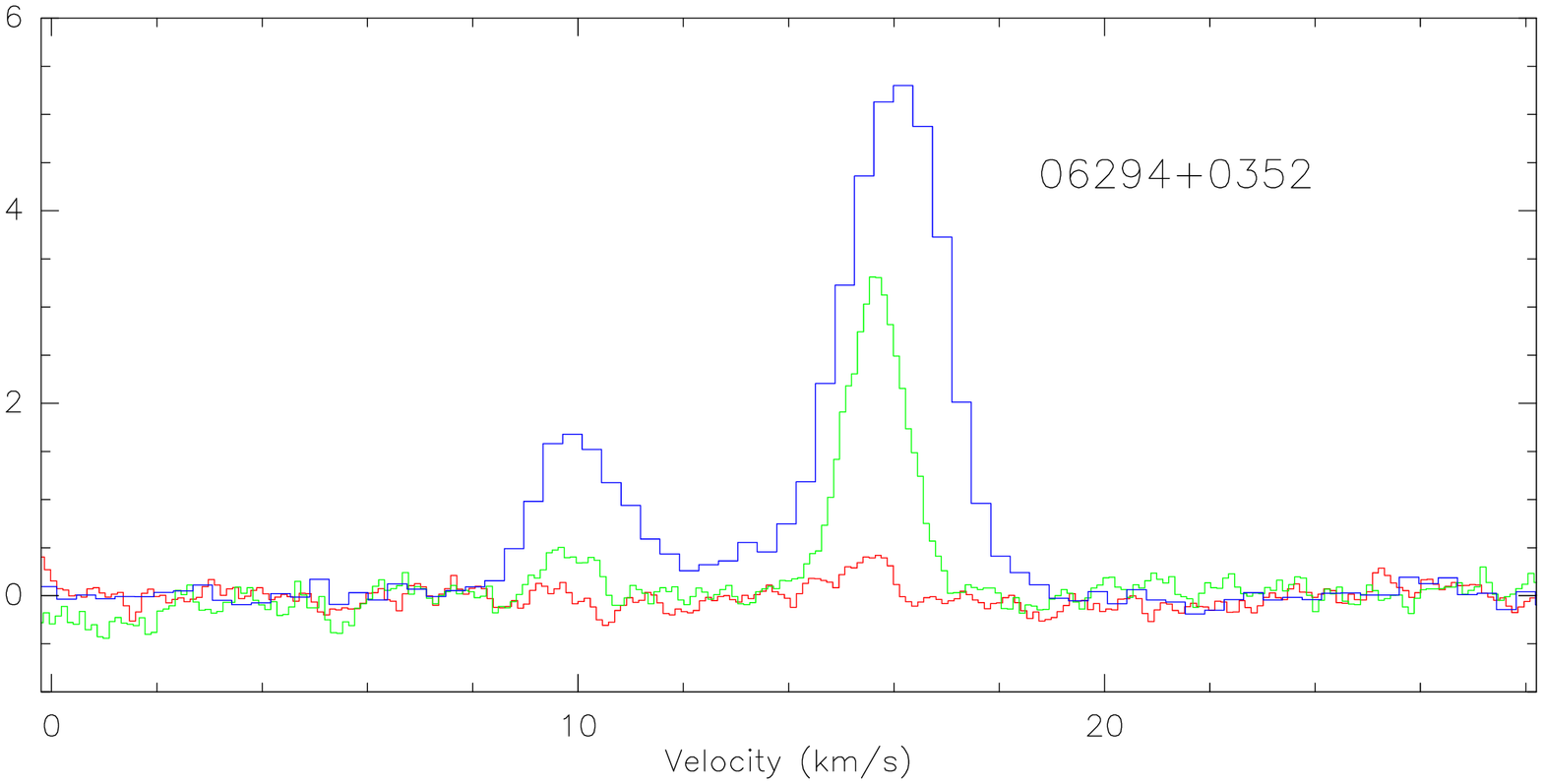}
\includegraphics[height=70mm,  angle=-90, clip, viewport=150 10 500 750]{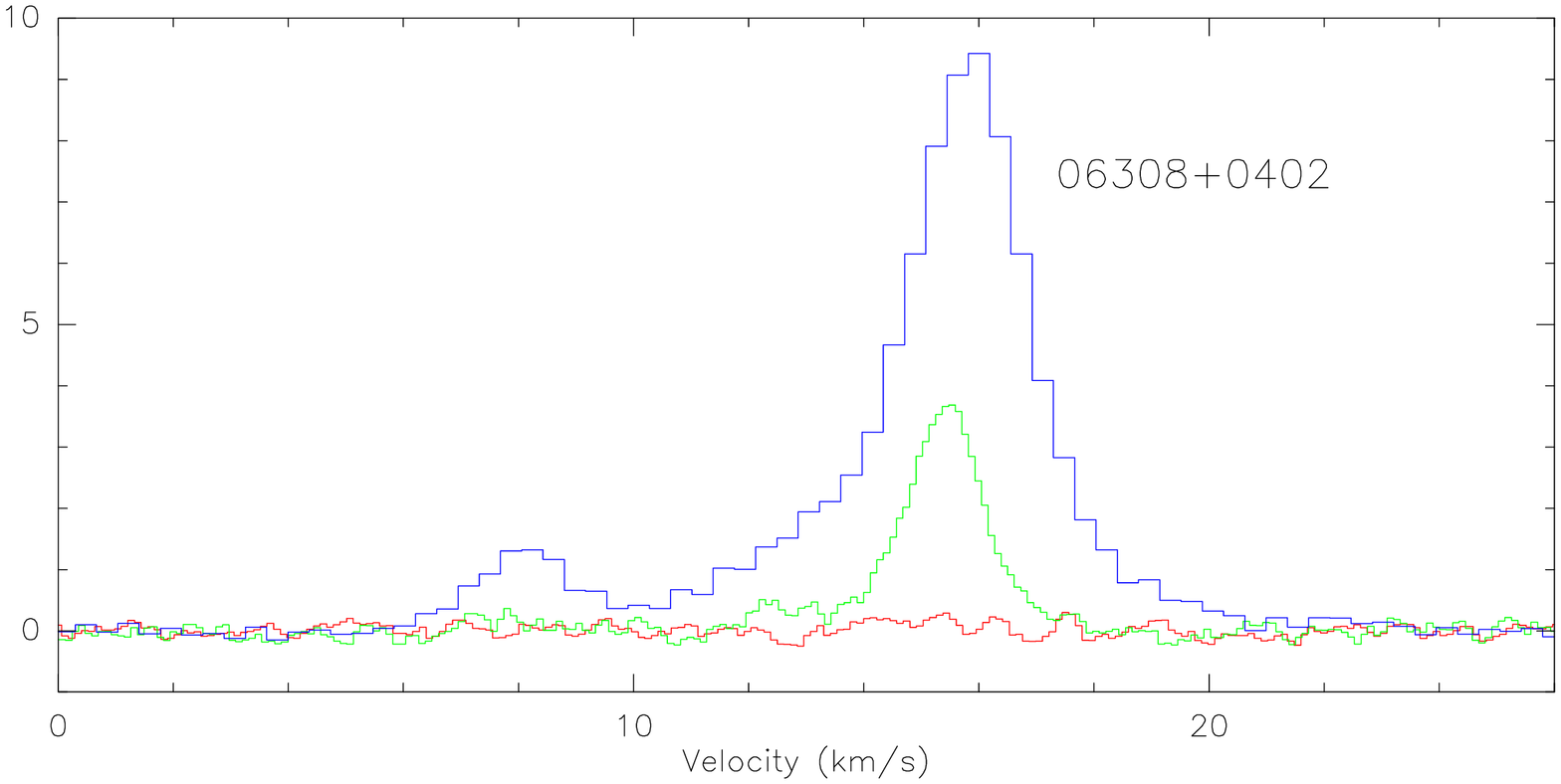}
\includegraphics[height=70mm,  angle=-90, clip, viewport=150 10 500 750]{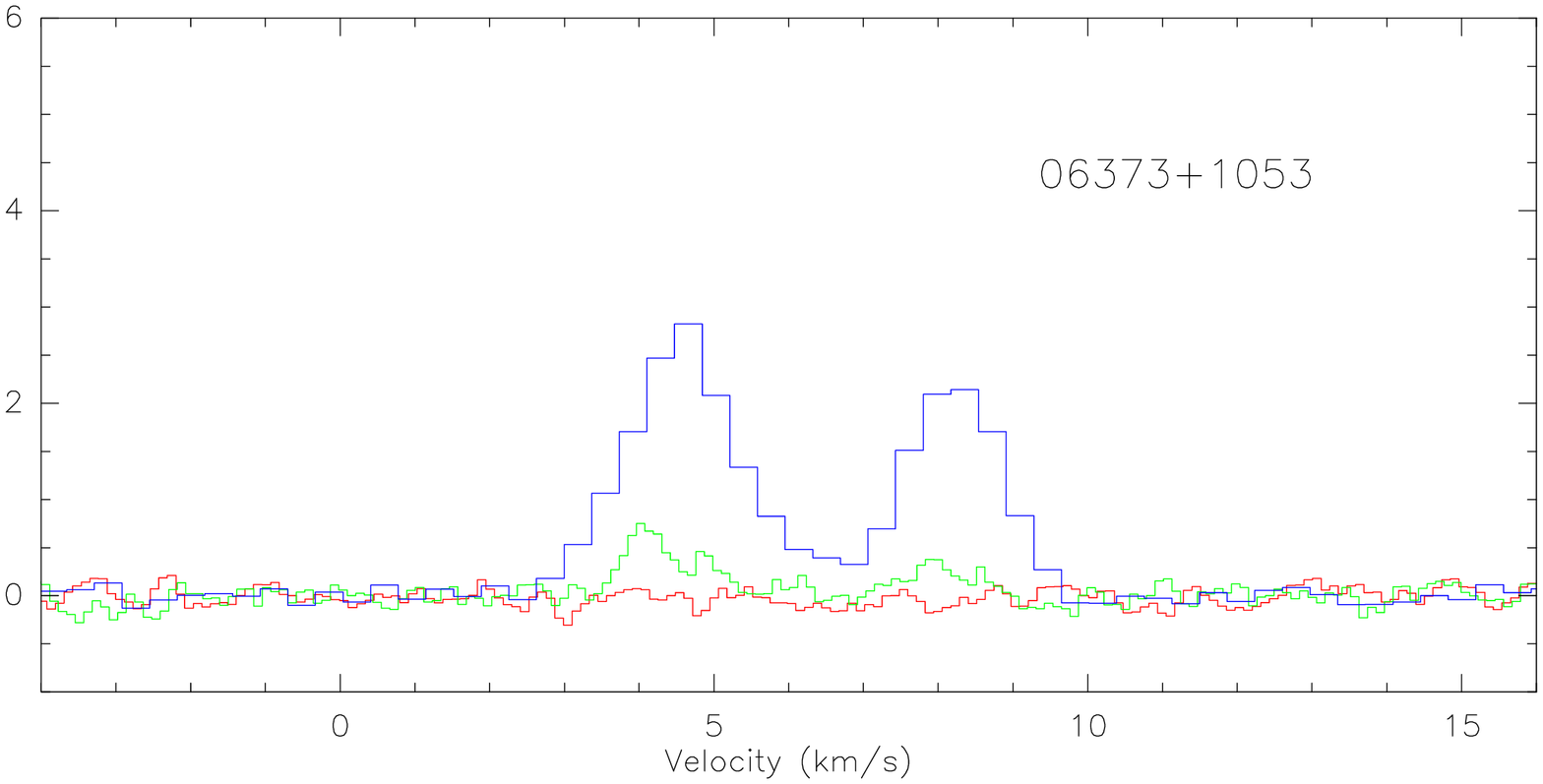}
\includegraphics[height=70mm,  angle=-90, clip, viewport=150 10 500 750]{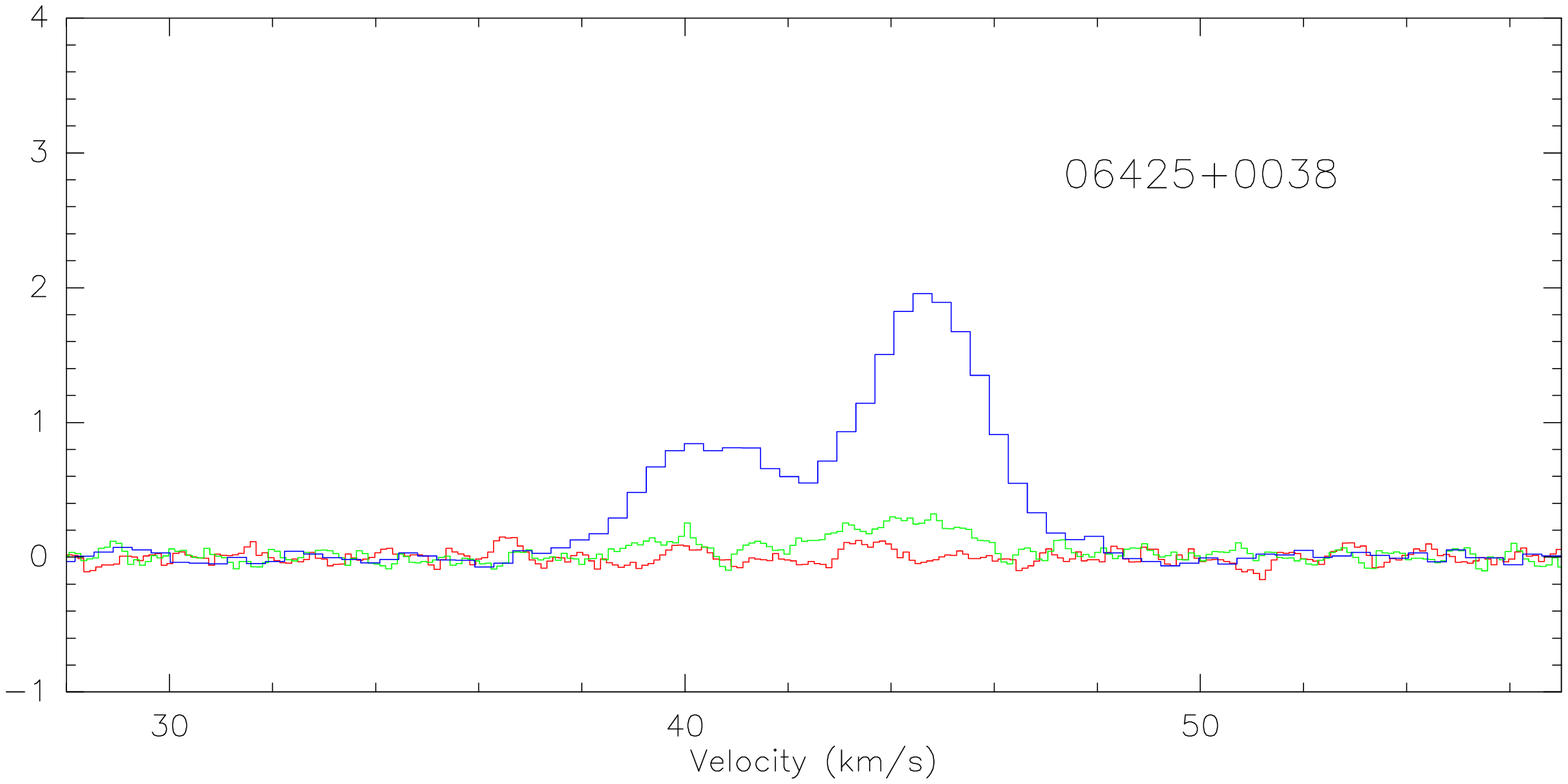}
\includegraphics[height=70mm,  angle=-90, clip, viewport=150 10 500 750]{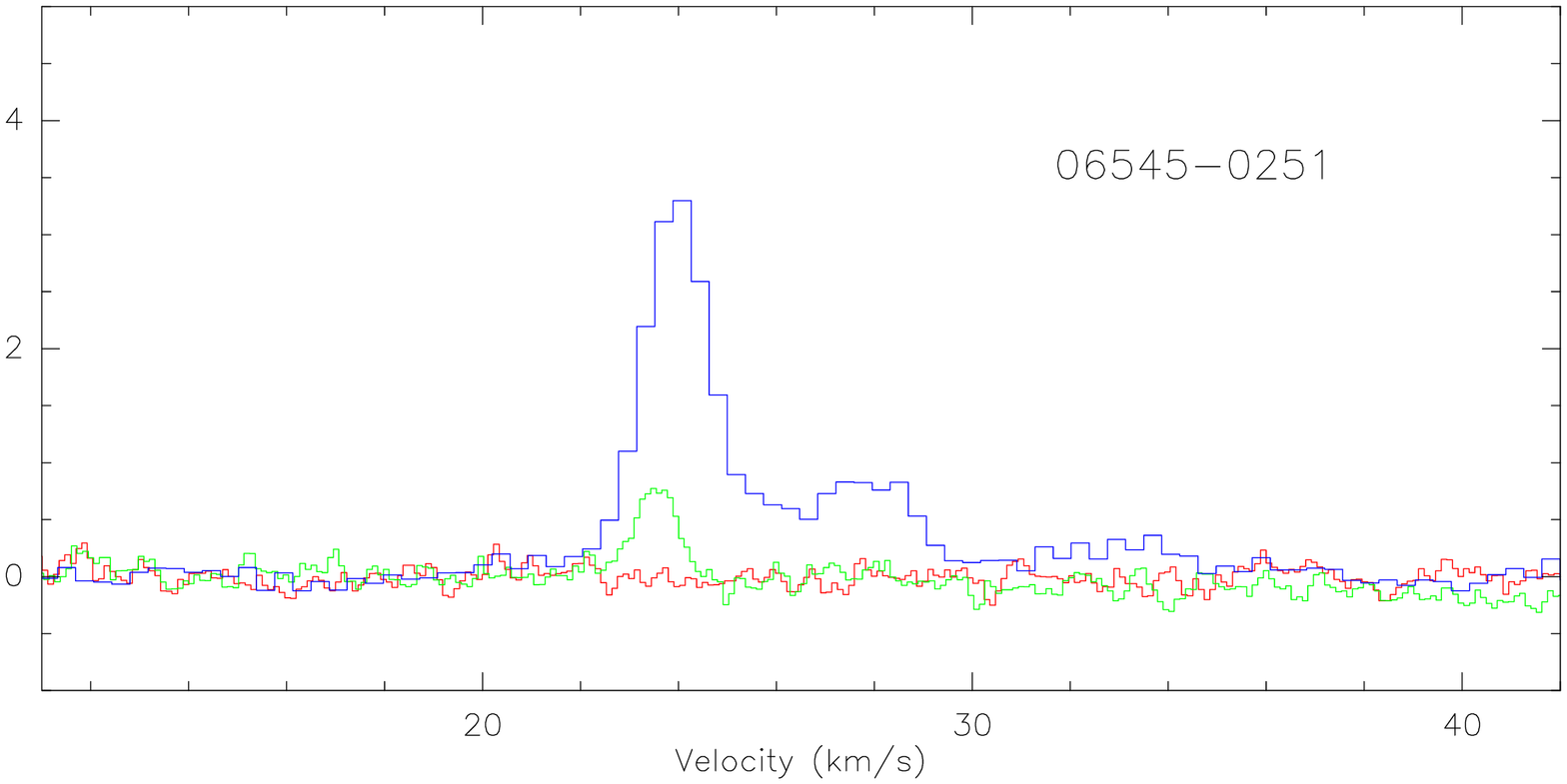}
\includegraphics[height=70mm,  angle=-90, clip, viewport=150 10 500 750]{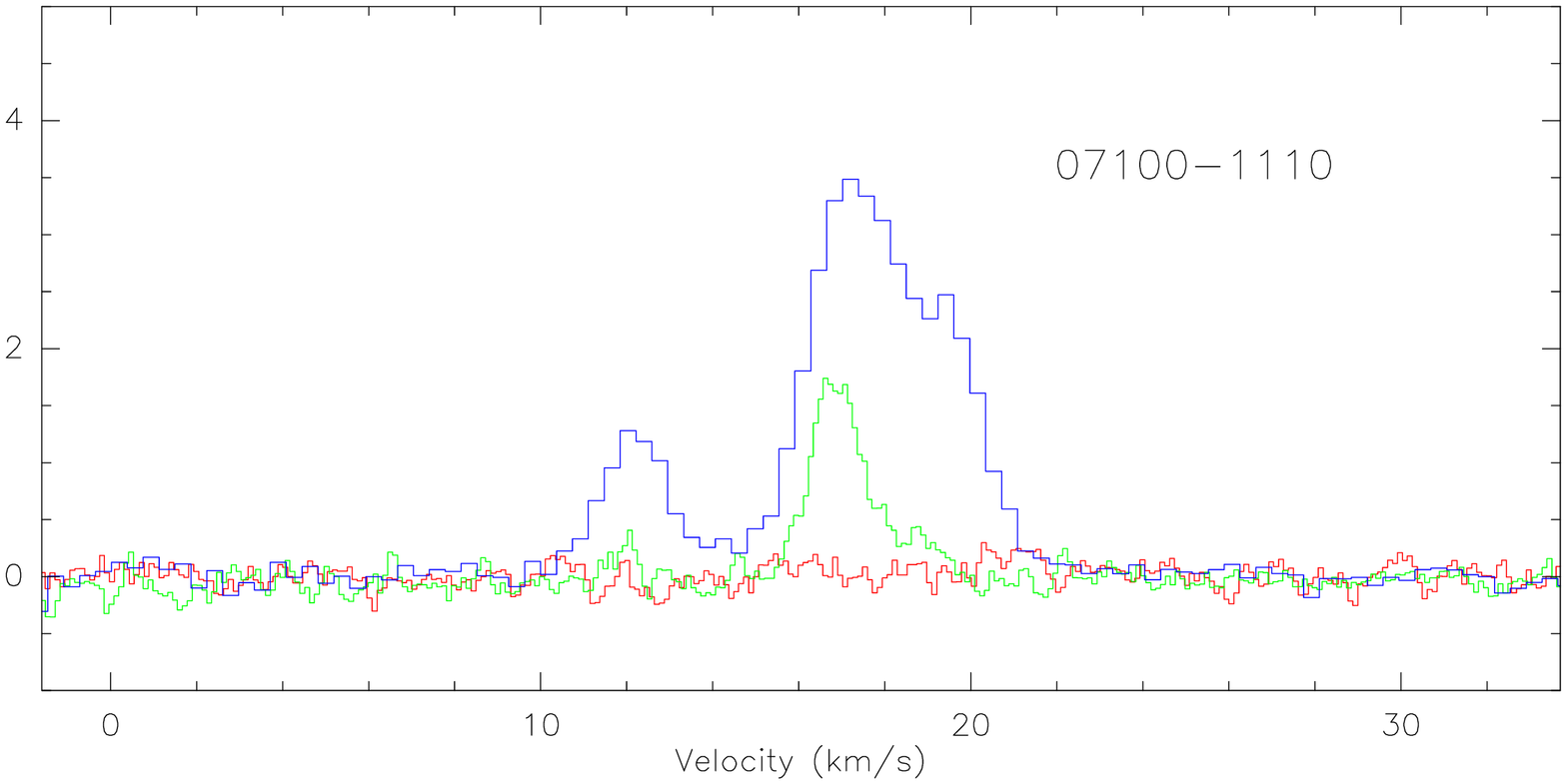}
\includegraphics[height=70mm,  angle=-90, clip, viewport=150 10 500 750]{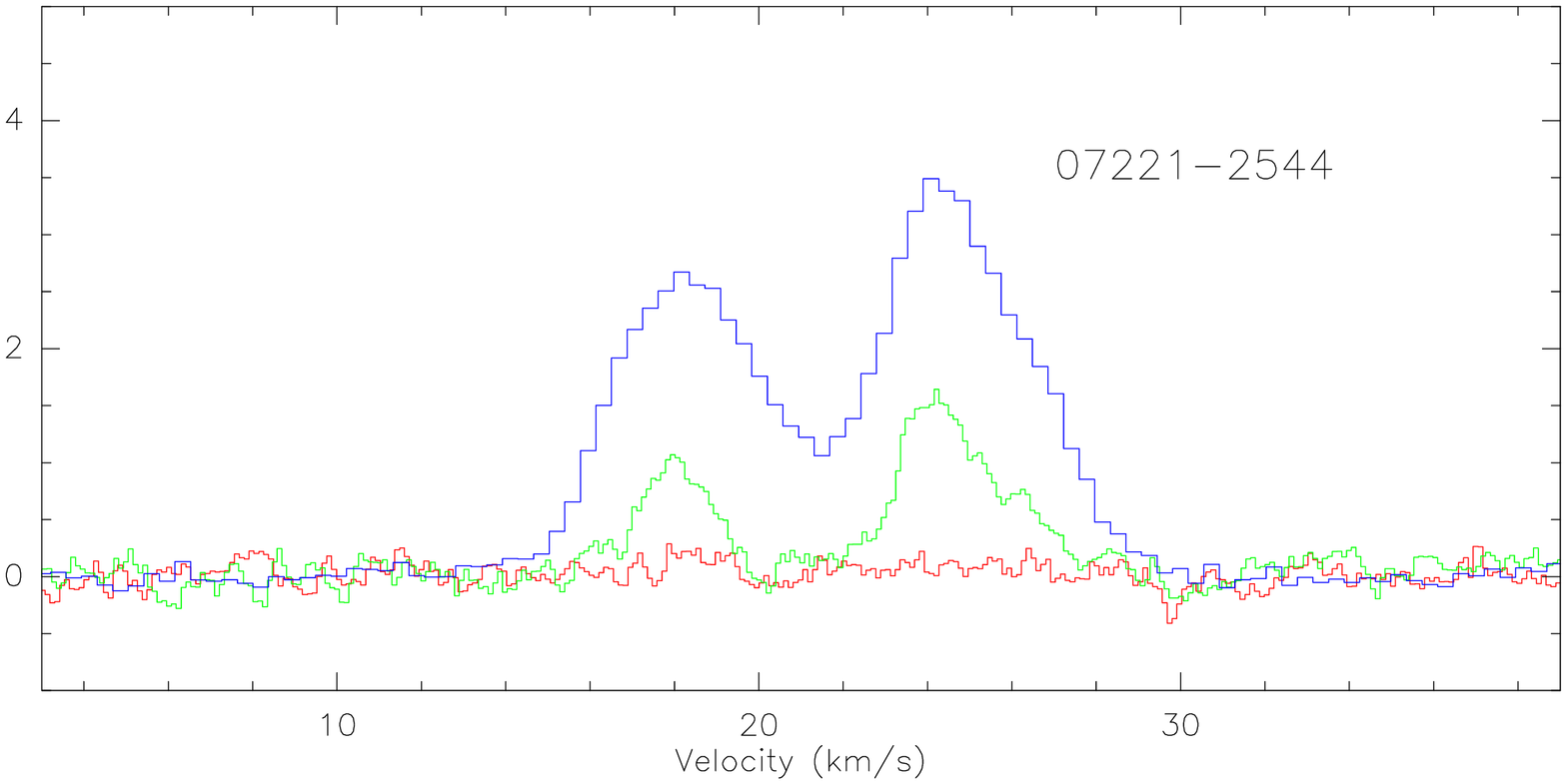}
\includegraphics[height=70mm,  angle=-90, clip, viewport=150 10 500 750]{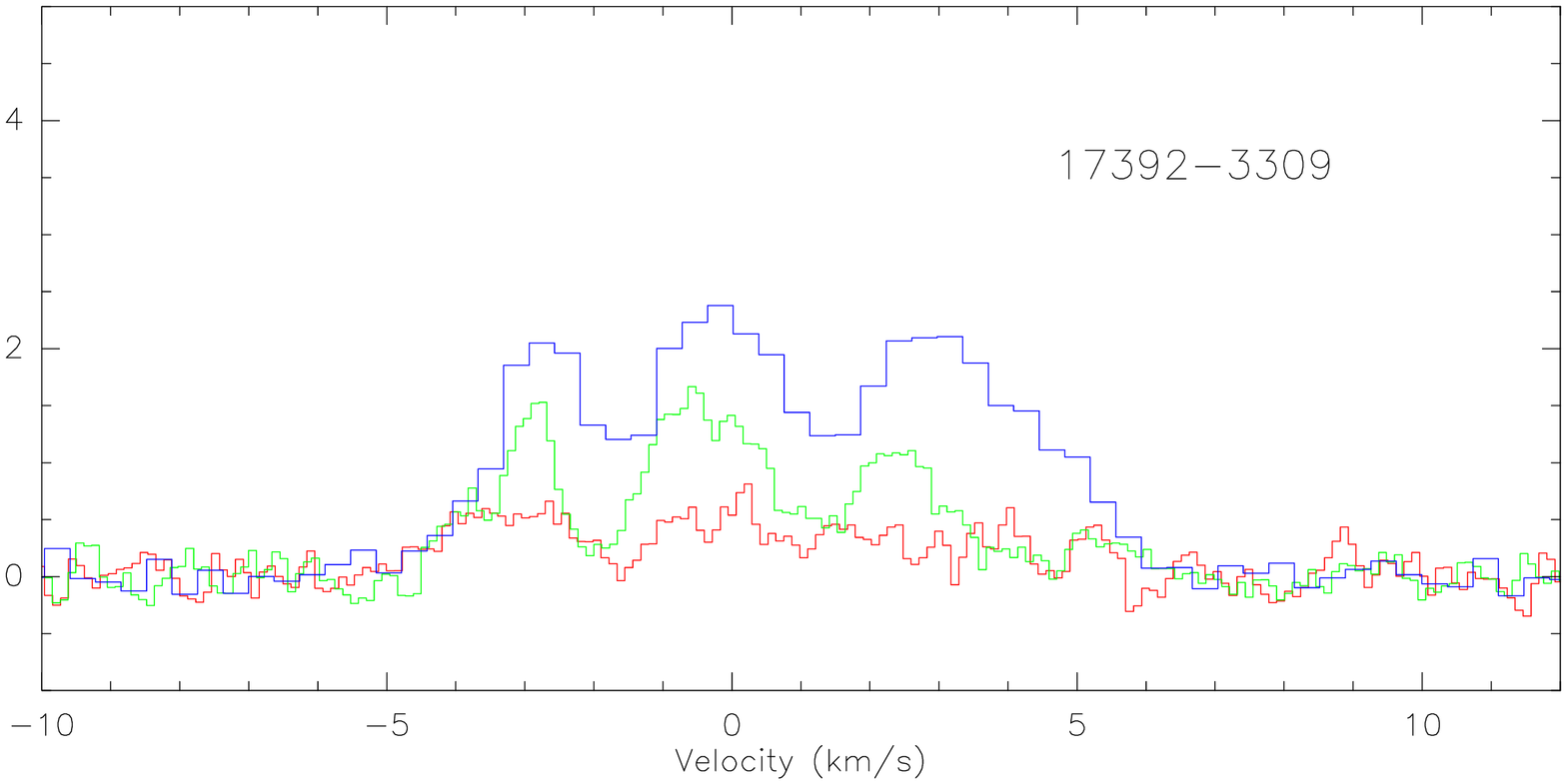}
\includegraphics[height=70mm,  angle=-90, clip, viewport=150 10 500 750]{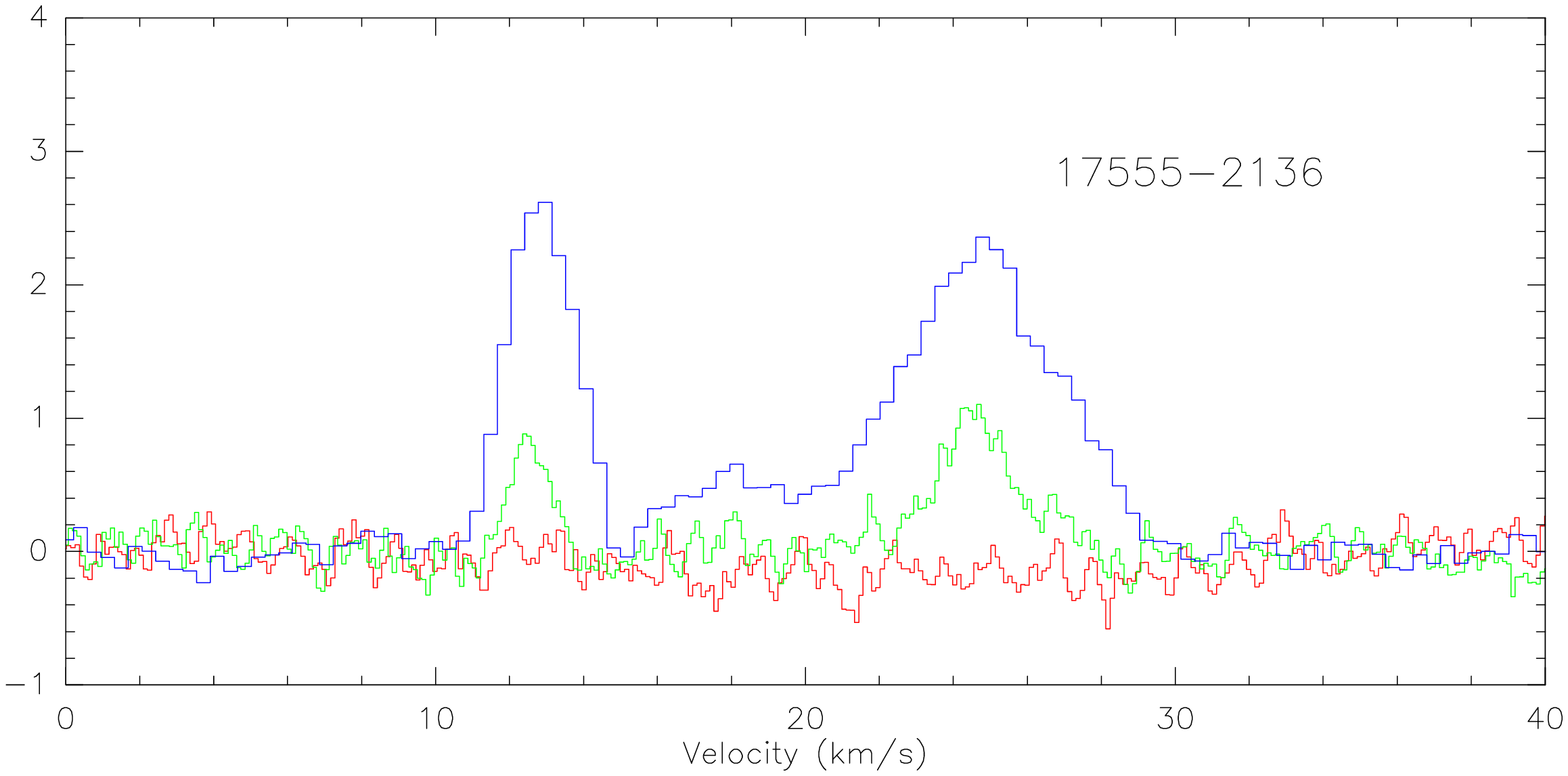}
\includegraphics[height=70mm,  angle=-90, clip, viewport=150 10 500 750]{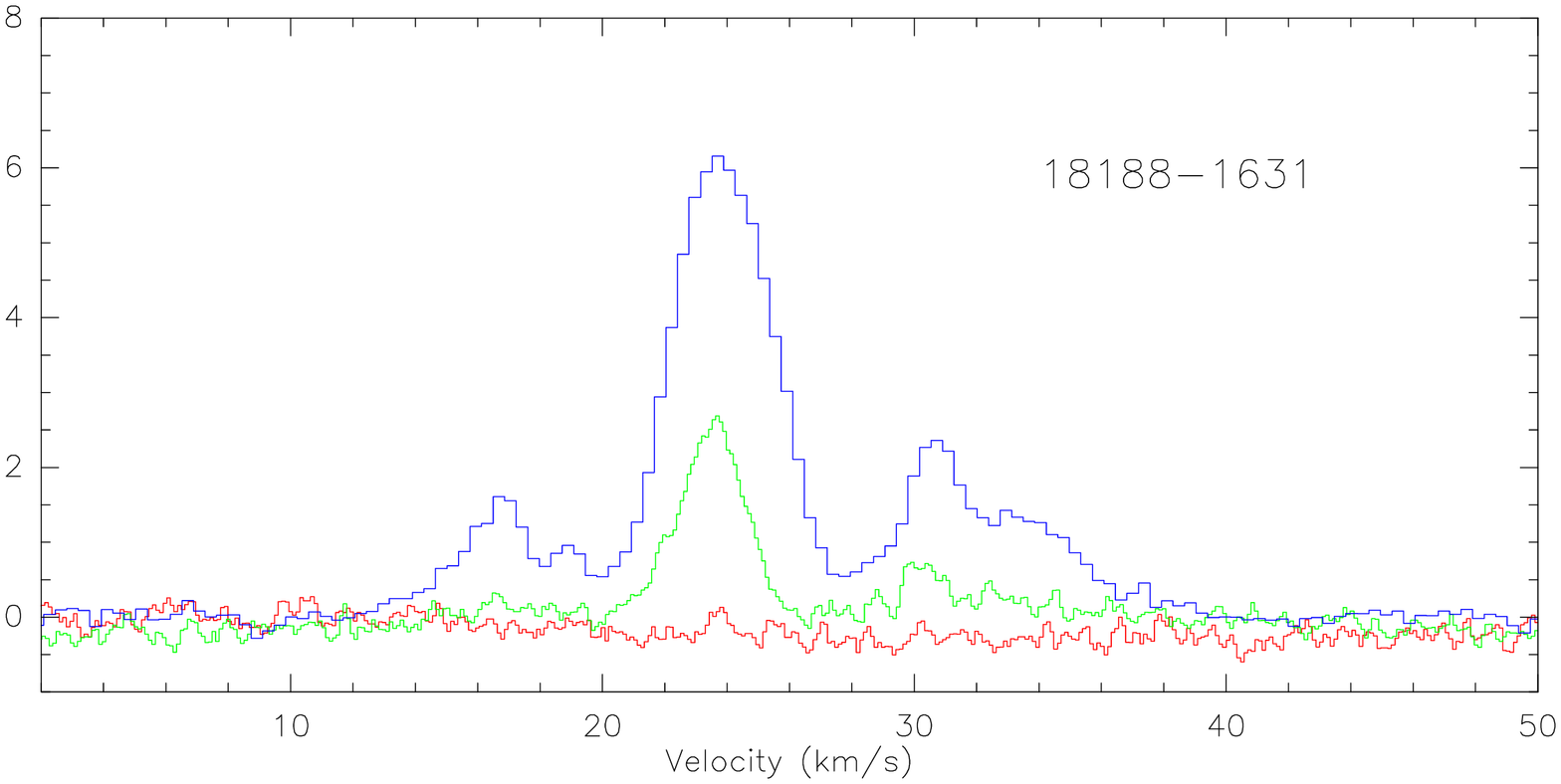}

\begin{minipage}[]{60mm}
  \caption{The sources of type 2
  }\end{minipage}
   \label{Fig7}
   \end{figure}

\addtocounter{figure}{-1}
\begin{figure}
\centering
\includegraphics[height=70mm,  angle=-90, clip, viewport=150 10 500 750]{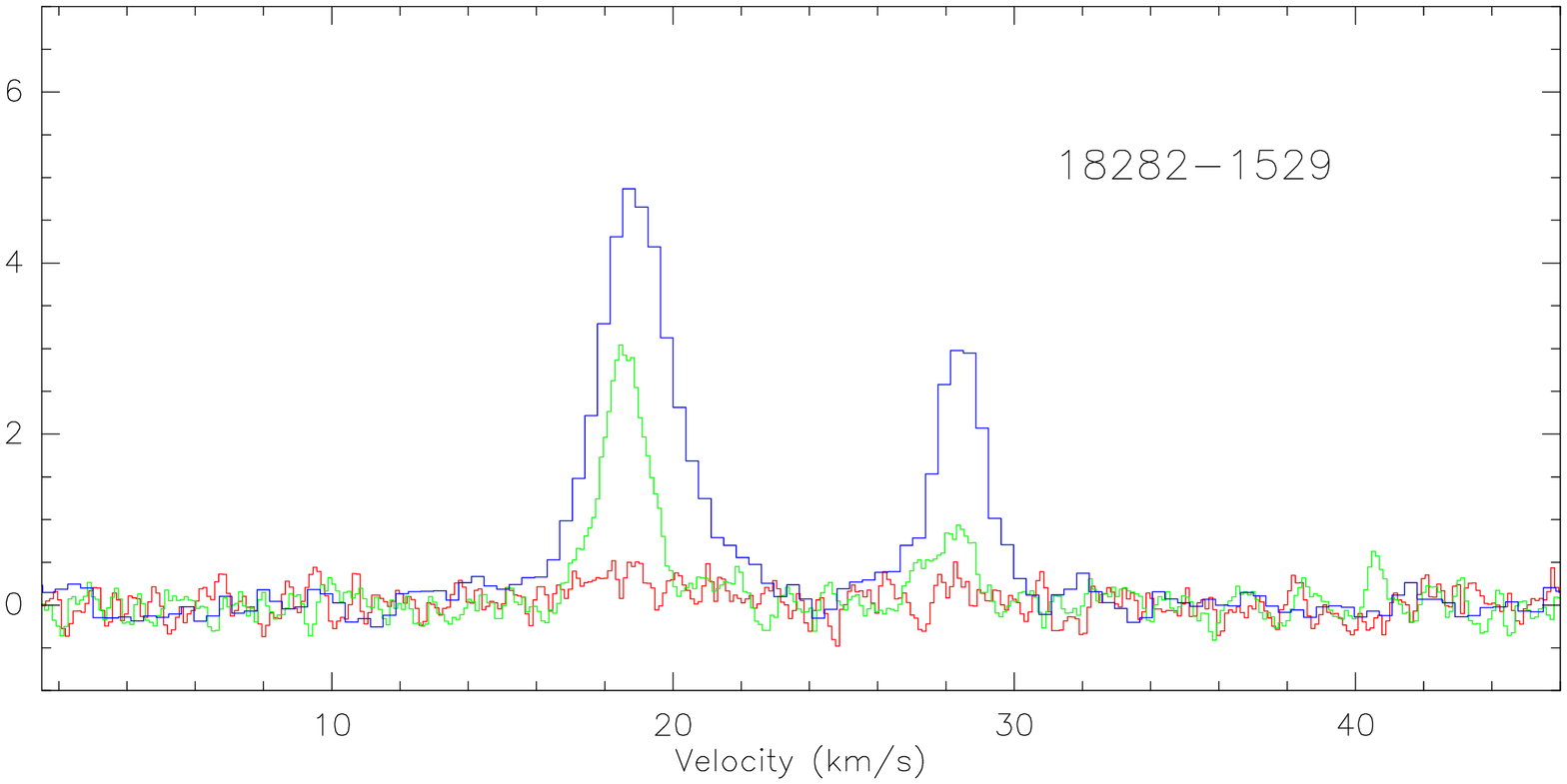}
\includegraphics[height=70mm,  angle=-90, clip, viewport=150 10 500 750]{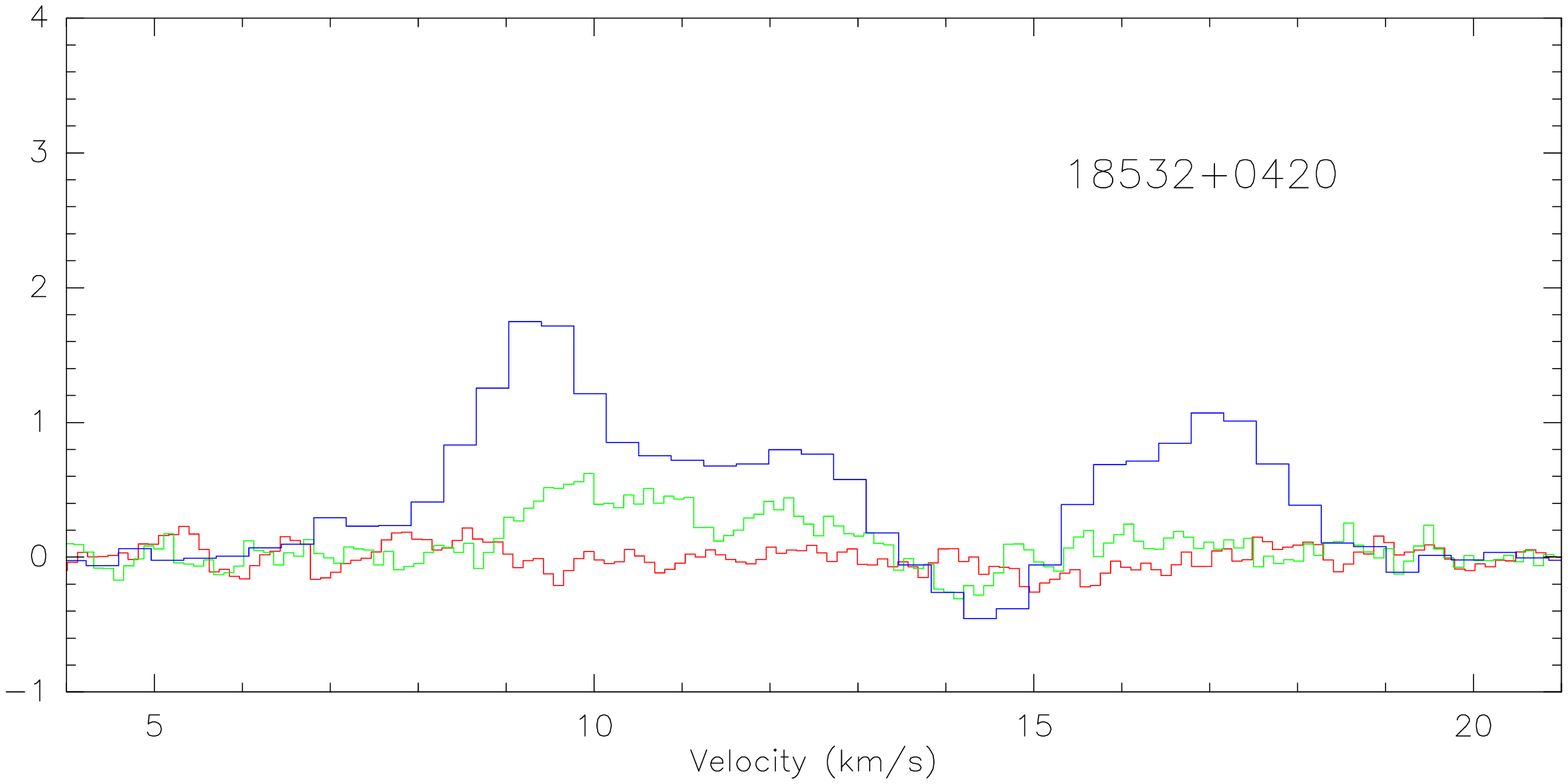}
\includegraphics[height=70mm,  angle=-90, clip, viewport=150 10 500 750]{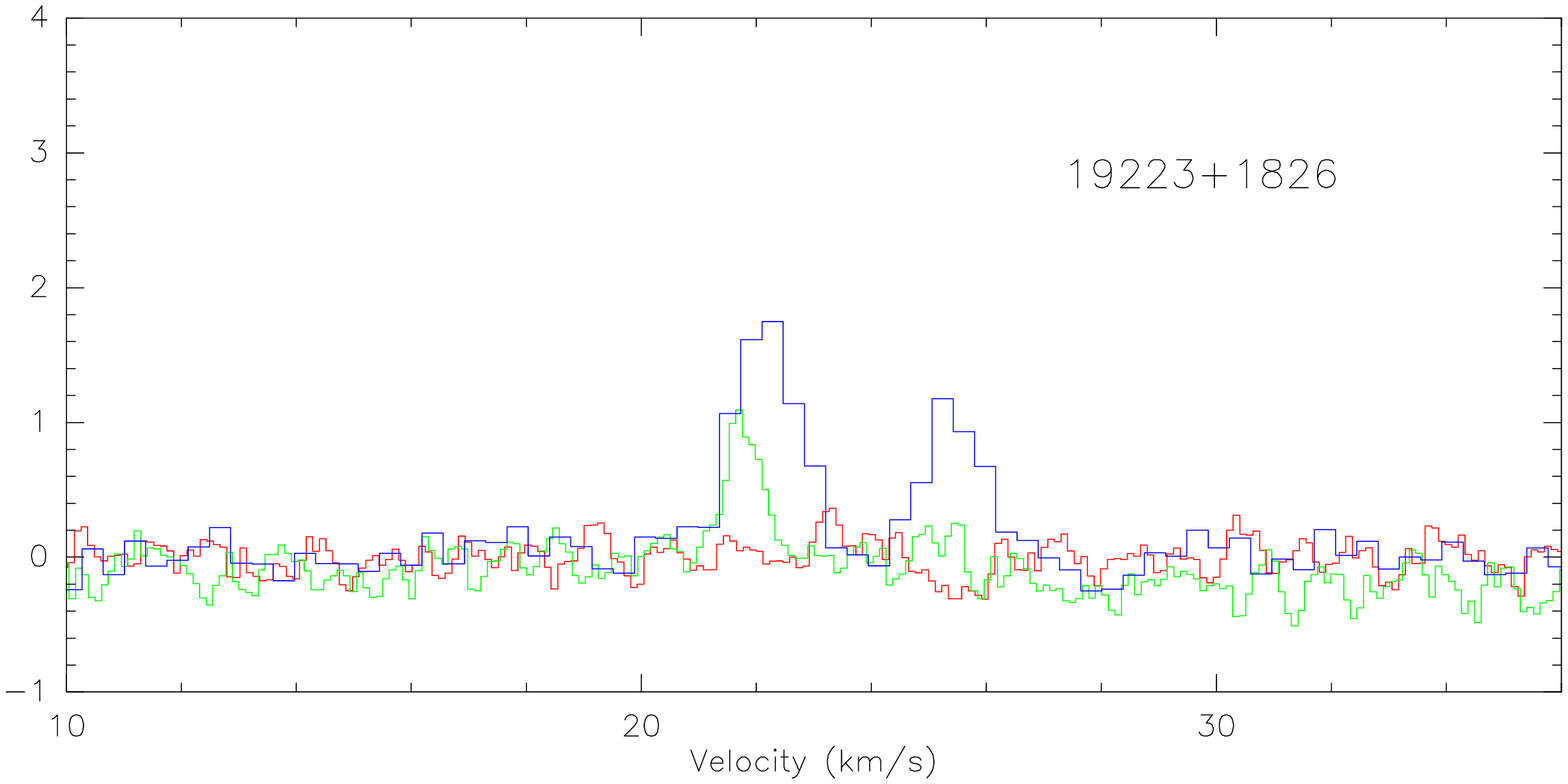}
\includegraphics[height=70mm,  angle=-90, clip, viewport=150 10 500 750]{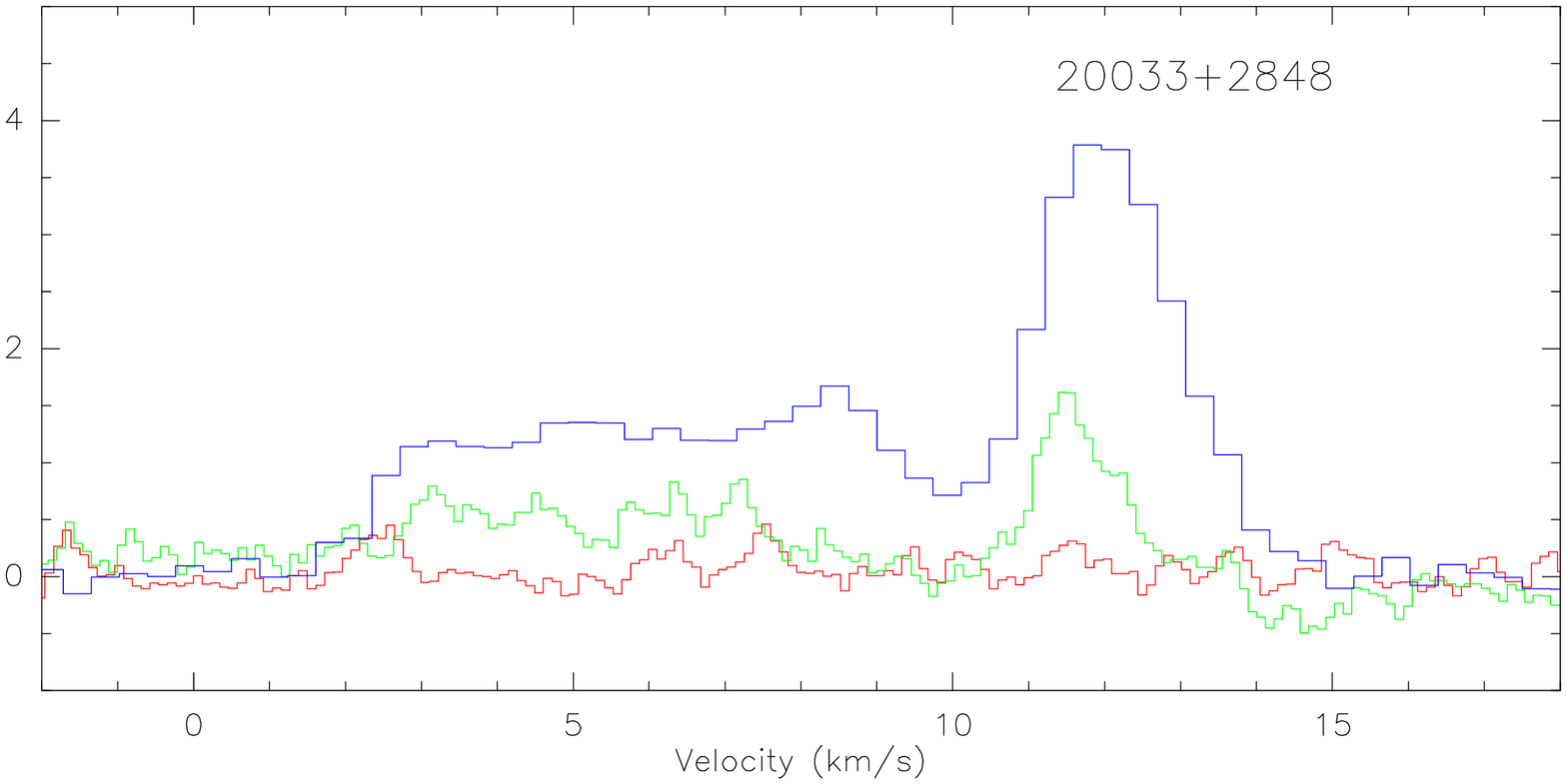}
\includegraphics[height=70mm,  angle=-90, clip, viewport=150 10 500 750]{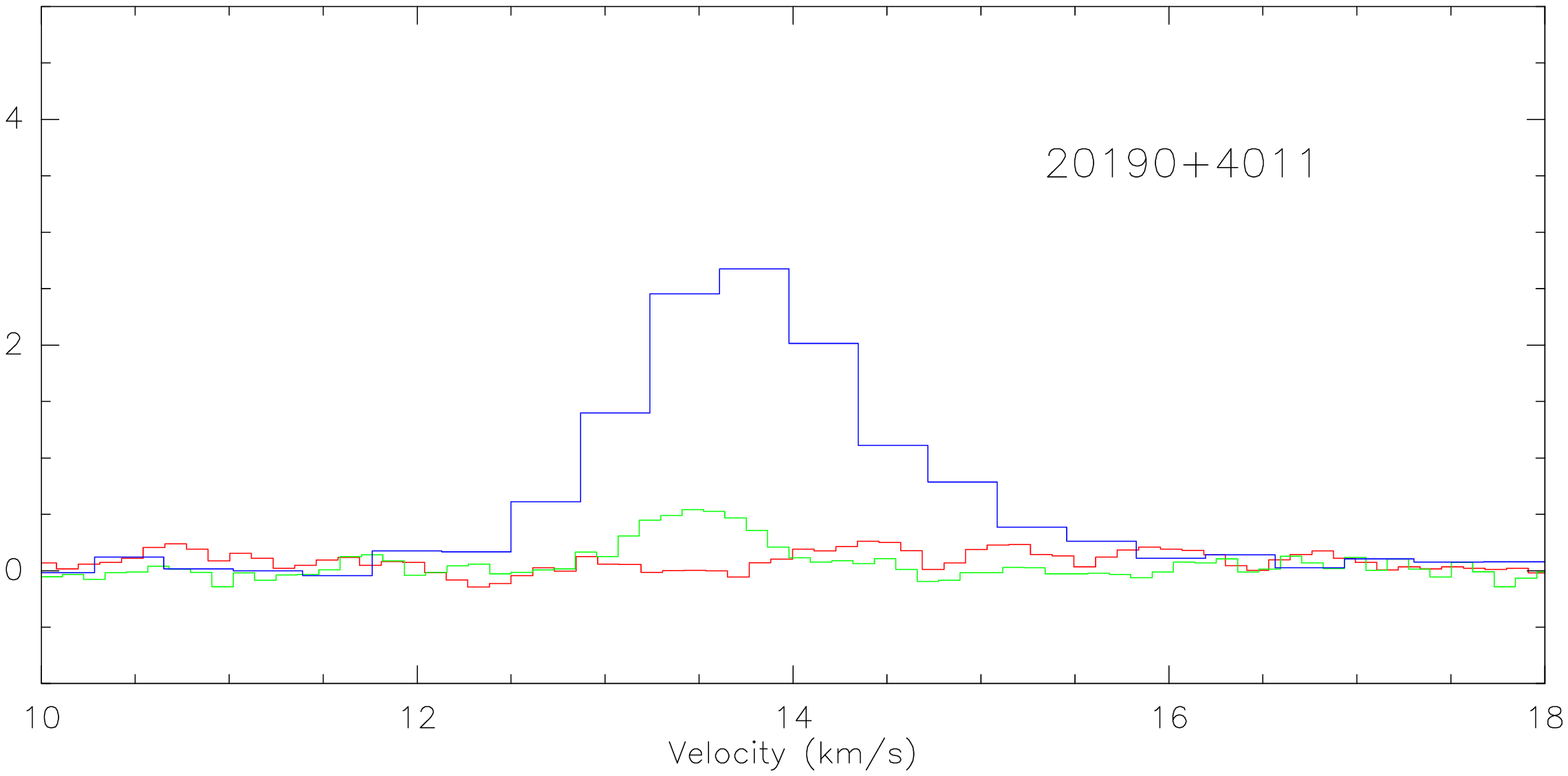}
\includegraphics[height=70mm,  angle=-90, clip, viewport=150 10 500 750]{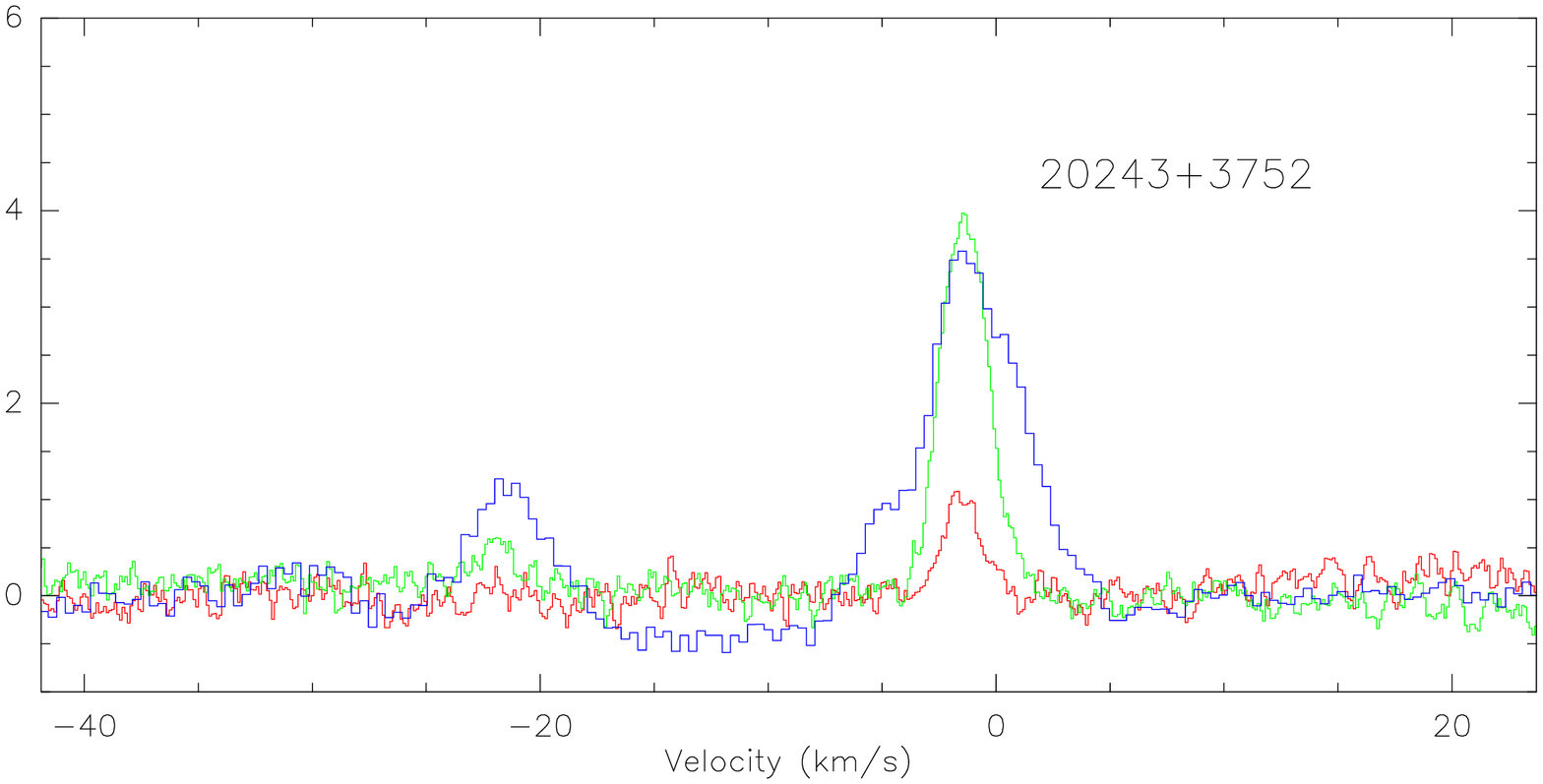}
\includegraphics[height=70mm,  angle=-90, clip, viewport=150 10 500 750]{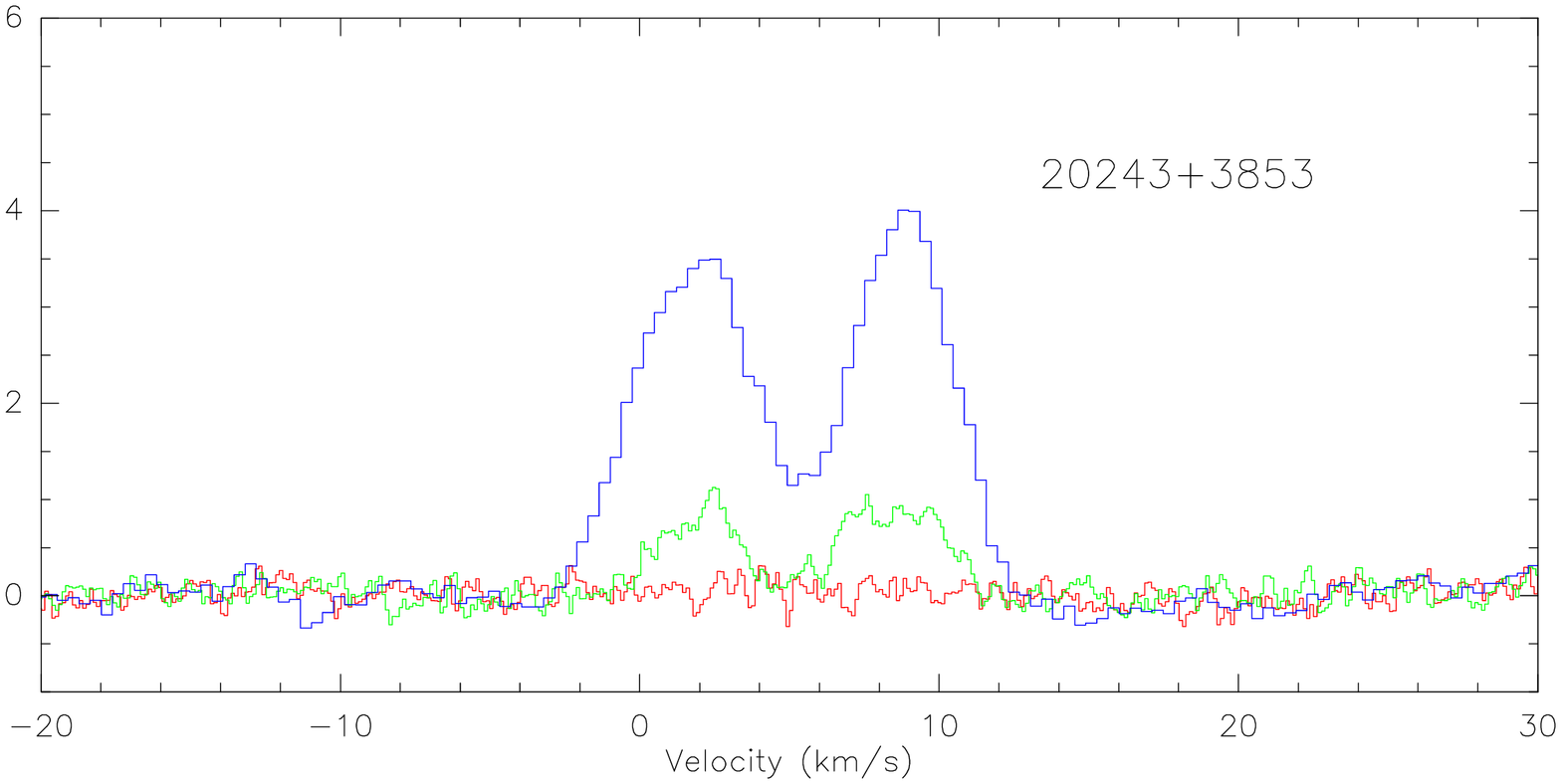}
\includegraphics[height=70mm,  angle=-90, clip, viewport=150 10 500 750]{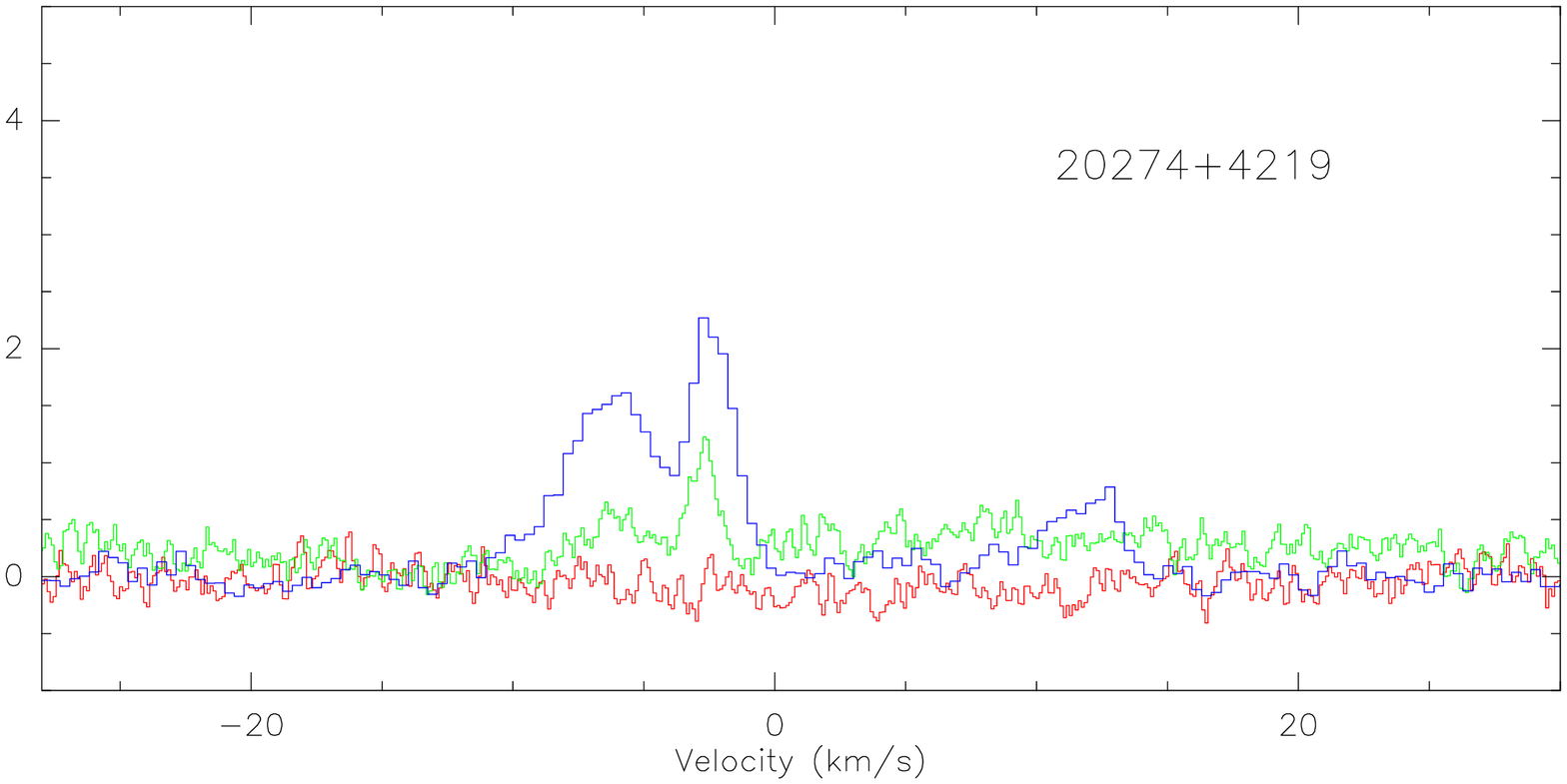}
\includegraphics[height=70mm,  angle=-90, clip, viewport=150 10 500 750]{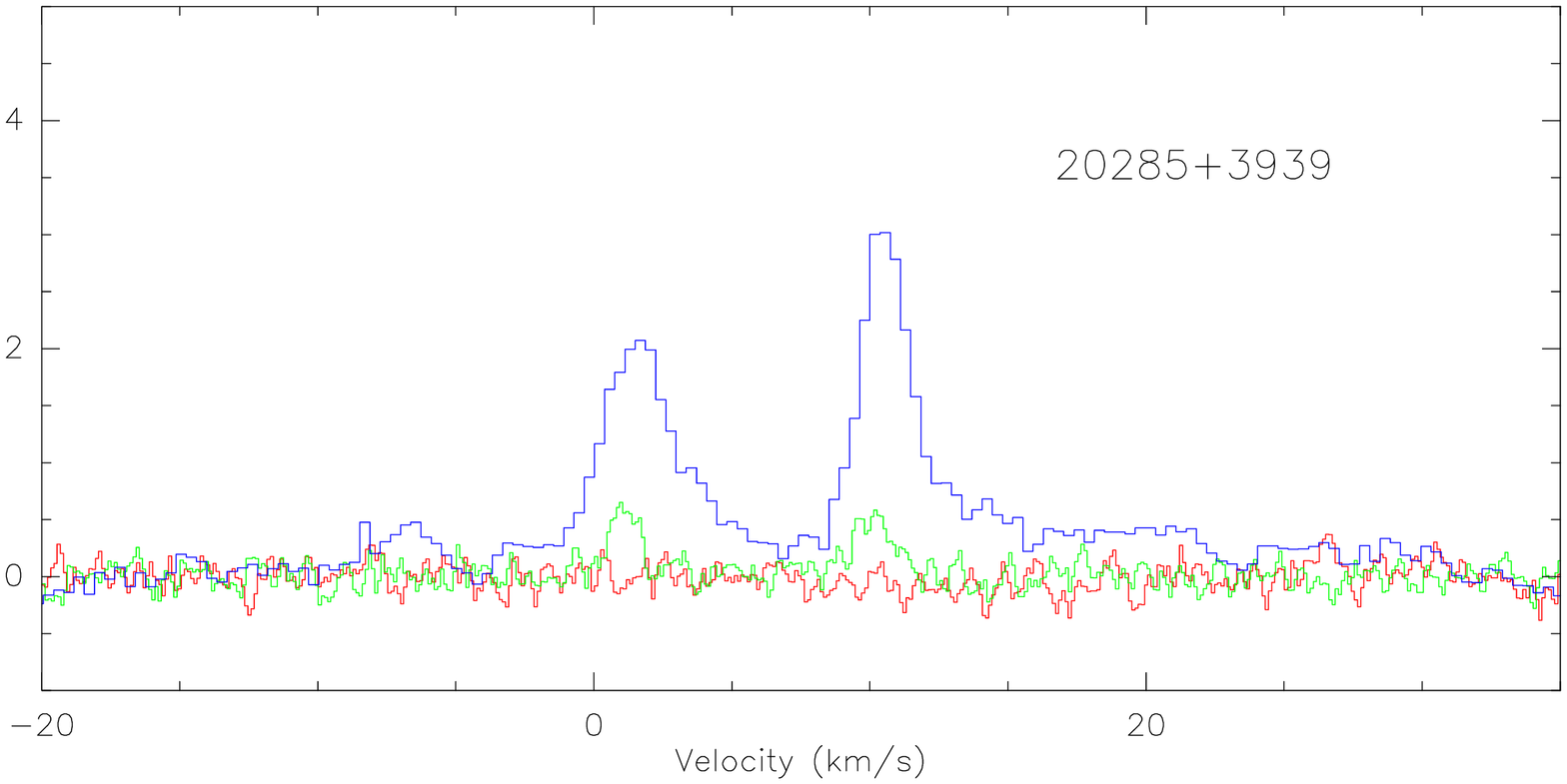}
\includegraphics[height=70mm,  angle=-90, clip, viewport=150 10 500 750]{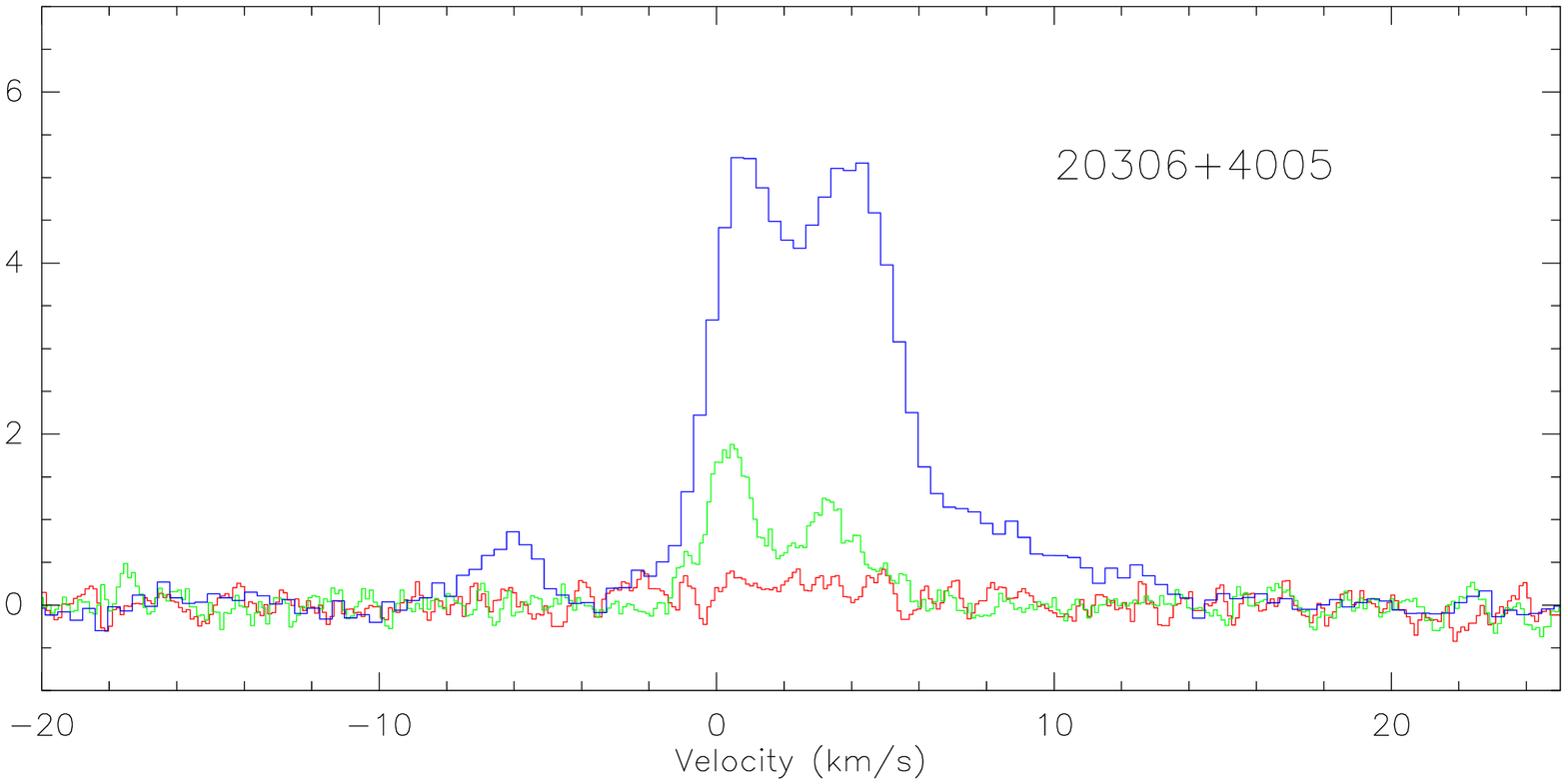}
\includegraphics[height=70mm,  angle=-90, clip, viewport=150 10 500 750]{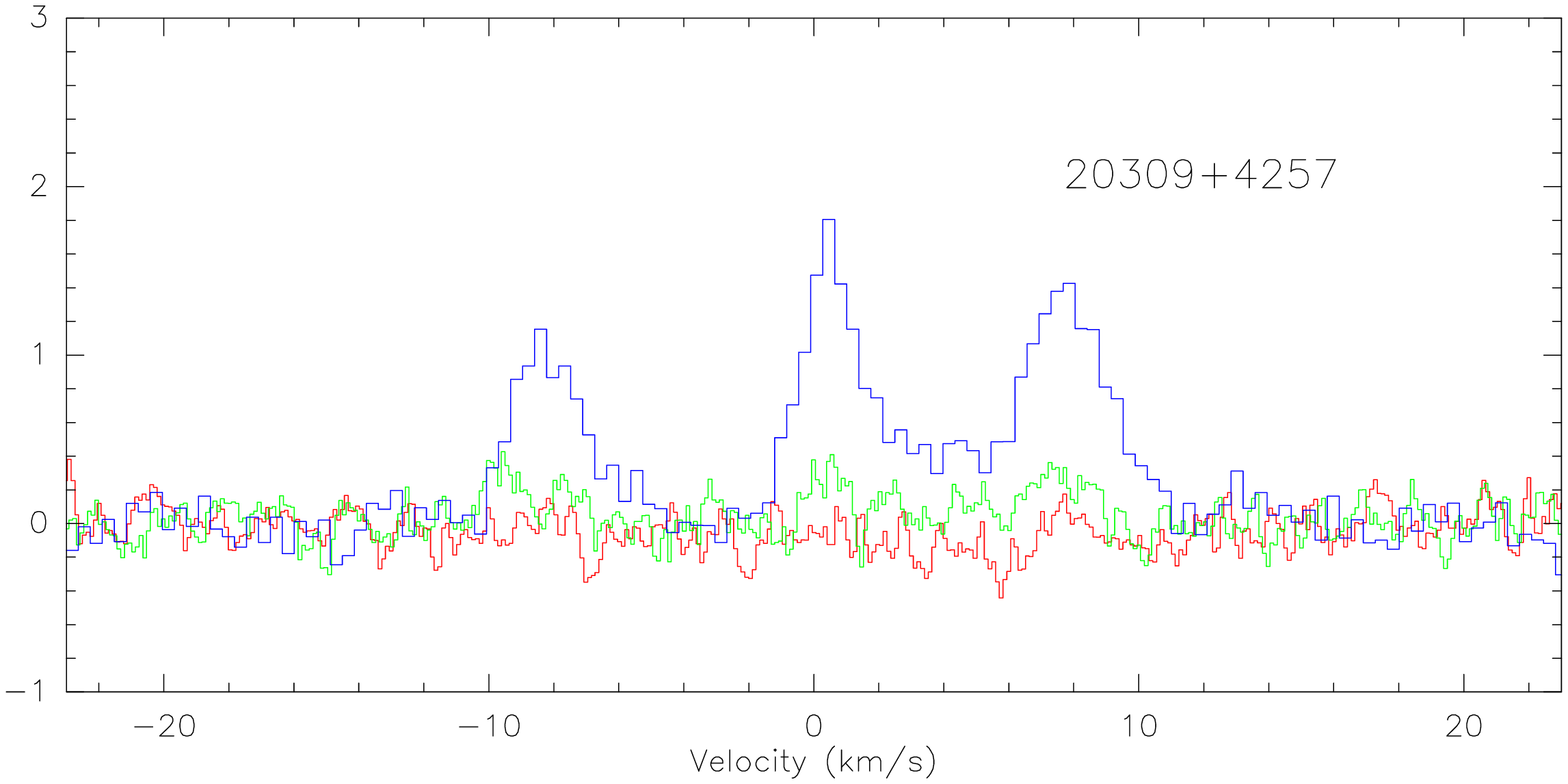}
\includegraphics[height=70mm,  angle=-90, clip, viewport=150 10 500 750]{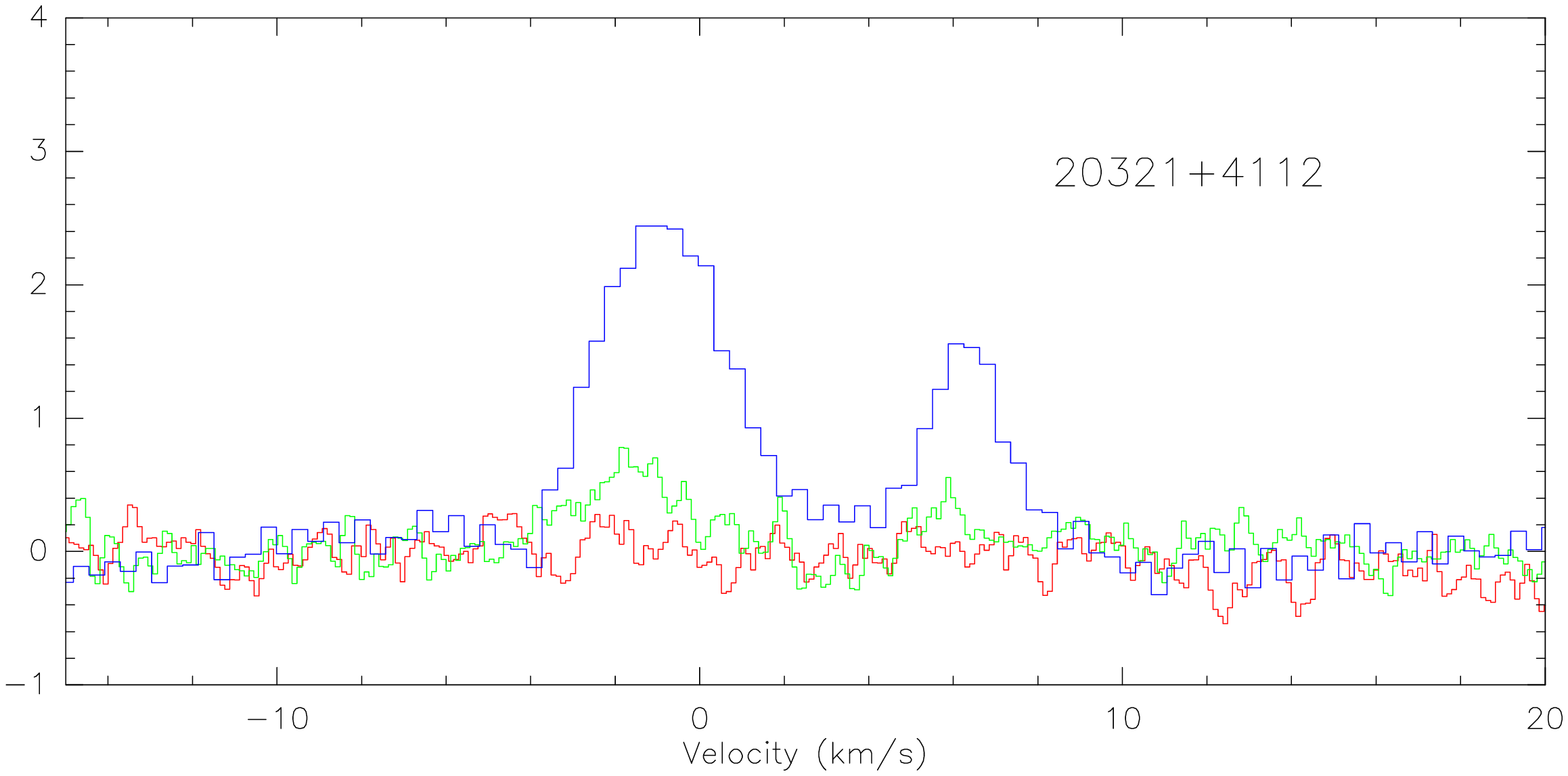}
\includegraphics[height=70mm,  angle=-90, clip, viewport=150 10 500 750]{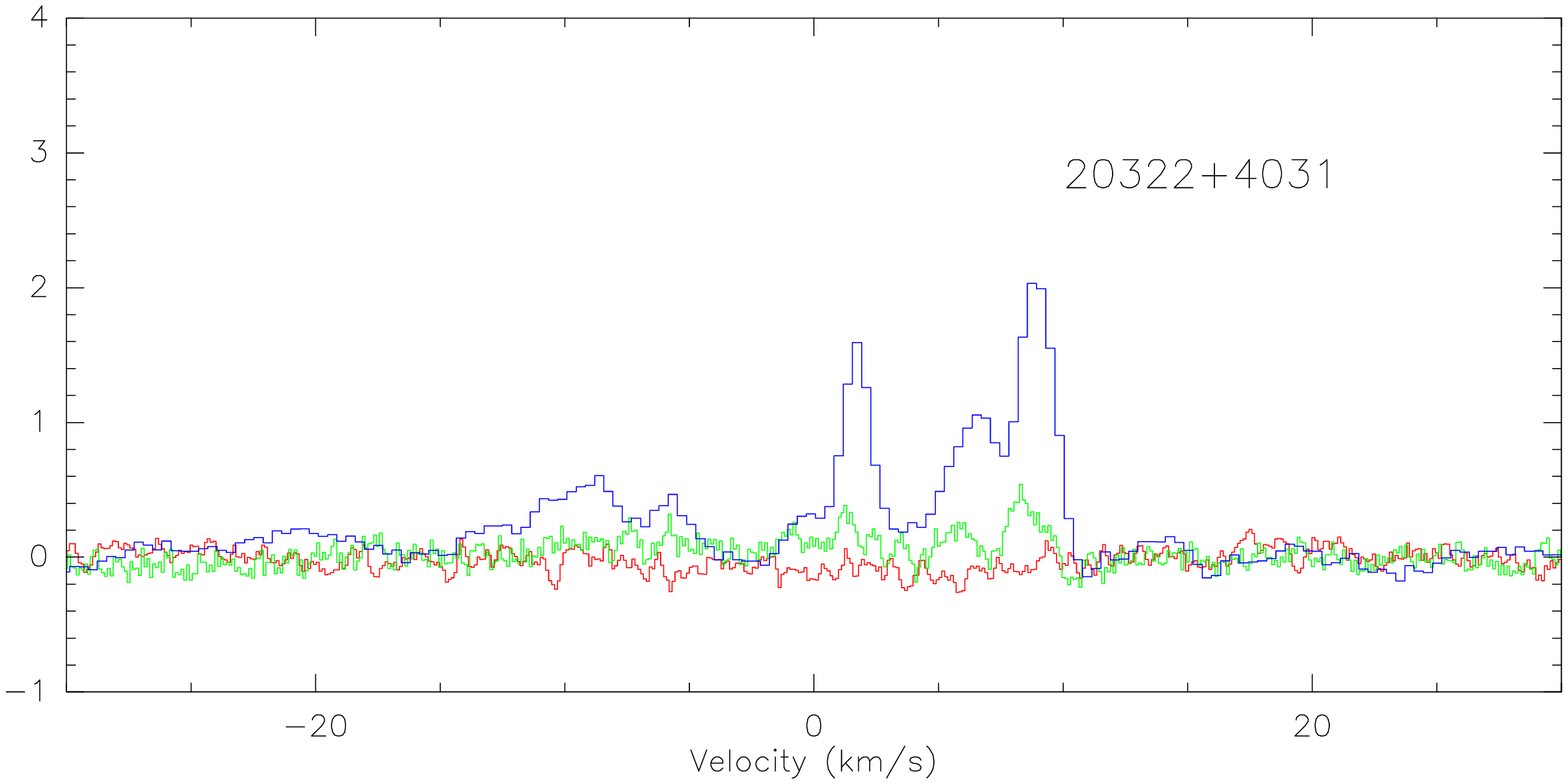}
\includegraphics[height=70mm,  angle=-90, clip, viewport=150 10 500 750]{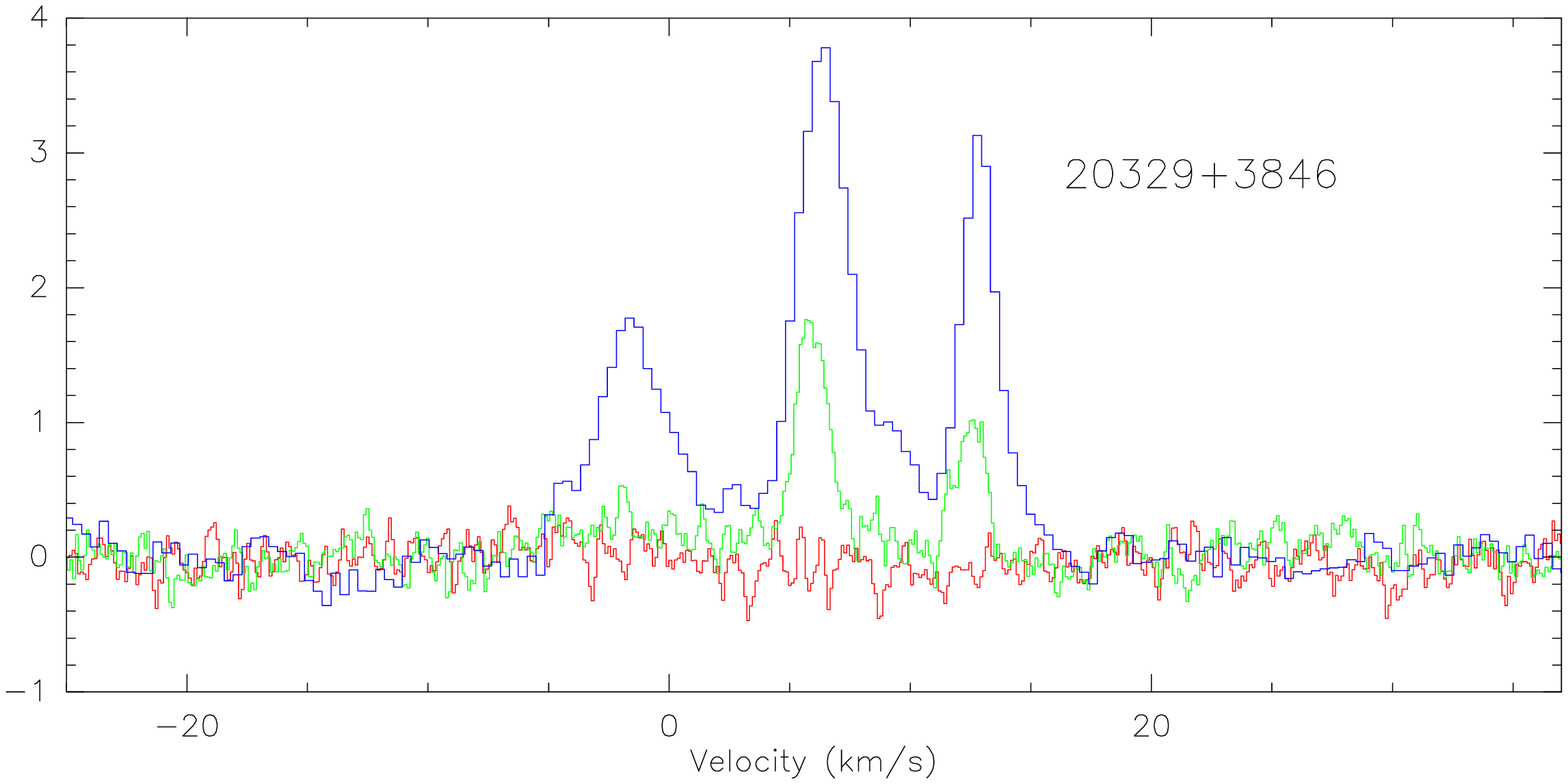}

\begin{minipage}[]{60mm}
  \caption{The sources of type 2
  }\end{minipage}
   \label{Fig7}
   \end{figure}

\addtocounter{figure}{-1}
\begin{figure}

\includegraphics[height=70mm,  angle=-90, clip, viewport=150 10 500 750]{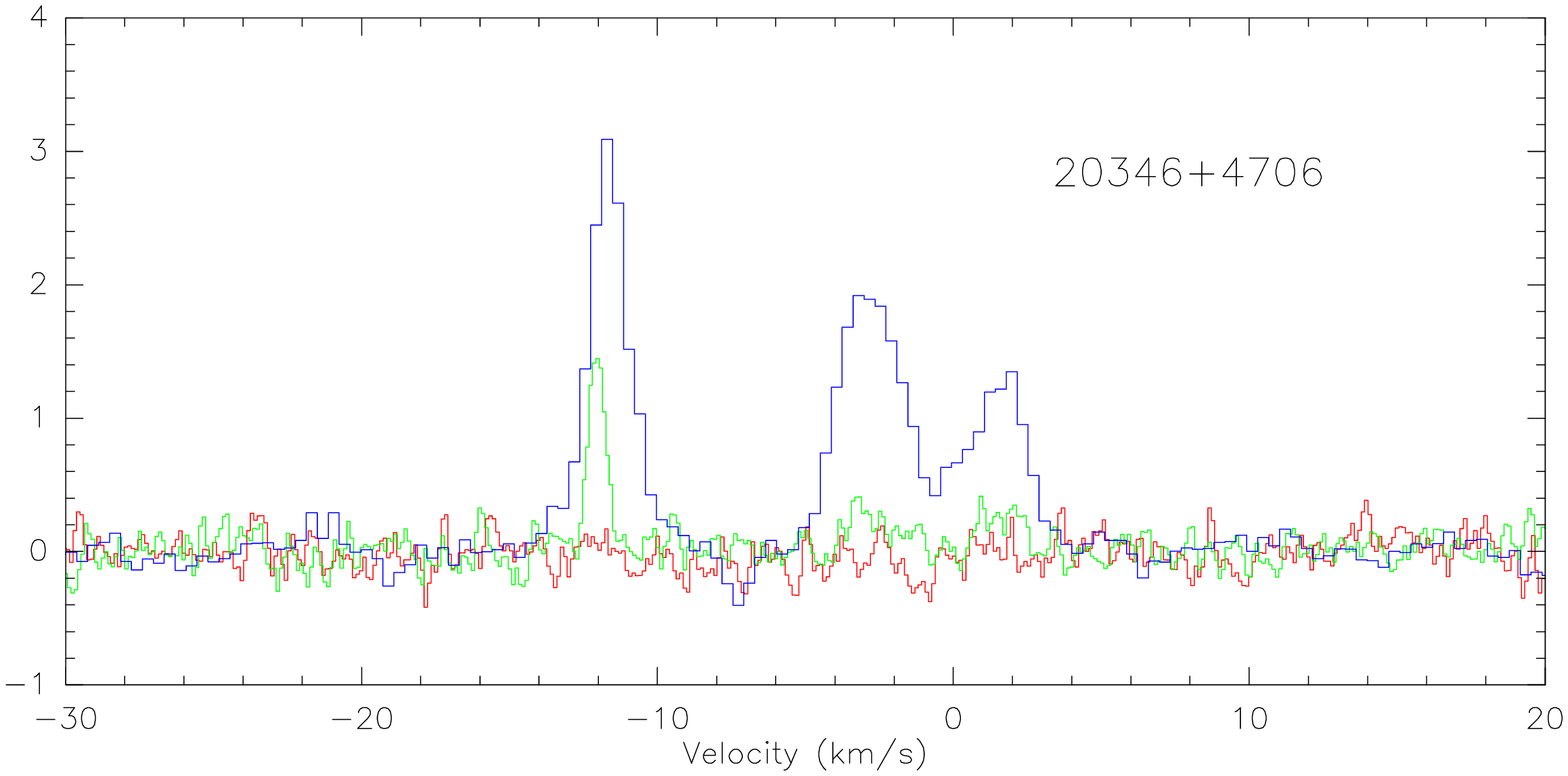}
\includegraphics[height=70mm,  angle=-90, clip, viewport=150 10 500 750]{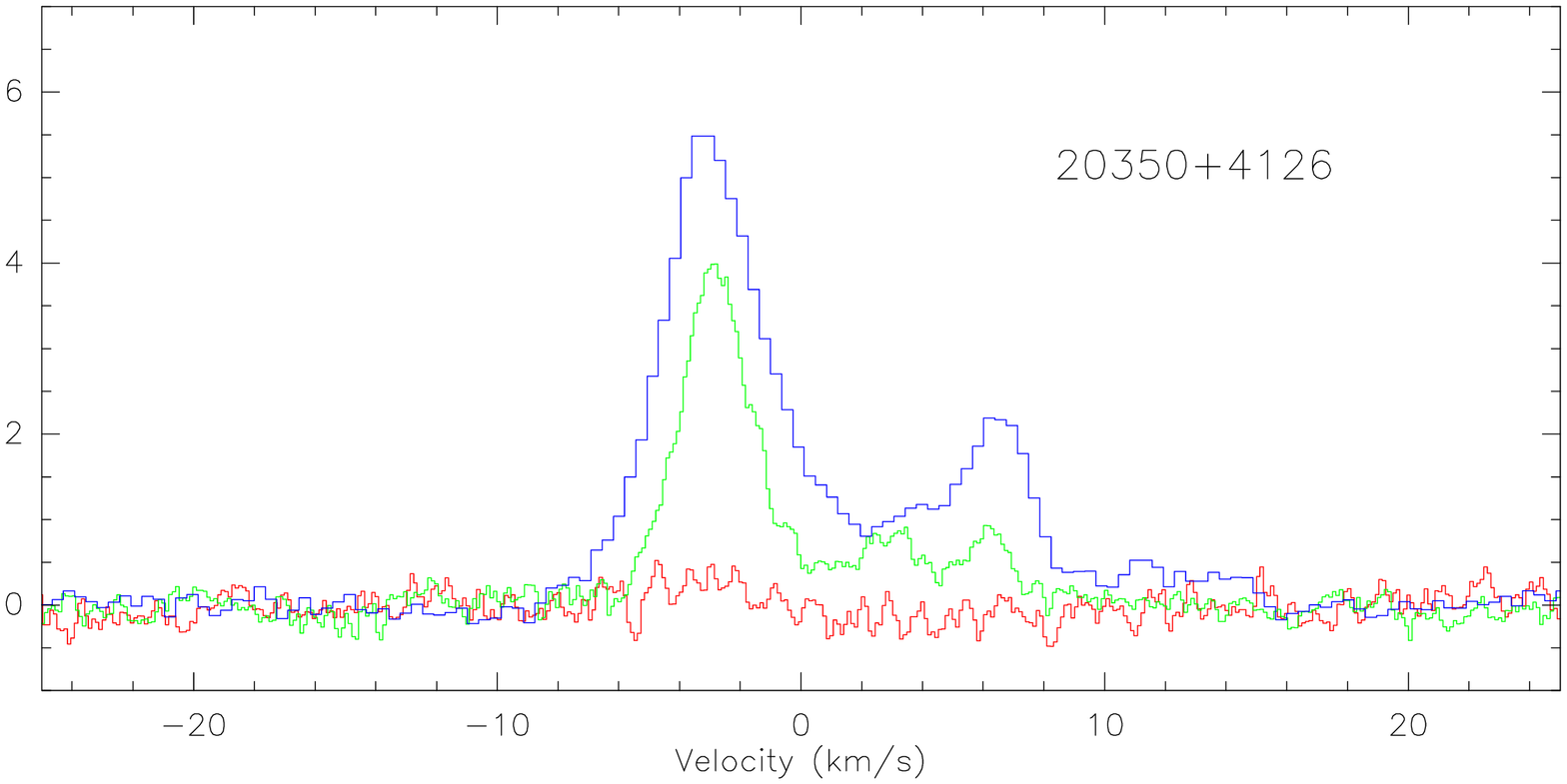}
\includegraphics[height=70mm,  angle=-90, clip, viewport=150 10 500 750]{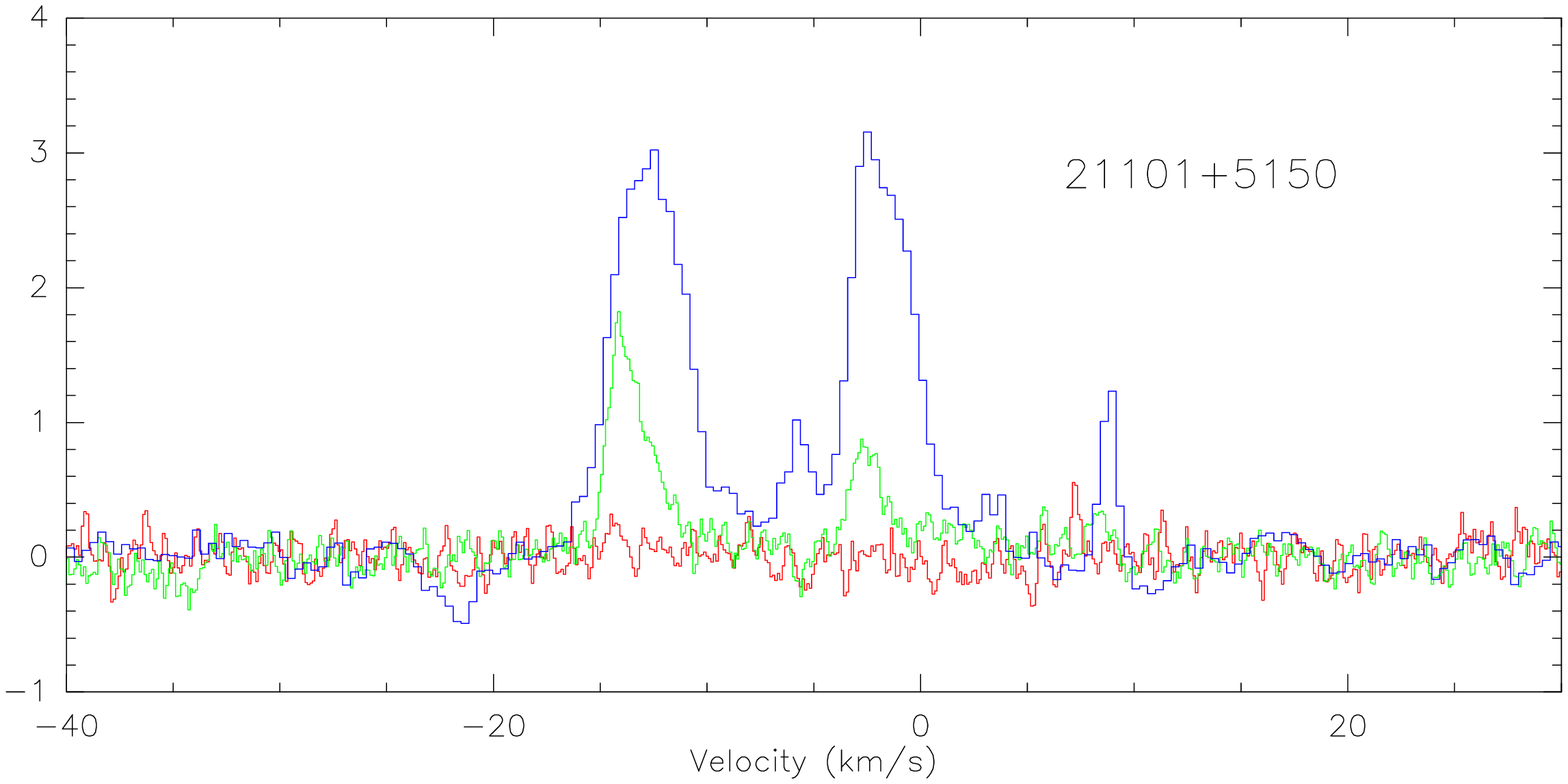}
\includegraphics[height=70mm,  angle=-90, clip, viewport=150 10 500 750]{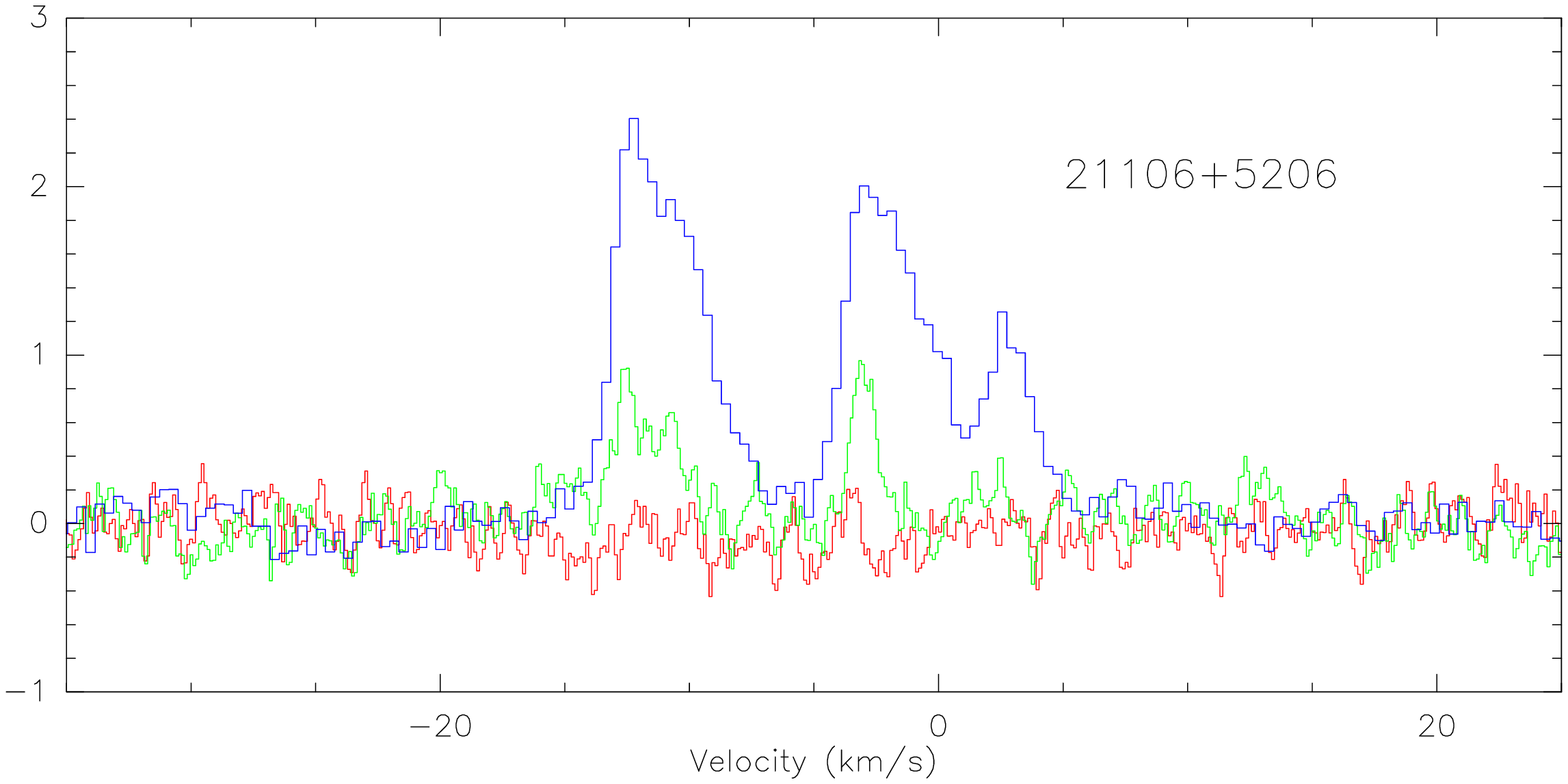}
\includegraphics[height=70mm,  angle=-90, clip, viewport=150 10 500 750]{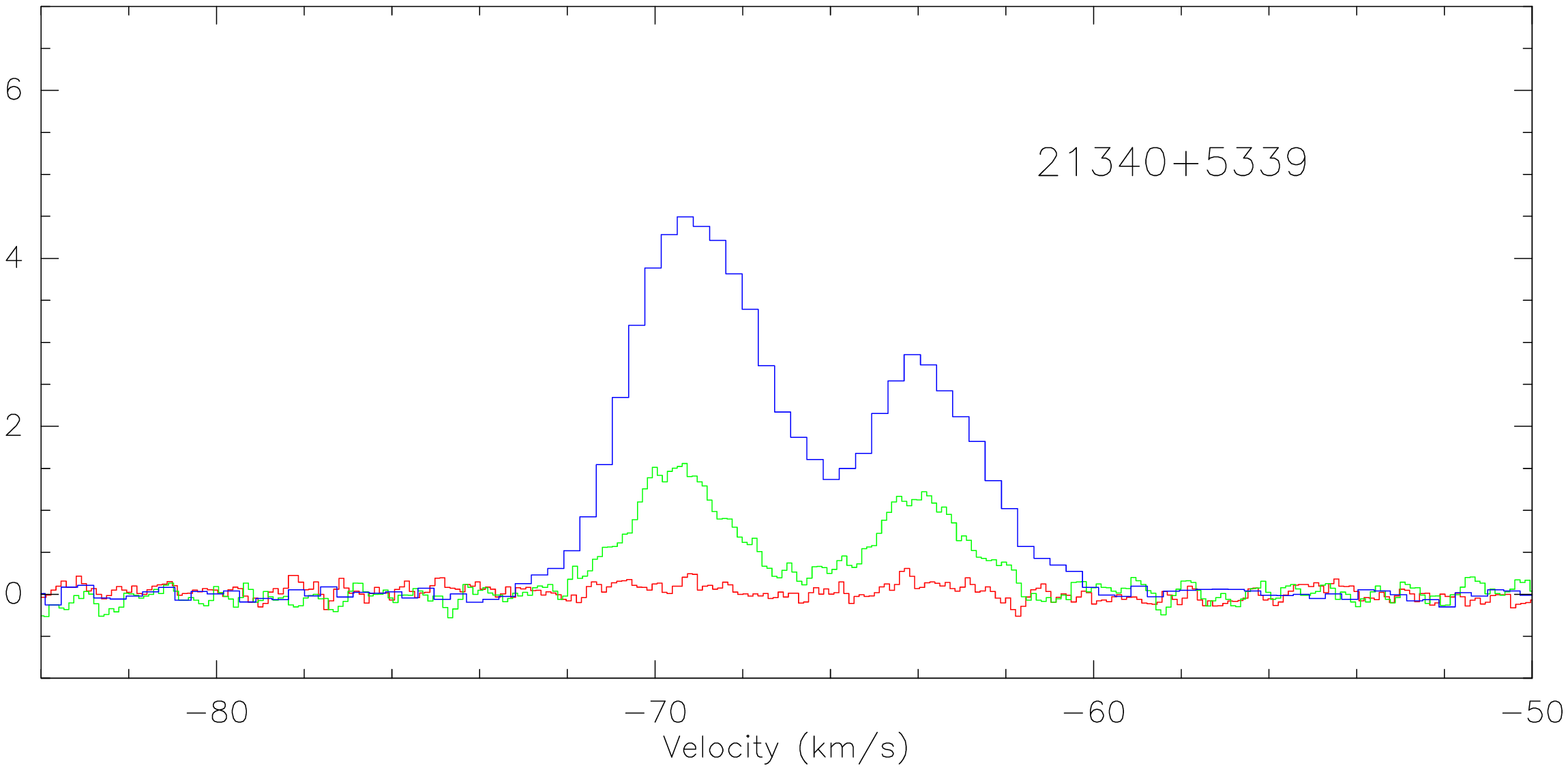}
\includegraphics[height=70mm,  angle=-90, clip, viewport=150 10 500 750]{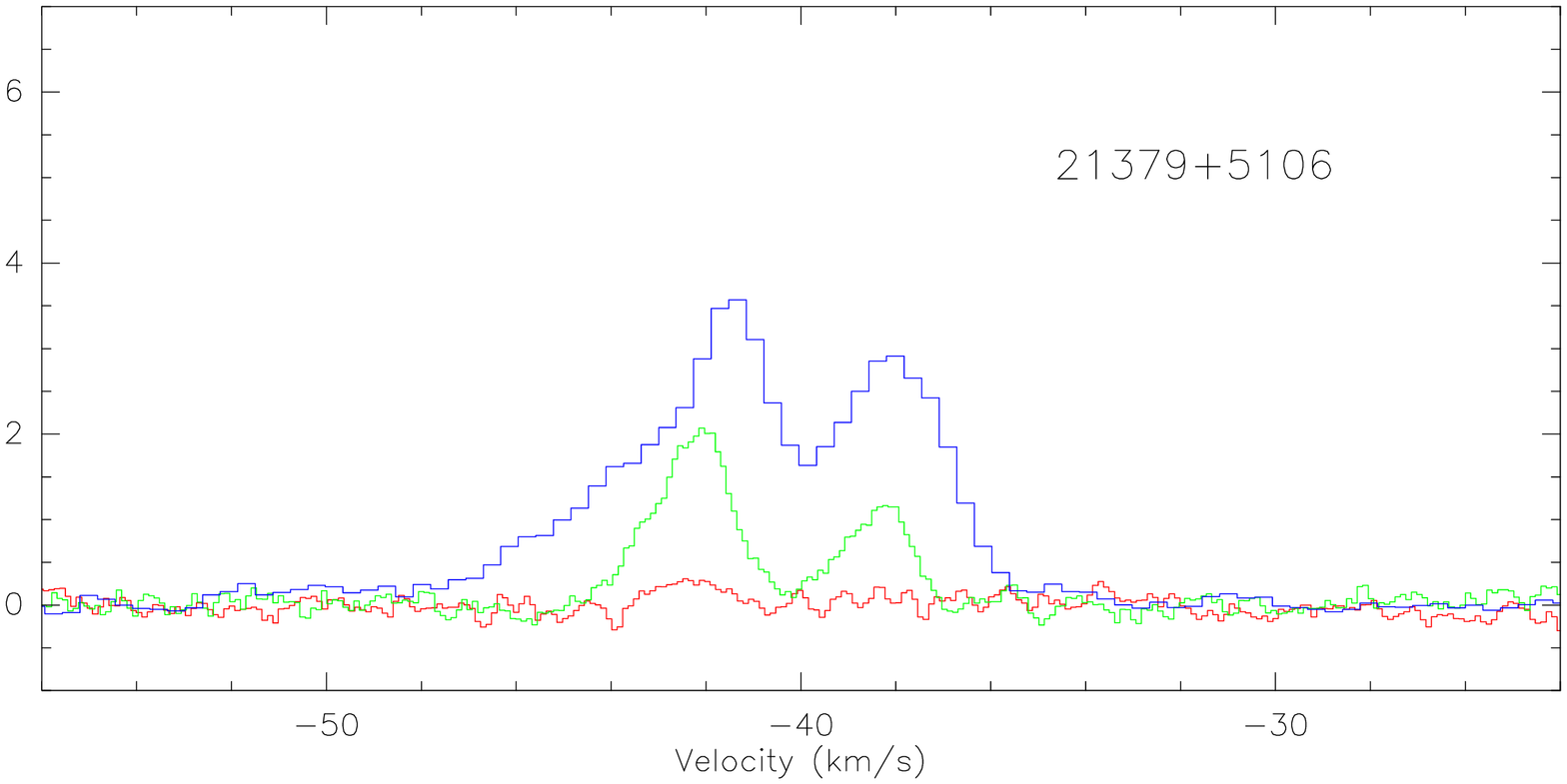}
\includegraphics[height=70mm,  angle=-90, clip, viewport=150 10 500 750]{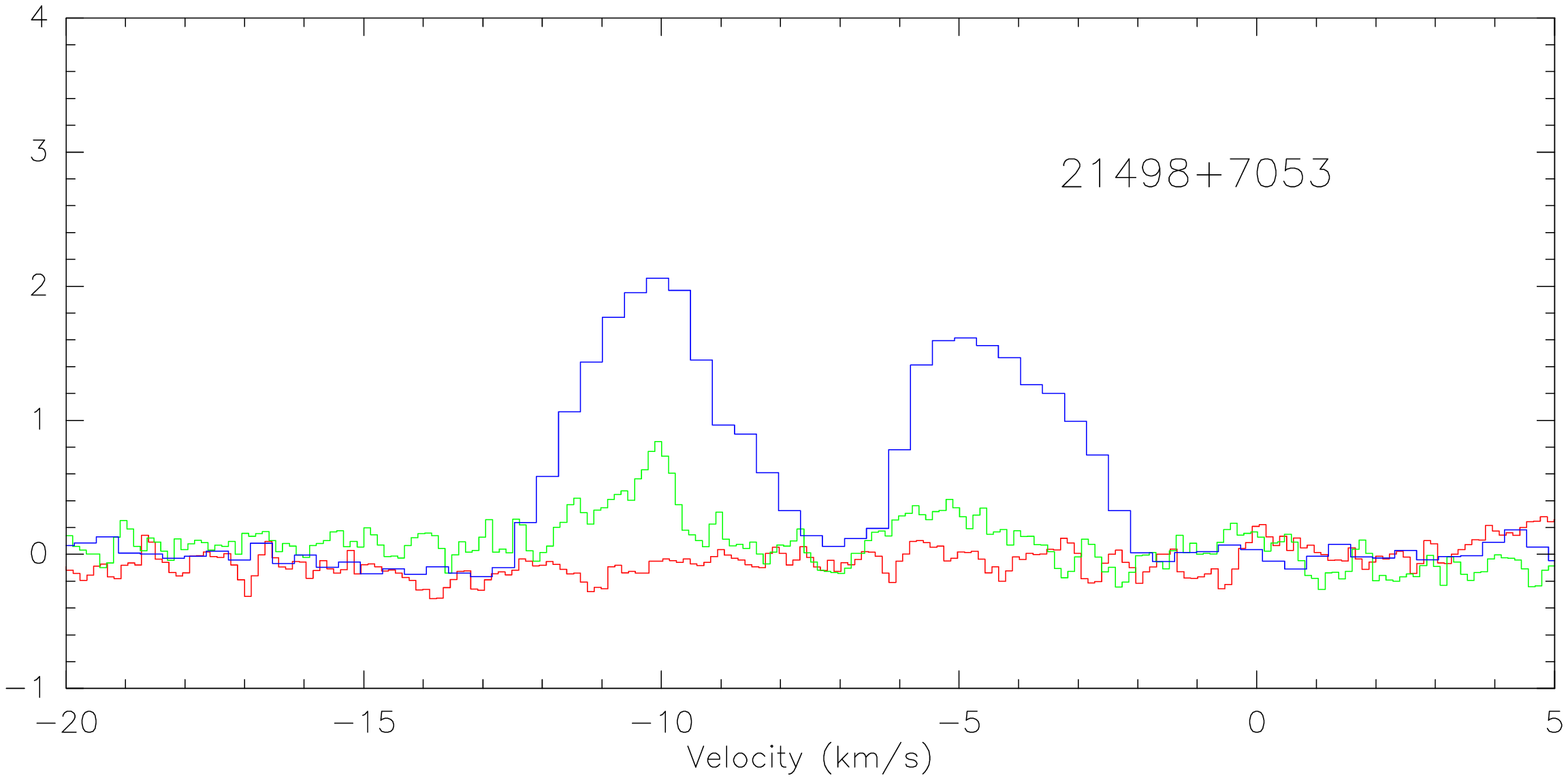}
\includegraphics[height=70mm,  angle=-90, clip, viewport=150 10 500 750]{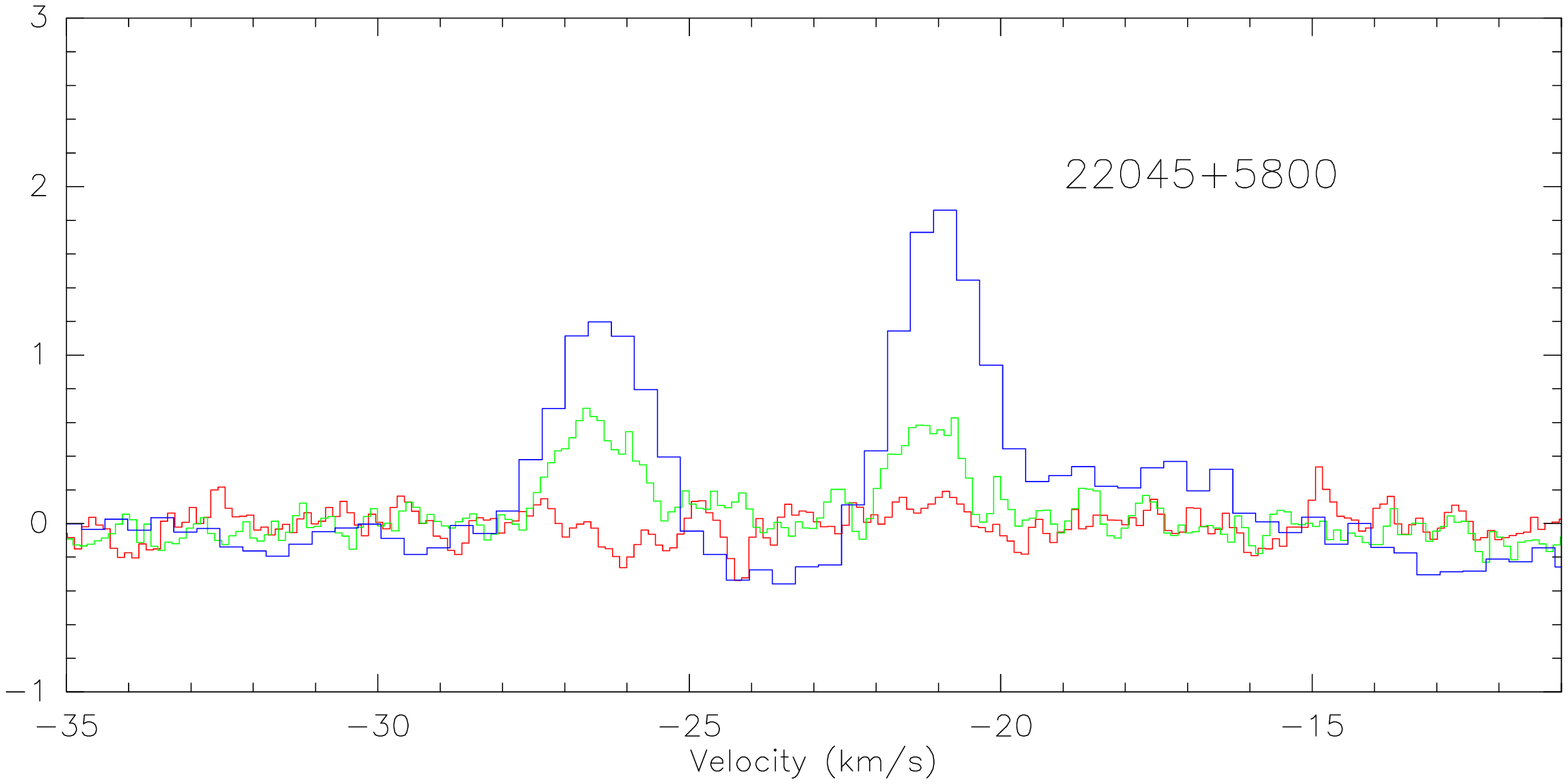}
\includegraphics[height=70mm,  angle=-90, clip, viewport=150 10 500 750]{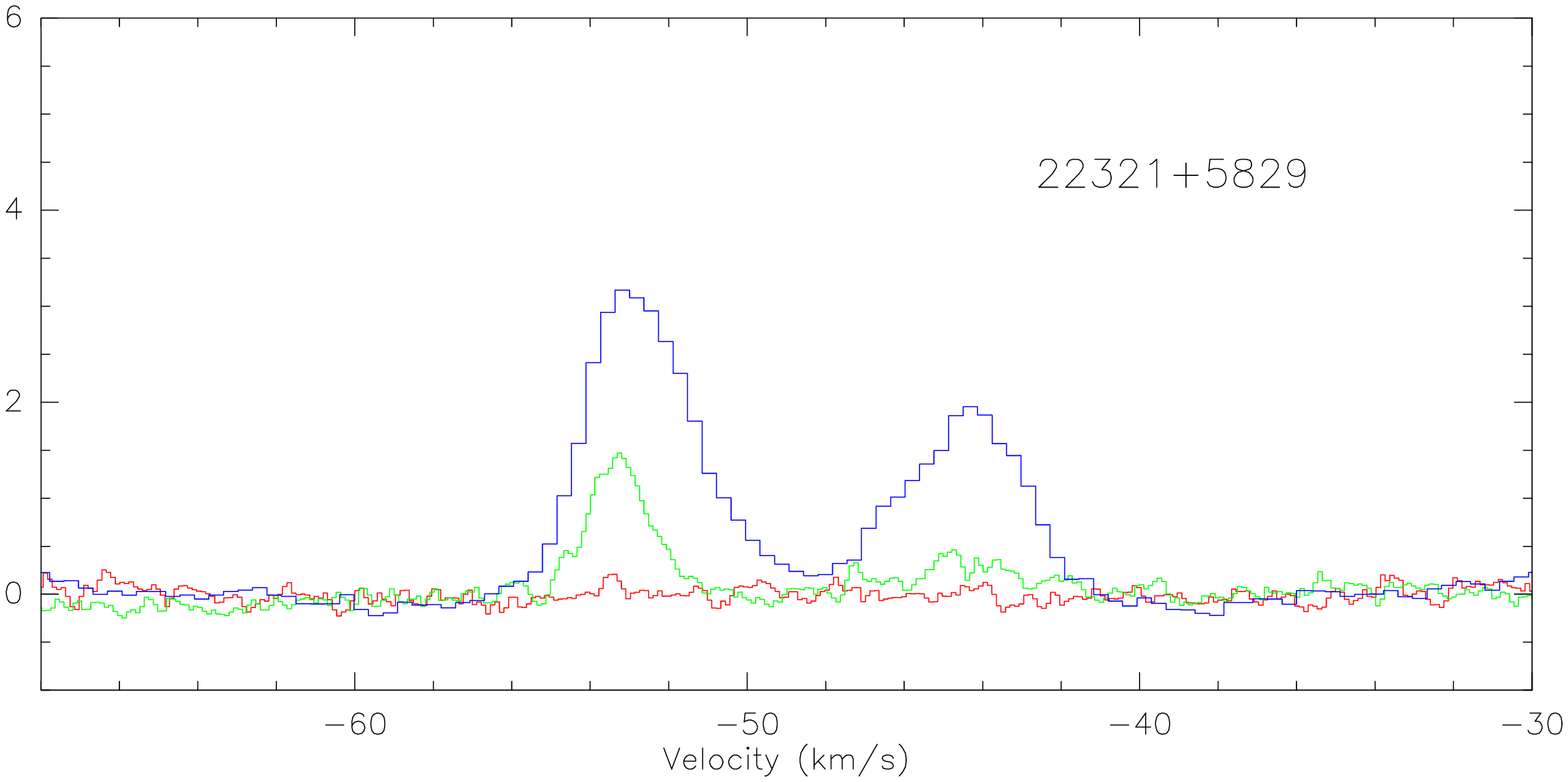}
\includegraphics[height=70mm,  angle=-90, clip, viewport=150 10 500 750]{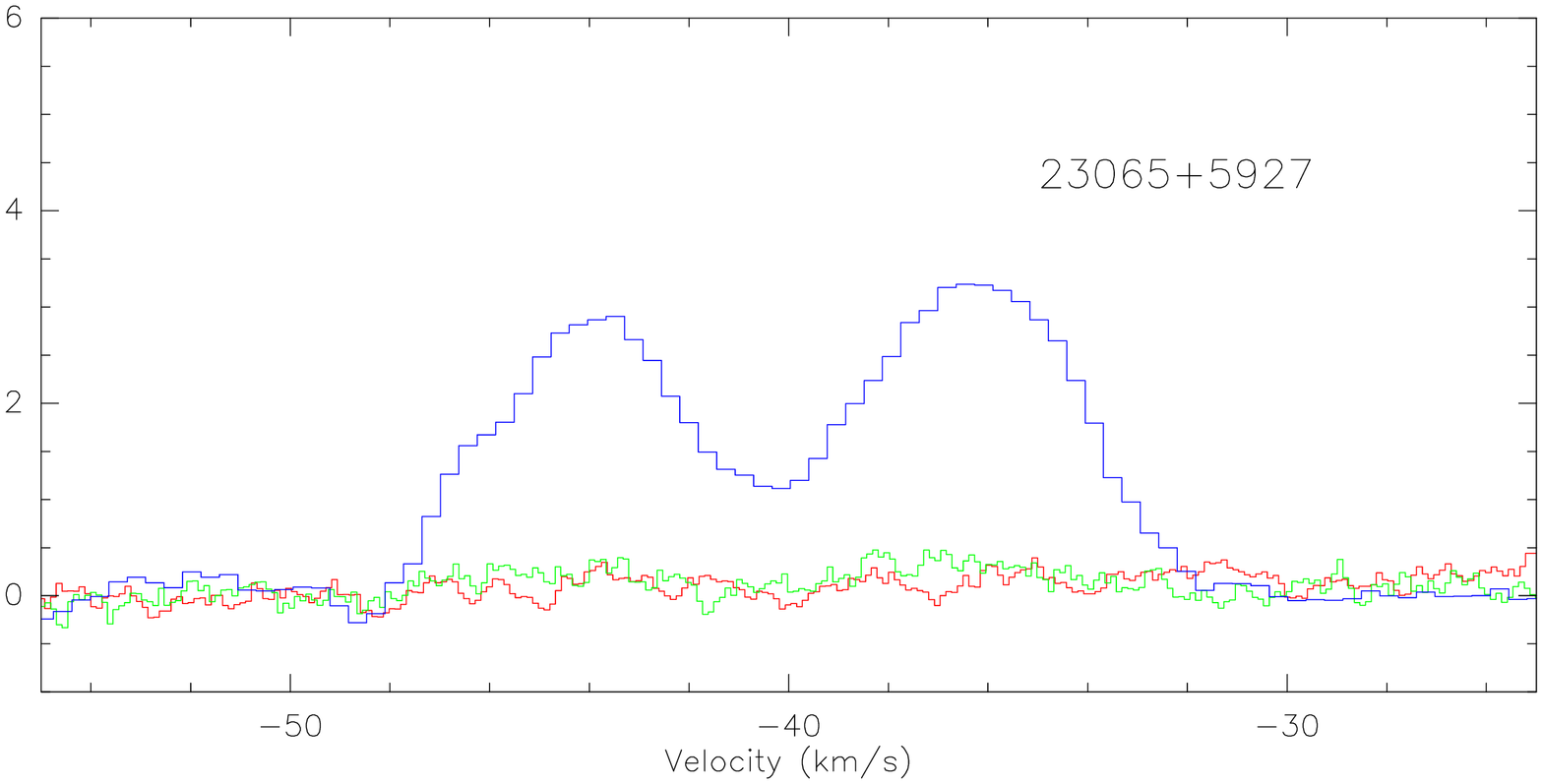}
\includegraphics[height=70mm,  angle=-90, clip, viewport=150 10 500 750]{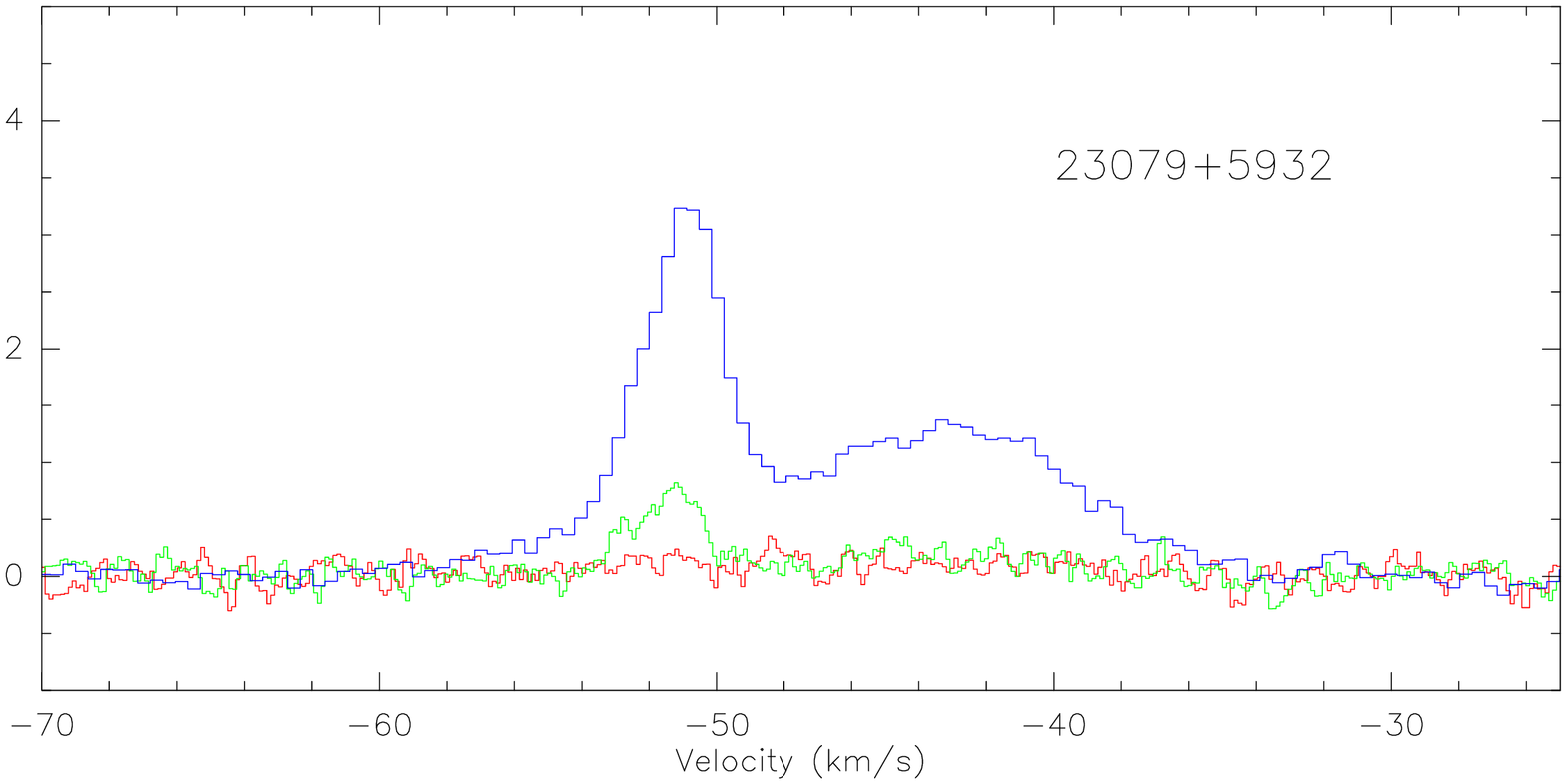}
\includegraphics[height=70mm,  angle=-90, clip, viewport=150 10 500 750]{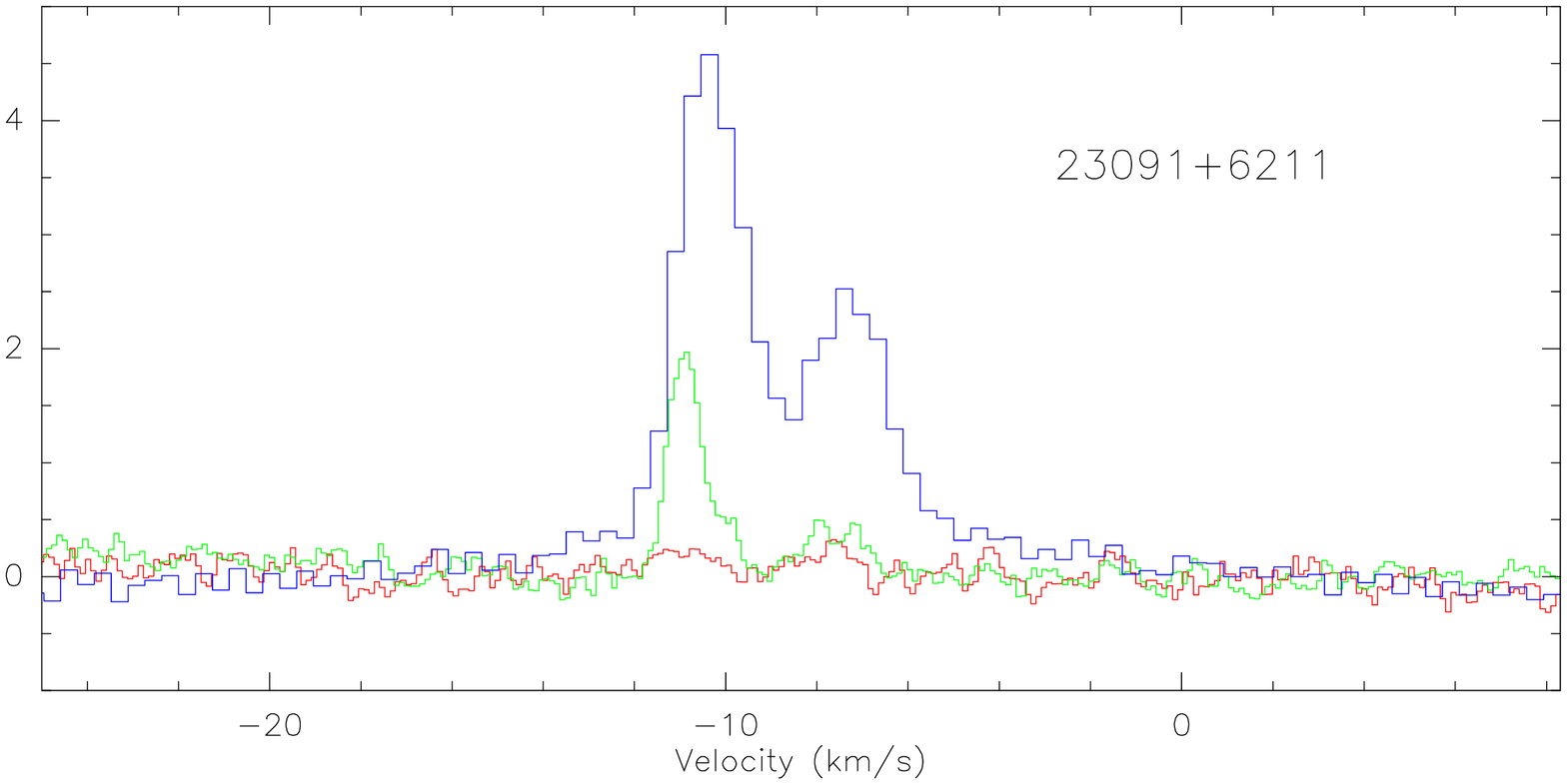}
\includegraphics[height=70mm,  angle=-90, clip, viewport=150 10 500 750]{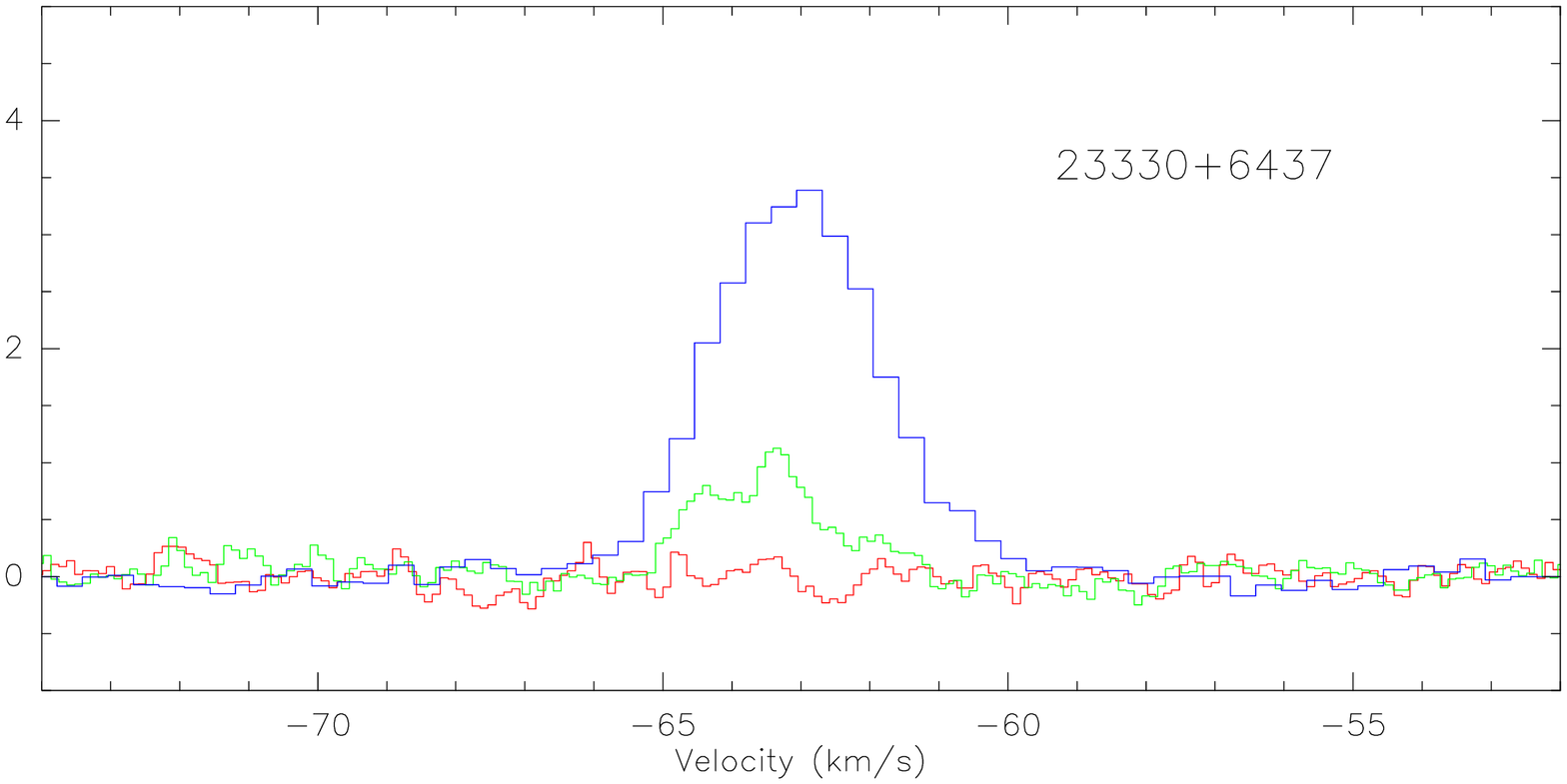}
\includegraphics[height=70mm,  angle=-90, clip, viewport=150 10 500 750]{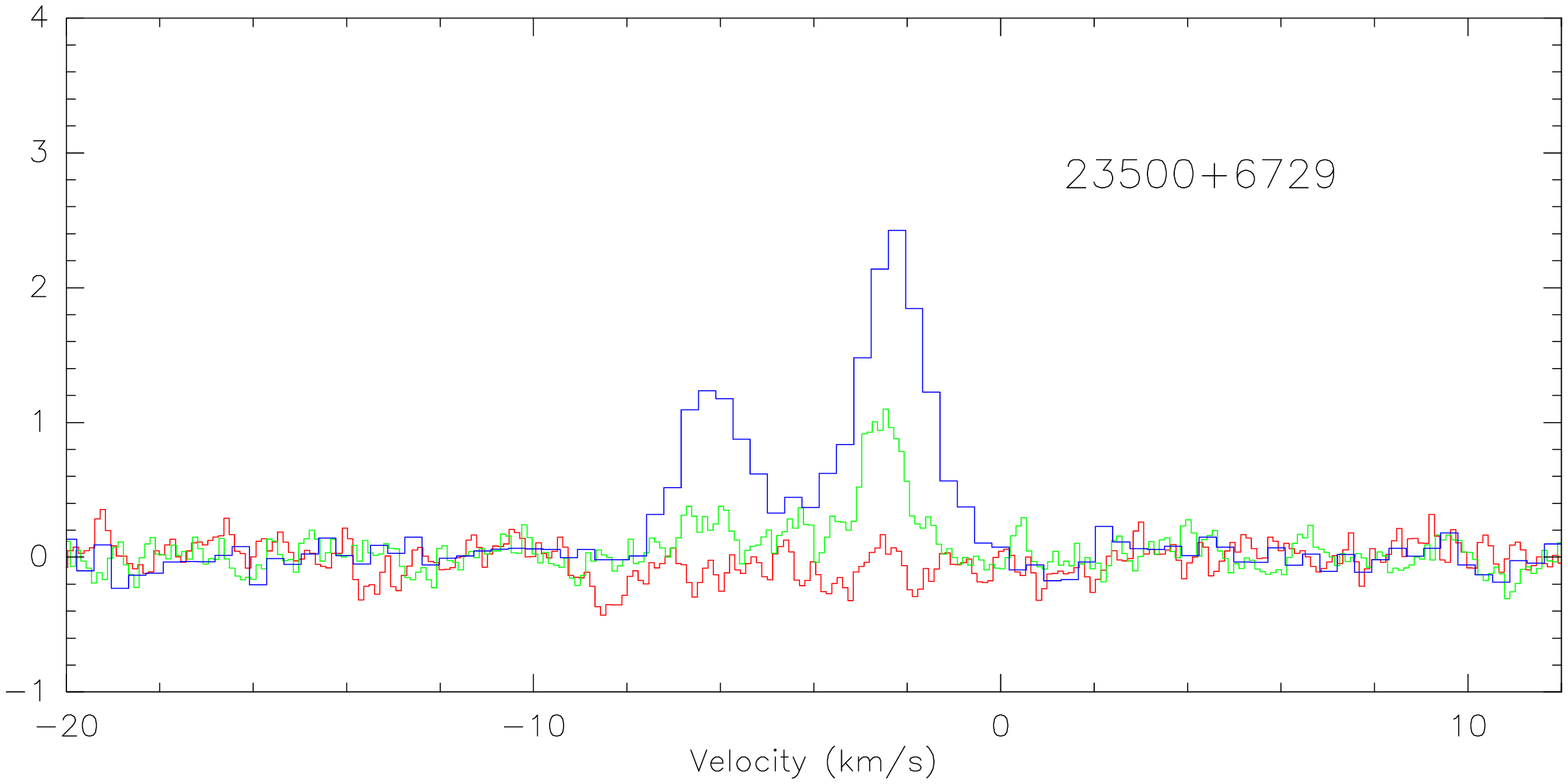}
\centering
\begin{minipage}[]{60mm}
  \caption{The sources of type 2
  }\end{minipage}
   \label{Fig7}
   \end{figure}

\addtocounter{figure}{-1}
\begin{figure}
   \centering
\includegraphics[height=70mm,  angle=-90, clip, viewport=150 10 500 750]{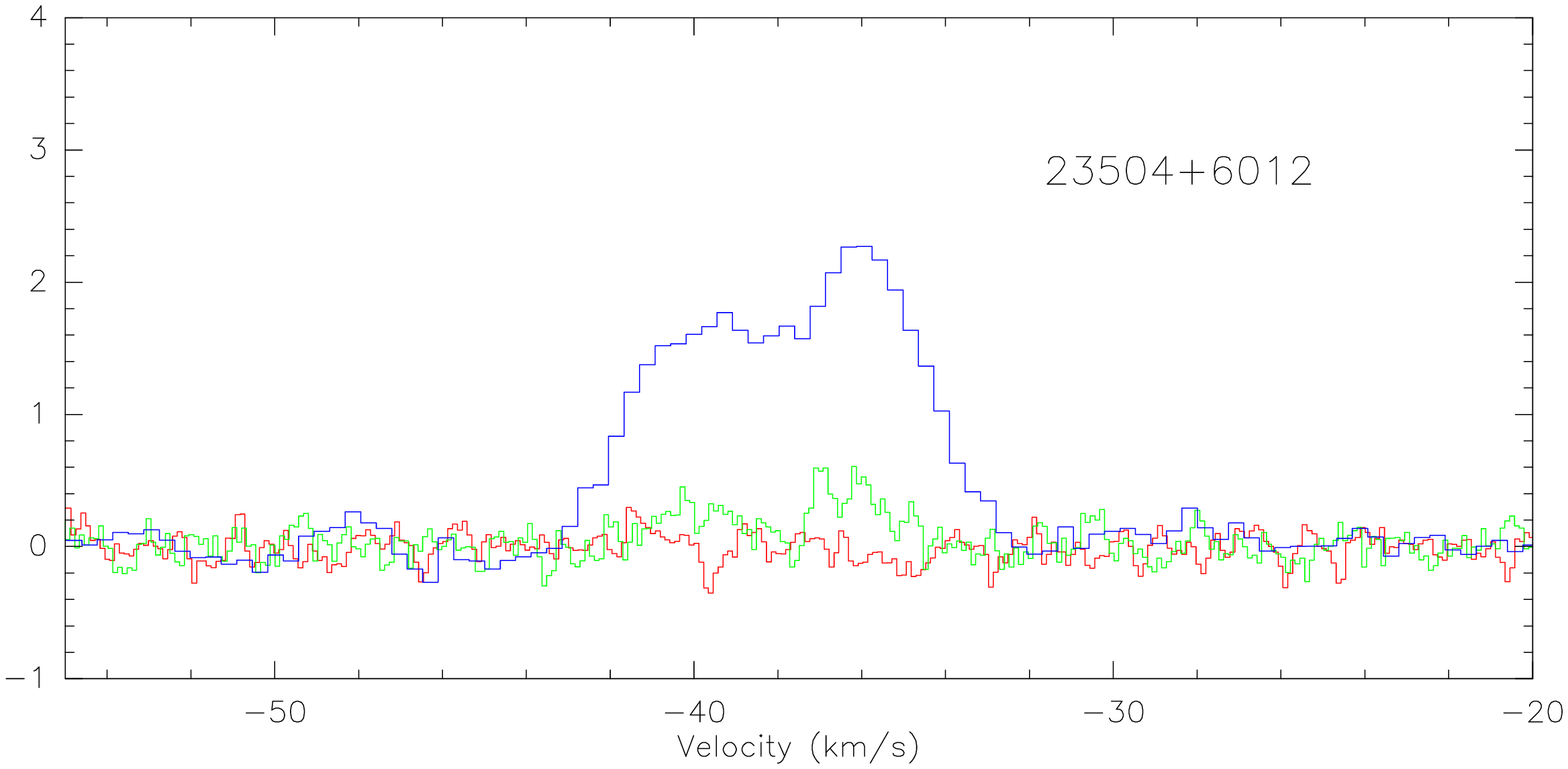}

\begin{minipage}[]{60mm}
   \caption{The sources of type 2
  }\end{minipage}
   \label{Fig7}
   \end{figure}

\begin{figure}
\centering
\includegraphics[height=70mm,  angle=-90, clip, viewport=150 10 500 750]{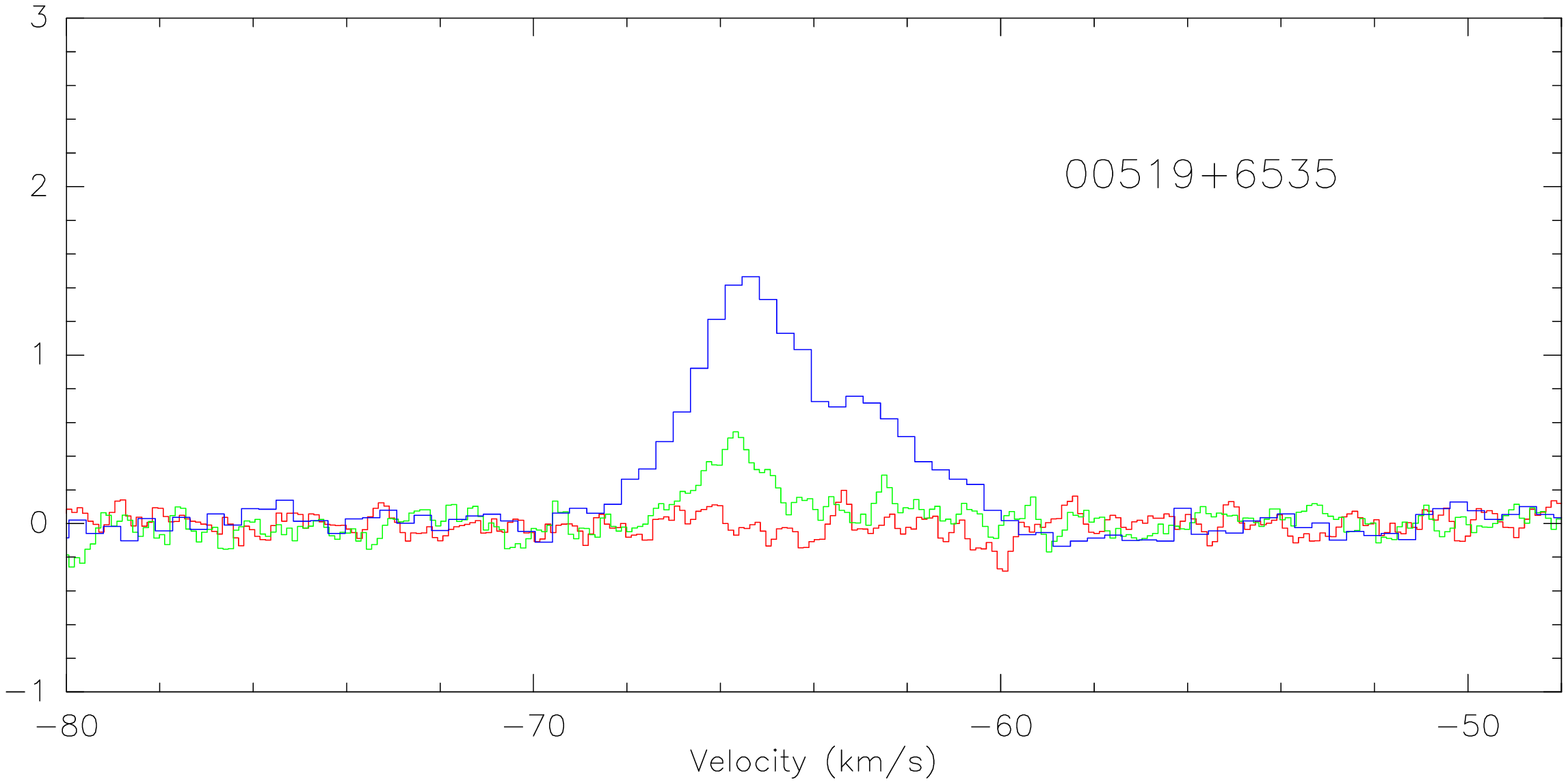}
\includegraphics[height=70mm,  angle=-90, clip, viewport=150 10 500 750]{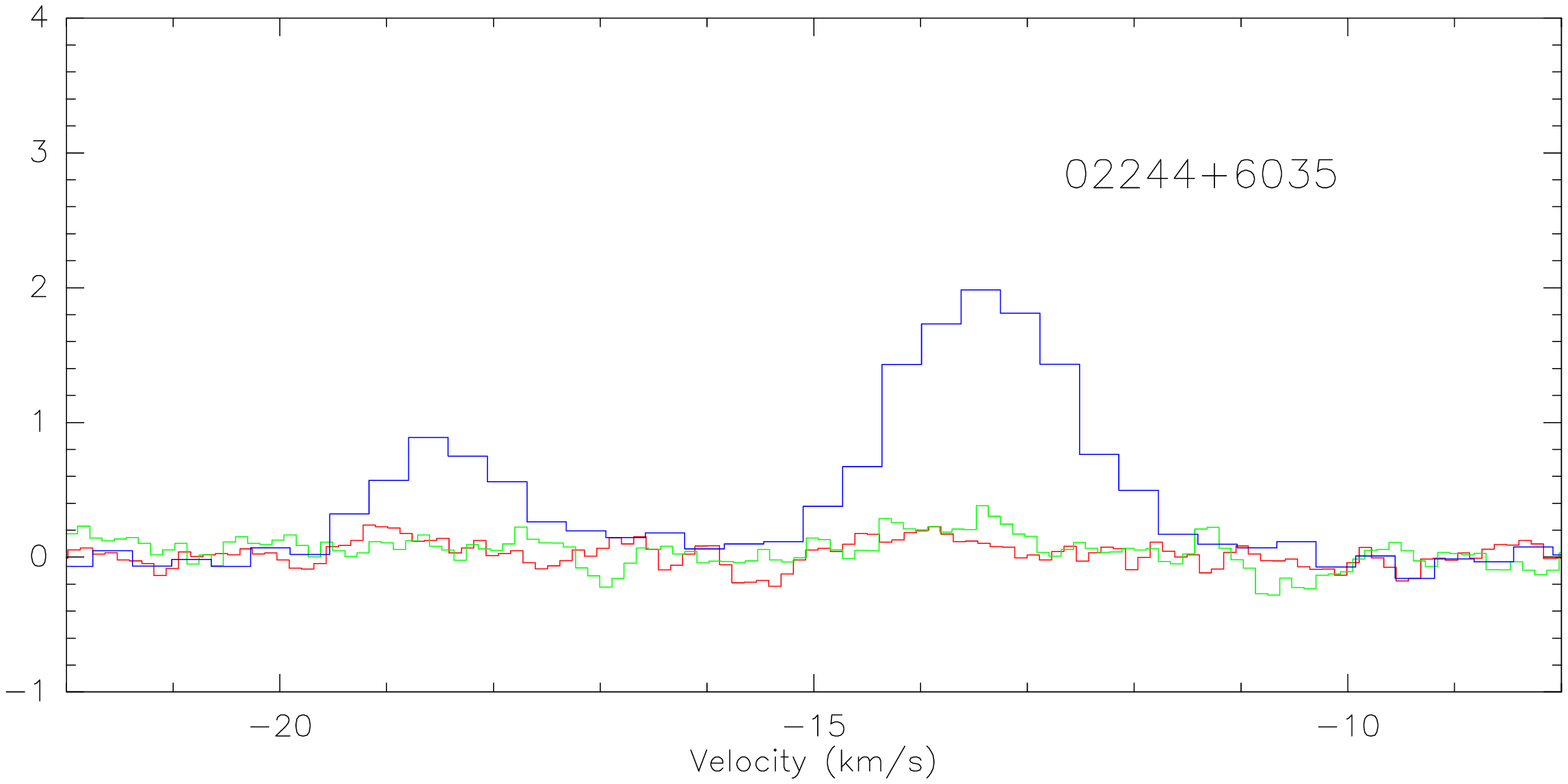}
\includegraphics[height=70mm,  angle=-90, clip, viewport=150 10 500 750]{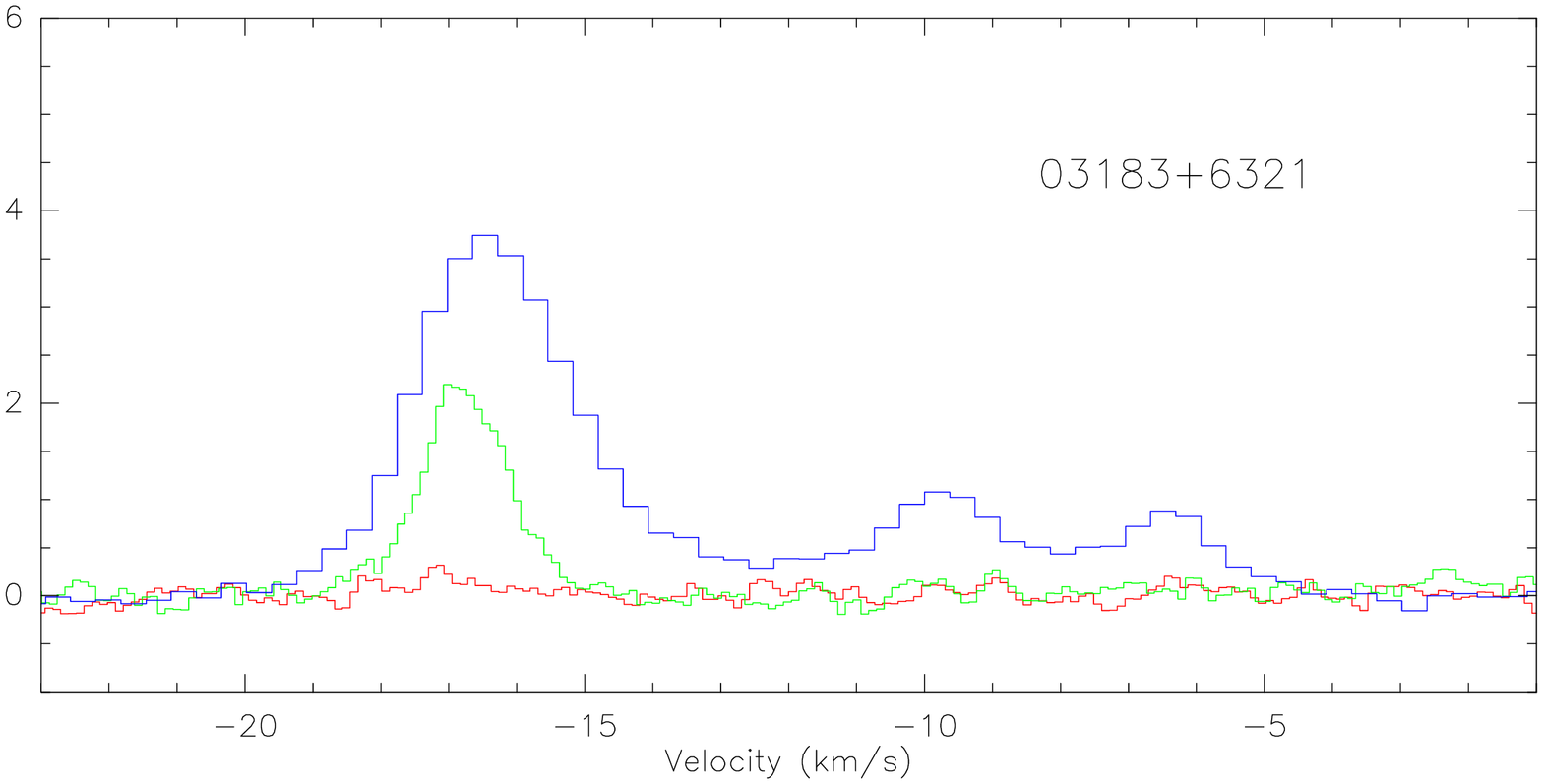}
\includegraphics[height=70mm,  angle=-90, clip, viewport=150 10 500 750]{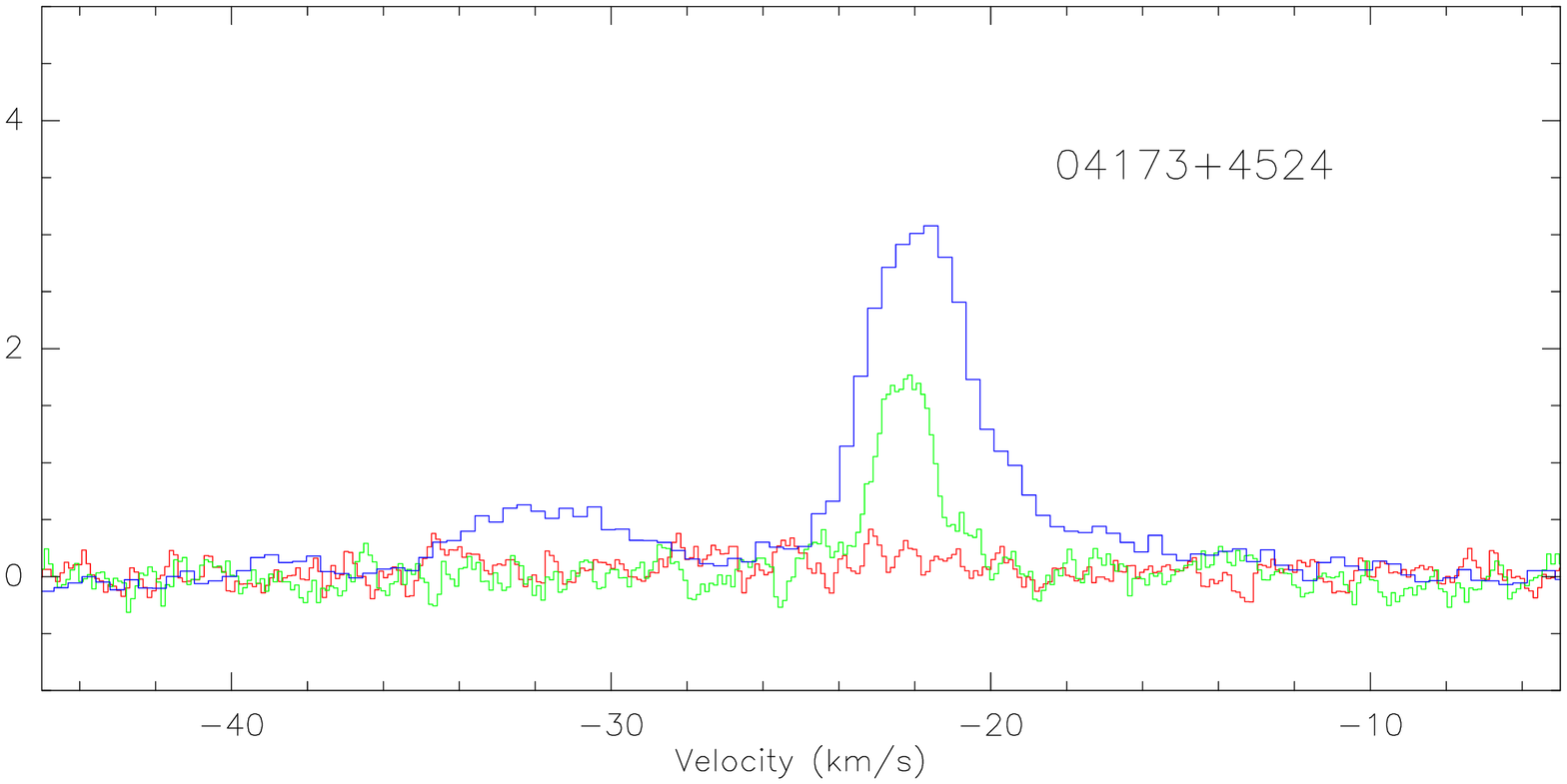}
\includegraphics[height=70mm,  angle=-90, clip, viewport=150 10 500 750]{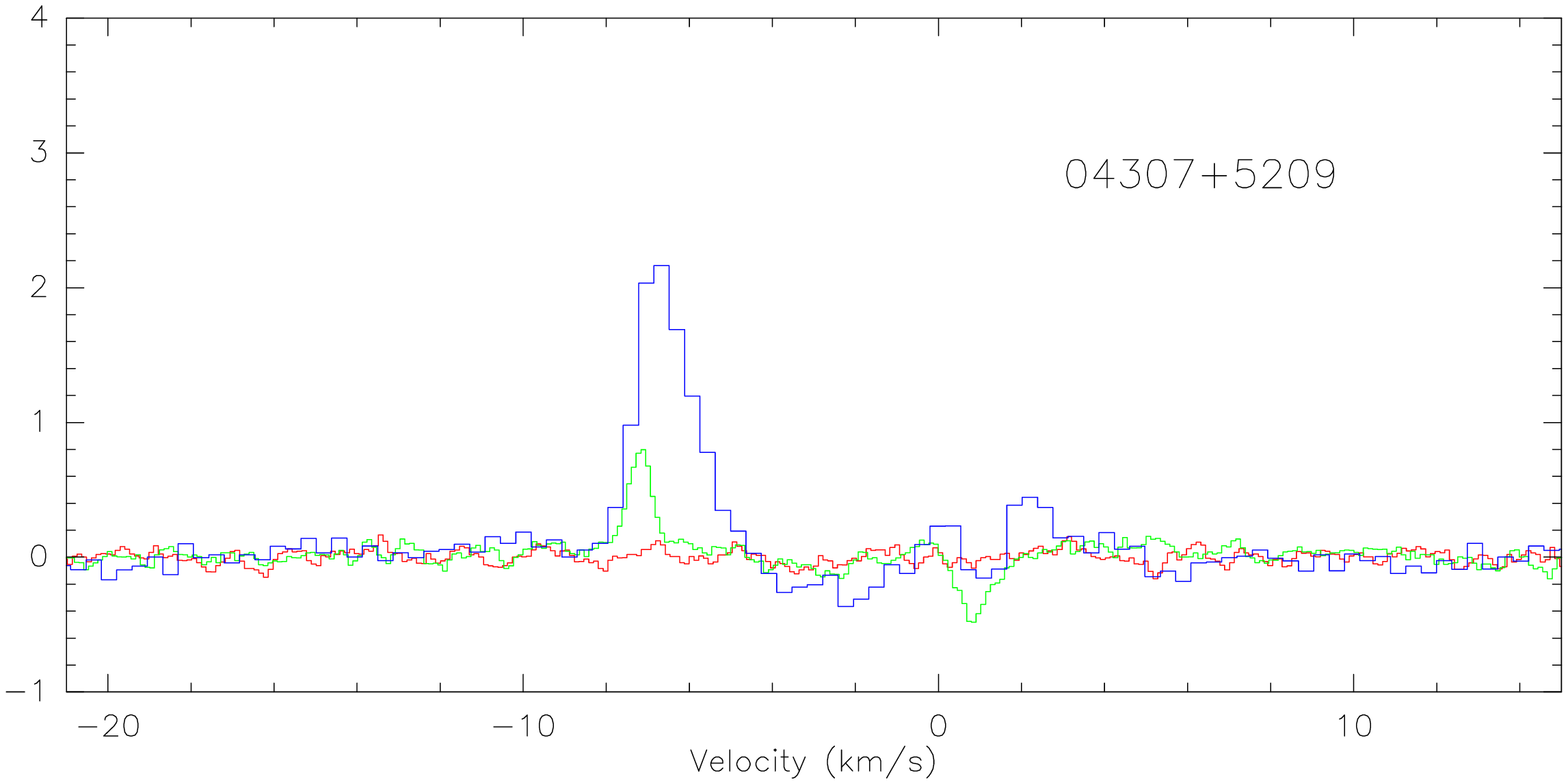}
\includegraphics[height=70mm,  angle=-90, clip, viewport=150 10 500 750]{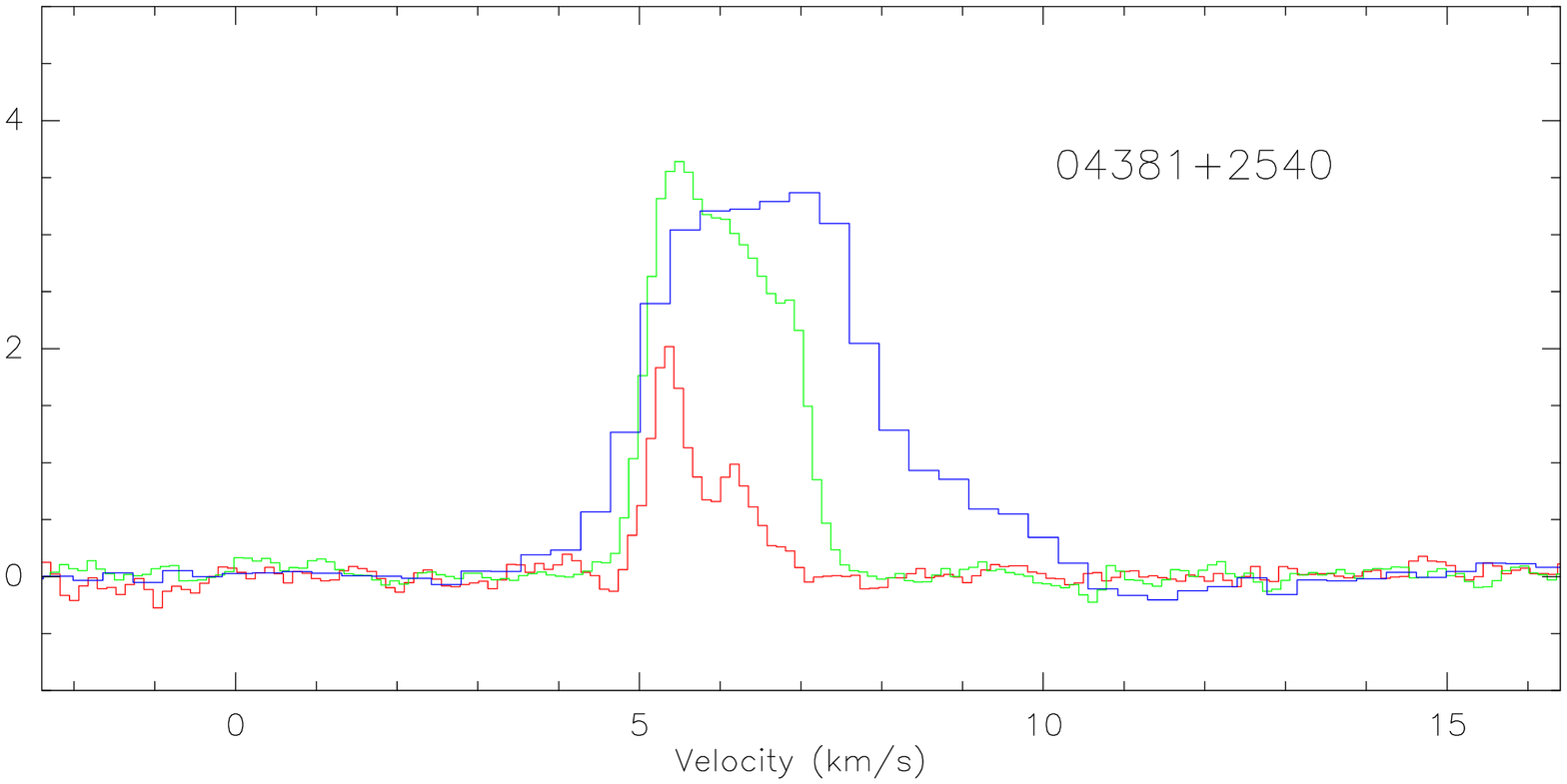}
\includegraphics[height=70mm,  angle=-90, clip, viewport=150 10 500 750]{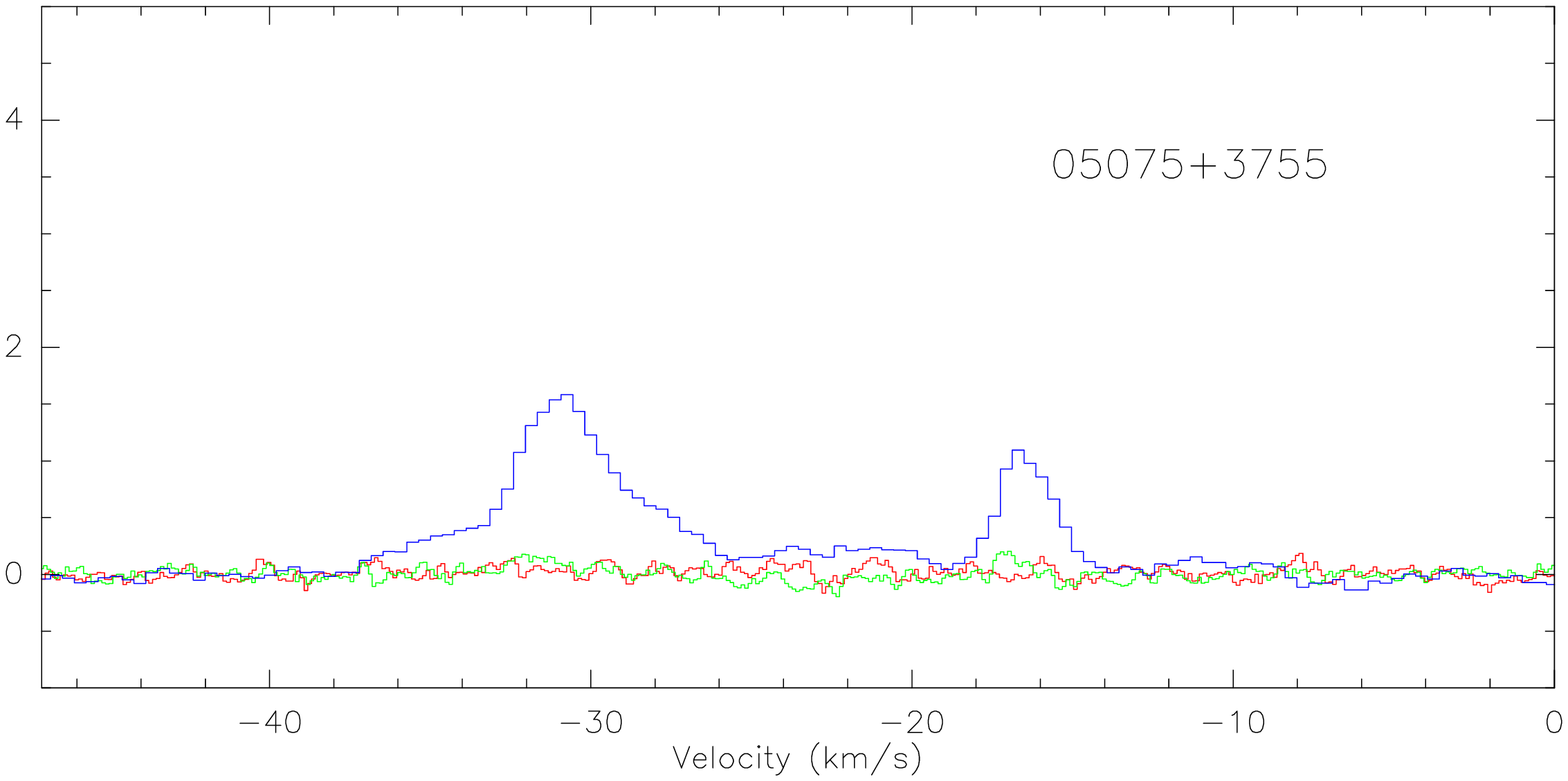}
\includegraphics[height=70mm,  angle=-90, clip, viewport=150 10 500 750]{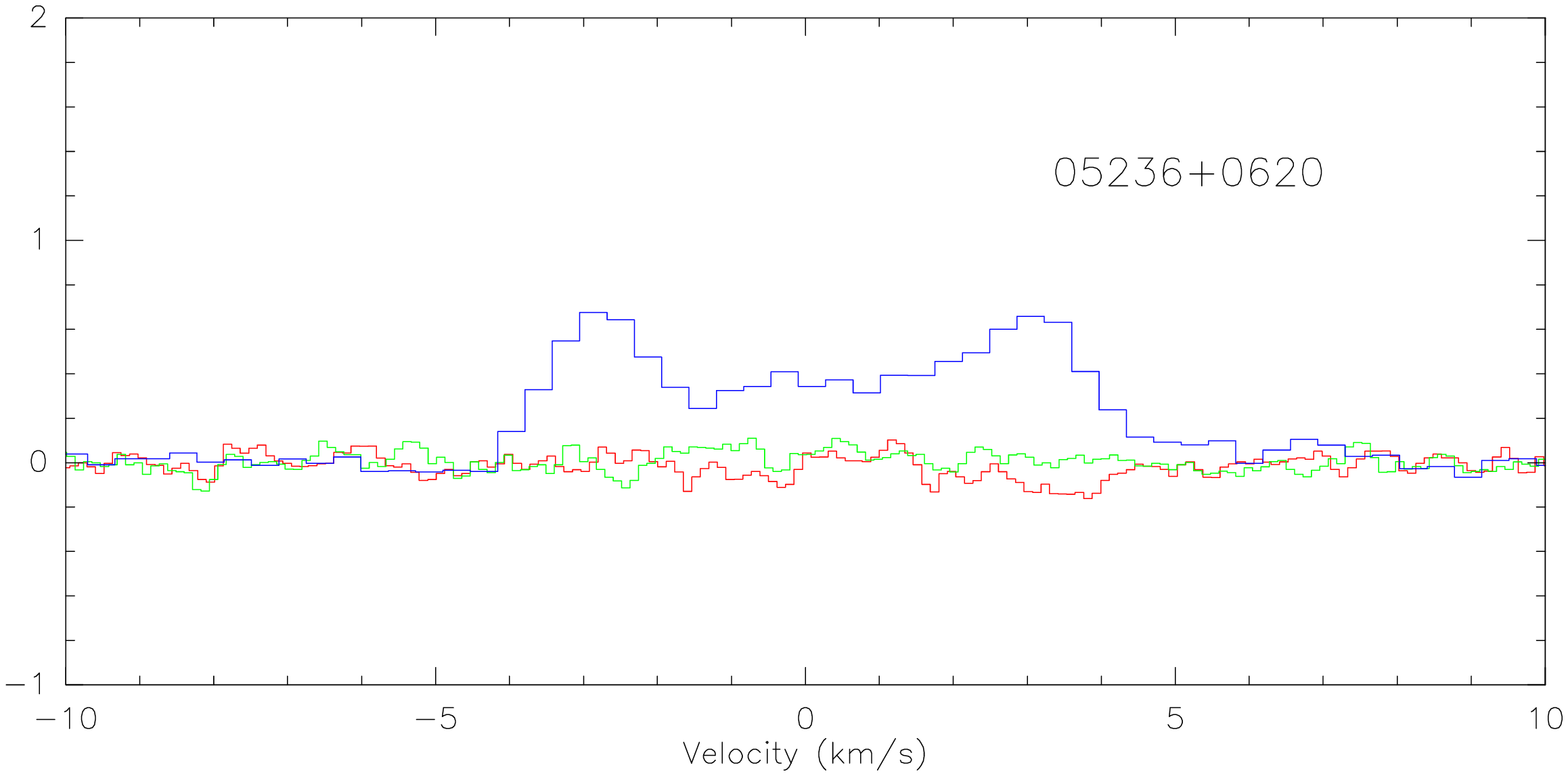}
\includegraphics[height=70mm,  angle=-90, clip, viewport=150 10 500 750]{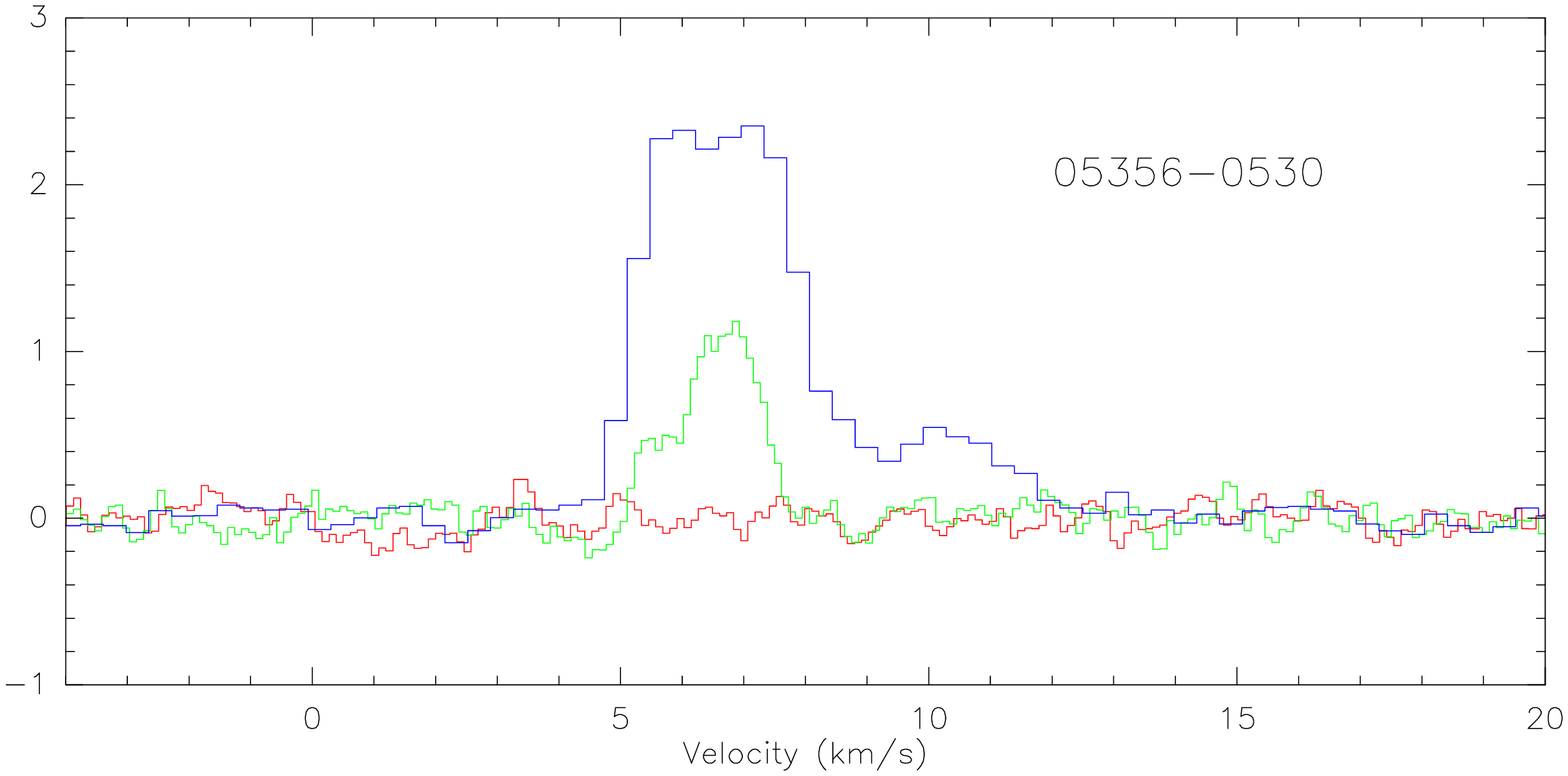}
\includegraphics[height=70mm,  angle=-90, clip, viewport=150 10 500 750]{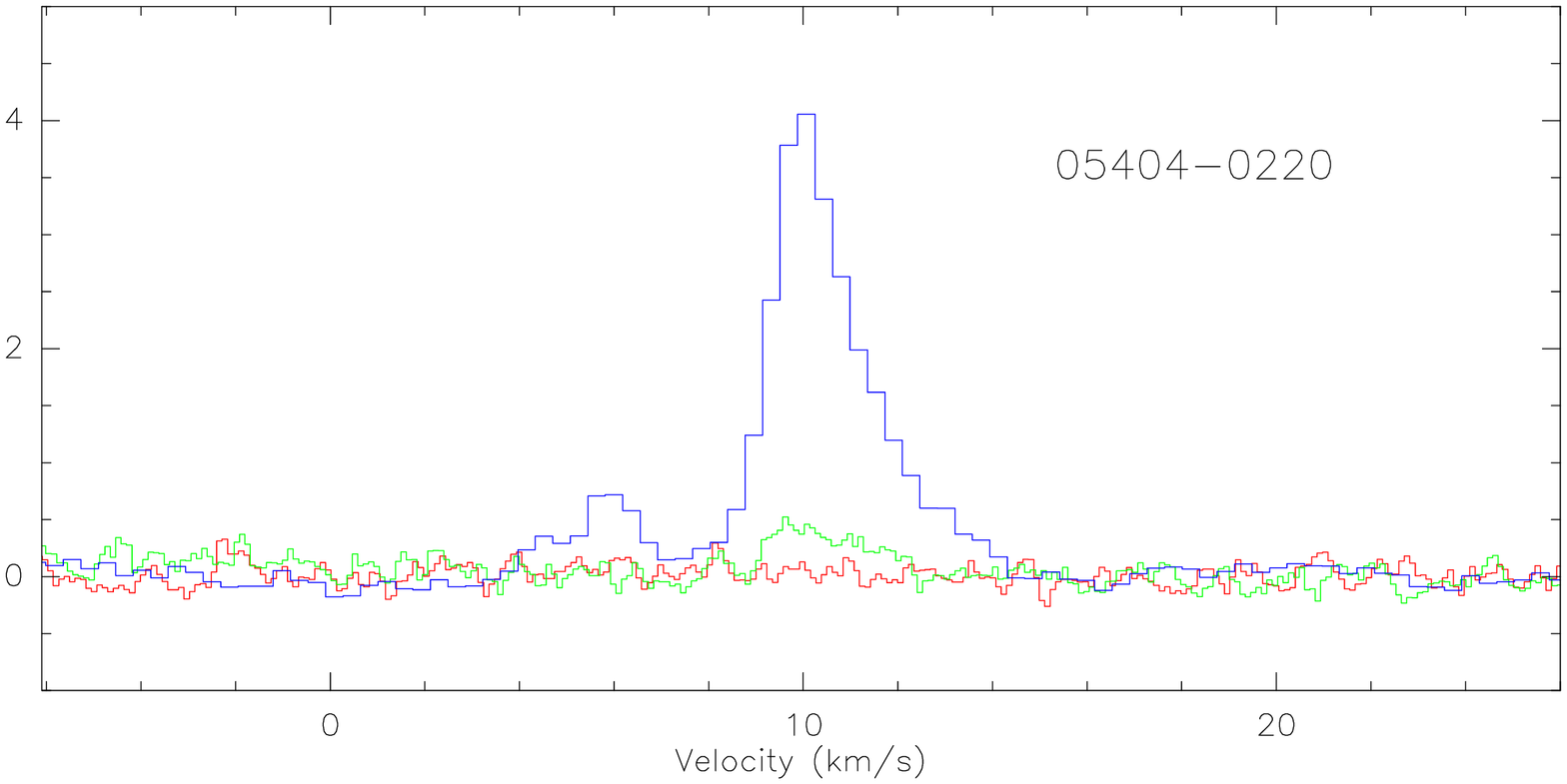}
\includegraphics[height=70mm,  angle=-90, clip, viewport=150 10 500 750]{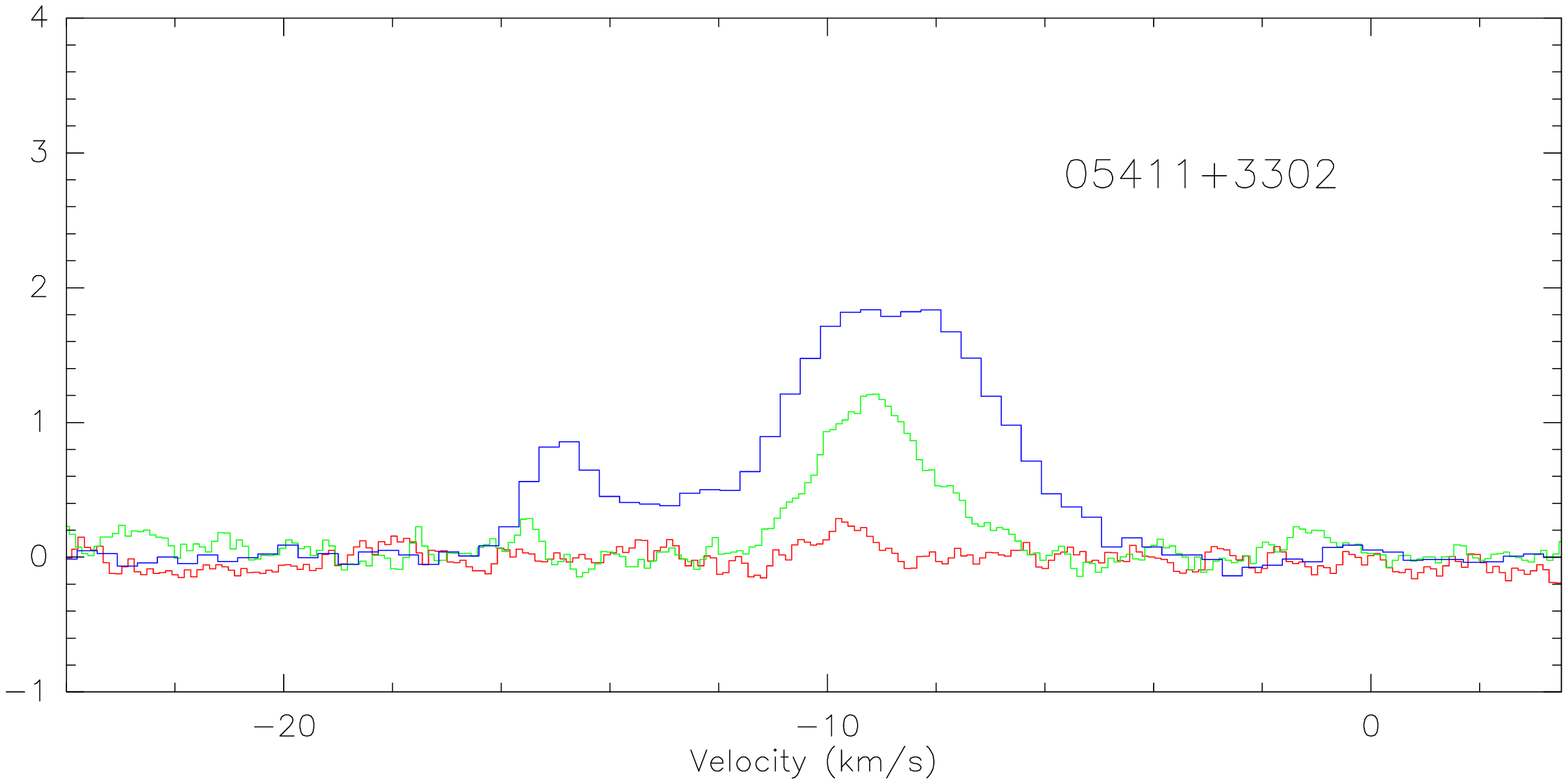}
\includegraphics[height=70mm,  angle=-90, clip, viewport=150 10 500 750]{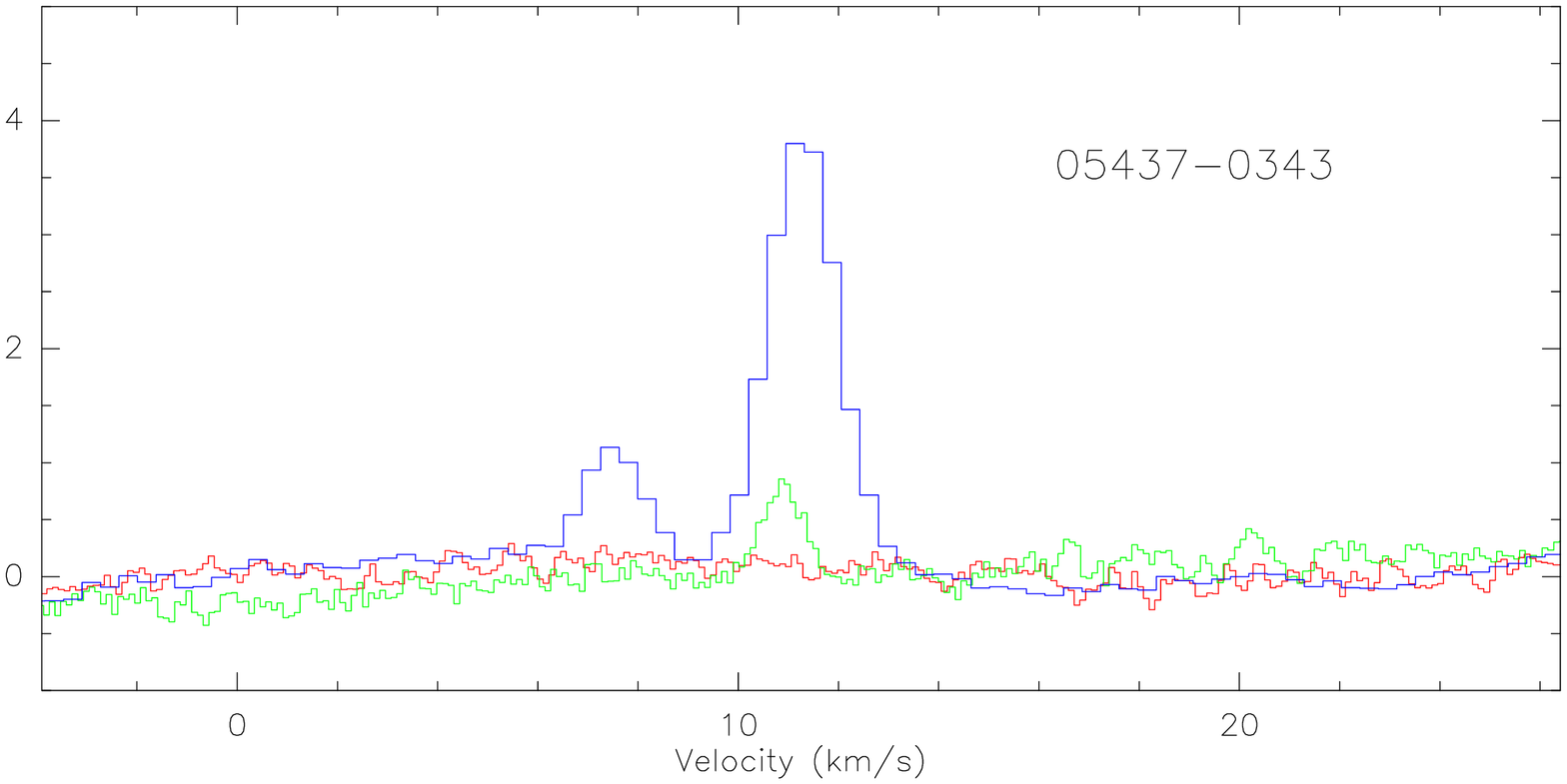}
\includegraphics[height=70mm,  angle=-90, clip, viewport=150 10 500 750]{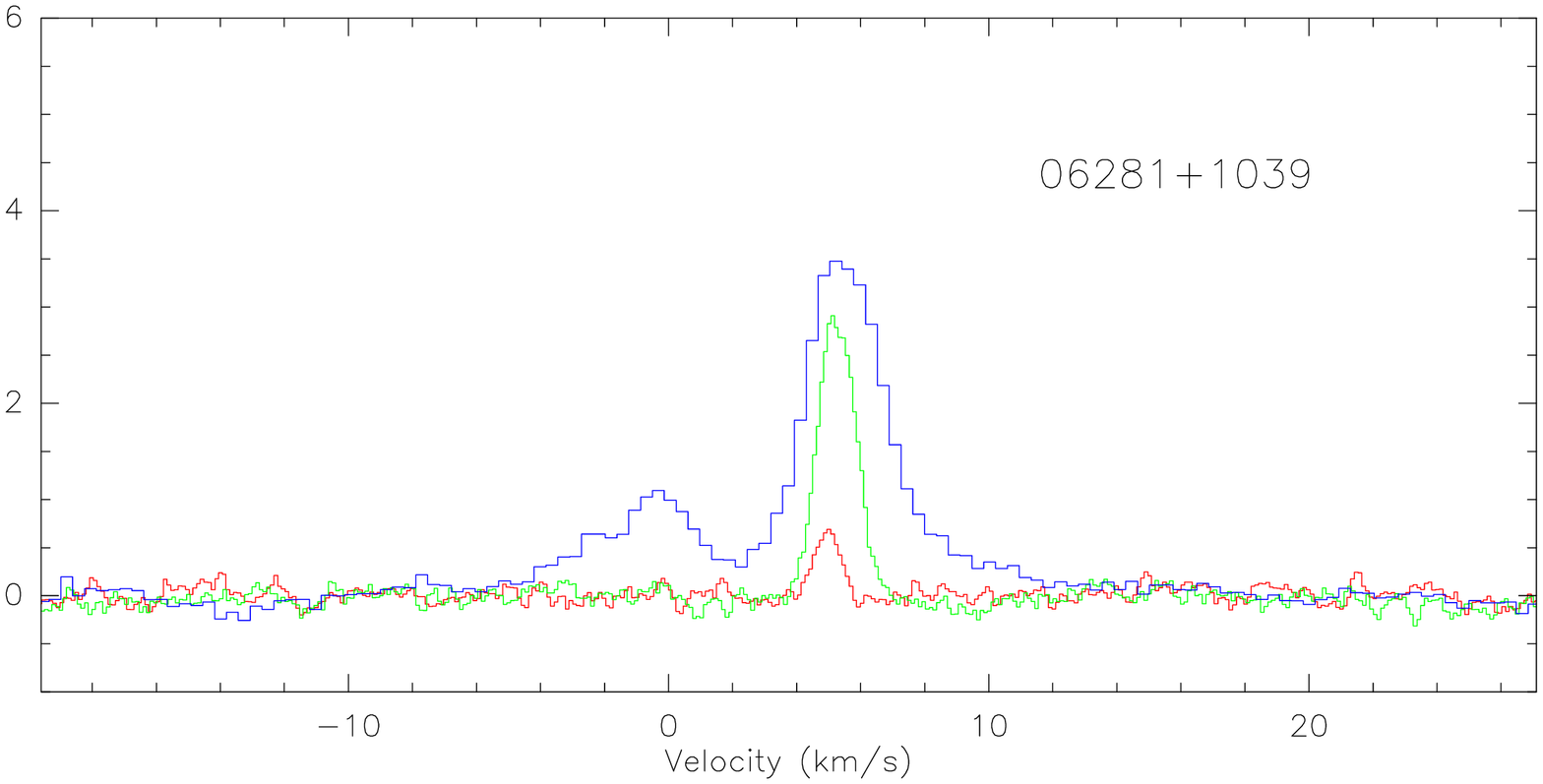}
\includegraphics[height=70mm,  angle=-90, clip, viewport=150 10 500 750]{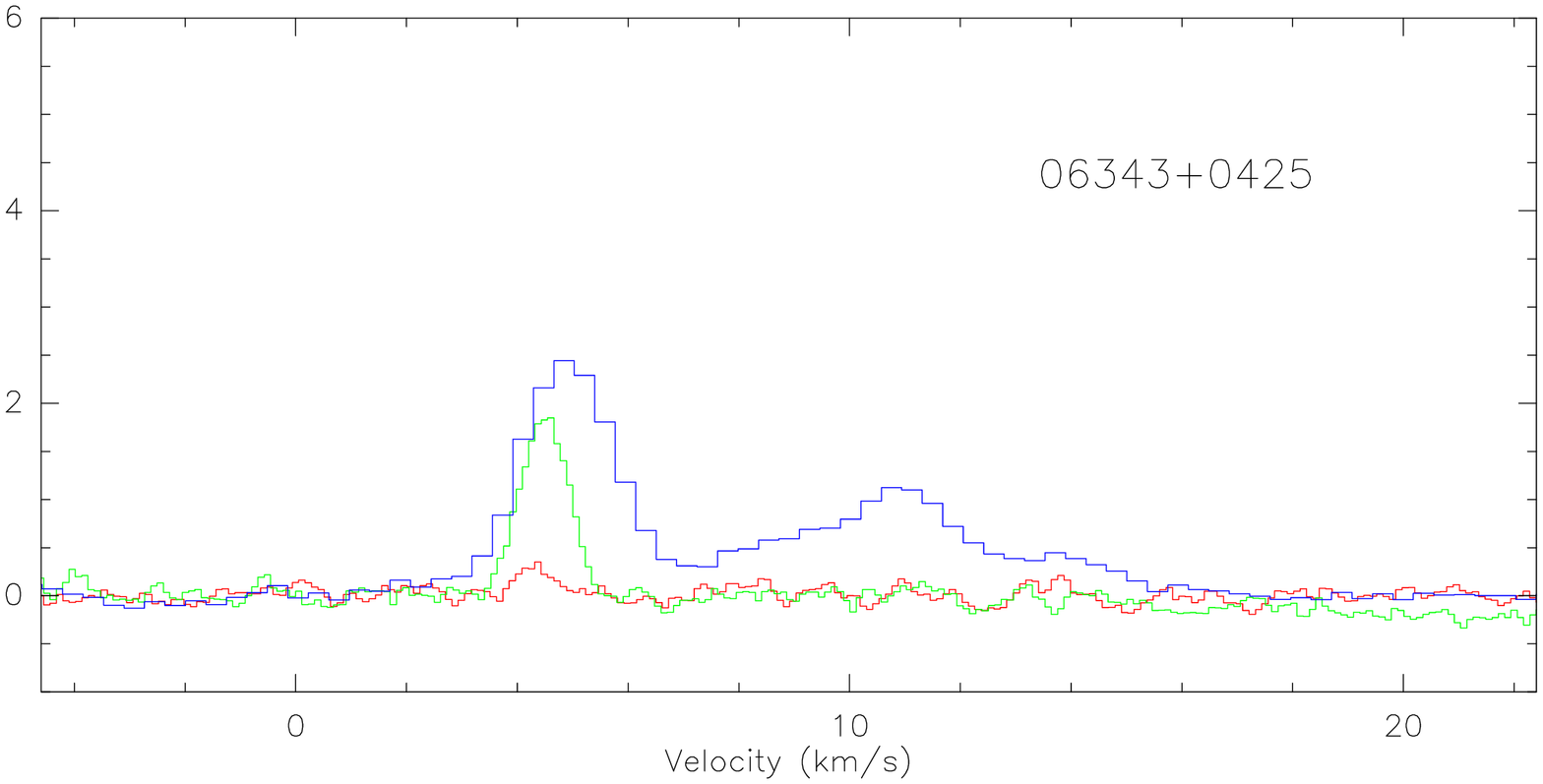}

\begin{minipage}[]{60mm}
  \caption{The sources of type 3
  }\end{minipage}
   \label{Fig8}
   \end{figure}

\addtocounter{figure}{-1}
\begin{figure}

\centering
\includegraphics[height=70mm,  angle=-90, clip, viewport=150 10 500 750]{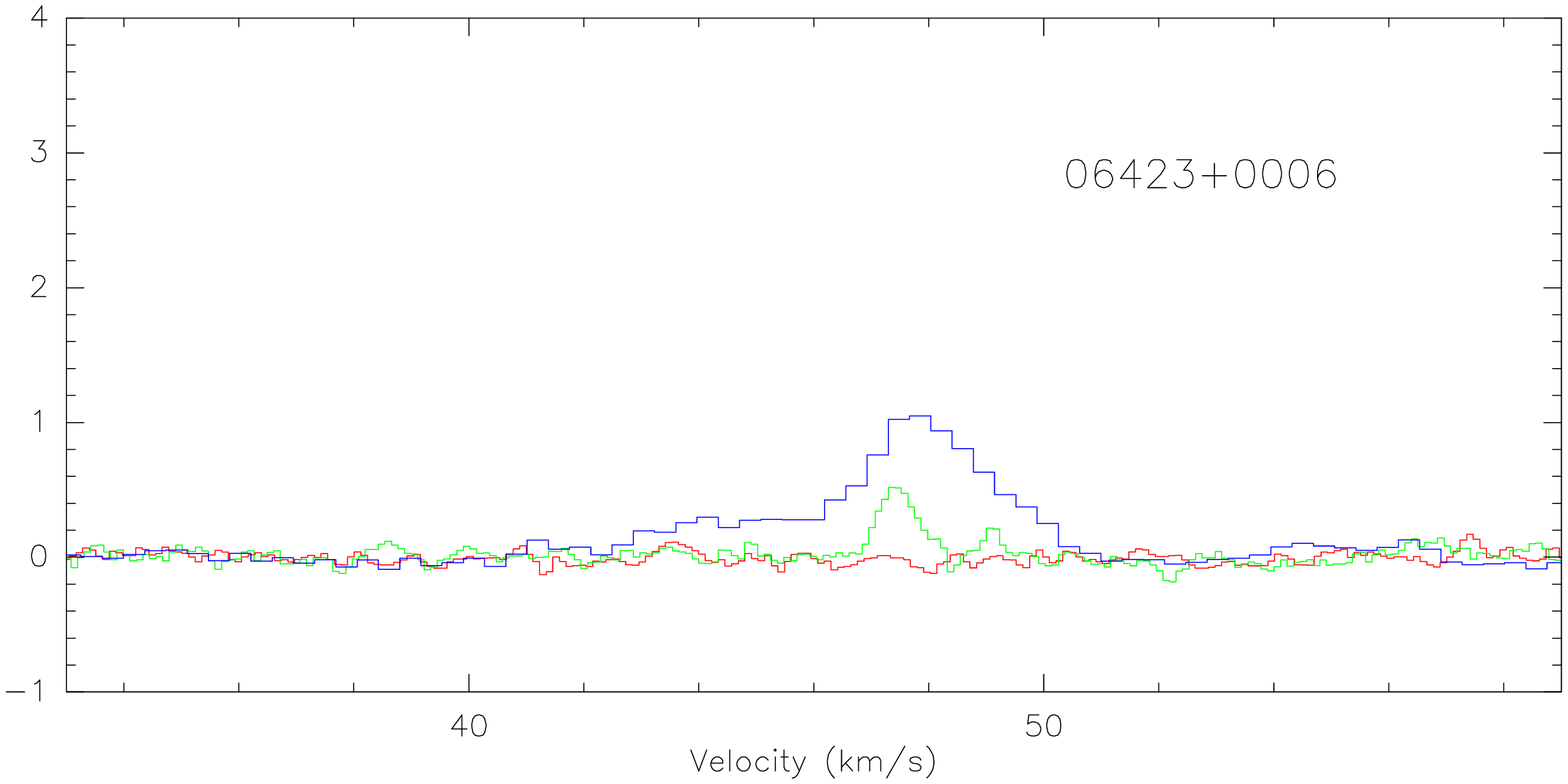}
\includegraphics[height=70mm,  angle=-90, clip, viewport=150 10 500 750]{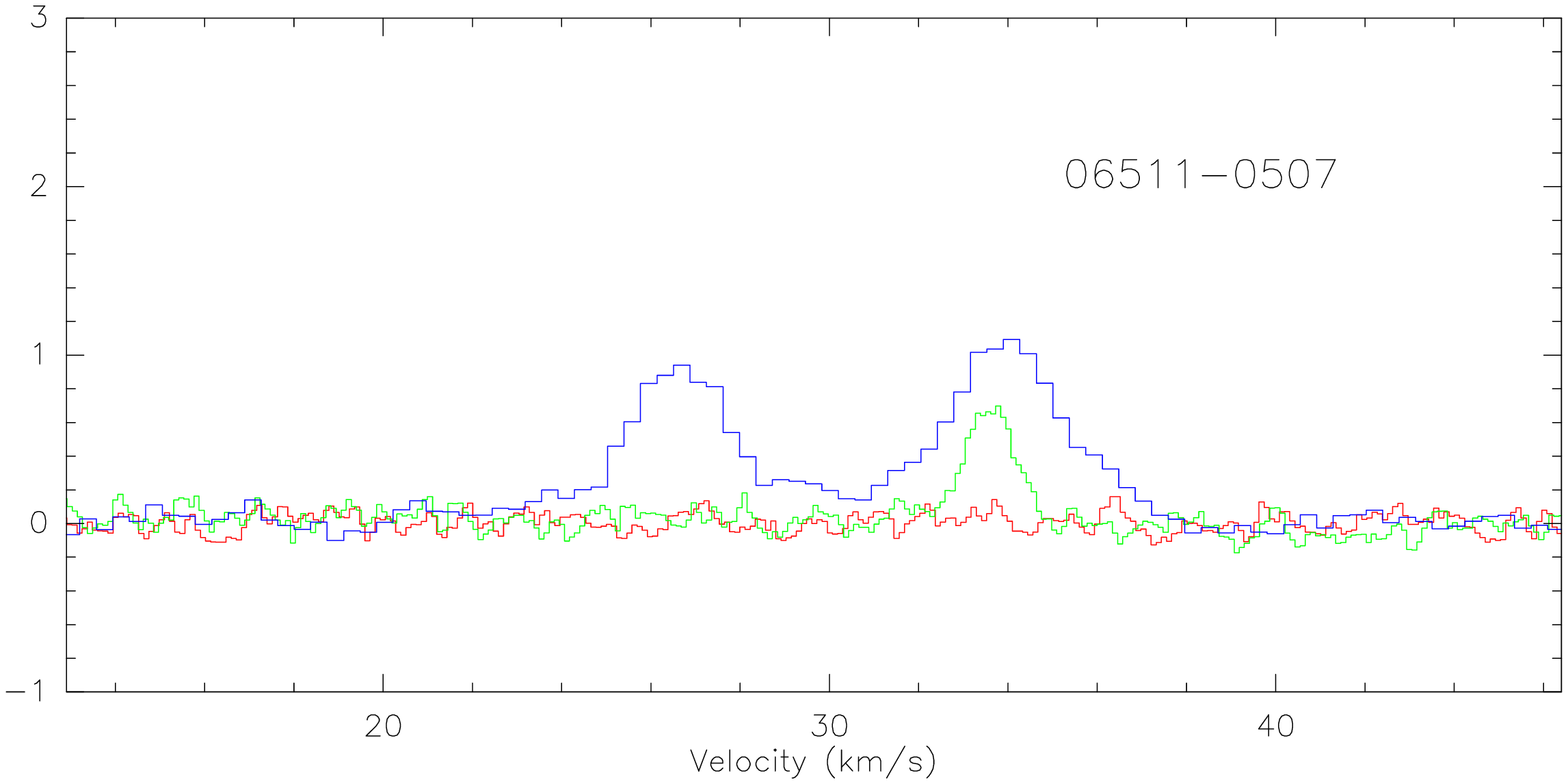}
\includegraphics[height=70mm,  angle=-90, clip, viewport=150 10 500 750]{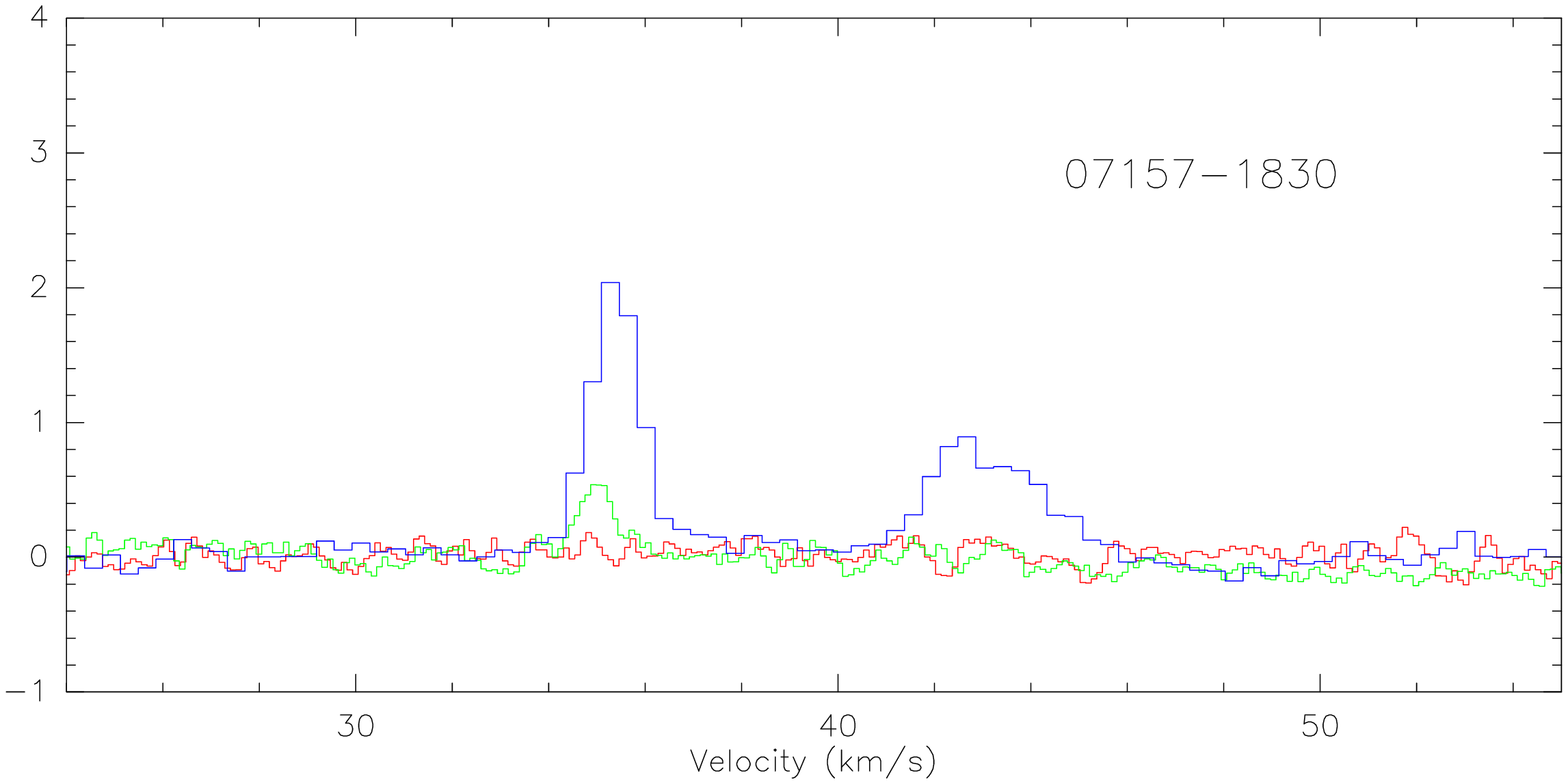}
\includegraphics[height=70mm,  angle=-90, clip, viewport=150 10 500 750]{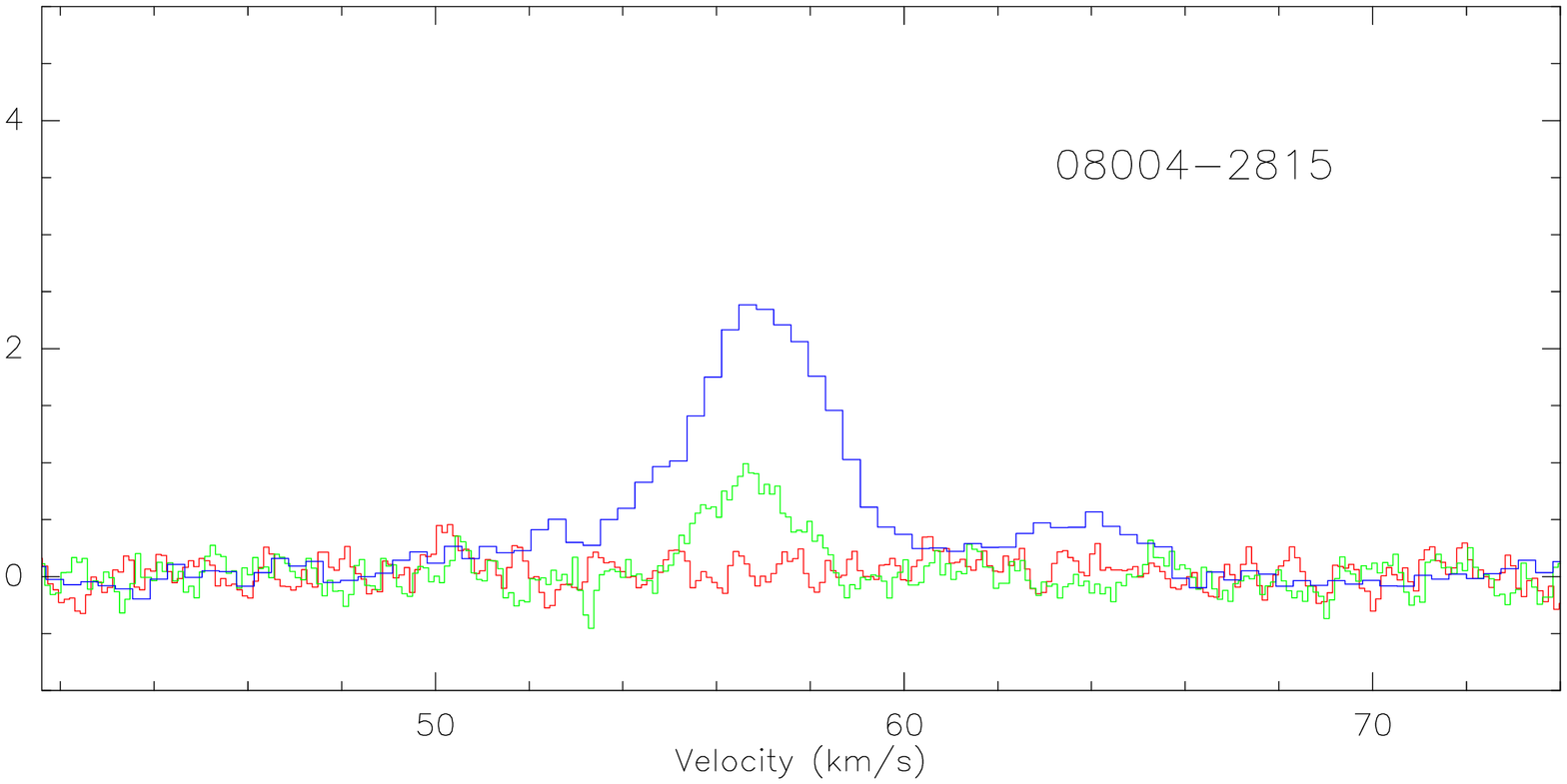}
\includegraphics[height=70mm,  angle=-90, clip, viewport=150 10 500 750]{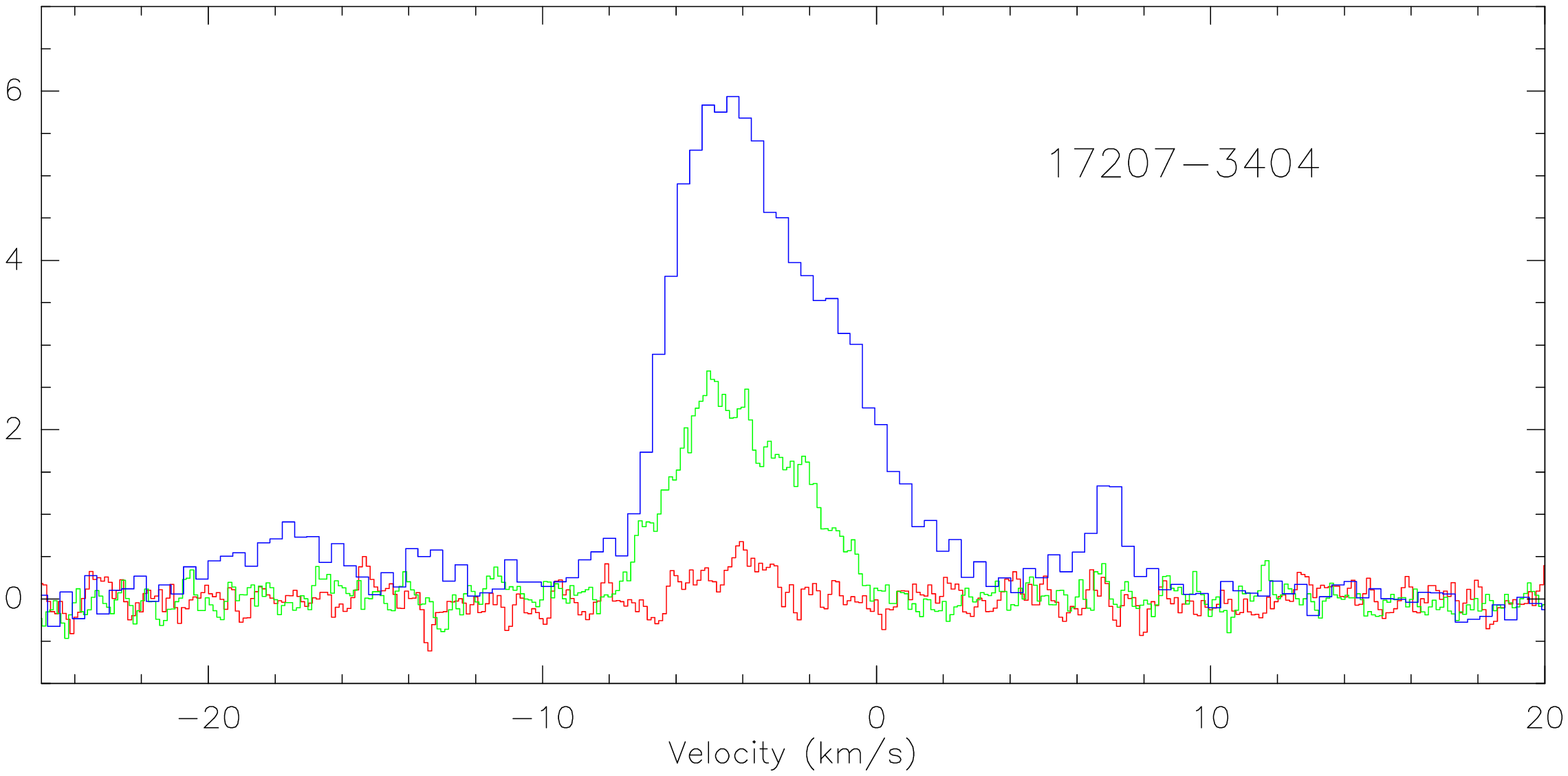}
\includegraphics[height=70mm,  angle=-90, clip, viewport=150 10 500 750]{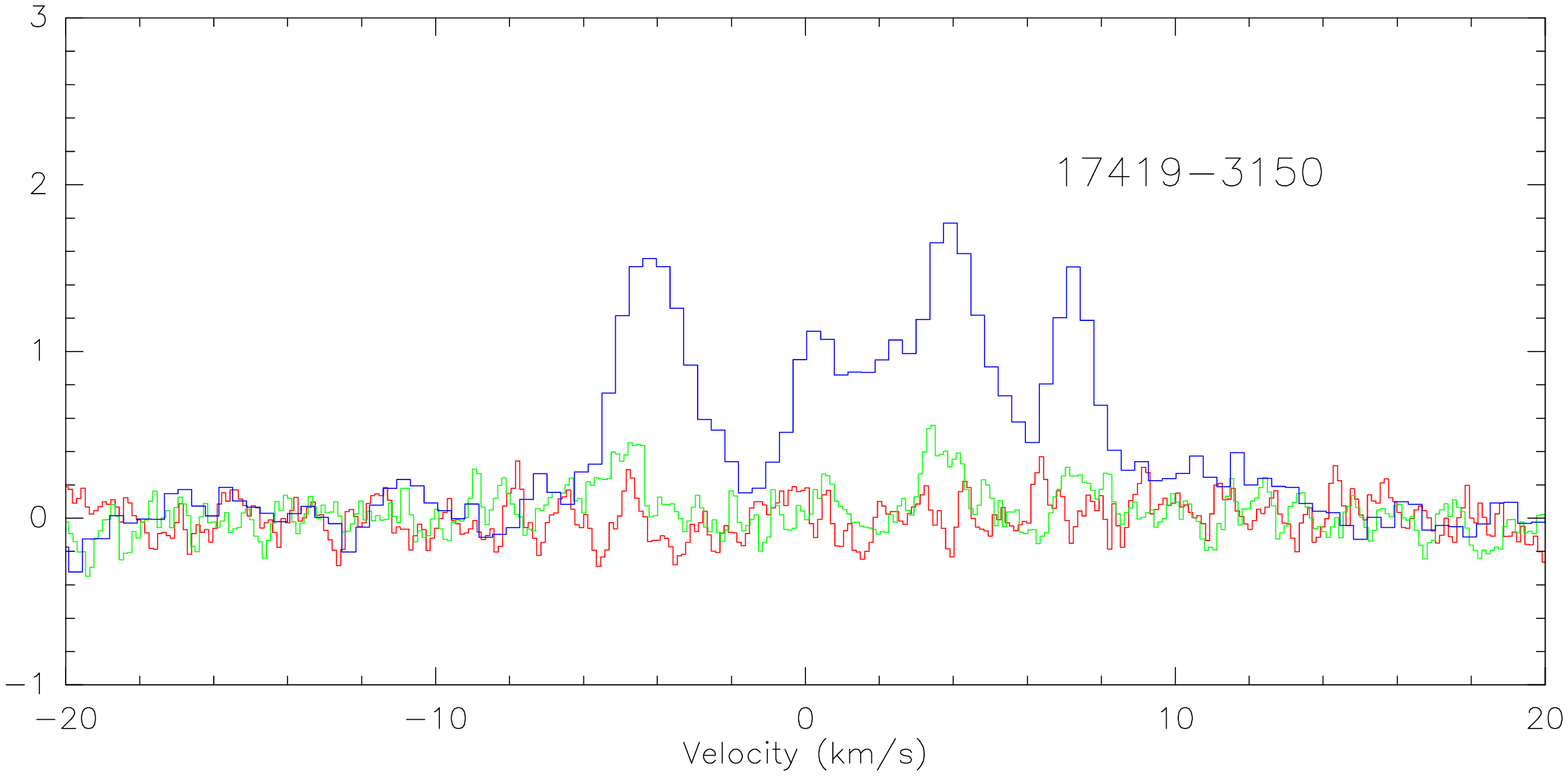}
\includegraphics[height=70mm,  angle=-90, clip, viewport=150 10 500 750]{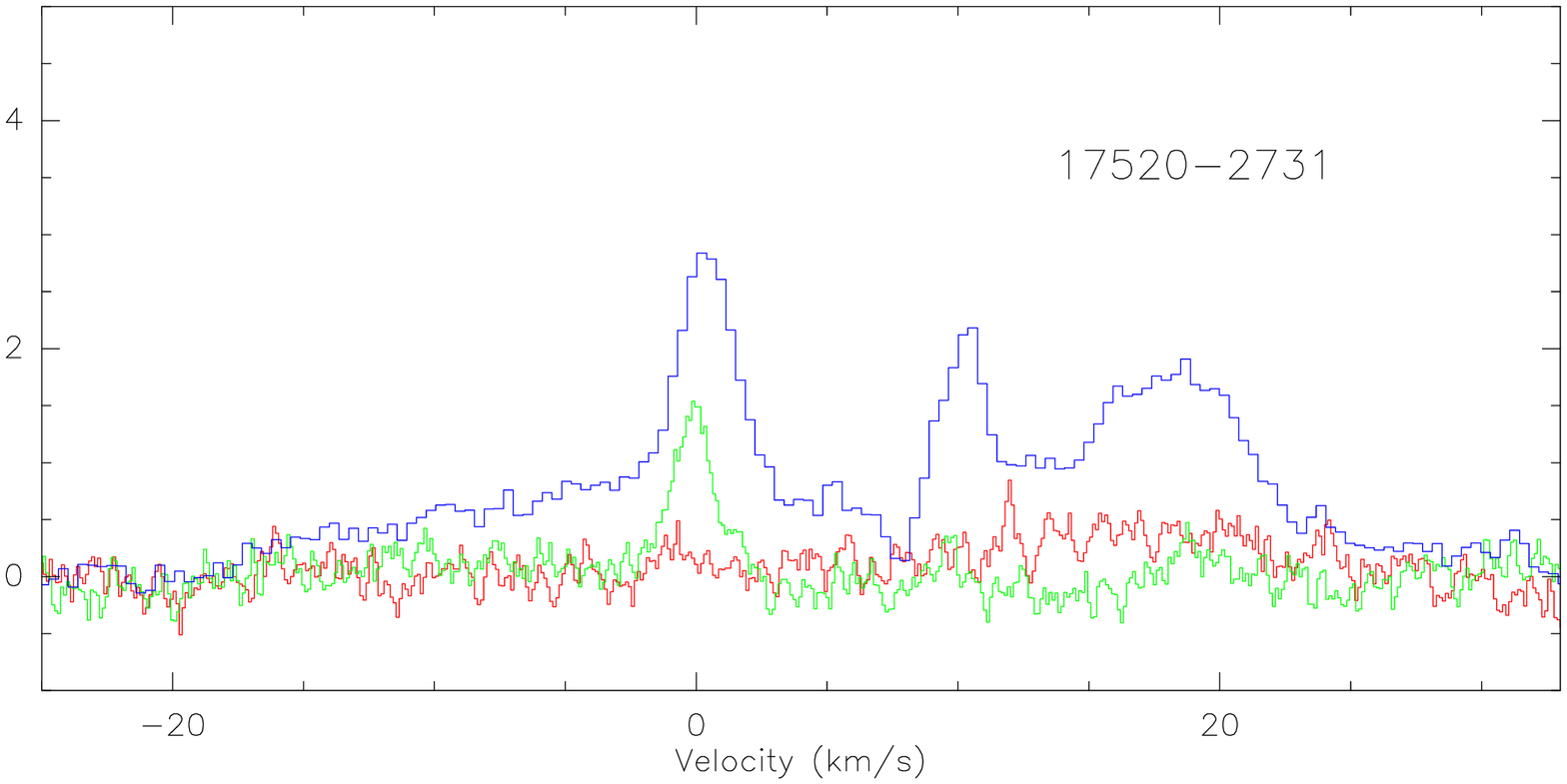}
\includegraphics[height=70mm,  angle=-90, clip, viewport=150 10 500 750]{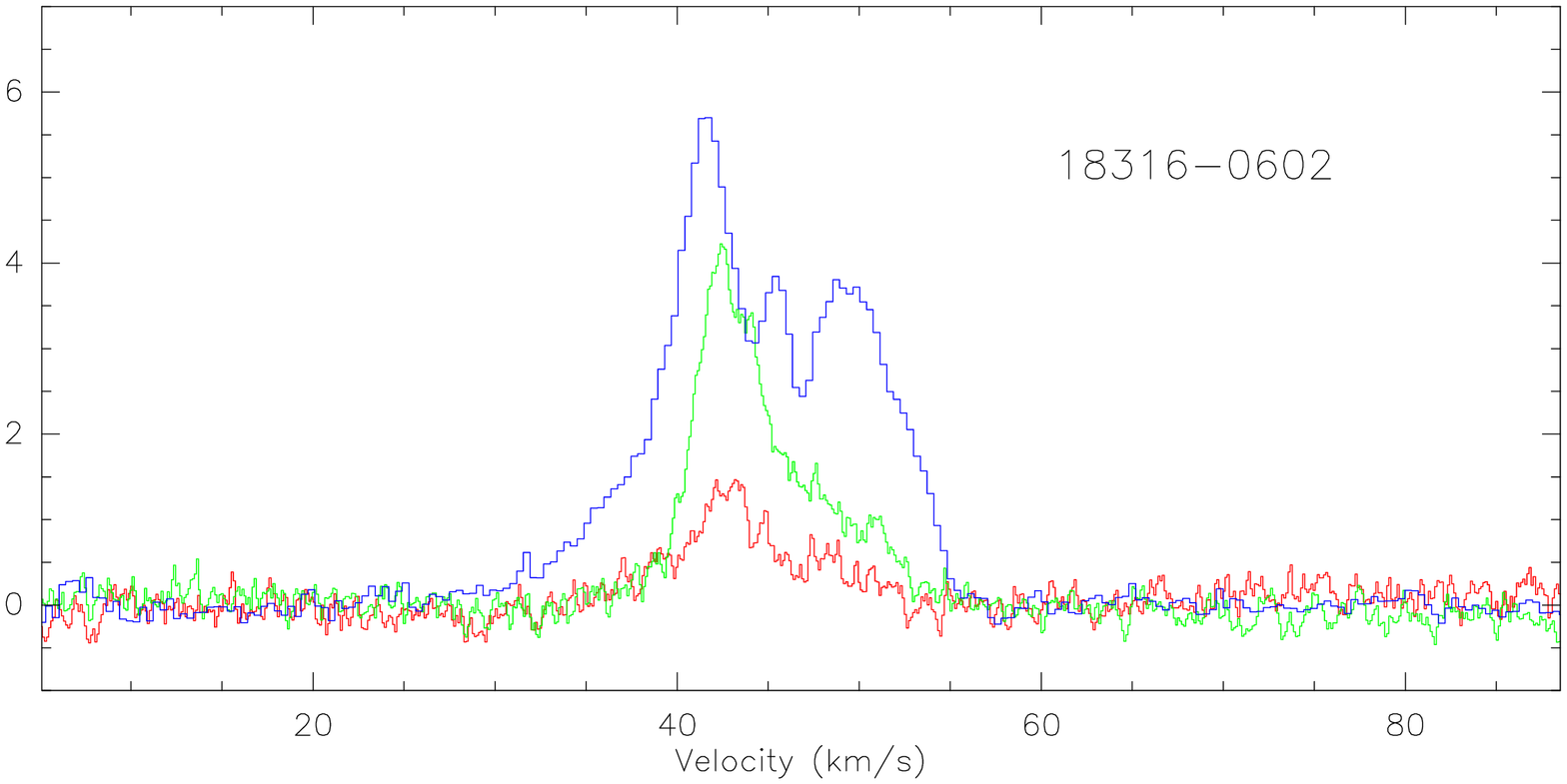}
\includegraphics[height=70mm,  angle=-90, clip, viewport=150 10 500 750]{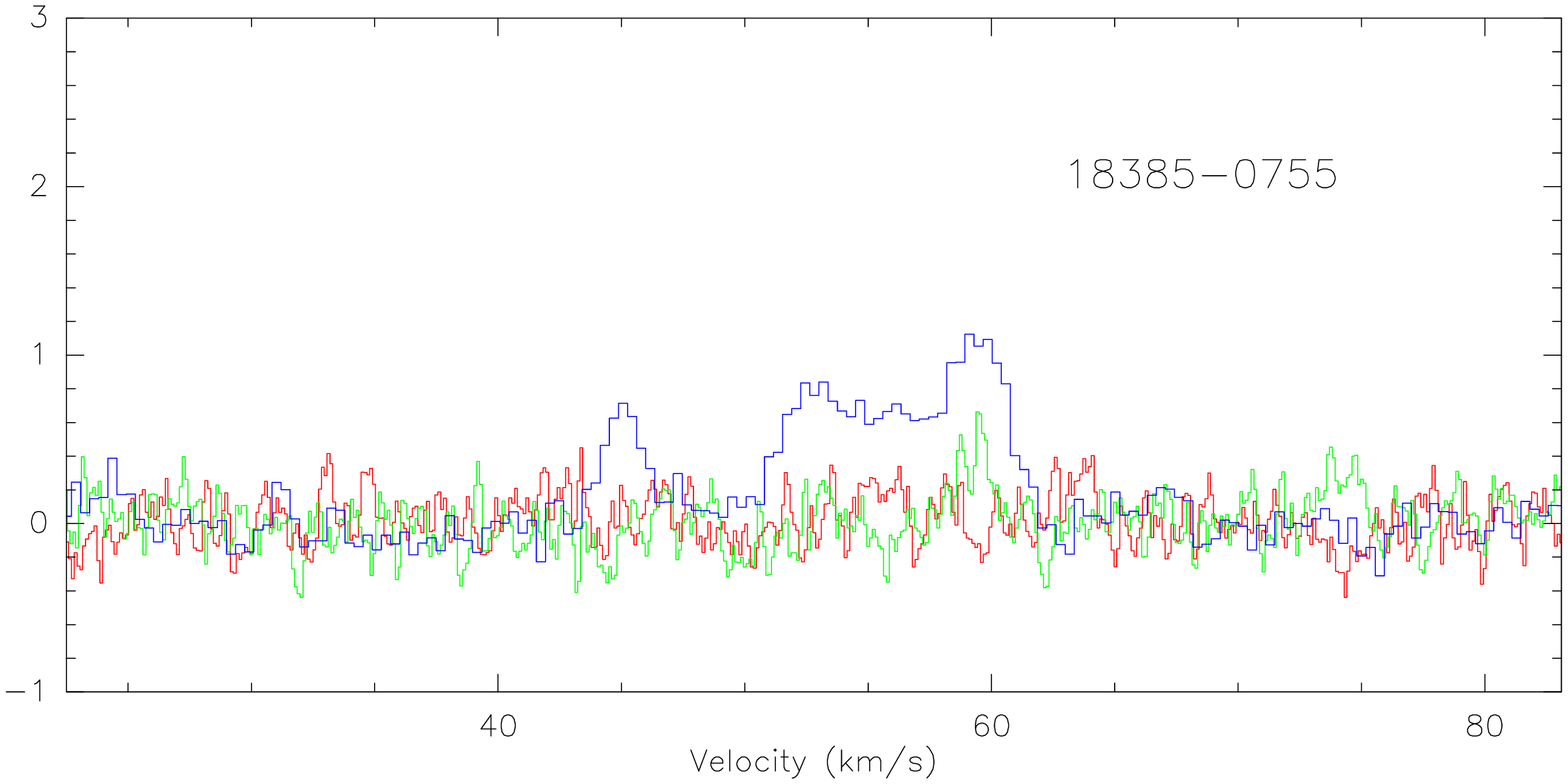}
\includegraphics[height=70mm,  angle=-90, clip, viewport=150 10 500 750]{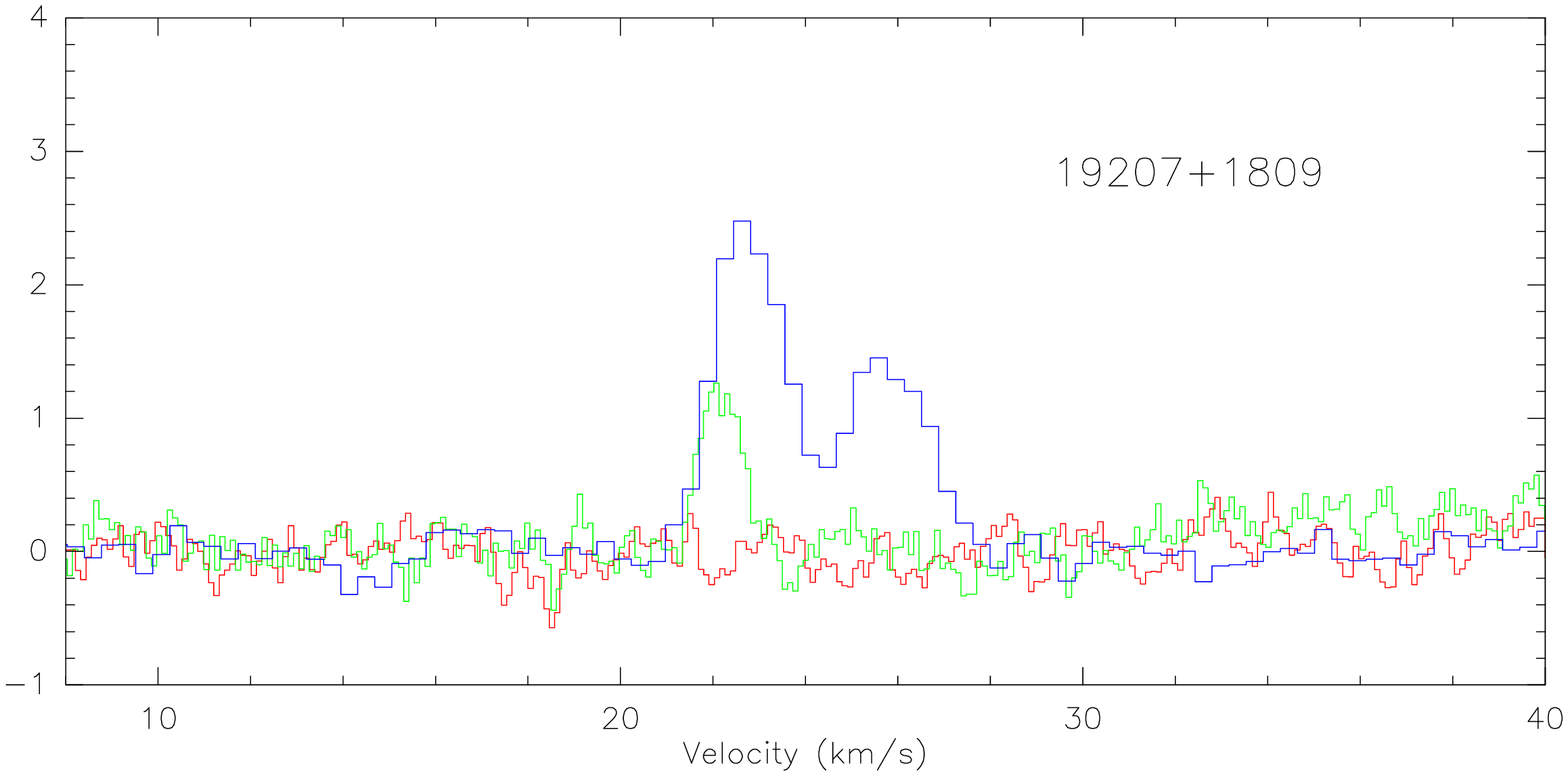}
\includegraphics[height=70mm,  angle=-90, clip, viewport=150 10 500 750]{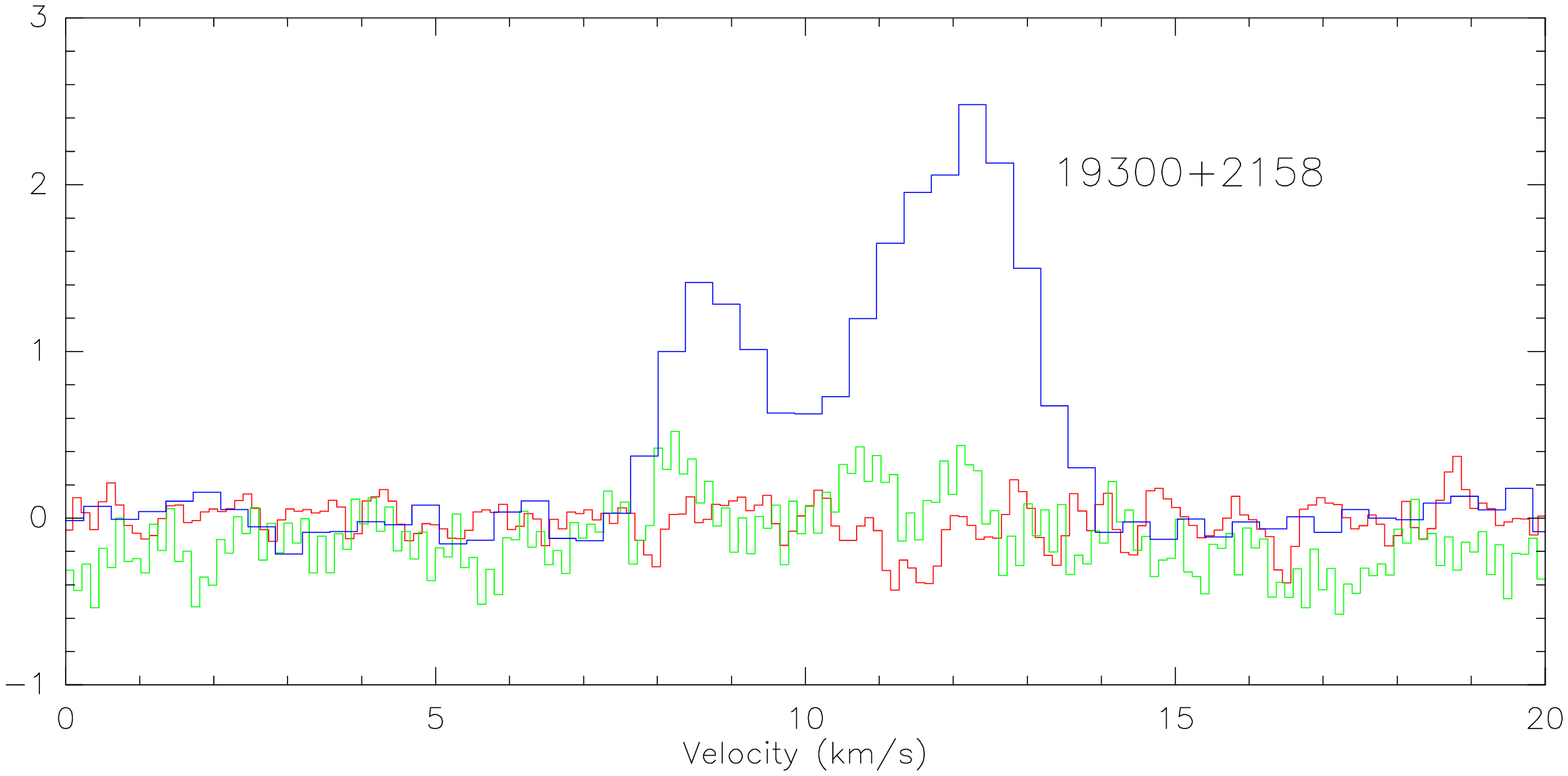}
\includegraphics[height=70mm,  angle=-90, clip, viewport=150 10 500 750]{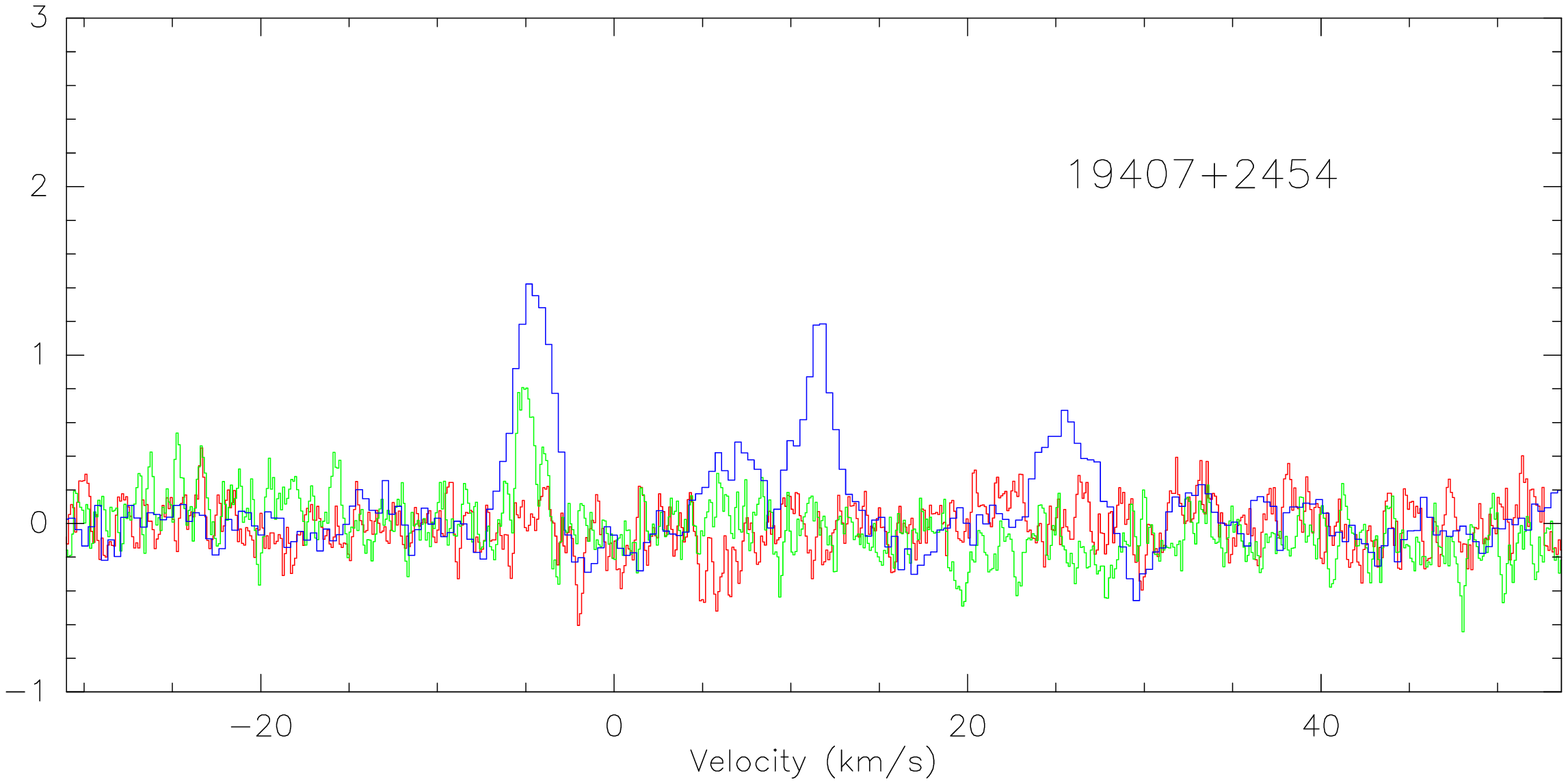}
\includegraphics[height=70mm,  angle=-90, clip, viewport=150 10 500 750]{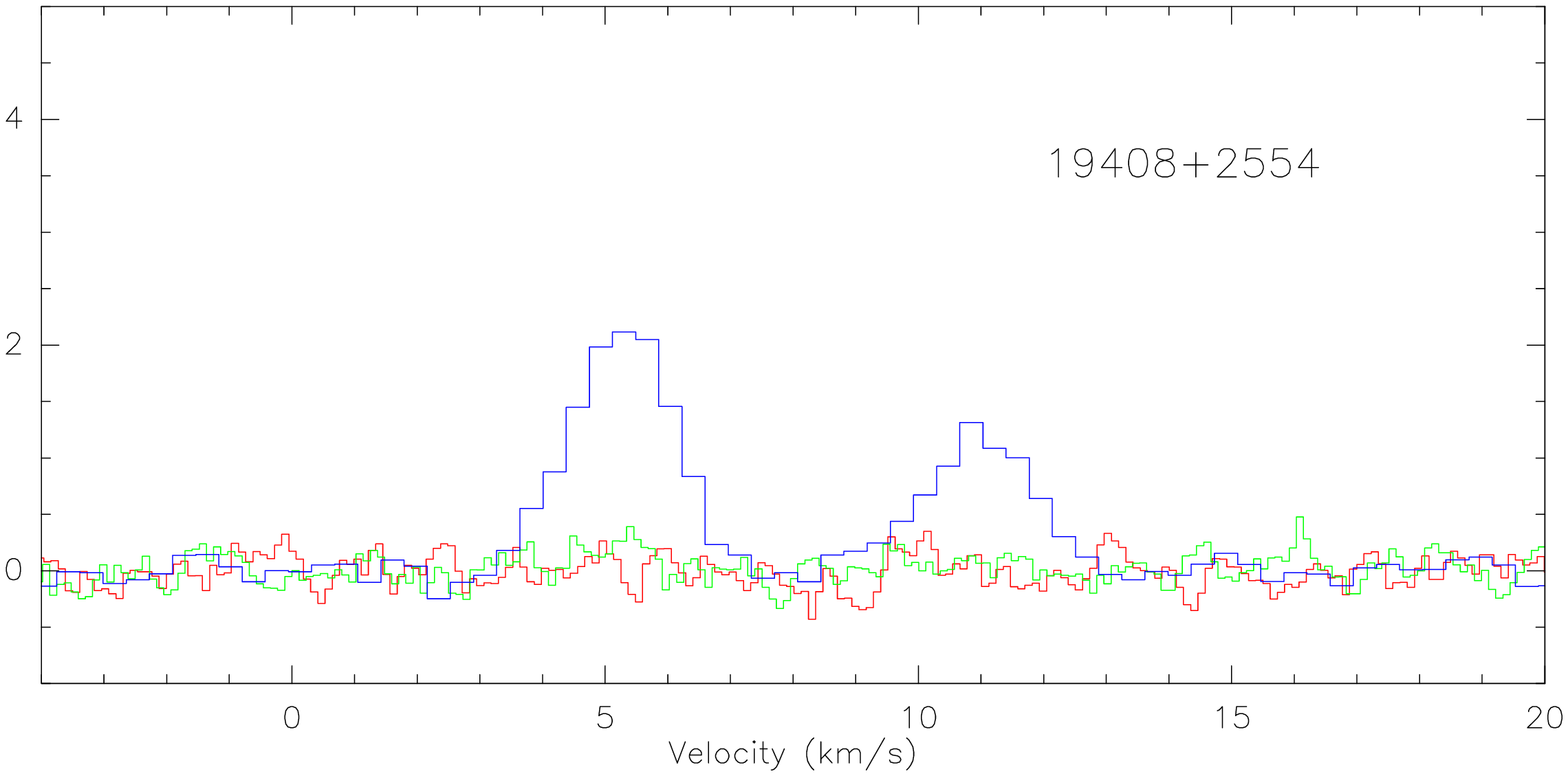}
\includegraphics[height=70mm,  angle=-90, clip, viewport=150 10 500 750]{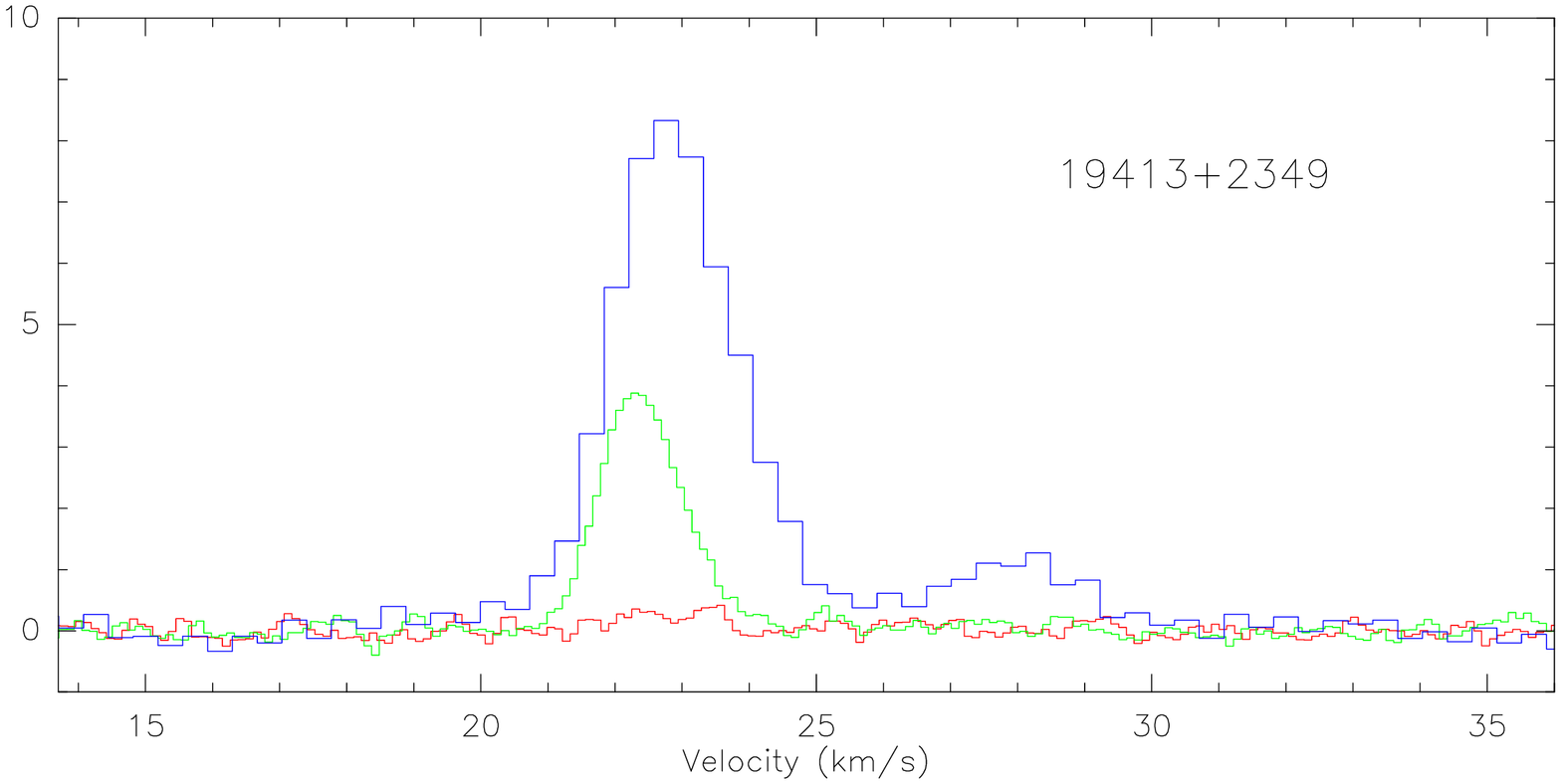}

\begin{minipage}[]{60mm}
  \caption{The sources of type 3
  }\end{minipage}
   \label{Fig8}
   \end{figure}

\addtocounter{figure}{-1}
\begin{figure}
\centering
\includegraphics[height=70mm,  angle=-90, clip, viewport=150 10 500 750]{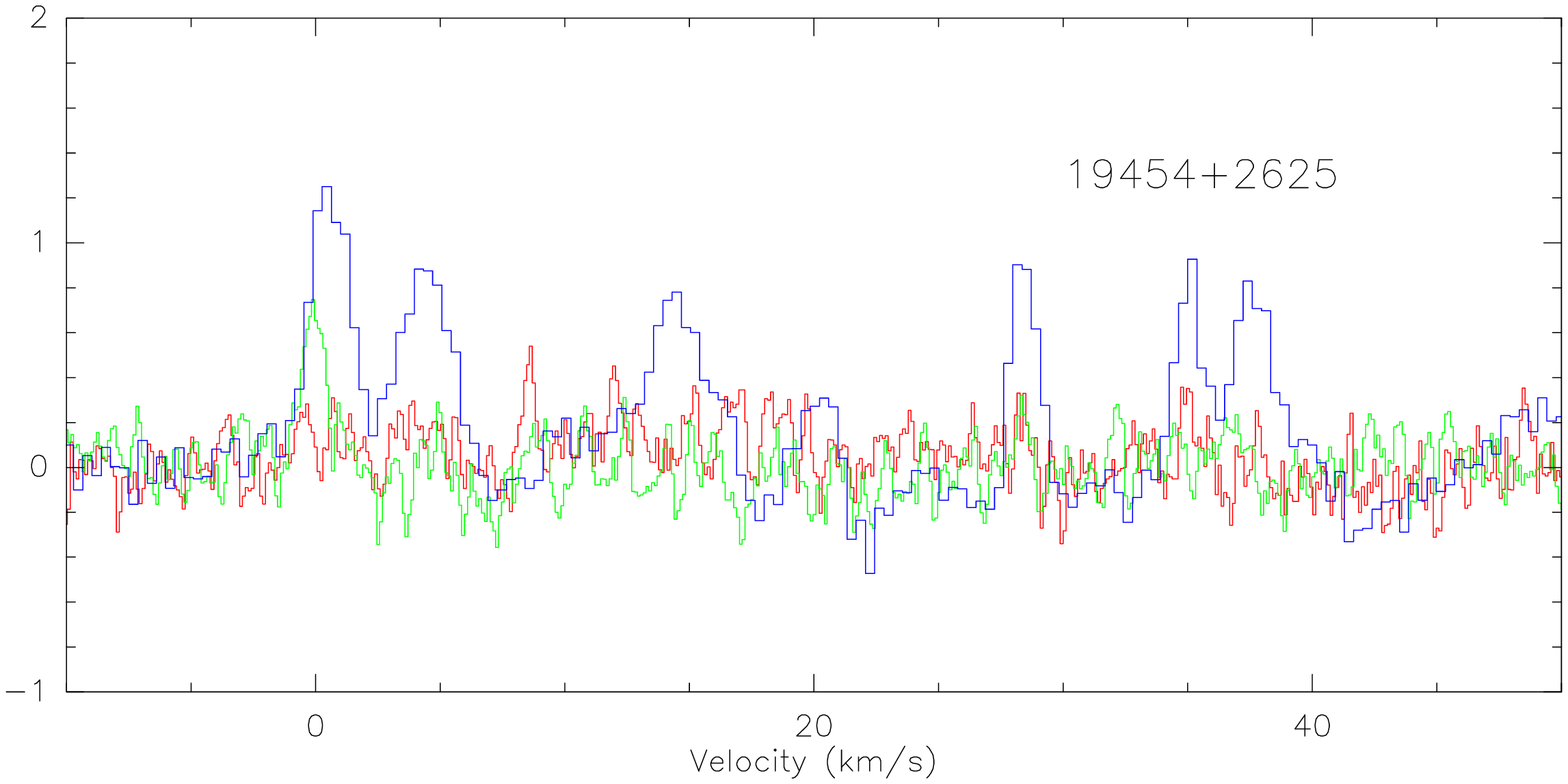}
\includegraphics[height=70mm,  angle=-90, clip, viewport=150 10 500 750]{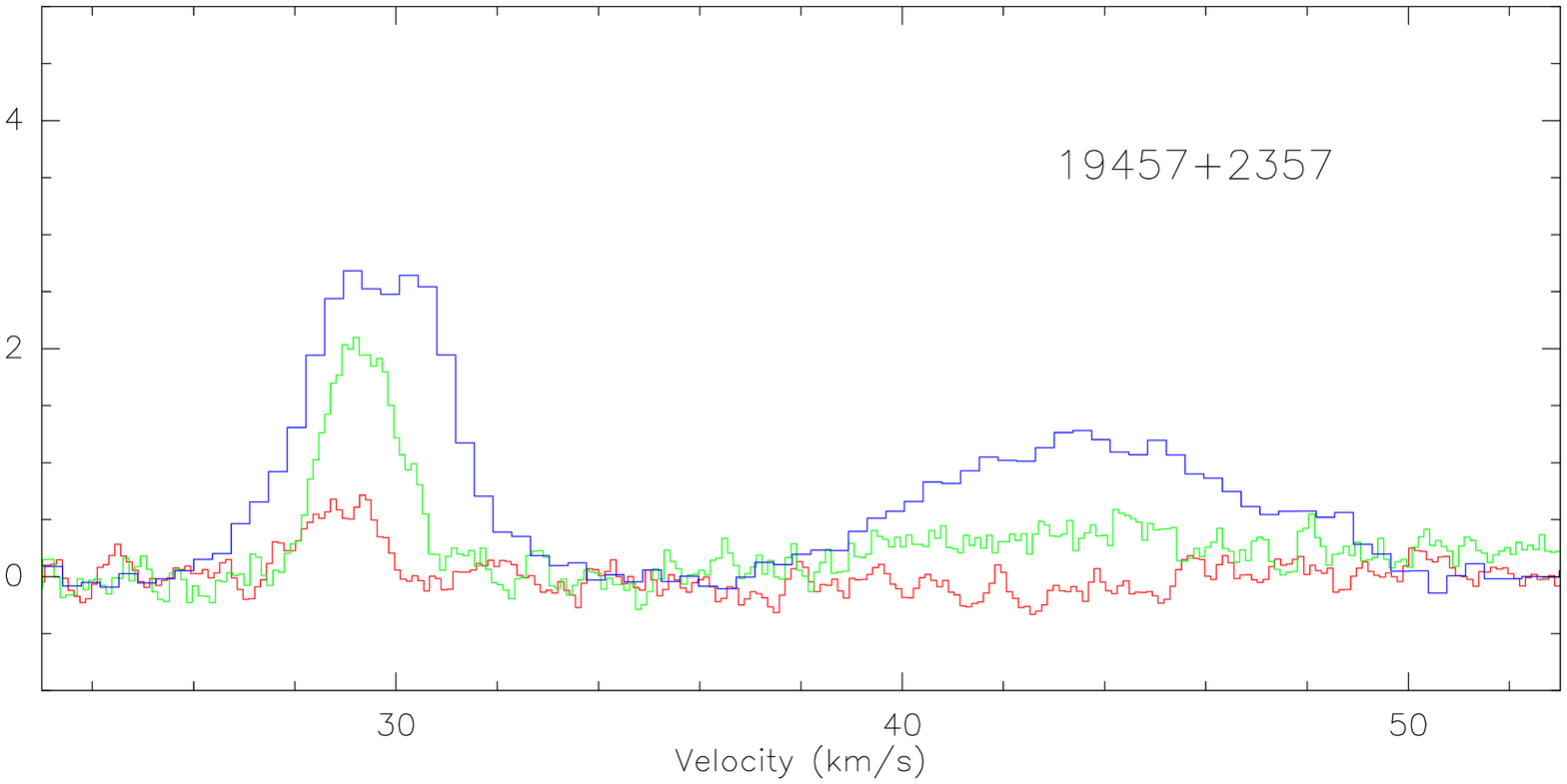}
\includegraphics[height=70mm,  angle=-90, clip, viewport=150 10 500 750]{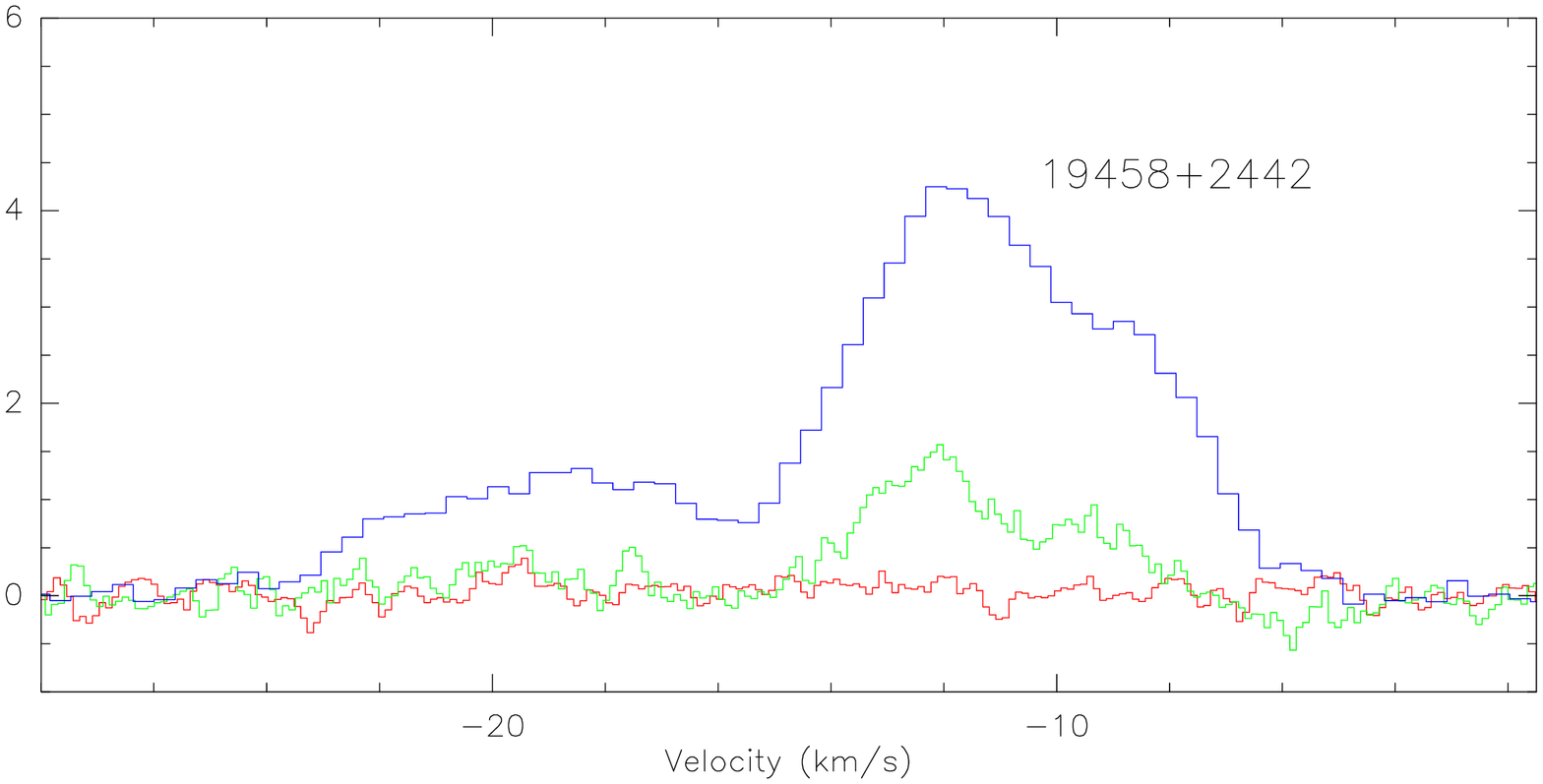}
\includegraphics[height=70mm,  angle=-90, clip, viewport=150 10 500 750]{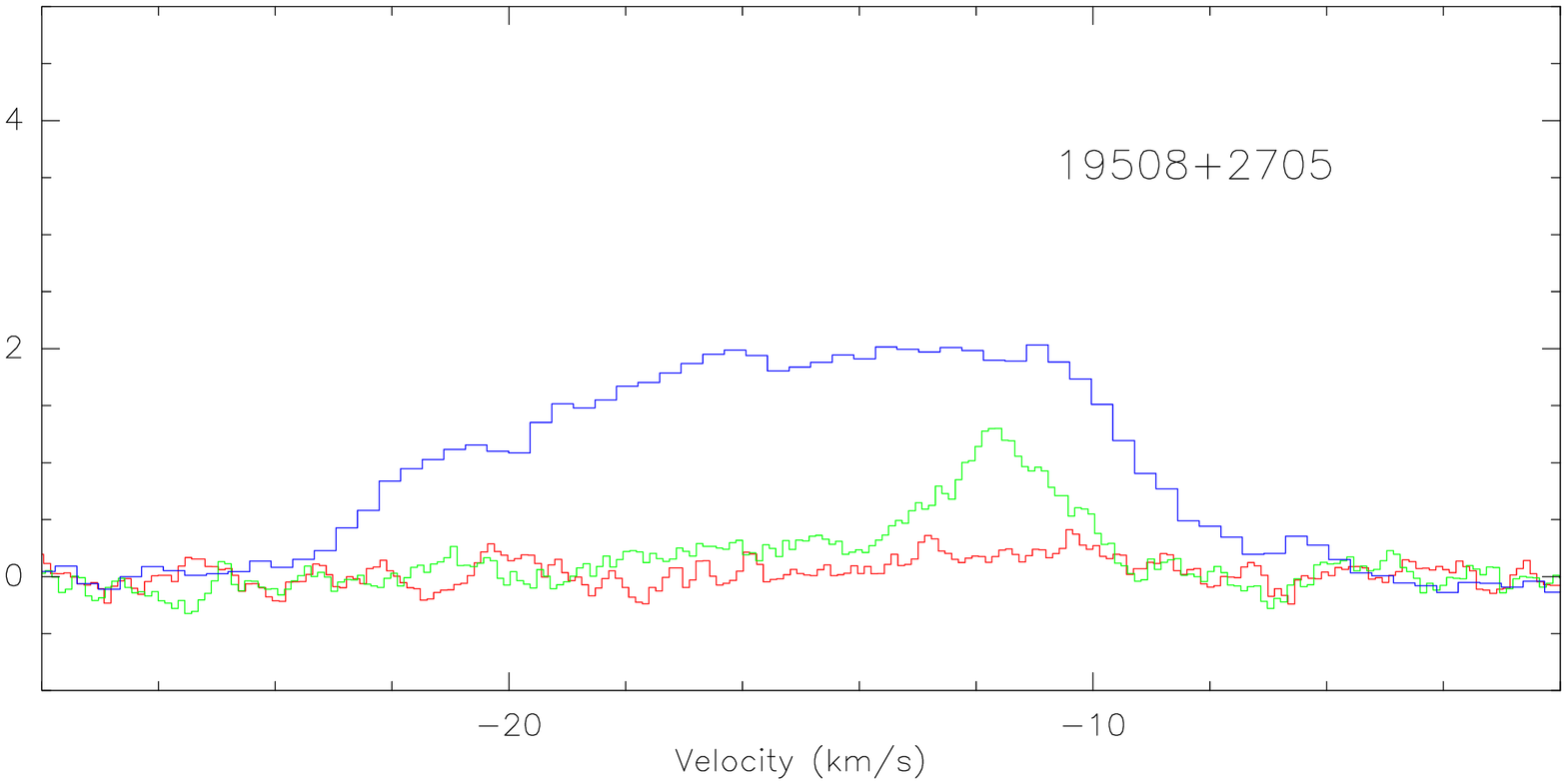}
\includegraphics[height=70mm,  angle=-90, clip, viewport=150 10 500 750]{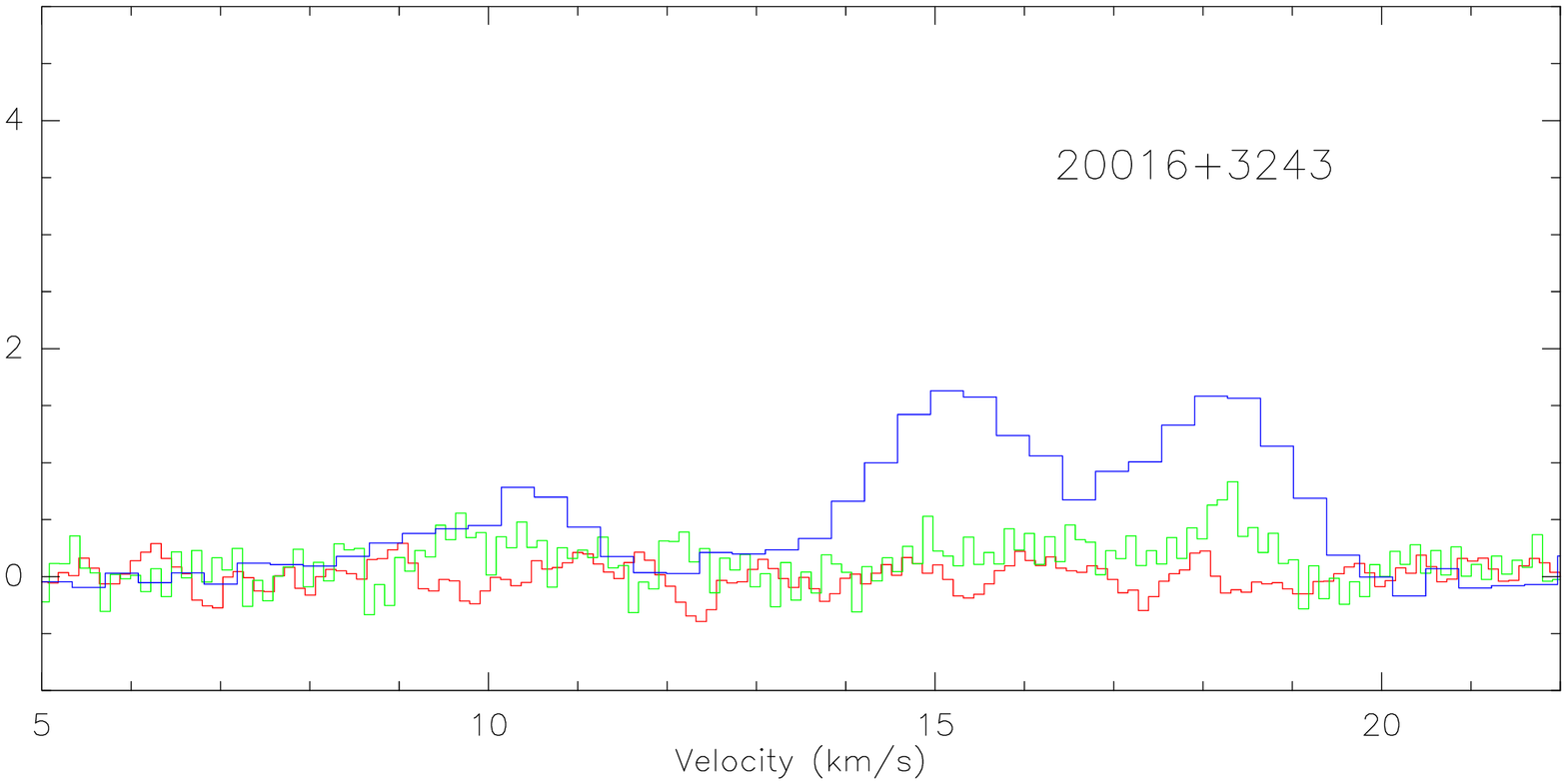}
\includegraphics[height=70mm,  angle=-90, clip, viewport=150 10 500 750]{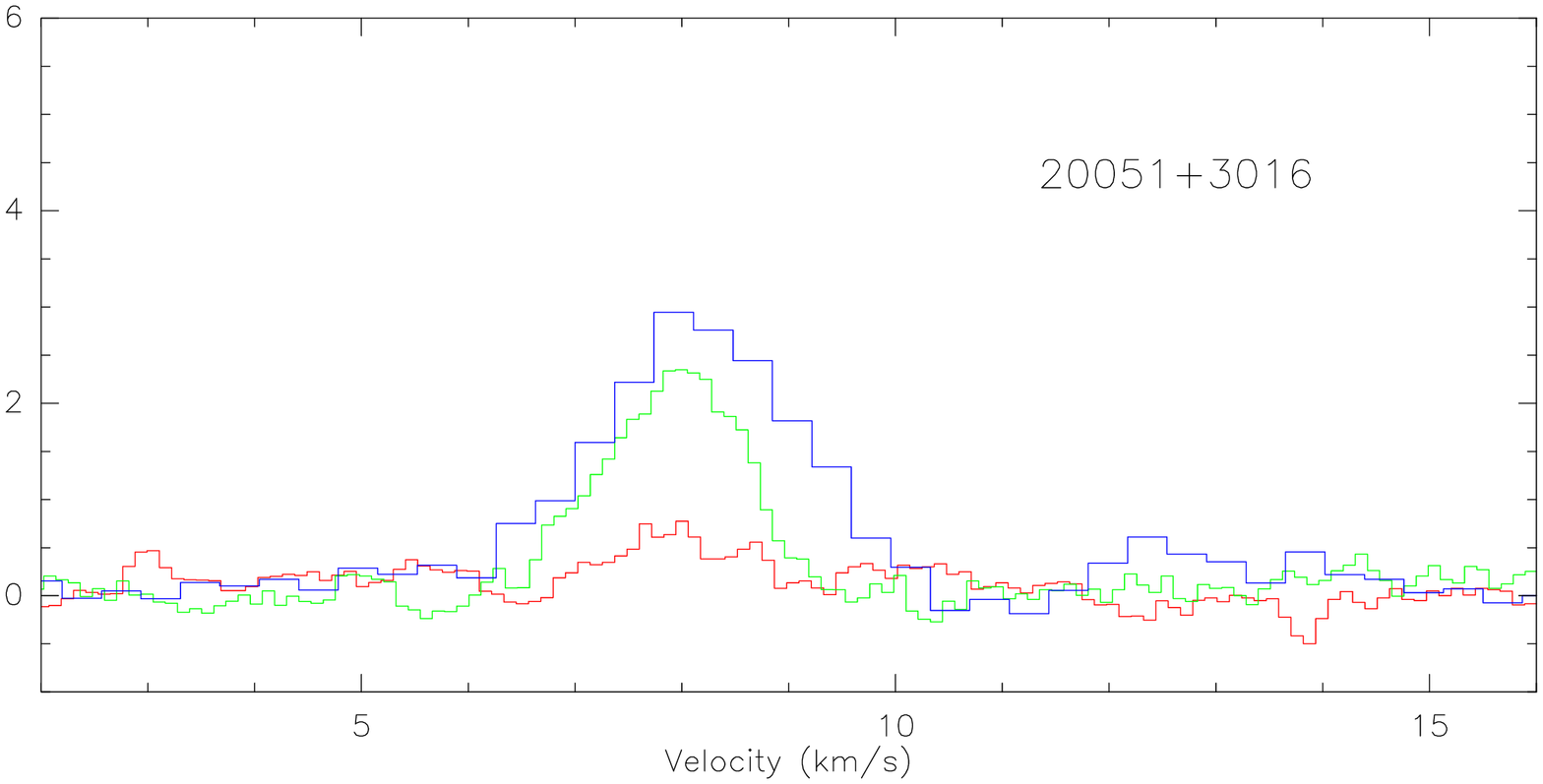}
\includegraphics[height=70mm,  angle=-90, clip, viewport=150 10 500 750]{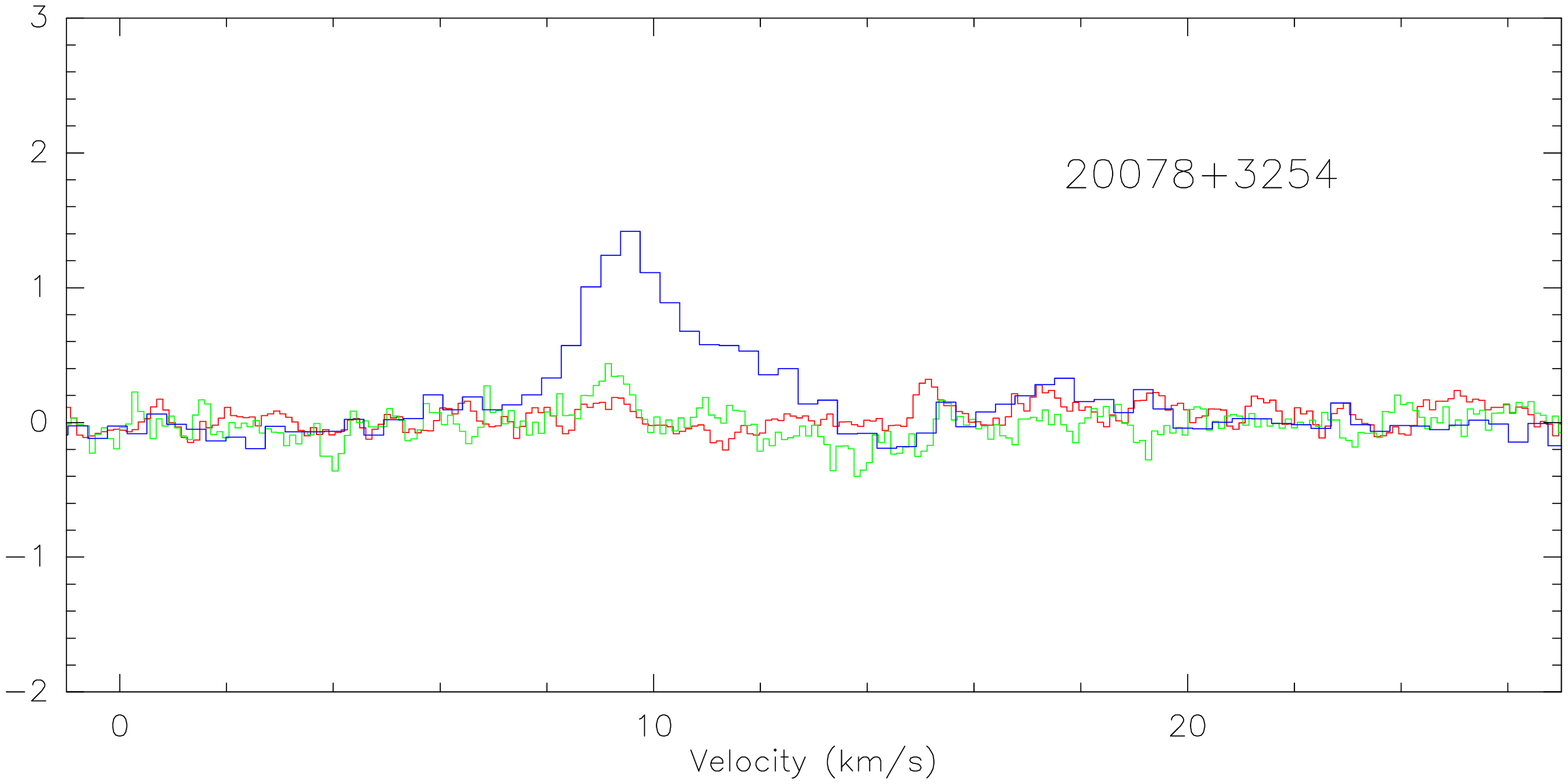}
\includegraphics[height=70mm,  angle=-90, clip, viewport=150 10 500 750]{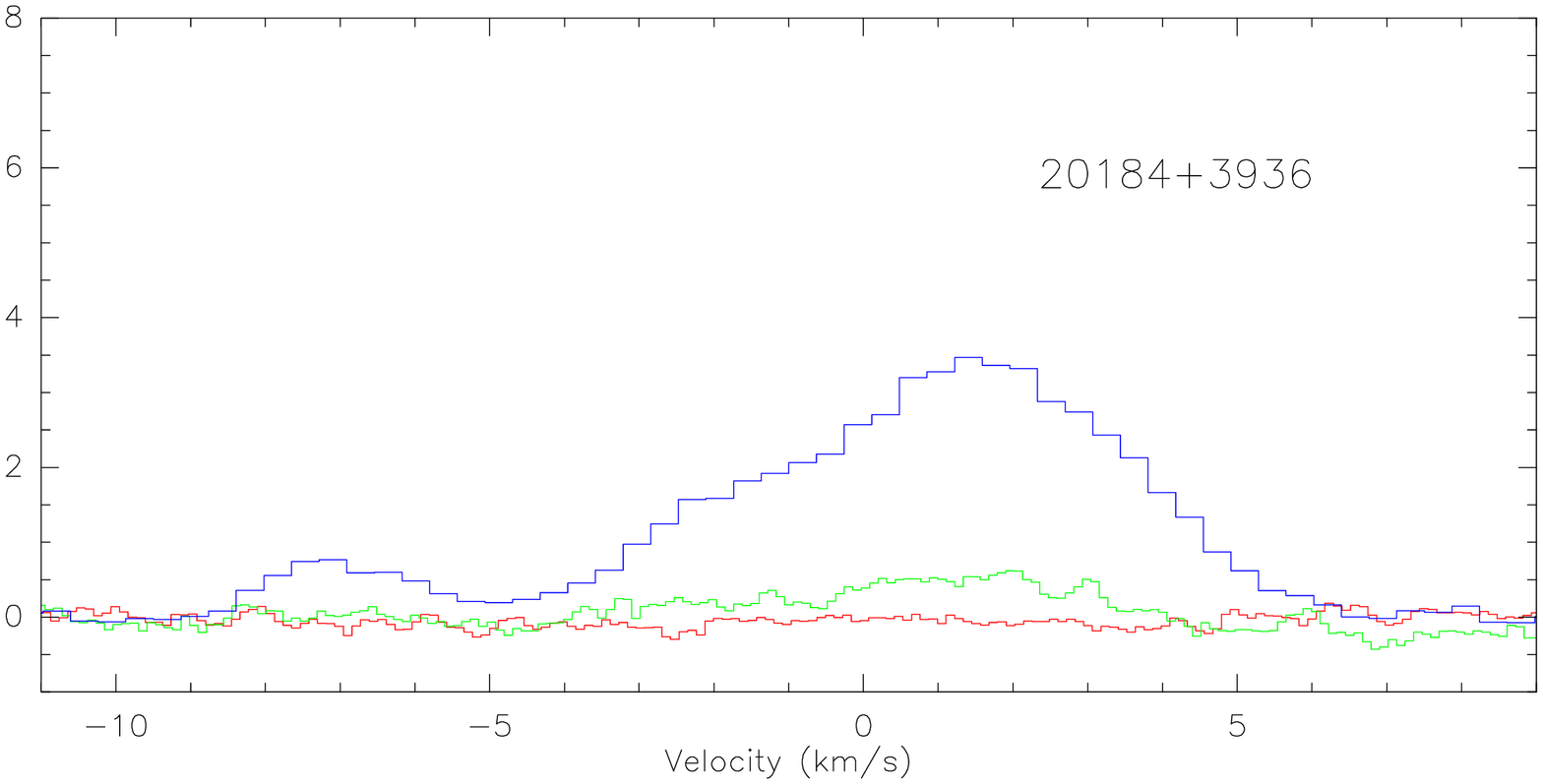}
\includegraphics[height=70mm,  angle=-90, clip, viewport=150 10 500 750]{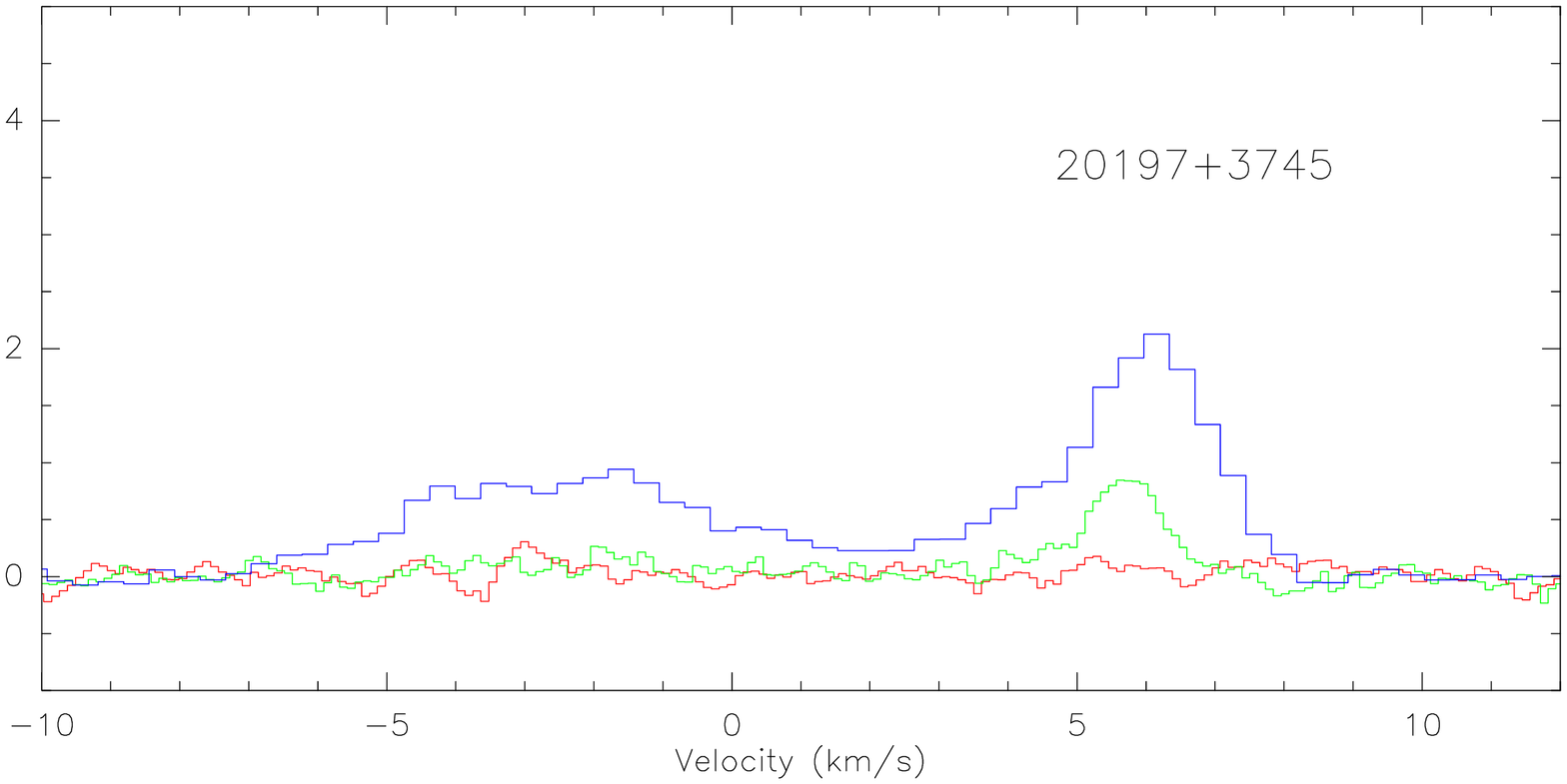}
\includegraphics[height=70mm,  angle=-90, clip, viewport=150 10 500 750]{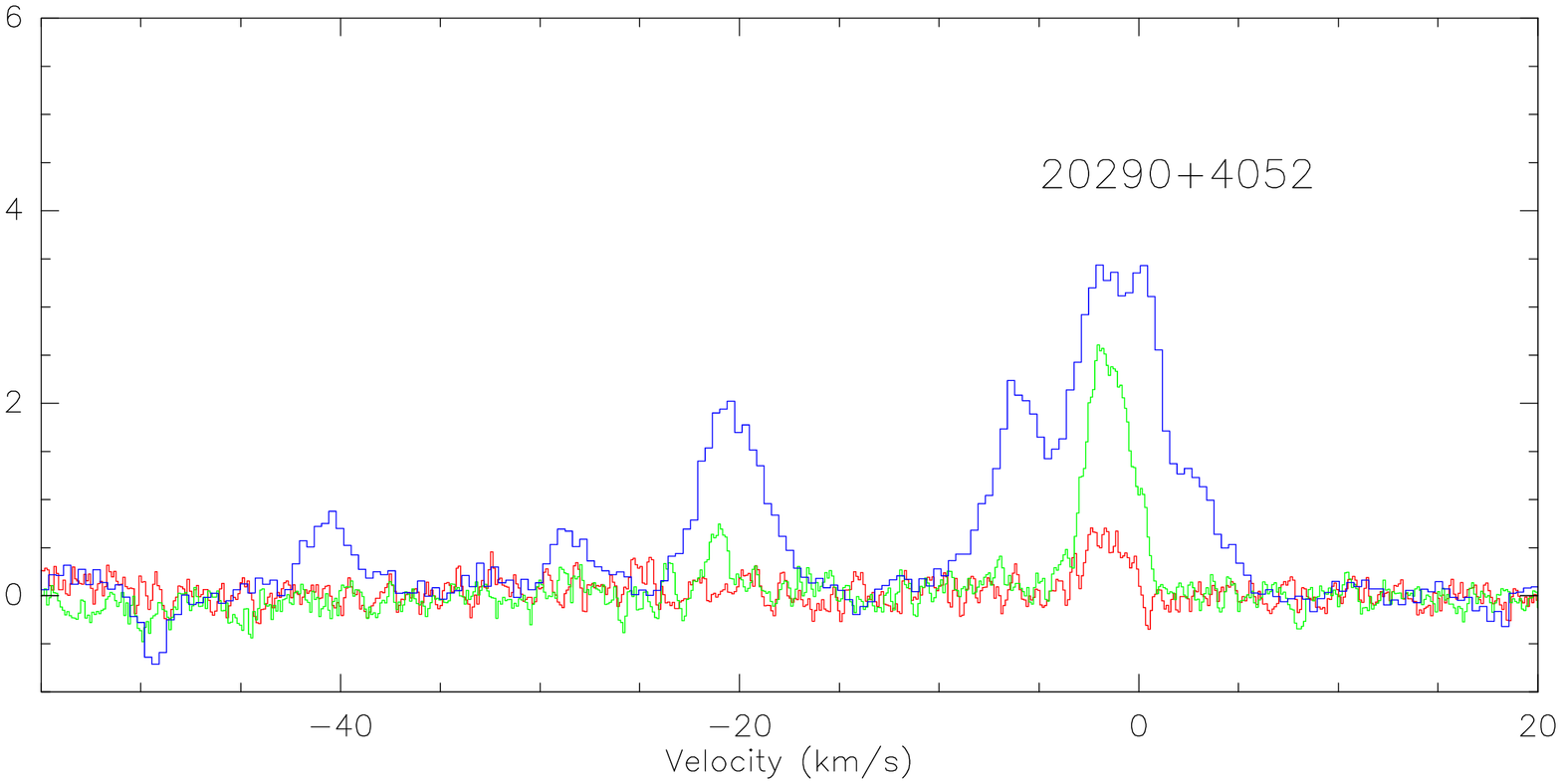}
\includegraphics[height=70mm,  angle=-90, clip, viewport=150 10 500 750]{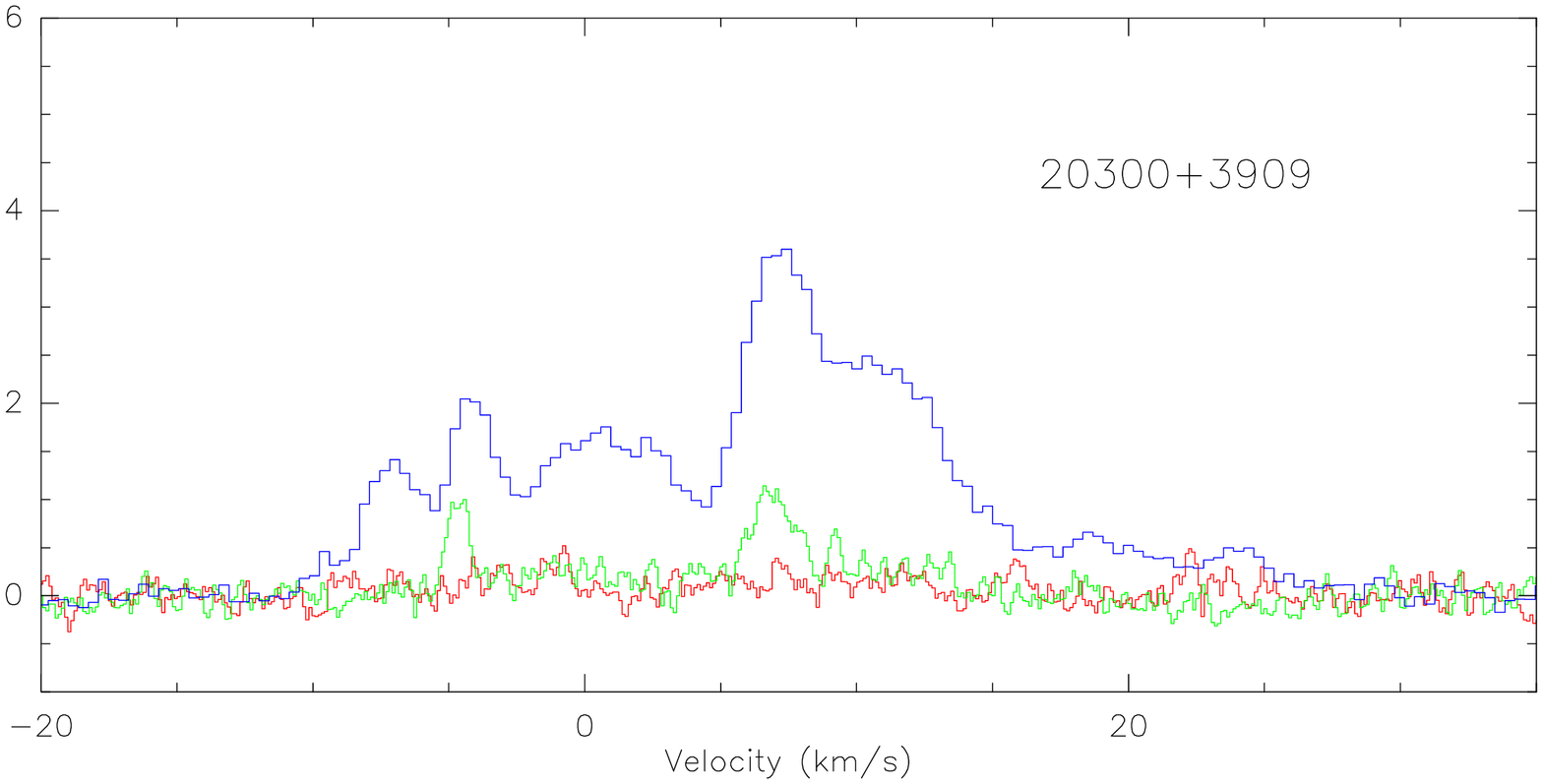}
\includegraphics[height=70mm,  angle=-90, clip, viewport=150 10 500 750]{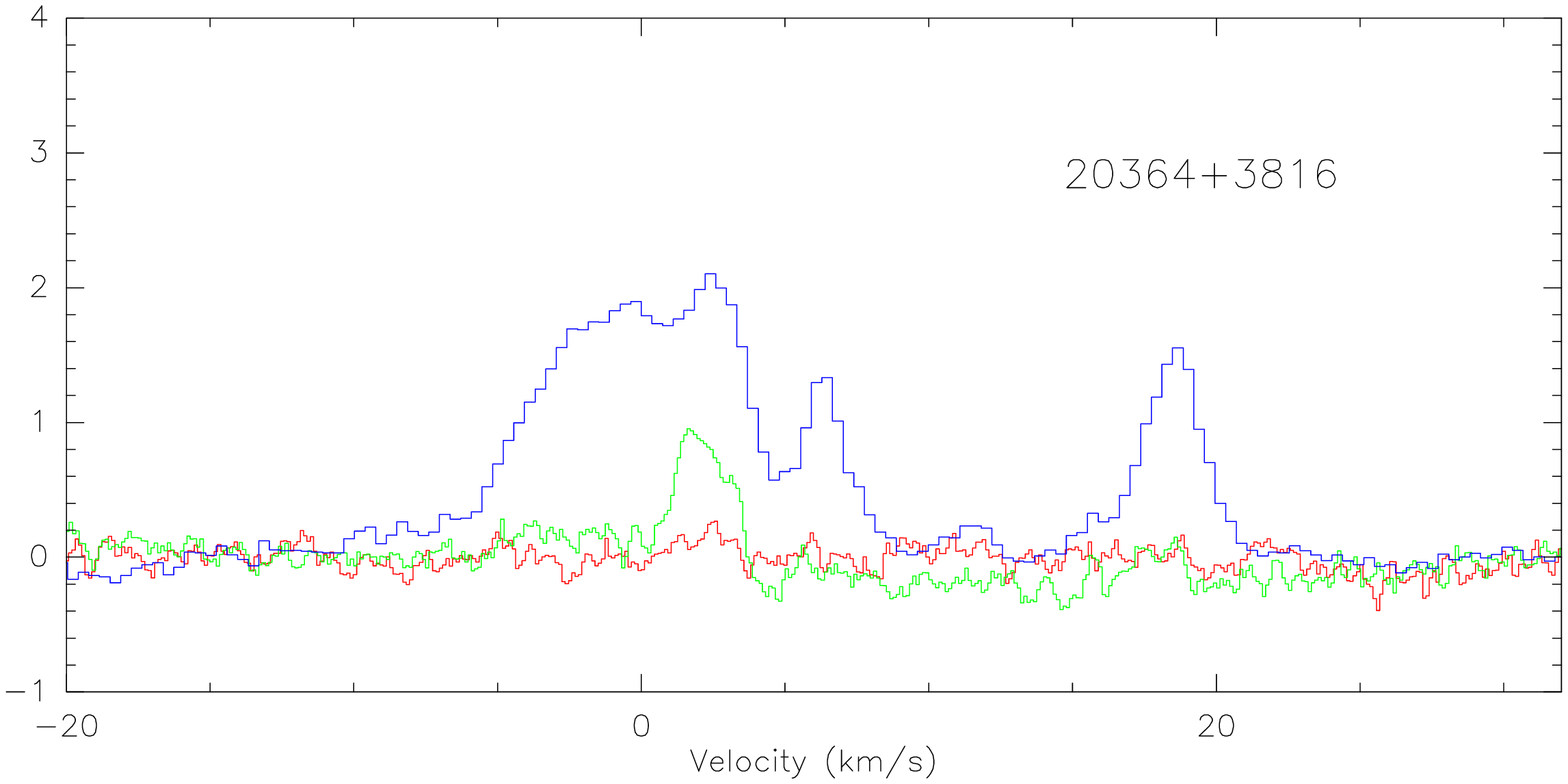}
\includegraphics[height=70mm,  angle=-90, clip, viewport=150 10 500 750]{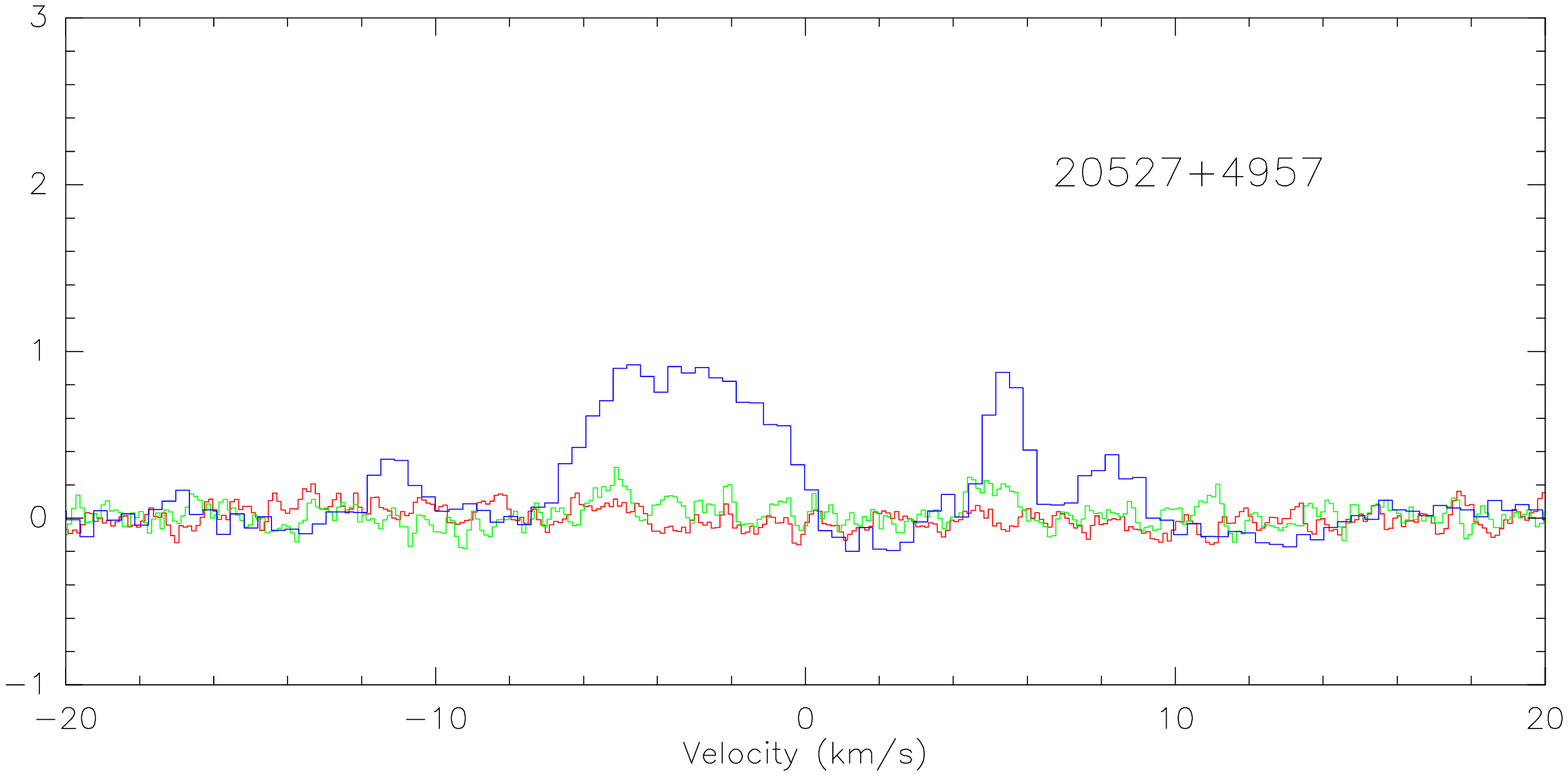}
\includegraphics[height=70mm,  angle=-90, clip, viewport=150 10 500 750]{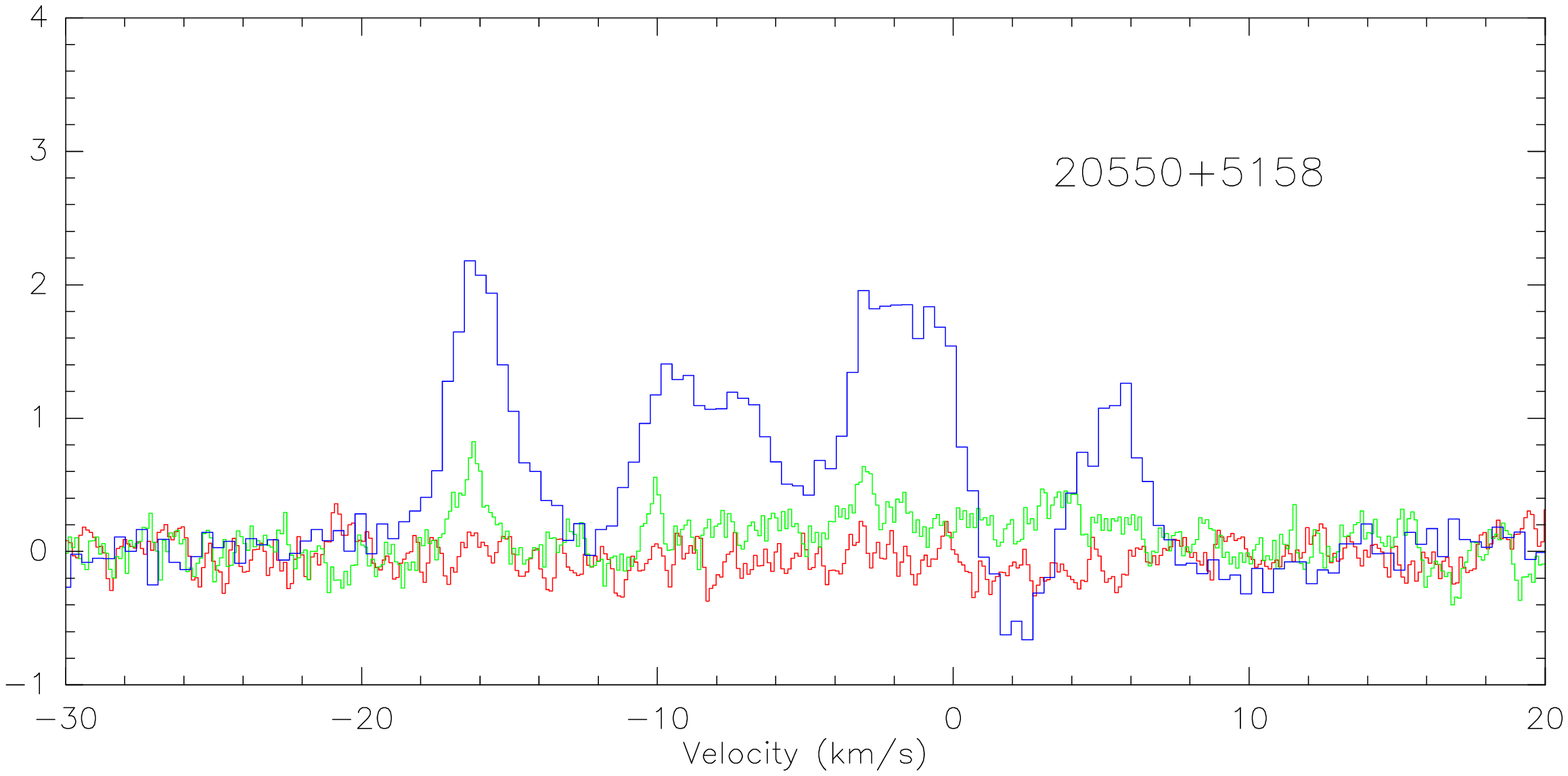}

\begin{minipage}[]{60mm}
  \caption{The sources of type 3
  }\end{minipage}
   \label{Fig8}
   \end{figure}

\addtocounter{figure}{-1}
\begin{figure}
\centering
\includegraphics[height=70mm,  angle=-90, clip, viewport=150 10 500 750]{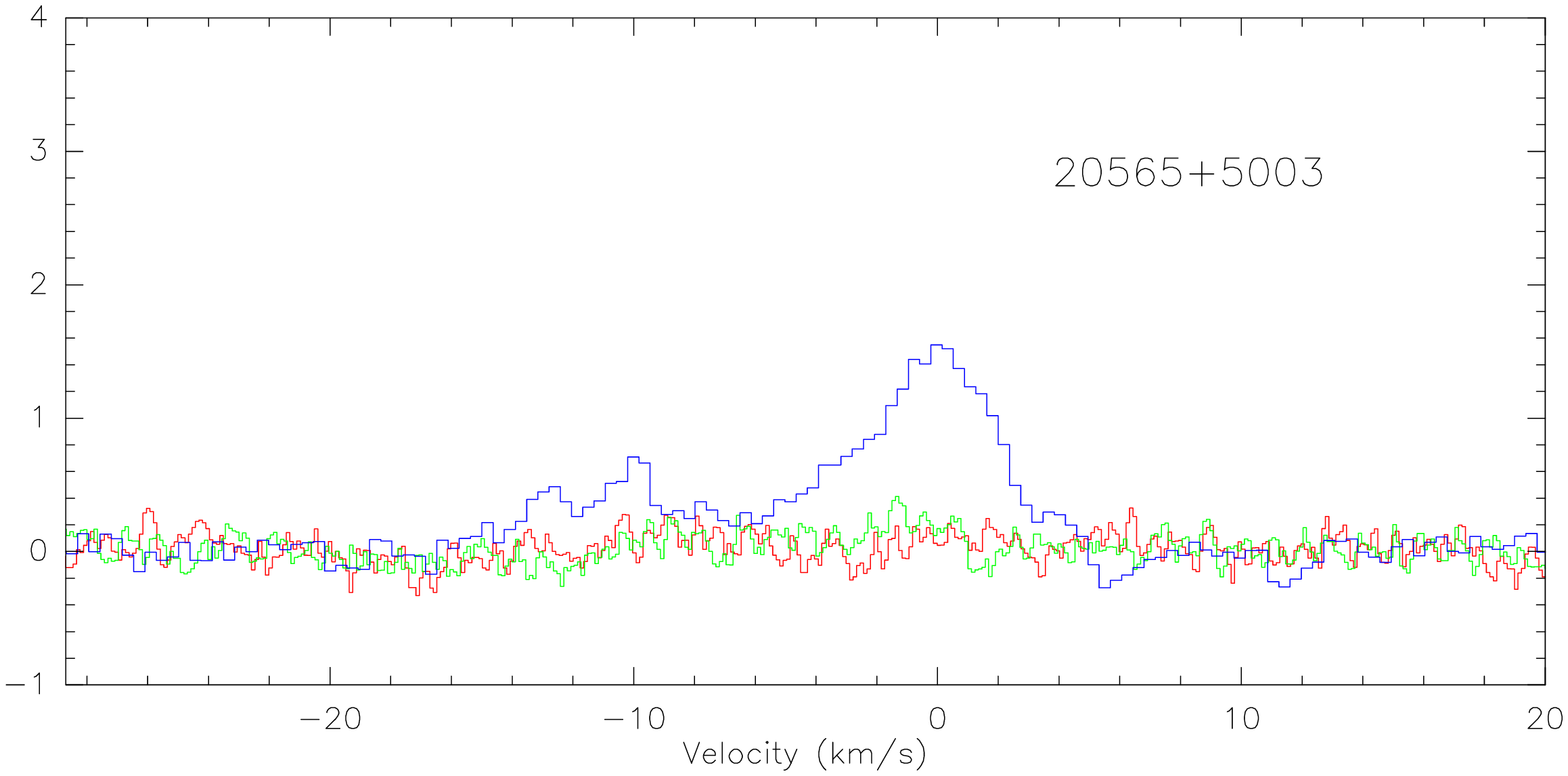}
\includegraphics[height=70mm,  angle=-90, clip, viewport=150 10 500 750]{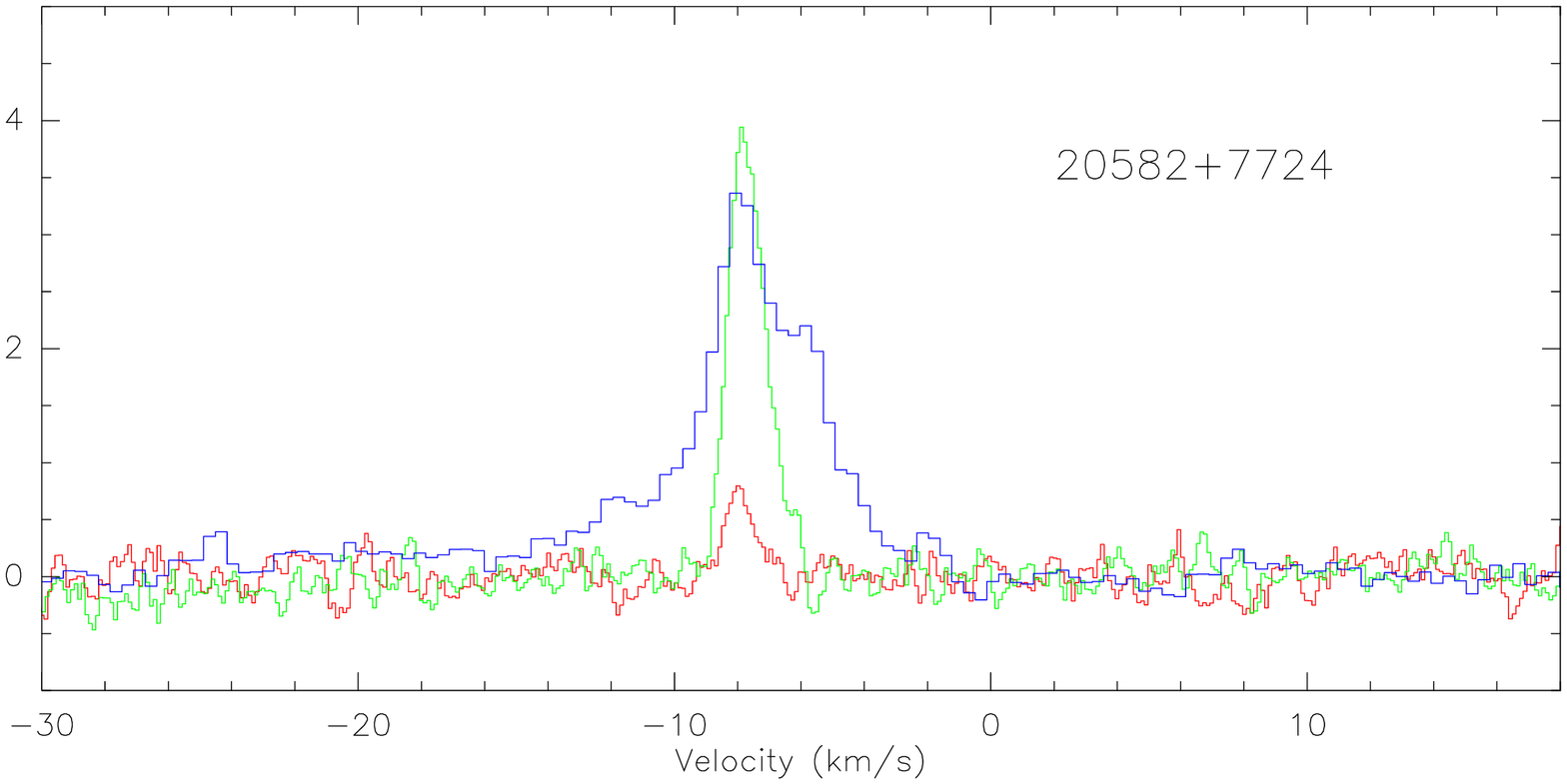}
\includegraphics[height=70mm,  angle=-90, clip, viewport=150 10 500 750]{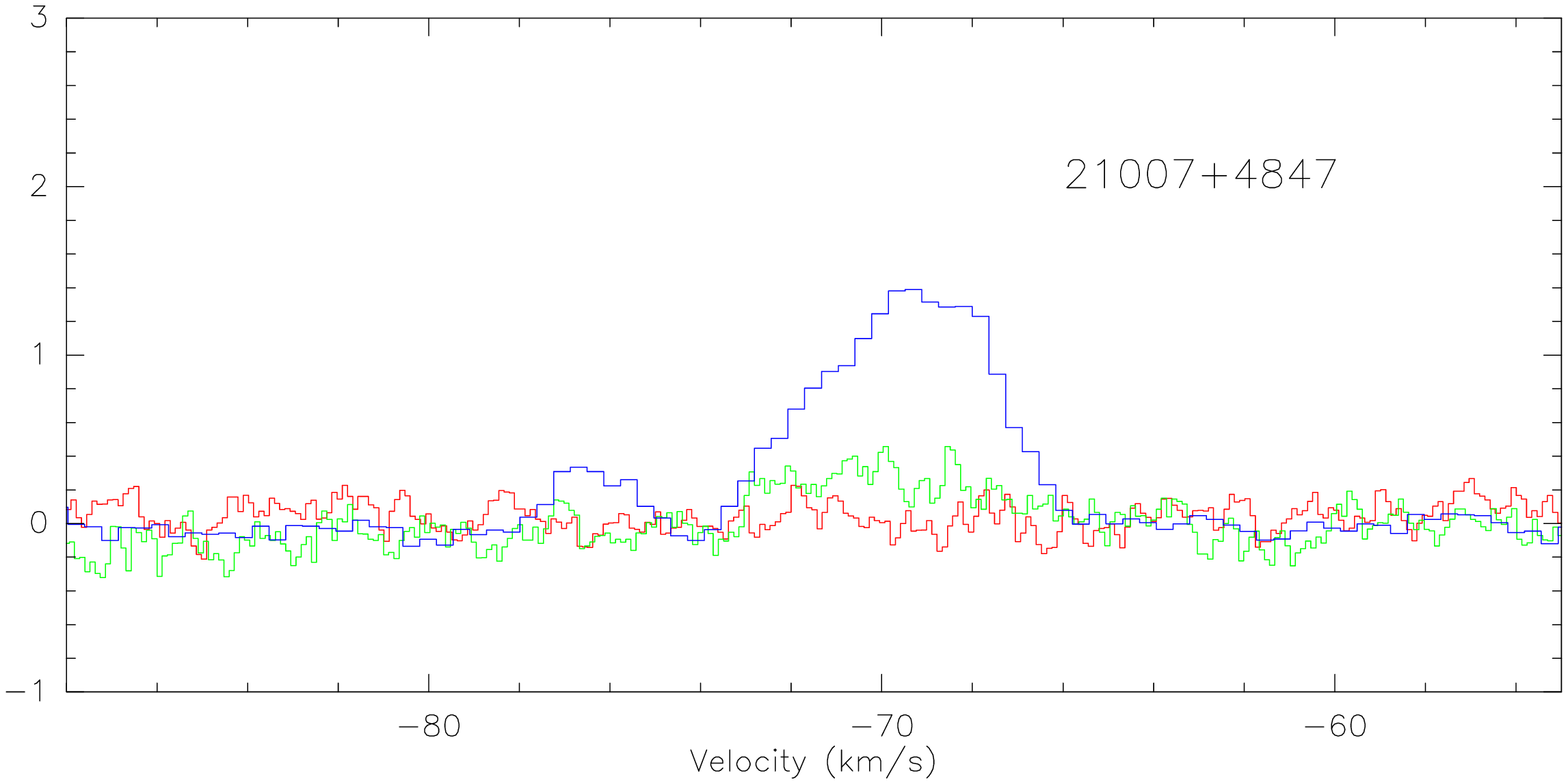}
\includegraphics[height=70mm,  angle=-90, clip, viewport=150 10 500 750]{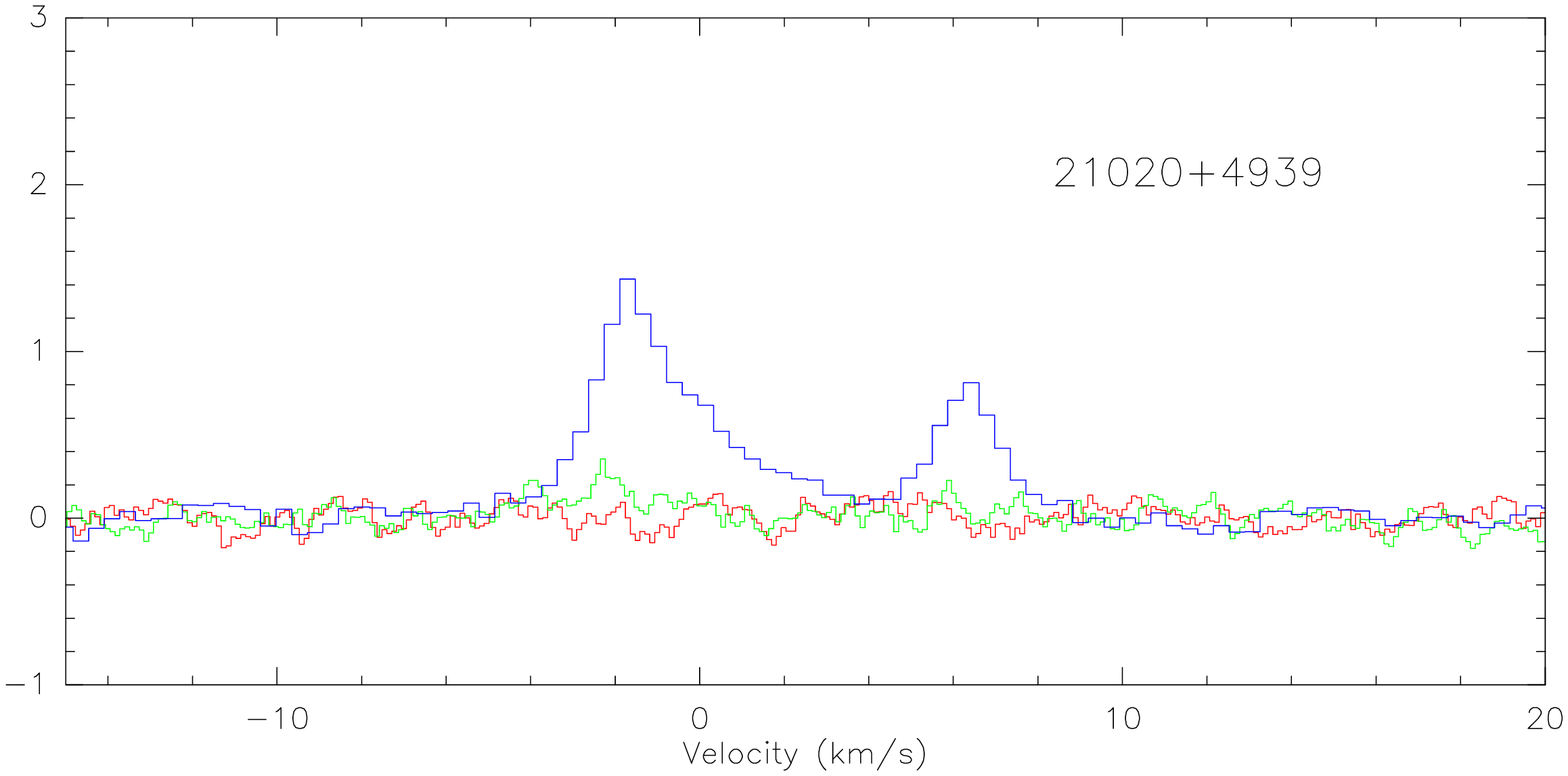}
\includegraphics[height=70mm,  angle=-90, clip, viewport=150 10 500 750]{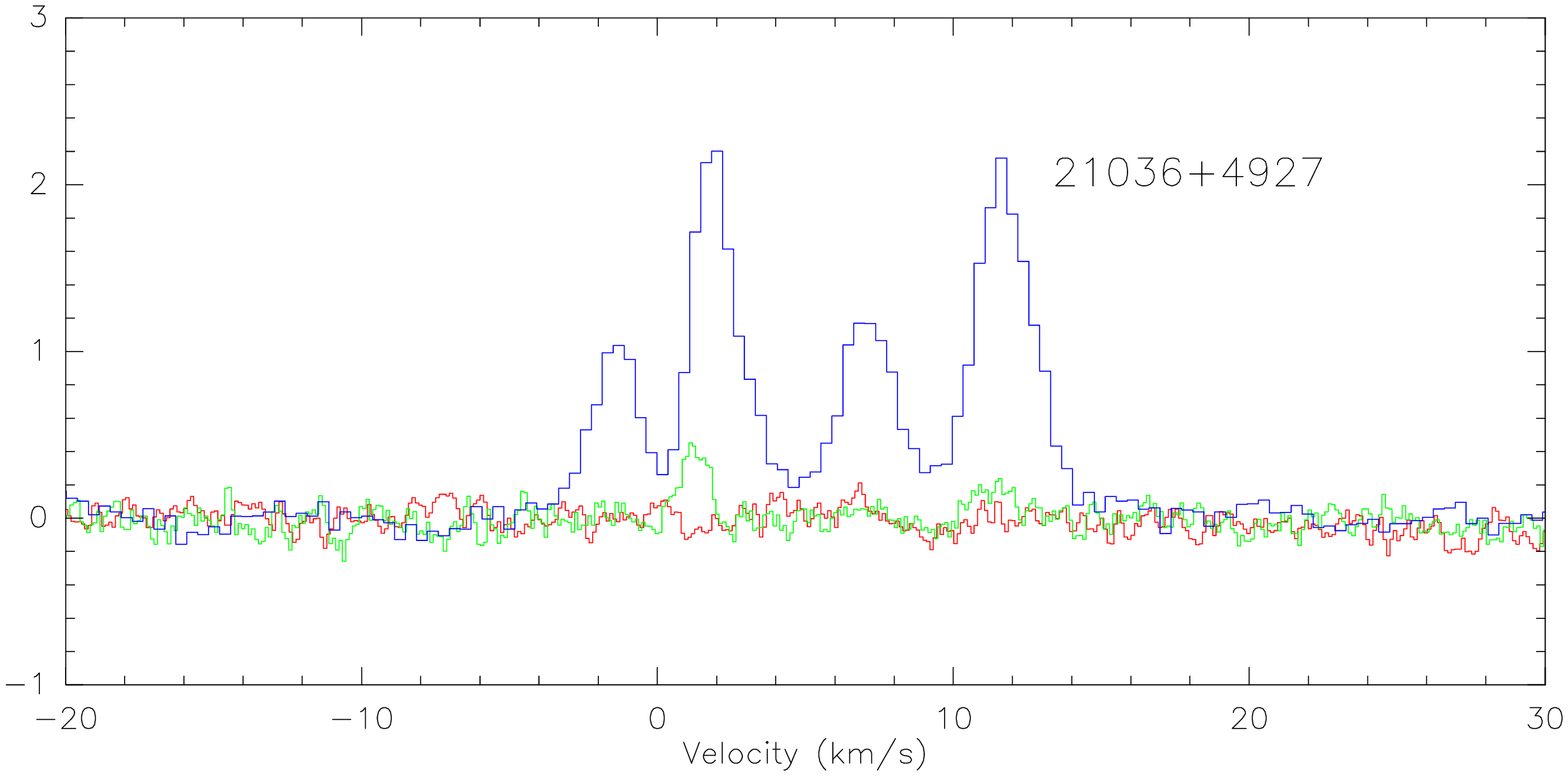}
\includegraphics[height=70mm,  angle=-90, clip, viewport=150 10 500 750]{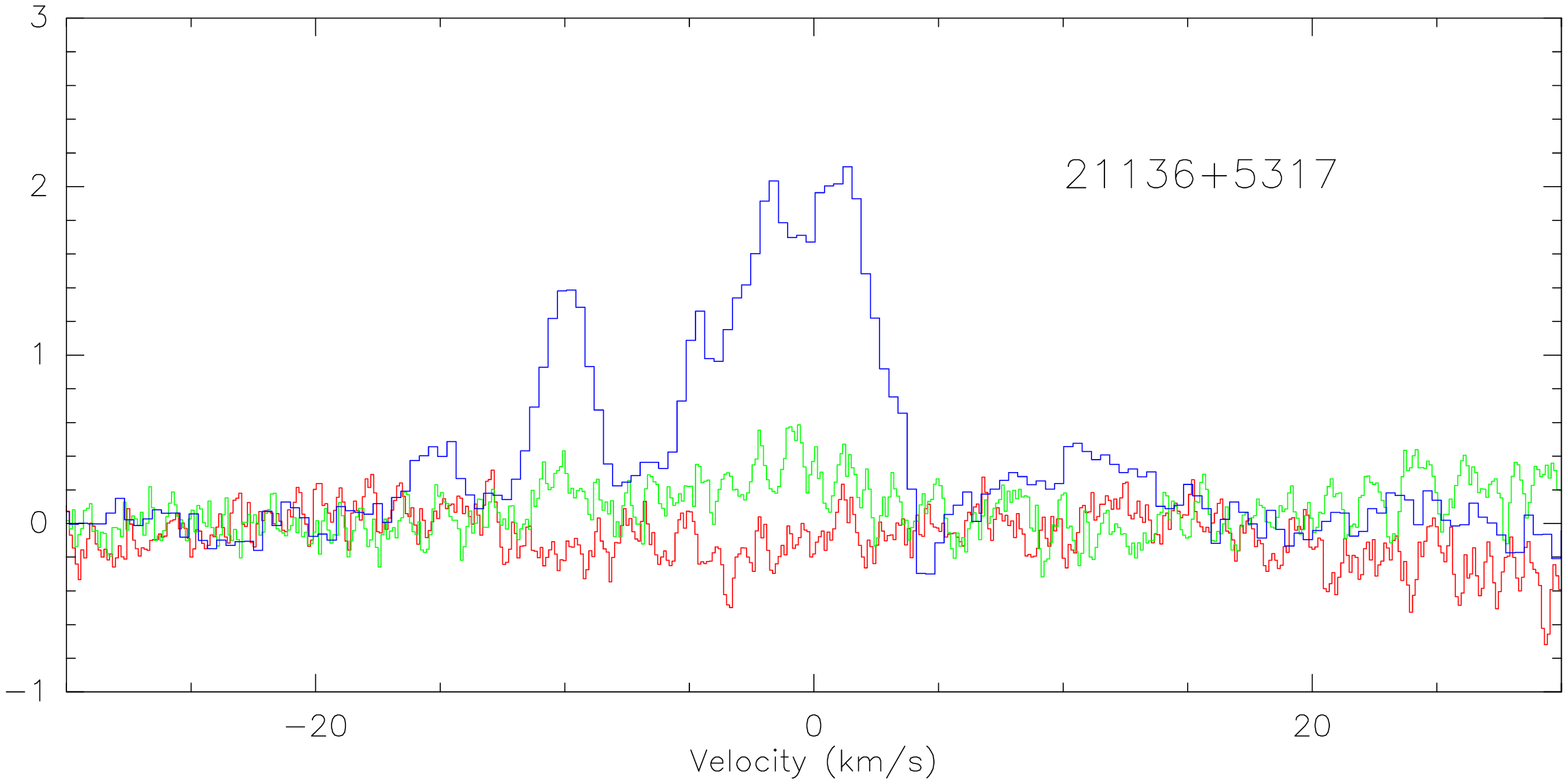}
\includegraphics[height=70mm,  angle=-90, clip, viewport=150 10 500 750]{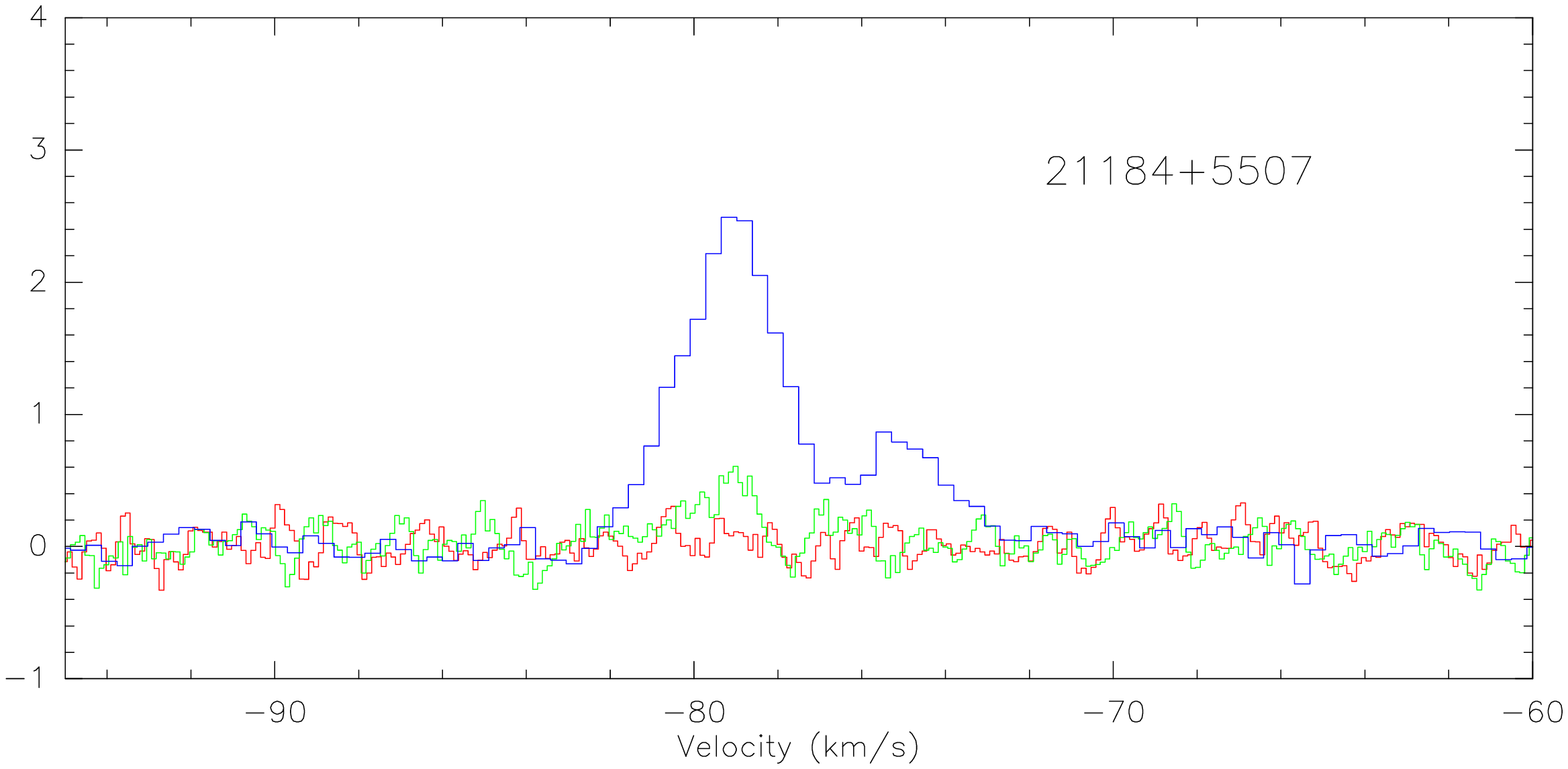}
\includegraphics[height=70mm,  angle=-90, clip, viewport=150 10 500 750]{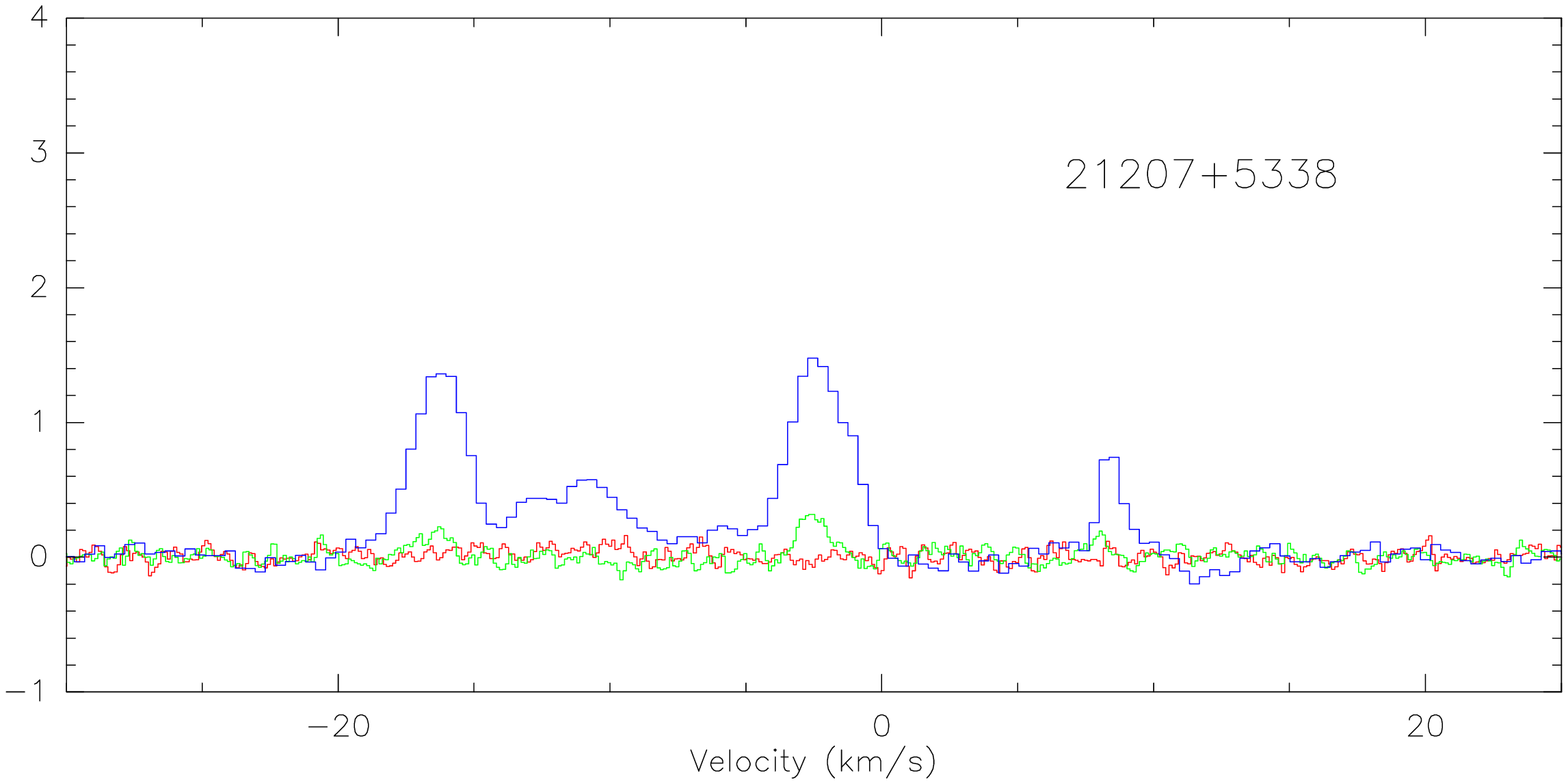}
\includegraphics[height=70mm,  angle=-90, clip, viewport=150 10 500 750]{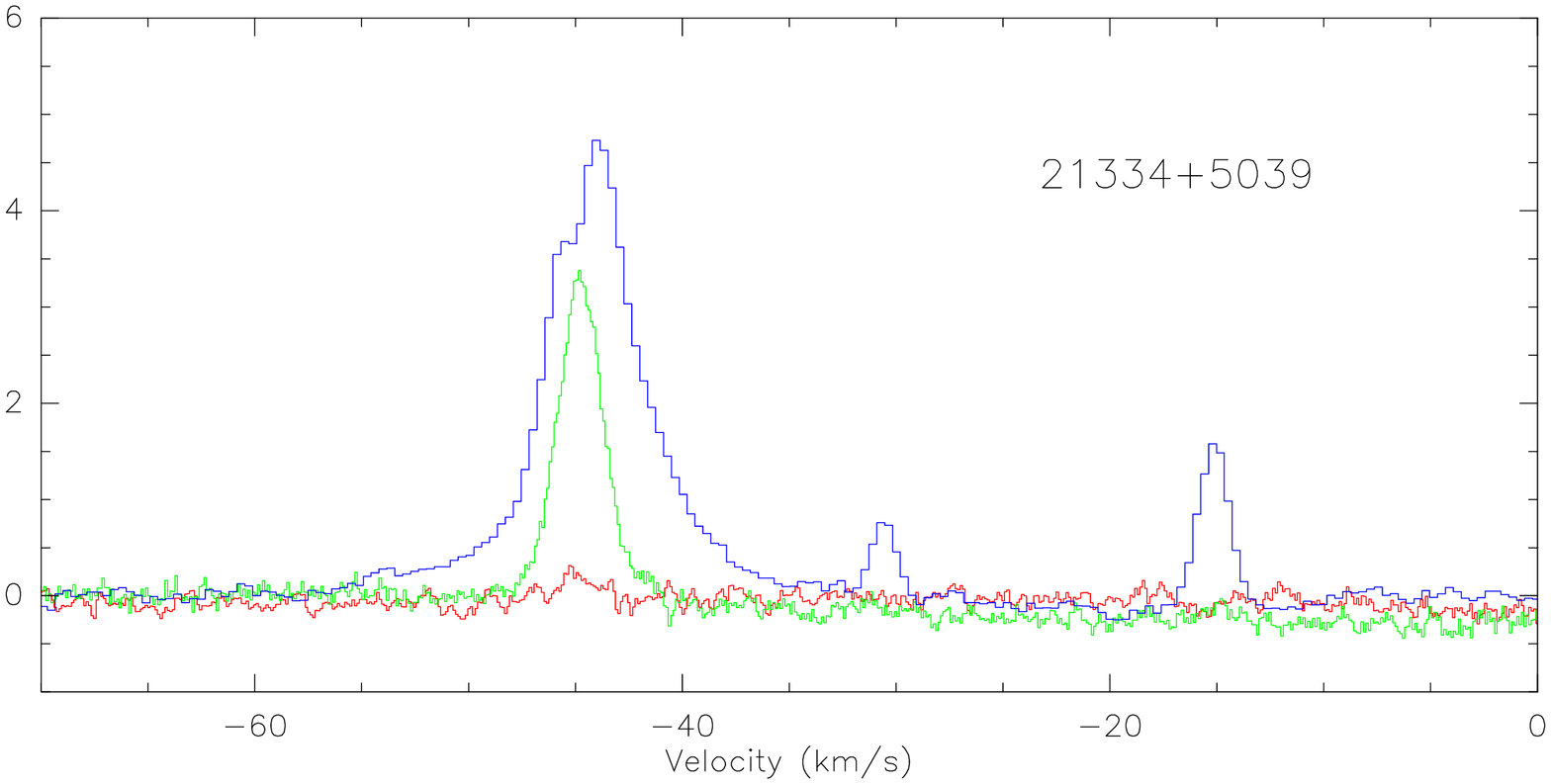}
\includegraphics[height=70mm,  angle=-90, clip, viewport=150 10 500 750]{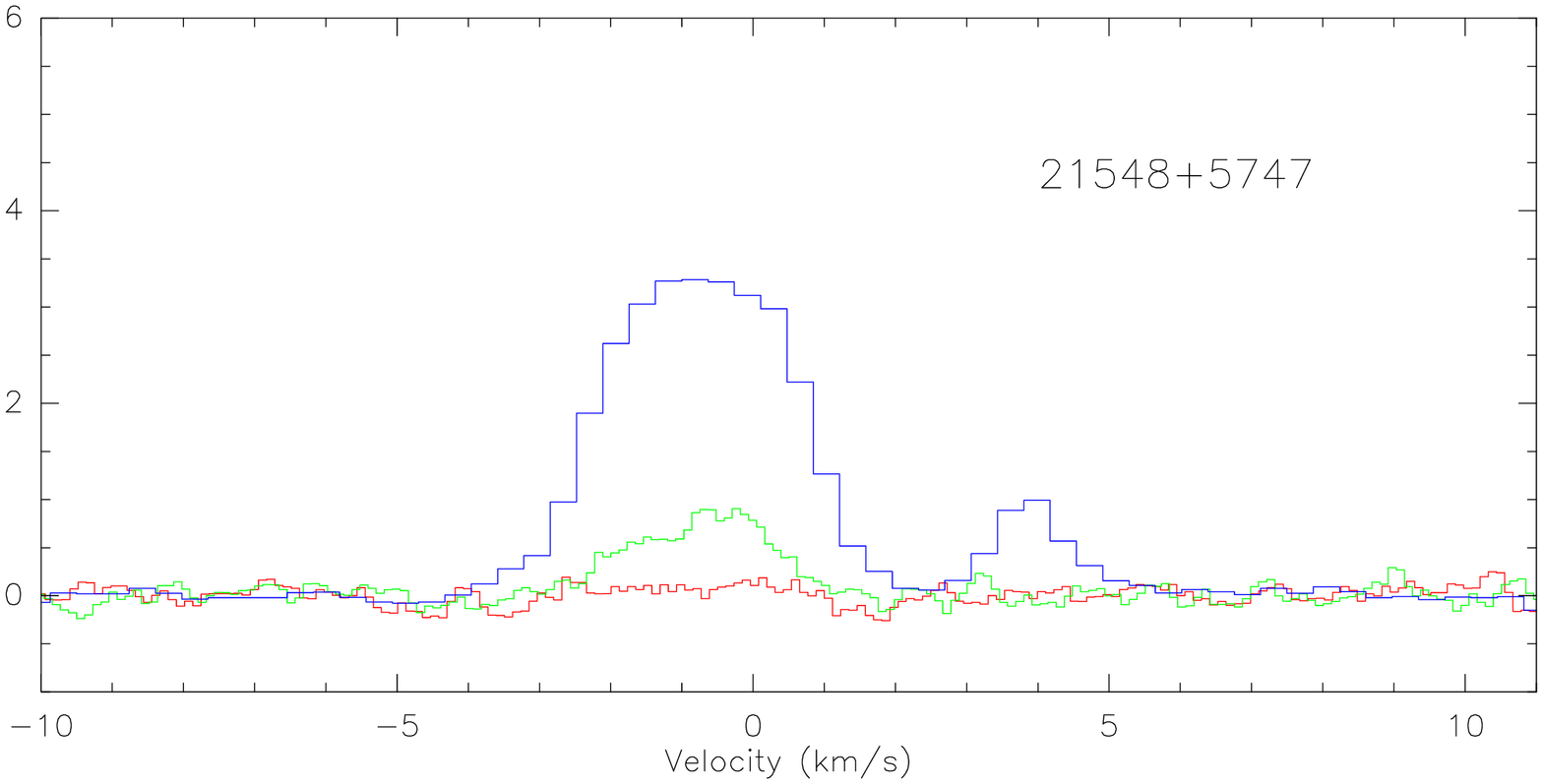}
\includegraphics[height=70mm,  angle=-90, clip, viewport=150 10 500 750]{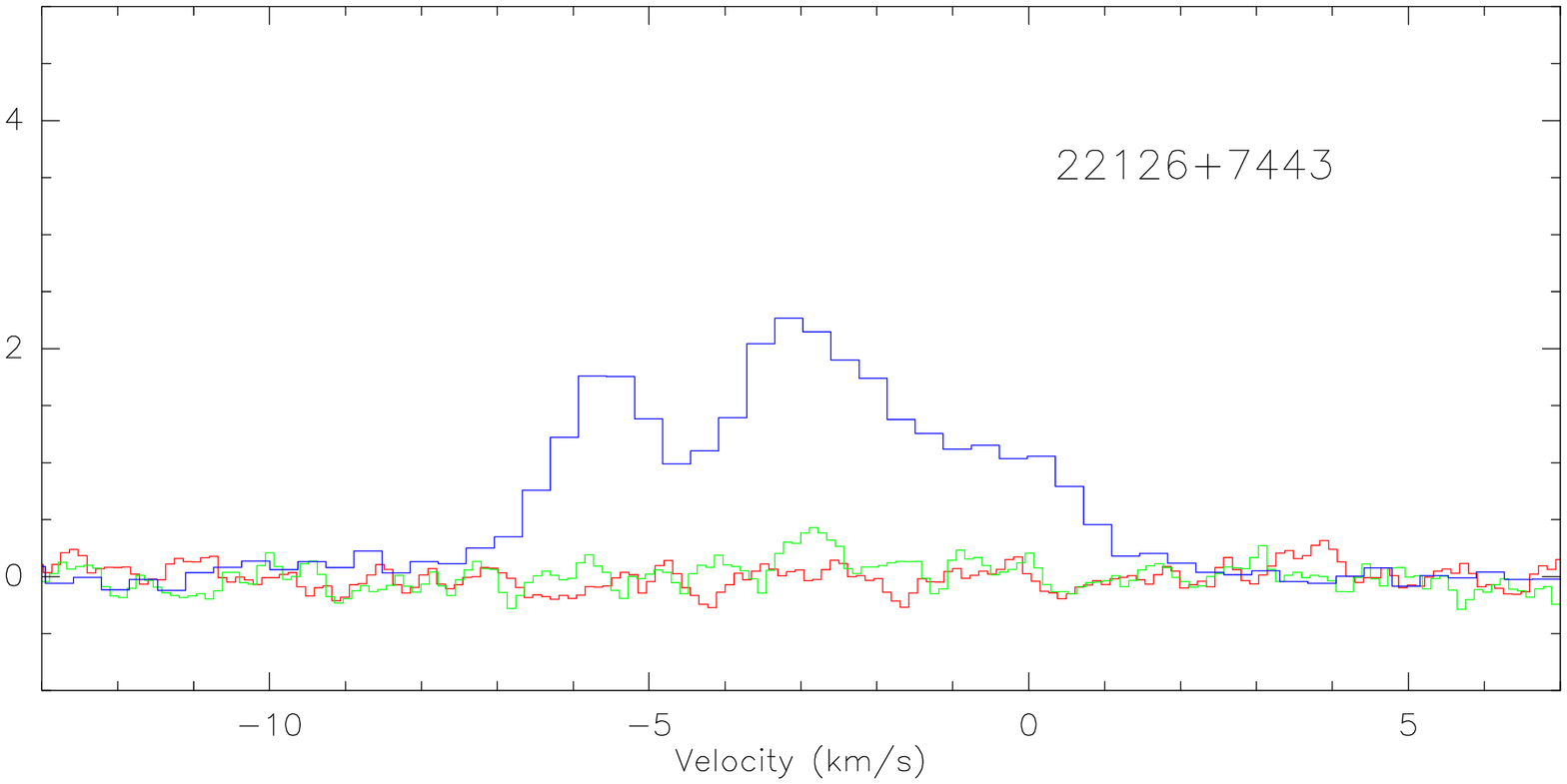}
\includegraphics[height=70mm,  angle=-90, clip, viewport=150 10 500 750]{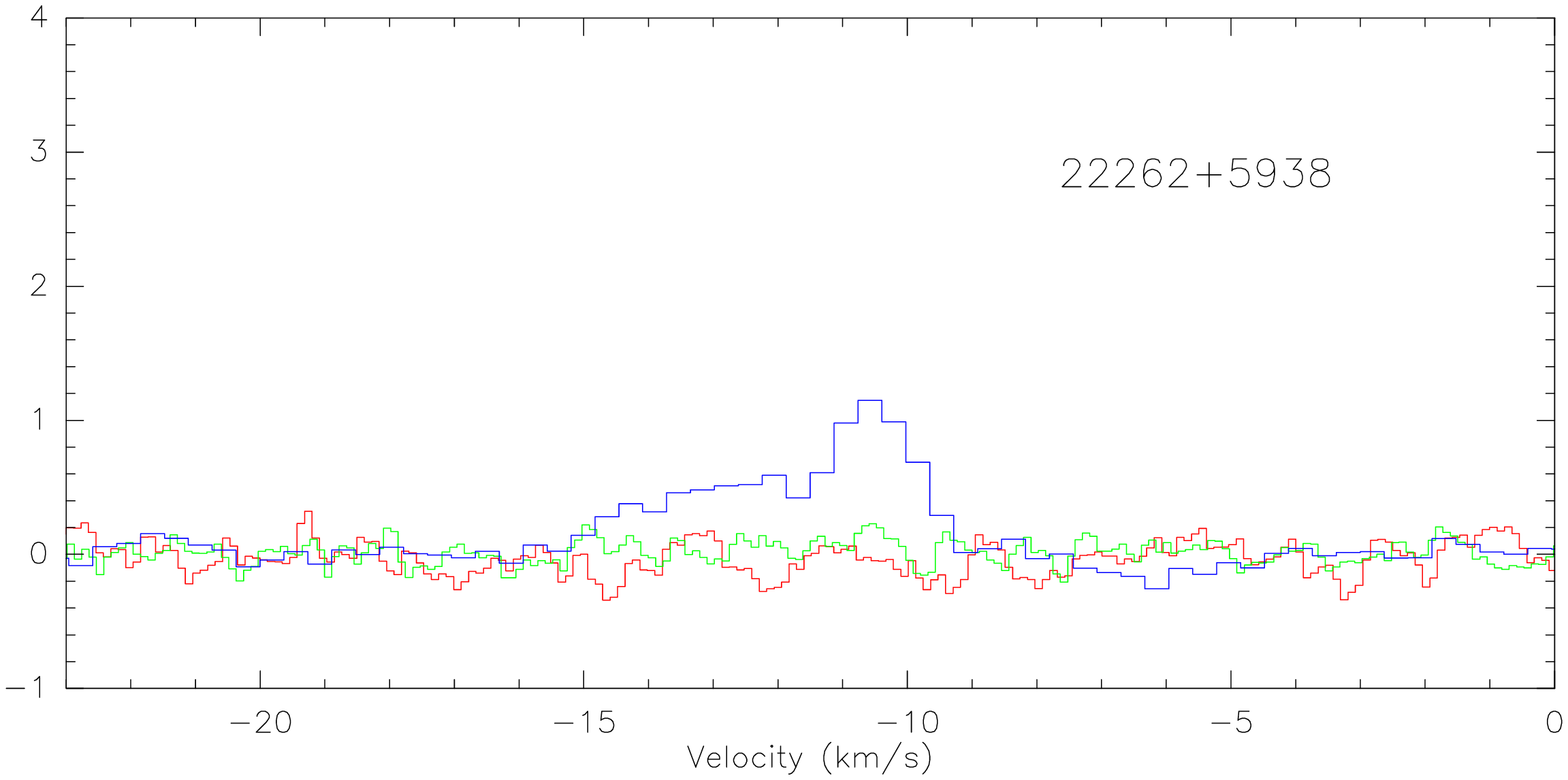}
\includegraphics[height=70mm,  angle=-90, clip, viewport=150 10 500 750]{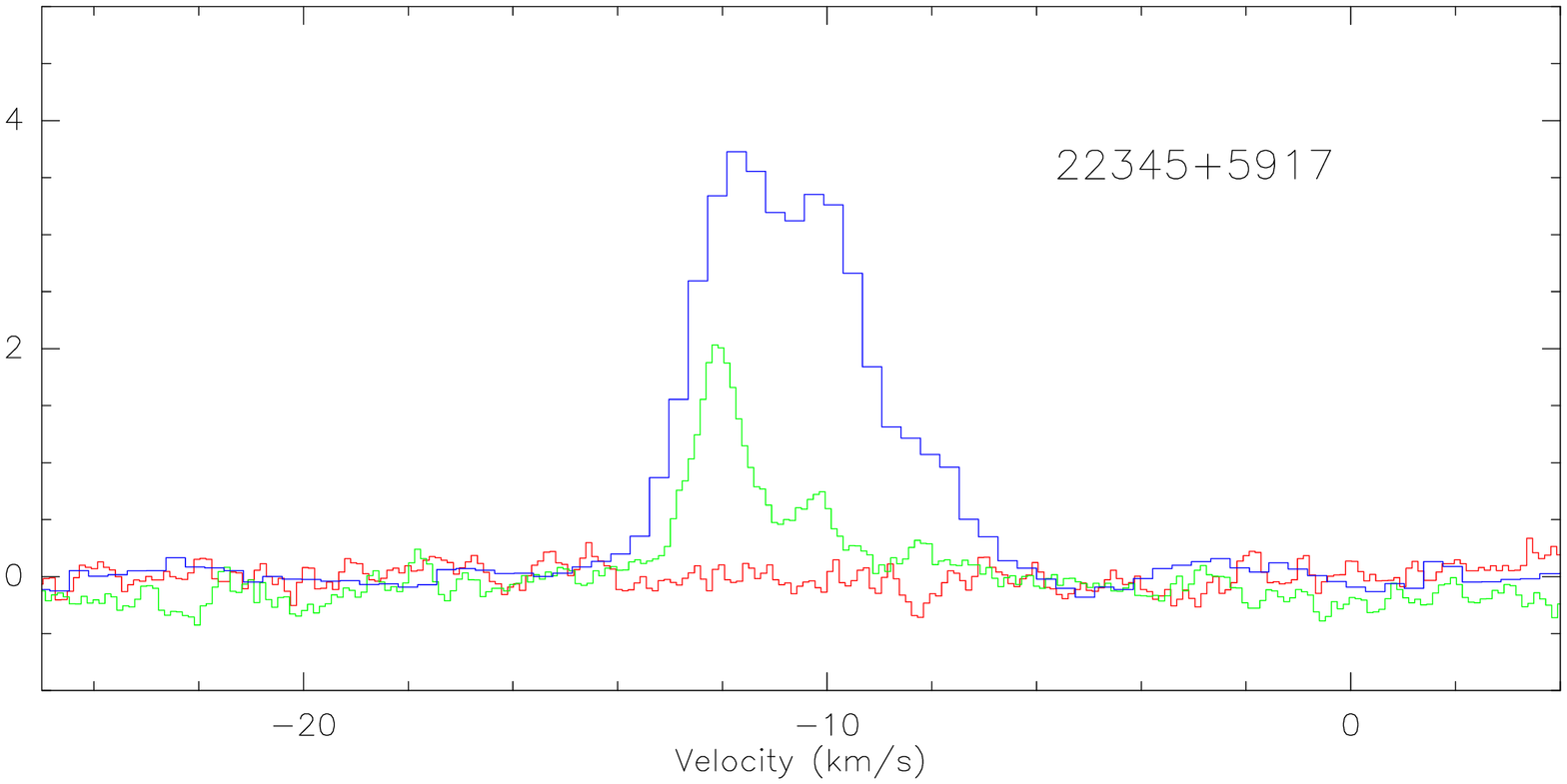}
\includegraphics[height=70mm,  angle=-90, clip, viewport=150 10 500 750]{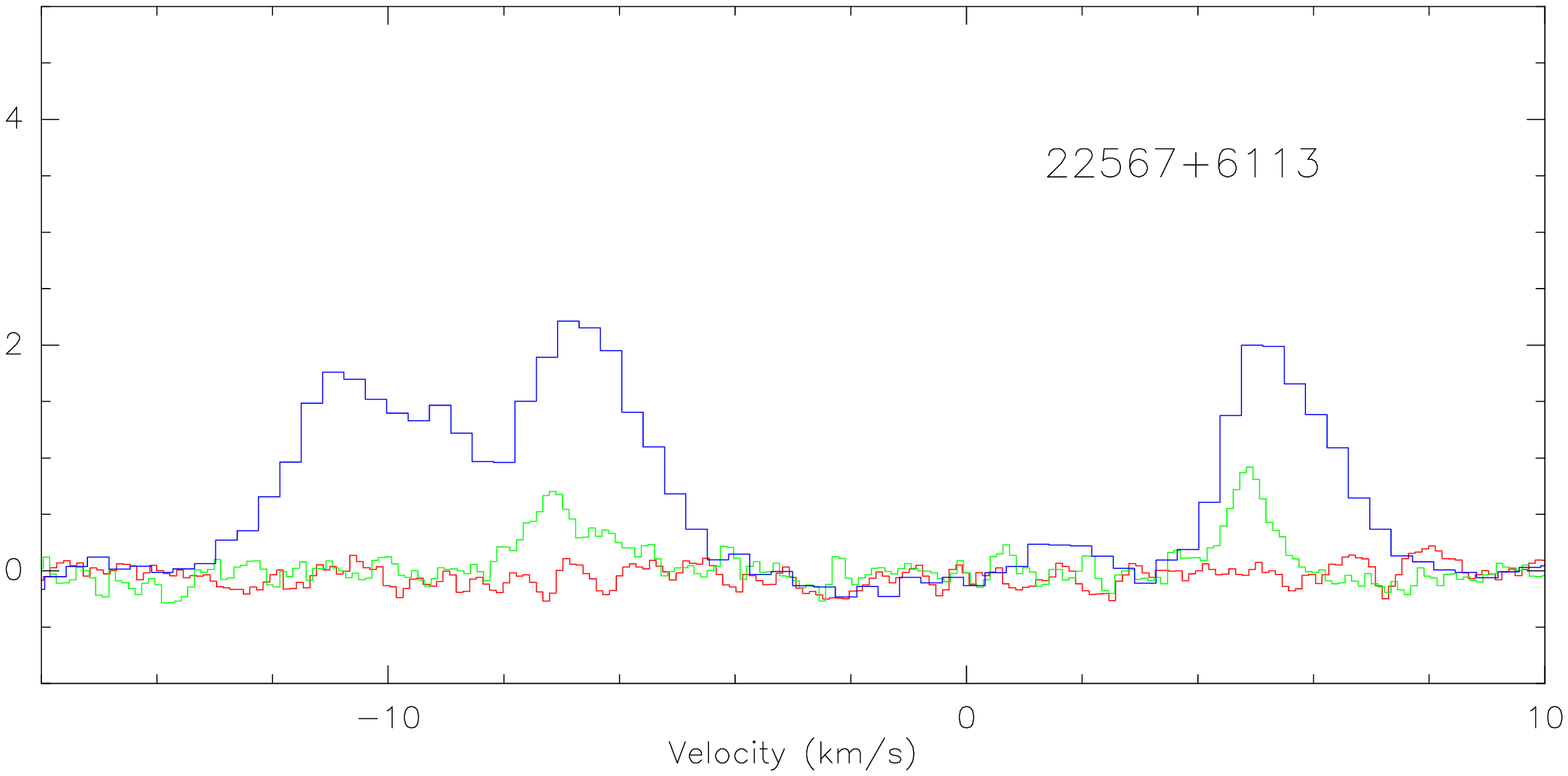}

\begin{minipage}[]{60mm}
  \caption{The sources of type 3
  }\end{minipage}
   \label{Fig8}
   \end{figure}

\addtocounter{figure}{-1}
\begin{figure}

\includegraphics[height=70mm,  angle=-90, clip, viewport=150 10 500 750]{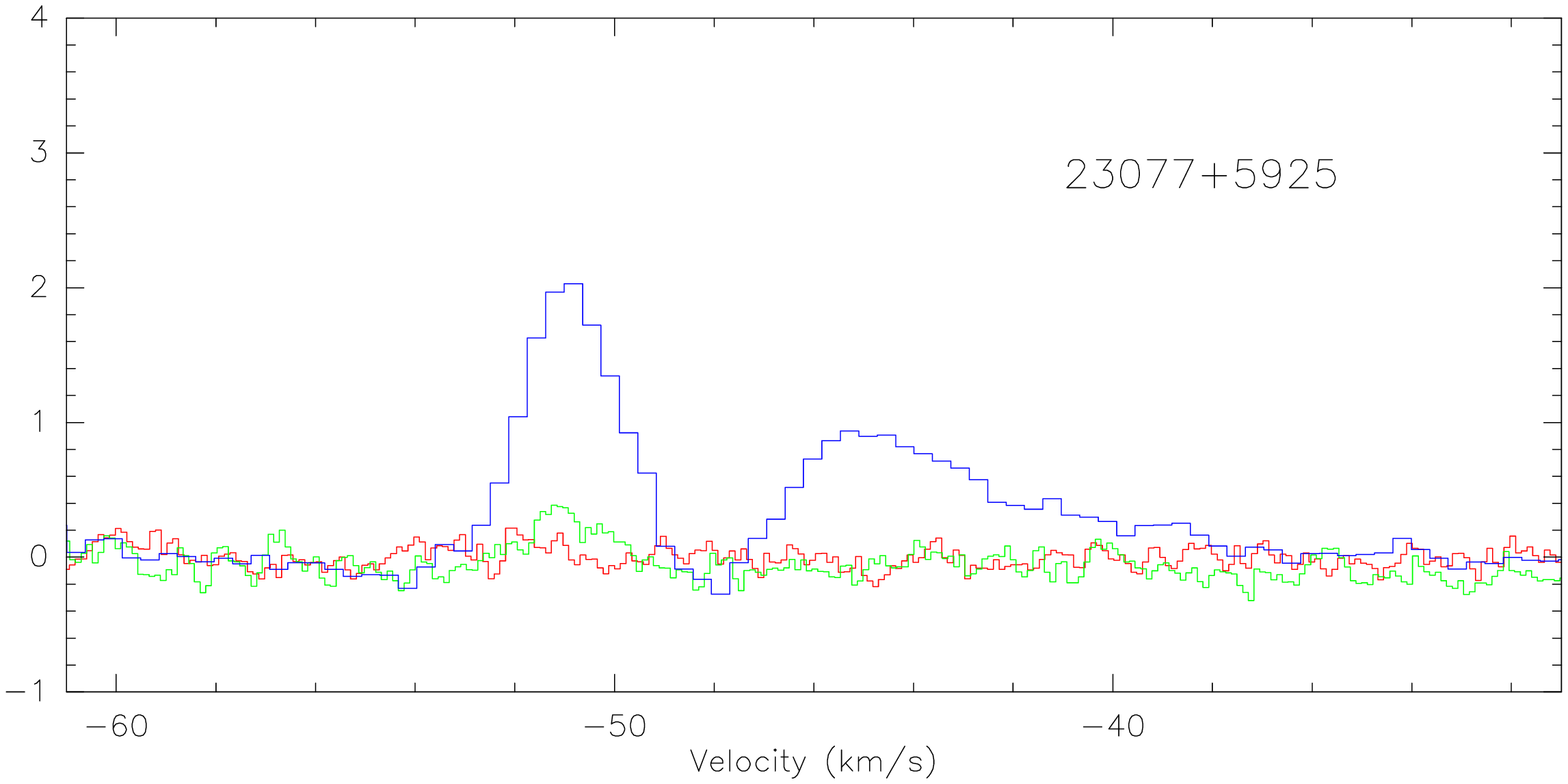}
\includegraphics[height=70mm,  angle=-90, clip, viewport=150 10 500 750]{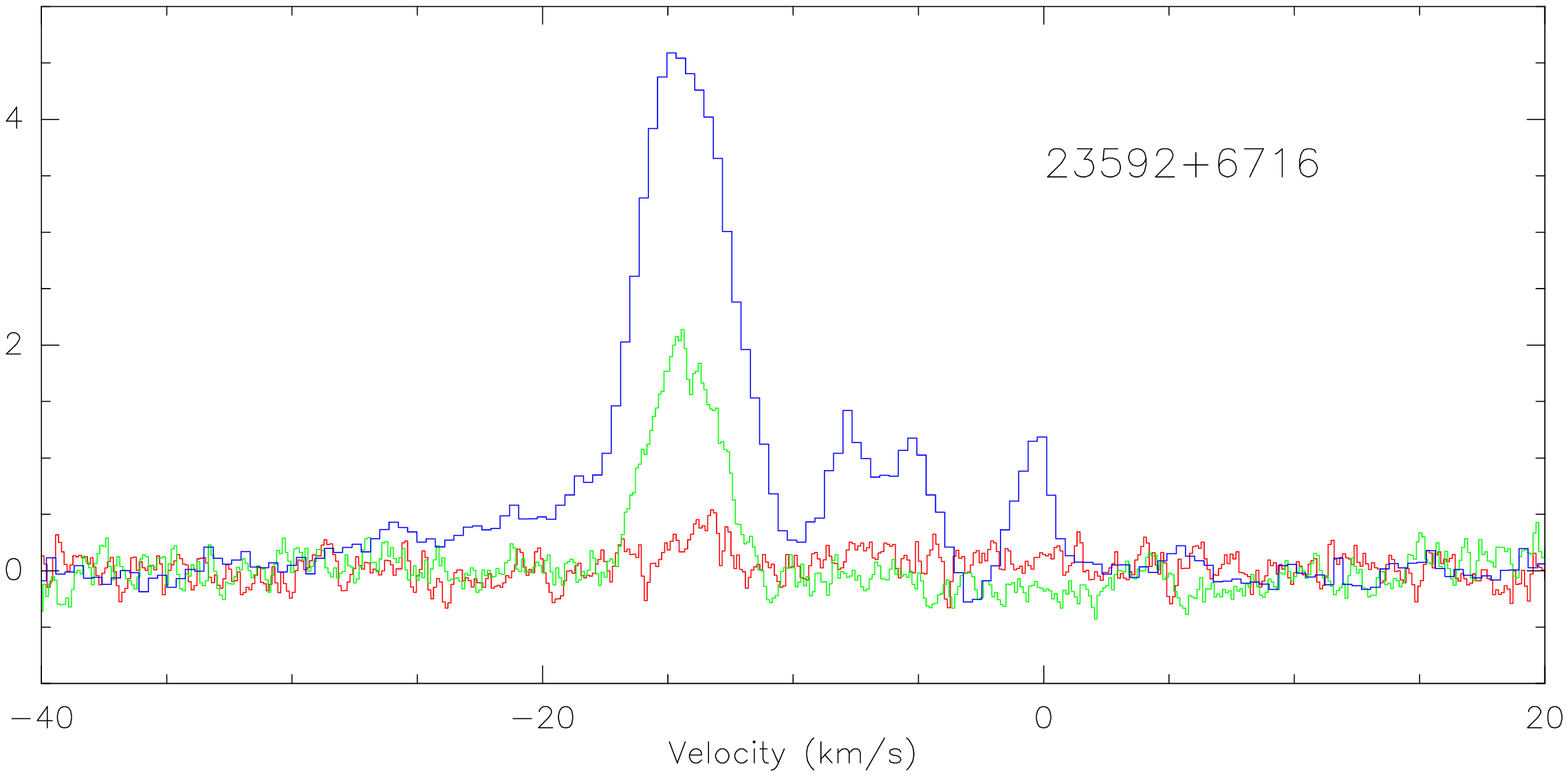}

\centering
\begin{minipage}[]{60mm}
   \caption{The sources of type 3
  }\end{minipage}
   \label{Fig8}
   \end{figure}

\begin{figure}
   \centering

\includegraphics[height=70mm,  angle=-90, clip, viewport=150 10 500 750]{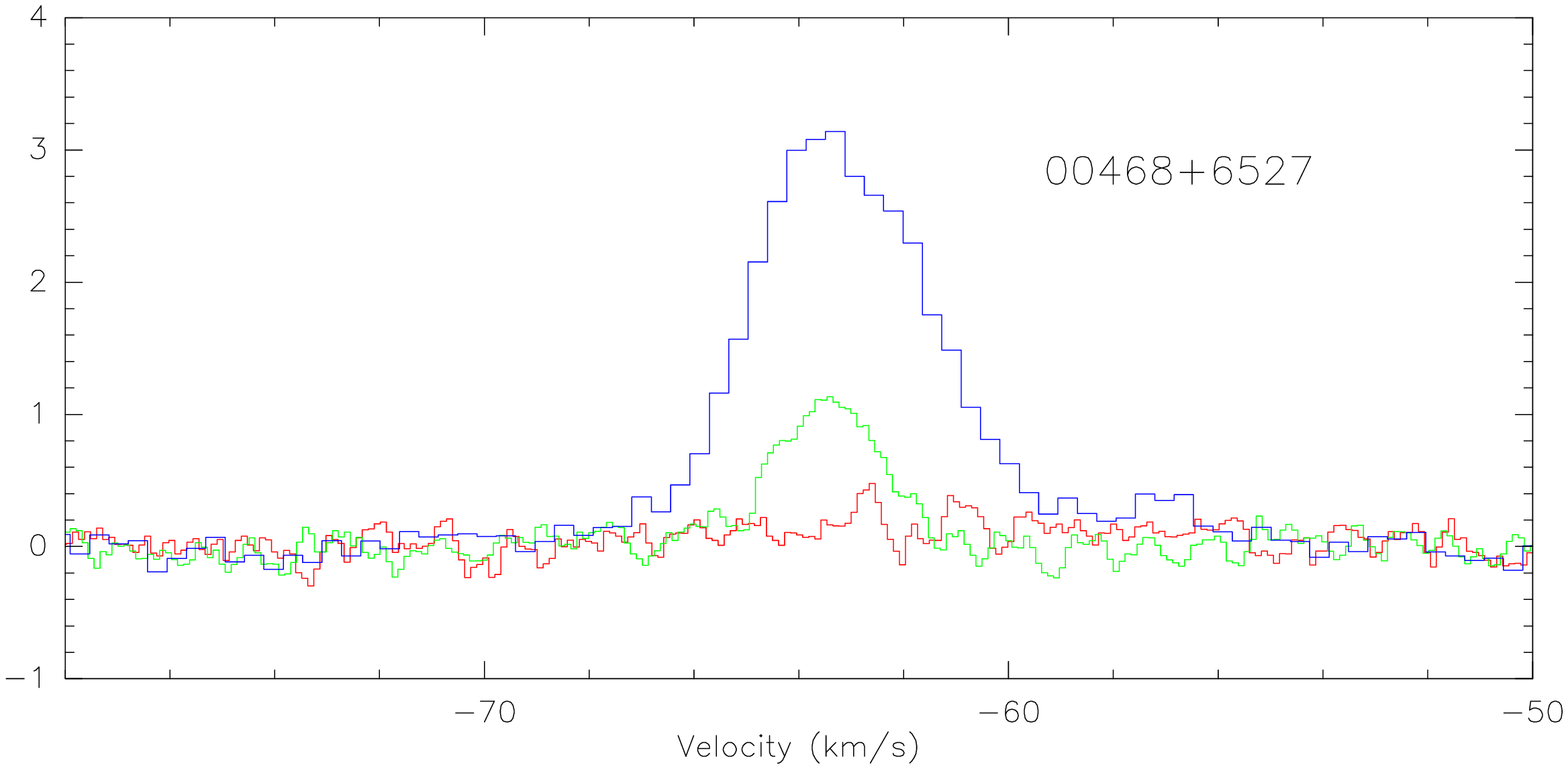}
\includegraphics[height=70mm,  angle=-90, clip, viewport=150 10 500 750]{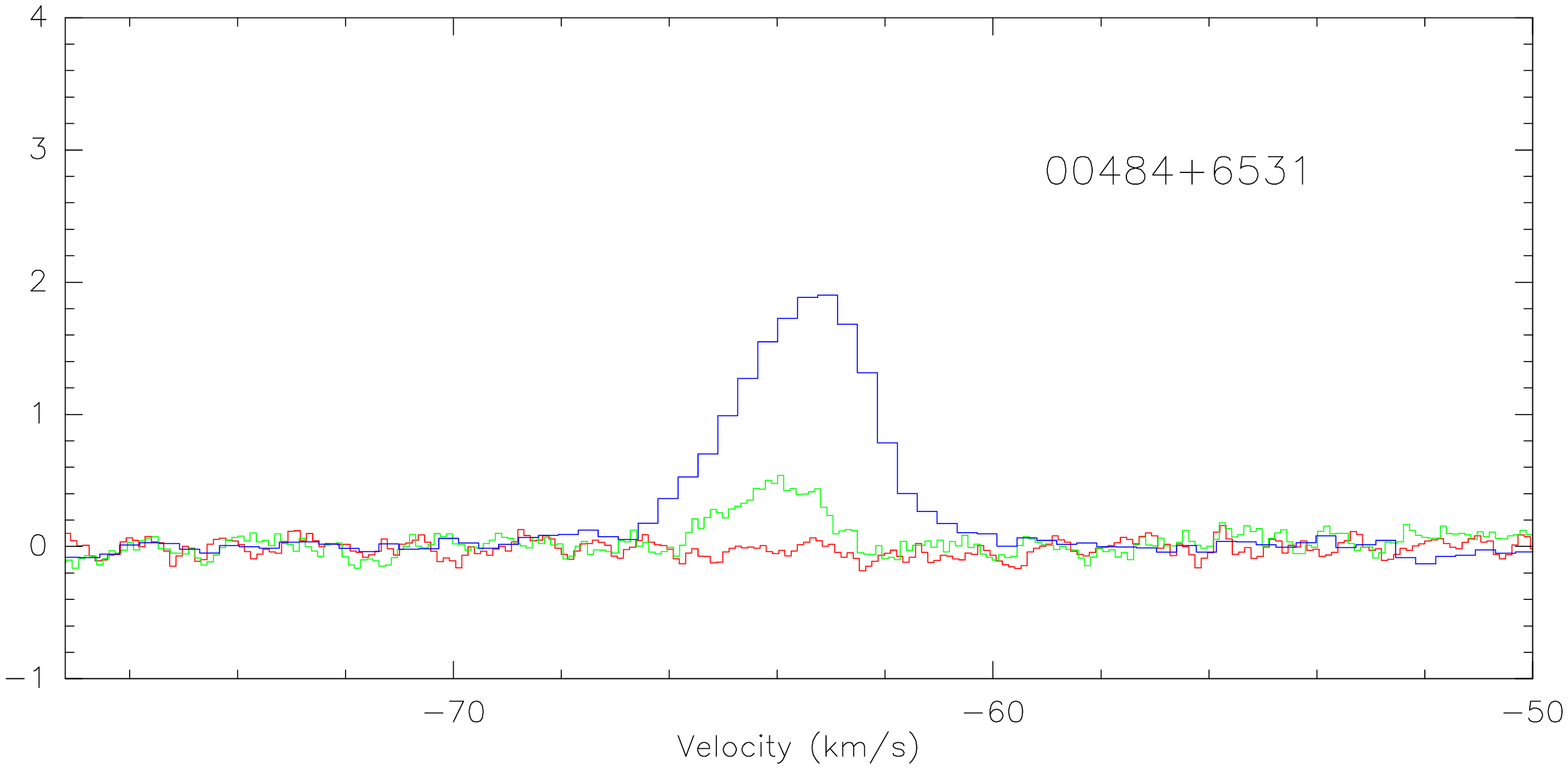}
\includegraphics[height=70mm,  angle=-90, clip, viewport=150 10 500 750]{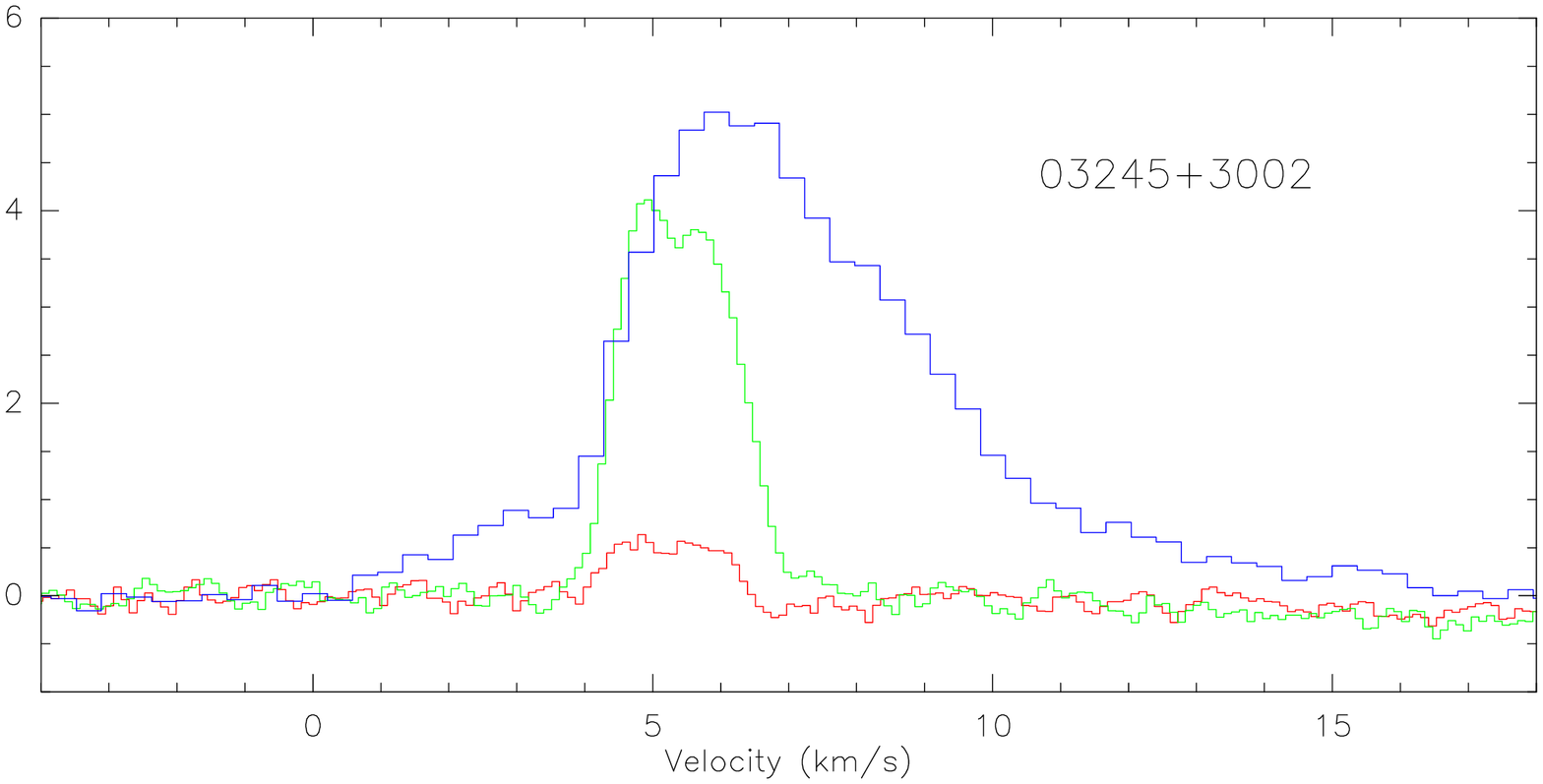}
\includegraphics[height=70mm,  angle=-90, clip, viewport=150 10 500 750]{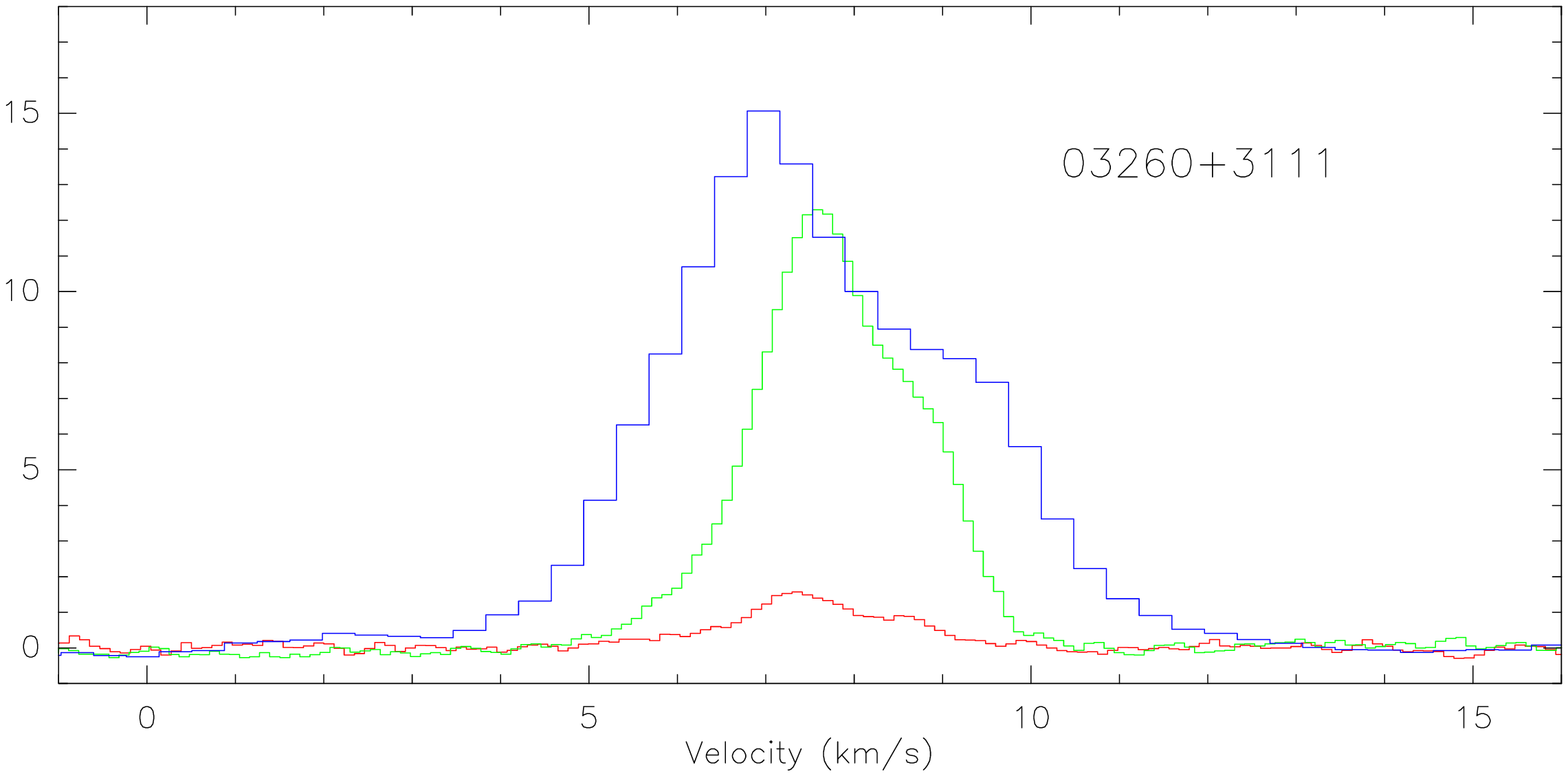}
\includegraphics[height=70mm,  angle=-90, clip, viewport=150 10 500 750]{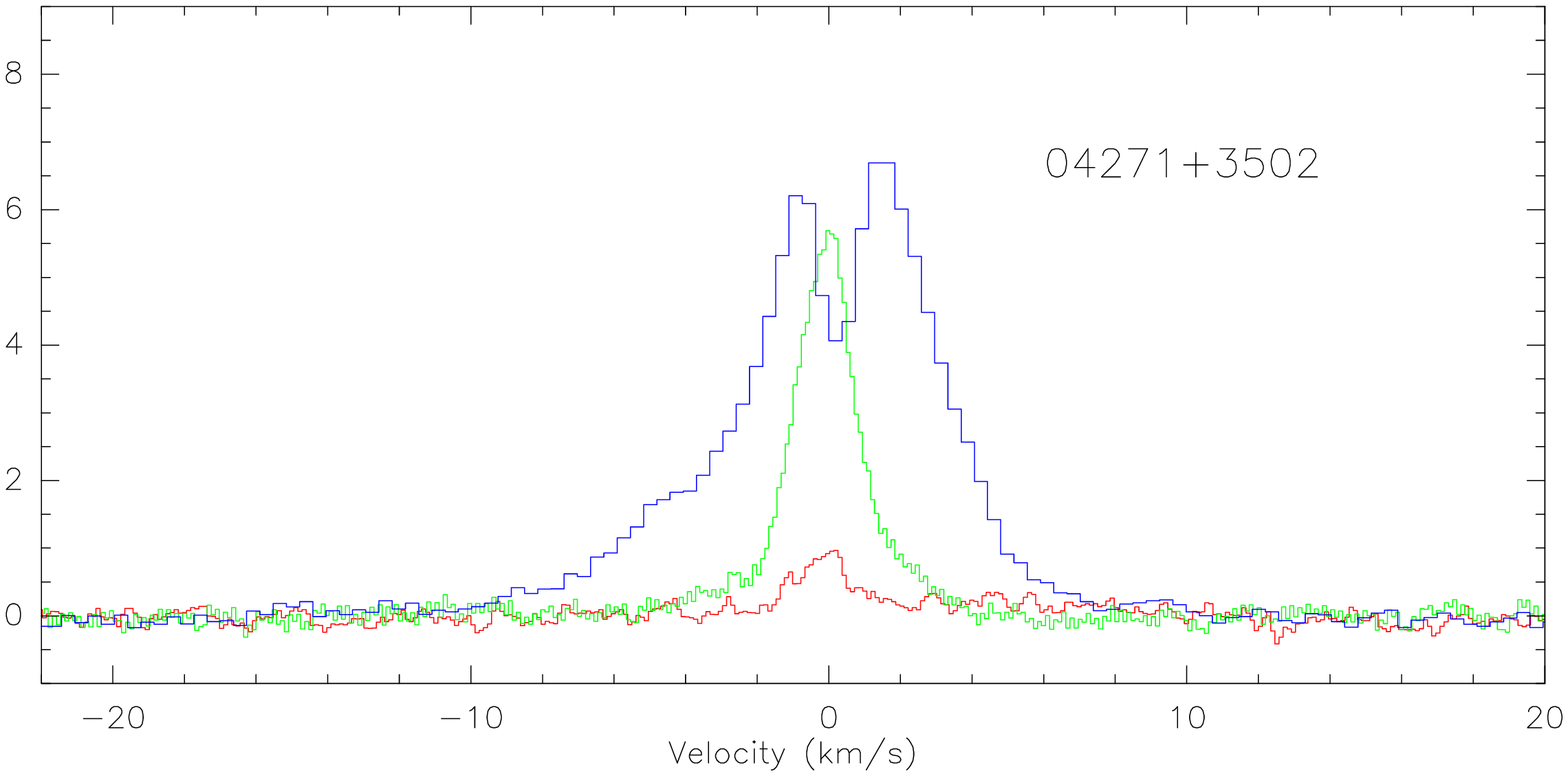}
\includegraphics[height=70mm,  angle=-90, clip, viewport=150 10 500 750]{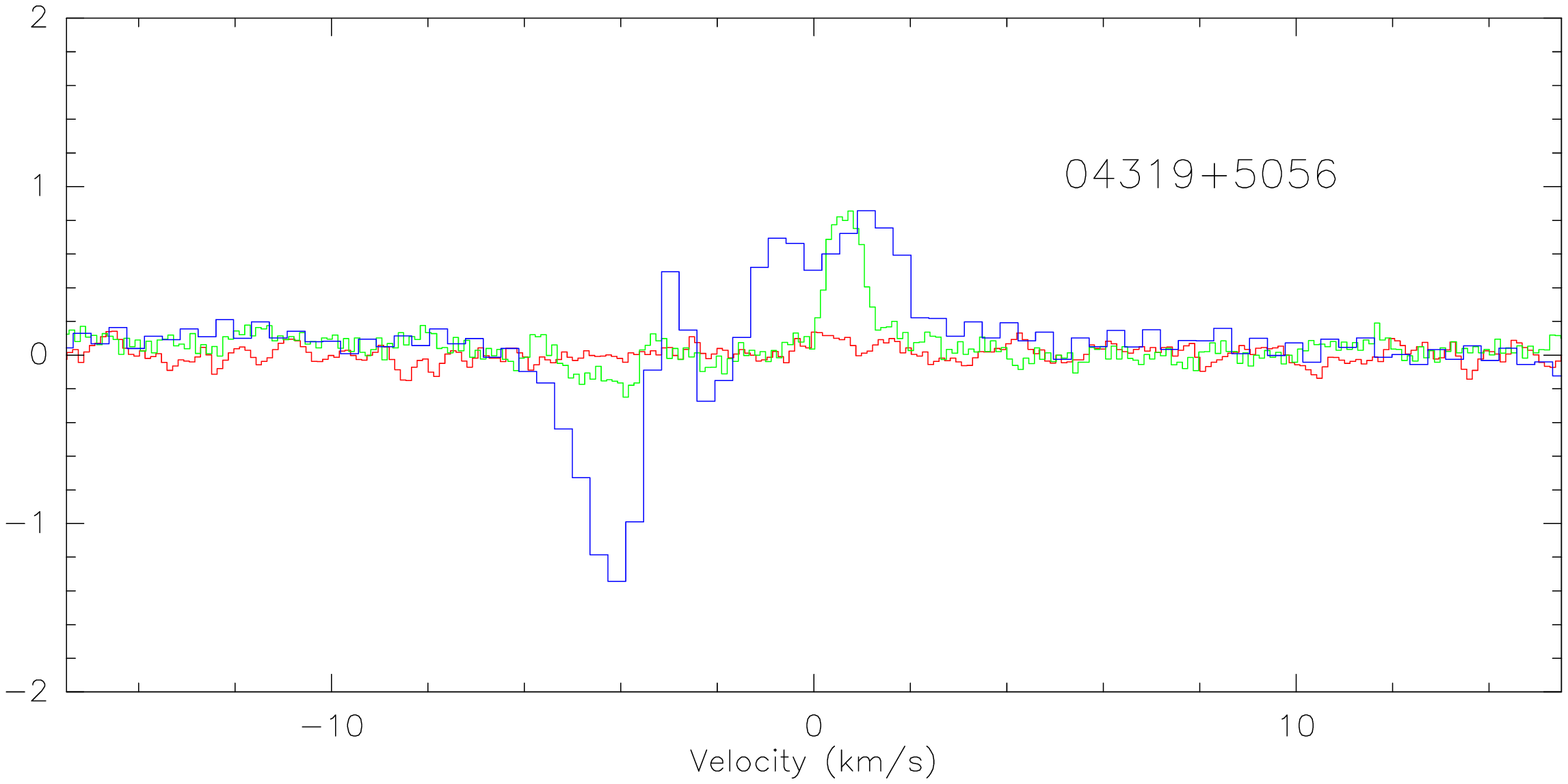}
\includegraphics[height=70mm,  angle=-90, clip, viewport=150 10 500 750]{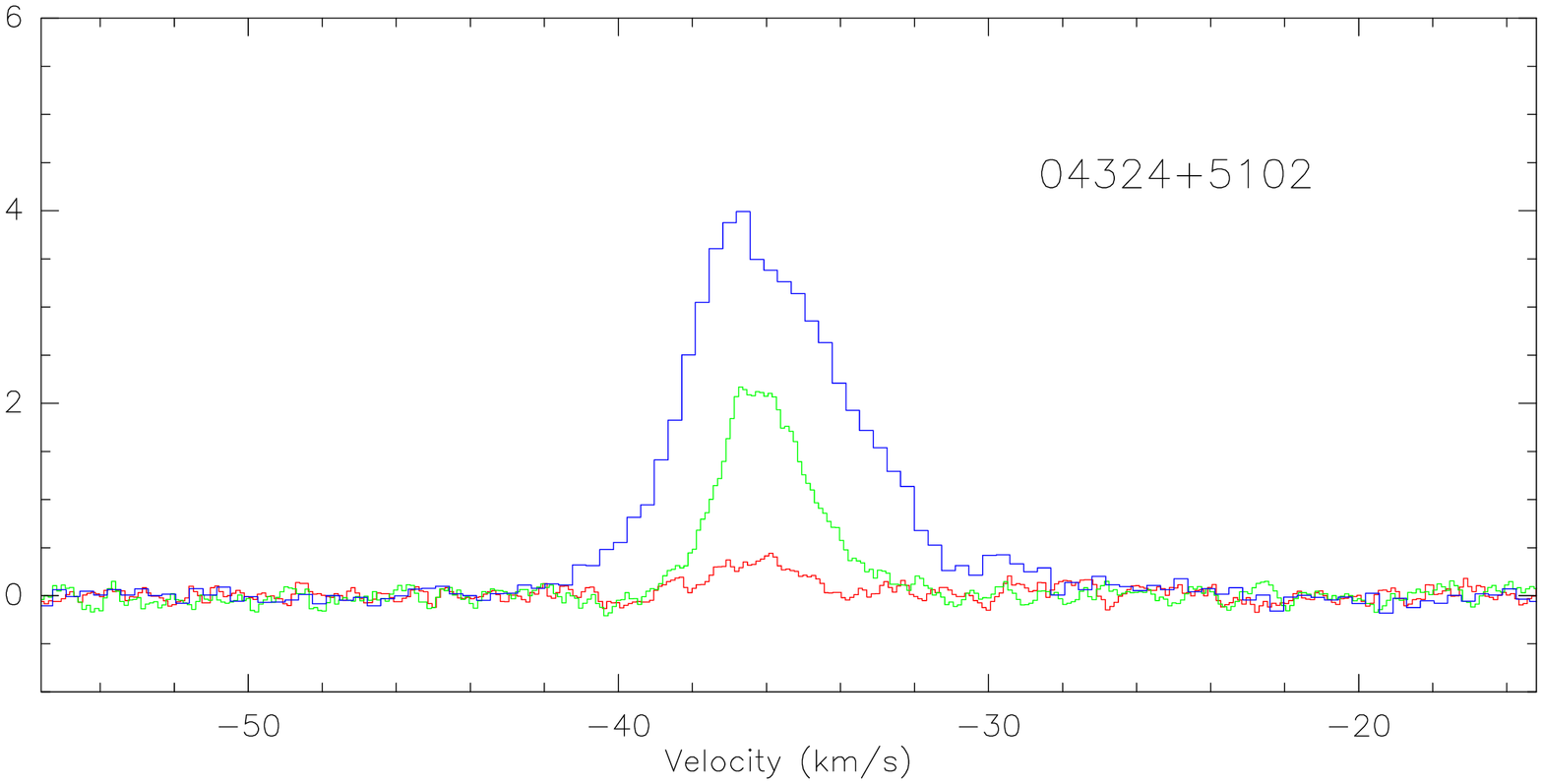}
\includegraphics[height=70mm,  angle=-90, clip, viewport=150 10 500 750]{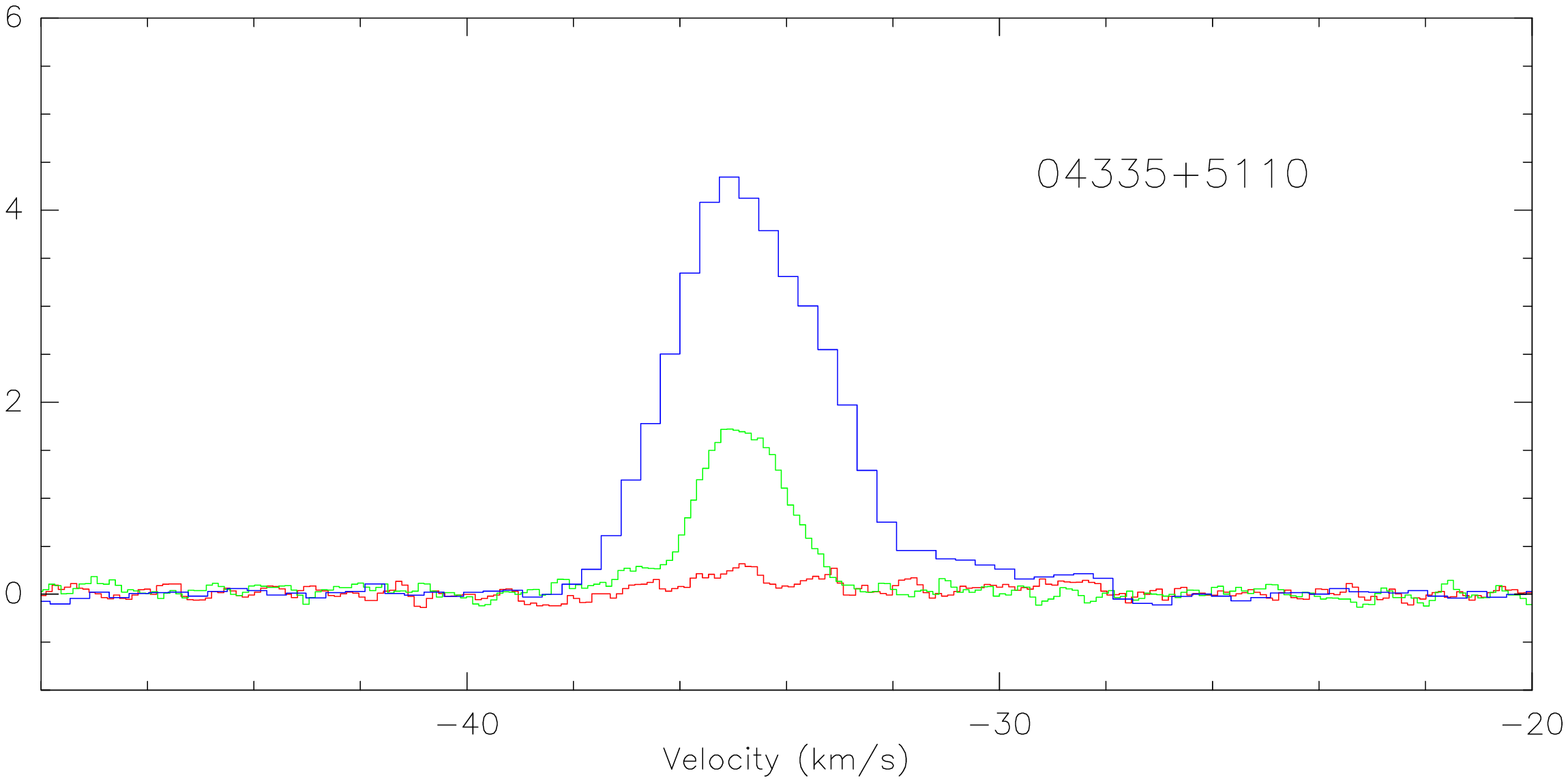}
\includegraphics[height=70mm,  angle=-90, clip, viewport=150 10 500 750]{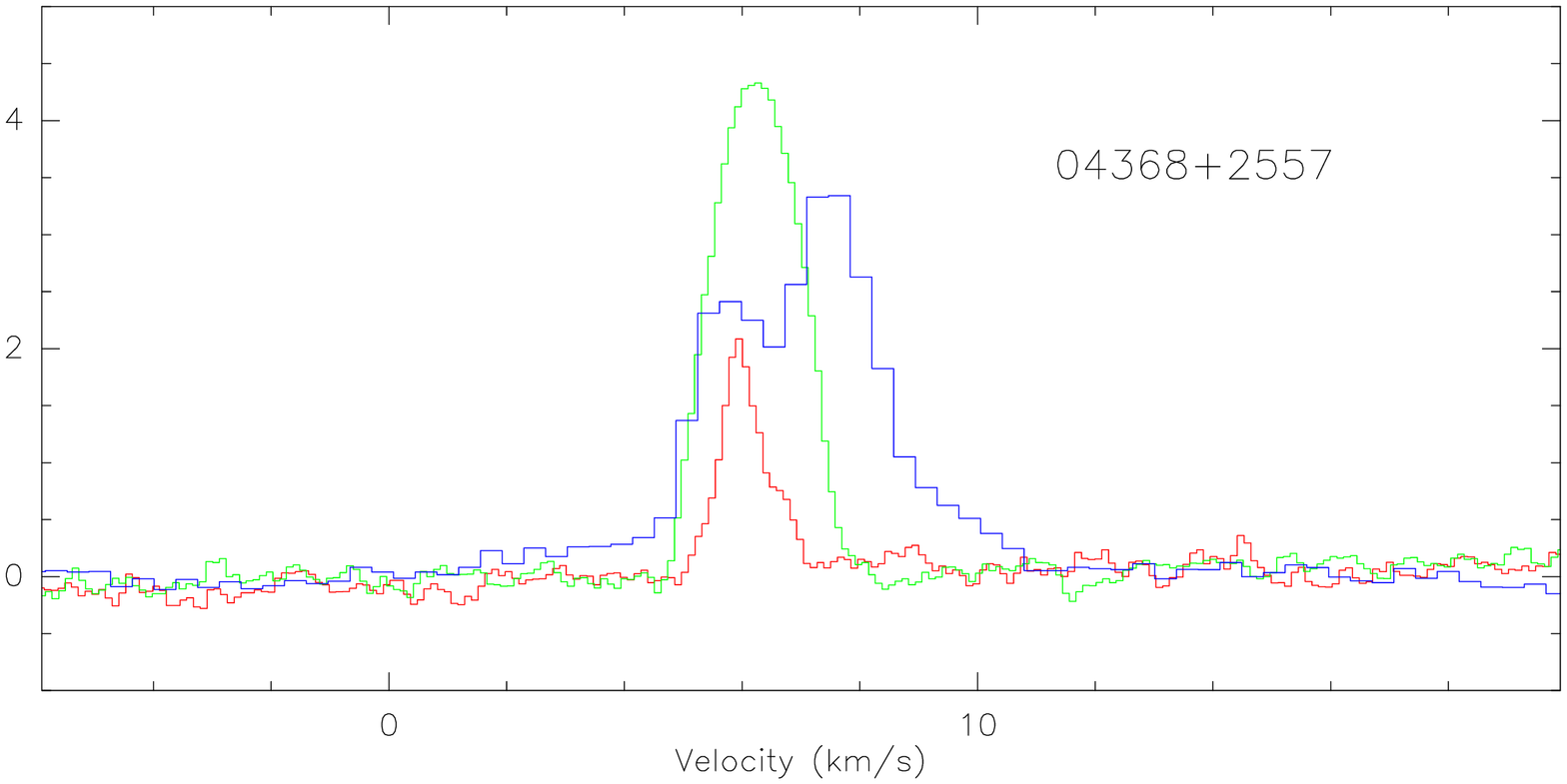}
\includegraphics[height=70mm,  angle=-90, clip, viewport=150 10 500 750]{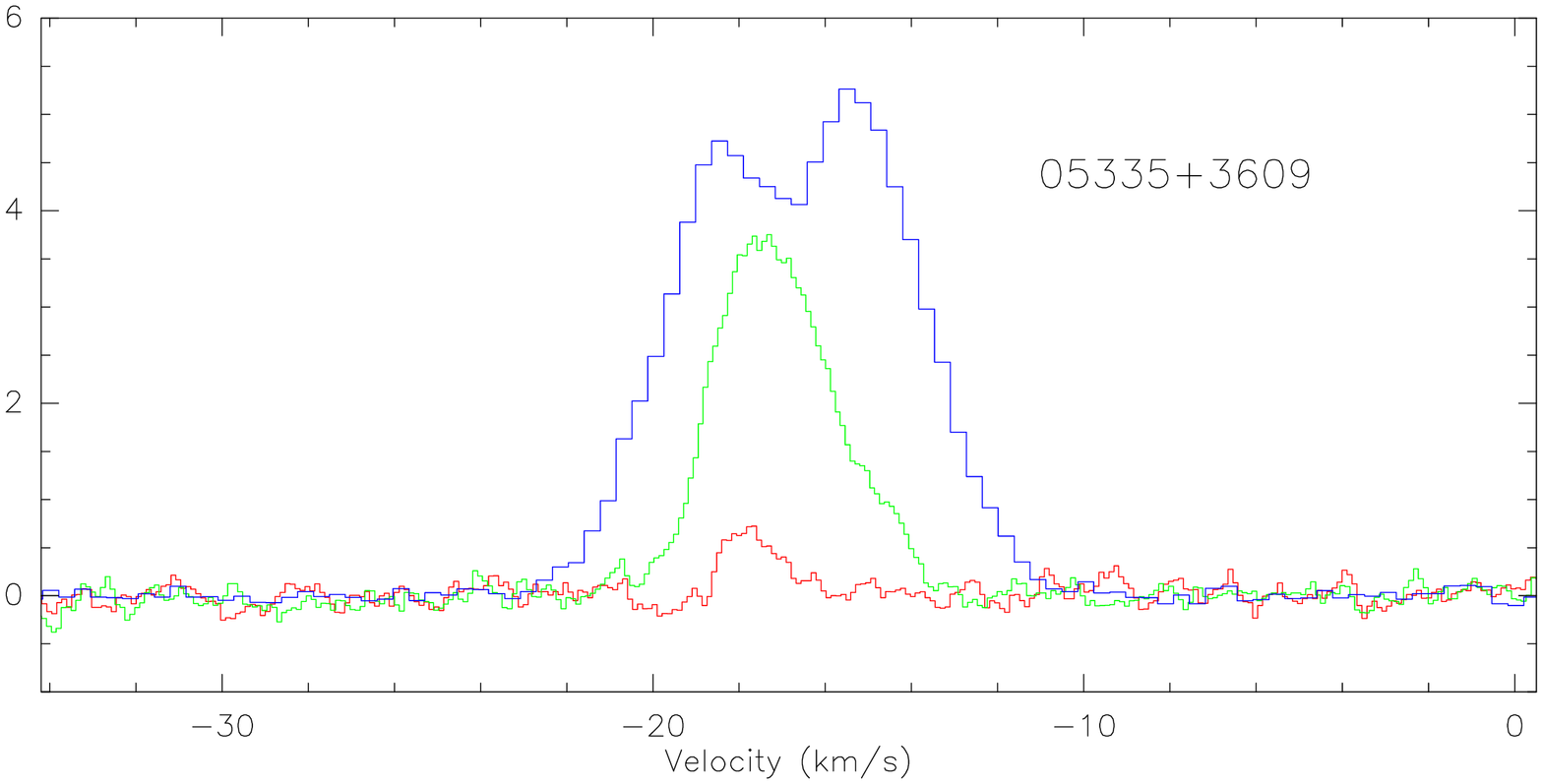}
\includegraphics[height=70mm,  angle=-90, clip, viewport=150 10 500 750]{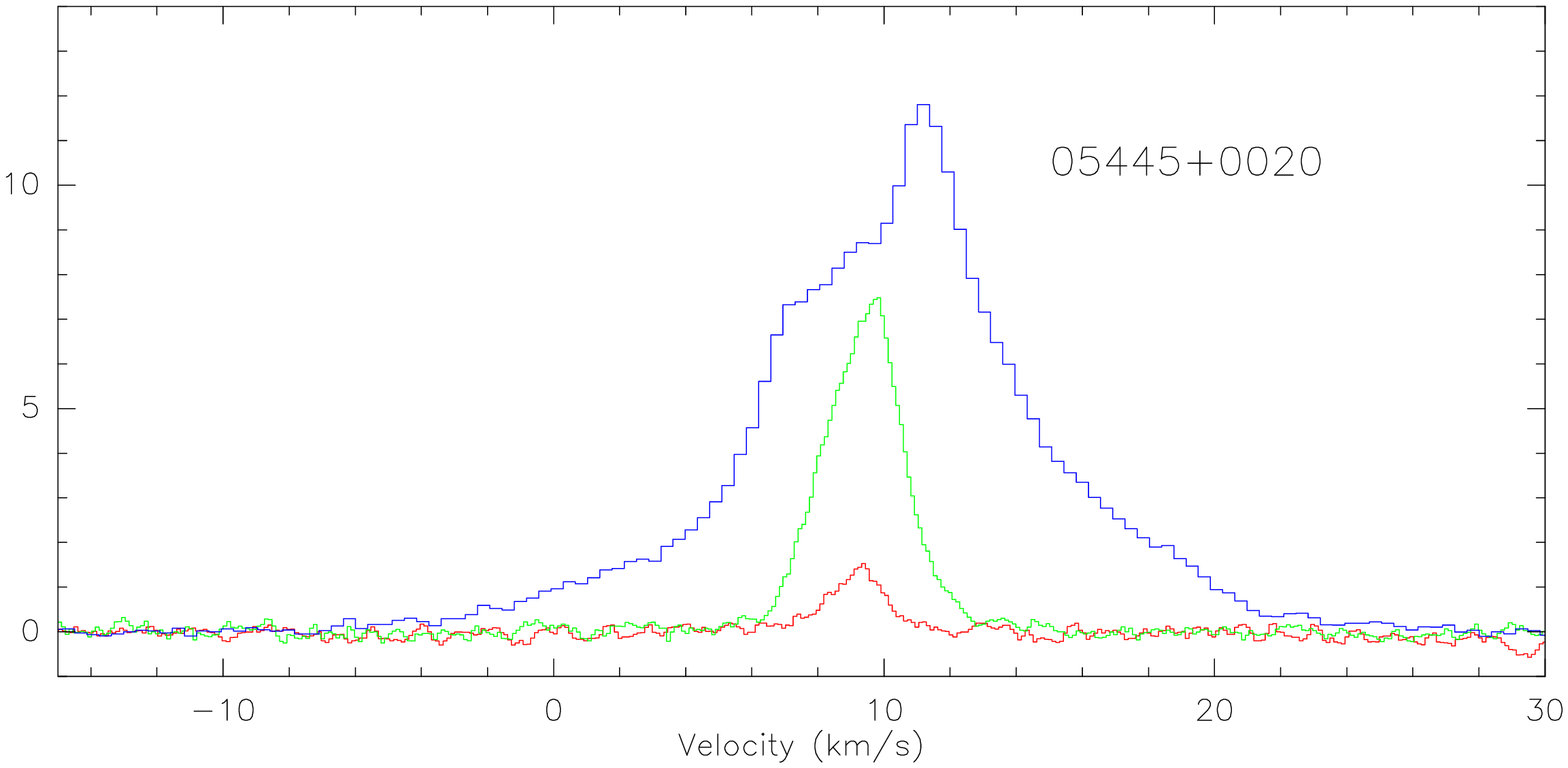}
\includegraphics[height=70mm,  angle=-90, clip, viewport=150 10 500 750]{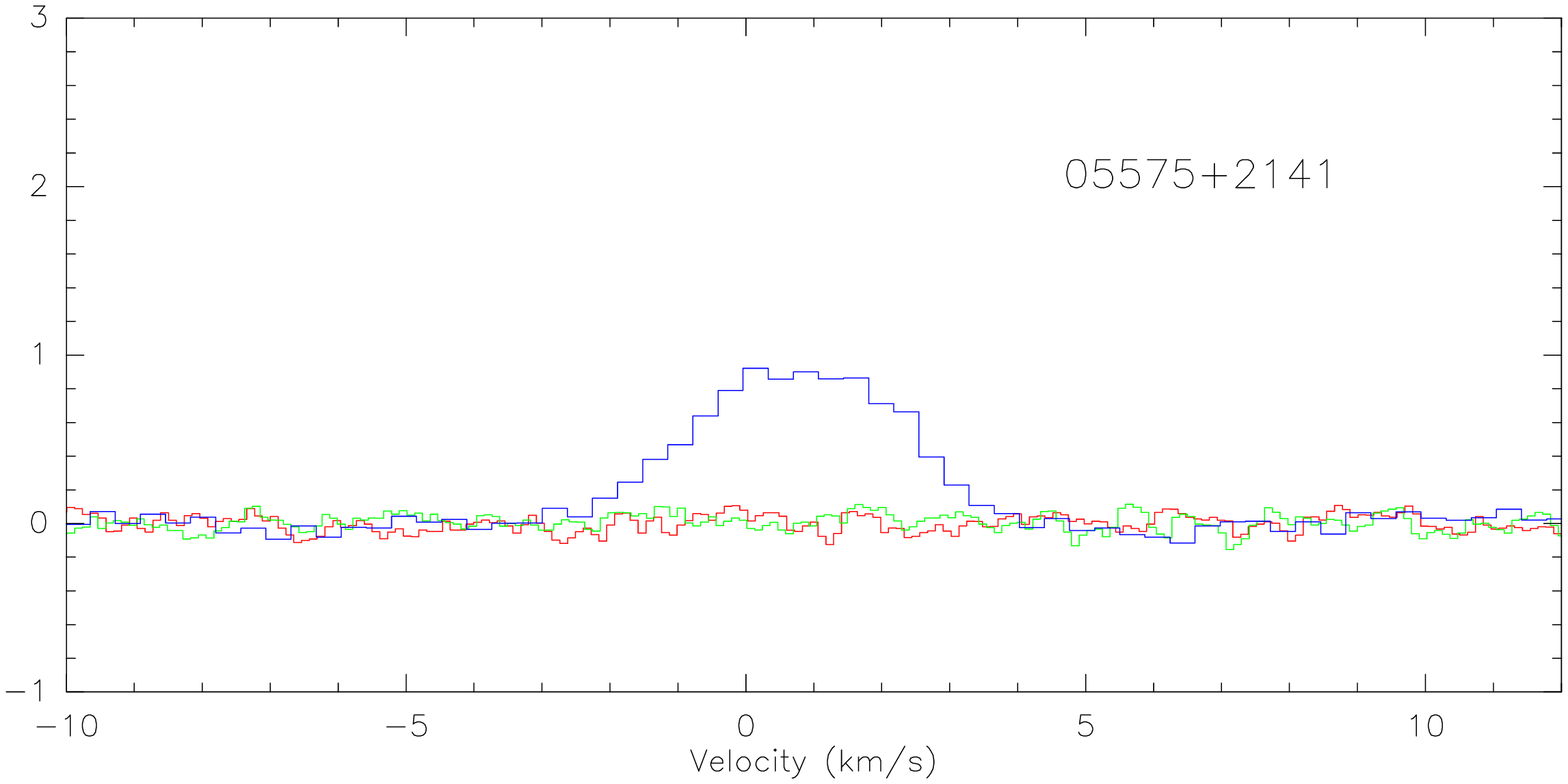}
\includegraphics[height=70mm,  angle=-90, clip, viewport=150 10 500 750]{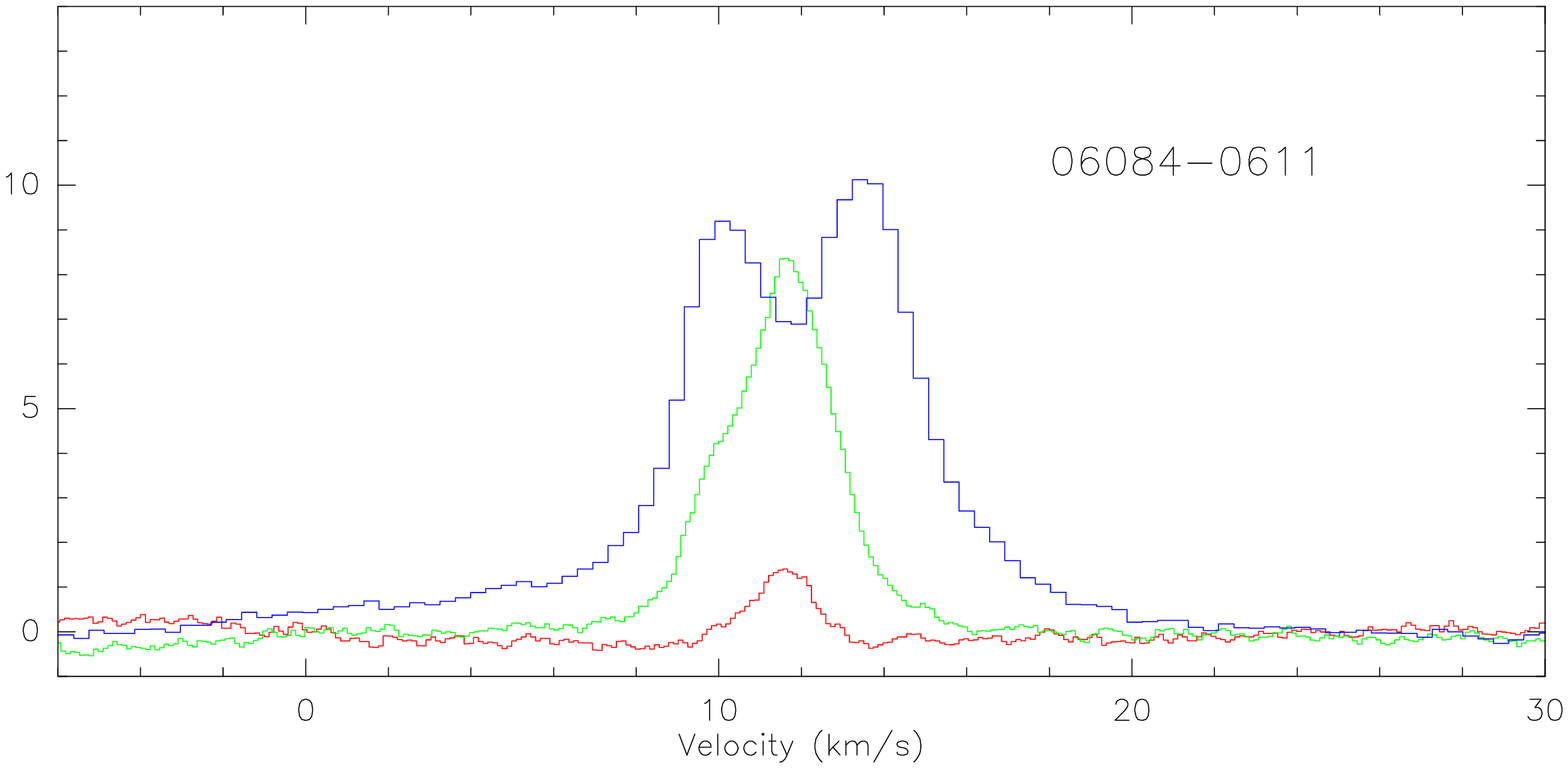}
\includegraphics[height=70mm,  angle=-90, clip, viewport=150 10 500 750]{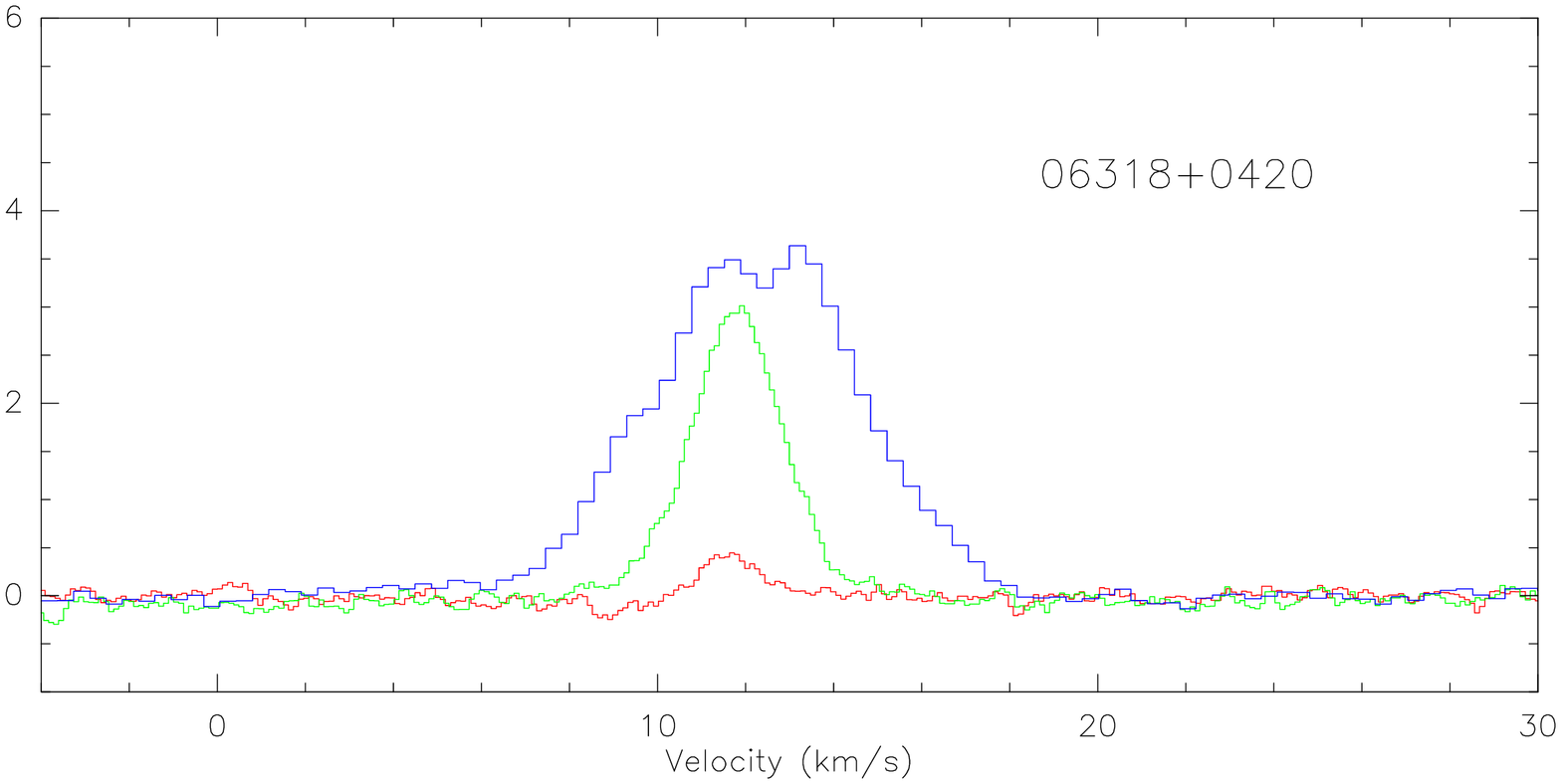}

\begin{minipage}[]{60mm}
   \caption{The sources of type 4
  }\end{minipage}
   \label{Fig9}
   \end{figure}

\addtocounter{figure}{-1}
\begin{figure}
   \centering

\includegraphics[height=70mm,  angle=-90, clip, viewport=150 10 500 750]{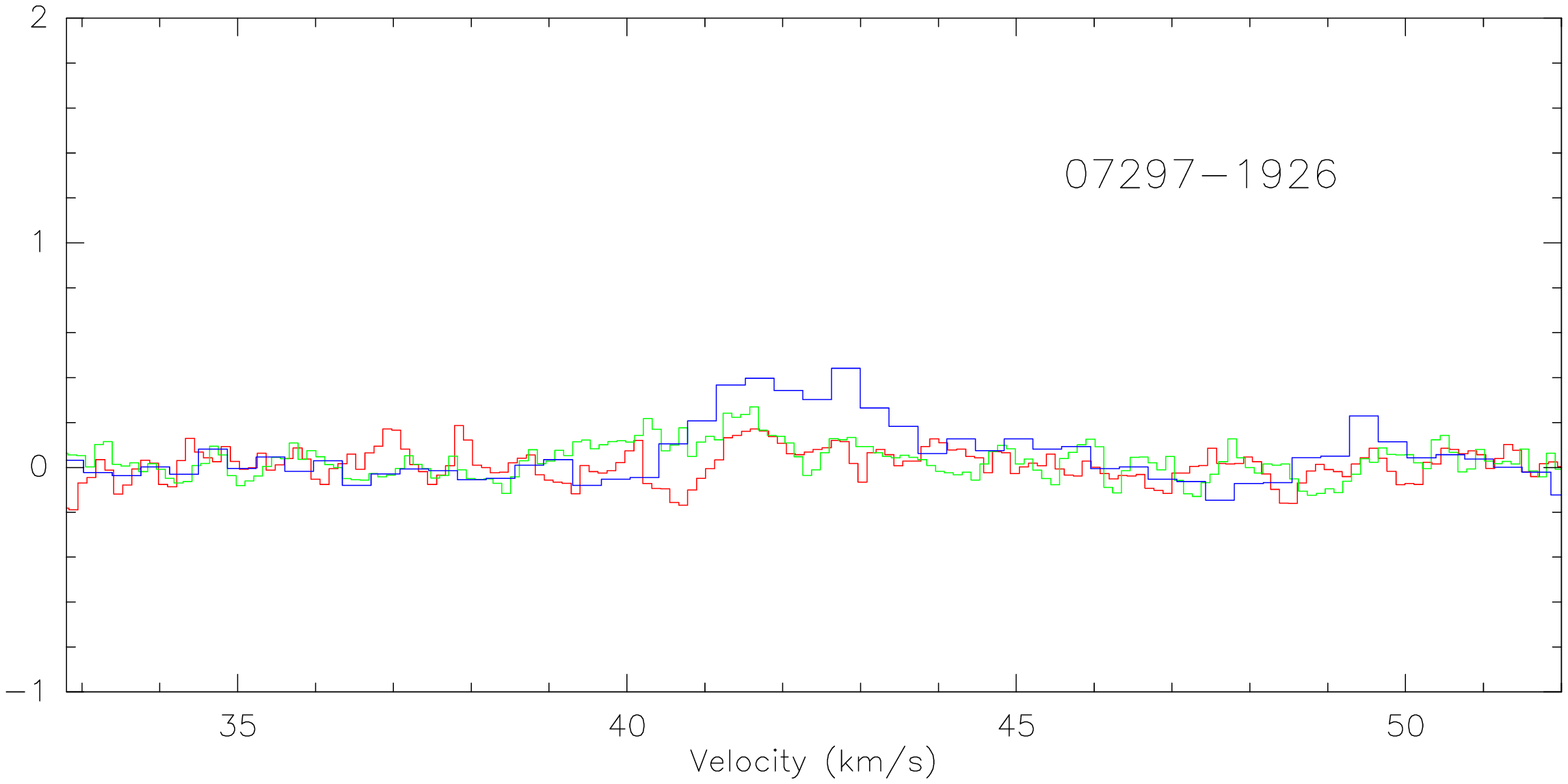}
\includegraphics[height=70mm,  angle=-90, clip, viewport=150 10 500 750]{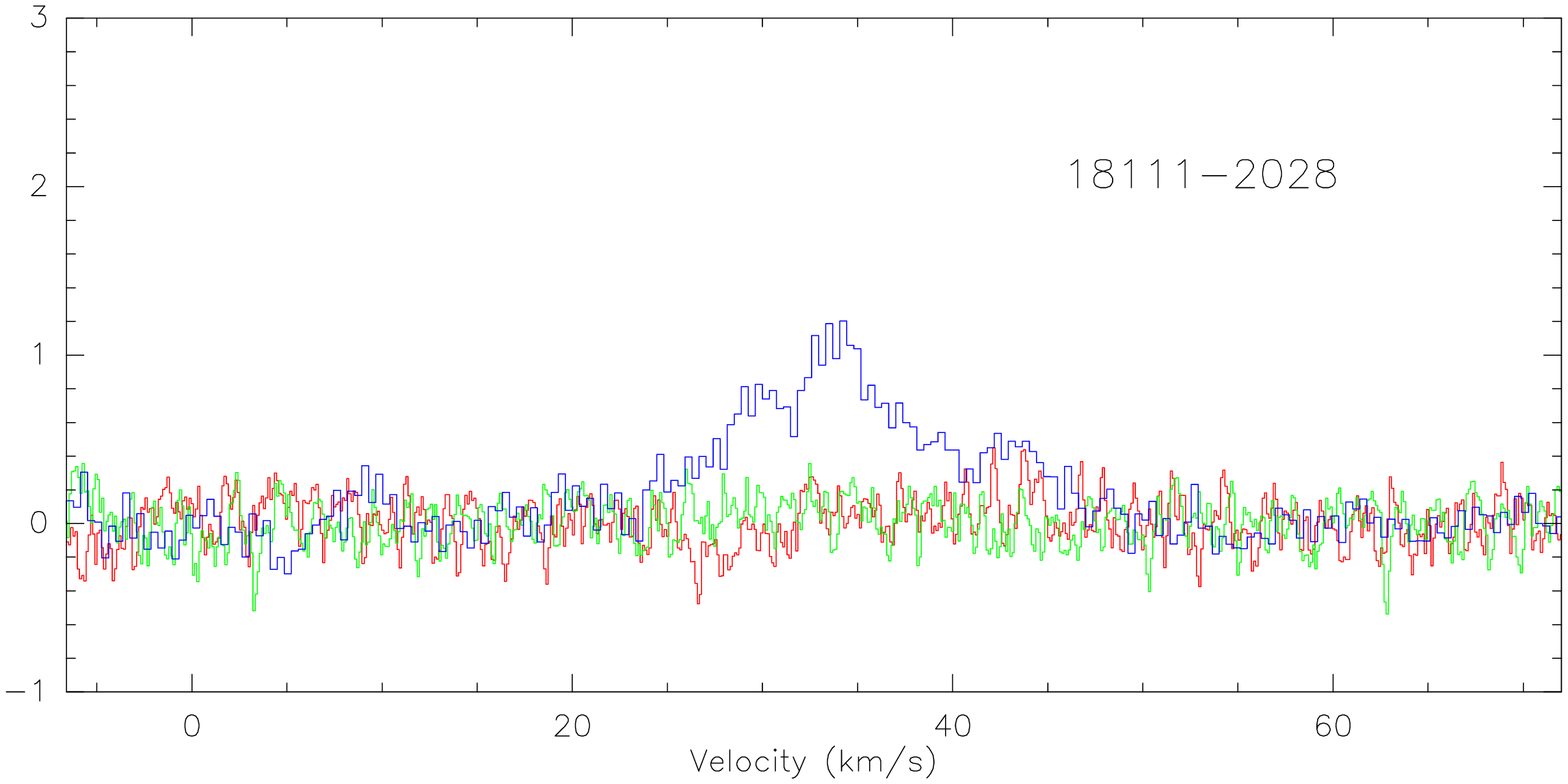}
\includegraphics[height=70mm,  angle=-90, clip, viewport=150 10 500 750]{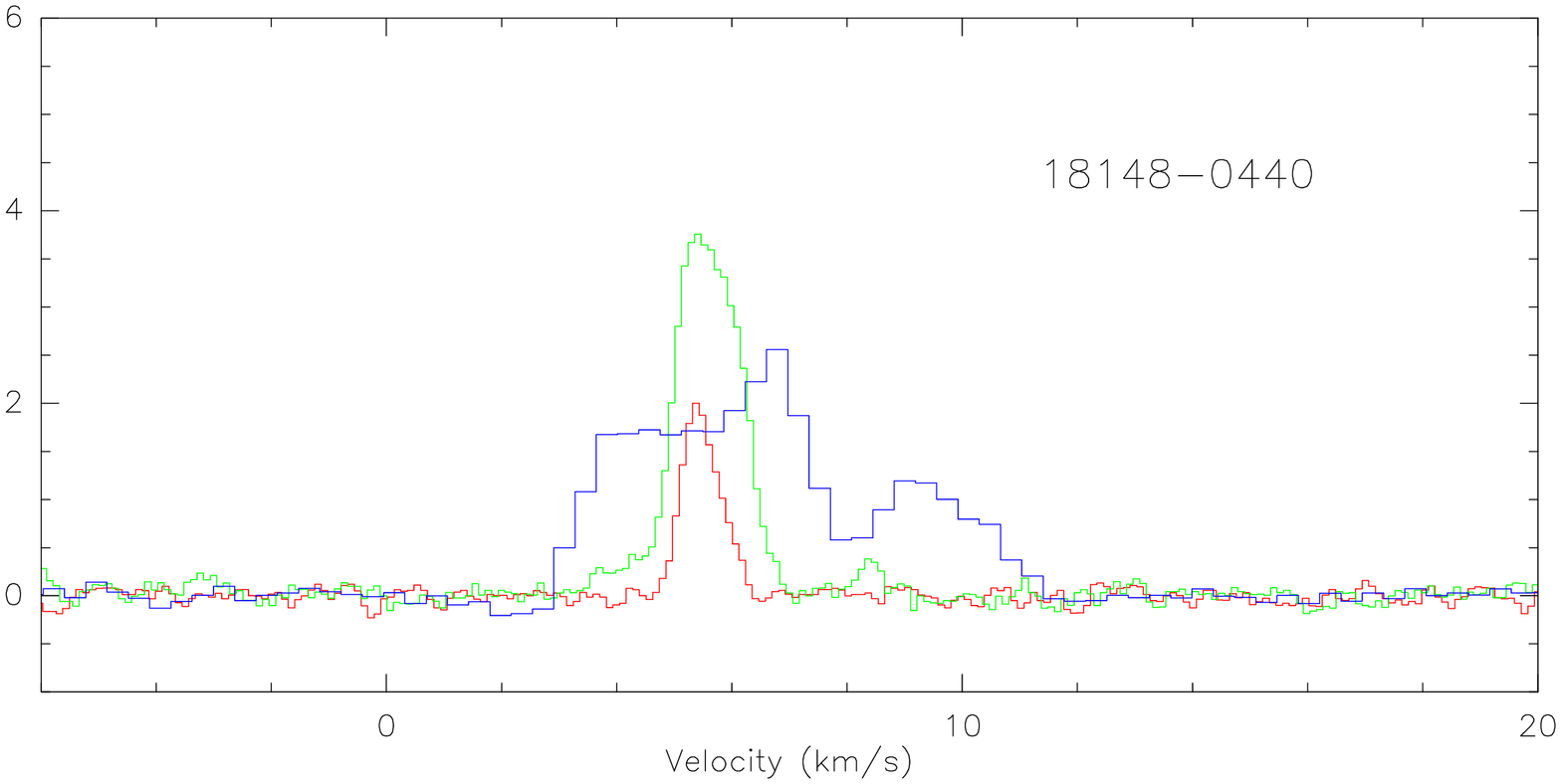}
\includegraphics[height=70mm,  angle=-90, clip, viewport=150 10 500 750]{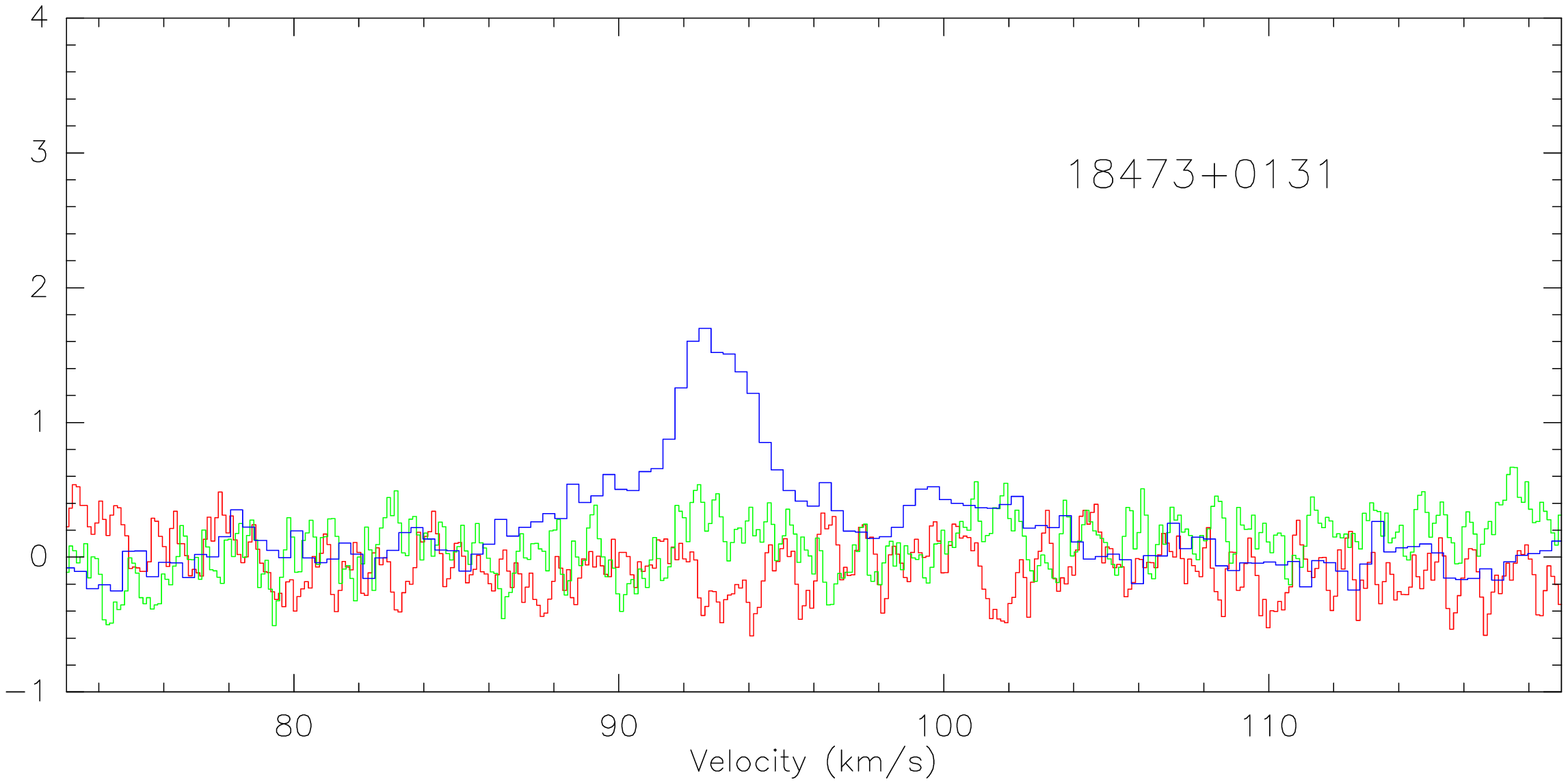}
\includegraphics[height=70mm,  angle=-90, clip, viewport=150 10 500 750]{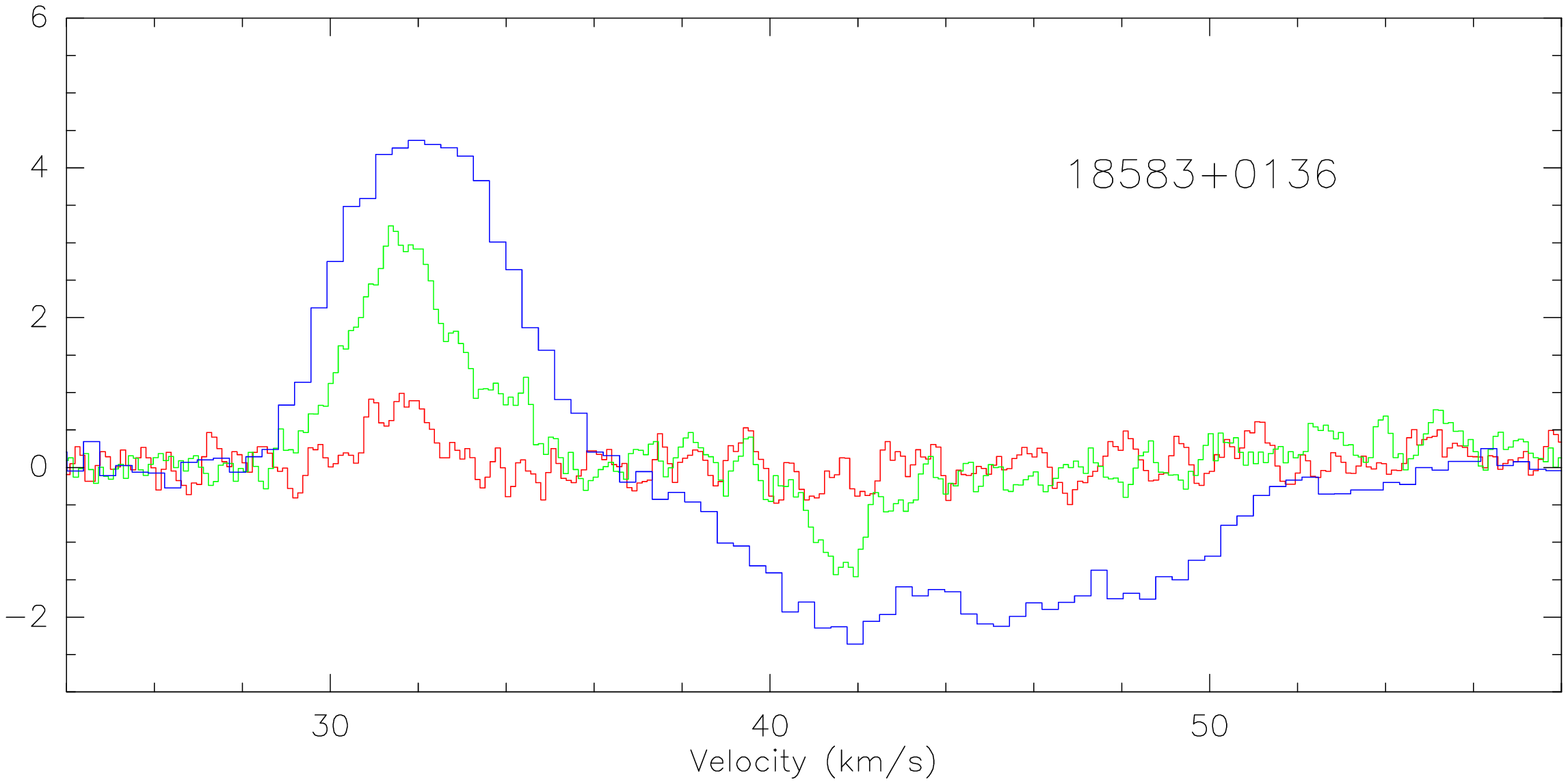}
\includegraphics[height=70mm,  angle=-90, clip, viewport=150 10 500 750]{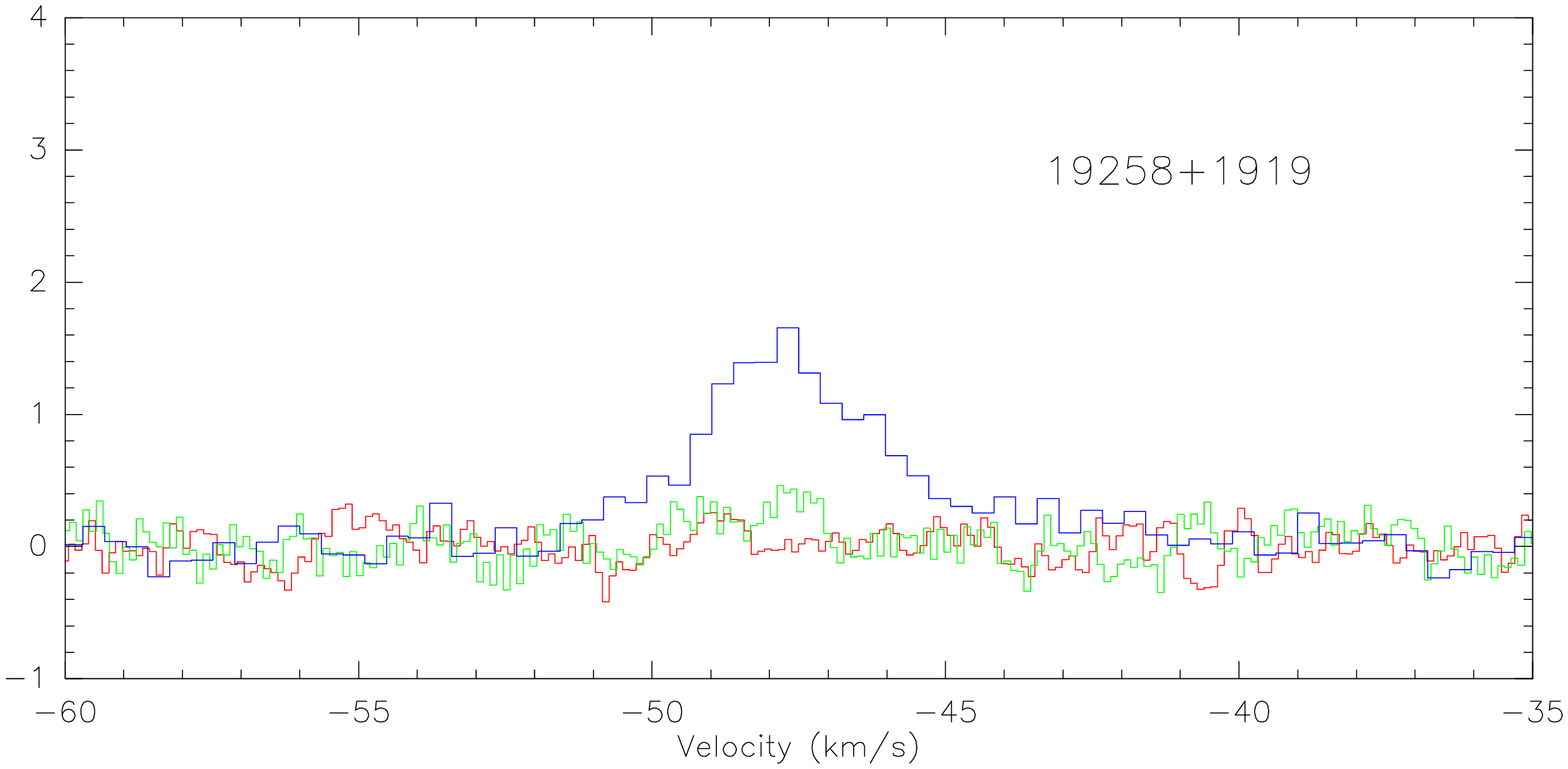}
\includegraphics[height=70mm,  angle=-90, clip, viewport=150 10 500 750]{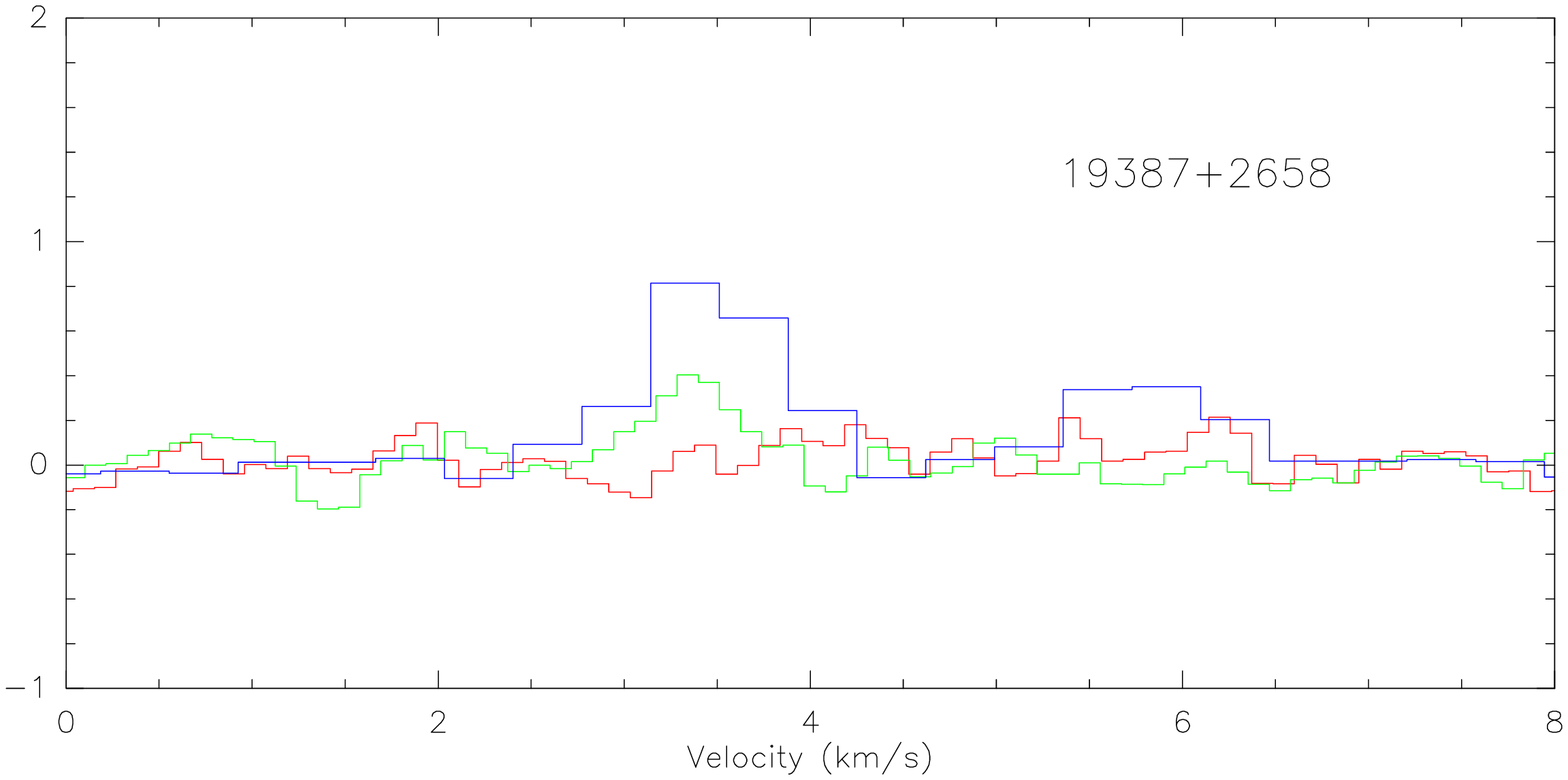}
\includegraphics[height=70mm,  angle=-90, clip, viewport=150 10 500 750]{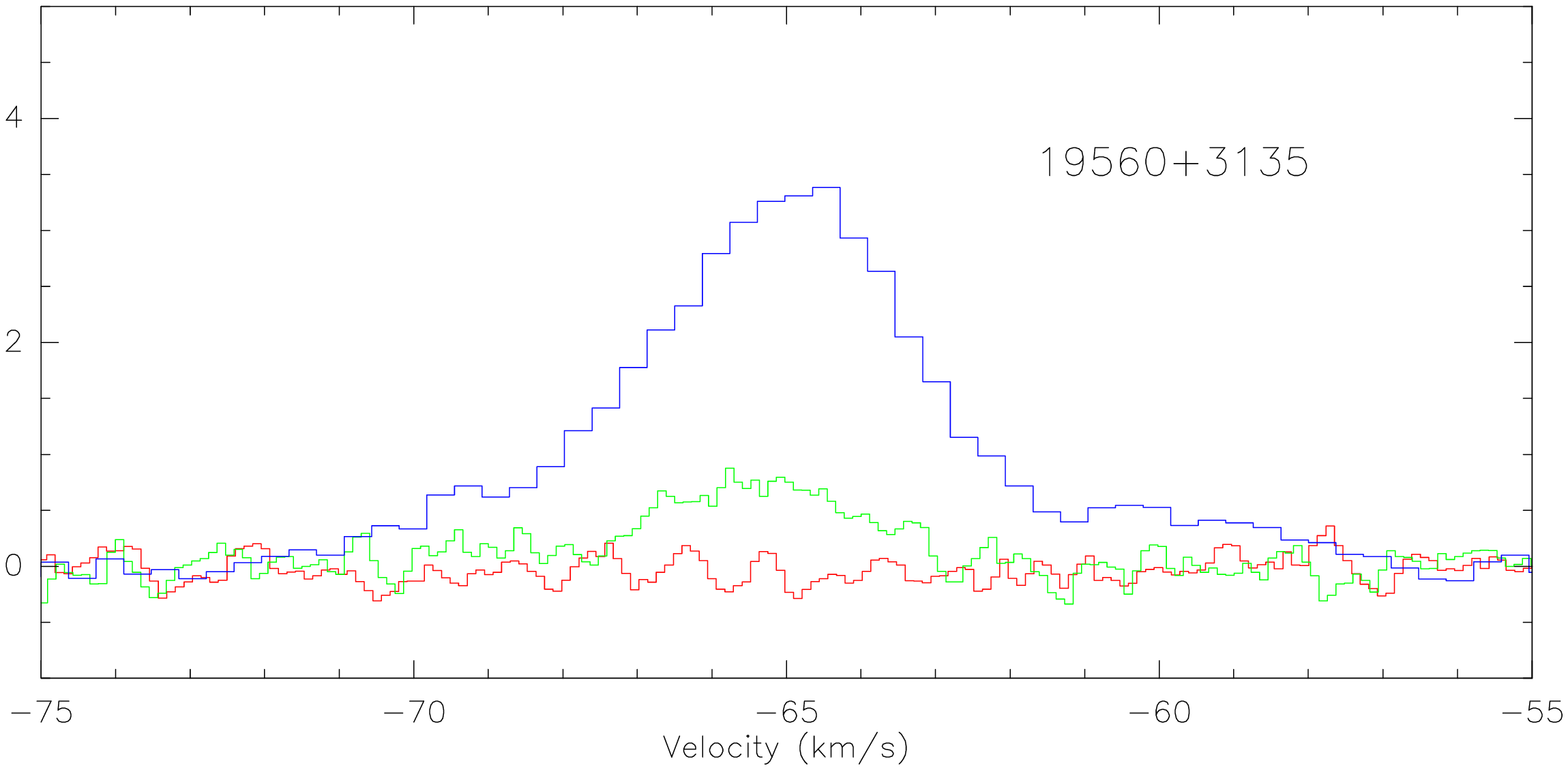}
\includegraphics[height=70mm,  angle=-90, clip, viewport=150 10 500 750]{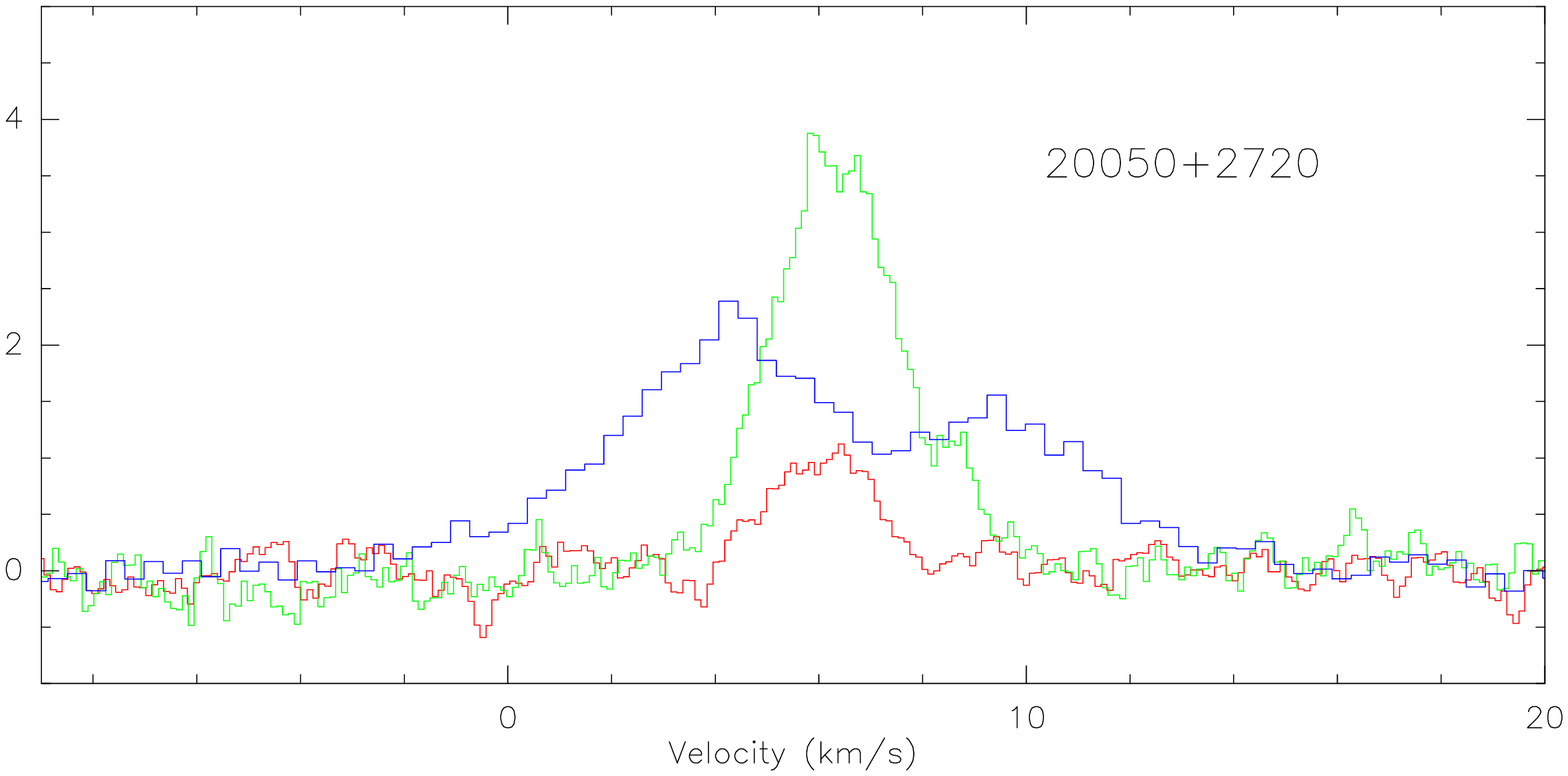}
\includegraphics[height=70mm,  angle=-90, clip, viewport=150 10 500 750]{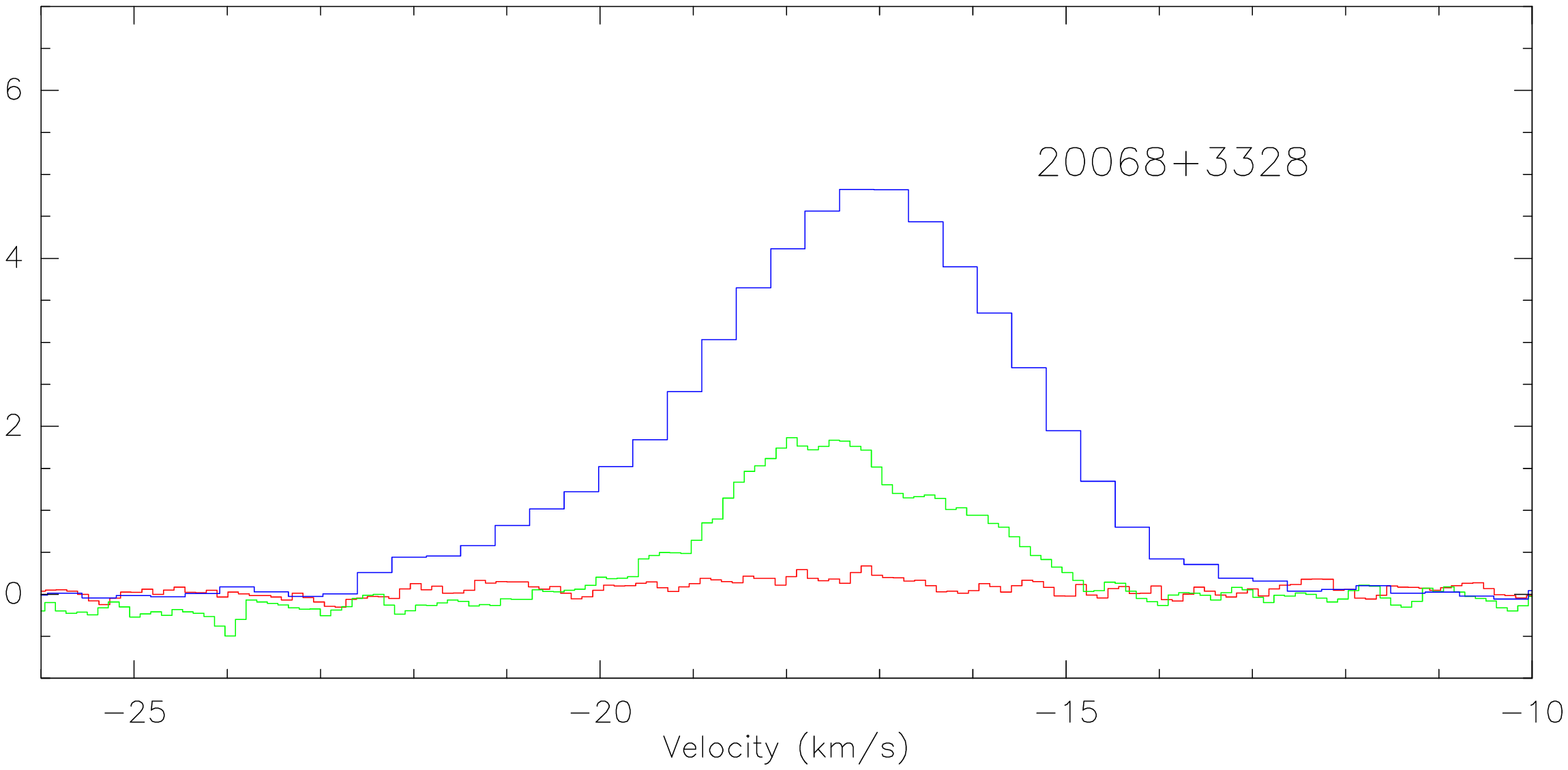}
\includegraphics[height=70mm,  angle=-90, clip, viewport=150 10 500 750]{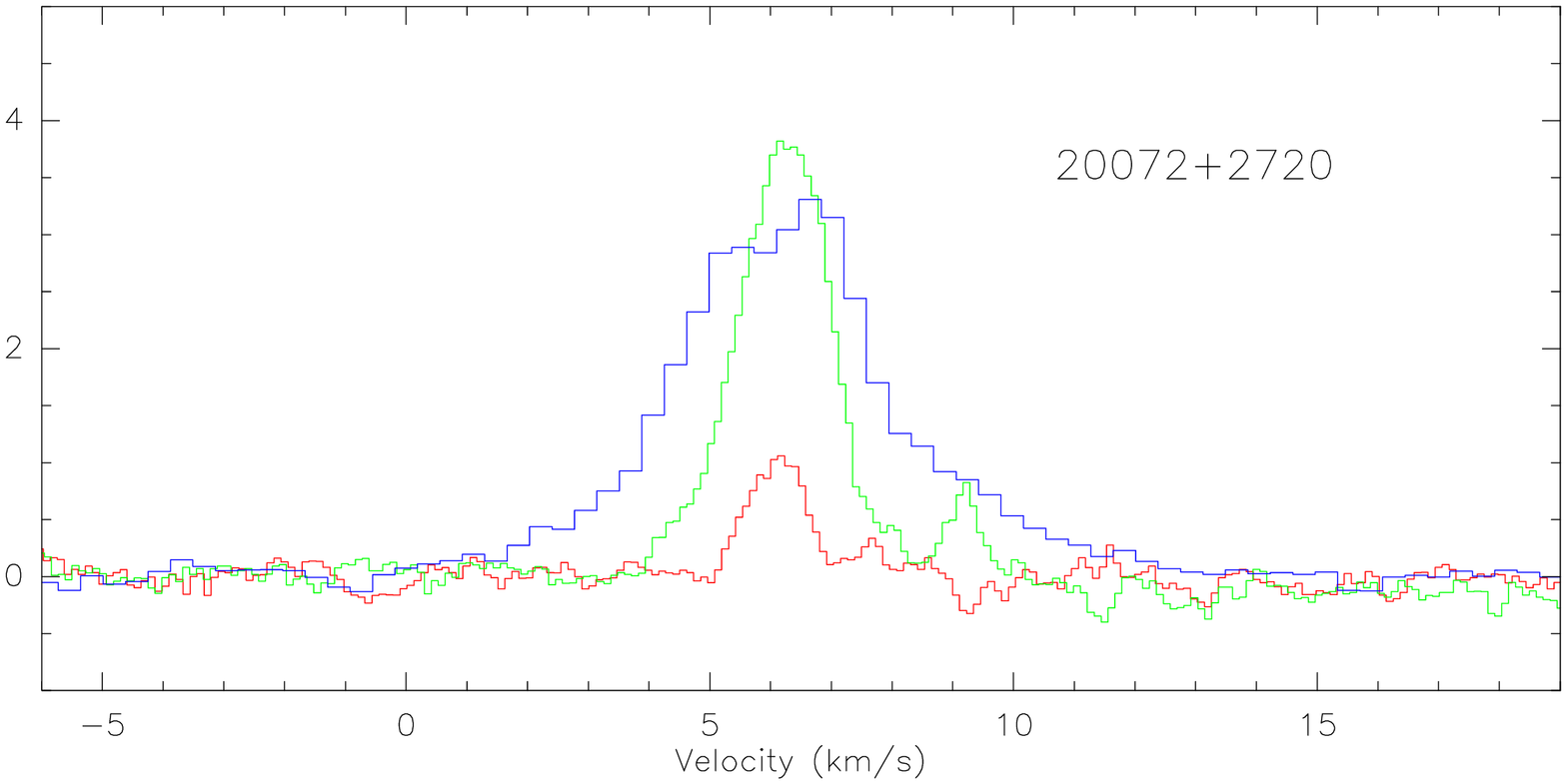}
\includegraphics[height=70mm,  angle=-90, clip, viewport=150 10 500 750]{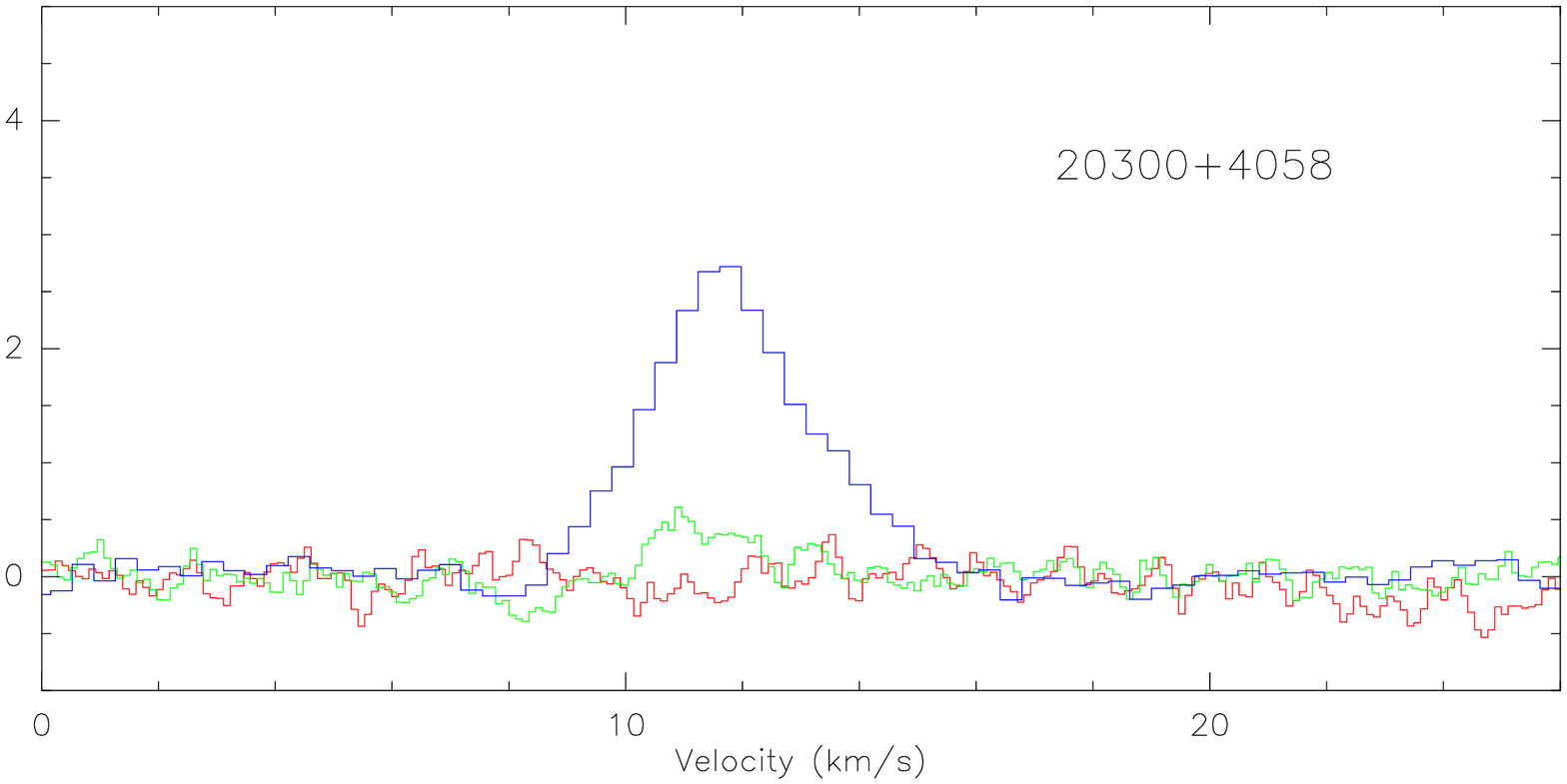}
\includegraphics[height=70mm,  angle=-90, clip, viewport=150 10 500 750]{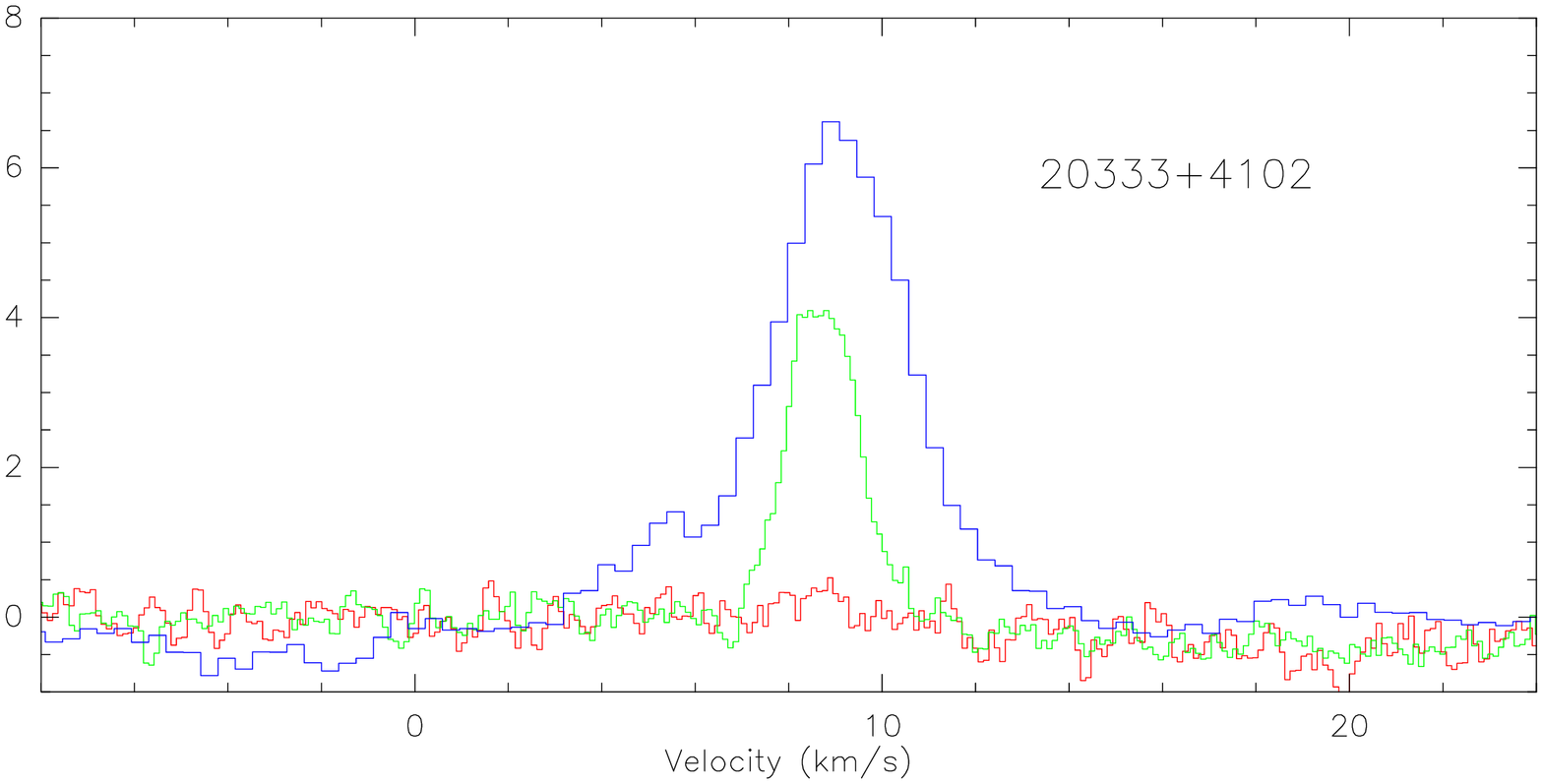}
\includegraphics[height=70mm,  angle=-90, clip, viewport=150 10 500 750]{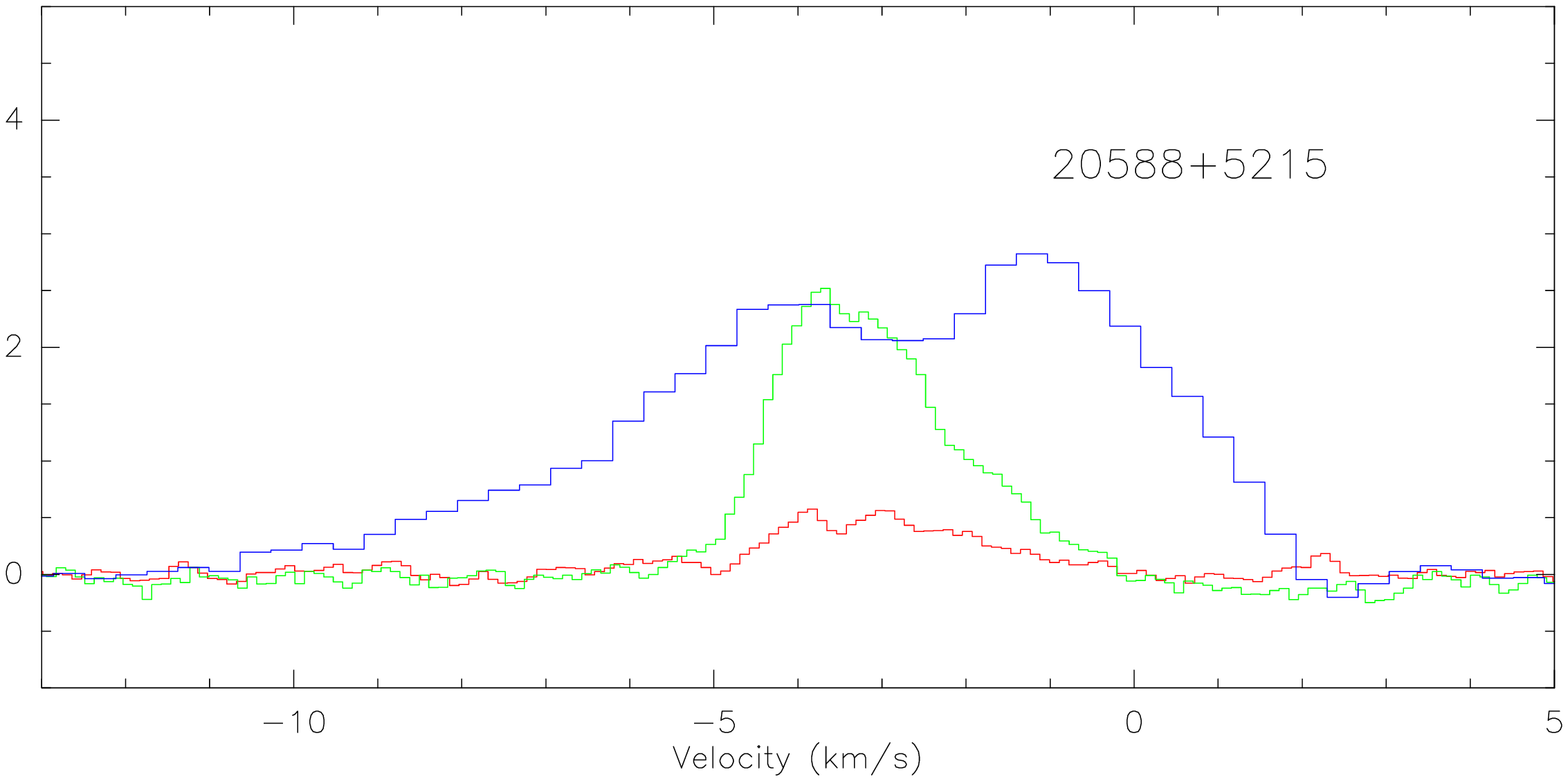}

\begin{minipage}[]{60mm}
   \caption{The sources of type 4
  }\end{minipage}
   \label{Fig9}
   \end{figure}

\addtocounter{figure}{-1}
\begin{figure}

\includegraphics[height=70mm,  angle=-90, clip, viewport=150 10 500 750]{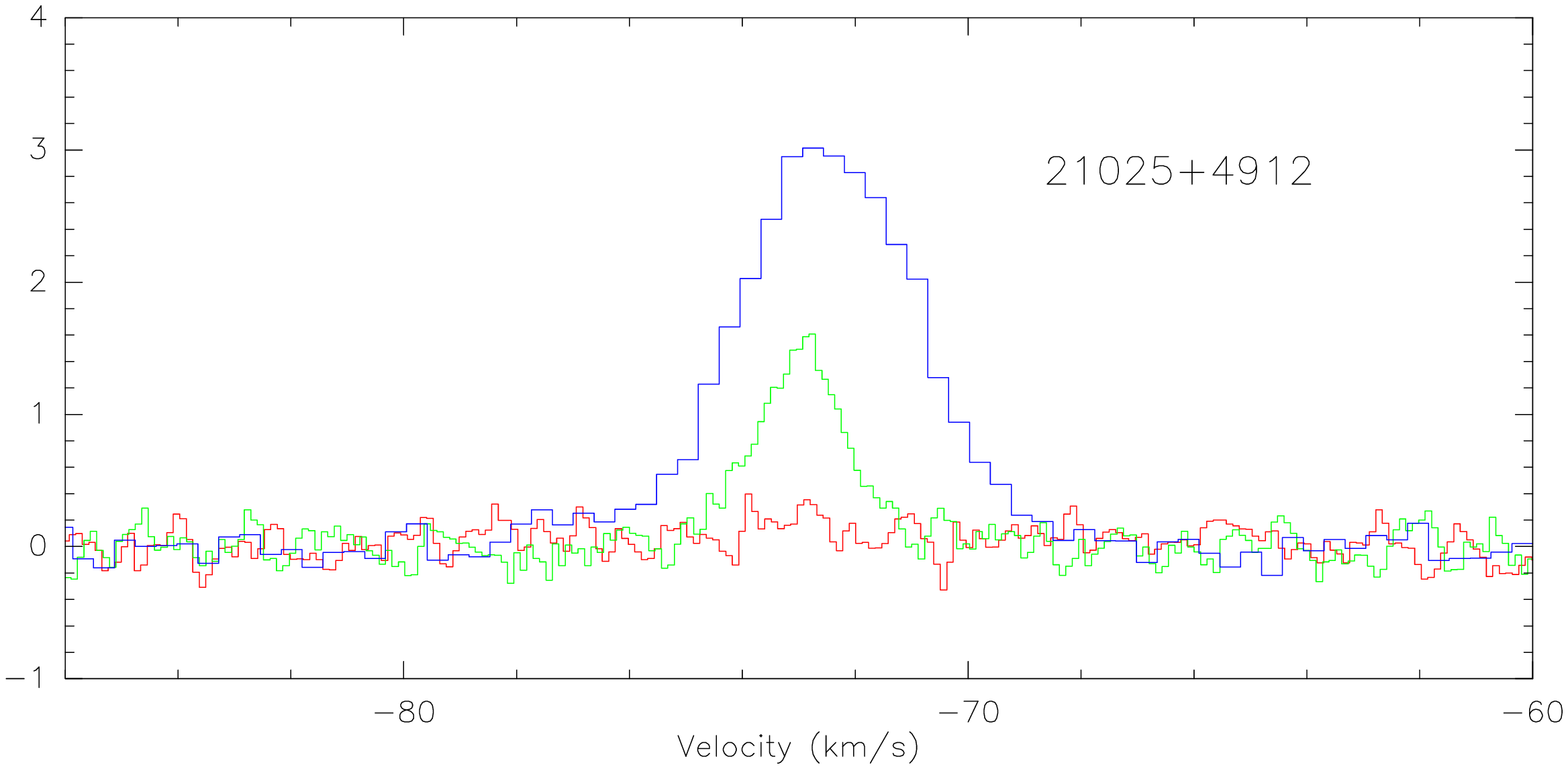}
\includegraphics[height=70mm,  angle=-90, clip, viewport=150 10 500 750]{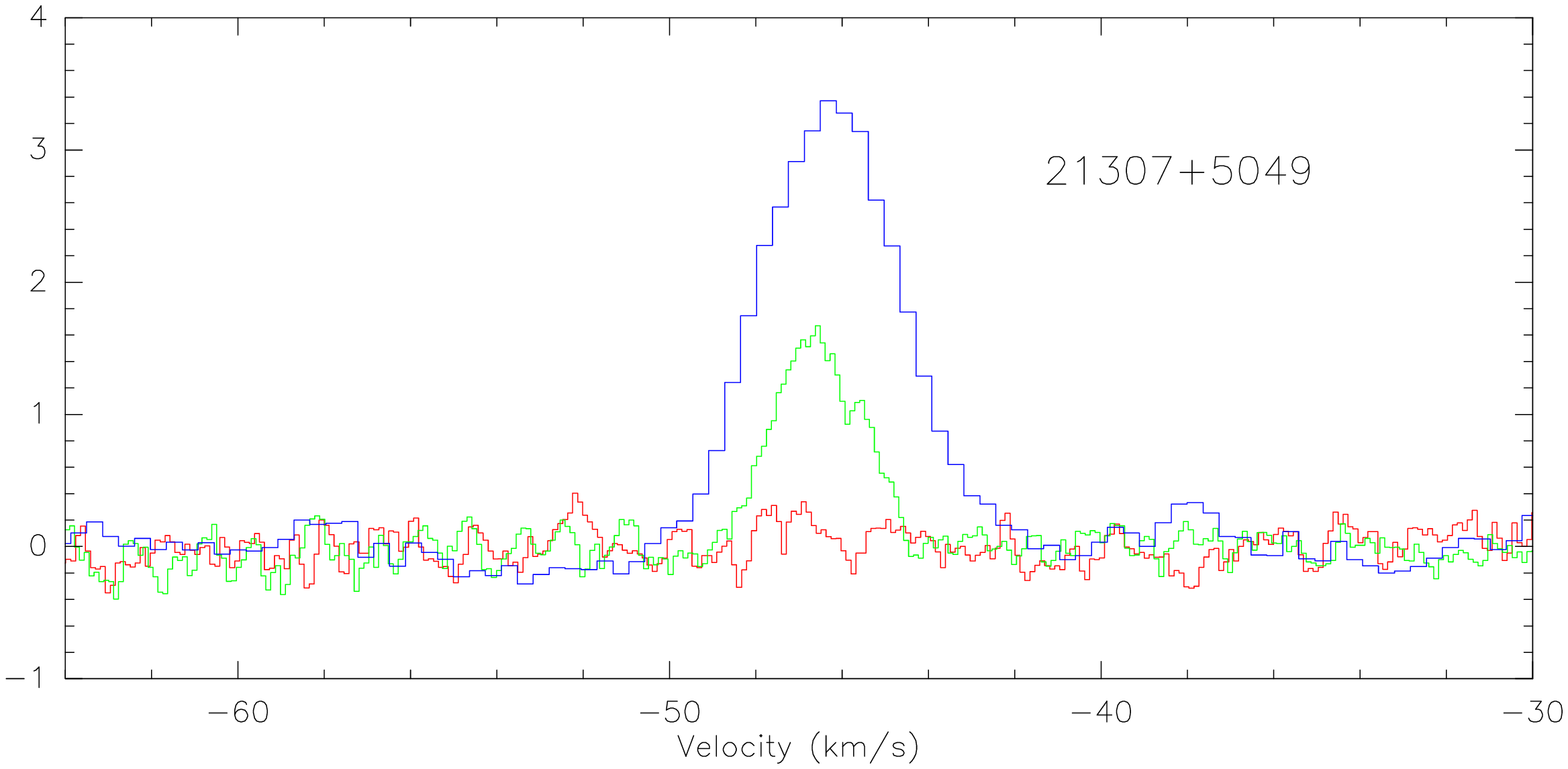}
\includegraphics[height=70mm,  angle=-90, clip, viewport=150 10 500 750]{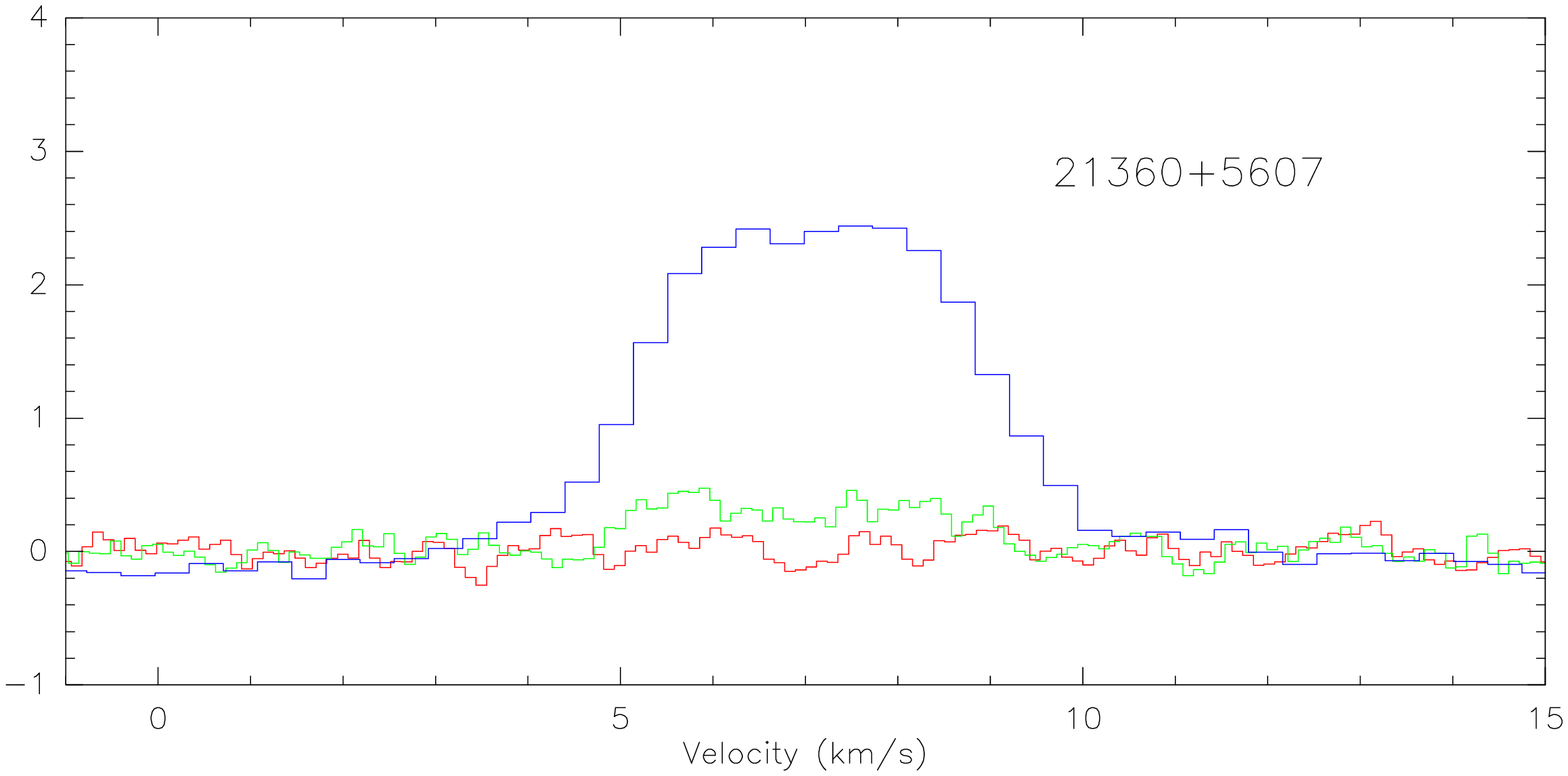}
\includegraphics[height=70mm,  angle=-90, clip, viewport=150 10 500 750]{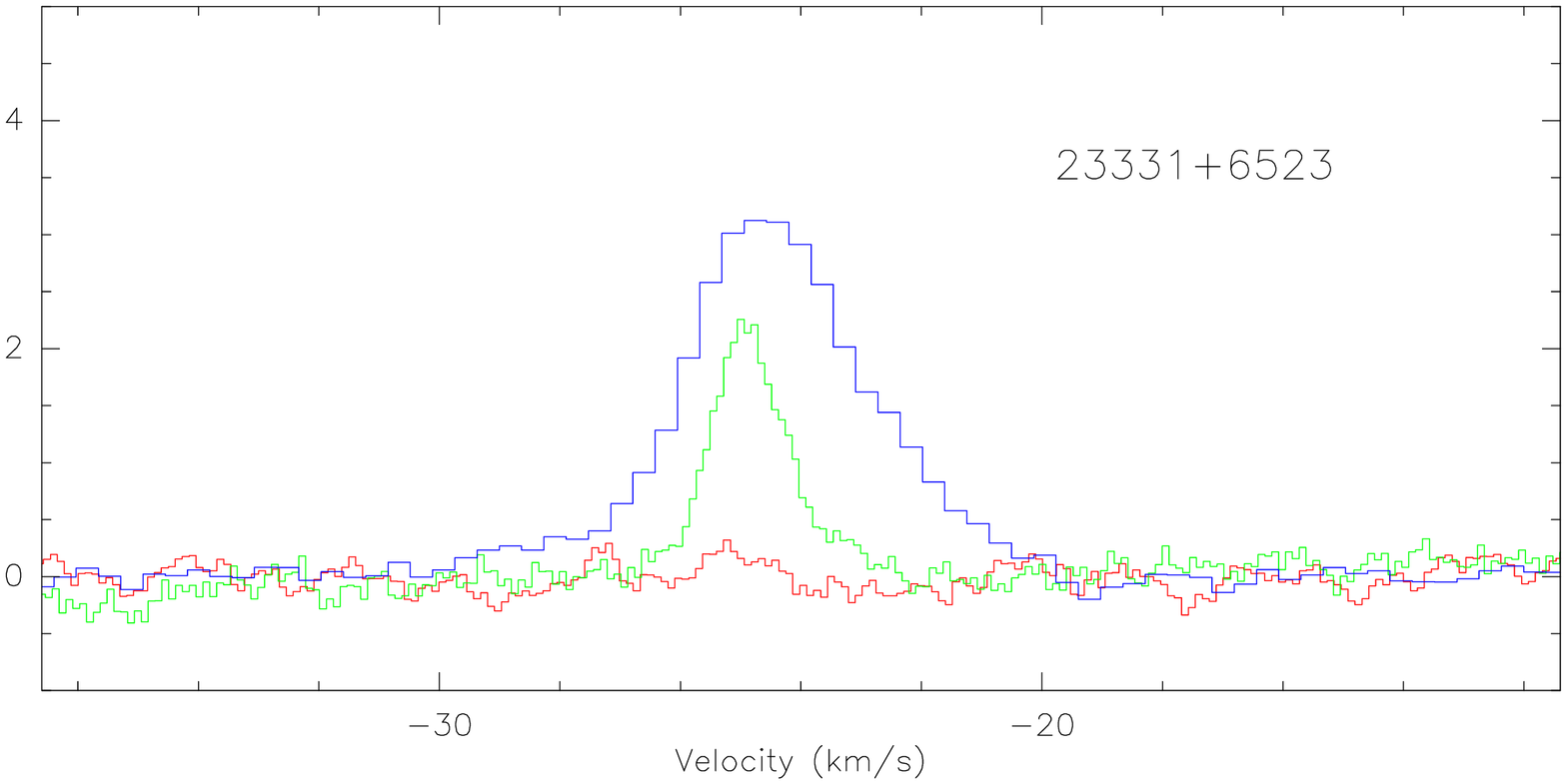}
\includegraphics[height=70mm,  angle=-90, clip, viewport=150 10 500 750]{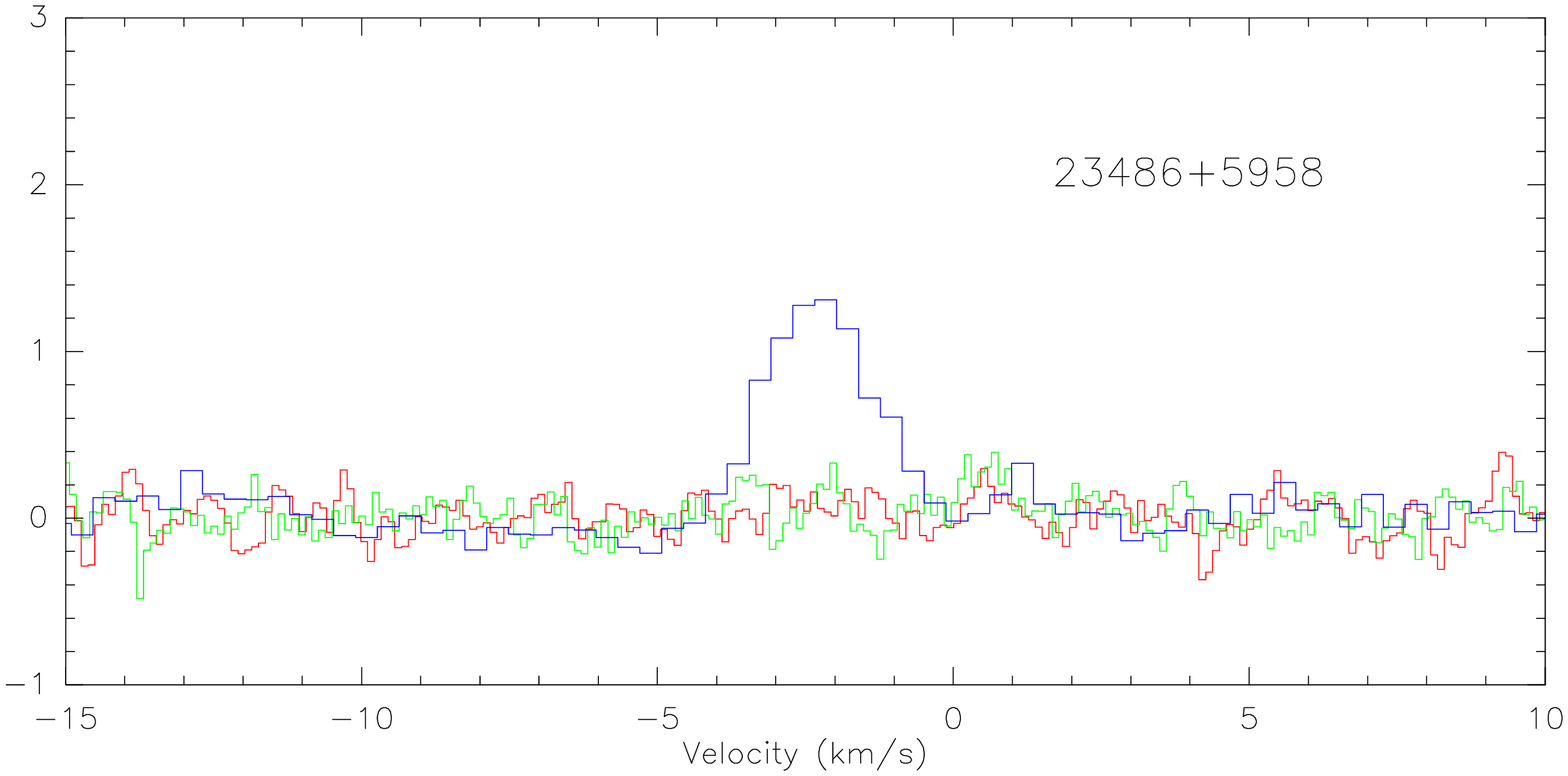}
\includegraphics[height=70mm,  angle=-90, clip, viewport=150 10 500 750]{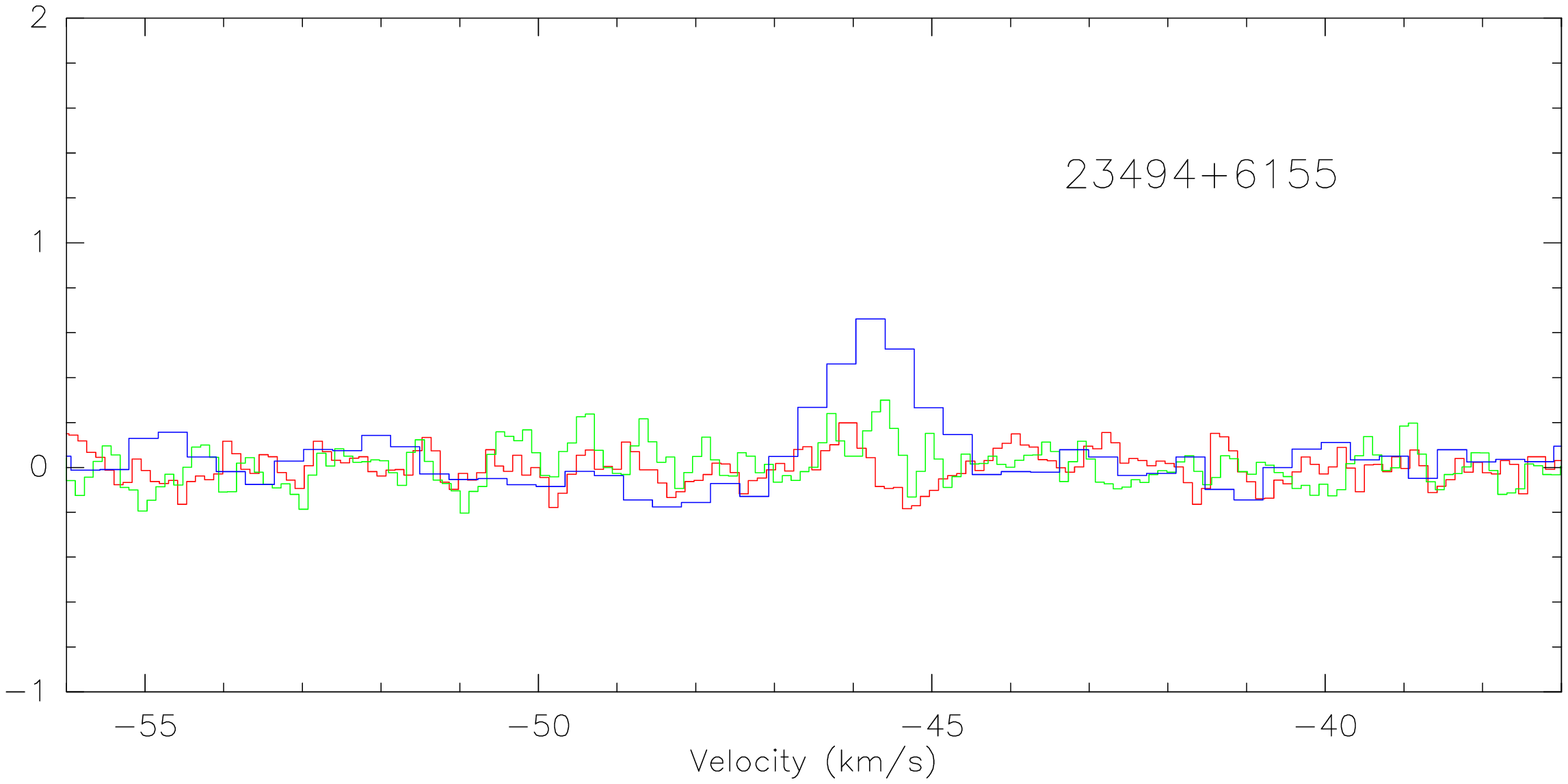}

   \centering
\begin{minipage}[]{60mm}
   \caption{The sources of type 4
  }\end{minipage}
   \label{Fig9}
   \end{figure}